\documentclass[
		twoside,openright,titlepage,numbers=noenddot,headinclude,
	 	footinclude=true,cleardoublepage=empty,
		dottedtoc, 
		BCOR=5mm,paper=a4,fontsize=11pt, 
		ngerman,american, 
		]{scrreprt} 
                
\PassOptionsToPackage{utf8}{inputenc} 
\usepackage{inputenc}

\PassOptionsToPackage{eulerchapternumbers,listings, pdfspacing, subfig,beramono,eulermath,parts}{classicthesis}

\newcommand{\myTitle}{Stochastic Processes in Mesoscale Physics \\ and the Early Universe\xspace}
\newcommand{\mySubtitle}{A work involving the Renormalisation Group and Primordial Black Holes\xspace}
\newcommand{\myDegree}{Doctor of Philosophy\xspace}
\newcommand{\myName}{Ashley Luke Wareham Wilkins\xspace}
\newcommand{\myProf}{Gerasimos Rigopoulos\xspace}
\newcommand{\myOtherProf}{Enrico Masoero\xspace}

\newcommand{\myFaculty}{Faculty of Science, Agriculture \& Engineering\xspace}
\newcommand{\myDepartment}{School of Mathematics, Statistics \& Physics\xspace}
\newcommand{\myUni}{Newcastle University\xspace}
\newcommand{\myLocation}{Newcastle upon Tyne \\ United Kingdom\xspace}
\newcommand{\myTime}{February 2023\xspace}
\newcommand{\myVersion}{version 1.0\xspace}

\newcommand{\lb}{\left(} 
\newcommand{\rb}{\right)}
\newcommand{\lsb}{\left[} 
\newcommand{\rsb}{\right]} 
\newcommand{\lcb}{\left\{} 
\newcommand{\rcb}{\right\}} 
\newcommand{\lan}{\left\langle} 
\newcommand{\ran}{\right\rangle}

\newcounter{dummy} 
\providecommand{\mLyX}{L\kern-.1667em\lower.25em\hbox{Y}\kern-.125emX\@}

\usepackage{lipsum} 

\usepackage{babel}

\usepackage{empheq}

\usepackage{csquotes}
\PassOptionsToPackage{%
backend=bibtex8,bibencoding=ascii,%
language=auto,%
style=numeric-comp,%
sorting = none, 
maxbibnames=10, 
backref=true,%
natbib=true 
}{biblatex}
\usepackage{biblatex}
 
\PassOptionsToPackage{fleqn}{amsmath} 
 \usepackage{amsmath}

 \usepackage{physics}
 \usepackage{float}
 \usepackage{rotating}
 
\PassOptionsToPackage{T1}{fontenc} 
\usepackage{fontenc}

\usepackage{textcomp} 

\usepackage{scrhack} 

\usepackage{xspace} 

\usepackage{mparhack} 

\usepackage{fixltx2e} 

\PassOptionsToPackage{printonlyused, smaller}{acronym} 
\usepackage{acronym} 

\PassOptionsToPackage{pdftex}{graphicx}
\usepackage{graphicx} 

\usepackage{tabularx} 
\usepackage{caption}
\usepackage{subfig}  

\usepackage{listings} 
\PassOptionsToPackage{pdftex,hyperfootnotes=false,pdfpagelabels}{hyperref}
\usepackage{hyperref}  
\hypersetup{
colorlinks=true, linktocpage=true, pdfstartpage=3, pdfstartview=FitV,
breaklinks=true, pdfpagemode=UseNone, pageanchor=true, pdfpagemode=UseOutlines,%
plainpages=false, bookmarksnumbered, bookmarksopen=true, bookmarksopenlevel=1,%
hypertexnames=true, pdfhighlight=/O,
urlcolor=webbrown, linkcolor=RoyalBlue, citecolor=webgreen, 
pdftitle={\myTitle},
pdfauthor={\textcopyright \myName, \myUni, \myFaculty},
pdfsubject={},
pdfkeywords={},
pdfcreator={pdfLaTeX},
pdfproducer={LaTeX with hyperref and classicthesis}
}

\makeatletter
\@ifpackageloaded{babel}
{
\addto\extrasamerican{

}
\addto\extrasngerman{

}
}{\relax}
\makeatother

\usepackage{classicthesis} 

\begin{document}

\frenchspacing 

\raggedbottom 

\selectlanguage{american} 


\pagenumbering{roman} 

\pagestyle{plain} 


\pdfbookmark[1]{Title Page}{Title Page} 

\begin{titlepage}

\begin{addmargin}[-1cm]{-3cm}
\begin{center}
\Large

\hfill
\vfill

\begingroup
\color{Maroon}\spacedallcaps{\myTitle} \\ \bigskip 
\endgroup

\spacedlowsmallcaps{\myName} 

\vfill
\large
Thesis submitted for the degree of\\
\myDegree \\
\bigskip
\includegraphics[width=6cm]{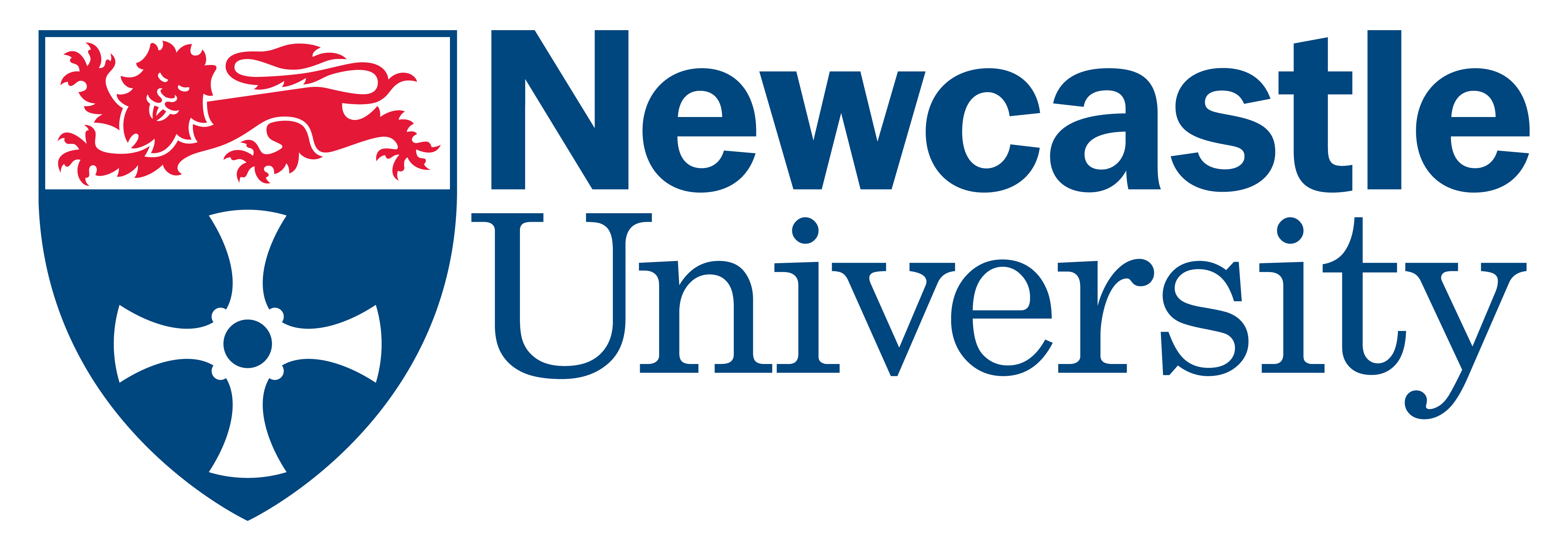} \\ \medskip 


\textit{\small \myDepartment \\
\myFaculty \\
\myUni \\ 
\myLocation \\ }\bigskip

\myTime 

\end{center}
\end{addmargin}

\end{titlepage} 


\thispagestyle{empty}

\hfill

\vfill

\noindent\myName: \\
\textit{\myTitle,}\\
\mySubtitle, 


\bigskip
\bigskip

\noindent\spacedlowsmallcaps{Supervisors}: \\
\myProf \\
\myOtherProf \\ 

\medskip 

\noindent\spacedlowsmallcaps{Location}: \\
\myLocation


\noindent\spacedlowsmallcaps{Time Frame}: \\
September 2018 -- \myTime

\cleardoublepage

\thispagestyle{empty}
\refstepcounter{dummy}

\pdfbookmark[1]{Dedication}{Dedication} 

\vspace*{3cm}

\begin{center}\slshape
All his life he tried to be a good person. \\
Many times, however, he failed. \\
For after all, he was only human. \\
He wasn't a dog. \\ \medskip
--- Charles M. Schulz
\end{center}

\medskip

\begin{center}
Dedicated to the best dog there ever was. \\ \smallskip
Lunaka May \\ \smallskip
2008\,--\,2021

\vfill

\includegraphics[width=6cm]{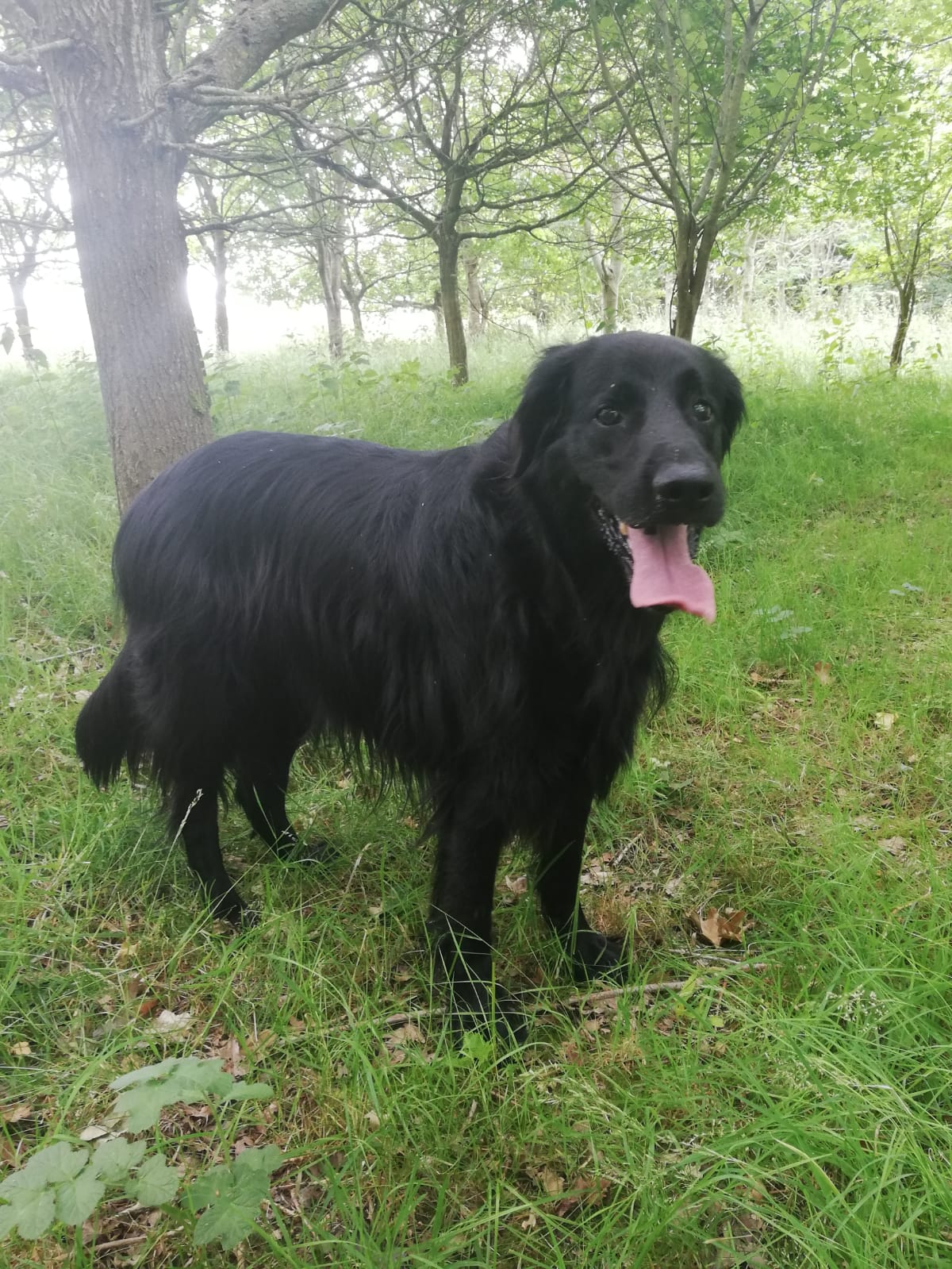}

\vfill
\end{center} 

\cleardoublepage\pdfbookmark[1]{Foreword}{Foreword} 

\begingroup
\let\clearpage\relax
\let\cleardoublepage\relax
\let\cleardoublepage\relax

\chapter*{Foreword}
This thesis represents work I have done over four years of my life and is therefore rather long. There are (I hope!) a lot of nice results and helpful pedagogical explanations in this thesis which the reader would find interesting. To that end I have done my utmost to make this thesis as readable as possible. \\

\noindent As many do, you are encouraged to start, at the end, with the summary chapter to help motivate the work in each of the chapters. There are many topics covered so there is not a singular background chapter but is instead present to greater and lesser extents within every chapter. The knowledgeable or busy reader is therefore directed to read the introduction of each chapter first, where I have included bullet points outlining the main results which are appropriately linked to. Important equations are boxed to aid the skimming of technical derivations. The reader is also encouraged to browse directly through the list of figures and tables to find ones that sound interesting and therefore jump right into the action. When one reads a document in such a non-linear order the meaning of common abbreviations can often become opaque. So all abbreviations used link to their definitions which can be found just after the list of tables. References in the bibliography also link to where they are cited in the text making it much easier to flip back and forth between them.\\

\noindent Hopefully you find this thesis much easier to read than I did to write!

\endgroup			

\vfill 

\cleardoublepage

\pdfbookmark[1]{Abstract}{Abstract} 

\begingroup
\let\clearpage\relax
\let\cleardoublepage\relax
\let\cleardoublepage\relax

\chapter*{Abstract}
This thesis is dedicated to the study of stochastic processes; non-deterministic physical phenomena that can be well described by classical physics. The stochastic processes we are interested in are akin to Brownian Motion and can be described by an overdamped Langevin equation comprised of a deterministic drift term and a random noise term. Because of the random noise term, one must solve the Langevin equation many times to make physical predictions for what would happen \emph{on average}. As Langevin equations have such wide applications this thesis is split into two parts where we examine them in two very different, yet connected, contexts.\\
In Part I we examine stochastic processes in the Mesoscale. For us this means that the Langevin equation is driven by thermal noise, with amplitude proportional to the temperature $T$ and there exists a genuine equilibrium thermal state. We outline how the Langevin equation can be reformulated using techniques from Quantum Field Theory as a path integral that is closely related to SuperSymmetric Quantum Mechanics. We apply a technique known as the Functional Renormalisation Group (FRG) which allows us to coarse-grain in temporal scales. We describe how the FRG can be used to compute correlation functions as the system relaxes towards equilibrium. In particular we describe how to obtain \emph{effective equations of motion} for the average position, variance and covariance of a particle evolving in highly non-trivial potentials and verify their accuracy by comparison to direct numerical simulations. In this way we outline a novel procedure for describing the behaviour of stochastic processes without having to resort to time consuming numerical simulations. \\
In Part II we turn to the Early Universe and in particular examine stochastic processes occurring during a period of accelerated expansion known as inflation. This inflationary period is driven by a scalar field called the inflaton which also obeys a Langevin equation. In this context however the noise term does not come from thermal fluctuations but from inherently quantum fluctuations that are stretched to cosmological scales by the expansion of the universe. These quantum fluctuations then act as the seed for the formation of all large scale structure. We review how to compute inflationary perturbations and focus on a formalism known as Stochastic Inflation. We outline how the Hamilton-Jacobi formulation of Stochastic Inflation allows one to move beyond the simplest inflationary approximation, slow-roll, and discuss an interesting period known as ultra-slow roll. A period of ultra-slow roll is generally needed to form very large density perturbations and we outline how to compute the full probability distribution function of curvature perturbations for a plateau in the inflationary potential using heat kernel techniques. We use this to study the formation of Primordial Black Holes. These extreme objects are formed in the early universe, before the first galaxies, and are a possible Dark Matter candidate. \\
We finish this thesis by applying the techniques developed in Part I to a spectator field during inflation. We confirm that the FRG techniques can compute cosmologically relevant observables such as the power spectrum and spectral tilt. We also extend the FRG formalism so that it can be used to solve first-passage time problems and verify that it gives the correct prediction for the average time taken for a field (or particle) to overcome a barrier in the potential. 

\endgroup			

\vfill 

\cleardoublepage

\refstepcounter{dummy}
\pdfbookmark[0]{Declaration}{declaration} 

\chapter*{Declaration} 

\thispagestyle{empty}

Whilst registered as a candidate for the above degree, I have not been registered for any other research award. The results and conclusions embodied in this thesis are the work of the named candidate and have not been submitted for any other academic award. \\

Part \ref{part:Meso} is based on two preprints \cite{Wilkins2020b,Wilkins2021} which have since been combined into one paper \cite{PhysRevE.106.054109} published in Physical Review E. Part \ref{part:Meso} is supplemented by lots of background information which was written by myself and drawn from many references, cited where appropriate. Chapters \ref{cha:Inflation} \& \ref{cha:PBH} are based on a publication in JCAP \cite{Rigopoulos2021} but again with lots more background information drawn from the literature and cited appropriately. The new work in chapters \ref{cha:Inflation} \& \ref{cha:PBH} begins in sections \ref{sec:stochastic_deltaN} \& \ref{sec: plateau} respectively. Chapter \ref{cha:RG_spect} is new research, done by myself, that has been published in JCAP \cite{Rigopoulos:2022gso} since the original submission of this thesis. \\

In the publications that entered this thesis all computations were done by myself, predominantly originally but a few as checks for my co-authors. I also contributed to and formulated many of the ideas developed. 

\bigskip
 
\noindent\textit{\myLocation, \myTime}

\smallskip

\begin{flushright}
\begin{tabular}{m{6cm}}
\\ \hline
\centering\myName \\
\end{tabular}
\end{flushright}


\cleardoublepage

\pdfbookmark[1]{Acknowledgements}{Acknowledgements} 

\begin{flushright}{\slshape    
It takes a village to raise a child.} \\ \medskip
--- African Proverb
\end{flushright}

\bigskip


\begingroup

\let\clearpage\relax
\let\cleardoublepage\relax
\let\cleardoublepage\relax

\chapter*{Acknowledgements}

And it took many people to make this thesis happen. Firstly I would like to thank my supervisors Gerasimos and Enrico for their guidance and support throughout a difficult journey. I am forever grateful for their reminders of why I loved research in the first place in difficult times. I would also like to thank all the members of the cosmology group at Newcastle for their kindness, insight and good fun. Many thanks also to my examiners Ian and Konstantinos for making my defence enjoyable and for their very fair critiques which have improved this thesis.\\

\noindent I also want to thank all my PhD colleagues, I have felt so privileged to fight for your rights and conditions. I'll never forget your support in giving me an award out of your own pocket for the work I was doing. I don't think I'll ever be able to express how much that meant to me. In particular I want to single out the (honorary) members of PhD i: Cristiana, Devika, Holly, Kate, Keaghan, Ryan, Sam, Sam and Stephen who have become dear friends and made going into Newcastle so much fun. Many thanks also to Archie for offering his thoughts on one of the drafts as well as being excellent fun to be with at conferences. I also would like to thank all those who I played water polo with in Nottingham, York and Newcastle. Doing a PhD can be very stressful and the catharsis of wrestling in the pool has proved invaluable to me. Thanks must also go to TransPennine Express for only making me late to Newcastle a mere most of the time.\\

\noindent My education did not begin at Newcastle and there are many people without whom I would not have been in a position to start this PhD. I had many great teachers and after a year trying to do it myself I have so much more appreciation for the work they did. At Nottingham I had many great lecturers and many great friends. In particular my success wouldn't have been possible without the support of Jamal and the knowledge of Andy, I am forever thankful I met you both. The person I want to thank most from Nottingham however is my personal tutor Ed Copeland. Your kindness during my degree and most importantly when I was in a very dark place after I had graduated was invaluable to me and I am eternally grateful for your support and faith in my abilities. \\

\noindent Finally I want to thank those who will perhaps understand the least of this thesis and yet are the biggest reasons it exists. I want to thank my parents Nick and Wendy and sister Hayley whose unwavering support and love throughout my life has been invaluable to me. Most importantly I want to thank my partner in all things Lucy who has been a constant source of strength for so many years. It is not an understatement to say that without your love, support and guidance I would not have completed this thesis. If this thesis had a co-author it would undoubtedly be you.

\endgroup 

\pagestyle{scrheadings} 

\cleardoublepage

\refstepcounter{dummy}

\pdfbookmark[1]{\contentsname}{tableofcontents} 

\setcounter{tocdepth}{2} 

\setcounter{secnumdepth}{2} 

\manualmark
\markboth{\spacedlowsmallcaps{\contentsname}}{\spacedlowsmallcaps{\contentsname}}
\tableofcontents 
\automark[section]{chapter}
\renewcommand{\chaptermark}[1]{\markboth{\spacedlowsmallcaps{#1}}{\spacedlowsmallcaps{#1}}}
\renewcommand{\sectionmark}[1]{\markright{\thesection\enspace\spacedlowsmallcaps{#1}}}

\clearpage

\begingroup 
\let\clearpage\relax
\let\cleardoublepage\relax
\let\cleardoublepage\relax


\refstepcounter{dummy}
\pdfbookmark[1]{\listfigurename}{lof} 

\listoffigures

\vspace{8ex}
\newpage


\refstepcounter{dummy}
\pdfbookmark[1]{\listtablename}{lot} 

\listoftables
        
\vspace{8ex}
\newpage
    



       

\refstepcounter{dummy}
\pdfbookmark[1]{Acronyms}{acronyms} 

\markboth{\spacedlowsmallcaps{Acronyms}}{\spacedlowsmallcaps{Acronyms}}

\chapter*{Acronyms}

\begin{acronym}[UML]
\acro{LO}{leading order}
\acro{NLO}{next-to leading order}
\acro{NNLO}{next-to-next-to leading order}
\acro{BM}{Brownian Motion}
\acro{BPI}{Brownian Motion Path Integral}
\acro{F-P}{Fokker-Planck}
\acro{RG}{Renormalisation Group}
\acro{FRG}{Functional Renormalisation Group}
\acro{LPA}{Local Potential Approximation}
\acro{WFR}{Wavefunction Renormalisation}
\acro{QFT}{Quantum Field Theory}
\acro{EFT}{Effective Field Theory}
\acro{SUSY}{SuperSymmetry}
\acro{EA}{Effective Action}
\acro{REA}{Regulated Effective Action}
\acro{EEOM}{Effective Equations of Motion}
\acro{CMB}{Cosmic Microwave Background}
\acro{GR}{General Relativity}
\acro{SR}{Slow-Roll}
\acro{USR}{Ultra Slow Roll}
\acro{H-J}{Hamilton-Jacobi}
\acro{DM}{Dark Matter}
\acro{PBHs}{Primordial Black Holes}
\acro{FPT}{First-Passage Time}

\end{acronym}  
                   
\endgroup 

\cleardoublepage

\pagenumbering{arabic} 

\cleardoublepage 


\ctparttext{\color{black}It is sometimes glibly said that modern physics is concerned about the very large -- astronomy, cosmology -- and the very small -- quantum phenomena -- with everything else left to the other sciences and engineering. In this thesis we reject this supposition and in part one discuss a small portion of the exciting physics to be done in \emph{The Mesoscale}.} 

\part{\label{part:Meso}The Mesoscale} 

\chapter{Stochastic Processes} 
\vspace{0.5cm}
\begin{flushright}{\slshape    
From where we stand the rain seems random. \\If we could stand somewhere else, we would see the order in it.} \\ \medskip
--- Tony Hillerman \cite{Hillerman1990}
\end{flushright}
\vspace{0.5cm}
\label{cha:stochastic_processes} 
\acresetall 
\section{Introduction}

Stochastic processes appear in all kinds of contexts in physics. From the \ac{BM} of small particles in a thermal bath \cite{VanKampen2007, gardiner2009stochastic} to scalar fields experiencing quantum fluctuations in the early inflationary universe \cite{Starobinsky1988}, many problems of interest can be described by the overdamped Langevin equation (\ref{eq:langevin}). However, the fluctuations (thermal or effectively thermal) occur very frequently and if one were to attempt to adequately simulate such a process a suitable small timestep size would have to be chosen. This means we only have an immediate understanding of the physics on small timescales. Understanding long-time behaviour and finding the equilibrium properties of the system from its initial out-of-equilibrium state requires following the stochastic process for times much longer than this fundamental timescale. We will describe these stochastic processes (a subset of the Wiener process) as realised in \ac{BM} so that the results are more physically intuitive.
\\

\ac{BM}, as originally described, is the random motion of large particles in a fluid of much smaller particles. The random motion is caused by collisions with the smaller, fast-moving particles in the fluid. There are a few different ways we can formulate this mathematically. In this chapter we will outline three different but equivalent descriptions:
  \begin{itemize}
  \item \textbf{The Langevin Equation} -- A stochastic differential equation
  \item \textbf{The Brownian Motion Path Integral} -- A path integral formulation that is equivalent to Euclidean SuperSymmetric Quantum Mechanics
  \item \textbf{The Fokker-Planck Equation} -- A non-linear PDE of a probability function
  \end{itemize}
Ultimately our goal will be to obtain a coarse-grained \emph{in time} theory for \ac{BM} such that one need not solve the Langevin equation many times to get a sense of the \emph{average} behaviour of the system. We will also seek to examine how the behaviour of this coarse-grained stochastic theory will change as the temperature of the system is varied. 

For the reader familiar with these topics we direct you to the main results of this chapter which will be used later:
\begin{itemize}
    \item The dimensionless Langevin equation (\ref{eq:langevindimless}) 
    \item The \ac{BPI} given by equation (\ref{eq:BM_path_int}) 
    \item The \ac{EA} is related to the \ac{BPI} by equation (\ref{eq:Gamma_S_relation}) 
    \item The standard and rescaled \ac{F-P} equations (\ref{eq: F-P}) and (\ref{eq:FP1}) respectively
\end{itemize}

 \section{\label{sec:langevin_eqn}The Langevin Equation}
\ac{BM} for a single particle of mass $m$ moving in a potential $V(x)$ coupled to an external heat bath with temperature $T$ can be described by the Langevin equation:
\begin{eqnarray}
m\ddot{x} + \gamma \dot{x} &=& -\partial_{x} V(x) + f(t) \label{eq:langevinfull} \\
\left\langle f(t)f(t') \right\rangle &=& 2D\gamma^2 \delta (t-t')\label{eq:f defn}
\end{eqnarray}
where $\gamma$ is a frictional term due to the surrounding fluid and $f(t)$ is a Gaussian ``noise" term. $D = k_bT/\gamma$ is the diffusion constant with equality given so as to match the Boltzmann equilibrium distribution (should it exist).  We will be concerned however with the overdamped limit which corresponds to $\varepsilon \equiv m/\gamma$ being a short timescale compared to the time scales of interest:
\begin{eqnarray}
\dot{x} &=& -\varepsilon \partial_{x} \bar{V}(x) + \eta(t) \label{eq:langevin} \\
\left\langle \eta(t)\eta(t') \right\rangle &=& 2D \delta (t-t')\label{eq:eta defn}
\end{eqnarray}
Here we have taking out a mass factor from the potential V so that mass appears explicitly in the equation (i.e. $V = m\bar{V}$). For convenience we will drop the overbar from here on in. In principle solving equation (\ref{eq:langevin}) once is not difficult numerically. The issue is that as $\eta$ is sampled randomly from a probability distribution one obtains a different answer every time (\ref{eq:langevin}) is solved. Therefore to get accurate average statistics one is required to solve it many times which can be computationally expensive. We will be examining the impact of changing the temperature, and hence changing the strength of the fluctuating force $\eta$, on the coarse-grained effective theory. Let us therefore introduce a reference temperature $T_0$ and a dimensionless parameter $\Upsilon$ which allows us to dial the temperature around $T_0$.  Writing $D=D_0\Upsilon$, we further define dimensionless variables 
\begin{eqnarray}
x=\sqrt{2D_0\varepsilon} \,\hat{x}\,,\quad t=\varepsilon \, \hat{t},\quad V(x)=\frac{2D_0}{\varepsilon} \, \hat{V}(\hat{x})\,,\quad \eta(t)=\sqrt{\frac{2D_0}{\varepsilon}} \, \hat{\eta}(\hat{t})
\end{eqnarray}
in terms of which the dynamical equation becomes
\begin{subequations}
\begin{empheq}[box = \fcolorbox{Maroon}{white}]{align}
\frac{d\hat{x}}{d\hat{t}} &= -\frac{\partial\hat{V}}{\partial\hat{x}} + \hat{\eta}(\hat{t}) \\
 \langle \hat{\eta}(\hat{t}) \hat{\eta}(\hat{t}')\rangle &= \Upsilon \delta(\hat{t} - \hat{t}') \label{eq:eta__therm_dimless}
 \end{empheq}\label{eq:langevindimless}
\end{subequations}  
From now on we will be dropping the hats for simplicity of notation but  generally refer to dimensionless quantities unless otherwise stated. 

\section{\label{sec:BM-PI}The Brownian Motion Path Integral}    

In order to bring the powerful tools of \ac{QFT} such as the \ac{FRG} to bear in chapter \ref{cha:fRG}, we will need to reformulate the stochastic differential equation (\ref{eq:langevindimless}) in terms of a \textit{path integral}. In this section we will outline one way to obtain this path integral, aiming to link this to Supersymmetric Quantum Mechanics. Our final expression, and the starting point of our subsequent analysis, is the \ac{BM} transition probability (\ref{eq:TransProb}), expressed in terms of an integral over possible histories weighted by the action (\ref{eq: PDF action}), to which the busy reader may progress if uninterested in the details of the derivation. We will be using a condensed functional notation of infinite dimensional functional integrals but all expressions can be considered as limits of large, finite dimensional ordinary integrals. This derivation is based on the path integral reformulation by De Dominicis, Peliti and Janssen \cite{DEDOMINICIS1976, Janssen1976, DeDominicis1978} of the well known Martin-Siggia-Rose approach for stochastic dynamics, first developed in \cite{Martin1973}.  More details on these path integrals, including the corresponding finite discretisation of the stochastic process can be found in \cite{Lau2007}.
\begin{figure}
    \centering
    \includegraphics[width=.75\textwidth]{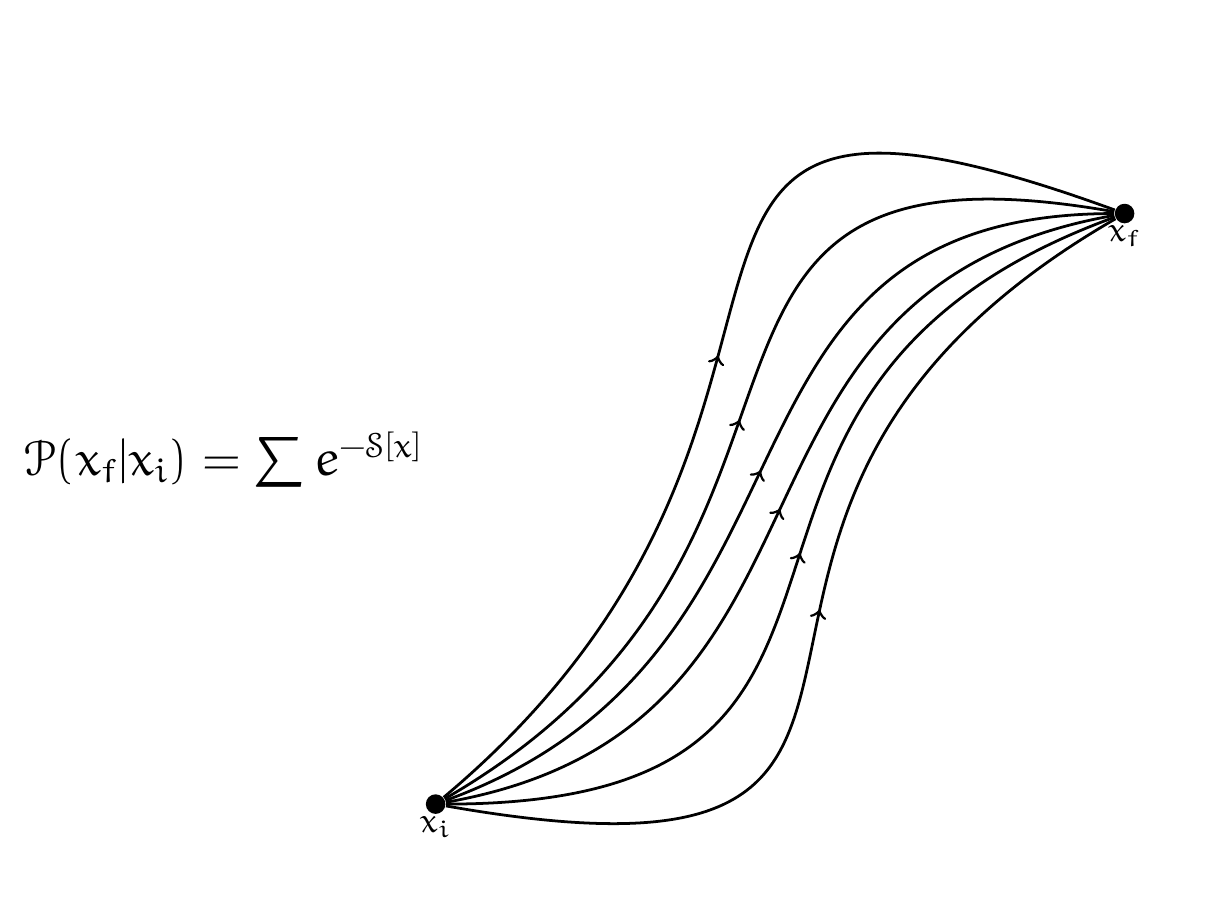}
    \caption[Representing a probability distribution function as a path integral]{A representation of the \textit{Probability Distribution Function} $\mathcal{P}(x_{f}\vert x_{i})$ as a sum over all possible trajectories between two points where each trajectory is weighted by the exponential of the appropriate action.}
    \label{fig:path_integral}
\end{figure}
The dynamics of the (dimensionless) Langevin equation (\ref{eq:langevindimless}) can be captured in terms of the \textit{Probability Distribution Function} (PDF) $\mathcal{P}(x_{f}\vert x_{i})$ of observing the particle at $x_f$ at time $t = t_f$ given that initially at $t = t_i$ the particle was at $x_i$. By definition this can be expressed as:
\begin{eqnarray}
\mathcal{P}(x_{f}\vert x_{i}) = \left\langle \delta\left( x(t_f)-x_f \right) \right\rangle \label{eq:PDF defn}
\end{eqnarray}
where the expectation value is taken over all possible realisations of the noise $\eta(t)$ and $\delta\left( x(t_f)-x_f \right) $ is the Dirac delta function. Put another way, $x(t_f)$ is the position at $t_f$ for a given noise history $\eta(t)$ and the brackets indicate averaging over all possible noise histories, or stochastic paths, which start at $x_i$ and end up at $x(t_f)=x_f$ at $t_f$. We demonstrate this idea in Fig.~\ref{fig:path_integral} by sketching out some of the possible trajectories that are summed over between $x_i$ and $x_f$. This sort of object is precisely a path integral so we can rewrite the PDF using a Gaussian measure for noise (\ref{eq:eta__therm_dimless}) and express the average as
\begin{eqnarray}
\mathcal{P}(x_{f}\vert x_{i}) = \int\mathcal{D}\eta(t)\delta\left( x(t_f)-x_f \right)\text{exp}\left[-\int \text{d}t \, \dfrac{\eta^2(t)}{2\Upsilon}\right]\nonumber \\
\label{eq: PDF first PI}
\end{eqnarray}
where each noise history is now appropriately weighted by the exponential factor in the above expression. In essence the $\lan \ran$ in (\ref{eq:PDF defn}) are shorthand for this integration over histories weighted by the exponential factor. In principle we are done: we have expressed the behaviour of this stochastic system as a path integral. Unfortunately it is not a very useful one! We will be interested primarily in correlation functions that depend on the position $x$ so it would be better if instead of our measure being $\mathcal{D}\eta(t)$ it was $\mathcal{D}x(t)$. To do this we will utilise some \ac{QFT} techniques. We now consider the identity (see e.g. \cite{zinn2002quantum}):
\begin{eqnarray}
1 &=& \int dx_f\int_{x_i}^{x_f}\mathcal{D}x(t)\,\delta\left( x(t)-x_\eta(t)\right)\\
& = & \int dx_f\int_{x_i}^{x_f}\mathcal{D}x(t)\,\delta\left( \dot{x} + V_{,x} -\eta (t)\right) \text{det}\textbf{M} \\
&=& \int_{x_i}\mathcal{D}x(t)\,\delta\left( \dot{x} + V_{,x} -\eta (t)\right) \text{det}\textbf{M} \label{eq: useful ident}
\end{eqnarray}
where the matrix $\textbf{M}(t,t')$ is:
\begin{eqnarray}
\textbf{M} &\equiv &\dfrac{\delta \eta(t)}{\delta x(t')} = \left(\dfrac{d}{dt} +  V_{,xx}\right) \delta(t-t')\label{eq: M defn}\,.
\end{eqnarray}
and $x_{\eta}(t)$ is the $x(t)$ resulting from a particular noise history $\eta (t)$. The identity (\ref{eq: useful ident}) expresses the obvious fact that, if the particle starts at some $x_i$ and follows a particular history $x_\eta(t)$ dictated by the Langevin equation without disappearing, it will end up somewhere after time $t_f$. We have used the standard subscript notation to denote derivative with respect to that variable e.g. $V_{,xx} = \partial_{xx}V$.
Note that the path integral in (\ref{eq: useful ident}) is over all paths starting at $x_i$ at $t_i$ and ending at any $x$ at $t_f$. Inserting our `fat unity' factor (\ref{eq: useful ident}) into (\ref{eq: PDF first PI}) and noting that the delta function there restricts $x(t_f)$ to be $x_f$ we obtain:
\begin{eqnarray}
\mathcal{P}(x_{f}\vert x_{i}) = \int\limits_{x(t_i)=x_i}^{x(t_f)=x_f}\mathcal{D}\eta\mathcal{D}x\,\delta\left( \dot{x} + V_{,x} -\eta \right) \text{ det}\textbf{M}\text{ exp}\left[-\int \text{d}t \,\dfrac{\eta^2(t)}{2\Upsilon}\right] \label{eq: PDF 2nd PI}
\end{eqnarray}
where the $\mathcal{D}x(t)$ integral is taken over all paths beginning at $x_i$ and ending at $x_f$. We now have a path integral over position which is good but in the process have had to introduce a new delta function as well as a determinant of a matrix. The delta function is simply restricting all possible paths in our path integral to satisfy the Langevin equation (\ref{eq:langevindimless}). We can rewrite the delta function as a functional Fourier transform using a new variable $\tilde{x}$ which is usually called the \textit{response field}:
\begin{eqnarray}
\delta\left( \dot{x} + V_{,x} -\eta \right) = \int \mathcal{D}\tilde{x}\text{ exp}\left[ i\int \text{d}t~\tilde{x}\left( \dot{x} + V_{,x} -\eta \right)\right]
\label{eq: tilde x PI}
\end{eqnarray}
Dealing with the determinant is a bit more tricky. There are a couple of standard ways we can incorporate $\mathrm{det} \textbf{M}$ into an exponential. Formally writing  
\begin{eqnarray}
\textbf{M}=\left(\frac{d}{dt}\right)\left(1+\left(\frac{d}{dt}\right)^{-1} V_{,xx}\right)\equiv\left(\frac{d}{dt}\right)\tilde{\textbf{M}}\,,
\end{eqnarray}
where the matrix $\left(\frac{d}{dt}\right)^{-1}(t,t')=\Theta(t-t')$, we see that 
\begin{eqnarray}
\text{det} \textbf{M} = \text{det}\left(\frac{d}{dt}\right) \times \text{det} \tilde{\textbf{M}} \propto\text{exp}\left[ \text{Tr}\,\text{log}\left(\tilde{\textbf{M}} \right)\right] \propto \text{exp}\left[\dfrac{1}{2}\int \text{d}t~V_{,xx}\right]\label{eq:det-Alg}
\end{eqnarray}
where we used the Stratonovich prescription $(\theta (0) = 1/2)$. On the face of it this appears desirable and we will later see this agrees with the \ac{F-P} description (\ref{eq:PI F-P}). However the use of Stratonovich is not a priori justified here and actually hides certain symmetries of the problem. Alternatively we can use anticommuting variables $c$ and $\bar{c}$  such that:
\begin{eqnarray}
\text{det } \textbf{M} = \int \mathcal{D}c\mathcal{D}\bar{c}\text{ exp}\left[\int\text{d}t \, \bar{c}\left( \partial_{t} +  V_{,xx} \right)c \right] \label{eq: ccbar PI}
\end{eqnarray}
The determination of $\text{det}\textbf{M}$ then requires appropriate boundary conditions for $\bar{c}$ and $c$. Here we will recall the computations of \cite{Marculescu1991} explicitly showing that
\begin{eqnarray}
	\text{det } \textbf{M} = \int \mathcal{D}c\mathcal{D}\bar{c}\text{ exp}\lsb\int\text{d}t \, \bar{c}\left( \partial_{t} +  V_{,xx} \right)c\rsb \propto \text{exp}\left[\dfrac{1}{2}\int \text{d}t~V_{,xx}\right] \label{eq: ccbar PI-Ap}
\end{eqnarray}
when the boundary conditions 
\begin{eqnarray}  
	c(t_{\mathrm in})=0\,,\quad \bar{c}(t_{\mathrm f})=0 \label{eq:ghost_bcs}
\end{eqnarray}
are chosen. The boundary condition (\ref{eq:ghost_bcs}) is implied by the discretised form of the path integral, see \cite{Hertz2017}. Roughly speaking the $c$ field is not present in the discretised path integral in the first integral and the $\bar{c}$ field is not present in the final integral, hence why they must vanish at these times. Similarly the discretised path integral can also be consulted to infer that we must further impose
\begin{eqnarray}
\tilde{x}(t_{\mathrm f}) = 0 . \label{eq:auxiliary_bc}
\end{eqnarray} 

Other choices are possible but lead to determinant values that are different from (\ref{eq:det-Alg}), corresponding to non-causal boundary conditions for the stochastic problem. \\

The Gaussian integral in (\ref{eq: ccbar PI}) can now be done explicitly leading to (\ref{eq:det-Alg}). However it pays to keep the determinant expressed in this form. Inserting equations (\ref{eq: tilde x PI}) \& (\ref{eq: ccbar PI}) into (\ref{eq: PDF 2nd PI}) we obtain:
\begin{eqnarray}
\mathcal{P}(x_{f}\vert x_{i}) = \int\mathcal{D}\eta\mathcal{D}x\mathcal{D}\tilde{x} \mathcal{D}c\mathcal{D}\bar{c} \text{ exp}\Bigg[\int\text{d}t \Big\{-\dfrac{\eta^2}{2\Upsilon} + i\tilde{x}\left( \dot{x} + V_{,x} -\eta \right) + \bar{c}\left( \partial_{t} +  V_{,xx} \right)c \Big\} \Bigg]\nonumber \\
\end{eqnarray}
We can now trivially perform the Gaussian integral over $\eta$ to obtain the path integral in terms of the \ac{BM} action $\mathcal{S}_{BM}(x,\tilde{x},\bar{c},c)$:
\begin{subequations}
\begin{empheq}[box=\fcolorbox{Maroon}{white}]{align}
\mathcal{P}(x_{f}\vert x_{i}) &= \int\mathcal{D}x\mathcal{D}\tilde{x} \mathcal{D}c\mathcal{D}\bar{c} \text{ exp}\left[-\mathcal{S}_{BM}(x,\tilde{x},\bar{c},c)\right] \label{eq:TransProb} \\
\mathcal{S}_{BM}(x,\tilde{x},\bar{c},c) &= \int \text{d}t\bigg[ \frac{\Upsilon}{2}\tilde{x}^2 - i\tilde{x}(\dot{x}+ V_{,x}) - \bar{c}\left( \partial_{t} +  V_{,xx} \right)c \bigg] \label{eq: PDF action}
\end{empheq}\label{eq:BM_path_int}
\end{subequations}

Computing this path integral, which henceforth shall be called the \ac{BPI}, is in general impossible analytically. At this stage the reader would be very justified to wonder why we went through all this trouble? Having started from a Gaussian path integral in one variable (\ref{eq:PDF defn}) we have arrived at a much more complicated one involving four different variables, why should this be computationally more tractable? To help explain that we look to \ac{SUSY}.
\subsection{Brownian Motion and SuperSymmetry}
\ac{SUSY} is a theoretical extension of the Standard Model of Particle Physics that proposes every fermion (e.g. electron) would have a supersymmetric partner that would be a boson, similarly every boson (e.g. photon) would have a supersymmetric partner that would be a fermion. This would essentially double the number of fundamental particles that exist in the Standard Model. There are many motivations for \ac{SUSY} not least of which is its requirement in String Theory. Regardless of whether \ac{SUSY} exists in nature there has been much work done analysing possible \ac{SUSY} theories. It has been known for a while that overdamped \ac{BM} (\ref{eq:langevindimless}) and Euclidean\footnote{What we mean by Euclidean in this context is that we are working in an imaginary time coordinate  e.g. $\tau = it$} SuperSymmetric Quantum Mechanics are in some sense equivalent. In this subsection we will demonstrate this explicitly.\\
To make the link with \ac{SUSY} we will suggestively redefine our fields as :
\begin{subequations}
\begin{align}
x(t) &\equiv  \sqrt{\Upsilon}\,\varphi(t) \\
	V(x) &\equiv  {\Upsilon} \, W(\varphi)  \\
	\tilde{x} &\equiv  \dfrac{1}{\sqrt{\Upsilon}}\,(i\dot{\varphi} - \tilde{F}) \\
	\bar{c}c &\equiv  i\bar{\rho}\rho   
\end{align}\label{eq:LPAidentifications}   
\end{subequations}

doing so allows us to relate the action for \ac{BM} to that of Euclidean SuperSymmetric Quantum Mechanics:
\begin{eqnarray}
	\mathcal{S}_{BM}[\varphi, \tilde{F}, \bar{\rho}, \rho] = \left[W(\varphi_f) -  W(\varphi_i)\right] + \mathcal{S}_{SUSY}
\end{eqnarray}
where
\begin{eqnarray}
	\mathcal{S}_{SUSY}[\varphi, \tilde{F}, \bar{\rho}, \rho] = \int dt\bigg[\dfrac{1}{2}\dot{\varphi}^2 + \dfrac{1}{2}\tilde{F}^2 + i\tilde{F}W_{,\varphi}(\varphi) -i \bar{\rho}(\partial_{t} +  W_{,\varphi\varphi}(\varphi))\rho\bigg]  \label{eq:SUSYClass}
\end{eqnarray} 
Action (\ref{eq:SUSYClass}) describes the dynamics of Euclidean Supersymmetric Quantum Mechanics where $\rho$ \& $\bar{\rho}$ are the fermionic fields and $\varphi$ \& $\tilde{F}$ are the bosonic fields \cite{Synatschke2009}. The same action also describes \ac{BM} and the \ac{BM} action is equivalent to the Supersymmetric Quantum Mechanics one up to a factor depending on the initial and final positions $x_i$ \& $x_f$; these terms can be simply taken outside the path integral as an exponential prefactor.

Variation of $\mathcal{S}_{SUSY}$ with respect to $\tilde{F}$ yields its ``equation of motion'' $\tilde{F}=-iW_{,\varphi}$ which when substituted back into $\mathcal{S}_{SUSY}$ yields the ``on mass-shell'' action
\begin{eqnarray}
	\mathcal{S}_{OM}[\varphi, \bar{\rho}, \rho] = \int dt\bigg[\dfrac{1}{2}\dot{\varphi}^2 + \dfrac{1}{2}W_{,\varphi}{}^2 -i \bar{\rho}(\partial_{t} +  W_{,\varphi\varphi})\rho\bigg]
\end{eqnarray} 
We will keep working with the auxiliary field $\tilde{F}$ and (\ref{eq:SUSYClass}) as it allows for the symmetry transformations to take on a simpler form, linear in all fields.     

It is illuminating to express the above action in terms of the original dimensional variables and perform the integration over $\bar{\rho}$ and $\rho$, leading to the alternative form of the term stemming from the determinant:
\begin{eqnarray}
	\mathcal{S}_{OM}[x] = \int \frac{dt}{2Dm}\bigg[\dfrac{1}{2}m\dot{x}^2 + \dfrac{1}{2}\varepsilon^2 mV_{,x}{}^2 - Dm\varepsilon V_{,xx}\bigg] \label{eq:dim-action}
\end{eqnarray} 
Note that $2Dm$ has the dimensions of action and therefore plays in the thermal problem a role analogous to $\hslash$ in quantum mechanics - see also section \ref{sec:F-P} in this respect. Unlike $\hslash$ of course, it can be varied by changing the temperature, therefore controlling the strength of fluctuations.

\subsection{\label{sec:gen_func}Correlation functions from the generating functional}
Now that we have made the link with \ac{SUSY} manifest we will exploit it to discuss how to compute correlation functions. One way to compute correlation functions is through the use of objects called \textit{generating functionals}. In this subsection we will outline how these generating functionals yield correlators in practice.

The first generating functional we examine is the partition functional $\mathcal{Z}(\mathcal{J})$ which depends on source terms ${\mathcal{J}}(t)$ (in analogy with a magnetic field source term $M(x)$ in spin systems):
\begin{eqnarray}
	\mathcal{Z}({\mathcal{J}}) = \int\mathcal{D}\Phi \text{ exp}\left[-\mathcal{S}_{BM}[\Phi]  + \int dt \,{\mathcal{J}} \Phi\right] \label{eq:partition functional} 
\end{eqnarray}
Variation of $\mathcal{Z}({\mathcal{J}})$ w.r.t. ${\mathcal{J}}$ will give any required correlator. In the above functional integral, $\Phi$ stands collectively for $(\varphi(t),\tilde{F}(t),\rho(t),\bar{\rho}(t))$ and ${\mathcal{J}}(t)$ for all the corresponding currents:
\begin{eqnarray}
 \int dt \,{\mathcal{J}} \Phi \equiv \int dt \left(  J_\varphi \varphi + J_{\tilde{F}} \tilde{F} + \bar{\rho}\theta + \bar{\theta}\rho    \right) \label{eq:currents}
\end{eqnarray}
The only constraint we will require of the currents is that they satisfy ${\mathcal{J}}(t_{\mathrm in}) ={\mathcal{J}}(t_{\mathrm f}) = 0$ at the initial and final times $t_{\mathrm in}$ and $t_{\mathrm f}$. The averages of the fields are defined by 
\begin{eqnarray}
	\left\langle \Phi(t)\right\rangle &\equiv&  \dfrac{\int\mathcal{D}\Phi\,\, \Phi(t)\,\,\text{ exp}\left[-\mathcal{S}_{BM}[\Phi] \right]}{\int\mathcal{D}\Phi \text{ exp}\left[-\mathcal{S}_{BM}[\Phi] \right]}\\
	&=&	\dfrac{\delta \mathcal{Z}[{\mathcal{J}}]}{\delta {\mathcal{J}}(t)}\Bigg|_{{\mathcal{J}}=0}
\end{eqnarray}
the two point correlation function is:
\begin{eqnarray}
	\left\langle \Phi(t_{1})\Phi(t_{2})\right\rangle &\equiv&  \frac{\int\mathcal{D}\Phi\,\, \Phi(t_1)\Phi(t_2)\,\,\text{ exp}\left[-\mathcal{S}_{BM}[\Phi] \right]}{\int\mathcal{D}\Phi \text{ exp}\left[-\mathcal{S}_{BM}[\Phi] \right]}\\
	&=& \dfrac{\delta^2\mathcal{Z}(\mathcal{J})}{\delta \mathcal{J}(t_{2})\delta \mathcal{J}(t_{1})}\bigg\rvert_{\mathcal{J} = 0}
\end{eqnarray}	
and similarly for higher correlation functions. N.B. that the usual prefactor of $\mathcal{Z}(0)$ is absent as it is equivalent to unity in this theory -- the reason for this will become apparent in section \ref{sec:symmetry}. This information can be similarly stored more compactly in another object $\mathcal{W}[{\mathcal{J}}]$. Defining 
\begin{eqnarray}
	\mathcal{W}[{\mathcal{J}}] \equiv \text{ln}\left(\mathcal{Z}({\mathcal{J}})\right) \label{eq:gen_func_W_defn}
\end{eqnarray}
allows us to generate connected correlation functions (or Ursell functions) as:
\begin{eqnarray}
	\left\langle \Phi(t_1)\dots\Phi(t_n)\right\rangle_{C} = \dfrac{\delta^n \mathcal{W}[{\mathcal{J}}]}{\delta {\mathcal{J}}(t_1)\dots\delta {\mathcal{J}}(t_n)}\Bigg|_{{\mathcal{J}}=0}
\end{eqnarray}
For instance the connected 2-point function (more commonly known as covariance) $G(t_1,t_2)$ is:
\begin{eqnarray}
	G(t_1,t_2) \equiv \left\langle \Phi(t_1)\Phi(t_2) \right\rangle_{C} &=& \left\langle \Phi(t_1)\Phi(t_2) \right\rangle - \left\langle \Phi(t_1) \right\rangle\left\langle \Phi(t_2) \right\rangle \nonumber \\
	&=& \dfrac{\delta^2 \mathcal{W}[{\mathcal{J}}]}{\delta {\mathcal{J}}(t_1)\delta {\mathcal{J}}(t_2)}\Bigg|_{{\mathcal{J}}=0}
\end{eqnarray}
However the most efficient object to work with is the \ac{EA} $\Gamma [\Phi]$ obtained by a Legendre transform of $\mathcal{W}[\mathcal{J}]$:
\begin{eqnarray}
\Gamma [\phi, \mathcal{\tilde{F}}, \psi, \bar{\psi}] = \int \mathcal{J} \Phi - \mathcal{W}[\mathcal{J}] \label{eq:Gamma_defn_SUSY}
\end{eqnarray}
where the right hand side is evaluated at $\mathcal{J}_{sup}$ corresponding to the supremum. From this definition it is clear that $\Gamma$ is guaranteed to be convex. What is nice about $\Gamma$ is that it treats the \textit{averaged} fields as the central objects of interest:
\begin{subequations}
\begin{align}
\phi &= \dfrac{\delta \mathcal{W}[\mathcal{J}]}{\delta J_{\varphi}} = \lan \varphi\ran_{J_{\varphi}} \\
\mathcal{\tilde{F}} &= \dfrac{\delta \mathcal{W}[\mathcal{J}]}{\delta J_{\tilde{F}}} = \lan \tilde{F}\ran_{J_{\tilde{F}}} \\
\psi &= \dfrac{\delta \mathcal{W}[\mathcal{J}]}{\delta \theta} = \lan \rho \ran_{\theta} \\
\bar{\psi} &= \dfrac{\delta \mathcal{W}[\mathcal{J}]}{\delta \bar{\theta}} = \lan \bar{\rho}\ran_{\bar{\theta}}    
\end{align}   \label{eq:SUSY_mean_fields} 
\end{subequations}
This points to how it is related to the original \ac{BM} action:
\begin{empheq}[box = \fcolorbox{Maroon}{white}]{equation}
e^{-\Gamma [\Phi]} = \mathcal{P}(x_{f}\vert x_{i}) = \int\mathcal{D}\varphi\mathcal{D}\tilde{\varphi} \mathcal{D}\psi\mathcal{D}\bar{\psi} \text{ exp}\lb-\mathcal{S}_{BM}[\varphi,\tilde{\varphi},\psi,\bar{\psi}]\rb \label{eq:Gamma_S_relation}
\end{empheq}
i.e. it resembles the classical action $\mathcal{S}$ but with all fluctuations integrated out. We will explore this in more detail in chapter \ref{cha:fRG EEOM}. \\
For completeness we can also define the currents in terms of the original physical variables $\mathbf{x} = (x,\tilde{x},c,\bar{c})$:
\begin{eqnarray}
 \int \mathrm{d}t \,{\mathcal{J}} \mathbf{x} \equiv \int \mathrm{d}t \left(  J_x x + J_{\tilde{x}} \tilde{x} + \bar{c}\vartheta + \bar{\vartheta}c    \right) \label{eq:currents_physical}
\end{eqnarray}
the corresponding mean fields:
\begin{subequations}
\begin{align}
\chi &= \dfrac{\delta \mathcal{W}[\mathcal{J}]}{\delta J_{x}} = \lan x\ran_{J_{x}} \\
\tilde{\chi} &= \dfrac{\delta \mathcal{W}[\mathcal{J}]}{\delta J_{\tilde{x}}} = \lan \tilde{x}\ran_{J_{\tilde{x}}} \\
C &= \dfrac{\delta \mathcal{W}[\mathcal{J}]}{\delta \vartheta} = \lan c \ran_{\vartheta} \\
\bar{C} &= \dfrac{\delta \mathcal{W}[\mathcal{J}]}{\delta \bar{\vartheta}} = \lan \bar{c}\ran_{\bar{\vartheta}}    
\end{align}   \label{eq:BM_mean_fields} 
\end{subequations}
and work directly with the generating functionals in terms of these physical variables. This means we can rewrite (\ref{eq:Gamma_S_relation}) like so:
\begin{eqnarray}
e^{-\Gamma [\mathbf{x}]} = \mathcal{P}(x_{f}\vert x_{i}) = \int\mathcal{D}x\mathcal{D}\tilde{x} \mathcal{D}c\mathcal{D}\bar{c} \text{ exp}\lb -\mathcal{S}_{BM}[x,\tilde{x},c,\bar{c}]\rb \label{eq:Gamma_S_relation_real}
\end{eqnarray}
In reality this makes computations much more difficult so we will generally work with the \ac{SUSY} variables and when appropriate go back to the physical variables using (\ref{eq:LPAidentifications}).

\subsection{\label{sec:symmetry}Symmetry transformations for \texorpdfstring{$\mathcal{S}_{BM}$}{} and Ward-Takahashi identities}
In this subsection we recall the transformations that leave the action  $\mathcal{S}_{BM}$ invariant, up to boundary terms. We comment on the implications of such symmetries, also paying attention to the boundary terms that are usually dropped under the assumption of equilibrium, or, equivalently, a corresponding infinite amount of elapsed time between initial and final states \cite{Mallick2011}. If we are to exploit \ac{SUSY} in chapter \ref{cha:fRG} it is crucial to verify the presence of this symmetry in an out-of-equilibrium context.

In general, invariances of the action imply relations between various correlation functions in field theory, generally known as Ward-Takahashi identities. Their derivation can be summarised as follows: A general infinitesimal transformation of the fields $\Phi\rightarrow \Phi'= \Phi+\Delta\Phi$ will generically change the action $S\rightarrow \mathcal{S}'=\mathcal{S}+\Delta \mathcal{S}$. Also shifting ${\mathcal{J}}\rightarrow {\mathcal{J}}+\Delta {\mathcal{J}}$ leads to 
\begin{eqnarray}
\Delta \mathcal{Z}=\int\!\! dt  \dfrac{\delta \mathcal{Z}}{\delta {\mathcal{J}}(t)}\Delta \mathcal{J}(t) =\! \int\!\!\mathcal{D}\Phi e^{-\mathcal{S}[\Phi]  + \int \!\!dt \,{\mathcal{J}} \Phi}\left(-\Delta S+{\mathcal{J}}\Delta\Phi + \Delta{\mathcal{J}}\,\Phi\right) 
\end{eqnarray}
where we used that a) $\Phi$ is simply an integration variable in (\ref{eq:partition functional}) and $\mathcal{Z}$ is not altered by a change in $\Phi$ but only via $\mathcal{J}$ and b) $\mathcal{D}\Phi = \mathcal{D}\Phi'$ i.e.~the transformation involves no non-trivial Jacobian determinant. Symmetries of the dynamical system comprise of transformations for which $\Delta S$ is, at most, a total derivative (or a total divergence for higher dimensions): $\Delta \mathcal{S}=\int dt \dfrac{d}{dt}\mathcal{A} = \mathcal{A}(t_f)-\mathcal{A}(t_i) = \left[\mathcal{A}\right]^{t_f}_{t_i}$. Further choosing $\Delta {\mathcal{J}}$ such that, for a given $\Delta\Phi$, ${\mathcal{J}}\Delta\Phi + \Delta{\mathcal{J}}\,\Phi=0$, leads to
\begin{eqnarray}
	\int dt  \dfrac{\delta \mathcal{Z}}{\delta {\mathcal{J}}(t)}\Delta {\mathcal{J}}(t) = \int\mathcal{D}\Phi \left[\mathcal{A}\right]^{t_f}_{t_i}e^{-\mathcal{S}[\Phi]  + \int dt \,{\mathcal{J}} \Phi} 
\end{eqnarray}
Differentiating this master equation w.r.t.~${\mathcal{J}}$ and setting ${\mathcal{J}}=0$, gives relations between correlations functions that are necessitated by the symmetry under $\Phi \rightarrow \Phi+\Delta\Phi$.

For our case, given two independent, infinitesimal Grassmann variables $\epsilon$ and $\bar{\epsilon}$, the following transformations of the fields \cite{Synatschke2009}
\begin{subequations}
\begin{align}
 \varphi &\rightarrow  \varphi + i \bar{\epsilon}\rho - i \bar{\rho}\epsilon \label{SUSY1}\\
	\tilde{F} &\rightarrow  \tilde{F} - \bar{\epsilon}\dot{\rho} - \dot{\bar{\rho}}\epsilon\\
	\rho &\rightarrow \rho +\left(\dot{\varphi}-i \tilde{F}\right)\epsilon \\
	\bar{\rho} &\rightarrow  \bar{\rho} + \bar{\epsilon}\left(\dot{\varphi} + i \tilde{F}\right)    
\end{align}\label{SUSY4}
\end{subequations}

leave $\mathcal{S}_{BM}$ invariant up to a boundary term at the initial time $t_{\mathrm in}$:
\begin{eqnarray}\label{eq:action change}
\mathcal{S}_{BM} \rightarrow \mathcal{S}_{BM} + \bar{\rho}_{\mathrm in}\left(i\dot{\varphi}+ \tilde{F} +2i W_{,\varphi}\right)_{\mathrm in}\epsilon
\end{eqnarray}
where a subscript $`{\mathrm in}$ 'denotes the initial time $t_{\mathrm in}$. The boundary term at $t_{\mathrm f}$ has been eliminated using the boundary condition  (\ref{eq:ghost_bcs}).
Note that the $\bar{\epsilon}$ transformation leaves $\mathcal{S}_{BM}$ invariant identically, irrespective of the boundary conditions.

Adding source currents $\left(J_\varphi,\, J_{\tilde{F}},\, \theta, \bar{\theta}\right)$ to the action \cite{Damgaard1987}
\begin{eqnarray}
	\mathcal{S}_{BM} \rightarrow \mathcal{S}_{BM} - \int dt \left(  J_\varphi \varphi + J_{\tilde{F}} \tilde{F} + \bar{\rho}\theta + \bar{\theta}\rho   \right)
\end{eqnarray}
and requiring appropriate transformations of those currents, 
\begin{subequations}

\begin{align}
J_\varphi &\rightarrow J_\varphi + \dot{\bar{\theta}} \epsilon + \bar{\epsilon} \dot{\theta}  \\
J_{\tilde{F}} &\rightarrow J_{\tilde{F}} + i\bar{\theta} \epsilon - i \bar{\epsilon} \theta \\
\theta &\rightarrow \theta + \epsilon \left(i J_\varphi -\dot{J}_{\tilde{F}}\right)\\
\bar{\theta} &\rightarrow \bar{\theta} - \bar{\epsilon} \left(i J_\varphi +\dot{J}_{\tilde{F}}\right)
\end{align}
\end{subequations}
we have
\begin{eqnarray}
	{\mathcal{J}}\Phi \rightarrow {\mathcal{J}}\Phi- \bar{\epsilon}\frac{d}{dt}\left(\rho J_{\tilde{F}}-\varphi\theta\right)- \frac{d}{dt}\left(\bar{\rho} J_{\tilde{F}}-\varphi\bar{\theta}\right){\epsilon}
\end{eqnarray}
We therefore see that the transformations result in     
\begin{eqnarray}
		\mathcal{S}_{BM} - 	{\mathcal{J}}\Phi \rightarrow  \mathcal{S}_{BM} - 	{\mathcal{J}}\Phi	+ \bar{\rho}_{\mathrm in}\left(i\dot{\varphi}+ \tilde{F} +2i W_{,\varphi}\right)_{\mathrm in} \!\! \epsilon
\end{eqnarray}
and the exponent in the integrand of (\ref{eq:partition functional}) only changes by a lower boundary term that is also independent of $\bar{\epsilon}$. 

The field transformations (\ref{SUSY4}) are linear shifts that leave the integration measure in the path integral invariant. Coupled with the shift in the currents we find, setting $\epsilon = 0$
\begin{eqnarray}
	\int dt \left[\frac{\delta \mathcal{Z}}{\delta J_{\varphi}(t)}\dot{\theta}-i\frac{\delta \mathcal{Z}}{\delta J_{\tilde{F}}(t)}\theta  -  \frac{\delta \mathcal{Z}}{\delta \bar{\theta}(t)}\left(iJ_\varphi+\dot{J}_{\tilde{F}}\right)\right] = 0 \label{BRST1}
\end{eqnarray}     
while for $\bar{\epsilon}=0$ we obtain
\begin{eqnarray}
	\int dt \left[\frac{\delta \mathcal{Z}}{\delta J_{\varphi}(t)}\dot{\bar{\theta}} + i\frac{\delta \mathcal{Z}}{\delta J_{\tilde{F}}(t)}\bar{\theta}  -  \frac{\delta \mathcal{Z}}{\delta {\theta}(t)}\left(iJ_\varphi-\dot{J}_{\tilde{F}}\right)\right] 	= \nonumber \\
	\int  \mathcal{D}\Phi  \left[-\bar{\rho}_{\mathrm in}\left(i\dot{\varphi}_{\mathrm in}+\tilde{F}_{\mathrm in} +2iW'_{\mathrm in}\right)\right]e^{-\mathcal{S}_{BM}+\int \!\! dt {\mathcal{J}}\Phi} \label{BRST2}
\end{eqnarray}	
These are the master equations from which so-called Ward-Takahashi identities between various correlators can be obtained. For example, differentiating (\ref{BRST1}) w.r.t.  $J_{\varphi}(t')$, $\theta(\tau)$ and setting ${\mathcal{J}}=0$ gives
\begin{eqnarray}\label{eq:WT1}
	\frac{d}{d\tau}\left\langle \varphi(t')\varphi(\tau)\right\rangle + \left\langle\varphi(t')W'(\varphi(\tau))\right\rangle - i \langle \rho(t')\bar{\rho}(\tau)\rangle = 0
\end{eqnarray}    
which, along with the original Langevin equation, allows us to infer that 
\begin{eqnarray}
i \langle \rho(t')\bar{\rho}(\tau)\rangle = \frac{1}{\sqrt{\Upsilon}}\left\langle  \varphi(t)\eta(\tau)\right\rangle\,,
\end{eqnarray}  
meaning that $\langle \rho(t')\bar{\rho}(\tau)\rangle$ is proportional to the response of $\varphi(t')$ to noise $\eta(\tau)$ (clearly a retarded quantity $\propto \Theta(t'-\tau)$). Furthermore, equation (\ref{eq:WT1}) can be rewritten as 
\begin{eqnarray}\label{eq:ghost=response}
	\sqrt{\Upsilon}\left\langle \tilde{x}(\tau)\varphi(t')\right\rangle =- \langle \bar{\rho}(\tau)\rho(t')\rangle
\end{eqnarray} 
which confirms that $\left\langle \varphi(t')\tilde{x}(\tau)\right\rangle$ is the retarded response function or propagator. Importantly, equation (\ref{eq:ghost=response}) also establishes that in a diagrammatic expansion closed ghost loops act to cancel closed loops involving the retarded propagator -- this can be seen explicitly in e.g. \cite{Bounakis2020}. This ensures $\mathcal{Z}[{\mathcal{J}}=0]=1$, which simply reflects conservation of probability, and furthermore that correlators do not depend on the ill-defined quantity $\Theta(0)$, reflecting the well-known fact that, for additive noise, the discretisation of the stochastic differential equation (Ito, Stratonovic etc) does not matter -- see e.g. \cite{Hertz2017, Bounakis2020}.         

Differentiating (\ref{BRST2}) w.r.t. $J_\varphi(t')$, $\bar{\theta}(\tau)$ and setting ${\mathcal{J}}=0$ gives, with the use of (\ref{eq:ghost=response}) and recalling that integration over $\tilde{F}$ gives $\tilde{F} \rightarrow -iW_{,\varphi}$, 
\begin{eqnarray}
	2\frac{d}{d\tau} \left\langle \varphi(t')\varphi(\tau)\right\rangle -i\sqrt{\Upsilon}\left\langle \tilde{x}(\tau)\varphi(t')\right\rangle + i\sqrt{\Upsilon}\left\langle \tilde{x}(t')\varphi(\tau)\right\rangle = -i\Upsilon \left\langle \tilde{x}_{\mathrm in} \varphi(\tau)\right\rangle\left\langle \tilde{x}_{\mathrm in} \varphi(t')\right\rangle
\end{eqnarray}  
This is a modified Fluctuation-Dissipation relation with the term on the RHS accounting for the initial condition. Sending $t_{\mathrm in} \rightarrow -\infty$ makes the RHS vanish and we recover the Fluctuation-Dissipation relation at equilibrium \cite{Mallick2011}:
\begin{eqnarray}
		\frac{d}{d\tau} \left\langle \varphi(t')\varphi(\tau)\right\rangle =  i\frac{\sqrt{\Upsilon}}{2}\left(\left\langle \tilde{x}(\tau)\varphi(t')\right\rangle - \left\langle \tilde{x}(t')\varphi(\tau)\right\rangle\right)
\end{eqnarray}
Before concluding this chapter we examine an alternative formulation of the Langevin problem.

\section{\label{sec:F-P}The Fokker-Planck Equation}
Instead of working with the Langevin equation directly one can deal with the probability distribution of position:
\begin{eqnarray}
	P(x,t) = \left\langle \delta (x-x_{\eta})\right\rangle \label{eq:pdf_pos_defn}
\end{eqnarray} 
where $x_{\eta}$ is the solution to (\ref{eq:langevindimless}) for a given noise function $\eta$ (i.e. a specific particle trajectory). To determine this probability distribution we consider an infinitesimal change in (\ref{eq:langevindimless}):
\begin{eqnarray}
\delta x = \dot{x}\delta t = - \partial_{x}V \delta t + \int_{t}^{t+\delta t}\mathrm{d}t' ~\eta(t') \label{eq:non_infinitesimal_x}
\end{eqnarray}
where we have assumed that we can evaluate the classical force $\partial_{x}V$ at the original position, $x$. This is a reasonable choice for classical dynamics but in stochastic dynamics there is ambiguity in this discretisation choice. Evaluating the force at the original time corresponds to the Ito convention whereas if we evaluated the force at $(t+\delta t)/2$ this would correspond to the Stratonovich convention and in general the two would give different predictions. Fortunately for us, in this system the two conventions agree -- see e.g. \cite{Hertz2017, Bounakis2020} -- and we will choose Ito for simplicity. We can take the stochastic average of (\ref{eq:non_infinitesimal_x}) and because the noise term has vanishing mean, $\lan \eta(t)\ran = 0$ we obtain:
\begin{eqnarray}
\lan \delta x\ran = -\partial_{x}V \delta t \label{eq:infinitesimal_x}
\end{eqnarray}
We can then compute the stochastic average of $\delta x_i \delta x_j$ in a similar manner:
\begin{eqnarray}
\lan \delta x_i \delta x_j\ran &=& \lan \partial_{x_i}V\partial_{x_j}V\ran \delta t^2 - \delta t \int_{t}^{t+\delta t}\mathrm{d}t' \lan \partial_{x_i}V\eta_{j}(t') + \partial_{x_j}V\eta_{i}(t') \ran \nonumber \\
&&+ \int_{t}^{t+\delta t}\mathrm{d}t'\int_{t}^{t+\delta t}\mathrm{d}t'' \underbrace{\lan \eta_{i}(t')\eta_{j}(t'')\ran}_{=\Upsilon \delta_{ij}\delta (t'-t'')} \\
\lan \delta x_i \delta x_j\ran &=& \delta_{ij} \Upsilon \delta t + \mathcal{O}(\delta t^2) \label{eq:infinitesimal_x^2}
\end{eqnarray}
As higher order moments are all higher order in $\delta t$ we will neglect them. Our goal then it to obtain the probability distribution that reproduces the correlations (\ref{eq:infinitesimal_x}) \& (\ref{eq:infinitesimal_x^2}). To achieve this we work with the conditional probability $P(x, t+ \delta t ; x',t )$, that the particle is at position $x$ at time $t + \delta t$ given that it was at position $x'$ a moment earlier at $t$. Using the definition (\ref{eq:pdf_pos_defn}) this can be written like so:
\begin{eqnarray}
P(x, t+ \delta t ; x',t ) = \lan \delta (x - x' -\delta x)\ran
\end{eqnarray}
and to the dismay of mathematicians everywhere we Taylor expand the delta function to obtain:
\begin{eqnarray}
P(x, t+ \delta t ; x',t ) = \lsb 1 + \lan \delta x_i \ran \dfrac{\partial}{\partial x_{i}'} + \dfrac{1}{2} \lan \delta x_i \delta x_j\ran \dfrac{\partial^2}{\partial x_{i}'\partial x_{j}' } + \dots\rsb \delta (x - x') \label{eq:Taylor_delta_func}
\end{eqnarray}
However we aren't interested in $P(x, t+ \delta t ; x',t ) $ but rather the probability $P(x,t; x_0, t_0)$ given some initial, arbitrary position. To do this we use the ``has to be somewhere" property discussed earlier in the form of the Chapman-Kolmogorov equation:
\begin{eqnarray}
P(x,t ; x_0, t_0) = \int_{-\infty}^{\infty}\mathrm{d}^3\Vec{x}' P(x,t; x', t')P(x',t'; x_0, t_0) \label{eq:Chap-Kolm}
\end{eqnarray}
We can now substitute (\ref{eq:Taylor_delta_func}) into (\ref{eq:Chap-Kolm}) so that the delta function kills the integral:
\begin{eqnarray}
P(x,t+\delta t; x_0, t_0) &=& P(x,t\; x_0, t_0) - \dfrac{\partial}{\partial x_{i}}\lsb \lan \delta x_i \ran P(x,t\; x_0, t_0) \rsb \nonumber \\
&& + \dfrac{1}{2}\dfrac{\partial^2}{\partial x_{i}\partial x_{j}} \lsb \lan \delta x_i \delta x_j\ran P(x,t\; x_0, t_0)\rsb + \dots \label{eq:deriv_2_FP}
\end{eqnarray}
The final simplification can be done by Taylor expanding the left hand side with respect to time:
\begin{eqnarray}
P(x,t+\delta t; x_0, t_0) = P(x,t; x_0, t_0) + \partial_t P(x,t; x_0, t_0) + \dots
\end{eqnarray}
and if we combine this with equations (\ref{eq:infinitesimal_x}) \& (\ref{eq:infinitesimal_x^2}) substituted into (\ref{eq:deriv_2_FP}) we find the probability obeys the following partial differential equation: 
\begin{empheq}[box = \fcolorbox{Maroon}{white}]{equation}
	\dfrac{\partial P(x,t)}{\partial t} = \partial_{x}(P(x,t)\partial_x V) + \dfrac{\Upsilon}{2}\partial_{xx} P(x,t) \label{eq: F-P}
\end{empheq}
which is known as the \ac{F-P} equation. It is usually more useful however to rescale the PDF like so:
\begin{eqnarray}
	P(x,t) = e^{-V/\Upsilon}\tilde{P}(x,t) \label{eq:Ptransform}
\end{eqnarray}
This leads to the \ac{F-P} equation taking the form:
\begin{subequations}
\begin{empheq}[box = \fcolorbox{Maroon}{white}]{align}
	\dfrac{\Upsilon}{2}\dfrac{\partial\tilde{P}(x,t)}{\partial t} &= \left(\dfrac{\Upsilon}{2}\right)^2\partial_{xx}\tilde{P}(x,t)  + \bar{U}\tilde{P}(x,t) \label{eq: rescaled F-P}\\
	\bar{U} &\equiv  \dfrac{\Upsilon}{4}\partial_{xx} V - \dfrac{1}{4}(\partial_{x} V)^2 \label{eq: Ubar=}
	\end{empheq}\label{eq:FP1}
\end{subequations}
which resembles a Euclidean Schr\"{o}dinger equation with $\Upsilon /2$ playing the role of $\hslash$ in controlling the fluctuation amplitude, as one might expect. The unusual form of the Schr\"{o}dinger potential $\bar{U}$ is because this is equivalent to a theory of \emph{SuperSymmmetric} Quantum Mechanics and $\bar{U}$ can therefore be expressed in terms of the superpotential $V$ through (\ref{eq: Ubar=}). Equation (\ref{eq: rescaled F-P}) can be solved in terms of a spectral expansion (see e.g. \cite{Gardiner2009,Markkanen2019}). Writing 
\begin{eqnarray}
	\tilde{P}(x,t) = \sum\limits_{n=0}^\infty c_n p_n(x)e^{-E_n t}
\end{eqnarray}  
we find that $p_n$ satisfy the corresponding, time independent  Euclidean Schr\"{o}dinger equation
\begin{eqnarray}
	-\frac{\Upsilon}{2}\dfrac{d^2 p_n}{dx^2}  +\frac{1}{2}\left(\frac{\left(V_{,x}\right)^2}{\Upsilon}- V_{,xx}\right)p_n=E_np_n \label{eq:F-P_spectral}
\end{eqnarray} 
The lowest eigenfunction with $E_0=0$ is 
\begin{eqnarray}
	p_0(x) =\mathcal{N} e^{-V(x)/\Upsilon} \label{eq:spec_lowest_eigfunc}
\end{eqnarray} 
corresponding to the equilibrium distribution $P_{\mathrm eq}(x)=p_0(x)^2$. The $p_n(x)$ eigenfunctions are complete and orthonormal  
\begin{eqnarray}
\int\limits_{-\infty}^{\infty} dx \, p_n(x)p_m(x)=\delta_{mn} \\
\sum\limits_{n=1}^{\infty} p_n(x)p_n(x_0)=\delta(x-x_0)
\end{eqnarray}
The conditional probability, a quantity akin to the evolution operator or propagator in quantum mechanics, can be expressed in terms of the spectral expansion as 
\begin{eqnarray}\label{eq:specral-prop1}
\tilde{P}(x,t|x_0,0) &=& \sum\limits_{n=0}^\infty p_n(x)p_n(x_0)e^{-E_n t}\\
P(x,t|x_0,0) &=& e^{-{V(x)}/{\Upsilon}}\tilde{P}(x,t|x_0,0)e^{+{V(x_0)}/{\Upsilon}} \label{eq:spectral-prop2}
\end{eqnarray}
Any correlation function can then be expressed by using (\ref{eq:spectral-prop2}). An economic notation can be achieved by using Dirac bra-ket notation in terms of which e.g.
\begin{eqnarray}
	\tilde{P}(t,0)=\sum\limits_{n=0}^{\infty}\left|n\rangle e^{-E_n t}\langle n \right|
\end{eqnarray} 
Correlation functions can then be expressed in the spectral expansion as:  
 \begin{eqnarray}
 \left\langle f(x(t)) g(x(0))\right\rangle = \sum\limits_{n=0}^{\infty}\left\langle 0 \right| f \left| n \right\rangle e^{-E_n t} \left\langle n \right|g \left| {\mathrm in} \right\rangle \label{eq:spectral_correlation}
 \end{eqnarray}
where, explicitly 
\begin{eqnarray}
\left\langle 0 \right| f \left| n \right\rangle &=& \int\limits_{-\infty}^{\infty}dx\, p_0(x)f(x)p_n(x) \\
\left\langle n \right| g \left| {\mathrm in} \right\rangle &=& \int\limits_{-\infty}^{\infty}dx\, p_n(x)\,g(x) \tilde{P}(x,0) 
\end{eqnarray}  
Note that the ``out state'' in the stochastic problem is always $\langle 0 |$ and the ``in state'' is defined in terms of $\tilde{P}(x,t=0)$. 

Following standard procedures from quantum mechanics, we can write the conditional probability
\begin{eqnarray}
P(x,t|x_0,0)= \left \langle x\right| e^{-{V(x)}/{\Upsilon}} \tilde{P}(t,0)e^{+{V(x_0)}/{\Upsilon}}\left|x_0\right\rangle 
\end{eqnarray}  
governed by the above Euclidean Schr\"{o}dinger equation, as a path integral
\begin{eqnarray}
P(x,t|x_0,0) &=& \mathcal{N}\text{ exp}\left(\dfrac{\varepsilon}{2D}\left[V(x)-V(x_0)\right]\right)  \nonumber \\
&&\times \int\limits_{x(0)=x_0}^{x(t)=x}\mathcal{D}x(\tau) \text{ exp}\left( -\int \dfrac{d\tau}{2Dm} \left\{\dfrac{1}{2}m(\partial_{\tau}{x})^2 - \bar{U}(x)\right\}\right) \label{eq:PI F-P}
\end{eqnarray}
where we have reinstated the dimensionful variables. We therefore recover the ``on mass-shell" path integral (\ref{eq:dim-action}) obtained earlier. Note the importance of including the determinant (\ref{eq:det-Alg}) in order to obtain the $\partial_{xx}V$ term in the Schr\"{o}dinger potential $\bar{U}$. 

\section{\label{sec:Stoch_conc}Conclusion}
\acresetall 
In this chapter we have accomplished several things. Firstly we introduced the fundamental problem that will be the focus of this thesis, namely the behaviour of the one-dimensional overdamped Langevin equation (\ref{eq:langevindimless}). We showed how this stochastic processes can be expressed in terms of an object called a path integral that can broadly be understood as a weighted sum over different possible stochastic trajectories. This path integral is equivalent to an action describing Euclidean SuperSymmetric Quantum Mechanics and we explored the impact this symmetry has on the behaviour of correlators. We also defined objects known as generating functionals which will later allow us to compute correlation functions. Most important of these is the \ac{EA} $\Gamma$ which is related to the classical action once all fluctuations have been integrated out (\ref{eq:Gamma_S_relation}). Finally we described an alternative formulation of the problem in terms of a \ac{F-P} equation (\ref{eq: F-P}) and demonstrated how correlators can be computed from this using a spectral expansion method.  
\chapter{The Renormalisation Group} 
\label{cha:fRG} 
\acresetall 
\vspace{0.5cm}
\begin{flushright}{\slshape    
How is it that our effective theories at different scales are so \\compelling as to
make physicists think they are gods?\\ The answer is that, like Aesop's mouse, they
walk in front of a lion,\\ and the lion is renormalisation.} \\ \medskip
--- \defcitealias{Huang1987}{Kerson Huang}\citetalias{Huang1987} \citep{Huang1987}
\end{flushright}
\vspace{0.5cm}
\section{Introduction}
The first area of physics that most people study is that of simple Newtonian Mechanics. Newton's three laws of motion describe very well many aspects of everyday life, from why it feels like your chair is pushing against you when you sit in it, to precisely how hard to hit the white ball with your cue to get the ball in the pocket. A strong candidate for the most famous equation in physics, $F=ma$, has incredibly wide applicability. Much of high school physics can be dealt with by careful use of this simple formula. Why should this be so? A well known complaint by many a student is that at the start of a new academic year their teacher will say how everything they learnt the previous year is actually wrong and this year they'll learn the ``real" physics. Why do we let teachers the world over lie to their students? The answer of course is that we don't. While certain justifications for why a particular equation is valid don't hold up to  scrutiny -- ``because the textbook says so" is unconvincing to anyone -- the equation itself is still valid \textit{in the right context}. It seems perfectly natural that we don't need the Schr\"{o}dinger equation to describe how a ball bounces off a wall but why should this be the case? We know our theories are incomplete -- we do not have a consistent theory of quantum gravity -- so why are we able to predict anything at all? Why are we able to make incremental improvements on our theories that we teach year on year? The answer is that all of physics is really made up of \textit{Effective Field Theories}.

An \ac{EFT} is a framework for describing physical phenomena in a certain range of validity. Typically this would correspond to being valid within a certain energy regime. More intuitively it might correspond to a certain lengthscale. In this way you could then see how Newtonian Physics is an \ac{EFT}\footnote{We are being a bit loose with terminology here as Newtonian mechanics is not usually formulated as a \textit{field} theory and for it to be a valid limit of a relativistic description one must assume small velocities.} that is valid over the length scales we typically associate with everyday life. For physics much smaller we must invoke quantum mechanics and for physics much larger we turn to \ac{GR}. One is then invited to ask how might we move between these descriptions? How might we view one as a limit of another? The \ac{RG} is a way of moving between these effective theories in a consistent way. Introducing it and applying it to stochastic processes is the subject of this chapter.

The \ac{RG} was brought to full force through the work of K. Wilson \cite{Wilson1983} who used it to understand phase transitions and since then the \ac{RG} has become a widely used technique in modern physics with many applications in both particle physics \cite{Peskin:1995ev} and condensed matter physics \cite{Chaikin1995}. 
The \ac{RG} is relevant whenever fluctuations significantly influence the state (static or dynamical) of a physical system. Its conceptual framework as applied in condensed matter physics is perhaps most apt for describing the goal in this work: the \ac{RG} interpolates between a small lattice size, where the underlying physics is known, to a much larger lattice size by including the effect of fluctuations on all intermediate length scales, providing an \textit{effective} picture that averages over all such fluctuations. In this chapter we apply this idea to the stochastic dynamics of a Brownian particle. For us the small lattice size corresponds to a small fundamental timescale over which the dynamics is adequately described by the Langevin equation (\ref{eq:langevin}). We seek an effective description, valid over much longer timescales, that captures the aggregate effect of fluctuations. The effective description is embodied in an \ac{EA} $\Gamma[\chi(t)]$ of the average position $\chi(t) \equiv \langle x(t) \rangle$. In particular, one can use the \ac{EA} to compute $n$-point correlation functions of the particle's position $\langle x(t_1) x(t_2)\ldots x(t_n) \rangle $, characterising the system's statistical properties. We will defer how to obtain the correlators from $\Gamma $ until chapter \ref{cha:fRG EEOM} and here focus on how to compute $\Gamma $ in the first place. To obtain this effective long-time behaviour we will use a version of the \ac{RG} known as the functional or exact or non-perturbative Renormalisation Group. 

In this chapter we will first introduce the \ac{FRG} in section \ref{sec:fRG intro} with a simple, one dimensional example. We will introduce the concept of a regulator and derive the well known Wetterich equation \cite{Wilson1983, Morris1994}. We will also cover the truncation scheme that will be used extensively in this work, namely the derivative expansion. Section \ref{sec:BM and fRG} is the start of our original research and discusses applying the \ac{FRG} technology to \ac{BM} and deriving the appropriate flow equations. We also include a discussion of the ``equilibrium flow''. In section \ref{sec:FRG SOL} we solve these flow equations for a few different potentials and comment on the behaviour seen. We summarise our results for this chapter in section \ref{sec:fRG_conc}. 

For those not interested in the technical details we refer the reader to the main results of this chapter:
\begin{itemize}
    \item Fig.~\ref{fig:fRG_fluctations} for a schematic picture of how the \ac{FRG} coarse-grains in time and Fig.~\ref{fig:DW_simple} for how this can smoothen out features in the potential. 
    \item Equations (\ref{eq:dV/dk}) and (\ref{eq:WFR_flow_eqn}) for the \ac{FRG} flow equations in the \ac{LPA} and \ac{WFR} approximations respectively for \ac{BM}.
    \item Fig.~\ref{fig: X2GaussMany_LPA_V_5T.png} for how the \ac{FRG} smoothens out highly complicated potentials for \ac{BM} and Fig.~\ref{fig: X2GaussMany_WFR_Zetax.png} for how in the \ac{WFR} it adds features to the parameter $\zeta_x$.
\end{itemize}

\section{\label{sec:fRG intro}The Functional Renormalisation Group}
In this section we will introduce a particular formulation of the \ac{RG} known as the \ac{FRG} \cite{Wetterich1993, Morris1994} -- see \cite{Berges2002} for a review and an entry point to the literature on the subject, \cite{Dupuis2020} for a comprehensive overview of applications as well as e.g. \cite{Gies2012, Delamotte2012} for more elementary introductions. It has many advantages over the original Wilsonian treatment, the most obvious is its ability to handle systems with strong couplings. As the name suggests, the \ac{FRG} uses \textit{functional methods} to describe the computation correlation functions of the fields. This is typically done through the use of \textit{generating functionals} which in principle should contain all relevant physical information about a theory. In section \ref{sec:gen_func} we introduced some examples of these for the \ac{BM} problem at hand. Before we apply the \ac{FRG} to \ac{BM} in section (\ref{sec:BM and fRG}) we will examine a much simpler one-dimensional system. \\

\begin{sidewaysfigure}
    \centering
    \includegraphics[width=.98\textwidth]{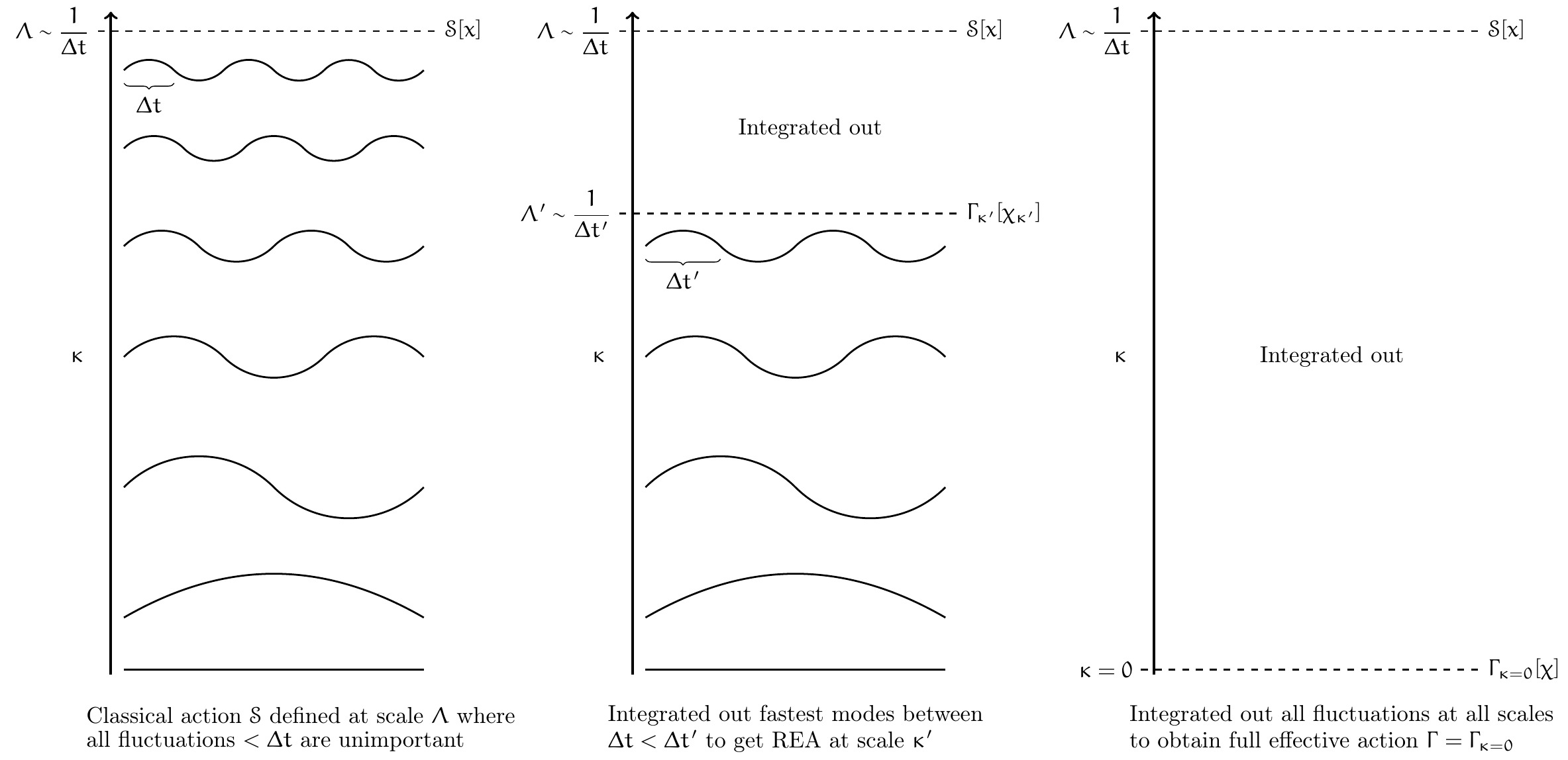}
    \caption[Schematic for how the \ac{FRG} works]{Schematic for how the \ac{FRG} works.}
    \label{fig:fRG_fluctations}
\end{sidewaysfigure}

We consider a very simple classical action corresponding to a particle evolving in a potential $V(x)$: 
\begin{eqnarray}
\mathcal{S}[x] = \int dt~\left[ \dfrac{1}{2}\dot{x}^2 - V(x)\right] \label{eq:classi_act_examp}
\end{eqnarray}
and recall that the generating functional we are most interested in is the \ac{EA} $\Gamma$:
\begin{eqnarray}
e^{-\Gamma[\chi]} = \int \mathcal{D}x~e^{-\mathcal{S}[x]} \label{eq:effect_class_comp}
\end{eqnarray}
i.e. it is equivalent to integrating out all fluctuations weighted by the classical action. Recall that the argument of $\Gamma$ is the mean field $\chi \equiv \lan x\ran$ averaged over all fluctuations. Computing this directly is impossible for all but the simplest potentials. Instead we ask ourselves first what we really mean by a \textit{classical} action $\mathcal{S}$ in the first place? The principle of least action states that the variational derivative of the action vanishing, i.e. $\delta\mathcal{S} = 0$, will yield the well known Euler-Lagrange equations:
\begin{eqnarray}
\dfrac{\delta S}{\delta x} = \dfrac{\partial L}{\partial x} - \dfrac{\mathrm{d}}{\mathrm{d}t}\dfrac{\partial L}{\partial \dot{x}} = 0
\end{eqnarray}
which for our very simple example (\ref{eq:classi_act_examp}) comes out to $F=ma$:
\begin{eqnarray}
\dfrac{\delta S[x]}{\delta x} = - \dfrac{dV(x)}{dx} - m \ddot{x} = 0 \Rightarrow  - \dfrac{dV(x)}{dx} = m \ddot{x} \label{eq:class_EOM}
\end{eqnarray}
However this equation of motion does not tell the full story, it is only valid at a certain energy scale $\Lambda$ often called the cutoff. Below this scale there are additional fluctuations\footnote{The exact nature of these fluctuations i.e. whether they are thermal or quantum is not important to this discussion.} which will modify the dynamics of (\ref{eq:class_EOM}) in some non-trivial way that must be taken into account. We can imagine therefore that the classical action fully describes the dynamics if one ignores all fluctuations with frequency/momentum $\kappa < \Lambda \sim 1/\Delta t$. Phrased another way we could say that all fluctuations that occur on a timescale $<\Delta t$ are already taken into account/don't importantly affect the dynamics of (\ref{eq:class_EOM}) -- this is what is represented in the leftmost plot of Fig.~\ref{fig:fRG_fluctations}. \\
With this in mind we introduce the \ac{REA} $\Gamma_{\kappa}[\chi]$ which modifies the \ac{EA} so that it depends on the momentum scale ${\kappa}$. It does this by modifying the lower bound of integration in (\ref{eq:effect_class_comp}) to be ${\kappa}$ meaning that not all fluctuations have yet been integrated over. Given what we have just discussed, at scale ${\kappa} = \Lambda$ there are no fluctuations to integrate over and equation (\ref{eq:effect_class_comp}) simply becomes:
\begin{eqnarray}
e^{-\Gamma_{{\kappa} = \Lambda}[\chi_{\kappa}]} = \int_{\Lambda} \mathcal{D}x~e^{-\mathcal{S}[x]} \approx e^{-\mathcal{S}[x]} \Rightarrow \Gamma_{{\kappa} = \Lambda}[\chi_{\kappa}] \approx \mathcal{S}[x]
\end{eqnarray}
telling us the equivalence of the \ac{REA} and the classical action at this scale. It is important to note that the mean field $\chi$ now has an implicit ${\kappa}$ dependence as the average is now only over fluctuations down to momentum scale ${\kappa}$, for ${\kappa} = \Lambda$ this means that $\chi  \approx x$. All this gives us an ``initial condition" for the \ac{REA}. Another way of expressing this ${\kappa}$ dependence of $\chi$ is by imagining we have split it into long timescale (slow) and short timescale (fast):
\begin{eqnarray}
\chi = \chi_{>} + \chi_{<} \label{eq:chi_slow_and_fast}
\end{eqnarray}
respectively and ${\kappa}$ tells us how we make this split. For instance at ${\kappa} = \Lambda$ the fast modes $\chi_{<}$ are those that occur on a timescale $<\Delta t$ and have already been incorporated into the action $\mathcal{S}[x]$ and the slow modes $\chi_{>}$ are the ones still present as shown on the left plot of Fig.~\ref{fig:fRG_fluctations}. As the \ac{REA} method involves computing the Gibbs free energy of the fast modes already accounted for it is clear that $\chi_{\kappa} \equiv \chi_{<}$ and why at ${\kappa}=\Lambda$ this is equivalent to $x$.\\

\begin{figure}[t!]
\centering
\includegraphics[width=.98\textwidth]{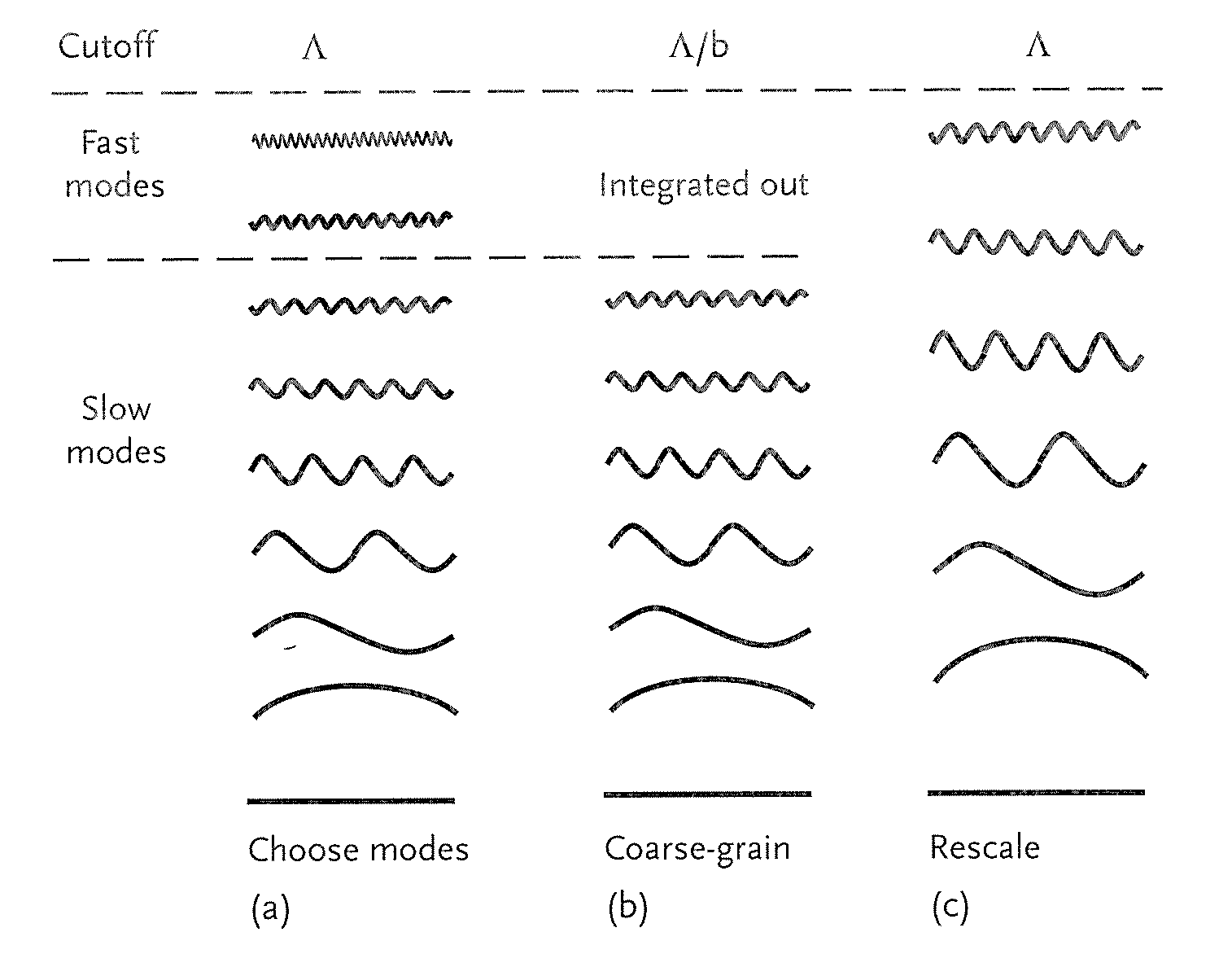}
\caption[The Kadanoff-Wilson `philosophy' of the Renormalisation Group]{An illustration of the Kadanoff-Wilson `philosophy' of the \ac{RG}, taken from \cite{Huang1987}.}
\label{fig: KW_fluctuations}
\end{figure}

We can imagine now wanting to determine the \ac{REA} at another momentum scale ${\kappa}' < {\kappa}$ corresponding to fluctuations occurring at timescale $\Delta t' > \Delta t$. In this way one would be capturing rarer and therefore slower fluctuations. The idea would then be to integrate out all fluctuations in the range ${\kappa}' < {\kappa}$ as shown in the middle plot of Fig.~\ref{fig:fRG_fluctations} which would then yield the \ac{REA} at scale ${\kappa}'$:
\begin{eqnarray}
e^{-\Gamma_{{\kappa}'}[\chi_{{\kappa}'}]} = \int_{{\kappa}'} \mathcal{D}x~e^{-\mathcal{S}[x]} 
\end{eqnarray}
where now the mean field has been averaged over all fluctuations down to ${\kappa}'$ and is distinct from $x$. At this scale the split between fast and slow modes now occurs at ${\kappa}'$ which means that modes which were previously in the ``slow" regime are now in the ``fast" regime and must be incorporated into the \ac{REA} $\Gamma$. This highlights the stark contrast between the \ac{FRG} approach and the original Kadanoff-Wilson idea where the object of interest is instead the Hamiltonian of the slow modes not yet integrated out $H[\chi_{>}]$. This is achieved by doing the split into short and fast modes, coarse-graining or integrating out the fast modes and then rescaling everything back to the original cutoff -- see Fig.~\ref{fig: KW_fluctuations}. This means that in the two descriptions ${\kappa}$ plays a subtly different role:
\begin{itemize}
\item In the Kadanoff-Wilson-Polchinski formulation ${\kappa}$ is a UV cutoff for the slow modes $\chi_{>}$
\item In the \ac{REA} method ${\kappa}$ is an IR cutoff for the fast modes $\chi_{<}$ since $\Gamma_{{\kappa}}$ is the free energy of the fast modes
\end{itemize}

If one continues this iterative process for the \ac{REA}, integrating out rarer and rarer fluctuations then eventually ${\kappa} = 0$ is reached and all fluctuations have been integrated out -- see rightmost plot of Fig.~\ref{fig:fRG_fluctations}. This means that the \ac{REA} at ${\kappa}=0$ is equivalent to the full \ac{EA} $\Gamma_{{\kappa}=0}[\chi] = \Gamma [\chi]$ and $\chi$ is the full mean field as expected.

\subsection{The regulator}
Having outlined the schematic picture of how the \ac{FRG} works let us make things a bit more precise. In order to obtain the \ac{REA} we must first regulate the other two generating functionals $\mathcal{Z}[J]$ \& $\mathcal{W}[J]$ like so:
\begin{eqnarray}
e^{\mathcal{W}_{\kappa}[J]} \equiv \mathcal{Z}_{\kappa}[J] \equiv \int \mathcal{D}x ~e^{-\mathcal{S}[x] -\Delta \mathcal{S}_{\kappa}[x] + \int Jx}
\end{eqnarray}
which we can see matches the definitions (\ref{eq:partition functional}) \& (\ref{eq:gen_func_W_defn}) from section (\ref{sec:gen_func}) except for the addition of a regulator term $\Delta \mathcal{S}_{\kappa}[x]$. This is usually defined to be quadratic in the field (in our case x):
\begin{eqnarray}
\Delta\mathcal{S}_{{\kappa}}[x] = \dfrac{1}{2}\int_{t,t'}x(t)R_{{\kappa}}(t,t')x(t') 
\end{eqnarray}
Crucially $R_{{\kappa}}$ is an IR regulator that depends on our Renormalisation scale ${\kappa}$ and the momentum $p$ of the modes
$R_{{\kappa}}$ is chosen in order to simplify following calculations. There are however some conditions imposed on the regulator term:
\begin{eqnarray}
R_{{\kappa}}(p) &\xrightarrow[p\gg {\kappa}]{\text{ }}& 0, \text{  leaves the UV modes unaffected}\nonumber\\
R_{{\kappa}}(p) &\xrightarrow[p\ll {\kappa}]{\text{ }}& {\kappa}^2, \text{  acts as large mass and freezes IR modes}\nonumber\\
R_{{\kappa}}(p) &\xrightarrow[{\kappa}\rightarrow \infty]{\text{ }}& \infty ,\text{  freezes all fluctuations when scale is large enough}\nonumber\\
R_{{\kappa}}(p) &\xrightarrow[{\kappa}\rightarrow 0]{\text{ }}& 0 ,\text{  Regulator vanishes allowing recovery of full theory}\nonumber\\ \label{eq:regulator conditions}
\end{eqnarray}
With these conditions in mind we can now more formally define the \ac{REA}:
\begin{eqnarray}
\Gamma_{{\kappa}}[\chi_{\kappa}] = \int J\chi_{\kappa} - \mathcal{W}_{{\kappa}}[J]-\Delta\mathcal{S}_{{\kappa}}[\chi_{\kappa}] \label{eq:REA_defn_formal}
\end{eqnarray}
where the regulated mean field $\chi_{\kappa}$ is defined analogously as in the full theory:
\begin{eqnarray}
\chi_{\kappa}(t) \equiv \lan x(t)\ran_{{\kappa},J} = \dfrac{\delta \mathcal{W}_{\kappa}[J]}{\delta J(t)} \label{eq:Regulated_mean_field}
\end{eqnarray}
It is worth mentioning that only the first two terms in (\ref{eq:REA_defn_formal}) are convex, for finite ${\kappa}$ any non-convexity of $\Gamma_{\kappa}$ is from the regulator term. Naturally in the limit ${\kappa}\rightarrow 0$, the regulator vanishes, $\Gamma_{{\kappa}=0} = \Gamma$ and the convexity is manifest.

\subsection{\label{sec:The Wetterich Equation}The Wetterich Equation}
\begin{figure}
    \centering
    \includegraphics[width = .75\linewidth]{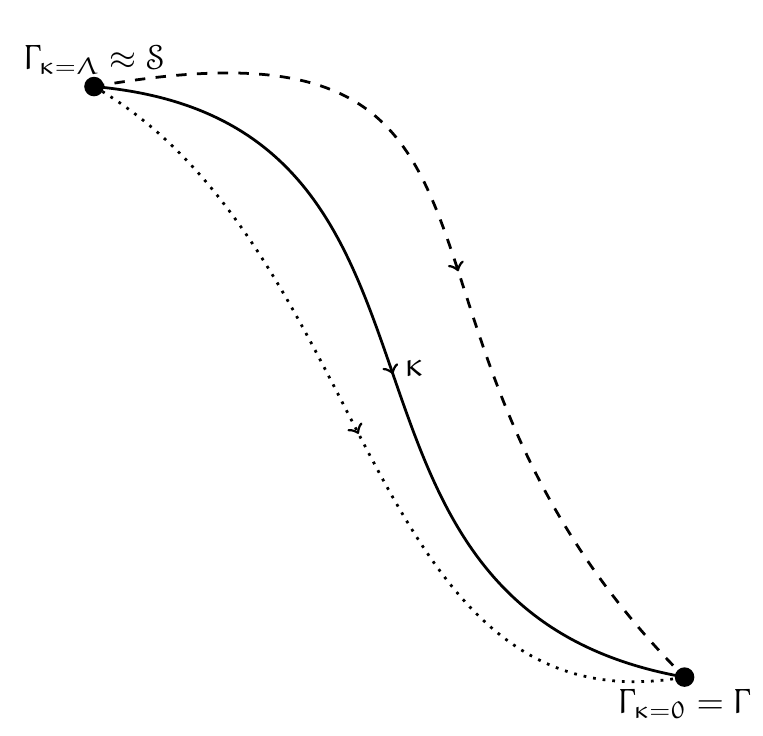}
    \caption[Trajectory of \ac{REA} in theory space]{The trajectory of the \ac{REA} according to the flow equation (\ref{eq:flow_equation}) in theory space for three different regulators $R_{\kappa}$. While the trajectories are different, they start and finish at the same point in theory space.}
    \label{fig:RG_flow_overview}
\end{figure}

We are now in a position to determine how the \ac{REA} varies with renormalisation scale ${\kappa}$. This will result in the well known Wetterich equation \cite{Wetterich1993,Morris1994} that describes the `flow' of the \ac{REA} between the microscopic and macroscopic scale. In our example microscopic corresponds to the small $\Delta t$ of the fluctuations at the cutoff $\Lambda$ and macroscopic corresponds to the increasing timescale of fluctuations as ${\kappa}\rightarrow 0$ and they are integrated out. The definition of $\Lambda \sim 1/\Delta t$ is analogous to the condensed matter interpretation of the cutoff being inversely proportional to the lattice size, the only difference here being that the condensed matter lattice is in space and ours is in time. \\
If we take functional derivatives of the \ac{EA} $\Gamma$ we get what is sometimes called the quantum equation of motion in analogy with the classical equations of motion\footnote{This suggests how we will later obtain \emph{effective} equations of motion in chapter \ref{cha:fRG EEOM}} (\ref{eq:class_EOM}) obtained by taking functional derivatives of the classical action. We can do the same thing to our \ac{REA} to get a modified equation of motion:
\begin{eqnarray}
J(t) = \dfrac{\delta \Gamma_{\kappa} [\chi]}{\delta \chi(t)} + (R_{\kappa}\chi )(t)
\end{eqnarray}
Here we have suppressed the ${\kappa}$ dependence of $\chi$ to avoid cluttered notation and will continue to do so in the rest of the text. We can take another functional derivative to obtain:
\begin{eqnarray}
\dfrac{\delta J (t)}{\delta \chi(\tau)} = \dfrac{\delta^2 \Gamma_{\kappa}[\chi]}{\delta \chi(t)\delta \chi(\tau)} + R_{\kappa}(t,\tau)
\end{eqnarray}
If however we take a functional derivative of (\ref{eq:Regulated_mean_field}) we can define the regulated propagator $G_{\kappa}$:
\begin{eqnarray}
\dfrac{\delta \chi(\tau )}{\delta J(t')} = \dfrac{\delta^2 \mathcal{W}_{\kappa}[J]}{\delta J (t') \delta J(\tau)} \equiv G_{\kappa} (\tau - t')
\end{eqnarray}
We can then combine the previous two equations to give the important identity:
\begin{eqnarray}
\delta (t-t') &=& \dfrac{\delta J(t)}{\delta J(t')} = \int \mathrm{d}\tau \dfrac{\delta J (t)}{\delta \chi(\tau)} \dfrac{\delta \chi(\tau )}{\delta J(t')} \nonumber \\
&=& \int\mathrm{d}\tau (\Gamma_{{\kappa}}^{(2)}[\chi] + R_{\kappa})(t,\tau)G_{\kappa}(\tau - t')
\end{eqnarray}
where we have introduced the notation:
\begin{eqnarray}
\Gamma_{{\kappa}}^{(n)}[\chi] = \dfrac{\delta^n \Gamma_{\kappa}[\chi]}{\delta \chi \dots\delta \chi}
\end{eqnarray}
In operator notation we can now write:
\begin{eqnarray}
\mathbb{1} = (\Gamma_{{\kappa}}^{(2)}[\chi] + R_{\kappa})G_{\kappa} \label{eq:operater_REA}
\end{eqnarray}
We now have all we need to derive the Wetterich or flow equation. We begin by taking a $\kappa$ derivative with respect to $e^{W_{\kappa}[J]}$ while holding $J$ constant:
\begin{eqnarray}
\partial_{\kappa}W_{\kappa}[J] e^{W_{\kappa}[J]} &=& -\int \mathcal{D}x ~\partial_{\kappa}\Delta \mathcal{S}_{\kappa}[x]~e^{-\mathcal{S}[x] -\Delta \mathcal{S}_{\kappa}[x] + \int Jx} \\
&=& \lsb -\dfrac{1}{2}\int_{t,t'}\partial_{\kappa}R_{\kappa}(t,t') \dfrac{\delta}{\delta J(t)}\dfrac{\delta}{\delta J(t')} \rsb e^{W_{\kappa}[J]} \\
&=&  -\dfrac{1}{2}\int_{t,t'}\partial_{\kappa}R_{\kappa}(t,t') \lb \dfrac{\delta W_{\kappa}[J]}{\delta J(t)}\dfrac{\delta W_{\kappa}[J]}{\delta J(t')} + \dfrac{\delta^2 W_{\kappa}[J]}{\delta J(t)\delta J(t')}\rb e^{W_{\kappa}[J]} \\
\Rightarrow \partial_{\kappa}W_{\kappa}[J] &=&  -\dfrac{1}{2}\int_{t,t'}\partial_{\kappa}R_{\kappa}(t,t') \lb \chi(t)\chi(t') + G_{\kappa}(t,t')\rb \label{eq:Polchinski}
\end{eqnarray}
To derive the Wetterich equation we then take a derivative of (\ref{eq:REA_defn_formal}) with respect to the renormalisation scale ${\kappa}$ for a fixed $\chi$ and at $J = J_{sup}$:
\begin{eqnarray}
\partial_{\kappa}\Gamma_{\kappa}[\chi] &=& \int (\underbrace{\partial_{\kappa} J}_{= 0})\chi - \partial_{\kappa} \mathcal{W}_{\kappa}[J]\vert_{\chi} - \partial_{\kappa}\Delta\mathcal{S}_{\kappa}[\chi] \\
&\underbrace{=}_{(\ref{eq:Polchinski})}& \dfrac{1}{2}\int_{t,t'}\partial_{\kappa}R_{\kappa}(t,t')G_{\kappa}\\
&\underbrace{=}_{(\ref{eq:operater_REA})}& \dfrac{1}{2}\mathbf{Tr}\lsb \partial_{\kappa} R_{\kappa}\lb \Gamma_{{\kappa}}^{(2)}[\chi] + R_{\kappa} \rb^{-1}\rsb \label{eq:flow_equation}
\end{eqnarray}
with $\mathbf{Tr}$ being a shorthand for a trace over any internal indices and integrating over time\footnote{In higher dimensional systems this would be a a full integral over all spatial dimensions as well.}. Equation (\ref{eq:flow_equation}) is the aforementioned Wetterich or flow equation. Crucially it allows us to describe a trajectory in \textit{theory space} -- see Fig.~\ref{fig:RG_flow_overview}. Here we can see how the flow equation (\ref{eq:flow_equation}) tells us how to move between the initial condition of $\Gamma_{{\kappa} = \Lambda }\approx \mathcal{S}$ down to our desired $\Gamma$. It is worth noting however that (\ref{eq:flow_equation}) does not fix a unique trajectory in theory space as the regulator conditions (\ref{eq:regulator conditions}) actually leave the choice of the regulator $R_{\kappa}$ very open. Therefore different choices of $R_{\kappa}$ will correspond to different trajectories in theory space -- in Fig.~\ref{fig:RG_flow_overview} we have shown how three different choices of regulators give different flows even though they start and end at the same point. This is why the \ac{REA} is heavily regulator dependent even though the full \ac{EA} $\Gamma$ is not. 

\subsection{\label{sec:deriv_expansion_intro}The derivative expansion}
Even though we now have a flow equation for the \ac{REA} we still in practice need to consider a functional form of $\Gamma_{\kappa}$ before we can make progress. In this work we will be focusing on the so-called derivative or operator expansion of $\Gamma_{\kappa}$. Bearing in mind the form of our classical action (\ref{eq:classi_act_examp}) in our simple example this would give us the following truncated expression for $\Gamma_{\kappa}$:
\begin{eqnarray}
\Gamma_{\kappa}[\chi] = \int \mathrm{d}t ~V_{\kappa}(\chi) + \dfrac{1}{2} (\partial_t Z_{\kappa}(\chi ))^2 + \mathcal{O}(\partial^4) \label{eq:deriv_expansion}
\end{eqnarray}
We can see how this has the same form as (\ref{eq:classi_act_examp}) except we now have objects which are ${\kappa}$ dependent. The \ac{LO} term in the derivative expansion is the potential so this is the first object which becomes ${\kappa}$ dependent: $V(x) \rightarrow V_{\kappa}(\chi)$. If this is the only thing that varies with ${\kappa}$ then one is working with the \ac{LPA}. The \ac{NLO} term involves renormalising the kinetic term and is known as \ac{WFR}. Clearly $Z_{{\kappa} = \Lambda}(\chi ) = \chi$ to match the classical action. It can also be rewritten as a coefficient of the standard kinetic term: $(\partial_t Z_{\kappa}(\chi ))^2 \rightarrow (\partial_{\chi}Z_{\kappa}(\chi))^2\dot{\chi}^2 $ which is a bit easier to keep track of as opposed to a redefinition of the field itself. It is worth noting at this stage that truncating the derivative expansion at finite order means that $\Gamma_{{\kappa}=0}$ will become regulator dependent and will not necessarily coincide with the full \ac{EA} $\Gamma$. However it is believed that the \ac{LO} term in the derivative expansion is sufficient for $\Gamma_{{\kappa}=0} \approx \Gamma$ regardless of the specific choice of regulator with the \ac{NLO} term reducing the regulator dependence further.\\

If we focus on just the \ac{LPA} for now then it is clear that
\begin{eqnarray}
\Gamma^{(2)}_{{\kappa}} = (-\partial_t^2 + \partial_{\chi\chi}V_{\kappa}(\chi))\delta (t-t')
\end{eqnarray}
We now chose a regulator of the following form:
\begin{eqnarray}
R_{\kappa}(p) &=& \lb {\kappa}^2-p^2\rb \Theta\lb {\kappa}^2 - p^2\rb \\
\Rightarrow\partial_{\kappa} R_{\kappa}(p) &=& 2{\kappa}\Theta\lb {\kappa}^2 - p^2\rb
\end{eqnarray}
where $\Theta $ is the Heaviside step function. We now have all the appropriate ingredients to compute the right hand side of (\ref{eq:flow_equation}):
\begin{eqnarray}
\dfrac{1}{2}\mathbf{Tr}\lsb \partial_{\kappa} R_{\kappa}\lb \Gamma_{{\kappa}}^{(2)}[\chi] + R_{\kappa} \rb^{-1}\rsb &=& \dfrac{1}{2}\int_{-\infty}^{\infty}\dfrac{\mathrm{d}p}{2\pi}\dfrac{2{\kappa}\Theta ({\kappa}^2-p^2)}{{\kappa}^2 + \partial_{\chi\chi}V_{\kappa}(\chi))}\\
&=& \dfrac{1}{\pi}\dfrac{{\kappa}^2}{{\kappa}^2 + \partial_{\chi\chi}V_{\kappa}(\chi))}
\end{eqnarray}
where we are performing the trace in momentum space. We can therefore write the flow of the potential as:
\begin{empheq}[box = \fcolorbox{Maroon}{white}]{equation}
\partial_{\kappa} V_{\kappa} (\chi) = \dfrac{1}{\pi}\dfrac{{\kappa}^2}{{\kappa}^2 + \partial_{\chi\chi}V_{\kappa}(\chi))} \label{eq:flow_V_basic}
\end{empheq}
Equation (\ref{eq:flow_V_basic}) is the \ac{LPA} flow equation for the simple problem at hand for our choice of regulator. All that remains to be done is to solve it.
\subsection{Renormalising the doublewell}
\begin{figure}[t!]
    \centering
    \includegraphics[width = .75\linewidth]{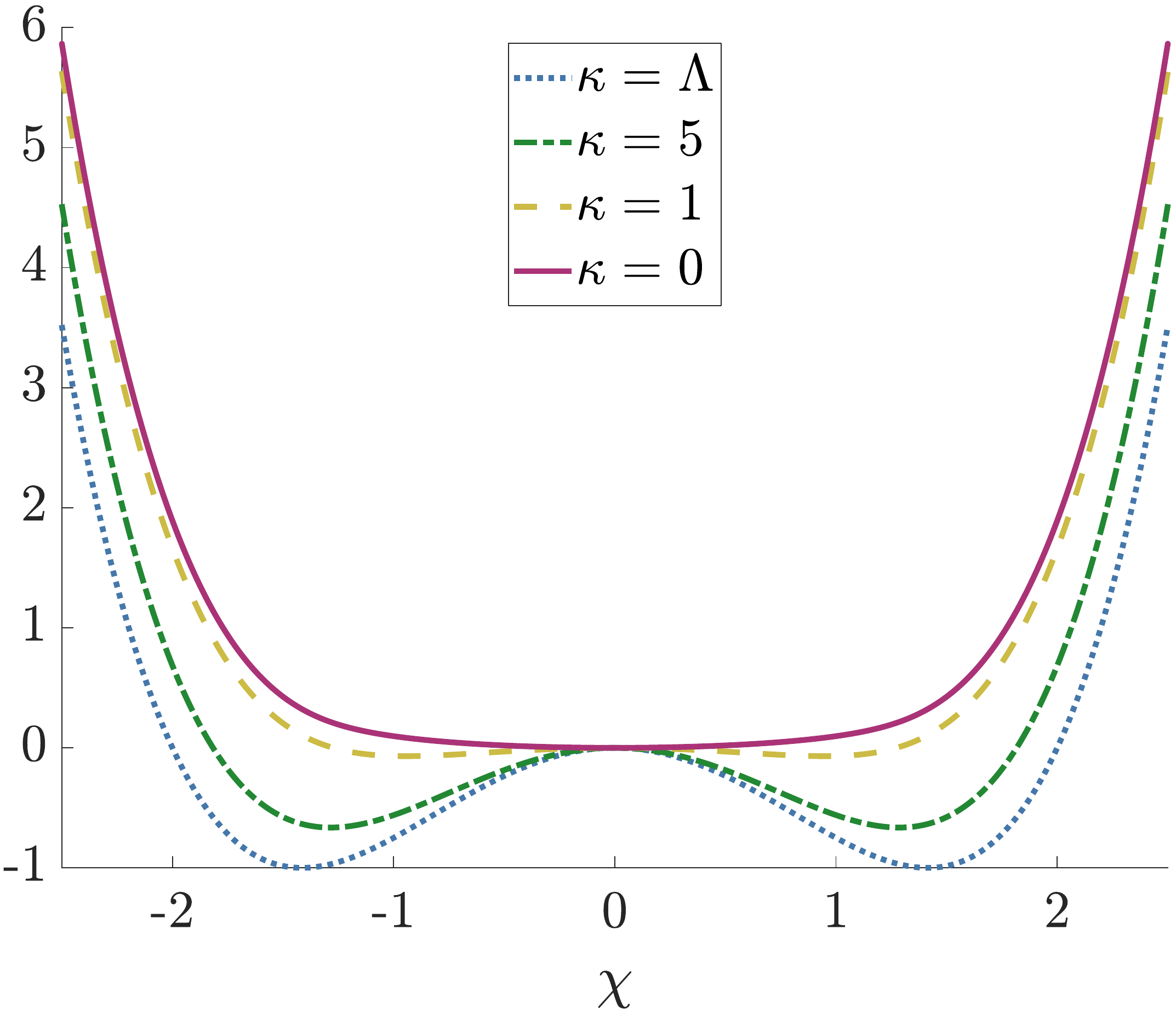}
    \caption[\ac{FRG} flow of doublewell in simple example]{The flow of the doublewell potential for our simple theory (\ref{eq:classi_act_examp}). The blue dotted curve corresponds to the ``bare" doublewell potential defined at the cutoff $\Lambda$. The \ac{FRG} flow equation (\ref{eq:flow_V_basic}) is solved down to $\kappa = 0$ (plotted by red solid line) and we have also plotted the flow at some intermediate values of $\kappa$. The curves have been vertically shifted by hand to coincide at $x = 0$ for visual clarity.}
    \label{fig:DW_simple}
\end{figure}
If we consider now a simple doublewell potential in our action (\ref{eq:classi_act_examp}):
\begin{eqnarray}
V(x) = -x^2 + \dfrac{x^4}{4}
\end{eqnarray}
then we can insert this as our initial condition for (\ref{eq:flow_V_basic}) i.e. $V_{{\kappa}=\Lambda}(\chi) = -\chi^2 + \chi^4/4$. The results from solving this numerically can be shown in Fig.~\ref{fig:DW_simple}. We can clearly see the original doublewell potential is given by the dotted blue ${\kappa} = \Lambda$ curve. As ${\kappa}$ is lowered we can see that the barrier in the potential gets smaller as rarer and rarer fluctuations are incorporated. These rarer fluctuations help to effectively reduce the size of the barrier until at ${\kappa} = 0$ (red solid curve) it has completely disappeared. This is as it should be as the convexity of the \ac{EA} $\Gamma$ ensures that the effective potential $V_{{\kappa} = 0}$ is also convex in the \ac{LPA}. 

\section{\label{sec:BM and fRG}Brownian Motion and the Functional Renormalisation Group}
Having demonstrated in section \ref{sec:fRG intro} how the \ac{FRG} works for a simple example we now see how it can be applied to our problem of \ac{BM}. As demonstrated in chapter \ref{cha:stochastic_processes}, our \ac{BM} problem is actually SuperSymmetric Quantum Mechanics. We can therefore apply the \ac{FRG} technology and incorporate the effect of thermal fluctuations by following the flow of the \ac{EA} $\Gamma_{\kappa}$ via the Wetterich equation which now involves a ``supertrace":
\begin{eqnarray}
	\partial_{{\kappa}}\Gamma_{{\kappa}}[X] = \dfrac{1}{2}\,{\mathrm STr}\left\{\int_{t,t'}\partial_{{\kappa}}R_{{\kappa}}(t,t')\left[R_{{\kappa}} + \Gamma_{{\kappa}}^{(2)}\right]^{-1}\right\} \label{eq:Wetterich functional}
\end{eqnarray}
which accounts for summing over the Bosonic and Fermionic degrees of freedom appropriately. Synatschke et. al have analysed a system with action $\mathcal{S}_{SUSY}$ in light of its underlying symmetries in \cite{Synatschke2009}. We adopt their results here. They find that from a SuperSymmetric perspective, the appropriate regulating term takes the form    
\begin{eqnarray}
\Delta\mathcal{S}_{\kappa}\! &=& \! \int_{\tau\tau'} \hspace{-0.2cm} r_2({\kappa},\Delta\tau)  \left[-\dot{\phi}(\tau)\dot{\phi}(\tau')+\mathcal{\tilde{F}}(\tau)\mathcal{\tilde{F}}(\tau') -i\bar{\psi}(\tau)\dot{\psi}(\tau')\right] \nonumber \\
&& \quad +~2ir_1({\kappa},\Delta\tau) \left[\phi(\tau)\mathcal{\tilde{F}}(\tau')-\bar{\psi}(\tau)\psi(\tau'))\right]
\end{eqnarray}
where $r_1$ and $r_2$ are two different regulators and $\Delta\tau \equiv \tau-\tau'$. Such a form was also suggested in \cite{Duclut2017}, however we will see that compatibility with the Boltzmann distribution suggests setting $r_{2} \rightarrow 0$. The flow equations of \cite{Synatschke2009} are discussed below. 

\subsection{\label{sec:LPA_BM_FLOW}Local Potential Approximation}
If we assume a derivative expansion as explained in section \ref{sec:deriv_expansion_intro} then the \ac{EA} takes the form:
\begin{eqnarray}
\Gamma_{{\kappa}}[\phi,\mathcal{\tilde{F}},\bar{\psi},\psi] = \int d\tau\bigg[\dfrac{1}{2}\dot{\phi}^2 + \dfrac{1}{2}\mathcal{\tilde{F}}^2 + i\mathcal{\tilde{F}}W_{{\kappa}, \phi}(\phi) - i\bar{\psi}\left(\partial_t + W_{{\kappa}, \phi\phi} \right)\psi \bigg] \label{eq:SUSYEffective}
\end{eqnarray}
such that $\Gamma_{{\kappa} = \Lambda} = \mathcal{S}_{SUSY}$ under the condition $W_{{\kappa} = \Lambda}(\phi) = W(\phi)$ with the mean fields $(\phi,\mathcal{\tilde{F}}, \psi, \bar{\psi})$ defined analogously to the full $\Gamma$ case (\ref{eq:SUSY_mean_fields}) although in a ${\kappa}$ dependent way as discussed around (\ref{eq:Regulated_mean_field}). In this approximation the only thing changing with ${\kappa}$ directly, progressively incorporating the effect of fluctuations on different timescales, is $W_{\kappa}$. This means we only have one flow equation to solve which turns out to be \cite{Synatschke2009}:
\begin{eqnarray}
\partial_{{\kappa}}W_{{\kappa}}(\phi) = \int_{-\infty}^{\infty}\dfrac{dp}{4\pi}\dfrac{(1+r_2)\partial_{{\kappa}}r_{1}- \partial_{{\kappa}}r_{2}~(r_{1} + \partial_{\phi\phi}W_{{\kappa}}(\phi))}{p^2+(r_{1} + \partial_{\phi\phi}W_{{\kappa}}(\phi))^2} \nonumber \\
\end{eqnarray}
We notice that if we set $r_2 = 0$ and choose a local-in-time $r_1({\kappa}, \delta\tau)={\kappa}\delta(\tau-\tau')$ the so-called Callan-Symanzik regulator then this choice\footnote{Physically speaking the final results should be independent of the regulator chosen. This is a subtlety we will not address in this work as it was shown in \cite{Synatschke2009} that even for other choice of regulators the difference in the final results was negligible, at least for the LPA.} effectively adds a quadratic term to the potential $W \rightarrow W + {\kappa}\phi^2$ and leads to a relatively simple flow equation: 
\begin{eqnarray}
\partial_{{\kappa}}W_{{\kappa}}(\phi) = \dfrac{1}{4}\cdot\dfrac{1}{{\kappa}+ \partial_{\phi\phi}W_{{\kappa}}(\phi)} \,.
\label{eq:dW/dk}
\end{eqnarray}
In terms of the physical variables we have 
\begin{empheq}[box = \fcolorbox{Maroon}{white}]{equation}
\partial_{{\kappa}}V_{{\kappa}}(\chi) = \dfrac{\Upsilon}{4}\cdot\dfrac{1}{{\kappa}+ \partial_{\chi\chi}V_{{\kappa}}(\chi)} \,,\label{eq:dV/dk}
\end{empheq}
which shows explicitly the effect of dialling the temperature $\Upsilon$: the higher the temperature the faster the flow as a result of stronger thermal fluctuations. Equation (\ref{eq:dV/dk}) can be discretised in the $\chi$ direction and become a set of coupled ODEs that can be solved in the ${\kappa}$ direction in order to obtain a numerical solution. The initial condition is, as discussed before, $V_{{\kappa}=\Lambda} = V$, and the boundary conditions at the edge of the $\chi$ is a one-sided derivative\footnote{For our bounded potentials we later solve for in section \ref{sec:FRG SOL} this is sufficient as the \ac{RG} flow effect is meant to become negligible as the gradient of the potential increases. If one wanted to work with an unbounded potential, e.g. Lennard-Jones, then the boundary conditions would need to be chosen more carefully.}.

It is important to note that equation (\ref{eq:dV/dk}) is identical to the flow of the effective potential that corresponds to the equilibrium Boltzmann distribution, see \cite{Guilleux2015} and section \ref{sec:equil-flow} with $R \rightarrow {\kappa}$. We  therefore see that the form of $\mathcal{S}_{SUSY}$ and deriving flow equations in a framework which respects its symmetries is crucial for establishing consistency with the equilibrium Boltzmann distribution. If one started  directly from the Onsager-Machlup functional (\ref{eq:dim-action}) and naively treated it as an $N=1$ Euclidean scalar theory in one-dimension with the combination $U = \frac{1}{2}\left(V_{,x}\right)^2 - \frac{\Upsilon}{2}V_{,xx}$ as the scalar potential to be evolved along the \ac{RG} flow, one would have obtained a different flow equation  
\begin{eqnarray}
\partial_{{\kappa}}U_{{\kappa}}(\phi) = \dfrac{1}{2}\int_{-\infty}^{\infty}\dfrac{dp}{2\pi}\dfrac{\partial_{{\kappa}}R_{{\kappa}}}{p^2+ R_{{\kappa}} + \partial_{\phi\phi}U_{{\kappa}}(\phi)}\,.
\end{eqnarray}
The corresponding Callan-Symanzik regulator would be  $R_{\kappa}={\kappa}^2$, giving
\begin{eqnarray}
\partial_{{\kappa}}U_{{\kappa}}(\phi) = \frac{1}{2} \frac{{\kappa}}{\sqrt{{\kappa}^2+\partial_{\phi\phi}U_{{\kappa}}(\phi)}}\,.
\end{eqnarray}
It is unclear how or if the end-of-the-flow potential $U_{{\kappa}=0}$ from this equation relates to the physical potential $V_{{\kappa}=0}$ and the flow appears a-priori incompatible with the Boltzmann distribution.   \\
\\
Before moving onto \ac{WFR} there is one final comment to make about the \ac{LPA} flow equation (\ref{eq:dV/dk}), namely the unphysical modification of the energy ground state. If one chose a constant potential $V = E$ then clearly the \ac{RG} flow should do nothing to the ground state energy $E$. However if we plug $V = E$ into (\ref{eq:dV/dk}) the flow is not zero:
\begin{eqnarray}
\partial_{\kappa} V_{\kappa} = \dfrac{\Upsilon}{4}\cdot\dfrac{1}{{\kappa}}
\end{eqnarray}
This can be solved analytically:
\begin{eqnarray}
V_{{\kappa} = 0} = E + \dfrac{\Upsilon}{4}\lsb \ln (\Lambda ) - \ln (0) \rsb
\end{eqnarray}
which gives not only a non-zero contribution but an infinite one! Therefore one should in principle regularise this unphysical divergence like so:
\begin{eqnarray}
\partial_{{\kappa}}V_{{\kappa}}(\chi) = \dfrac{\Upsilon}{4}\lsb \dfrac{1}{{\kappa}+ \partial_{\chi\chi}V_{{\kappa}}(\chi)} -\dfrac{1}{{\kappa}}\rsb \label{eq:dV/dk_regularised}
\end{eqnarray}
In reality however for our potentials of interest the \ac{RG} flow only introduces a small minor vertical shift which no quantities of interest in this work will depend upon. Given the difficulty of solving numerically the regularised version (\ref{eq:dV/dk_regularised}) we instead work with the original flow equation (\ref{eq:dV/dk}) and will later on vertically shift the curves by hand for legibility.
\subsection{\label{sec:WFR_BM_FLOW}Wave Function Renormalisation}

In the previous subsection we assumed that the \ac{REA} $\Gamma_{\kappa}$ only depends on the renormalisation scale through the form of the potential. We now allow for the field $\varphi$ itself to be renormalised which results in a scaling of the kinetic term. The new \ac{REA} in the \ac{SUSY} formalism is \cite{Synatschke2009}:
\begin{eqnarray}
\Gamma_{{\kappa}}[\phi,\bar{\psi},\psi] &=& \int dt~\dfrac{1}{2}Z_{,\phi}^{2}\dot{\phi}^{2} + \dfrac{1}{2}\left(\dfrac{W_{,\phi}}{Z_{,\phi}}\right)^2 \nonumber \\
&&- i\bar{\psi}\left(Z_{,\phi}^2 \partial_{t} + Z_{,\phi} Z_{,\phi\phi}\dot{\phi} - Z_{,\phi\phi} \dfrac{W_{,\phi}}{Z_{,\phi}} + W_{,\phi\phi}\right)\psi \nonumber \\
\end{eqnarray}
where we have suppressed the explicit dependence on ${\kappa}$ of $W$ \& $Z$ to avoid overly cluttered notation. From now on we will in general drop this explicit dependence on ${\kappa}$ for $W$, $V$, $Z$ \& $\zeta$, defined below, only restoring it when we are directly comparing it to the original cutoff value. We introduce an additional identification in addition to (\ref{eq:LPAidentifications}):
\begin{eqnarray}
\zeta(x) = \sqrt{\Upsilon} Z(\phi) \Rightarrow \zeta_{,x} = Z_{,\phi}, \quad \bar{c}c = -i\zeta_{,x}\bar{\psi}\psi
\end{eqnarray}
such that the (on-shell) \ac{REA} for \ac{BM} is now written as:
\begin{eqnarray}
\Gamma_{{\kappa}}[\chi,\bar{c},c] = \int dt~\dfrac{1}{2\Upsilon}\zeta_{,\chi}^{2}\dot{\chi}^{2} + \dfrac{1}{2\Upsilon}\left(\dfrac{V_{,\chi}}{\zeta_{,\chi}}\right)^2 - \bar{c}\left(\zeta_{,\chi}^2 \partial_{t} + \zeta_{,\chi} \zeta_{,\chi\chi}\dot{\chi} - \zeta_{,\chi\chi} \dfrac{V_{,\chi}}{\zeta_{,\chi}} + \cdot V_{,\chi\chi}\right)c \nonumber \\
\end{eqnarray}
The regulator term becomes more complicated for this action and we do not reproduce it here, see \cite{Synatschke2009} for details of this. Following their approach one arrives at the \ac{LPA} + \ac{WFR} flow equations: 
\begin{subequations}
\begin{empheq}[box = \fcolorbox{Maroon}{white}]{align}
\partial_{\kappa }V_{\kappa}(\chi ) &= \dfrac{\Upsilon}{4}\cdot\dfrac{1}{\kappa+ \partial_{\chi \chi}V_{\kappa}(\chi)} \label{eq:dV/dktilde2} \\
\partial_{\kappa}\zeta_{,\chi} &= \dfrac{\Upsilon}{4}\cdot\dfrac{\mathcal{P}}{\zeta_{,\chi}\cdot\mathcal{D}^2 } \label{eq:dzetax/dktilde}\\
\mathcal{D} &\equiv  V_{,\chi\chi} + {\kappa}\,\zeta_{,\chi}^{2}, \quad \mathcal{P} \equiv \dfrac{4\zeta_{,\chi\chi} V_{,\chi\chi\chi}}{\mathcal{D}} - \left( \zeta_{,\chi\chi}\zeta_{\chi}\right)_{,\chi} - \dfrac{3\zeta_{,\chi}^{2}V_{,\chi\chi\chi}^{2}}{4\mathcal{D}^2} 
\end{empheq}\label{eq:WFR_flow_eqn}
\end{subequations}
which now consist of the previous \ac{LPA} equation for the effective potential (\ref{eq:dV/dk}) as expected, augmented by one more flow equation for the wavefunction renormalisation $\zeta_{,\chi}$. 

As before we will integrate the \ac{LPA} equation (\ref{eq:dV/dk}) by discretising along the $\chi$ direction and solving the resulting set of coupled ODEs in ${\kappa}$. Once the effective potential $V_{\kappa}(\chi)$ has been obtained the second PDE can be solved for $\zeta_{,\chi}$ in a similar way. It is worth pointing out here that our approach differs slightly from \cite{Synatschke2009} in that the effective potential obeys the same equation as in the \ac{LPA} approximation even with the inclusion of \ac{WFR}.\footnote{For the \ac{WFR} approximation the authors of \cite{Synatschke2009} use a spectrally adjusted regulator which is evaluated on a background field $\bar{\phi}$. They make the simple choice of identifying this background field with the fluctuation field (i.e. $\bar{\phi} = \phi$). This approach however modifies the flow of $V_{\kappa}$ -- i.e. equation (\ref{eq:dV/dktilde2}) differs from the \ac{LPA} version (\ref{eq:dV/dk}) -- which means the flow no longer correctly approaches the Boltzmann equilibrium distribution's effective potential and leads to deviations from the correct equilibrium position and variance. The only choice of $\bar{\phi}$ that prevents this from happening is one where $Z_{{\kappa}}'(\bar{\phi}) = 1$ for all ${\kappa}$ which is what we have done here.} This is because the equilibrium state is described exactly by the \ac{LPA} equation \cite{Guilleux2015,Guilleux2017, Moreau2020a}, as we mentioned above and explicitly recall in section \ref{sec:equil-flow}. The \ac{LPA} flow equation was first solved in \cite{Synatschke2009, Guilleux2015, Guilleux2017}, while more recently \ac{WFR} was included for a double well potential in \cite{Moreau2020a}.  \\
Before we conclude this chapter with some specific solutions to the flow equations (\ref{eq:dV/dk}) \& (\ref{eq:WFR_flow_eqn}) we outline how the \ac{RG} can tell us explicit details of equilibrium quantities. 

\subsection{\label{sec:equil-flow} The equilibrium flow equation} 

In equilibrium, all equal-time expectation values can be generated by the generating function \\
\begin{eqnarray}\label{equil1}
\mathcal{Z}(J)=\int dx  \,  e^{-2 V(x)/\Upsilon + Jx} 
\end{eqnarray}
in a manner directly analogous to that described in section \ref{sec:gen_func} but with functional derivatives replaced by ordinary derivatives w.r.t. $J$. In a spirit identical to the renormalisation group but in the simpler setting of one degree of freedom, we can define a modified generating functional \cite{Guilleux2015}
\begin{eqnarray}
\mathcal{Z}_{{\kappa}}(J)=\int dx ~ e^{-2 V(x)/\Upsilon - \frac{1}{2}R({\kappa})x^2+Jx}
\end{eqnarray}    
with an additional quadratic term controlled by an arbitrary function $R({\kappa})$ of a parameter ${\kappa}$, satisfying $\lim\limits_{{\kappa} \rightarrow 0} R({\kappa}) = 0$, giving back the original $Z(J)$. Correlation functions are generated by $W_{\kappa}(J)=\ln Z_{\kappa}(J)$ via
\begin{eqnarray}\label{app:correlations}
\chi_{\kappa} \equiv \langle x\rangle_{\kappa}=\frac{\partial W_{\kappa}(J)}{\partial J}\,,\quad \langle x^2\rangle_{\kappa} - \chi^2_{\kappa}=\frac{\partial^2 W_{\kappa}(J)}{\partial J^2}
\end{eqnarray}     
e.t.c. In the limit ${\kappa}=0$ and after setting $J=0$ the usual predictions of the equilibrium Boltzmann distribution are recovered.   

The source $J$ has been considered as an external, independent variable controlling expectation values such as $\chi$ and higher correlators. One could also consider $\chi$ as the independent variable, solving $\chi = \partial W/\partial J$ for $J(\chi)$ and defining the effective potential $U(\chi)$ via a Legendre transform 
\begin{eqnarray}
\Gamma_{\kappa}(\chi) + W_{\kappa}(J) = J\chi -\frac{1}{2}R({\kappa})\chi^2
\end{eqnarray}  
with
\begin{eqnarray}
\Gamma(\chi) \equiv 2 U(\chi)/\Upsilon \label{eq:equi_effective_potential}
\end{eqnarray}            
Note that 
\begin{eqnarray}\label{Legendre}
\frac{\partial \Gamma_{\kappa}}{\partial \chi } = J_{\kappa} - R({\kappa})\chi
\end{eqnarray}
implying that the minimum of the effective potential defines the equilibrium expectation value of $x$ (at $J=0$ and ${\kappa}=0$).

The dependence of the generating function $W_{\kappa}(J)$ on ${\kappa}$ can be easily obtained as   
\begin{eqnarray}
\partial_{\kappa} W_{\kappa}(J)=-\frac{1}{2}\partial_{\kappa}R\left[\frac{\partial^2 W_{\kappa}(J)}{\partial J^2}+\left(\frac{\partial W_{\kappa}(J)}{\partial J}\right)^2\right]
\end{eqnarray}
which is an ``\ac{RG} equation'' for $W_{\kappa}(J)$. We can also obtain an equation determining how $\Gamma_{\kappa}(\chi)$ runs with ${\kappa}$. Reciprocally, taking $\chi$ as the independent variable, $J$ becomes a function of $\chi$ and ${\kappa}$. Taking a ${\kappa}$ derivative of (\ref{Legendre}) at fixed $\chi$ we obtain 
\begin{eqnarray}
\partial_{\kappa} \Gamma_{\kappa}(\chi) = \frac{1}{2}\partial_{\kappa} R \frac{\partial^2 W_{\kappa}}{\partial J^2} 
\end{eqnarray}            
To express the RHS in terms of $\Gamma_{\kappa}(\chi)$, consider the first relation of (\ref{app:correlations}). Taking a $\chi$ derivative we find 
\begin{eqnarray}
\left(\frac{\partial^2\Gamma_{\kappa}}{\partial\chi^2}+R\right)\frac{\partial^2 W_{\kappa}}{\partial J^2} =1
\end{eqnarray}   
Hence, the ``\ac{RG} flow'' of $\Gamma$ is determined by 
\begin{eqnarray}\label{app:Gamma-flow}
\partial_{\kappa}\Gamma_{\kappa}(\chi)=\frac{1}{2}\partial_{\kappa}R\left(\frac{\partial^2\Gamma}{\partial\chi^2}+R \right)^{-1}
\end{eqnarray} 
Note also that, at ${\kappa} \rightarrow 0$ 
\begin{eqnarray}
\dfrac{\delta^2 \Gamma}{\delta \chi \delta \chi} = \dfrac{2\partial_{\chi\chi}U}{\Upsilon}
\end{eqnarray}
from (\ref{eq:equi_effective_potential}). Also as $\Gamma^{(2)}$ and the two point function $G$ are inverse to each other we obtain:
\begin{eqnarray}
\langle x^2 \rangle - \chi^2=\frac{\Upsilon}{2 \, \partial_{\chi\chi} U(\chi_{eq})} \label{eq:equi_flow_var}
\end{eqnarray}  
and hence the variance at equilibrium is determined by the curvature of the effective potential around its minimum. 

All the above manipulations can be generalised to many or even infinite degrees of freedom and continuum actions, leading to the Wetterich equation (\ref{eq:Wetterich functional}), which is directly equivalent to (\ref{app:Gamma-flow}), and the relations of section  (\ref{sec:1-point function}). For this work it is important to note that the equilibrium effective potential $U(\chi)$ discussed here obeys the \ac{LPA} flow equation \emph{exactly} if we choose $R({\kappa})={\kappa}$.         
\section{\label{sec:FRG SOL}Solutions to the Brownian Motion flow equations}
In this section we will solve the flow equations (\ref{eq:dV/dk}) \& (\ref{eq:WFR_flow_eqn}) down to ${\kappa}=0$ for a range of interesting potentials. We will consider five different potentials:\\
{\color{RoyalBlue}$i)$} A simple polynomial:
\begin{eqnarray}
V(x) = x + \dfrac{x^2}{2} + \dfrac{2x^3}{3} + \dfrac{x^4}{4} \label{eq:Vpoly}
\end{eqnarray}
{\color{RoyalBlue}$ii)$} The doublewell with unit depth:
\begin{eqnarray}
V(x) = -x^2 + \dfrac{1}{4}x^4 \label{eq:Vdoublewell}
\end{eqnarray}
{\color{RoyalBlue}$iii)$}
A doublewell made by two Lennard-Jones (LJ) potentials back to back:
\begin{eqnarray}
V(x) = 4\left(\dfrac{1}{(x+3)^{12}}-\dfrac{1}{(x+3)^6} \right) + 40\left(\dfrac{1}{(x-3)^{12}}-\dfrac{1}{(x-3)^6}\right)
\end{eqnarray} 
where the left well is one unit deep and the right well is ten units deep. Clearly here the domain of interest is $x \in (-3,3)$ as the potential diverges at $x = \pm 3$. This potential could represent the interaction of two different particles.\\
We will also consider the scenario of a simple $x^2$ potential with additional Gaussian bumps (or dips):
\begin{eqnarray}
V(x) &=& x^2 + \sum_{i=1}^{n} a_i ~\text{exp} \left[ -\dfrac{(x-b_i)^2}{\mu}\right]
\end{eqnarray}
where there are $n$ bumps or dips with the prefactor $a_i$ being positive or negative respectively. $b_i$ marks the location of each bump and $\mu$ the width of each bump which for simplicity we take to be $ = 0.06$ for all. For our purposes we will focus on two variants of this setup: {\color{RoyalBlue}$iv)$} An $x^2$ plus two bumps placed symmetrically away from the origin and {\color{RoyalBlue}$v)$} an $x^2$ plus 3 bumps and 3 dips in an asymmetrical setup. Concretely the parameters we will use are:  
\begin{eqnarray}
&& x^2 + 2\text{ bumps: }  a_1 = a_2 = 1.5, b_1 = -b_2 = 1 \label{eq:x^2_2bumpsdefn}\\
&& x^2 + 6\text{ bumps/dips: }   a_1 = a_4 = a_5 = -1.5 , a_2 = a_3 = a_6 =1.5 \nonumber \\ 
&& \quad \quad b_1 = -b_2 = 0.7, b_3 = -b_4 = 1.4, b_5 = -b_6 = 2.1  \label{eq:x^2_6bumpsdefn}
\end{eqnarray}
This potential represents a rudimentary toy model for motion over a ``potential energy landscape'' with a series of local energy minima. The last two cases clearly demonstrate the effect of local extrema on the final shape of the effective potential since the underlying $x^2$ potential does not alter its shape under the \ac{RG} flow\footnote{To see this examine (\ref{eq:dV/dk}) and notice that at ${\kappa}=\Lambda$ the RHS becomes $\Upsilon/4\lb {\kappa} + 1\rb$ which has no $x$ dependence. This means the \ac{RG} flow will only serve to translate the potential vertically but not the overall shape.}.

\subsection{Polynomial Truncation}
Before solving the full PDE (\ref{eq:dV/dk}) it is instructive to consider an approximation, focusing on the double well potential (\ref{eq:Vdoublewell}) for illustration. We consider a truncated polynomial ansatz for the effective potential $V_{{\kappa}}(\chi)$ of the form
\begin{eqnarray}\label{eq:polytruncation}
V_{{\kappa}}(\chi) = E({\kappa}) + \sum_{i = 1}^{N}\alpha_{i}({\kappa})\chi^{2i}
\end{eqnarray}
with initial conditions defined such that it matches the original doublewell potential (\ref{eq:Vdoublewell}) at the cutoff:
\begin{eqnarray}
E({\kappa} = \Lambda) = 0,\quad \alpha_{1}({\kappa}=\Lambda) =  -1,\quad \alpha_{2}({\kappa}=\Lambda) =  1/4
\end{eqnarray}
and all coefficients of higher powers vanishing. We can then expand the RHS of (\ref{eq:dV/dk}) in powers of $\chi$, truncate the series at $2N$ and therefore write a set of N + 1 coupled ODEs in terms of the couplings that can then be solved numerically. Below we write the set of ODEs for the $\mathcal{O}(4)$ truncation as it only concerns coupling coefficients up to order $x^4$:
\begin{eqnarray}
\dfrac{dE({\kappa})}{d{\kappa}} &=& \dfrac{\Upsilon}{4}\cdot\left(\dfrac{1}{{\kappa}+ 2\cdot \alpha_1({\kappa})}-\dfrac{1}{{\kappa}}\right) \label{eq:de/dk}\\
\dfrac{d\alpha_1({\kappa})}{d{\kappa}} &=& -\dfrac{3\Upsilon\cdot\alpha_2({\kappa})}{({\kappa}+ 2\cdot\alpha_1({\kappa}))^2} \label{eq:da/dk}\\
\dfrac{d\alpha_2({\kappa})}{d{\kappa}} &=& \dfrac{36\Upsilon\cdot\alpha_{2}^{2}({\kappa})}{({\kappa}+ 2\cdot\alpha_1({\kappa}))^3} \label{eq:db/dk}
\end{eqnarray}
\begin{figure}
		\centering
		\includegraphics[width=0.65\textwidth]{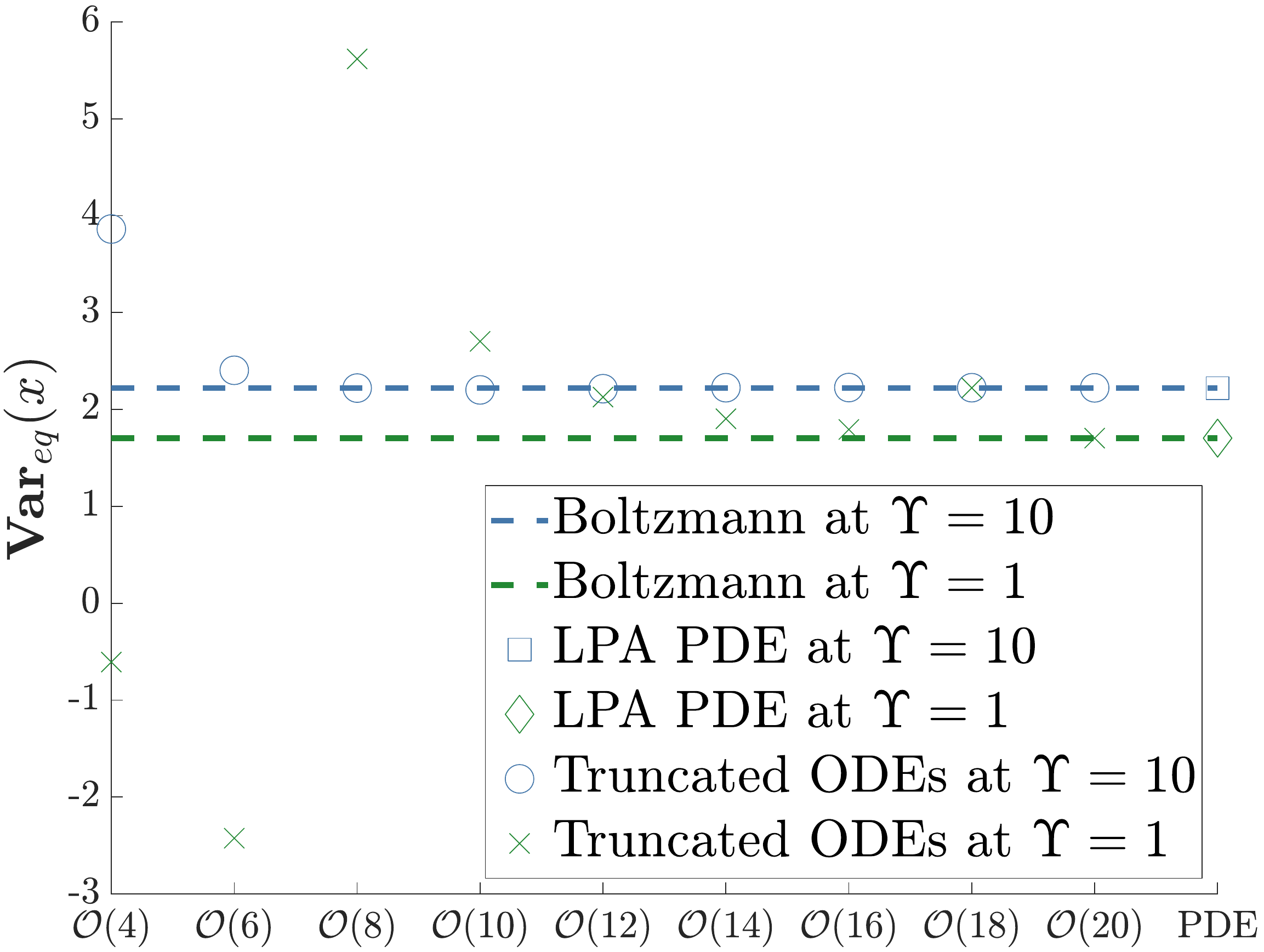}
		\caption[Truncated \ac{LPA} solution to flow equation for a doublewell potential]{The convergence of the truncated system of ODEs to the full \ac{LPA} PDE value for $\textbf{Var}(x)$ at equilibrium for $\Upsilon = 10$ (blue) and $\Upsilon = 1$ (green). $\textbf{Var}(x)$ as calculated from the Boltzmann distribution is also included for reference.}\label{fig: truncation_DW.png}
		\end{figure}
These equations show how the coefficients in the polynomial ansatz for the potential evolve when fluctuations of lower and lower frequencies are averaged over. Keeping more terms in the polynomial truncation is straightforward, leading to a hierarchy of flow equations for the different coefficients that can be easily obtained via a computer algebra software. Solving such polynomial flow equations is numerically much easier than solving the full PDE (\ref{eq:dV/dk}) and the solution to the full PDE should be approached as $N \rightarrow \infty$. However, this method is only well suited to initial potentials of polynomial form of small degree (e.g. the doublewell $-x^2 + x^4/4$). For potentials with more complex shapes the full PDE must be solved.  

The system of ODEs at each truncation was solved using Matlab's built in ode23s function which is based on a modified Rosenbrock formula of order 2. Focusing for concreteness on the variance and noting that it is related to the second derivative of the effective potential at the minima through (\ref{eq:equi_flow_var})\footnote{It may not be obvious that this is equation is still valid but we will later show in section \ref{sec:observables} explicitly that it is resulting in (\ref{eq:equal2pt}).}, we can rewrite it as: \textbf{Var}(x) = $\Upsilon /4\alpha_1({\kappa}=0)$, showing naturally how the different coupling constants relate to physical quantities -- the variance is inversely proportional to $\alpha_1({\kappa}=0)$.

The results of this truncation for the doublewell potential are displayed in Fig.~\ref{fig: truncation_DW.png}. Here we can see that the lowest order truncations match poorly with the correct value as given by the Boltzmann distribution. This discrepancy being particularly noticeable for $\Upsilon = 1$ with predictions of negative variance which is unphysical. However the value calculated by solving the full \ac{LPA} PDE (\ref{eq:dV/dk}) is approached by including more terms with the $\mathcal{O}(20)$ truncation matching the full PDE at both temperatures. The available thermal kinetic energy is $E_{th}=\Upsilon/2$ so $\Upsilon = 2$ corresponds to a thermal energy equal to the height of the doublewell barrier. We can therefore also think of $\Upsilon > 2$ corresponding to a high temperature regime, where the barrier can be overcome, and $\Upsilon < 2$ to a low temperature regime where trapping in one of the two minima occurs.   

We see that, in this example at least, the polynomial approximation to the flow equations offers a viable option to solving the flow, with the added bonus that it can be solved much quicker than the full \ac{LPA} PDE. However, if the initial potential is not well approximated by a polynomial such as the Unequal-Lennard Jones, or our bumpy potentials, then one is forced to solve the full \ac{LPA} PDE. Furthermore, going beyond the \ac{LPA} and include \ac{WFR}, which is sometimes required for accuracy doubles the complexity of any polynomial truncation. We therefore now turn to the full PDEs, the numerical solution to which is both feasible and accurate as we demonstrate.   

 \subsection{\label{sec:LPA-flow} Solutions to the full LPA flow equation}
 \begin{figure}[t!]
	\centering
		\includegraphics[width=0.45\textwidth]{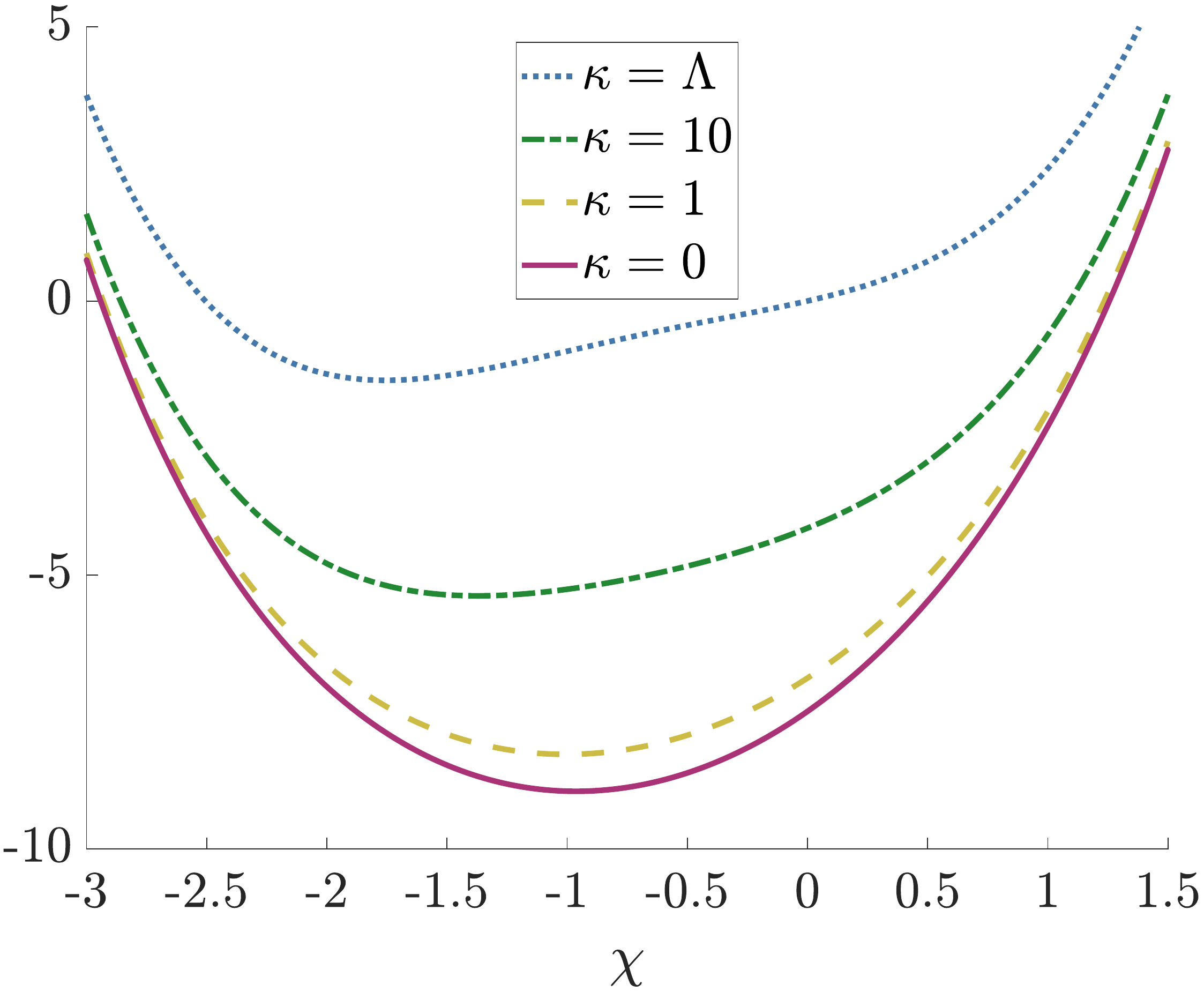}
		\includegraphics[width=0.45\textwidth]{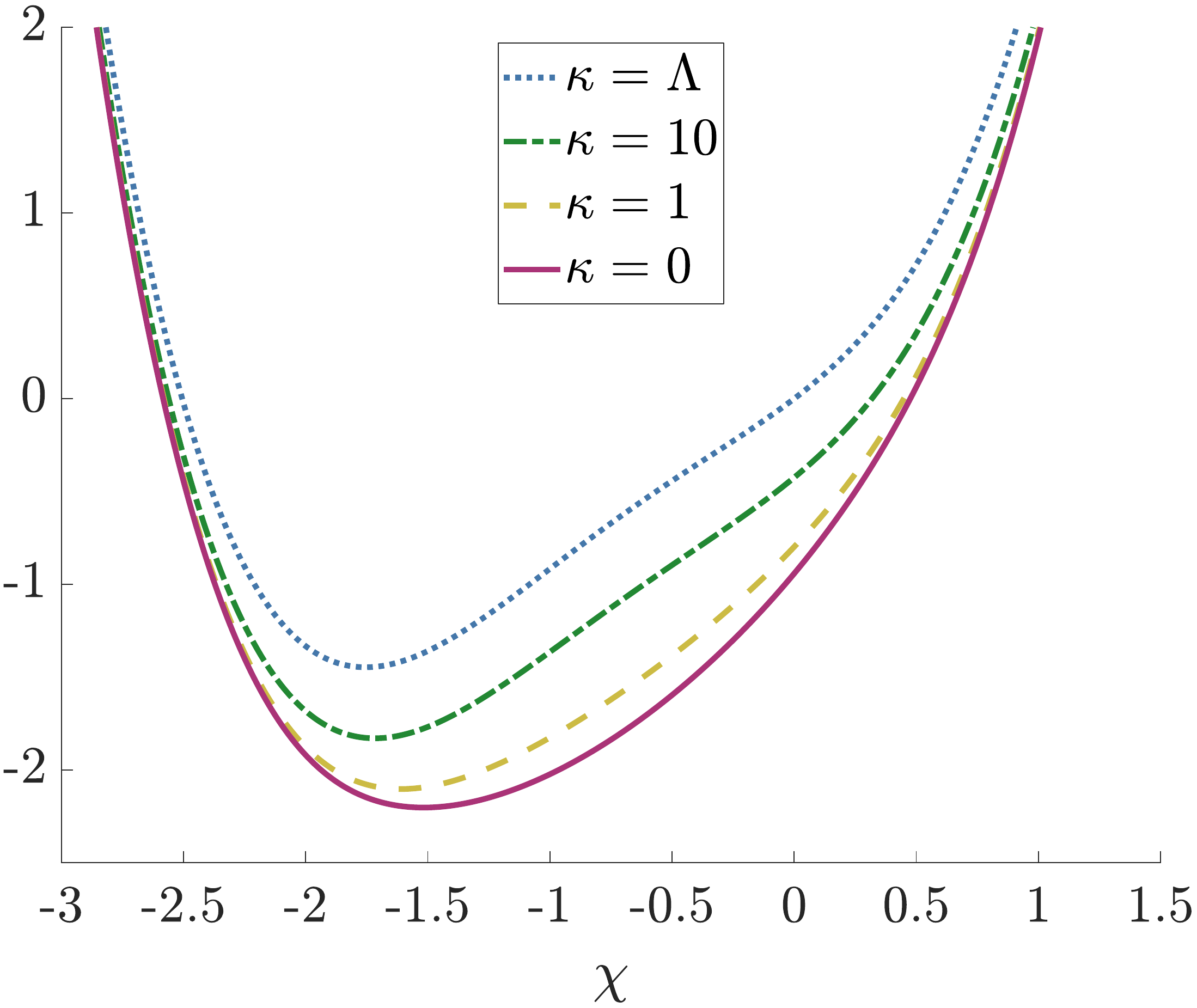}
		\caption[\ac{LPA} solution to flow equation for a polynomial potential]{The flow of the polynomial Langevin potential V in the \ac{LPA} for $\Upsilon = 10$ (High temperature/strong fluctuations - left) and $\Upsilon = 1$ (Low temperature/Weak fluctuations - right). The blue dotted curve indicates the bare potential which is progressively changed, through green dot-dashed and yellow dashed curves, into the red solid curve for the effective potential, as fluctuations are integrated out.}\label{fig: Gies_LPA_V_5T.png}
\end{figure}
\begin{figure}[t!]
		\centering
		\includegraphics[width=0.45\textwidth]{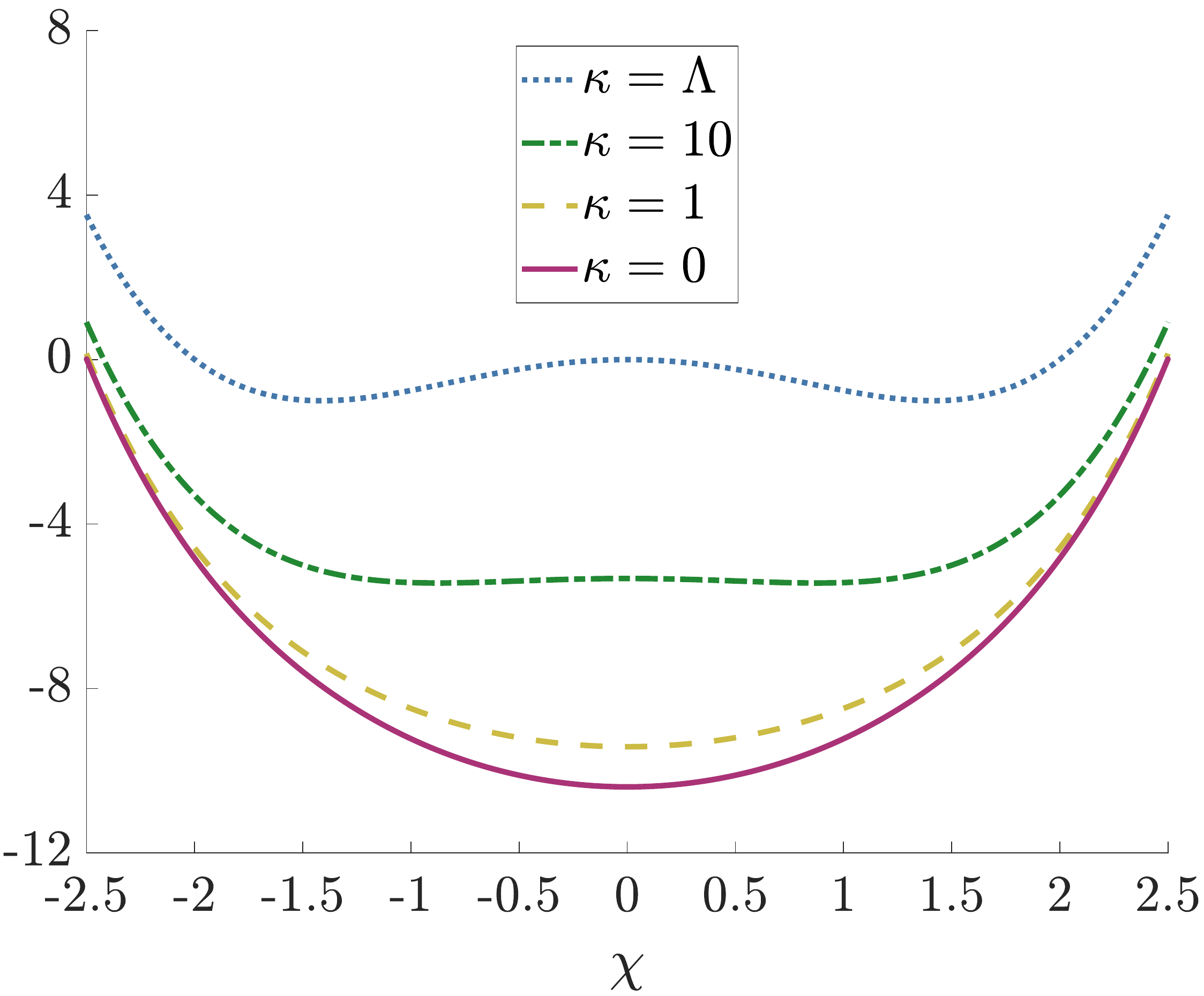}		
		\includegraphics[width=0.45\textwidth]{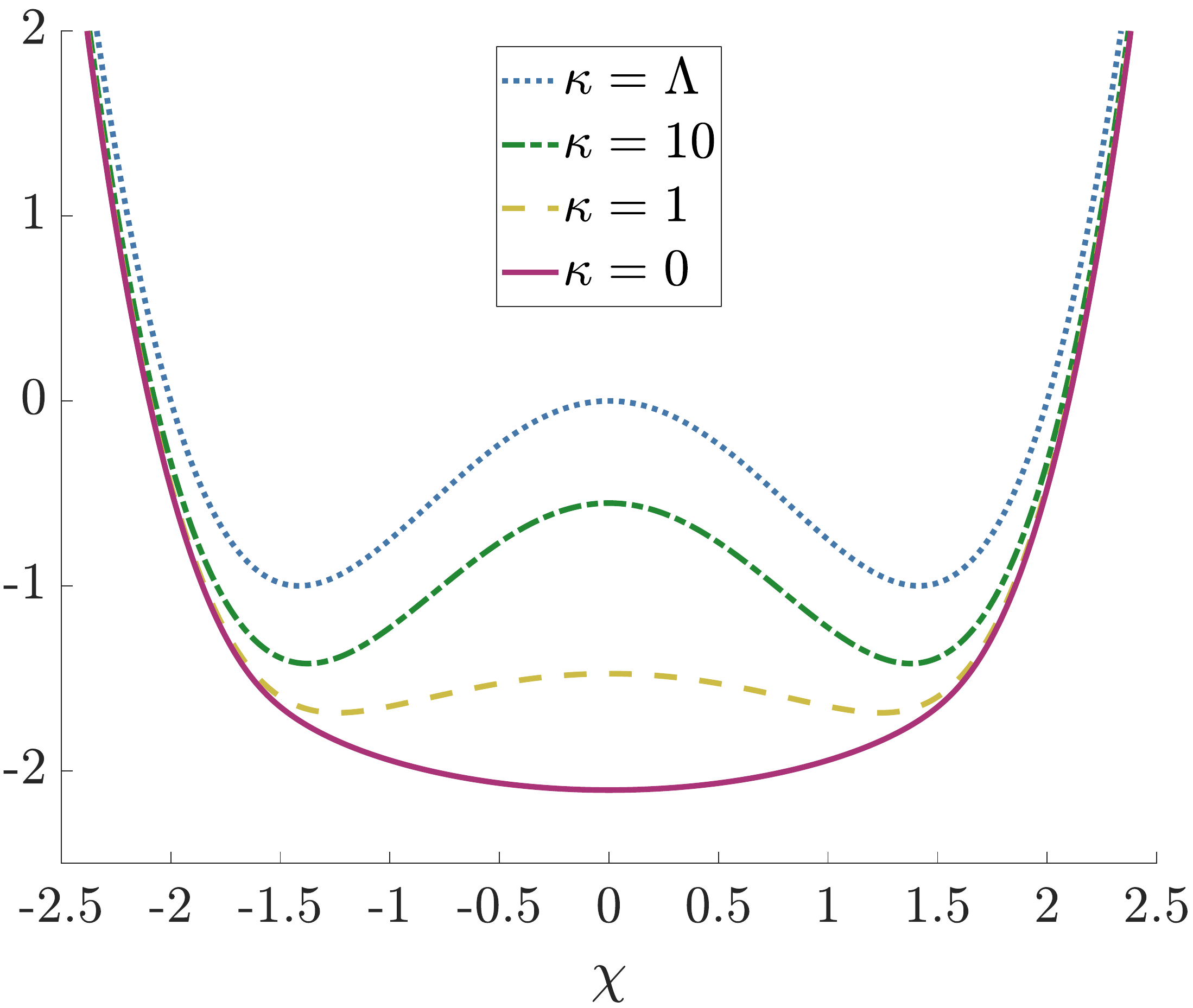}
		\caption[\ac{LPA} solution to flow equation for a doublewell potential]{The flow of the doublewell Langevin potential $V$ in the \ac{LPA} for $\Upsilon = 10$ (left) and $\Upsilon = 1$ (right). Again, the bare potential is denoted by the blue dotted curve and the ${\kappa}=0$ effective potential by the red solid one.}
		\label{fig: DW_LPA_V_5T.png}
\end{figure}
\begin{figure}[t!]
	\centering
		\includegraphics[width=0.45\textwidth]{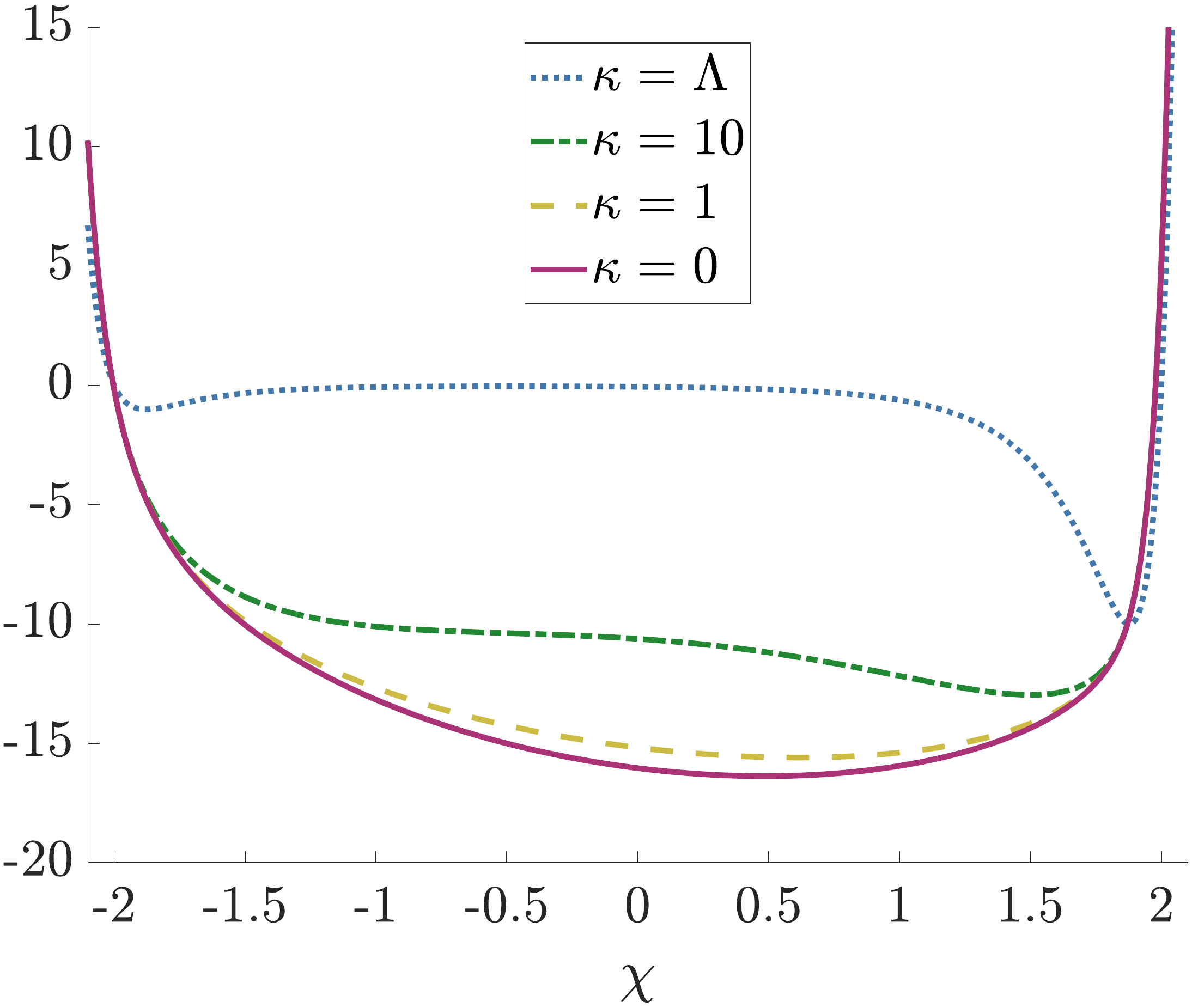}
		\includegraphics[width=0.45\textwidth]{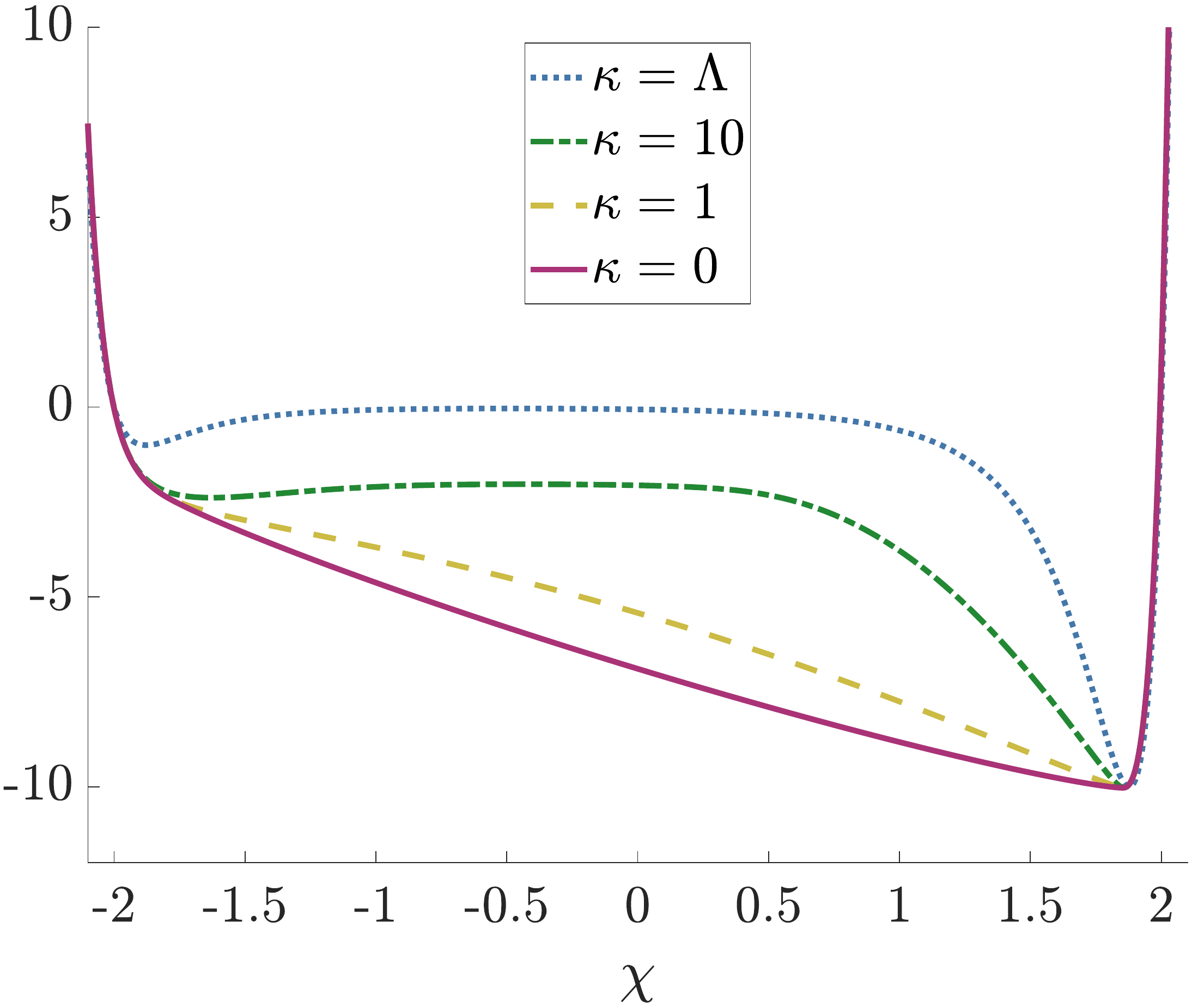}
		\caption[\ac{LPA} solution to flow equation for a Lennard-Jones type potential]{The flow of the unequal L-J Langevin potential V in the \ac{LPA} for $\Upsilon = 10$ (left) and $\Upsilon = 2$ (right). As before, the bare potential is denoted by the blue dotted curve and the ${\kappa}=0$ effective potential by the red solid one.}
		\label{fig: ULJ_LPA_V_5T.png}
\end{figure}
\begin{figure}[t!]
	\centering
		\includegraphics[width=0.45\textwidth]{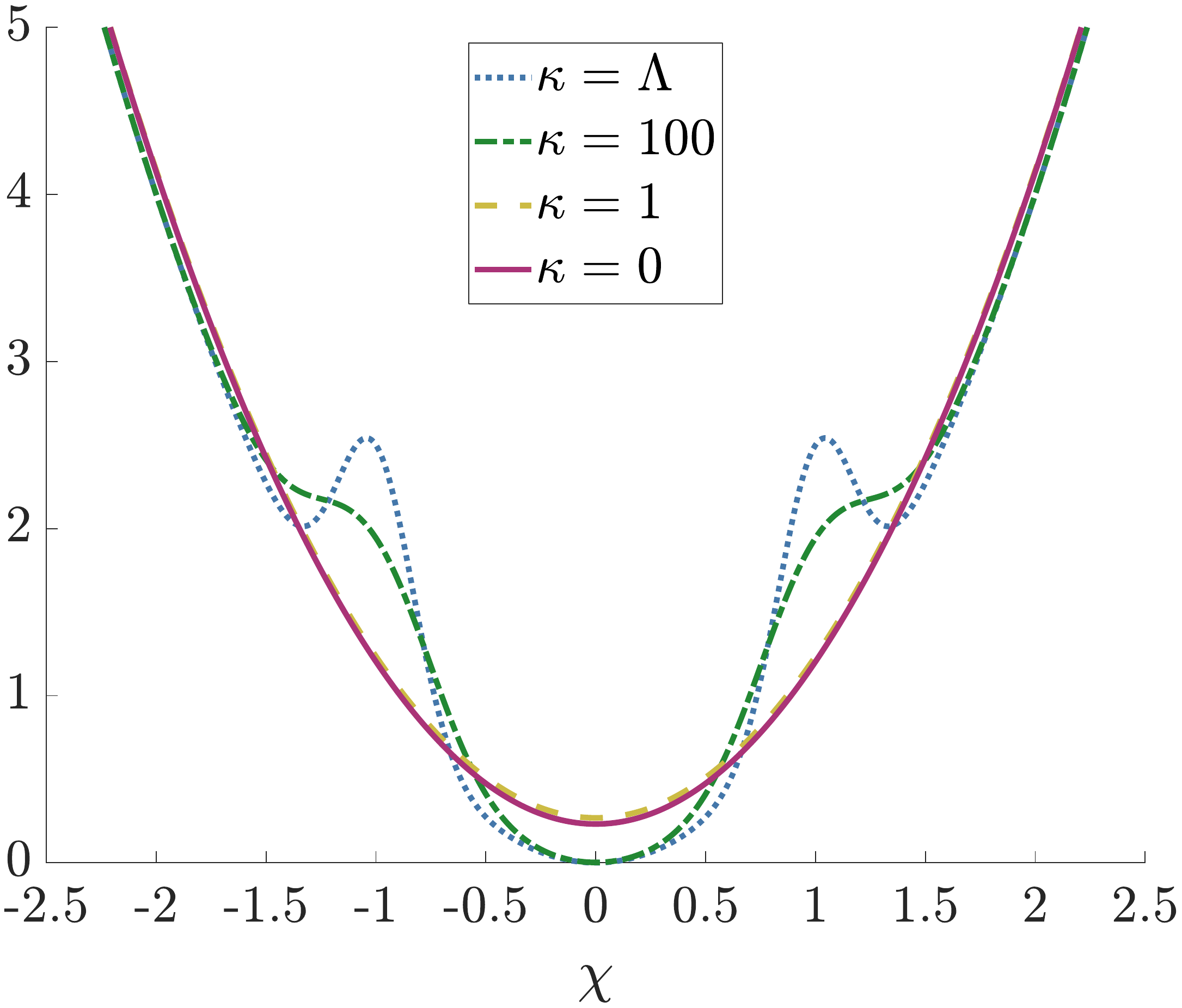}
		\includegraphics[width=0.45\textwidth]{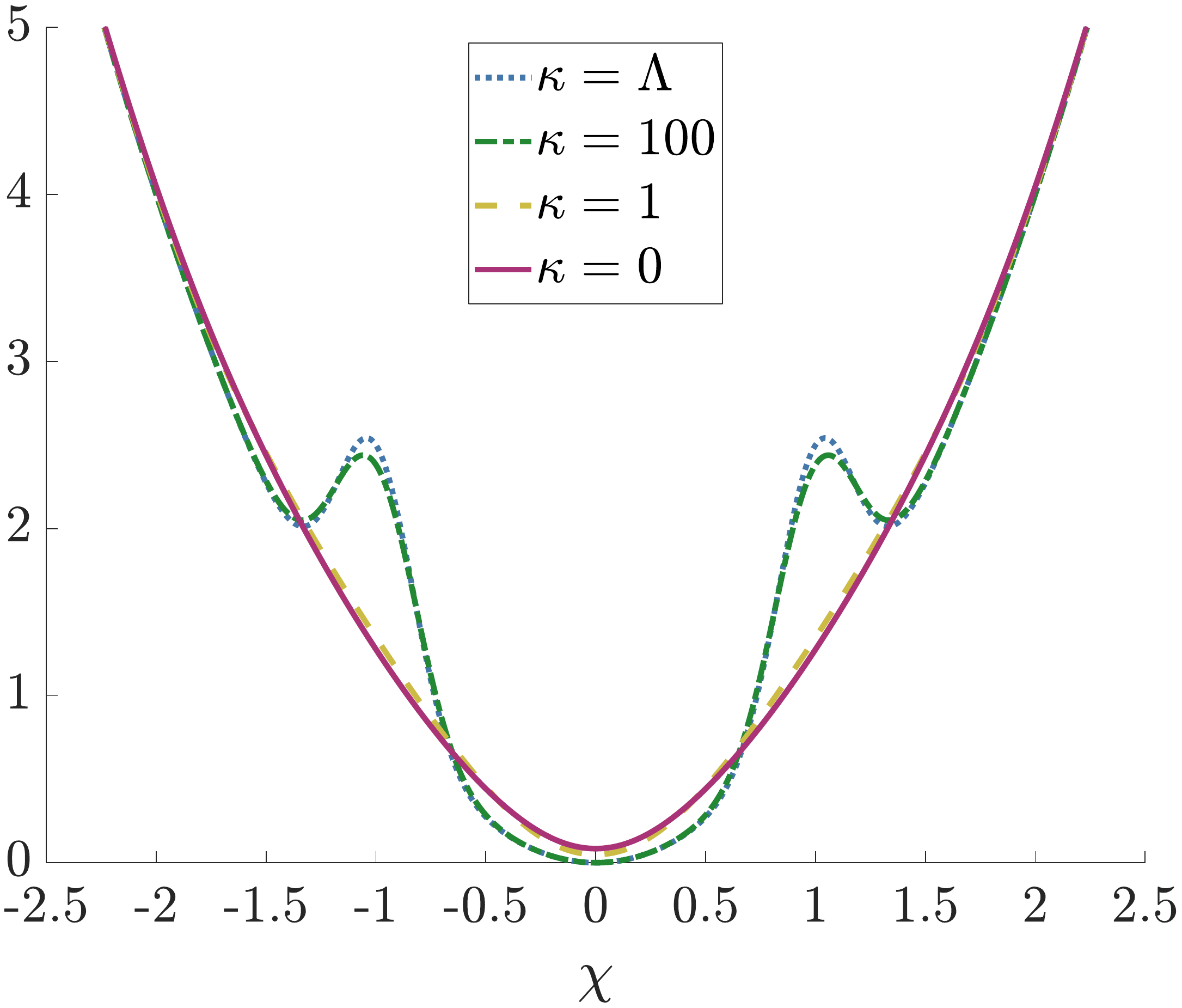}
		\caption[\ac{LPA} solution to flow equation for $x^2$ plus two bumps potential]{The flow of the $x^2$ potential with two additional bumps for $\Upsilon = 10$ (left) and $\Upsilon= 1 $ (right). As before, the bare potential is denoted by the blue dotted curve and the ${\kappa}=0$ effective potential by the red solid one.}
		\label{fig: X2Gauss_LPA_V_5T.png}
\end{figure}
\begin{figure}[t!]
	\centering
		\includegraphics[width=0.45\textwidth]{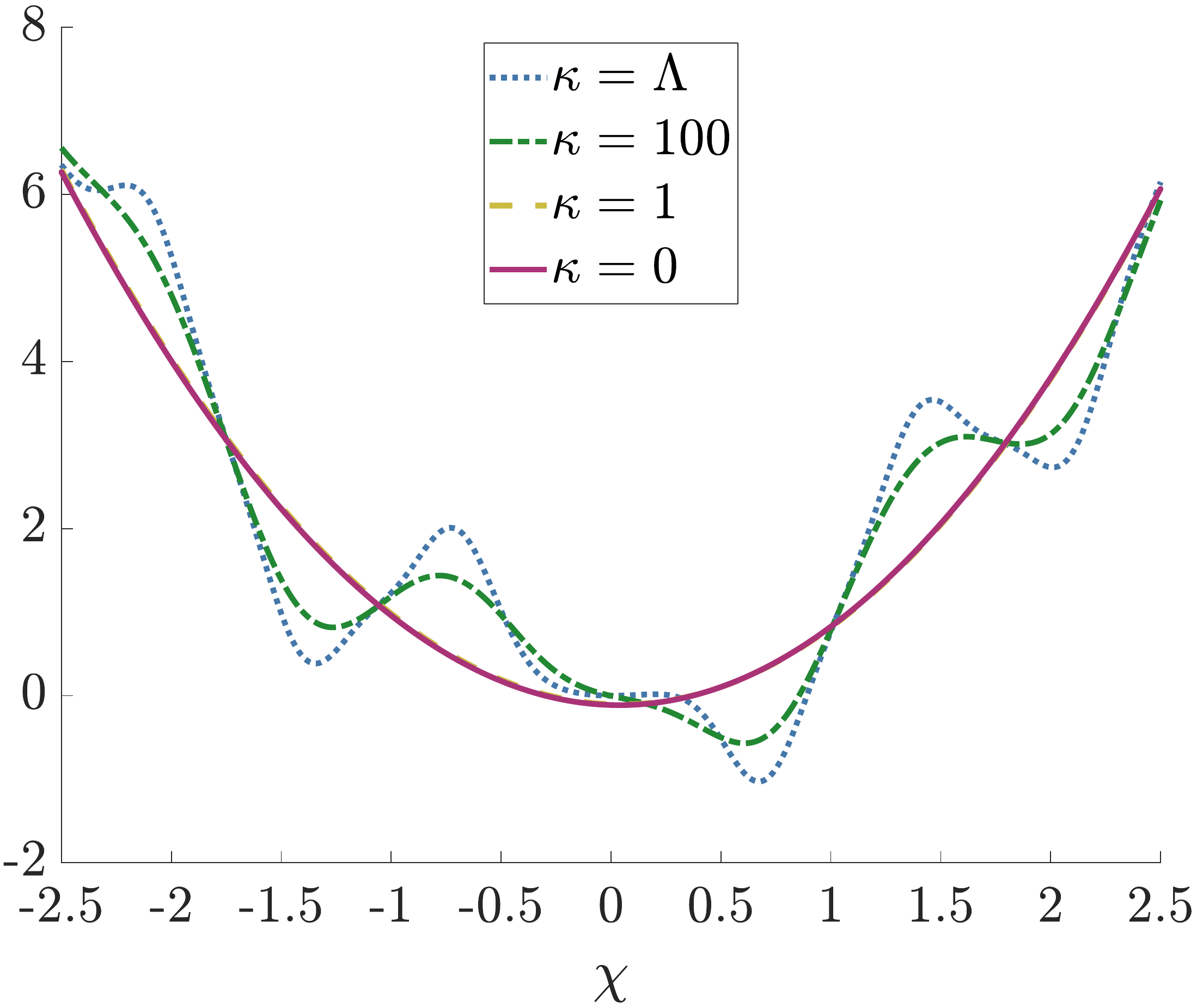}
		\includegraphics[width=0.45\textwidth]{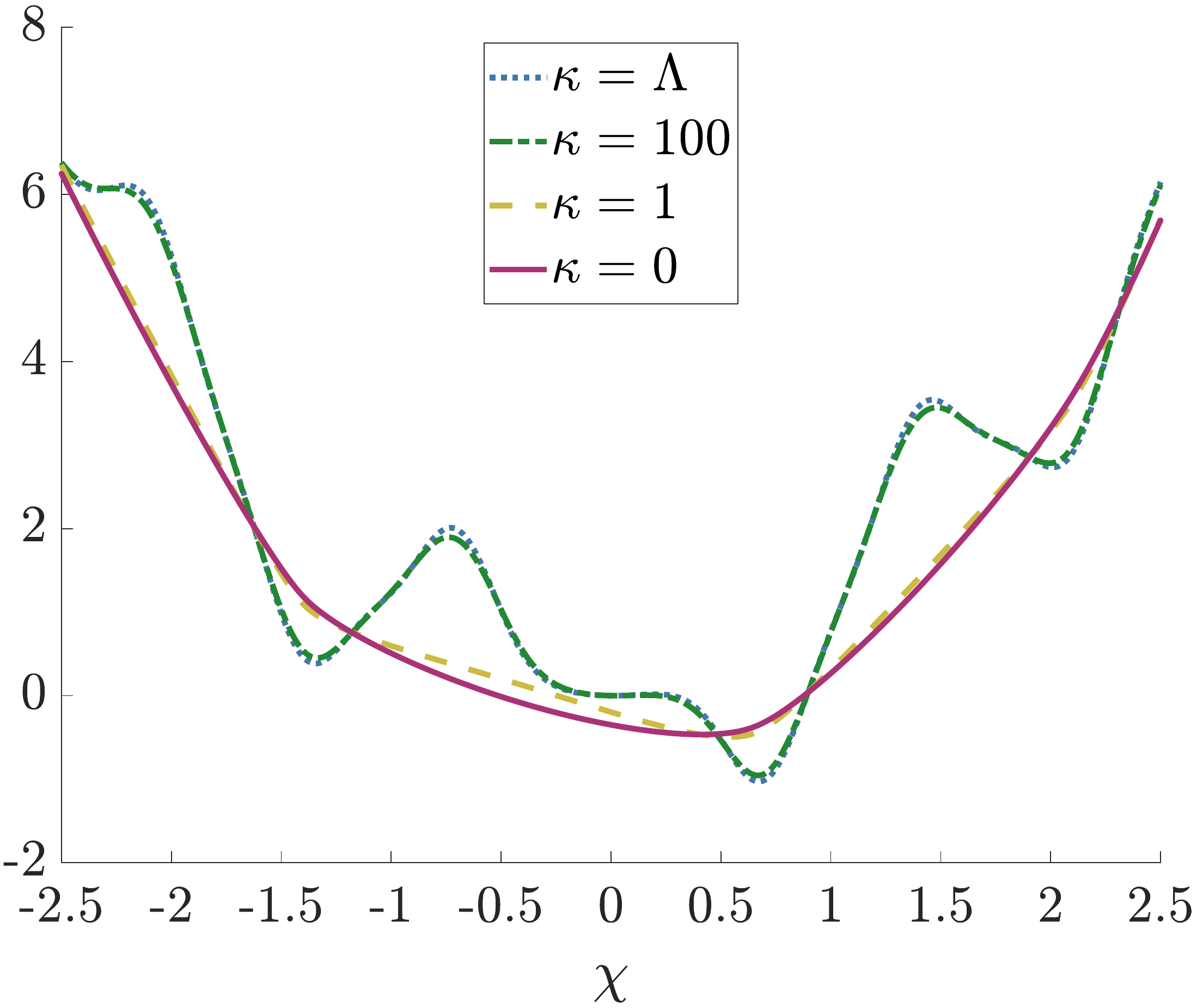}
		\caption[\ac{LPA} solution to flow equation for $x^2$ plus six bumps potential]{The flow of the $x^2$ potential with three additional Gaussian bumps and three dips in the \ac{LPA} for $\Upsilon = 10$ (left) and $\Upsilon = 1$ (right). As before, the bare potential is denoted by the blue dotted curve and the ${\kappa}=0$ effective potential by the red solid one.}
		\label{fig: X2GaussMany_LPA_V_5T.png}
\end{figure}
 We solve the \ac{LPA} flow equation (\ref{eq:dV/dk}) on a grid in the $\chi$ direction, using Matlab's built in ode45 or ode15s function to evolve in the ${\kappa}$ direction, depending on the potential. For most potentials ode45 -- which is based on an adaptive step size Runge-Kutta method -- was sufficient. The numerical derivatives in the $\chi$ direction were based on a finite difference scheme using the Fornberg method with a stencil size of 5 for the potentials under study. While increasing the grid size improves the accuracy of the numerical derivative it also increases the number of coupled ODEs to be solved, making the integration much more computationally expensive. A balance must be drawn depending on the potential in question. We considered 1001 points with $x \in (-3,3)$ for the unequal L-J potential and $x \in (-5,5)$ for the others. 

Our first example of a flow from the bare to the effective potential for high and low temperature, $\Upsilon = 10$ and $\Upsilon = 1$ respectively, is shown in Fig.~\ref{fig: Gies_LPA_V_5T.png} involving a polynomial potential. The flow in the range ${\kappa} \in (10, 600)$ is rather inconsequential and there is not much change in the shape of the potential. As ${\kappa}\rightarrow 0$ is approached however, a distinct single minimum develops indicating the average position of the particle. As expected, the lower the temperature, the closer the effective potential's minimum is to the bare potential's minimum, indicating the relative weakness of fluctuations to force the particle to spend time away from it.    

A perhaps more interesting case is shown in Fig.~\ref{fig: DW_LPA_V_5T.png}, displaying how the double well potential flows with renormalisation scale ${\kappa}$ to its effective incarnation for high $\Upsilon = 10$ and low $\Upsilon = 1$ temperature. Again, for the high temperature in the range ${\kappa} \in (10, 600)$ there is not much change in the shape of the potential. Physically this means that the fluctuations we have integrated out in this range do not contribute significantly to the particle moving between the two minima, only displacing the particle about each of the two distinct minima. However, by ${\kappa} = 1$ the energy barrier has gotten significantly smaller meaning that we have started to integrate over fluctuations that drive the particle over the barrier. Naturally, when ${\kappa} = 0$ is reached the potential is fully convex (as it must be by definition of $\Gamma$) with no barriers to overcome. Similar behaviour is obtained where again we consider the lower temperature, $\Upsilon = 1$. As one might expect it takes `longer' in ${\kappa}$ evolution for the barrier to disappear as fluctuations at each ${\kappa}$ scale have less energy than their equivalent for the $\Upsilon = 10$ case. Of note is that not only is the evolution different but the final shape of $V_{{\kappa}=0}(x)$ is different for the two different temperature regimes. For $\Upsilon = 1$ it is clear that the potential is much flatter around the origin than for $\Upsilon = 10$. This is suggestive of longer time scales required at lower temperatures to overcome the energy barrier and reach equilibrium. It also indicates longer times for the connected 2-point function to decay, as we later discuss in section \ref{sec:observables}.    

Also of note is that for both cases the global minimum shifts from its degenerate values at $\pm \sqrt{2}$ to $x = 0$. This makes physical sense as one expects that the particle will spend most of its time at the bottom of each well so that its average position will be in the middle i.e. at the origin. We know that the minimum of the fully flowed potential $V_{{\kappa}=0}(x)$ should correspond to the equilibrium position of the particle so we can intuitively see already that we are getting the correct behaviour.

As our third example we turn to a non-symmetric non-polynomial potential. Fig.~\ref{fig: ULJ_LPA_V_5T.png} displays the evolution with ${\kappa}$ for an unequal Lennard-Jones (L-J) potential under the \ac{LPA} for high ($\Upsilon = 10$) and low ($\Upsilon = 2$) temperature. Similarly to the double well case, the energy barriers get smaller and eventually disappear as ${\kappa}$ is lowered and $V_{{\kappa}=0}$ is fully convex. As one might expect however, $V_{{\kappa}=0}$ is not symmetric. Furthermore, the minimum of $V_{{\kappa}=0}$ does not match the global minimum of the bare potential. At high temperature, $\Upsilon=10$, the effective global minimum is located at $x > 0$ which is suggestive that the particle spends most of its time in the deeper well on the right but still spends a significant amount of time in the smaller well such that the average position lies in between the two. This is no longer the case for low temperature $\Upsilon = 2$, as shown in the bottom plot of Fig.~\ref{fig: ULJ_LPA_V_5T.png}. Here the minimum at ${\kappa} = 0$ is very close to the bare potential's global minimum, suggesting that the particle is nearly always found here at equilibrium. We see that the form of the effective potential clearly reflects the physical fact that, as the temperature is lowered, the particle is more likely to be found in the global minimum as it has less energy to escape and explore its surroundings.

We finally turn to our last two example potentials, consisting of a simple $x^2$ potential with the addition of two Gaussian bumps placed symmetrically at $x = \pm 1$, shown in Fig.~\ref{fig: X2Gauss_LPA_V_5T.png}, and six Gaussian bumps/dips, shown in Fig.~\ref{fig: X2GaussMany_LPA_V_5T.png}, representing the most complicated landscape we deal with in this thesis. These examples clearly demonstrate how the flow of the effective potential is driven by the local curvature (the Gaussian features imposed here) since for an $x^2$ potential the \ac{FRG} flow equation (\ref{eq:dV/dk}) yields no change beyond an unphysical shift by an overall additive constant. Again, the difference between the low and high temperature cases is evident in the asymmetric case with the high temperature flow eradicating the potential's substructure, while the low temperature flow ends up with a preferred equilibrium position, indicating the particle being more likely to be found near the global minimum.
\subsection{\label{sec:WFR-flow} Solutions to the WFR flow equation} 

\begin{figure}[t!]
	\centering
	\includegraphics[width=0.45\textwidth]{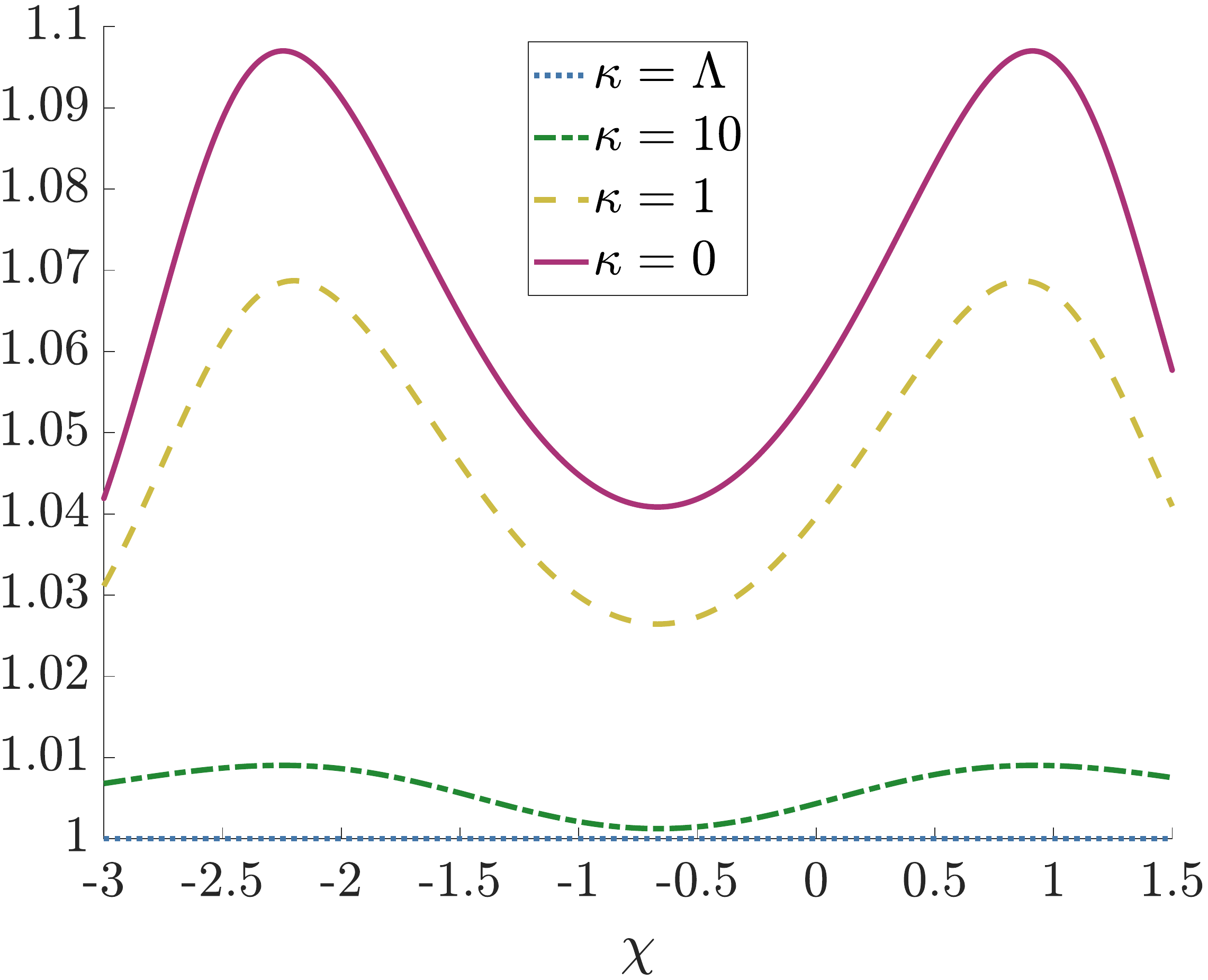}	
	\includegraphics[width=0.45\textwidth]{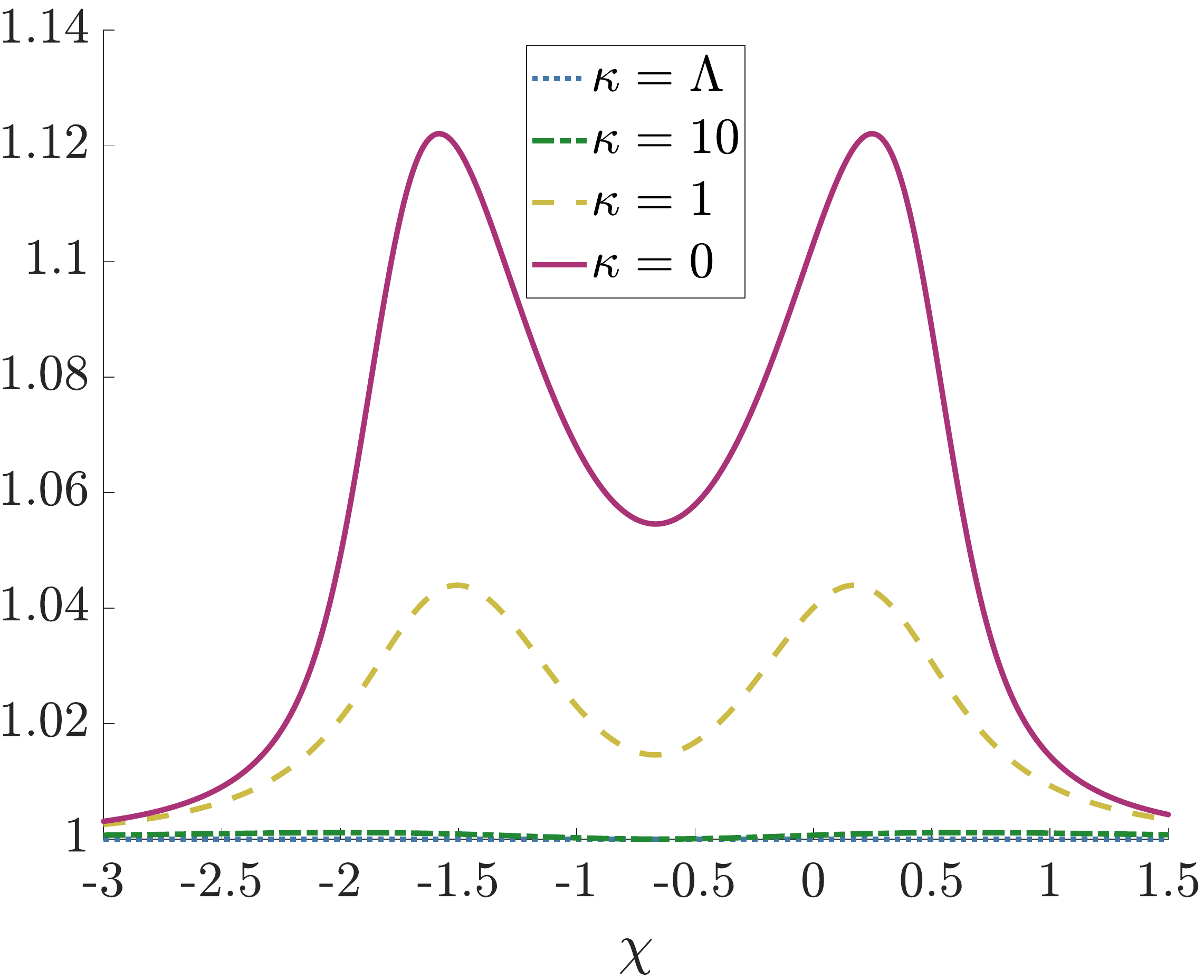}
	\caption[\ac{WFR} solution to the flow equation for a polynomial potential]{The flow of $\zeta_{x}$ for the polynomial potential for $\Upsilon = 10$ (left) and $\Upsilon = 1$ (right).}
	\label{fig: Poly_WFR_N_5T.png}. 
\end{figure}
\begin{figure}[t!]
	\centering
	\includegraphics[width=0.45\textwidth]{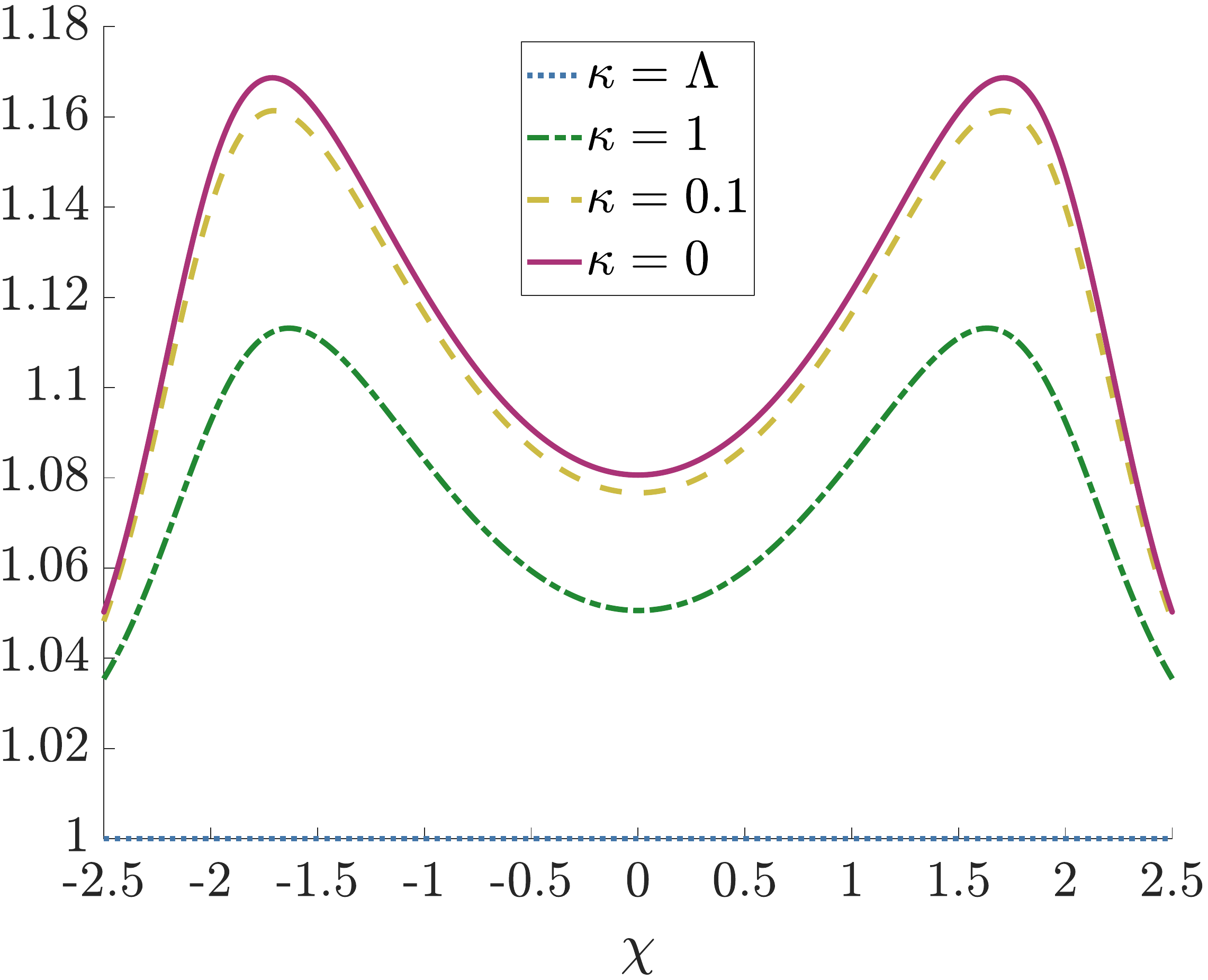}	
	\includegraphics[width=0.45\textwidth]{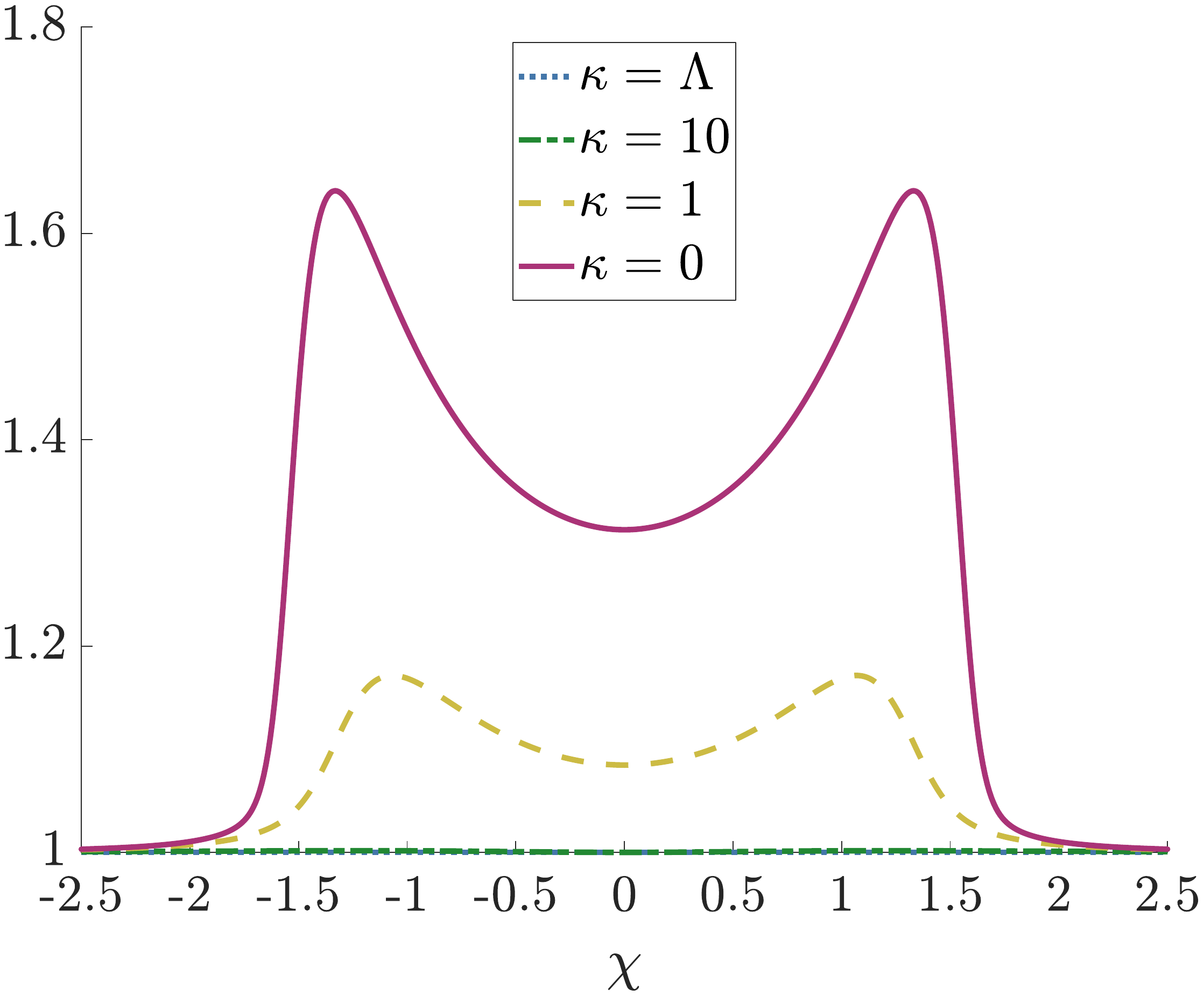}
	\caption[\ac{WFR} solution to the flow equation for a doublewell potential]{The flow of $\zeta_{x}$ for the double well potential for $\Upsilon = 10$ (left) and $\Upsilon = 1$ (right).}
	\label{fig: DW_WFR_N_5T.png}
\end{figure}
\begin{figure}[t!]
	\centering
	\includegraphics[width=0.45\textwidth]{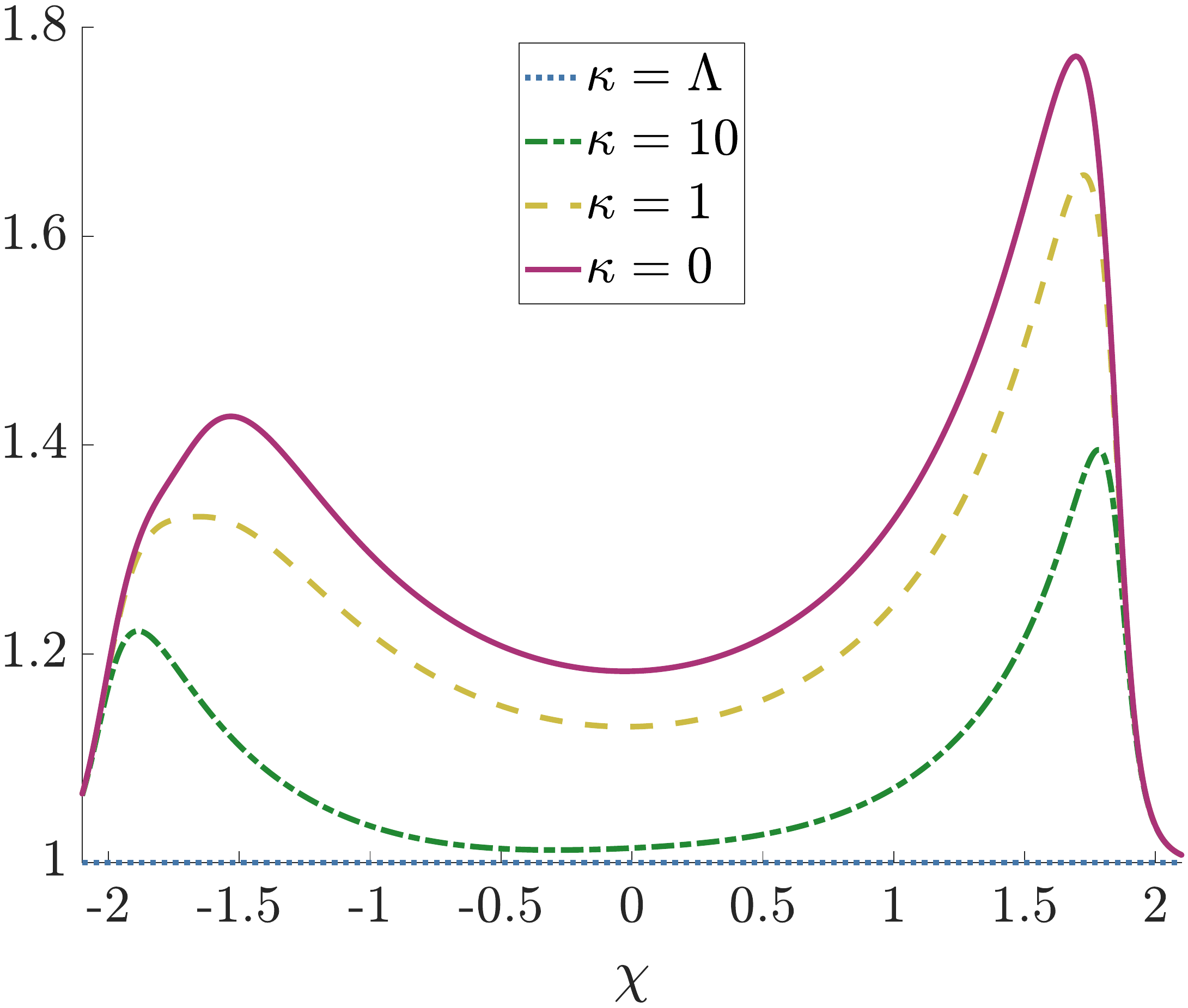}
	\includegraphics[width=0.45\textwidth]{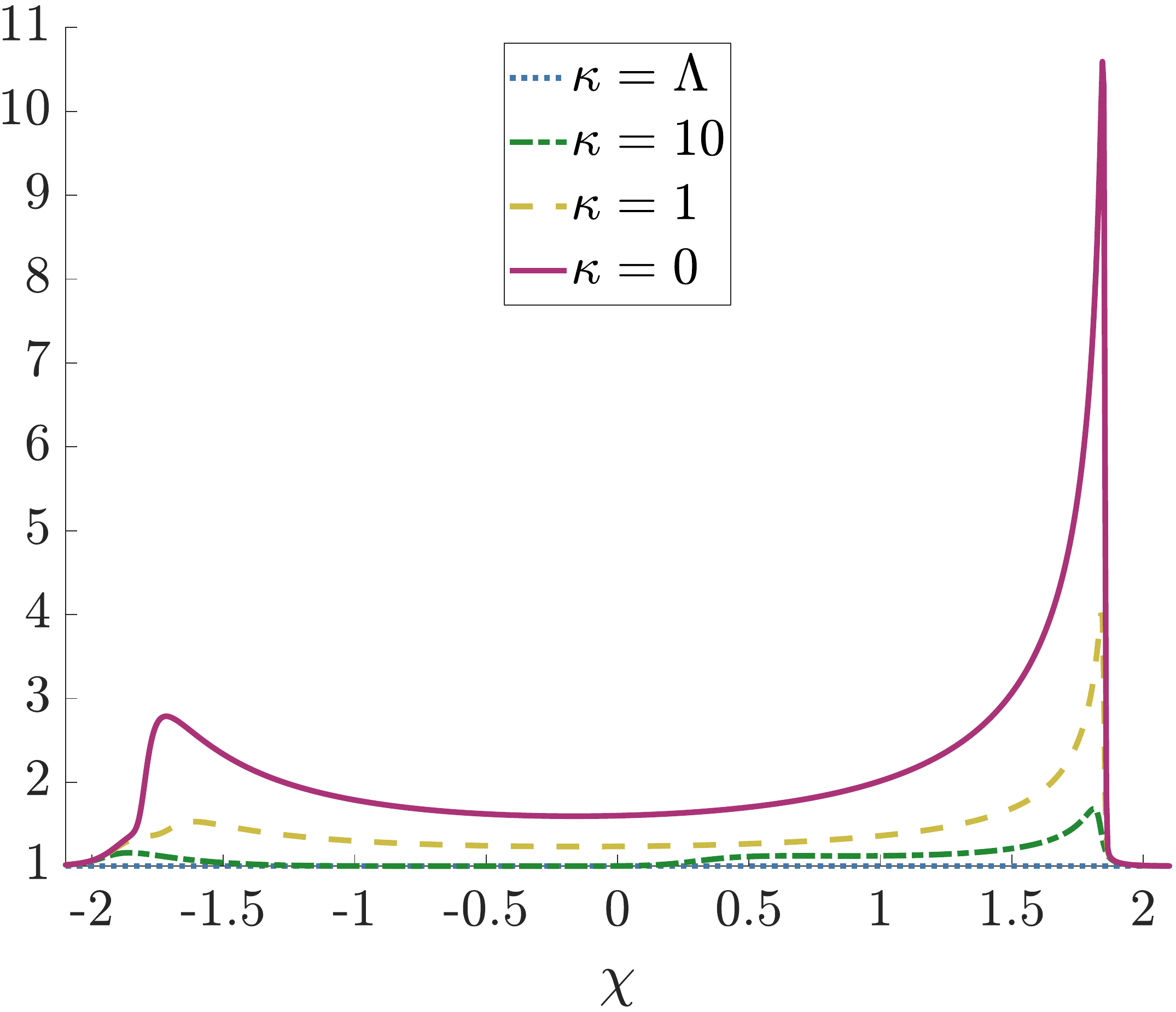}
	\caption[\ac{WFR} solution to the flow equation for a Lennard-Jones type potential]{The flow of $\zeta_{x}$ in an unequal L-J potential under \ac{WFR} for $\Upsilon = 10$ (left) and $\Upsilon = 2$ (right). }
	\label{fig: ULJ_WFR_N_5T.png}
\end{figure}
\begin{figure}[t!]
	\centering
	\includegraphics[width=0.45\textwidth]{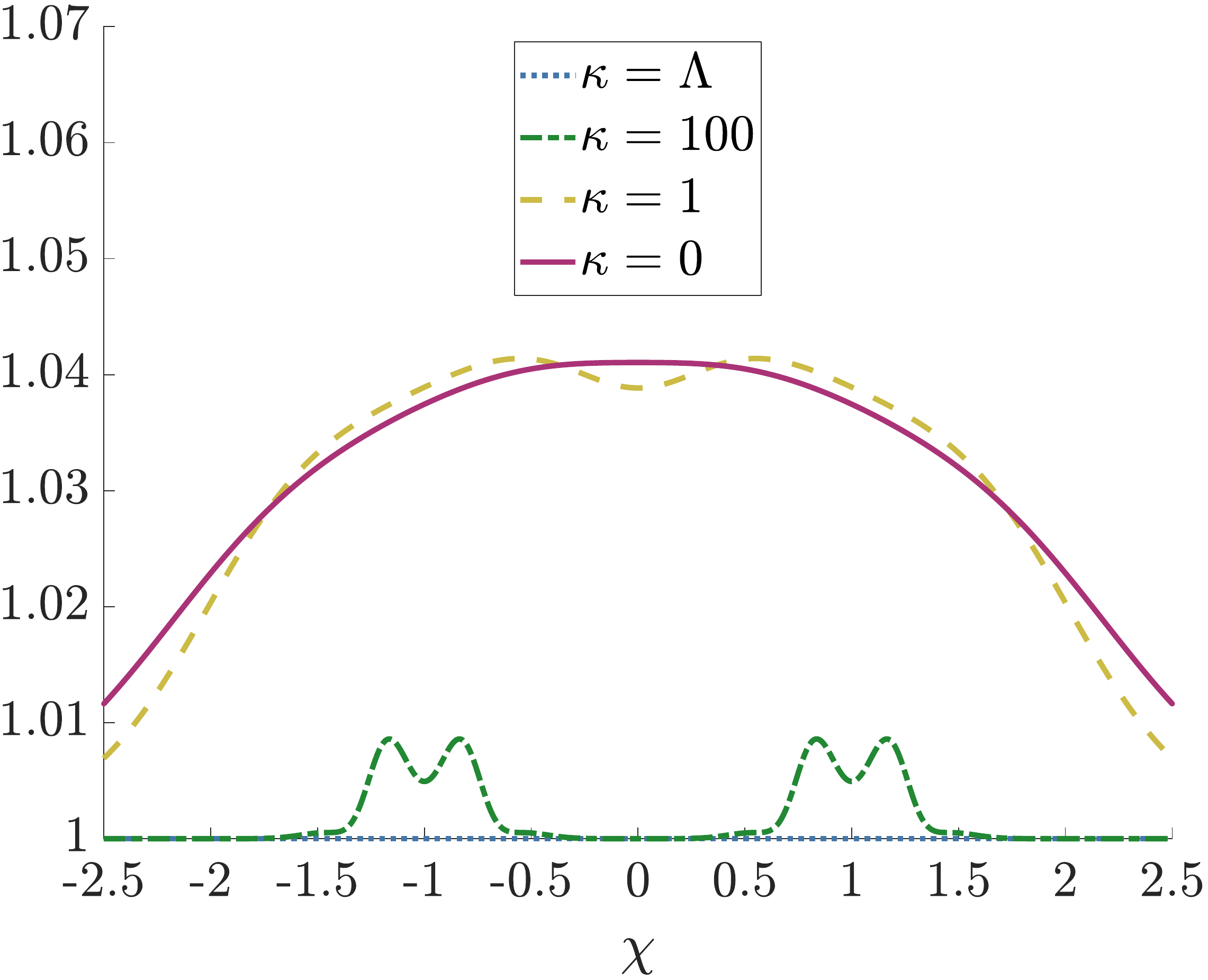}	
	\includegraphics[width=0.45\textwidth]{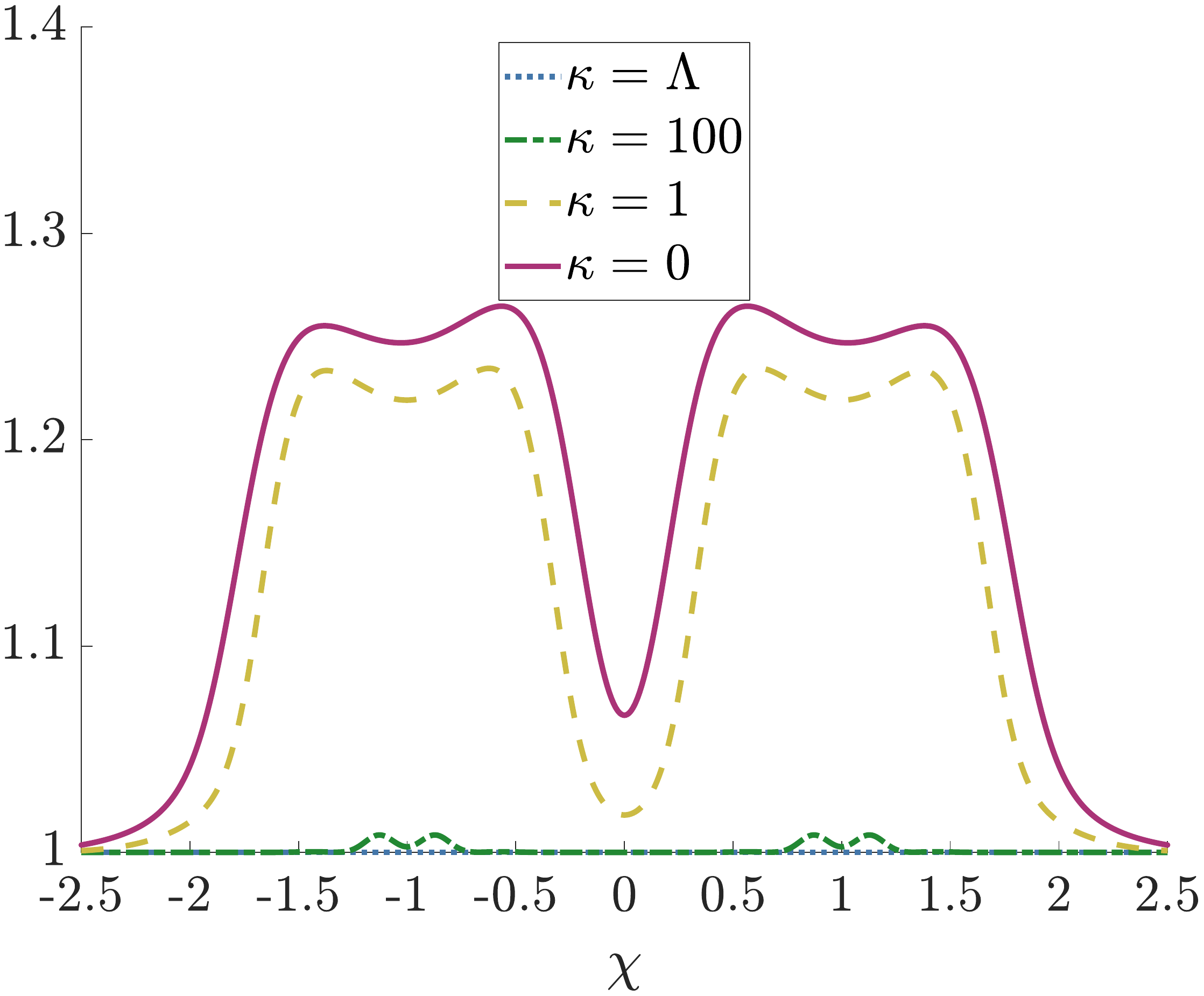}
	\caption[\ac{WFR} solution to the flow equation for $x^2$ plus two bumps potential]{The flow of $\zeta_{x}$ for an $x^2$ potential with two Gaussian bumps for $\Upsilon = 4$ (left) and $\Upsilon = 1$ (right).}
	\label{fig: X2Gauss_WFR_N_5T.png}
\end{figure}
\begin{figure}[t!]
	\centering
	\includegraphics[width=0.45\textwidth]{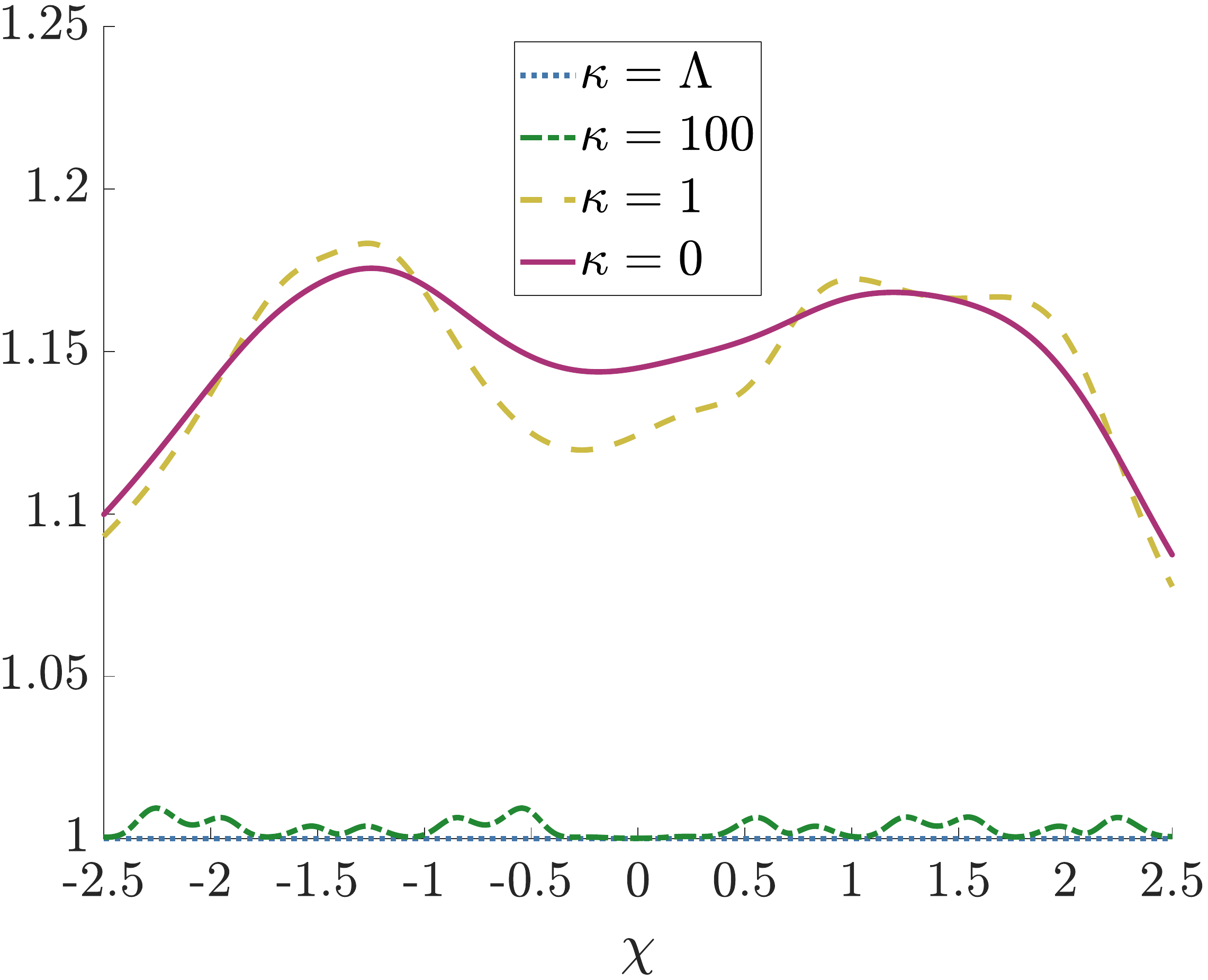}	
	\includegraphics[width=0.45\textwidth]{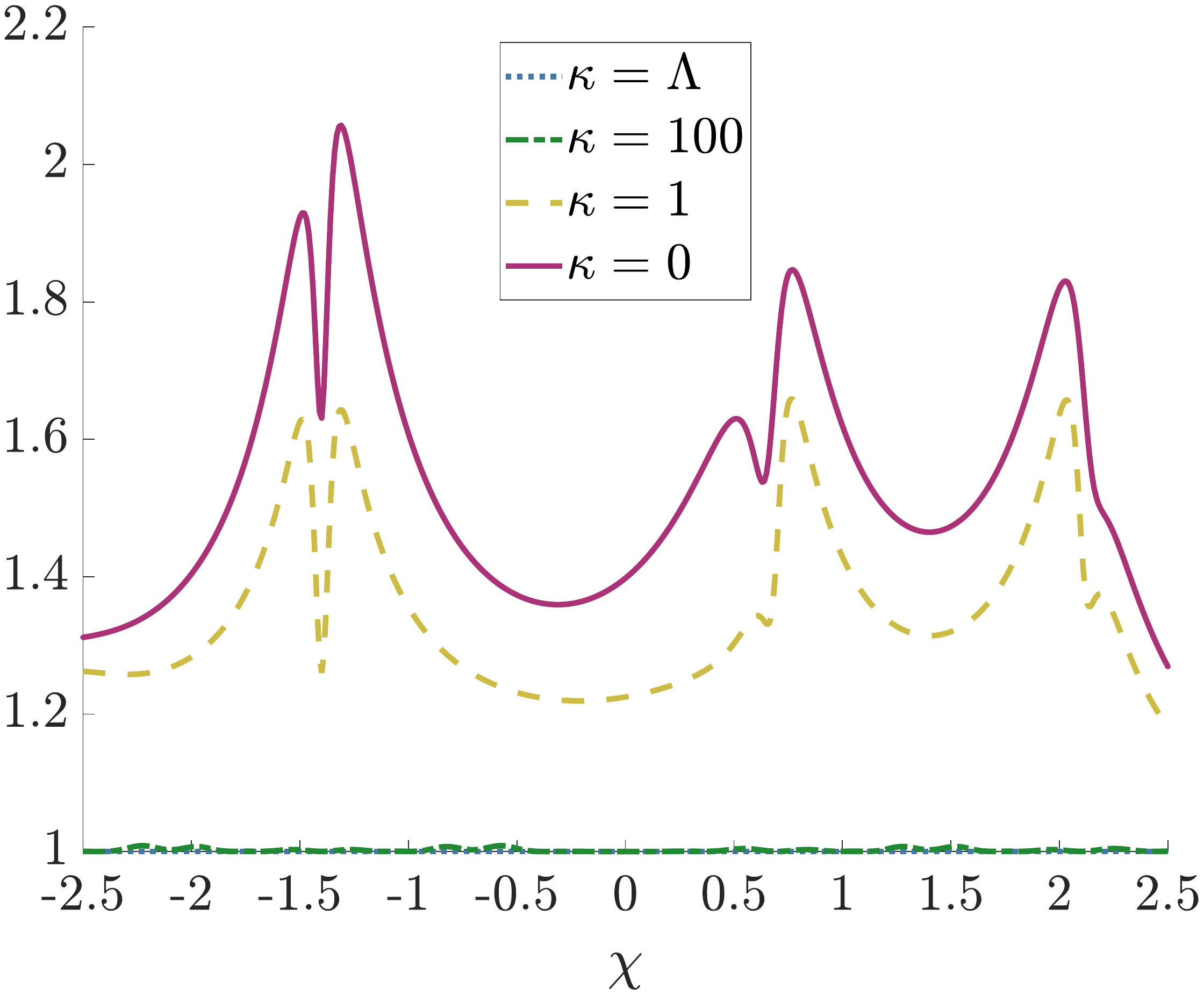}
	\caption[\ac{WFR} solution to the flow equation for $x^2$ plus six bumps potential]{The flow of $\zeta_{x}$ for an $x^2$ potential with six Gaussian bumps/dips for $\Upsilon = 3$ (left) and $\Upsilon = 1$ (right).}
	\label{fig: X2GaussMany_WFR_Zetax.png}
\end{figure}

Including \ac{WFR} does not change the evolution or final shape of $V_{\kappa}(x)$ making it irrelevant for time independent equilibrium quantities -- see section \ref{sec:observables}. However many other quantities of interest will depend on the inclusion of one more function $\zeta_{,x}$ to be evolved in ${\kappa}$. Its evolution with ${\kappa}$ for the potentials is shown in Figs.~\ref{fig: Poly_WFR_N_5T.png}, \ref{fig: DW_WFR_N_5T.png}, \ref{fig: ULJ_WFR_N_5T.png}, \ref{fig: X2Gauss_WFR_N_5T.png} \& \ref{fig: X2GaussMany_WFR_Zetax.png}. At the start of the flow (blue dotted curves) $\zeta_{,\chi}=1$ with a non-trivial $\chi$ dependence developing as ${\kappa} \rightarrow 0$, shown by the red curve. Similar to the potential, as $\Upsilon$ is lowered it takes longer for changes to happen. $\zeta_x(x)$ at ${\kappa} = 0$ for $\Upsilon = 1$ is much flatter than for $\Upsilon = 10$. The evolution of $\zeta_x(x)$ with ${\kappa}$ for the unequal L-J potential is shown in Fig.~\ref{fig: ULJ_WFR_N_5T.png} for both $\Upsilon = 10$ and $\Upsilon = 2$. The behaviour is similar to the doublewell case except now it is not symmetric with a larger peak for $x > 0$ as one might expect considering the initial shape of V. The  origin of these peaks is clear for the doublewell and unequal L-J potential. In both cases they form around the local minima of the bare V potential, e.g. for the doublewell this is at $x = \pm \sqrt{2}$ which we can see matches the peak of the red curves in Fig.~\ref{fig: DW_WFR_N_5T.png}. As the temperature is lowered for the unequal L-J potential the size of the peak around the deeper well increases. The structure of $\zeta_x(x)$ is even more complicated for $x^2$ with two bumps as shown in Fig.~\ref{fig: X2Gauss_WFR_N_5T.png}. Interestingly highly complicated structure appears part way through the flow (by ${\kappa} = 100$) before being smoothed out as $ {\kappa}\rightarrow 0$ for $\Upsilon = 4$. Some of this structure still remains at $\Upsilon = 1$ around the origin for ${\kappa} = 0$ as similarly observed in Figs.~\ref{fig: Poly_WFR_N_5T.png}, \ref{fig: DW_WFR_N_5T.png} \& \ref{fig: ULJ_WFR_N_5T.png}. Qualitatively similar behaviour is also observed for the flow of $\zeta_x(x)$ for $x^2$ plus six Gaussian bumps/dips -- see Fig.~\ref{fig: X2GaussMany_WFR_Zetax.png}.

It is perhaps not surprising that including the running of $\zeta_x$ seriously complicates the numerics of the problem, equation (\ref{eq:WFR_flow_eqn}) is more complicated than (\ref{eq:dV/dk}), however the effect is significant. The calculations for $\zeta_x$ also take significantly longer than for $V_{\kappa}$ alone as much smaller timesteps are required to be within acceptable numerical tolerances. As an example the calculation of $V_{\kappa}$ at $\Upsilon = 10$ for the doublewell took $\sim$ 5 minutes on a simple desktop machine whereas also calculating $\zeta_x$ on the same machine took $\sim $ 30 minutes. With more specialist numerical integrators tailor made for these equations it is conceivable that computations could be done quicker and $\zeta_x$ could be calculated for potentials and temperatures currently inaccessible using proprietary software. The advantage of this approach compared to competing methods is that optimising solving equation (\ref{eq:WFR_flow_eqn}) is independent of the potential.

\section{\label{sec:fRG_conc}Conclusion}
\acresetall 
In this chapter we have successfully introduced the concept of the \ac{RG} and focused on the formalism of the \ac{FRG}. We derived, based on the works of \cite{Wetterich1993, Morris1994}, the Wetterich equation (\ref{eq:flow_equation}) and applied it to a simple case of a doublewell potential. We then took this \ac{FRG} technology and applied it with full force to the problem of \ac{BM} outlined in chapter \ref{cha:stochastic_processes}, writing down the first two orders of the widely used derivative expansion of the \ac{EA}, referred to as the \ac{LPA} and \ac{WFR}. This built off the work of Synatschke et. al \cite{Synatschke2009} but connected it with the physical problem of \ac{BM} as opposed to \ac{SUSY}. We used a particular type of regulator, the frequency independent Callan-Symanzik regulator, for which the flow equations take on a relatively simple form, and further recalled that obtaining flow equations within the supersymmetric framework is convenient for ensuring compatibility with the Boltzmann equilibrium distribution, something that is not a priori obvious or guaranteed if one starts with the Onsager-Machlup form of the action (\ref{eq:dim-action}) and considers it a Euclidean $N=1$ scalar theory in one dimension with the Schr\"{o}dinger potential $\bar{U} = \Upsilon/4 \, V'' - 1/4\,(V')^2$. This has important consequences for certain aspects of the \ac{FRG} including the restriction to the derivative expansion that higher order terms (e.g. \ac{WFR}) must not modify the flow of the potential (\ref{eq:dV/dk}) as this would no longer reproduce the correct equilibrium result. We then took a variety of potentials which could represent various different complicated physical phenomena and solved the appropriate \ac{FRG} flow equations for them explicitly demonstrating how even potentials with lots of features are smoothed out by the \ac{FRG}. On the other hand solving for \ac{WFR} induces features in the $\zeta_x$ parameter where before there were none. 
\chapter{Effective Equations of Motion} 
\label{cha:fRG EEOM} 
\acresetall 
\vspace{0.5cm}
\begin{flushright}{\slshape    
People take the longest possible paths,\\ digress to numerous dead ends,\\ and make all kinds of mistakes.\\ Then historians come along and write summaries\\ of this messy, nonlinear process and\\ make it appear like a simple, straight line} \\ \medskip
--- Dean Kamen
\end{flushright}
\vspace{0.5cm}
\section{Introduction}
In chapter \ref{cha:fRG} we successfully applied the \ac{FRG} formalism to the problem of \ac{BM} and solved the corresponding flow equations for a variety of potentials. We have claimed that the effective potential $V_{{\kappa}=0}$ now fully incorporates the effect of all fluctuations but what does this mean? In this chapter we will derive \ac{EEOM} so called because they are derived from the \ac{EA} $\Gamma$. \\

This chapter is all new research and begins by deriving the \ac{EEOM} for the one- and two-point function in section \ref{sec:acc eom} which in our \ac{BM} example corresponds to the average position of the particle and variance/covariance of position respectively. In section \ref{sec:observables} we then examine in detail the limit of these \ac{EEOM} as the system approaches equilibrium. We identify the position of the minimum and the second derivative at the minimum of the effective potential as the equilibrium position and variance respectively and verify the \ac{FRG}'s ability to correctly capture this information for the potentials considered in chapter \ref{cha:fRG}. We also see how the \ac{FRG} can predict the decay of covariance at equilibrium. In section \ref{sec:Accelerated} we solve the \ac{EEOM} derived in section \ref{sec:acc eom} for some potentials of interest. We compare the \ac{FRG} results with direct numerical simulation of (\ref{eq:langevindimless}) and where possible solving the full \ac{F-P} PDE (\ref{eq:FP1}) and spectral expansion (\ref{eq:F-P_spectral}). We are therefore able to identify a regime of validity for the \ac{FRG} approach. We summarise this chapter with our conclusions in section \ref{sec:EEOM_conc}.\\

The busy reader is directed to the main results of this chapter:
\begin{itemize}
    \item Fig.~\ref{fig:fRG_flow_schematic_EEOM} for the schematic idea behind obtaining the \ac{EEOM} from the Langevin equation.
    \item The \ac{EEOM} for the average position $\lan x \ran$ given by equation (\ref{eq:QEOM}).
    \item The \ac{EEOM} for the variance $\lan x(t)x(t)\ran_{C}$ (\ref{eq:EEOM Variance}) and for the covariance $\lan x(t_1)x(t_2)\ran_{C}$ (\ref{eq:EEOM Covariance}).
    \item Fig.~\ref{fig: ChiT_X2GaussMany} for how well the \ac{FRG} can predict $\lan x \ran$ as the system relaxes to equilibrium in a complicated potential.
    \item Fig.~\ref{fig: VarT_X2Gaussmany.png} for how well the \ac{FRG} can predict $\lan x(t)x(t)\ran_{C}$ as the system relaxes to equilibrium in a complicated potential.
\end{itemize}

\section{\label{sec:acc eom}The Effective Equations of Motion}
A standard formulation of classical mechanics involves the principle of least action. As discussed around equation (\ref{eq:class_EOM}) if one considers the classical action $\mathcal{S}$:
\begin{eqnarray}
\mathcal{S} = \int dt~ L(x,\dot{x})
\end{eqnarray}
where $L(x,\dot{x})$ is the Lagrangian, then one can obtain the equations of motion by taking the variational derivative and setting it equal to zero:
\begin{eqnarray}
\dfrac{\delta \mathcal{S}}{\delta x} = 0 \label{eq:delta S = 0}
\end{eqnarray}
The \ac{EA} $\Gamma$ is so named because its definition makes it look like a standard classical action once fluctuations have been integrated out (\ref{eq:Gamma_S_relation}), (\ref{eq:effect_class_comp}):
\begin{eqnarray}
e^{-\Gamma} = \int\mathcal{D}x~e^{-\mathcal{S}}
\end{eqnarray}
We showed in section \ref{sec:The Wetterich Equation} how taking functional derivatives of the \ac{REA} ultimately lead to the Wetterich or flow equation (\ref{eq:Wetterich functional}) needed to compute $\Gamma$. However what about taking functional derivatives of the full \ac{EA}?
It is natural to ask whether we can extend the variational principle used to obtain the classical equations of motion from S to obtain \textit{effective}\footnote{Typically in the literature these would be called \textit{quantum} equations of motion. However the fluctuations we integrate over are not quantum, they are thermal, and so this would be misleading for our treatment of \ac{BM} hence \textit{effective}.} \textit{equations of motion} from $\Gamma$. As the \ac{FRG} has $\Gamma$ as its central object it is ideally placed to calculate these \ac{EEOM}. Also of importance is the fact that $\Gamma$ is written in terms of the mean fields (e.g. $\chi = \lan x\ran$) directly. This is what we will demonstrate in the rest of this section.
\subsection{\label{sec:EEOM_one}The EEOM for the one point function}
\begin{figure}
    \centering
    \includegraphics[width = .95\linewidth]{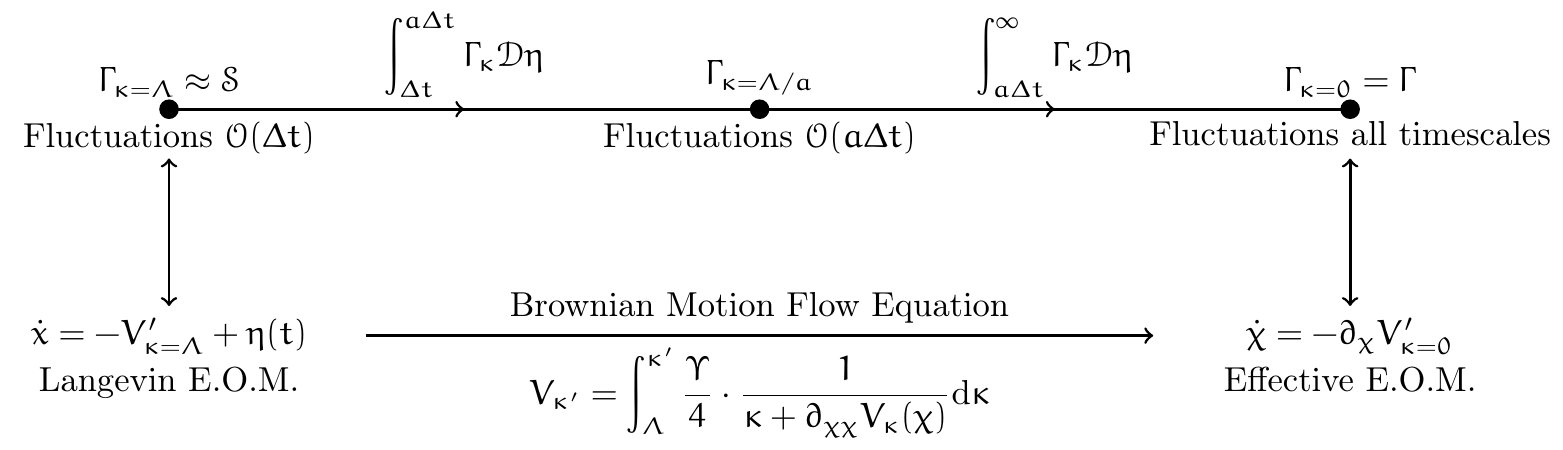}
    \caption[Conceptual look at how the \ac{FRG} yields the \ac{EEOM}]{A schematic diagram of how the \ac{FRG} takes the Langevin equation (\ref{eq:langevindimless}) and creates an effective theory that incorporates the effect of fluctuations.}
    \label{fig:fRG_flow_schematic_EEOM}
\end{figure}

In a similar way to how the classical action $\mathcal{S}(x)$ yields the famous Euler-Lagrange equations through one functional derivative, so too does $\Gamma [\chi]$ yield the \ac{EEOM} for the one point function (or average position) $\chi = \lan x\ran$:
\begin{eqnarray}
\dfrac{\delta\Gamma}{\delta \chi(t)} = 0 \label{eq:gen QEOM}
\end{eqnarray}
Here we have assumed there are no external sources\footnote{N.B. this is not the same as assuming that the noise term (\ref{eq:eta defn}) is zero as this is true for $\Gamma$ by definition} (J = 0).  \\
If we consider the \ac{LPA} truncation\footnote{Chosen so as to appropriately match (\ref{eq: PDF action}).} of $\Gamma$ written in terms of the physical variables:
\begin{eqnarray}
\Gamma_{\kappa}[\chi,\tilde{\chi},C,\bar{C}] = \int\mathrm{d}t~\dfrac{\Upsilon}{2}\tilde{\chi}^2 -i\tilde{\chi}\lb \dot{\chi} + \partial_{\chi}V_{{\kappa}}\rb  -\bar{C}\dot{C} -\bar{C}C\partial_{\chi\chi}V_{\kappa} \label{eq:Gamma_k_LPA_BM}
\end{eqnarray}
then we can obtain the \ac{EEOM} for each variable in turn. If we start with the \ac{EEOM} for $\bar{C}$:
\begin{eqnarray}
\dfrac{\delta\Gamma_{\kappa}}{\delta \bar{C}} &=& \dot{C} + C\partial_{\chi\chi}V_{{\kappa}} = 0 \\
\Rightarrow \dot{C} &=& - C\partial_{\chi\chi}V_{{\kappa}} \label{eq:EEOM_Cbar}
\end{eqnarray}
and substitute this back into (\ref{eq:Gamma_k_LPA_BM}) we see that the terms involving $\bar{C}$ cancel. We can then derive the \ac{EEOM} for $\tilde{\chi}$:
\begin{eqnarray}
\dfrac{\delta\Gamma_{\kappa}}{\delta \tilde{\chi}} &=& -\Upsilon\tilde{\chi} + i\lb \dot{\chi} + \partial_{\chi}V_{{\kappa}}\rb = 0 \\
\Rightarrow -i\tilde{\chi} &=& \dot{\chi} + \partial_{\chi}V_{{\kappa}} \label{eq:EEOM_tilde_chi}
\end{eqnarray}
which if we also substitute back into the \ac{EA} leaves us with:
\begin{eqnarray}
\Gamma_{{\kappa}}[\chi] &=& \dfrac{1}{\Upsilon}\int\mathrm{d}t~\dfrac{1}{2}\dot{\chi}^2 + \dot{\chi}\partial_{\chi}V_{{\kappa}} + \dfrac{1}{2}\lb\partial_{\chi}V_{{\kappa}}\rb^2 \\
\Rightarrow \Gamma_{{\kappa}}[\chi] &=& \dfrac{1}{\Upsilon}\lsb V_{\kappa} (\chi_f) - V_{\kappa}(\chi_i)\rsb + \dfrac{1}{\Upsilon} \int\mathrm{d}t~\dfrac{1}{2}\dot{\chi}^2 + \dfrac{1}{2}\lb\partial_{\chi}V_{{\kappa}}\rb^2 \label{eq:Gamma_k_OM}
\end{eqnarray}
where in the second line we have integrated out the $\dot{\chi}\partial_{\chi}V_{{\kappa}}$ term to give us the boundary terms outside the integral. These will play no part in the \ac{EEOM} so we will ignore them from here on in. Applying (\ref{eq:gen QEOM}) to (\ref{eq:Gamma_k_OM}) we obtain the \ac{EEOM} for the average position:
\begin{eqnarray}
\dfrac{\delta\Gamma_{{\kappa} = 0}}{\delta \chi(t)} &=& \ddot{\chi} -  \partial_\chi V_{{\kappa}=0}(\chi)\,\partial_{\chi\chi}V_{{\kappa}=0}(\chi) = 0 \\
\Rightarrow \dot{\chi} &=& - \partial_\chi V_{{\kappa}=0}(\chi) \label{eq:EEOM_LPA_chi}
\end{eqnarray} 
where we have noted from section \ref{sec:equil-flow} that the equilibrium position corresponds to the minimum of the effective potential (\ref{eq:equi_effective_potential}) in order to fix a constant of integration. \\
Equation (\ref{eq:EEOM_LPA_chi}) provides the final step in our conceptual journey as to how the \ac{FRG} works. We saw earlier in Fig.~\ref{fig:fRG_fluctations} how the \ac{FRG} integrates out fluctuations of increasing rarity until it recovers the full \ac{EA} $\Gamma$. In Fig.~\ref{fig:fRG_flow_schematic_EEOM} we show how this procedure gives us the \ac{EEOM}. If one starts with the Langevin equation (\ref{eq:langevindimless}) in the bottom left, we say that this can be described by some classical action (\ref{eq: PDF action}) which we identify the \ac{REA} with at some cutoff scale $\Lambda$. This cutoff corresponds to fluctuations that occur over some timescale $\mathcal{O}(\Delta t)$, if one wanted to simulate the Langevin equation (\ref{eq:langevindimless}) this is the timestep they should use in their numerical scheme. The \ac{FRG} then moves across the top line of Fig.~\ref{fig:fRG_flow_schematic_EEOM} integrating out fluctuations that occur over ever-increasing timescales until they are all integrated over and $\Gamma_{{\kappa}=0}$ is reached. One can then use (\ref{eq:gen QEOM}) to obtain the \ac{EEOM} for the average position (\ref{eq:EEOM_LPA_chi}) bringing us to the bottom right of Fig.~\ref{fig:fRG_flow_schematic_EEOM}. The flow equation (\ref{eq:flow_V_basic}) derived in section \ref{sec:LPA_BM_FLOW} is shown at the bottom of Fig.~\ref{fig:fRG_flow_schematic_EEOM} as a straightforward way of moving between the Langevin equation (\ref{eq:langevindimless}) and the \ac{EEOM} (\ref{eq:EEOM_LPA_chi}). In this way it is clear that the effective potential is the result of incorporating the fluctuating degrees of freedom hidden in the noise term $\eta$ and that the symmetries we worked so hard to enforce in section $\ref{sec:BM-PI}$ are exactly what is needed for the forms of the two equations to match. \\

The story is not spoiled if we include \ac{WFR} as (\ref{eq:gen QEOM}) then becomes:
\begin{eqnarray}
\left(\zeta_{,\chi}\dot{\chi}\right)\dot{} - \frac{\partial_\chi V_{{\kappa}=0}}{\zeta_{,\chi}^2}\left(\partial_{\chi\chi}V_{{\kappa}=0}-\frac{\zeta_{,\chi\chi}}{\zeta_{,\chi}}\partial_\chi V_{{\kappa}=0}\right) = 0
\end{eqnarray} 
where $\zeta_{\chi}$ and $\zeta_{\chi\chi}$ are also evaluated at ${\kappa}=0$.
Like the \ac{LPA} \ac{EEOM} this can be reduced to a first order differential equation and we can express both of them in the following form: 
\begin{empheq}[box = \fcolorbox{Maroon}{white}]{equation}
\dot{\chi} = -\tilde{V}_{,\chi}(\chi) \label{eq:QEOM}
\end{empheq} 
where we have introduced the \textit{effective dynamical potential} $\tilde{V}$ defined by 
\begin{eqnarray}
\tilde{V}_{\chi}(\chi) \equiv 
\begin{cases}
V_{\chi}({\kappa} = 0, \chi), & \text{ for \ac{LPA}}  \\[10pt]
\dfrac{V_{\chi}({\kappa} = 0, \chi)}{\zeta_{\chi}^{2}({\kappa} = 0, \chi)}, & \text{ for \ac{WFR}} 
\end{cases}\label{eq: Vtilde}
\end{eqnarray}
Here we can clearly see that for \ac{LPA} the \textit{effective} and \textit{effective dynamical} potentials are equivalent whereas \ac{WFR} receives an additional factor. 

Equation (\ref{eq:QEOM}) tells us that the equation of motion for the average position $\chi$ is an extremely simple first order differential equation that appears like a Langevin equation (\ref{eq:langevindimless}) with no noise. This means that once you have obtained the \textit{effective dynamical} potential you can compute the evolution of the average position $\chi$ trivially from any starting position.
\subsection{\label{sec:EEOM_2point}The EEOM for the two point function}
As discussed in section \ref{sec:The Wetterich Equation} the connected 2-point function, $G(t,t')=\left\langle x(t)x(t')\right\rangle_{C}=\delta^2\mathcal{W}/\delta J(t_1)\delta J(t_2)$, and the second functional derivative of the \ac{EA} $\Gamma_{{\kappa}=0}$ are inverse to each other:
\begin{eqnarray}
\int d\tau~\dfrac{\delta^2\Gamma_{{\kappa}=0}}{\delta \chi(t)\delta \chi(\tau)}\dfrac{\delta^2\mathcal{W}_{{\kappa}=0}}{\delta J(\tau)\delta J(t')} = \delta(t-t')\label{eq:2ptW and gamma}
\end{eqnarray} 
Concretely this means that the connected 2-point function $G(t,t')$ satisfies the following equation:
\begin{eqnarray}
\left(\dfrac{d^2}{dt^2} -  \mathcal{U}(\chi) \right)G(t,t') &=& -2\Delta\delta(t-t')  \label{eq:2pointfuncdef} 
\end{eqnarray}
where $\mathcal{U}(\chi)$ is:
\begin{eqnarray}
\mathcal{U}(\chi)= 
\begin{cases}
    V_{,\chi\chi}^{2} + V_{,\chi}V_{,\chi\chi\chi}, & \text{for \ac{LPA}}\\[10pt]
 \dfrac{V_{,\chi\chi}^{2}}{\zeta_{,\chi}^{4}}+ \dfrac{V_{,\chi}V_{,\chi\chi\chi}}{\zeta_{,\chi}^{4}} -\dfrac{4V_{,\chi}V_{,\chi\chi}\zeta_{,\chi\chi}}{\zeta_{,\chi}^{5}}    + \dfrac{4V_{,\chi}^{2}\zeta_{,\chi\chi}^{2}}{\zeta_{,\chi}^{6}} , & \text{for \ac{WFR}}
  \end{cases} \label{eq:U = }
\end{eqnarray}
and
\begin{eqnarray}
\Delta\equiv
\begin{cases}
\dfrac{\Upsilon}{2}, & \text{for \ac{LPA}}\\[10pt]
\dfrac{\Upsilon}{2\zeta_{,\chi}^{2}}, & \text{for \ac{WFR}}
\end{cases}
\label{eq:Deltadef}
\end{eqnarray}
In order to get the general solution for the two-point function we will rewrite (\ref{eq:2pointfuncdef}) in terms of some general functions:
\begin{eqnarray}
\left(\dfrac{d^2}{dt^2} -  Q(t) \right)G(t,t') &=& -\dfrac{\Upsilon}{P(t)}\delta(t-t')  \label{eq:appendix2pointfuncdef} 
\end{eqnarray}
Where for us $Q(t) = U(\chi(t))$ is given by (\ref{eq:U = }) and $P(t) = 1$ or $\zeta_{\chi}^2(\chi(t))$ for \ac{LPA} and \ac{WFR} respectively. We now consider the homogeneous version of (\ref{eq:appendix2pointfuncdef}):
\begin{eqnarray}
\ddot{f}(t) -  Q(t)f(t) = 0 \label{eq:appendhomo2ptfunc}
\end{eqnarray}
which will generically have two independent solutions $Y_1(t)$ and $Y_2(t)$ that we would like to obtain. In order to do this we consider what these solutions asymptote to at late times. We know for large t (denoted by T) the system will reach the equilibrium distribution (or at least will be asymptotically close to it) for which (\ref{eq:appendhomo2ptfunc}) becomes:
\begin{eqnarray}
\ddot{f}(T) - \lambda^2 f(T) = 0 \label{eq:appendhomo2ptfuncequi}
\end{eqnarray}
as $Q(t)$ asymptotes to a time independent quantity $\lambda^2$ as the system approaches equilibrium. For us $\lambda^2$ is defined as in the equilibrium limit of (\ref{eq:U = }):
\begin{eqnarray}\label{eq:lambda}
\lambda^2 \equiv  
\begin{cases}
V_{,\chi\chi}^{2}\vert, & \text{for \ac{LPA}}\\[10pt]
\dfrac{V_{,\chi\chi}^{2}\vert}{\zeta_{,\chi}^{4}\vert}, & \text{for \ac{WFR}}
\end{cases} 
\end{eqnarray}
The notation $\vert$ means we have evaluated the function at ${\kappa} = 0$ and at equilibrium $\chi = \chi_{eq}$. 
Equation (\ref{eq:appendhomo2ptfuncequi}) has two solutions, one growing and one decaying:
\begin{eqnarray}
Y_1(T) &=& A\exp (\lambda T) \label{eq:appendY1equi}\\
Y_2(T) &=& B\exp (-\lambda T)\label{eq:appendY2equi}
\end{eqnarray}
We can now consider the Wronskian $\mathcal{\mathcal{W}}$ which in our case must be constant for all time:
\begin{eqnarray}
\mathcal{W}(t) &\equiv & Y_1(t)\dot{Y}_2(t) - \dot{Y}_1(t)Y_2(t) = \text{constant} \\
\mathcal{W}(T) &=& -2AB\lambda \Rightarrow \mathcal{W}(t) = -2AB\lambda \label{eq:appendWronskianconst}
\end{eqnarray}
We will make use of this fact later. \\
Substituting the ansatz $G(t,t') = Y_1(t)F(t,t')$, where F is some function to be determined, into (\ref{eq:appendix2pointfuncdef}) we obtain:
\begin{eqnarray}
\dot{F}(t,t') = \dfrac{1}{Y_{1}^{2}(t)}\left[-\Upsilon\dfrac{Y_1(t')}{P(t')}\Theta (t-t') + C_1(t')\right] \nonumber \\
\label{eq:append Fdot}
\end{eqnarray}
where $\Theta (t-t')$ is the Heaviside step function as before and $C_1(t')$ is a `constant' of integration function to be determined. If we now integrate (\ref{eq:append Fdot}) we obtain the following expression for $G(t,t')$:
\begin{eqnarray}
G(t,t') = -\Upsilon\dfrac{Y_1(t)}{P(t')}\left[\Theta (t-t') \int_{t'}^{t} \dfrac{Y_1(t')}{Y_{1}^{2}(u)} du + C_2(t')\right]  +~C_1(t')Y_1(t)\int^{t} \dfrac{du}{Y_{1}^{2}(u)} \label{eq:appendGwithints}
\end{eqnarray}
where $C_2(t')$ is another `constant' of integration function to be determined. To compute the integrals in (\ref{eq:appendGwithints}) we note that by the definition of the Wronskian:
\begin{eqnarray}
Y_1(t)\int^t \dfrac{\mathcal{W}(u)}{Y_{1}^{2}(u)}du = \mu Y_1(t) +  Y_2(t)
\end{eqnarray}
where $\mu$ is simply a constant of integration. As the Wronskian is constant we simply write:
\begin{eqnarray}
Y_1(t)\int^t \dfrac{du}{Y_{1}^{2}(u)} = \dfrac{1}{-2AB\lambda}\left[\mu Y_1(t) + Y_2(t)\right]
\end{eqnarray}
Such that (\ref{eq:appendGwithints}) becomes:
\begin{eqnarray}
G(t,t') &=& \dfrac{\Upsilon}{2AB\lambda P(t')}\Big\lbrace\bar{C}_1(t')Y_2(t) + \bar{C}_2(t')Y_1(t) \nonumber \\
&&\qquad \qquad \qquad+~\Theta (t-t')\left[Y_1(t')Y_2(t) - Y_1(t)Y_2(t')\right]\Big\rbrace   \label{eq:append GCbar}
\end{eqnarray}
where $C_1$ and $C_2$ have been rescaled to $\bar{C}_1$ and $\bar{C}_2$ in order to absorb some irrelevant constant factors. We note that the functions $\bar{C}_i$ can only be linear combinations of $Y_1$ and $Y_2$:
\begin{eqnarray}
\bar{C}_1(t') &\equiv &  \alpha ~Y_1(t') + \beta ~Y_2 (t') \\
\bar{C}_2(t') &\equiv &  \gamma ~Y_1(t') + \delta ~Y_2 (t') 
\end{eqnarray}
where the constants $\alpha$, $\beta$, $\gamma$ and $\delta$ will be determined in a moment.\footnote{N.B. the $\delta$ here should not to be confused with the Dirac delta function}. Combining all this together we obtain the most general solution:
\begin{eqnarray}
G(t,t') &=& \dfrac{\Upsilon}{2\lambda P(t')}\dfrac{1}{AB}\Big\lbrace \left[\alpha + \Theta (t-t') \right] Y_1(t')Y_2(t) +~\beta ~Y_2(t')Y_2(t) \nonumber \\
&& \qquad \qquad \qquad +~\gamma ~Y_1(t')Y_1(t) +~\left[\delta - \Theta (t-t')\right]Y_2(t')Y_1(t)\Big\rbrace  \label{eq:append G general}
\end{eqnarray}
To obtain the values of the constants we must impose physical conditions:
\begin{enumerate}
\item The variance, $G(t,t)$, should remain finite as $t\rightarrow \infty$ \\
i.e. an equilibrium distribution exists at late times \\
$\Rightarrow \gamma = 0$
\item The variance should approach the correct equilibrium distribution, $G_{eq}$, at late times T \\
$\Rightarrow \alpha = 0$
\item The covariance, $G(t,0)$, should remain finite as $t\rightarrow \infty$  \\
$\Rightarrow \delta = 1$
\item The initial condition is $G(0,0) \equiv G_{00}$ \\
$\Rightarrow \dfrac{\beta\Upsilon}{2AB\lambda} = \dfrac{P(0)}{Y_2(0)Y_2(0)}\left[G_{00} - \dfrac{Y_1(0)Y_2(0)}{AB}\dfrac{\Upsilon}{2\lambda P(0)}\right]$
\end{enumerate}
Which gives us the two point function:
\begin{eqnarray}
G(t,t') &=& \dfrac{\Upsilon}{2\lambda P(t')}\left[ \Theta (t-t')\tilde{Y}_1(t')\tilde{Y}_2(t) + \Theta (t'-t)\tilde{Y}_2(t')\tilde{Y}_1(t)\right] \nonumber \\
&& \quad +~ \dfrac{P(0)}{P(t')}\left[G_{00} - \dfrac{\Upsilon}{2\lambda P(0)}\right]\tilde{Y}_2(t')\tilde{Y}_2(t) \label{eq:append finalG}
\end{eqnarray}
where $\tilde{Y}_i(t) \equiv Y_i(t)/Y_i(0)$ are the `normalised' solutions to the homogeneous equation (\ref{eq:appendhomo2ptfunc}). We have also set $A = Y_1(0)$ and $B = Y_2(0)$ which we are free to do.
Equation (\ref{eq:append finalG}) has two important limits: \\
The \textit{Variance} $t'\rightarrow t$:
\begin{empheq}[box = \fcolorbox{Maroon}{white}]{equation}
\textbf{Var}(x) \equiv G(t,t) = \dfrac{\Upsilon}{2\lambda P(t)} \tilde{Y}_1(t)\tilde{Y}_2(t)  +~ \dfrac{P(0)}{P(t)}\left[G_{00} - \dfrac{\Upsilon}{2\lambda P(0)}\right]\tilde{Y}_{2}^{2}(t)
 \label{eq:EEOM Variance}
\end{empheq}
and the \textit{Covariance} $t'\rightarrow 0$, $t > 0$:
\begin{empheq}[box = \fcolorbox{Maroon}{white}]{equation}
\textbf{Cov}(x(0)x(t)) \equiv G(t,0) = G_{00}\tilde{Y}_2(t) \label{eq:EEOM Covariance}
\end{empheq}
Equations (\ref{eq:EEOM Variance}) \& (\ref{eq:EEOM Covariance}) are the main results of this section.
\section{\label{sec:observables}The equilibrium limit}
While the \ac{EEOM} derived in Section \ref{sec:acc eom} are valid for non-equilibrium evolution it is important to ensure that they converge to the correct equilibrium limit. At equilibrium the equations are greatly simplified resulting in the equilibrium position $\chi_{eq}$ and variance $\textbf{Var}_{eq}(x)$ becoming static quantities as expected. We will also show how the covariance at equilibrium is given by an exponential decay with exponent predicted by the \ac{FRG}.


\subsection{Equilibrium 1-point function}
\label{sec:1-point function}
At equilibrium the average position of the particle should not change, this means that $\dot{\chi} = 0$. It naturally follows from this condition and the \ac{EEOM} for $\chi$ (\ref{eq:QEOM}) that equilibrium is defined for both \ac{LPA} \& \ac{WFR} by the condition 
\begin{eqnarray}
\partial_\chi V_{{\kappa} = 0}(\chi_{eq}) = 0 \label{eq:equilbirum position}
\end{eqnarray}  
As the potential $V_{{\kappa}=0}(\chi)$ should be convex (by definition of $\Gamma$) equation (\ref{eq:equilbirum position}) tells us that $\chi_{eq}$ corresponds to the minimum of  $V_{{\kappa}=0}(\chi)$. Or more concretely: 
\begin{eqnarray}
\lim\limits_{t \to \infty} \left\langle x(t) \right\rangle =  x \text{ that minimises } V_{{\kappa}=0}(x)
\end{eqnarray}
The equilibrium position is obviously the same for both \ac{LPA} and \ac{WFR} as they both lead to the same effective potential. As the equilibrium position is straightforwardly computed from the Boltzmann distribution verifying that the minimum of the effective potential matches the predicted equilibrium position is a good first test that the procedure to obtain the numerical solution to (\ref{eq:dV/dk}) we have outlined here is valid.

\begin{table}[t!]
	\centering
	\begin{tabular}{l l l l l}
		\toprule
		Potentials & $\Upsilon$ & Boltz & \ac{LPA} \\
		\midrule
		Polynomial 	& 10   & -0.9618 & -0.96   \\
		& 2 & -1.3227 & -1.33   \\
		& 1 & -1.5170 & -1.52   \\
		Unequal L-J & 10   & 0.4854  & 0.485   \\
		& 2 & 1.8522 & 1.85   \\
		& 1 & 1.8684  & 1.87    \\
		$x^2$ plus six bumps/dips	& 5 & 0.0531 & 0.055 \\
		& 2 & 0.1597 & 0.16 \\
		\bottomrule 
	\end{tabular}\captionof{table}{$\chi_{eq}$ as calculated from the Boltzmann distribution and the \ac{LPA} effective potential.} 
	\label{tabel:X0 for different potentials and methods}
\end{table}

Let us consider the (normalised) equilibrium Boltzmann distribution defined in the standard way:
\begin{eqnarray}
P(x) = N\text{exp}\left(-\dfrac{2 V(x)}{\Upsilon}\right)
\end{eqnarray}
where $N$ is chosen so that $\int_{-\infty}^{\infty}P(x) = 1$. We can then compute $\chi_{eq}$ from the equilibrium probability distribution function:
\begin{eqnarray}
\int_{-\infty}^{\infty}x\cdot P(x) = \chi_{eq}
\end{eqnarray}
\begin{figure}[t!]
	\centering
		\includegraphics[width=0.45\textwidth]{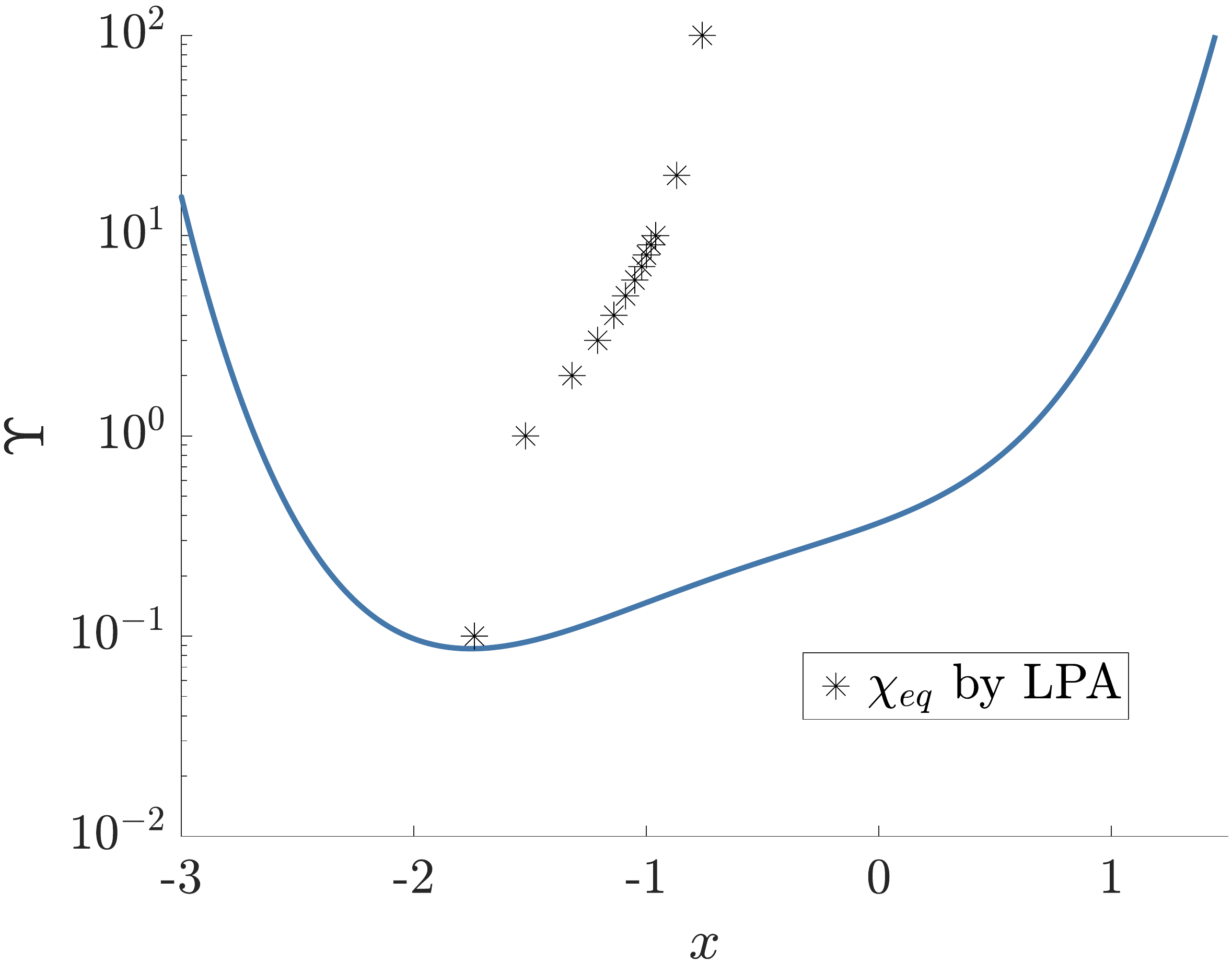}
		\includegraphics[width=0.45\textwidth]{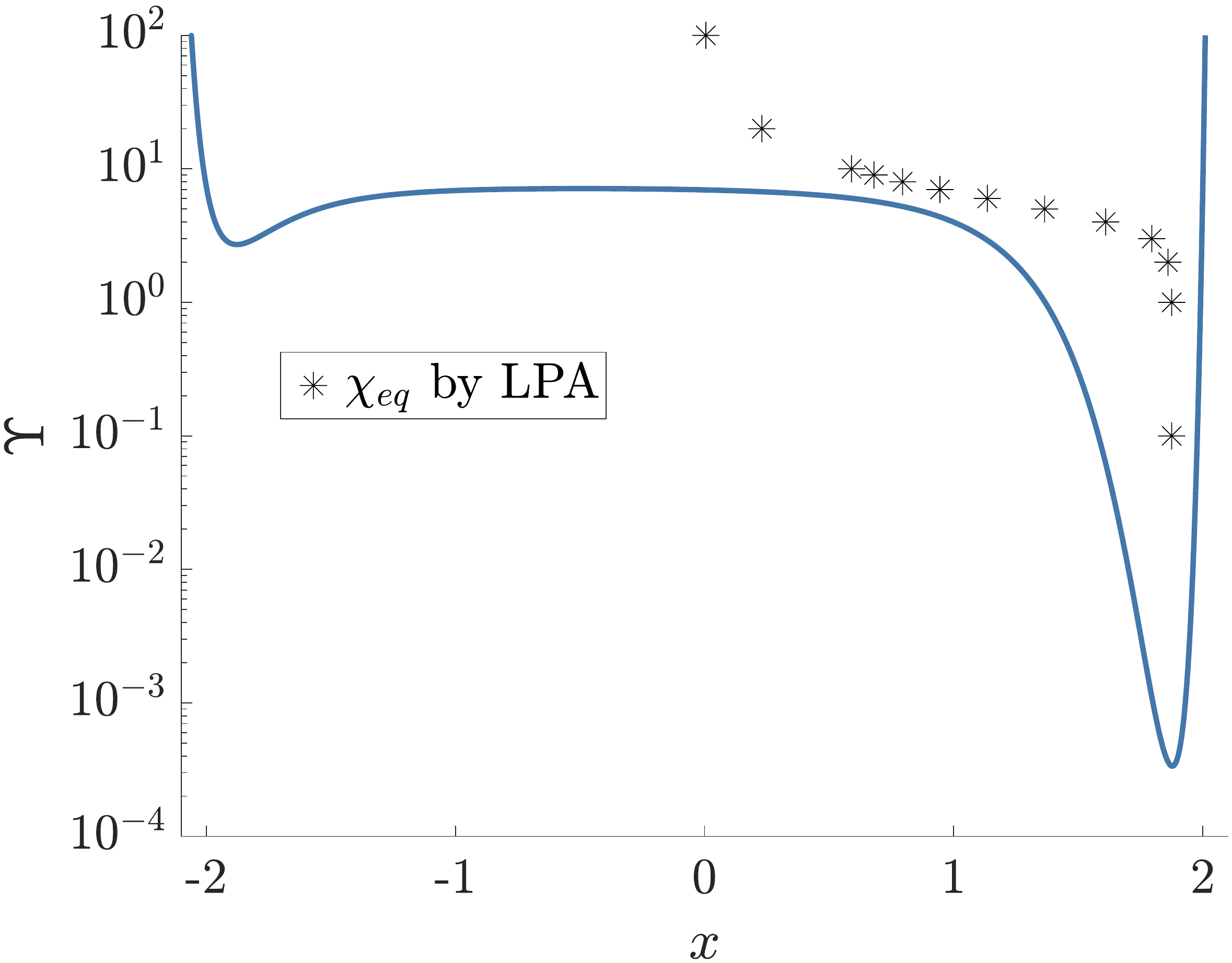}
		\caption[\ac{LPA} prediction for equilibrium position in polynomial and Lennard-Jones type potentials]{The value of $\chi_{eq}$ for different values of the thermal energy $\Upsilon$ in the polynomial potential (left) and unequal Lennard-Jones type potential (right) as calculated via the \ac{LPA}. The original bare polynomial Langevin potential is plotted (not to scale) in blue for context}
		\label{fig: X0 for Poly and ULJ}
\end{figure}
Looking at Table.~\ref{tabel:X0 for different potentials and methods} we can see that the \ac{LPA} matches the Boltzmann distribution extremely well for a wide range of different potentials across the range of temperatures we examined. In Fig.~\ref{fig: X0 for Poly and ULJ} we have plotted the \ac{LPA} prediction for the average position as the thermal energy $\Upsilon$ of the system is lowered for the polynomial potential (left) and unequal L-J (right). As $\Upsilon$ is lowered the equilibrium position shifts closer to the original potential's minimum. This is particularly stark in the right plot of Fig.~\ref{fig: X0 for Poly and ULJ} as it is clear at high temperature the equilibrium position is in the middle of the two wells suggesting a roughly symmetric Boltzmann distribution. However as temperature is lowered the $\chi_{eq}$ moves into the deeper well indicating that particles at equilibrium at low temperatures would nearly always be in this region as one would expect. Table \ref{tabel:X0 for different potentials and methods} verifies that the numerical solution to the \ac{LPA} flow equation (\ref{eq:dV/dk}) is accurate in capturing this crucial physical aspect of the system at equilibrium. 
\subsection{\label{sec:equi_2point}Equilibrium 2-point function}
If we now take the equilibrium limit $\chi \rightarrow \chi_{eq}$ of the full \ac{EEOM} for the 2-point function (\ref{eq:2pointfuncdef}) we find that it simplifies to:
\begin{eqnarray}
\left(\dfrac{d^2}{dt^2} -  \lambda^2\right)G_{eq}(t_1,t_2) =  -2\Delta \vert\delta(t_2-t_1)  \label{eq:2pointfuncequi} 
\end{eqnarray}
where, as before,
\begin{eqnarray}\label{eq:lambdadef}
\lambda^2 \equiv  
\begin{cases}
V_{,\chi\chi}^{2}\vert, & \text{for \ac{LPA}}\\[10pt]
\dfrac{V_{,\chi\chi}^{2}\vert}{\zeta_{,\chi}^{4}\vert}, & \text{for \ac{WFR}}
\end{cases} 
\end{eqnarray}
and $\Delta$ is defined as in (\ref{eq:Deltadef}). The notation $\vert$ means we have evaluated the function at ${\kappa} = 0$ and at equilibrium $\chi = \chi_{eq}$. \\
The appropriate solution to (\ref{eq:2pointfuncequi}) providing the connected correlation function at equilibrium is  
\begin{eqnarray}
G_{eq}(t_1,t_2) = \textbf{Cov}_{eq}(x(t_1)x(t_2))  &=& \dfrac{\Upsilon}{2V_{,\chi\chi}\vert} e^{-\lambda|t_1-t_2|} \label{eq:2-pointfunc} \\
\Rightarrow G_{eq}(t,t) = \textbf{Var}_{eq}(x) &=& \dfrac{\Upsilon}{2V_{,\chi\chi}\vert}
\label{eq:equal2pt}
\end{eqnarray}
As the equilibrium variance is also easily computed from the Boltzmann distribution, equation (\ref{eq:equal2pt}) gives us a second test to verify that the effective potential and by extension the \ac{FRG} recipe we have outlined has physical significance. 
In the \ac{LPA} approximation the variance and the decay rate of the autocorrelation function are both directly given by the curvature of the effective potential at its minimum. The inclusion of \ac{WFR} however alters the decay rate without changing the equilibrium variance. This is as it should since the latter is fixed by the equilibrium Boltzmann distribution. 
\subsubsection{Variance}
\begin{table}[t!]
	\centering
	\begin{tabular}{l l l l l}
		\toprule
		Potentials & $\Upsilon$ & Boltz & \ac{LPA}  \\
		\midrule
		Polynomial 	& 10 & 1.5690 & 1.5695 \\
					& 2 & 0.5931 & 0.5894  \\
					& 1 & 0.2938 & 0.2922 \\
		Doublewell 	& 10 & 2.2198 & 2.2199 \\
					& 2 & 1.6655 & 1.6655  \\
					& 1 & 1.7043 & 1.7042  \\
		Unequal L-J & 10 & 1.6858 & 1.6860 \\
					& 2 & 0.01426 & 0.01963  \\
		$x^2$ plus two bumps & 10 & 2.5317 & 2.5759 \\
					& 2 & 0.4145 & 0.4145  \\
					& 1 & 0.1554 & 0.1554 \\
		$x^2$ plus six bumps/dips & 5 & 1.2550 & 1.2551  \\
			& 3 & 0.7824 & 0.7813  \\
			& 2 & 0.5497 & 0.5499  \\
		
		\bottomrule 
	\end{tabular} \captionof{table}{Variance as calculated from the Boltzmann distribution and the \ac{LPA} effective potential. } 
	\label{tabel:VarX for different potentials and methods}
\end{table}
We begin by examining the time-independent quantity at equilibrium, the variance, for the potentials studied in section \ref{sec:LPA-flow}. Equation (\ref{eq:equal2pt}) tells us clearly that the variance is related to how flat the ${\kappa} = 0$ potential is near the minimum, controlled by $V_{,\chi\chi}$ at the equilibrium point. Unsurprisingly, the bigger the curvature of the effective potential, the smaller the variance for a fixed temperature. As the temperature is lowered the equilibrium distribution is confined to a smaller and smaller region of the potential energy surface with a smaller variance. If variance changed linearly with temperature we can see from (\ref{eq:equal2pt}) that $V_{,\chi\chi}\vert$ would not change as temperature was varied. However this variance does not generically scale linearly with temperature which is why the ${\kappa} = 0$ curves in Figs.~\ref{fig: Gies_LPA_V_5T.png} \& \ref{fig: DW_LPA_V_5T.png} are generically flatter about the equilibrium point for $\Upsilon = 1$ than for $\Upsilon = 10$ in order to accommodate the fact that the equilibrium variance decreases by less than a factor of 10 for $\Upsilon = 10 \rightarrow 1$; $V_{,\chi\chi}$ near the equilibrium point must \textit{decrease} as temperature is lowered. However in Fig.~\ref{fig: X2Gauss_LPA_V_5T.png} the $\Upsilon = 10 \rightarrow 1$ transition marks a bigger transition in equilibrium behaviour due to the particle at equilibrium being now mostly trapped at the origin instead of spread out over the bumps. This means that the variance decreases by more than a factor of 10 as $\Upsilon = 10 \rightarrow 1$ and therefore $V_{,\chi\chi}$ near the equilibrium point must \textit{increase} as temperature is lowered resulting in a steeper curve at ${\kappa} = 0$. The takeaway point is that lowering temperature in a particular range can generically make the effective potential flatter or steeper around the minimum depending on the scaling of variance with temperature in that regime.

Either way, once the \ac{FRG} flow equations have been solved, calculating the effective potential's curvature at the minimum is very straightforward. Our results are summarised in Table.~\ref{tabel:VarX for different potentials and methods} and it is clear that the \ac{LPA} offers very good agreement for the variance of the equilibrium distribution for all the potentials examined. 

\subsubsection{\label{sec:equi_cov}Covariance}

In addition to the static variance at equilibrium, the curvature of the effective potential around the minimum also determines the time dependence of correlations in equilibrium, quantified by the time dependent covariance or connected 2-point function (\ref{eq:2-pointfunc}). Furthermore, now the solution to the \ac{WFR} flow equation (\ref{eq:dzetax/dktilde}) for $\zeta_{,\chi}$ also contributes, providing a correction to the decay rate $\lambda$.    

\begin{figure}[t!]
	\centering
		\includegraphics[width=0.45\textwidth]{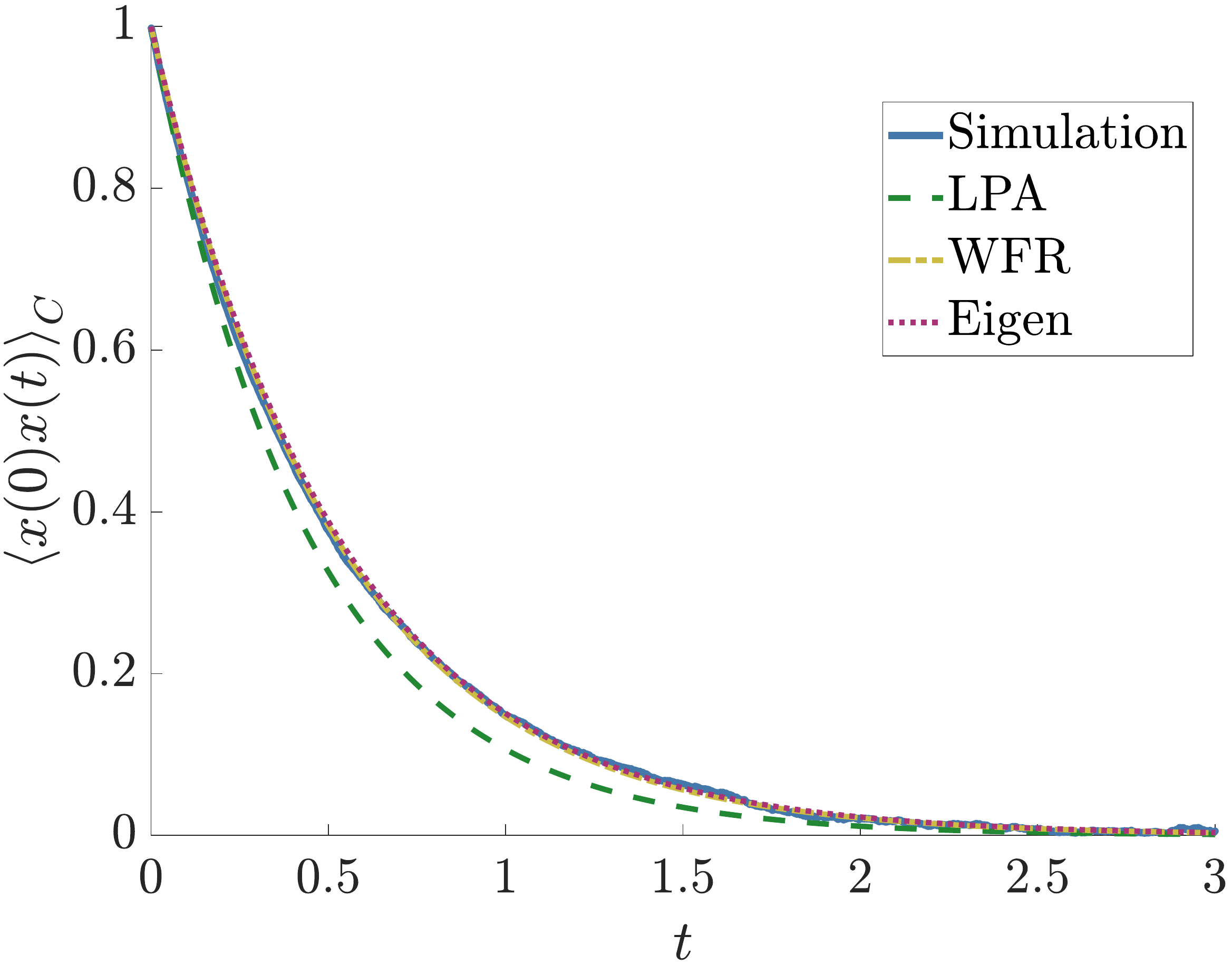}		
		\includegraphics[width=0.45\textwidth]{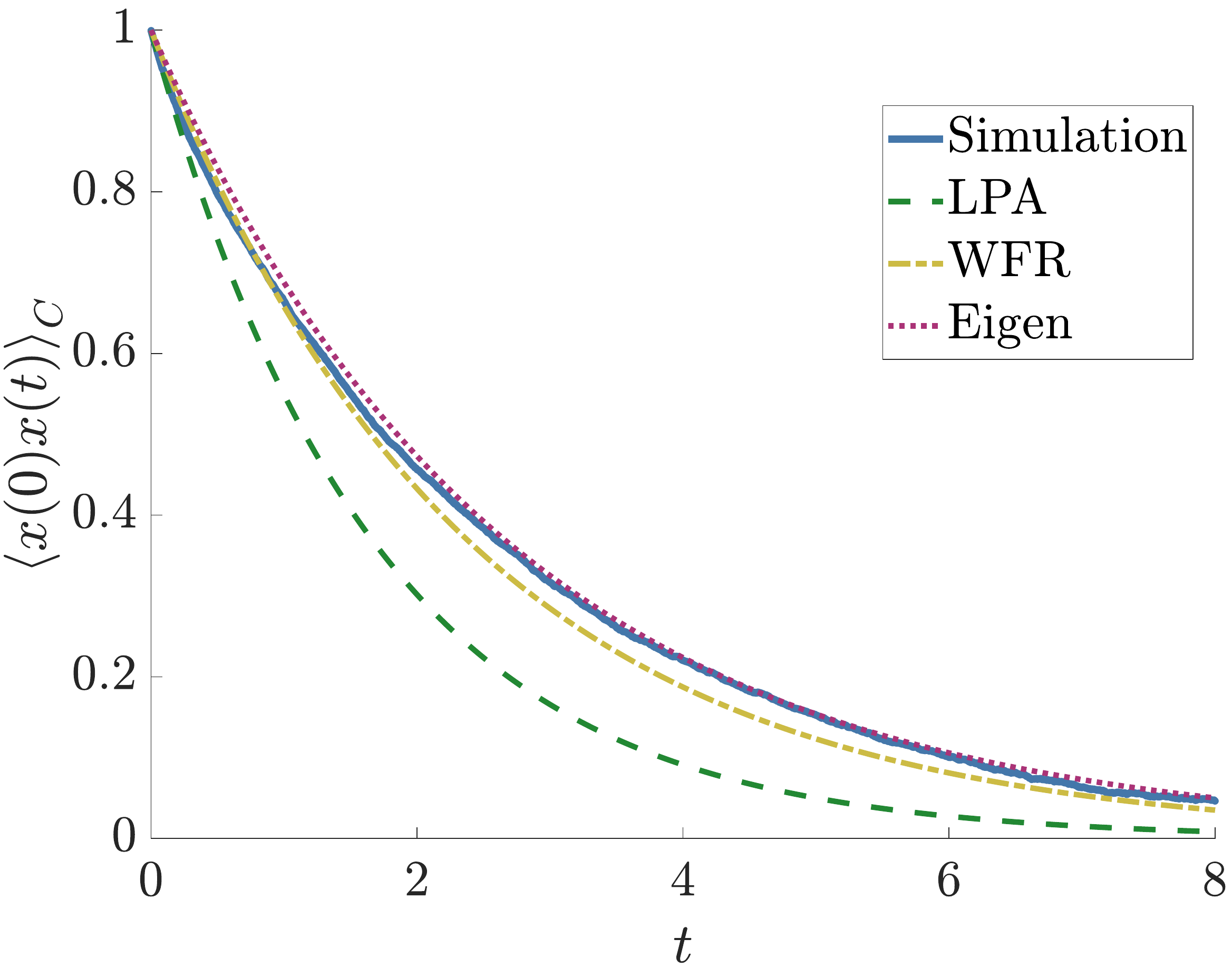}
		\caption[Equilibrium covariance in a doublewell potential]{The decay of the (normalised) covariance $\left\langle x(0)x(t)\right \rangle_C$ at equilibrium in a doublewell potential for $\Upsilon = 10$ (left) and $\Upsilon = 2$ (right). }
		\label{fig: 2pt_eq_DW.png}
\end{figure}

\begin{figure}[t!]
	\centering
		\includegraphics[width=0.45\textwidth]{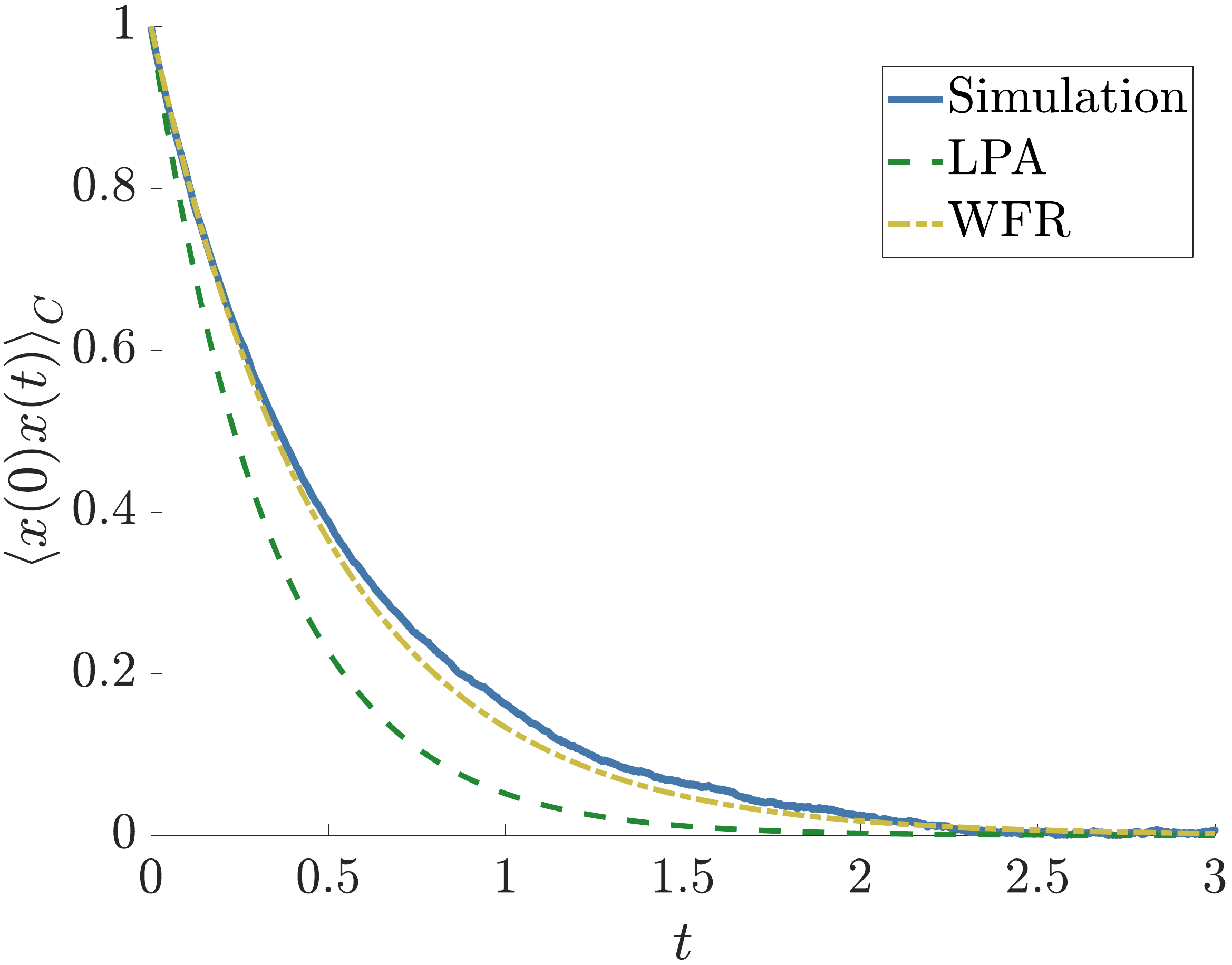}
		\includegraphics[width=0.45\textwidth]{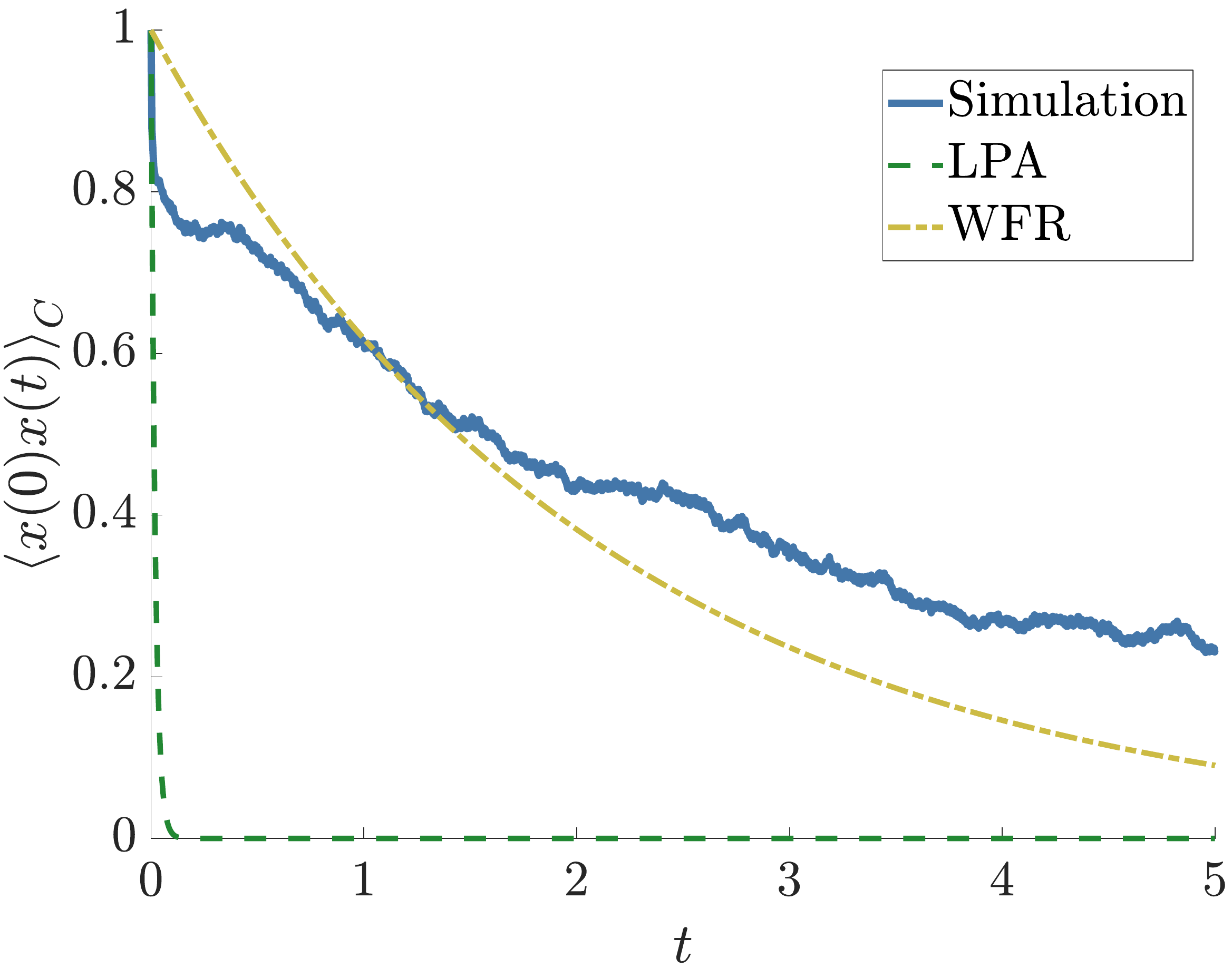}
		\caption[Equilibrium covariance in a Lennard-Jones type potential]{The decay of the (normalised) covariance $\left\langle x(0)x(t)\right \rangle_C$ at equilibrium in the unequal L-J potential for $\Upsilon = 10$ (left) and $\Upsilon = 1$ (right).}
		\label{fig: 2pt_eq_ULJ.png}
\end{figure}

\begin{figure}[t!]
	\centering
		\includegraphics[width=0.45\textwidth]{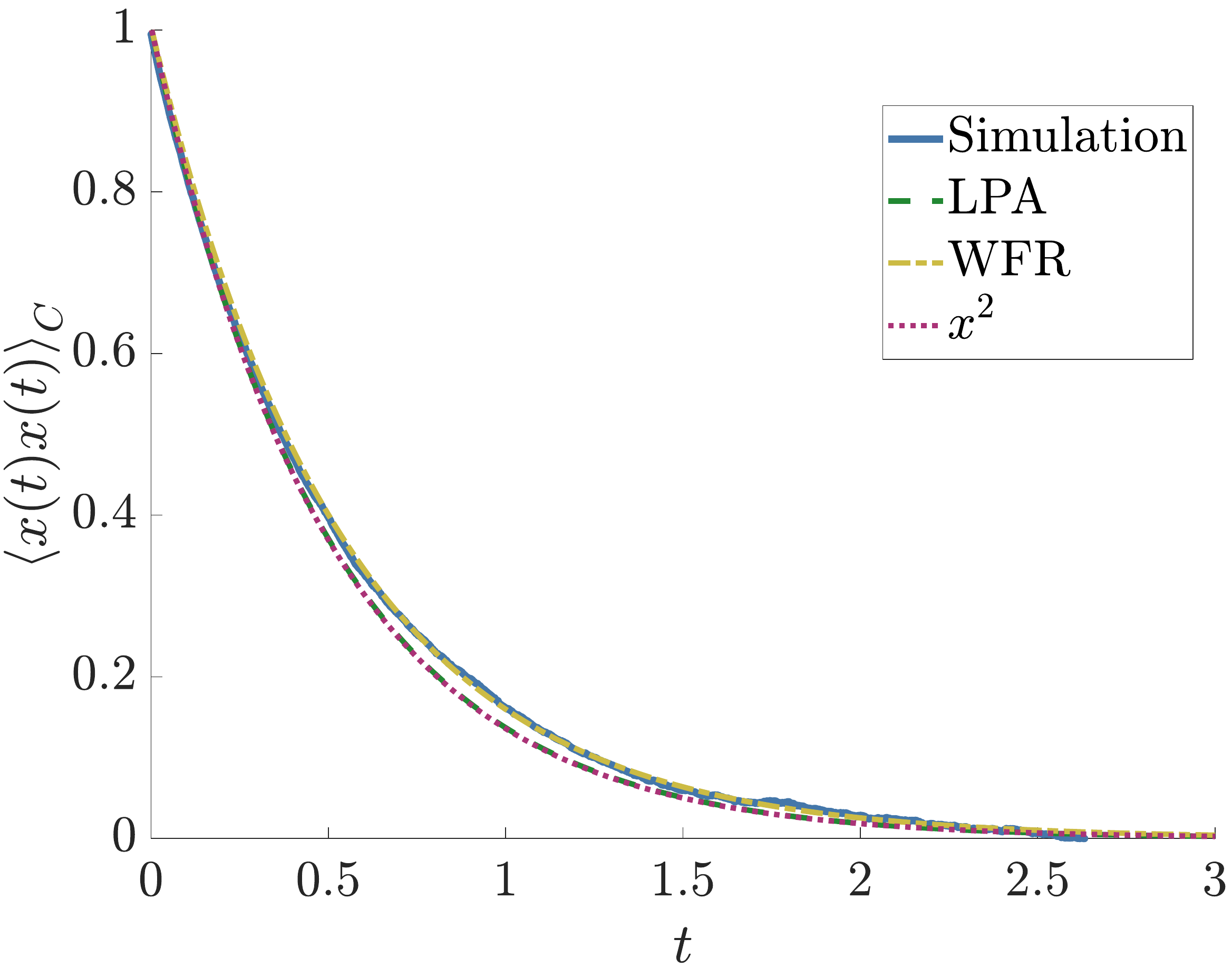}		
		\includegraphics[width=0.45\textwidth]{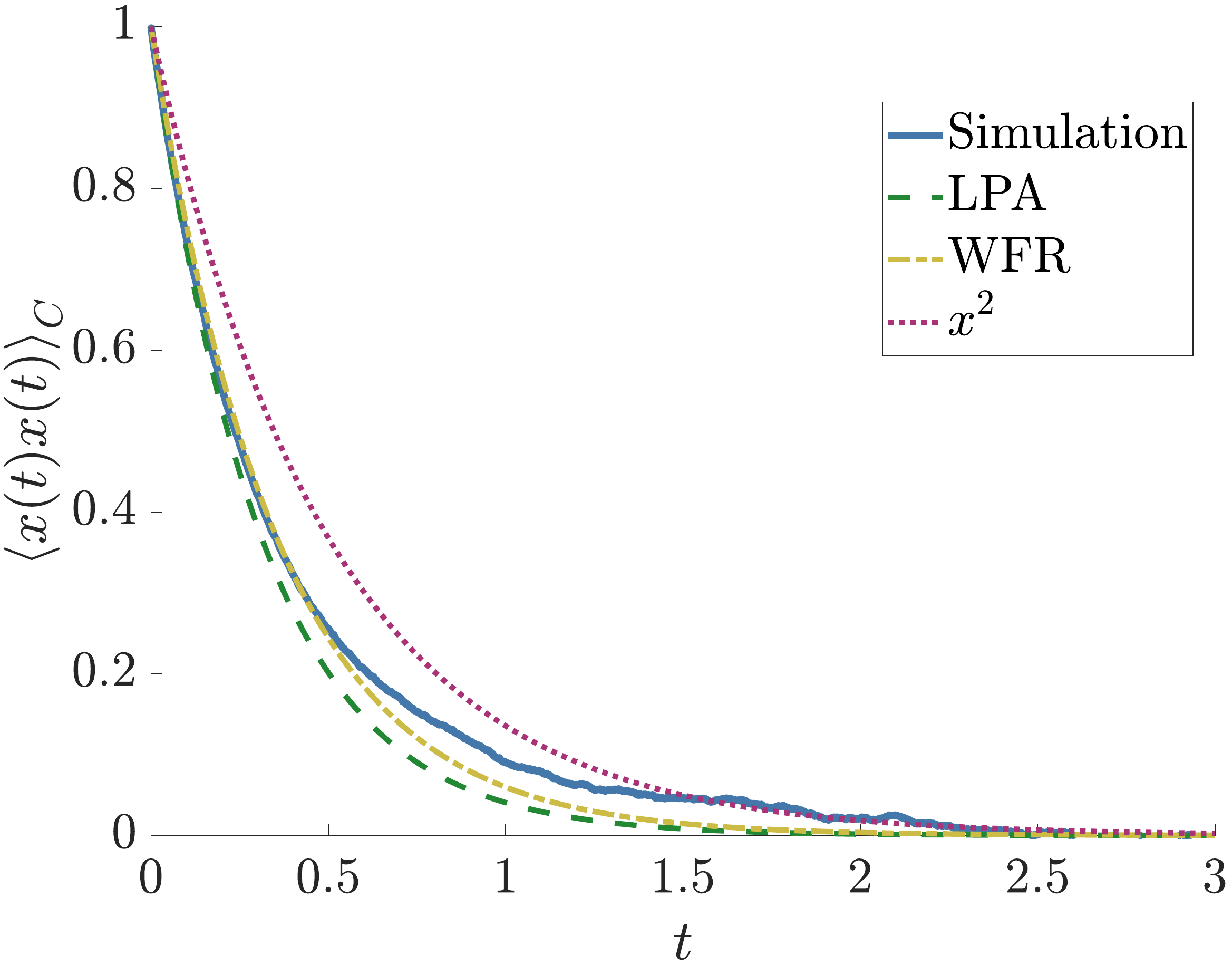}
		\caption[Equilibrium covariance in a $x^2$ plus two bumps potential]{The decay of the (normalised) covariance $\left\langle x(0)x(t)\right \rangle_C$ at equilibrium in an $x^2$ plus two Gaussian bumps potential for $\Upsilon = 4$ (left) and $\Upsilon = 1$ (right).}
		\label{fig: 2pt_eq_X2Gauss.png}
\end{figure}

\begin{figure}[t!]
	\centering
		\includegraphics[width=0.45\textwidth]{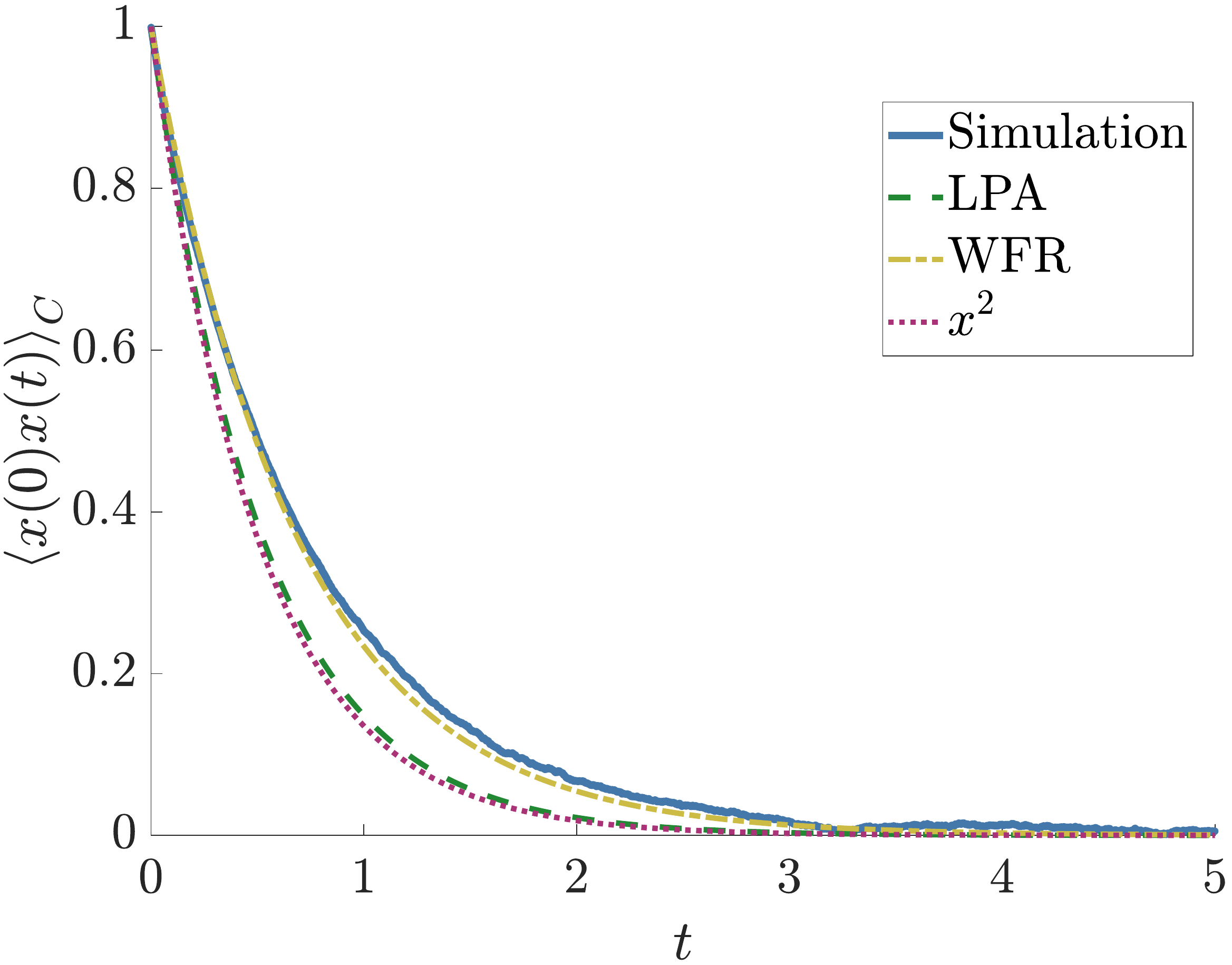}		
		\includegraphics[width=0.45\textwidth]{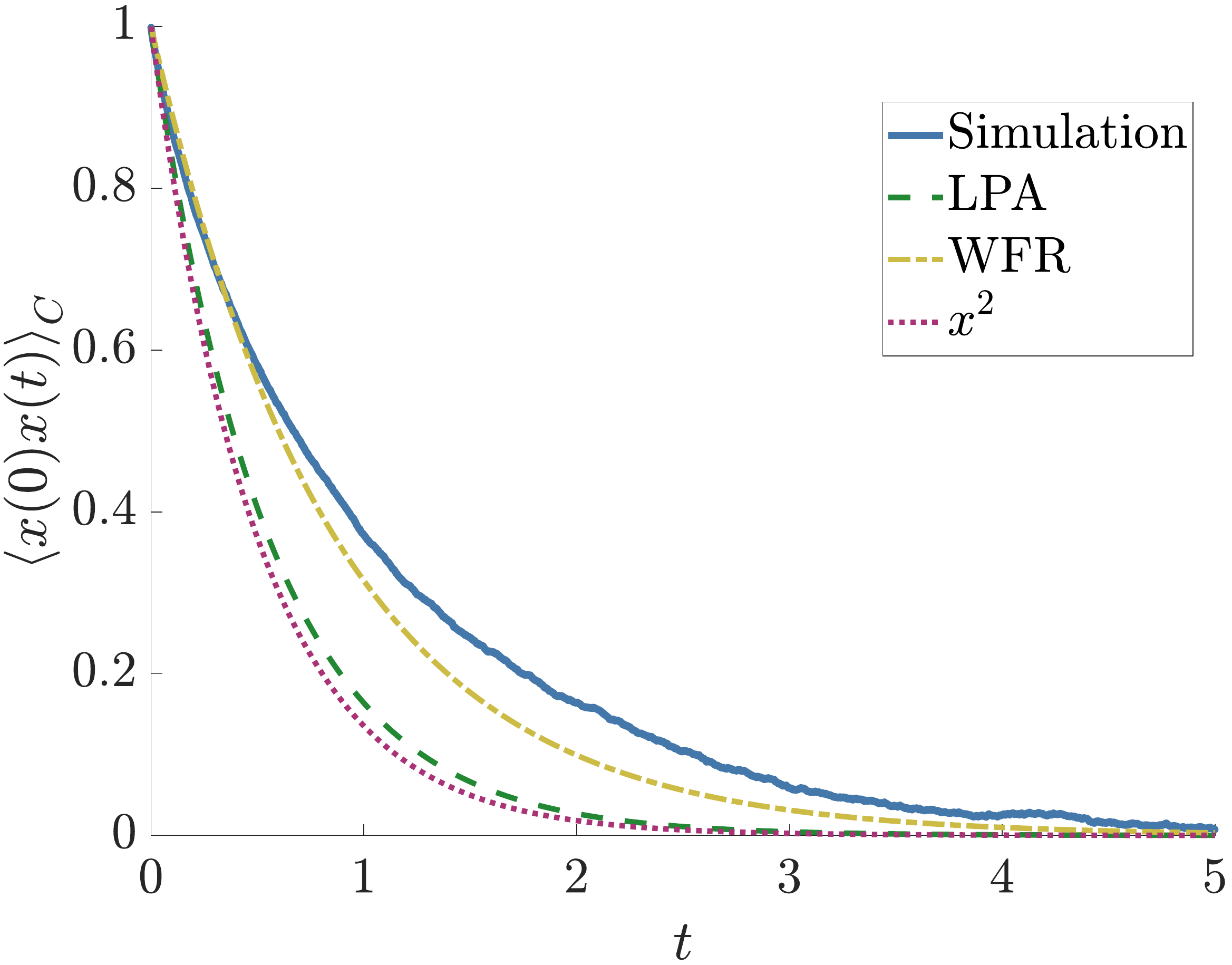}
		\caption[Equilibrium covariance in a $x^2$ plus six bumps potential]{The decay of the (normalised) covariance $\left\langle x(0)x(t)\right \rangle_C$ at equilibrium in an $x^2$ plus six Gaussian bumps/dips potential for $\Upsilon = 3$ (left) and $\Upsilon = 2$ (right).}
		\label{fig: 2pt_eq_X2GaussMany.png}
\end{figure}

In Table.~\ref{tabel:effective mass for different potentials and methods} we collect the values of $\lambda$ obtained using the \ac{FRG} under \ac{LPA} \& \ac{WFR} for different  $\Upsilon $ values, higher or comparable to the typical depth or barrier heights of the different potentials, and compare this directly to numerical simulations of the Langevin equation (\ref{eq:langevindimless}). Where possible we also computed the first non-zero eigenvalue $E_1$ by diagonalising the Hamiltonian from the Schr\"{o}dinger (or rescaled \ac{F-P}) equation in (\ref{eq:FP1}). We can clearly see from Table. \ref{tabel:effective mass for different potentials and methods} that the \ac{LPA} can have good agreement with the simulation value for simple potentials at high temperature but can deviate drastically more drastically as temperature is lowered. Inclusion of the \ac{WFR} factor $\zeta_{,\chi}$ reduces the deviation error from the value obtained in the simulations substantially to $\sim 1\% $ for the simplest cases and order of magnitude agreement for the most complicated, low temperature systems.\\
\begin{figure}[t!]
	\centering
		\includegraphics[width=0.45\textwidth]{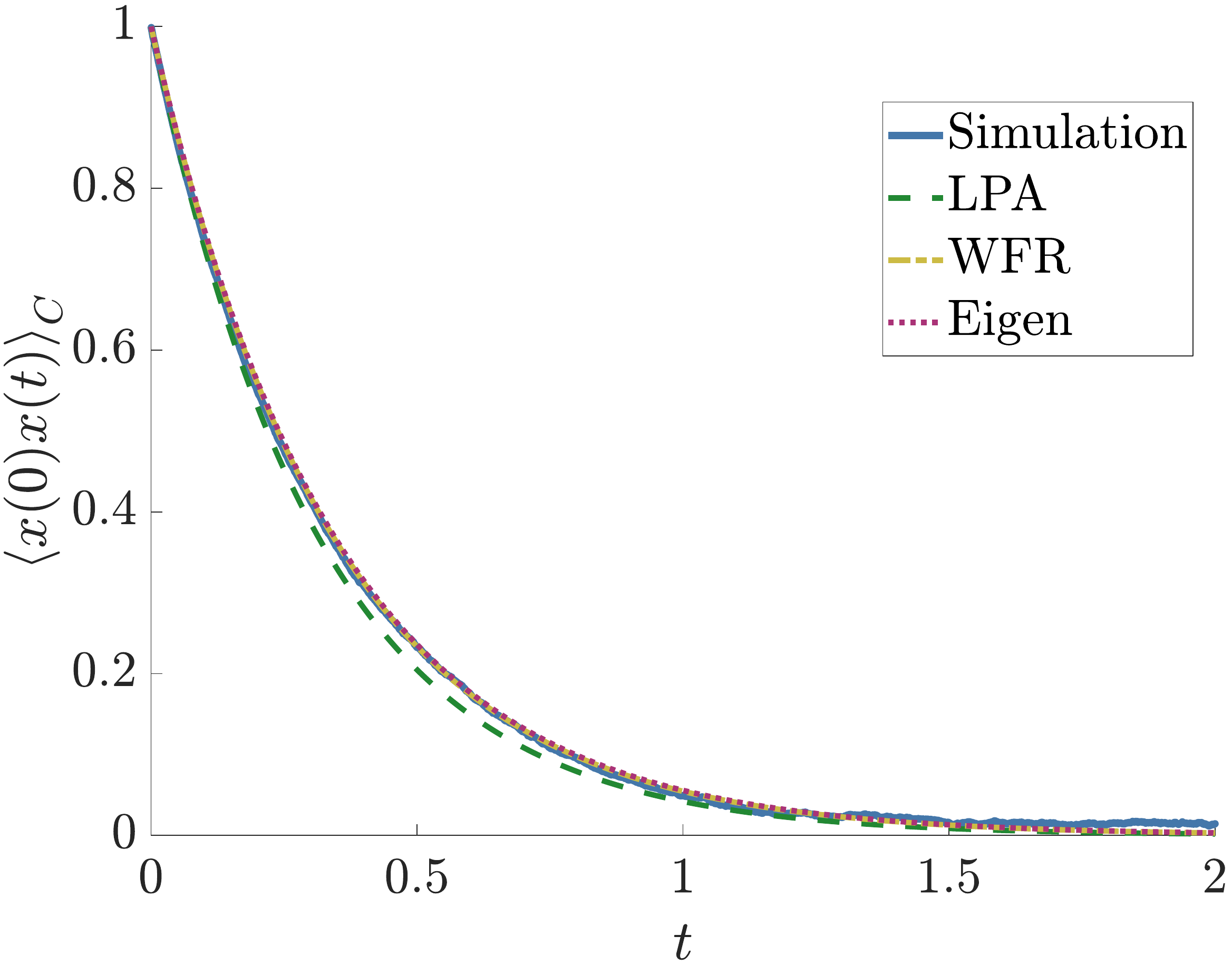}		
		\includegraphics[width=0.45\textwidth]{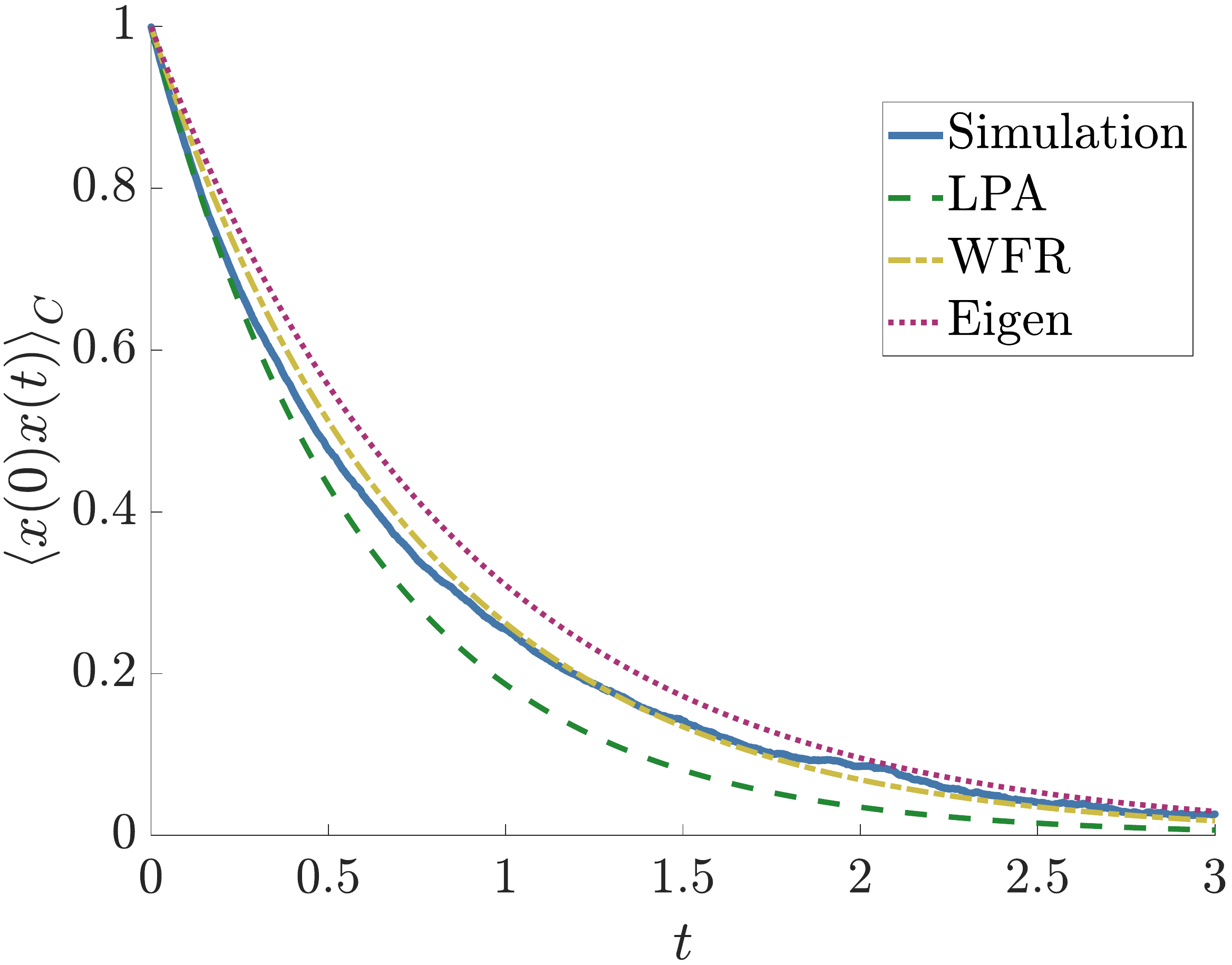}
		\caption[Equilibrium covariance in a polynomial potential]{The decay of the (normalised) covariance $\left\langle x(0)x(t)\right \rangle_C$ at equilibrium in a polynomial potential for $\Upsilon = 10$ (left) and $\Upsilon = 1$ (right).}
		\label{fig: 2pt_eq_Poly.png}
\end{figure}
We plot the decay of the covariance at equilibrium for our five potentials of interest in Figs. \ref{fig: 2pt_eq_Poly.png}, \ref{fig: 2pt_eq_DW.png}, \ref{fig: 2pt_eq_ULJ.png}, \ref{fig: 2pt_eq_X2Gauss.png} \& \ref{fig: 2pt_eq_X2GaussMany.png}  as calculated by \ac{FRG} techniques compared to direct numerical simulations of the Langevin equation. For a polynomial potential, as shown in Fig.~\ref{fig: 2pt_eq_Poly.png}, we can see how the decay rate as calculated via the \ac{FRG} for both \ac{LPA} and \ac{LPA} + \ac{WFR} closely matches the simulations at both high and low temperature. Fig.~\ref{fig: 2pt_eq_DW.png} shows the decay in the doublewell which at $\Upsilon = 10$ (left) shows great agreement with the simulation and Schr\"{o}dinger calculation of $\lambda$ with \ac{FRG} methods. At $\Upsilon = 2$ (right) of Fig.~\ref{fig: 2pt_eq_DW.png} however we can see that the \ac{LPA} is poorly capturing the correct decay rate and the improvement gained by including \ac{WFR} offers is much more dramatic. \\
\begin{table}[t!]
	\centering
		\begin{tabular}{l l l l l l}
			\toprule
			Potentials & $\Upsilon$ & \ac{LPA} & \ac{WFR} & Sim & $E_1$ \\
			\midrule
			Poly & 10 &  3.1857 & 2.9101 & 2.9191 & 2.8882\\
			& 4 & 2.0842 & 1.8664 & 1.8767  & 1.8260  \\
			& 1 & 1.7112 & 1.3381 & 1.3585 & 1.1733 \\
			Doublewell & 10 & 2.2252 & 1.9192 & 1.9274  & 1.8918\\
			& 2 & 0.6004 & 0.4188 & 0.3851 & 0.3744 \\
			& 1 & 0.2934 & 0.1700 & 0.1199 & 0.1136 \\
			Unequal L-J & 10 & 2.9655 & 2.0146 & 1.8783 & ---\\
			& 2 & 50.9312 & 0.4806 & 0.3691  & ---\\
			$x^2$ + 2 bumps & 10 & 1.9411 & 1.9140 & 1.9367 & ---\\
			& 2 & 2.4125 & 1.9881 & 1.9778   & ---\\
			& 1 & 3.2175 & 2.8184 & 2.7213 & ---\\
			$x^2$ + 6 b/d & 3 & 1.9199 & 1.4529 & 1.3977 & ---\\
			& 2& 1.8185 & 1.1552 & 0.9710 & ---\\
			& 1& 1.9667 & 0.7769 & 0.3725 & --- \\
			\bottomrule 
		\end{tabular}
	 \caption[Value of $\lambda $ as computed by \ac{FRG}, simulations and solving the Schr\"{o}dinger equation]{Value of the autocorrelation decay rate obtained for various potentials at different temperatures by different methods. The \ac{LPA} \& \ac{WFR} columns display $\lambda $ as calculated from the \ac{FRG} flow. The simulation values were generated by averaging over 50,000 runs. 
	 }
	  	 \label{tabel:effective mass for different potentials and methods}
\end{table}

The calculation of $E_1$ from the Schr\"{o}dinger equation has proved a non-trivial numerical exercise for the non-polynomial potentials, hence its omission from Table.~\ref{tabel:effective mass for different potentials and methods} and Figs.~\ref{fig: 2pt_eq_ULJ.png}, \ref{fig: 2pt_eq_X2Gauss.png} \& \ref{fig: 2pt_eq_X2GaussMany.png}. Here the \ac{FRG} offers a very real advantage over more conventional methods to calculating this decay rate as we do not have to develop special numerical routines for every potential of interest, we simply solve the same two flow equations (\ref{eq:dV/dk}) \& (\ref{eq:WFR_flow_eqn}). We can see in the left plot of Fig.~\ref{fig: 2pt_eq_ULJ.png} for $\Upsilon = 10$ how the \ac{LPA} + \ac{WFR} decay rate closely matches the simulated decay at high temperature with the advantage of being calculated much more quickly than the direct simulation. At $\Upsilon = 2$ we can see that while the \ac{WFR} agreement with simulation is relatively poor, it is a massive improvement over the \ac{LPA} prediction. 

For our $x^2$-plus-bumps potentials the decay rate is shown for two and six bumps/dips in Figs.~\ref{fig: 2pt_eq_X2Gauss.png} \& \ref{fig: 2pt_eq_X2GaussMany.png} respectively and as in the unequal L-J case the computation of eigenvalues for these potentials is a non-trivial exercise. We can see in the left plot ($\Upsilon = 10$) of Fig.~\ref{fig: 2pt_eq_X2Gauss.png} that the \ac{LPA} and \ac{WFR} both in good agreement with simulations and in the right plot ($\Upsilon = 1$) the two decays correctly bound the simulated decay -- we will discuss this more in a moment. In Fig.~\ref{fig: 2pt_eq_X2GaussMany.png} we can see that for $\Upsilon = 3$ the \ac{LPA} and \ac{WFR} predictions appropriately bound the simulated decay with the simulations asymptoting to the \ac{WFR} decay at late times.  This indicates that even for highly non-trivial systems where the simulated decay is vastly different from the bare $x^2$ potential -- see Table. \ref{tabel:effective mass for different potentials and methods} and compare to the $x^2$ prediction for $\lambda /2$ which is 1 -- the \ac{FRG} can appropriately capture these effects.
 
It is also worth pointing out that the simulated decay rate does not appear to follow a pure exponential at all times in all cases. This can be best seen in the right plot of Fig.~\ref{fig: 2pt_eq_X2Gauss.png} -- also Fig.~\ref{fig: 2pt_eq_X2GaussMany.png} --  where the decay initially closely follows the \ac{LPA} decay before moving towards the \ac{WFR} decay rate at later times. This sort of behaviour has been identified in similar systems in the early universe \cite{Markkanen2020} where it was noticed that the smallest non-zero eigenvalue's spectral coefficient was sufficiently small for higher order eigenvalues to dominate the decay at earlier times. We speculate that this is the reason for the simulated decay not following a true exponential in all cases and emphasises the inadequacy of $E_1$ to accurately describe covariance for all systems of interest. As \ac{LPA} matches the decay rate predicted by the Boltzmann distribution and \ac{WFR} is closer to the decay predicted by $E_1$ it is apparent why having both is highly useful and why it is nice to be able to get both in the same framework.

\section{\label{sec:Accelerated}Accelerated dynamics out of equilibrium}
In order to solve the equations of motion for the one point function $\chi (t)$ and two point function $G(t,t')$ we must first solve the PDEs for the \ac{LPA} \& \ac{WFR} to obtain the \textit{dynamical effective potential} $\tilde{V}$ and the function $\mathcal{U}$. We will use the solutions obtained in section \ref{sec:FRG SOL} in order to compute these parameters and then solve the appropriate \ac{EEOM}.  
\subsection{The dynamical effective potentials}

\begin{figure}[t!]
\centering
\includegraphics[width=0.45\textwidth]{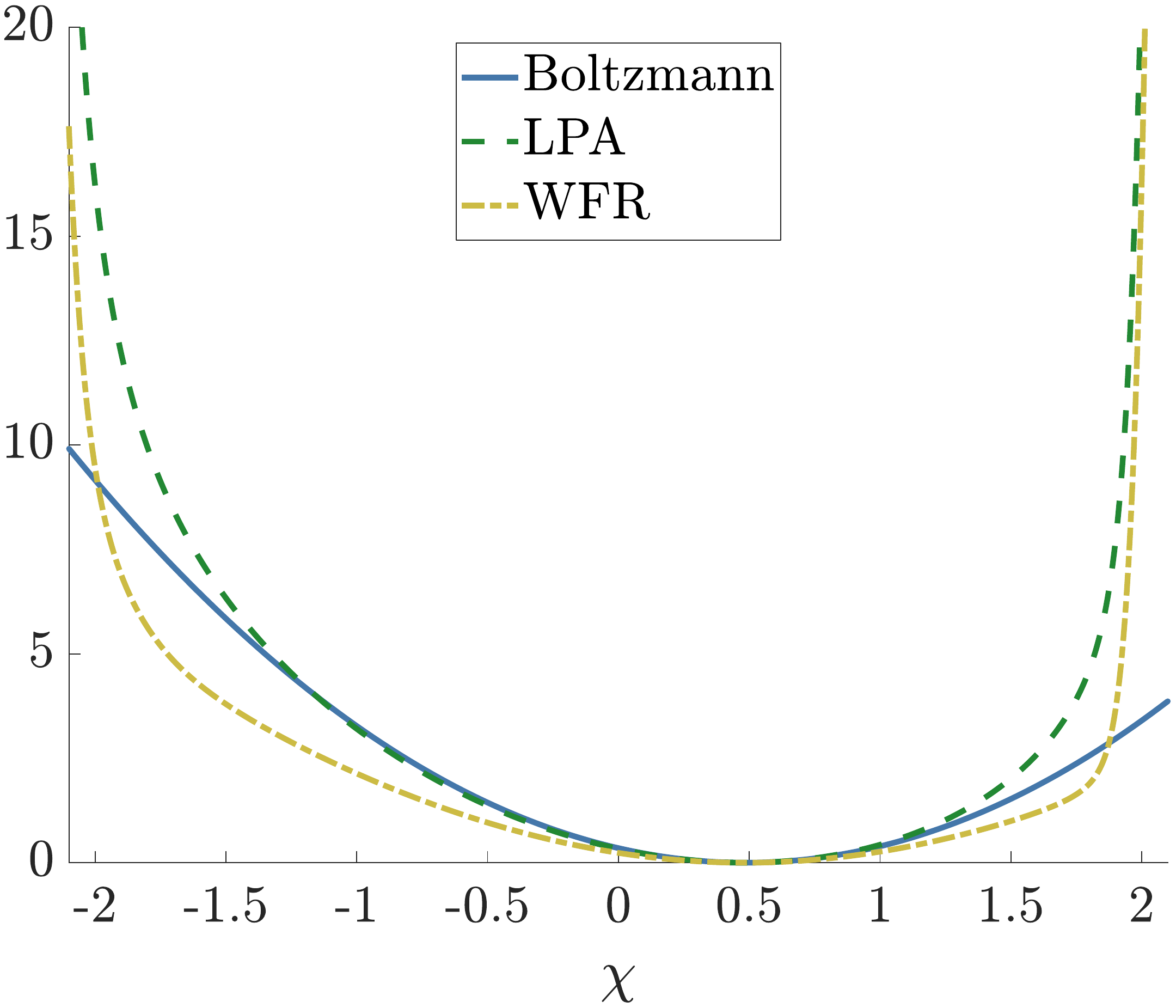}
\includegraphics[width=0.45\textwidth]{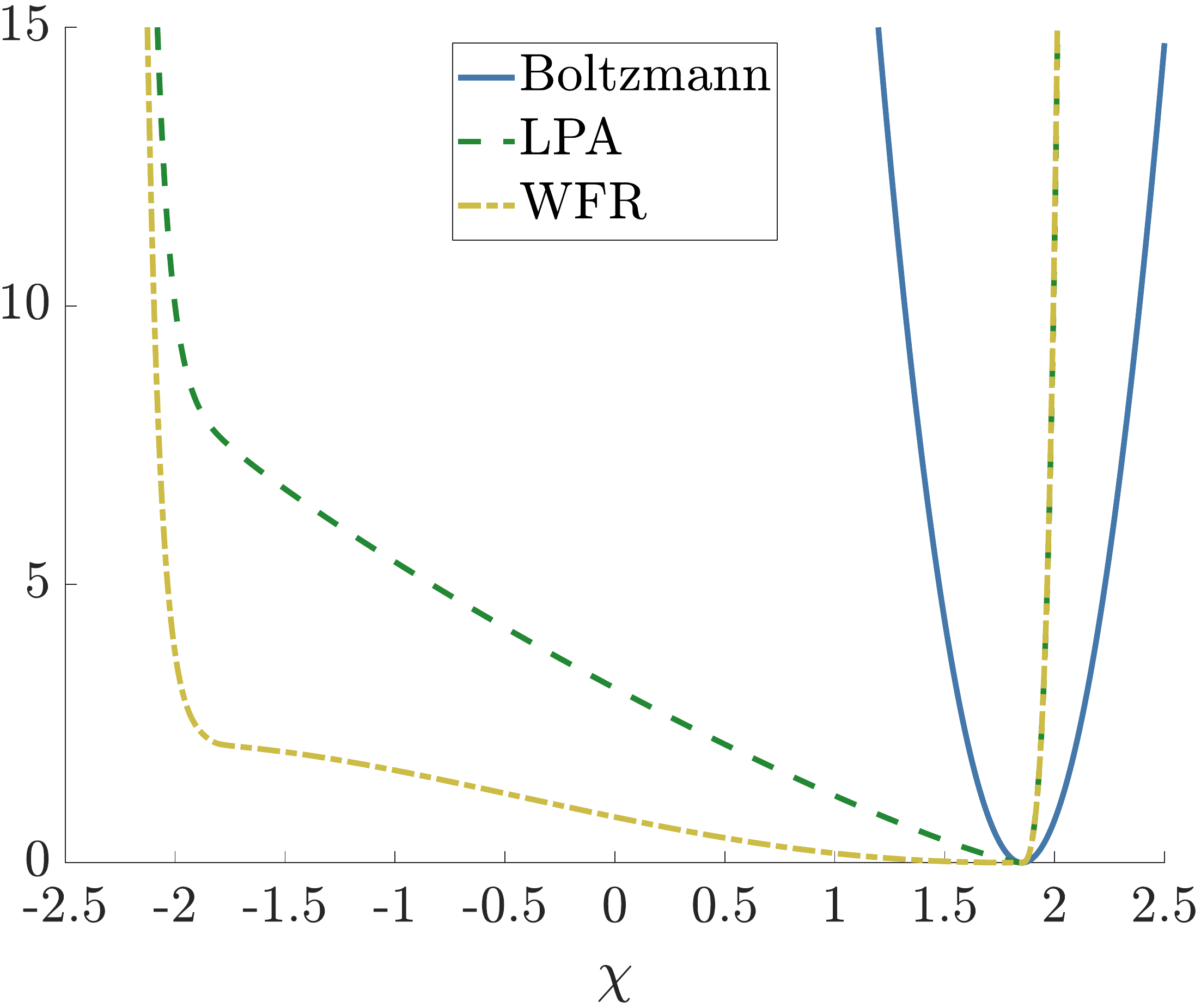}
\caption[\ac{LPA} vs \ac{WFR} vs Boltzmann dynamical effective potentials]{Comparison of $\tilde{V}$ for the unequal L-J potential at $\Upsilon = 10$ (left) and $\Upsilon = 2$ (right) as calculated using \ac{FRG} methods \ac{LPA} and \ac{WFR} compared to the Boltzmann ``near-equilibrium" approximation given by equation (\ref{eq: BoltztildeV}). All potentials have been vertically shifted so that their minima (corresponding to the equilibrium position) coincide. }\label{fig: ULJ_LPAvsBoltz_V_5T}
\end{figure}

In section \ref{sec:EEOM_one} we introduced the notion of the \textit{dynamical effective potential} $\tilde{V}$ given by equation (\ref{eq: Vtilde}). As the \ac{FRG} guarantees that the fully effective potential $V_{{\kappa}=0}$ will be convex this implies that the dynamical effective potential $\tilde{V}$ will also be either fully or extremely close to fully convex for \ac{LPA} and \ac{WFR} respectively thus greatly simplifying dynamical calculations. In the previous section we emphasised how the \ac{FRG} \ac{LPA} effective potential gives us the Boltzmann equilibrium quantities such as equilibrium position and variance. What we would like to emphasise now however is that away from the minimum of the effective potential the \ac{FRG} gives us information that the near equilibrium Boltzmann assumption does not. To be concrete the (Gaussian) Boltzmann distribution assumes that the potential is of the form:
\begin{eqnarray}
\tilde{V}_{Boltz}(\chi) = \dfrac{\Upsilon}{4\cdot \text{Var}_{eq}}\left( \chi - \chi_{eq}\right)^2 \label{eq: BoltztildeV}
\end{eqnarray}
where $\chi_{eq}$ and $\text{Var}_{eq}$ are the equilibrium position and variance respectively. We show in Fig.~\ref{fig: ULJ_LPAvsBoltz_V_5T} how this approximation can break down dramatically as one moves away from the equilibrium position suggesting that the \ac{FRG} captures the far away from equilibrium dynamics well. While this deviation is noticeable for $\Upsilon = 10$ (left plot), the deviation is so dramatic at $\Upsilon = 2$ (right plot) that the region of validity of (\ref{eq: BoltztildeV}) is tiny. In principle one could attempt to include higher order cumulants of the Boltzmann distribution such as skewness and kurtosis into the effective potential, however the relationship between these cumulants and higher derivatives of the effective potential is highly non-trivial and is cumbersome to include. In any case it is not expected including these corrections would lead to significant improvement away from equilibrium.
\begin{figure}[t!]
\centering
\includegraphics[width=0.45\textwidth]{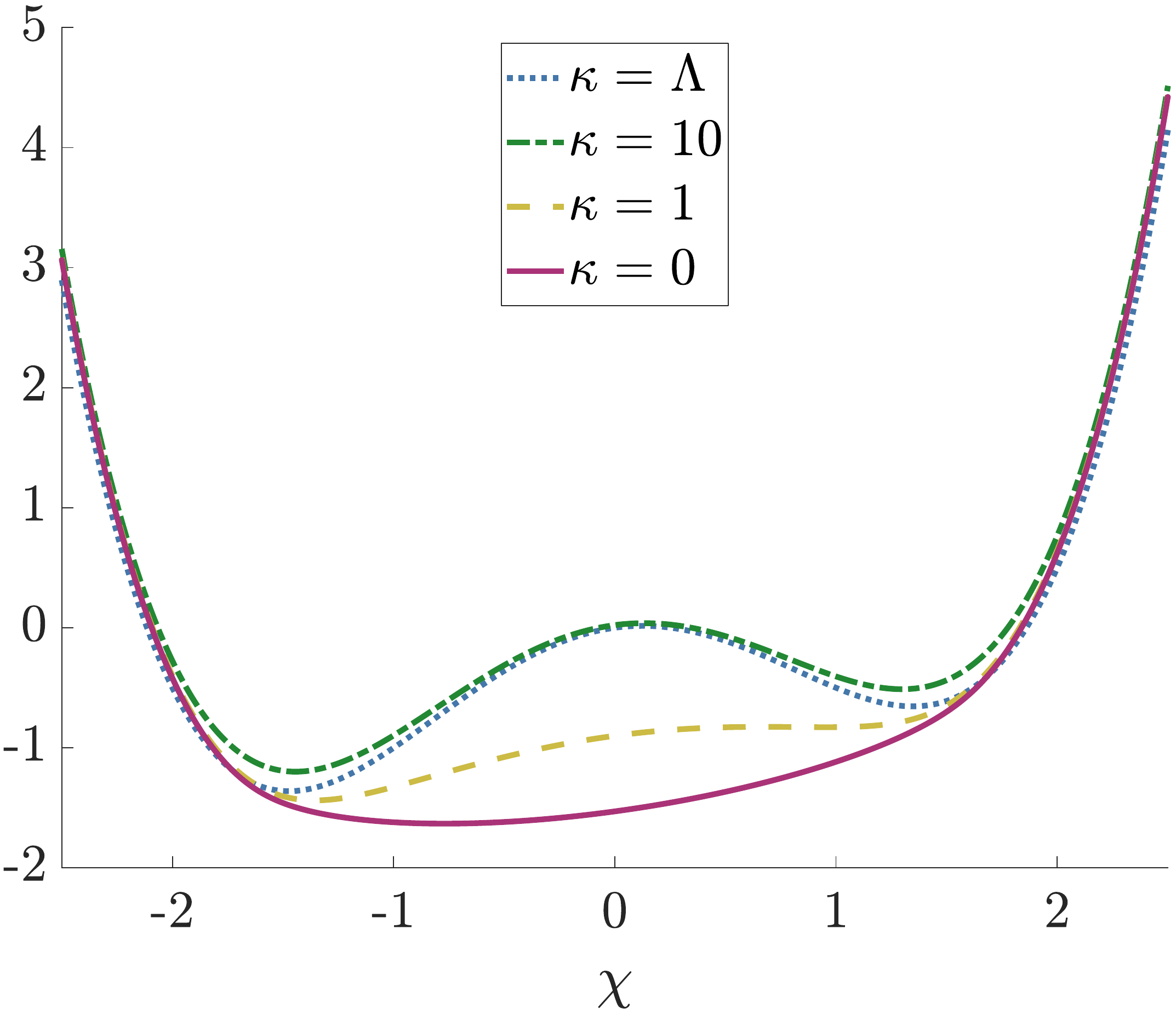}
\includegraphics[width=0.45\textwidth]{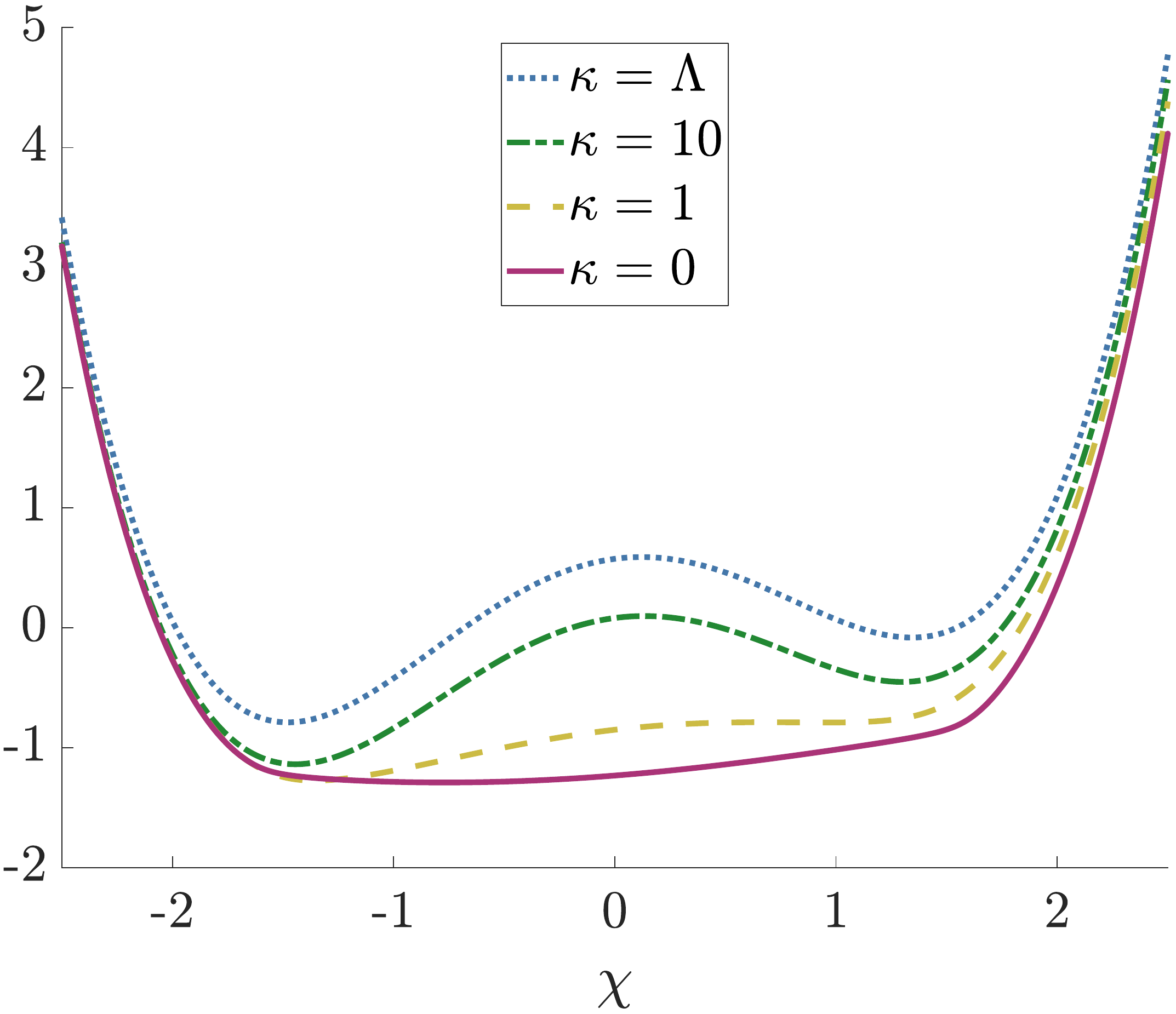}
\caption[Flow of dynamical effective potential for asymmetric doublewell potential]{The flow of the \textit{dynamical effective potential} $\tilde{V}$ for an initially asymmetric doublewell potential defined as $V = V_{DW} + x/4$ for $\Upsilon$ = 1 using \ac{LPA} method  (left) and \ac{WFR} method (right).}\label{fig: ADW_tildeV}
\end{figure}

In Fig.~\ref{fig: ADW_tildeV} we show the evolution of $\tilde{V}$ as ${\kappa}$ is lowered to zero -- or equivalently as all the fluctuations are integrated out -- for the asymmetric doublewell. We can see for both the \ac{LPA} (left) and \ac{WFR} (right) how the barrier gets smaller as fluctuations are integrated out until it completely disappears. The equilibrium position is represented by the global minimum of the red curve (${\kappa} = 0$) and we can infer the speed of the evolution to this equilibrium by the slope of the curve to it. Similar behaviour can be seen for all the other potentials we consider in this thesis. The fact that the fully flowed potential (red curve) is guaranteed to be (near) convex by definition of the \ac{EA} $\Gamma$ ensures that the dynamics we perform in it will be trivial to solve. This is what we cover in the following subsection. 

Where possible we will compare our results with those obtained by direct simulation of the Langevin  equation (\ref{eq:langevindimless}) and by solving the \ac{F-P} equation (\ref{eq: rescaled F-P}).
\subsection{Accelerated trajectories}
\begin{figure}[t!]
\centering
\includegraphics[width=0.65\textwidth]{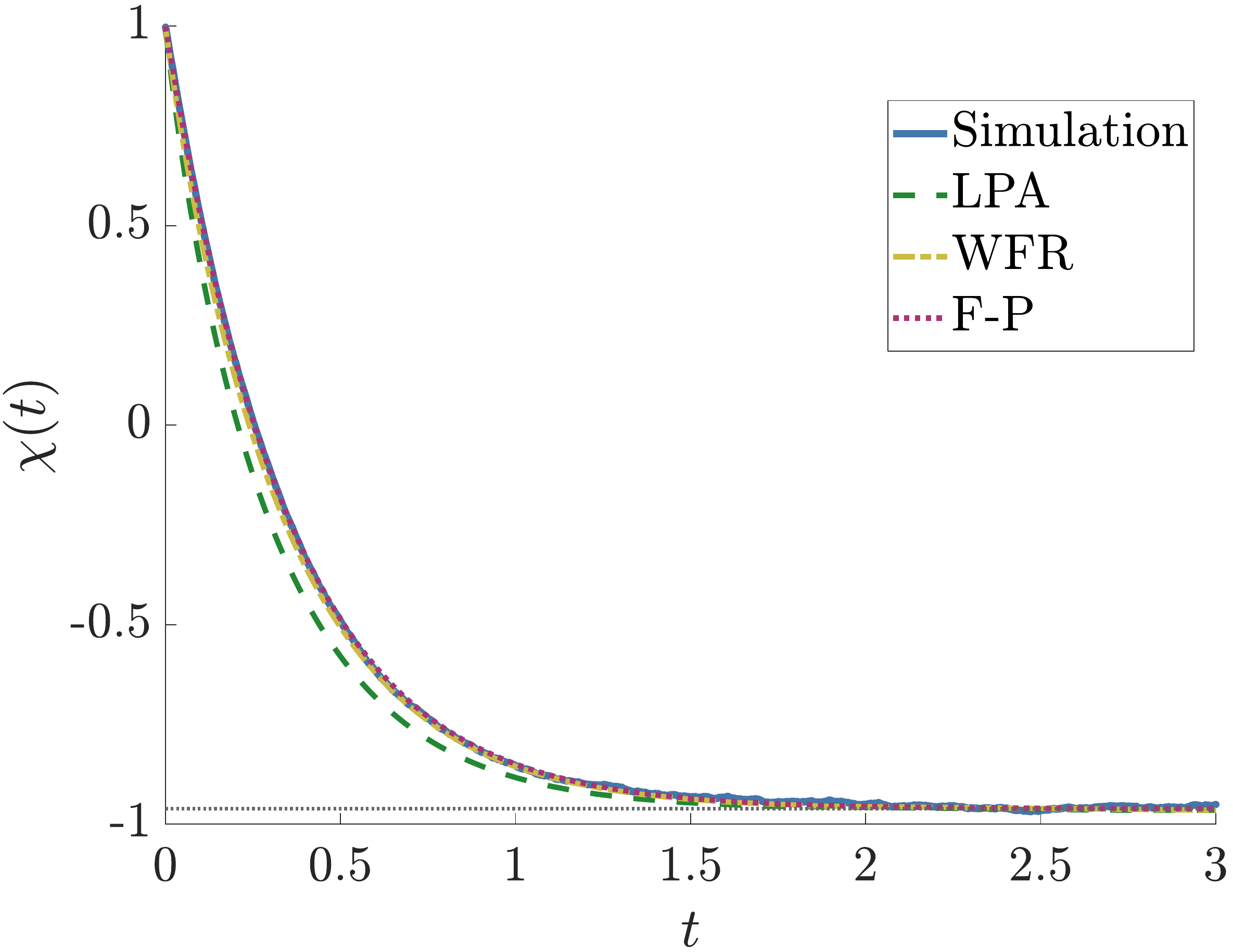}
\caption[Evolution of average position for a polynomial potential]{The trajectory of the average position $\chi$ in a polynomial potential $\tilde{V}$ by direct simulation \& solving the \ac{EEOM} (\ref{eq:QEOM}) using \ac{LPA} and \ac{WFR} for $\Upsilon = 10$.}\label{fig: Chi_Gies}
\end{figure}
\begin{figure}[t!]
\centering
\includegraphics[width=0.65\textwidth]{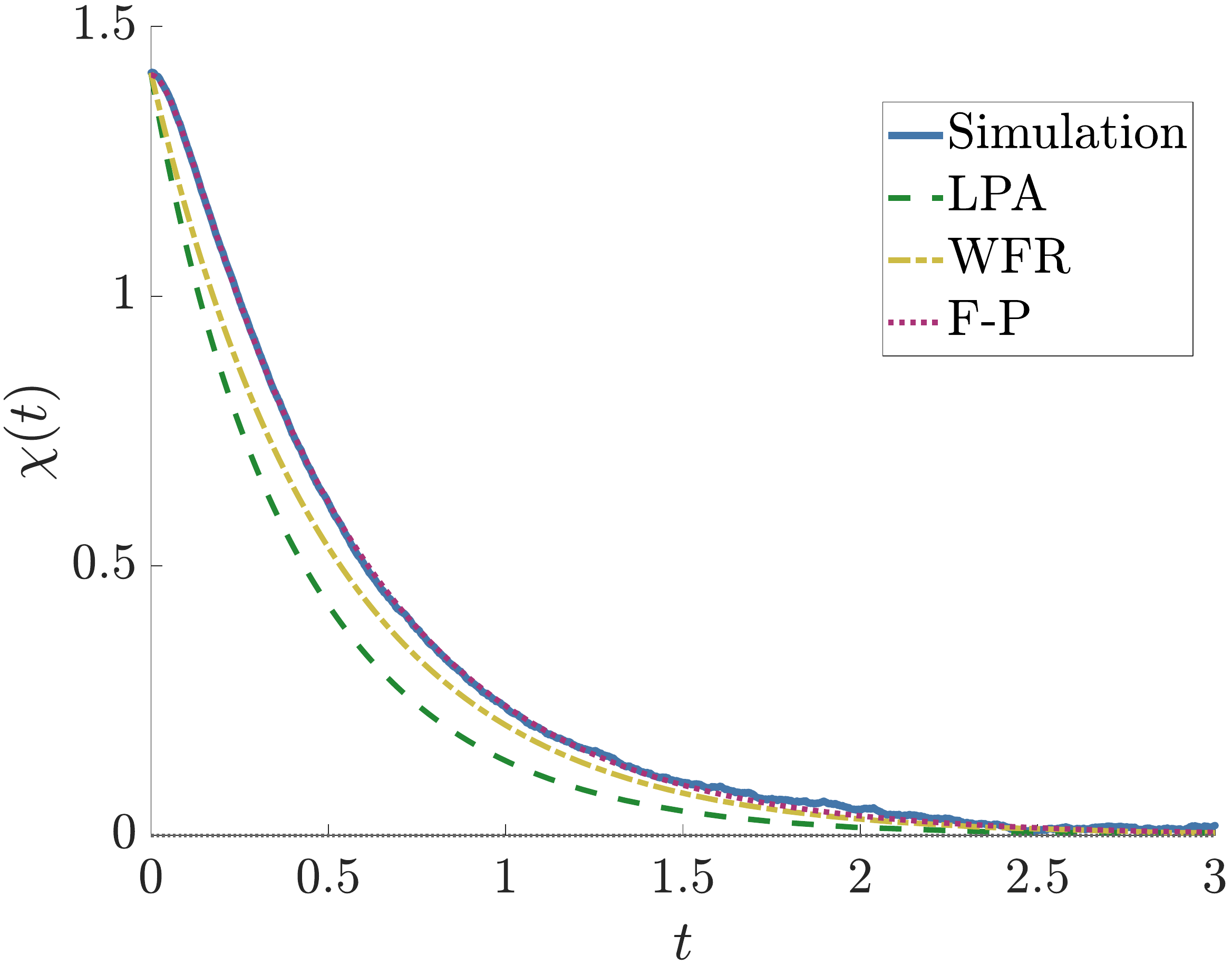}
\caption[Evolution of average position for a doublewell potential]{The trajectory of the average position $\chi$ in a doublewell potential $\tilde{V}$ by direct simulation \& solving the \ac{EEOM} (\ref{eq:QEOM}) using \ac{LPA} and \ac{WFR} for $\Upsilon = 10$.  }\label{fig: Chi_doublewell}
\end{figure}
\begin{figure}[t!]
\centering
\includegraphics[width=0.65\textwidth]{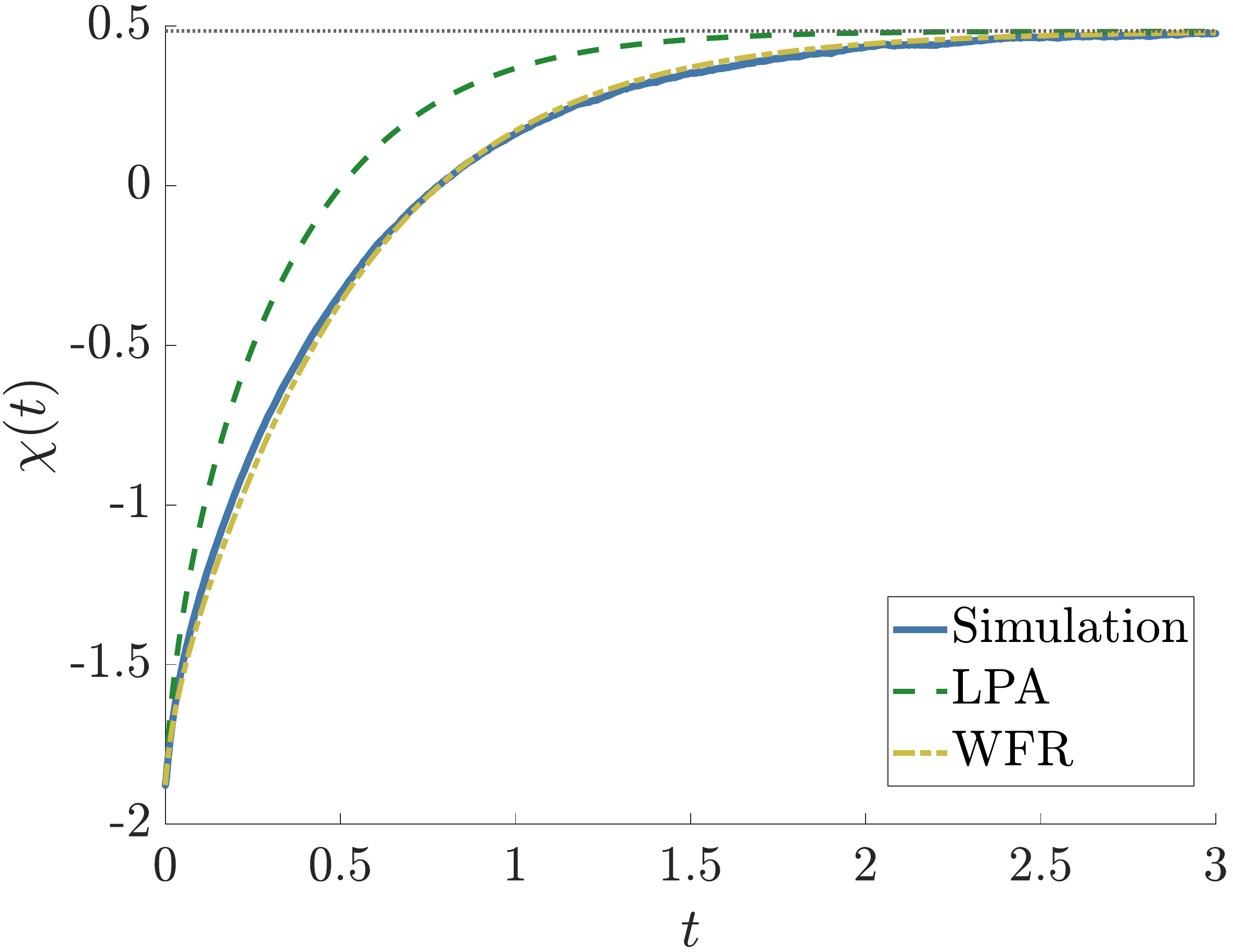}
\caption[Evolution of average position for a Lennard-Jones type potential]{The trajectory of the average position $\chi$ in a unequal L-J potential $\tilde{V}$ by direct simulation \& solving the \ac{EEOM} (\ref{eq:QEOM}) using \ac{LPA} and \ac{WFR} for $\Upsilon = 10$. }\label{fig: Chi5T_ULJ}
\end{figure}
\begin{figure}[t!]
\centering
\includegraphics[width=0.65\textwidth]{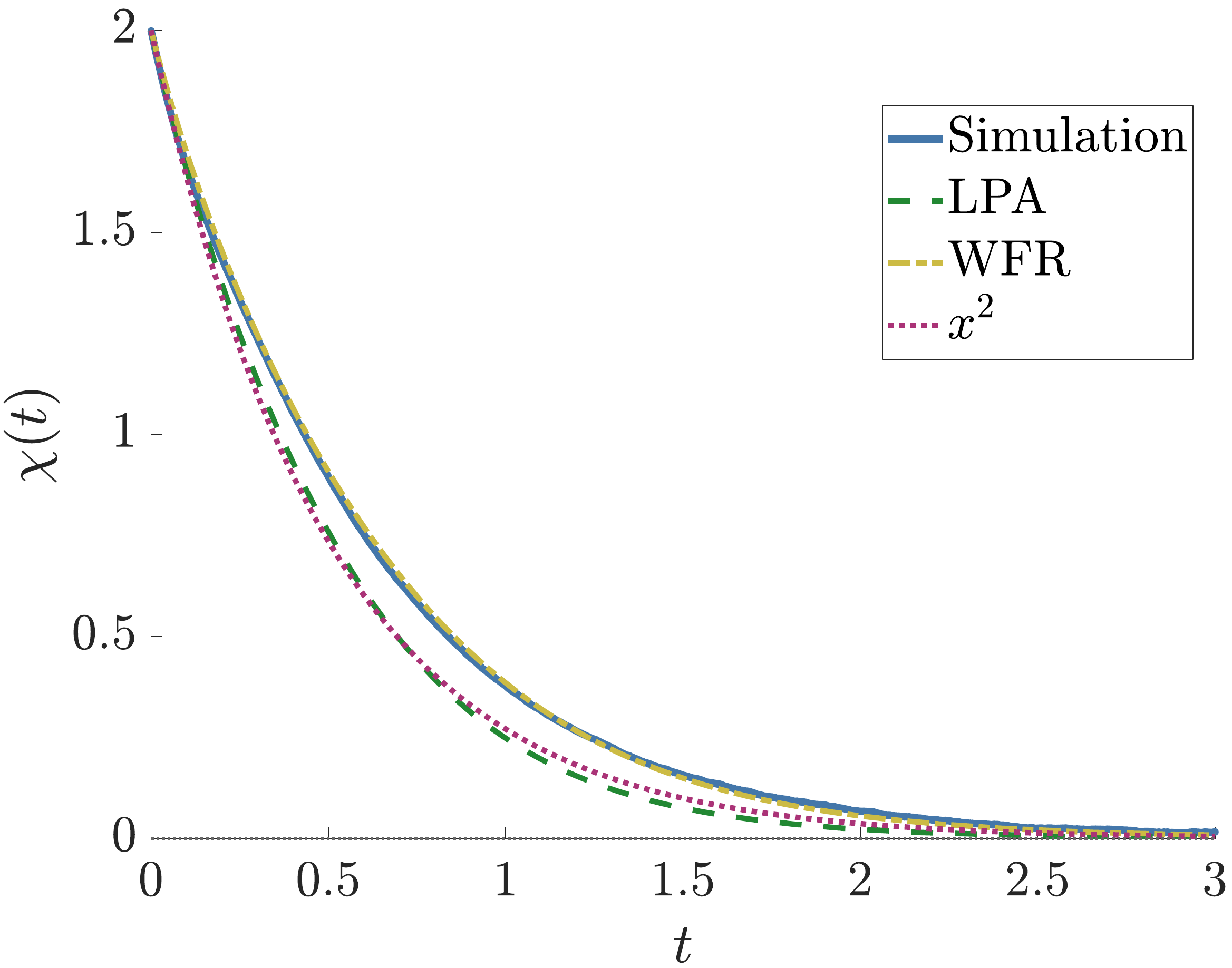}
\caption[Evolution of average position for a $x^2$ plus two bumps potential]{The trajectory of the average position $\chi$ for an $x^2$ potential plus two Gaussian bumps potential $\tilde{V}$ by direct simulation \& solving the \ac{EEOM} (\ref{eq:QEOM}) using \ac{LPA} and \ac{WFR} for $\Upsilon = 2$. The average position $\chi$ predicted for a simple $x^2$ potential is also displayed to highlight the non-trivial behaviour the \ac{FRG} is capturing.}\label{fig: Chi1T_X2Gauss}
\end{figure}
\begin{figure}[t!]
\centering
\includegraphics[width=0.65\textwidth]{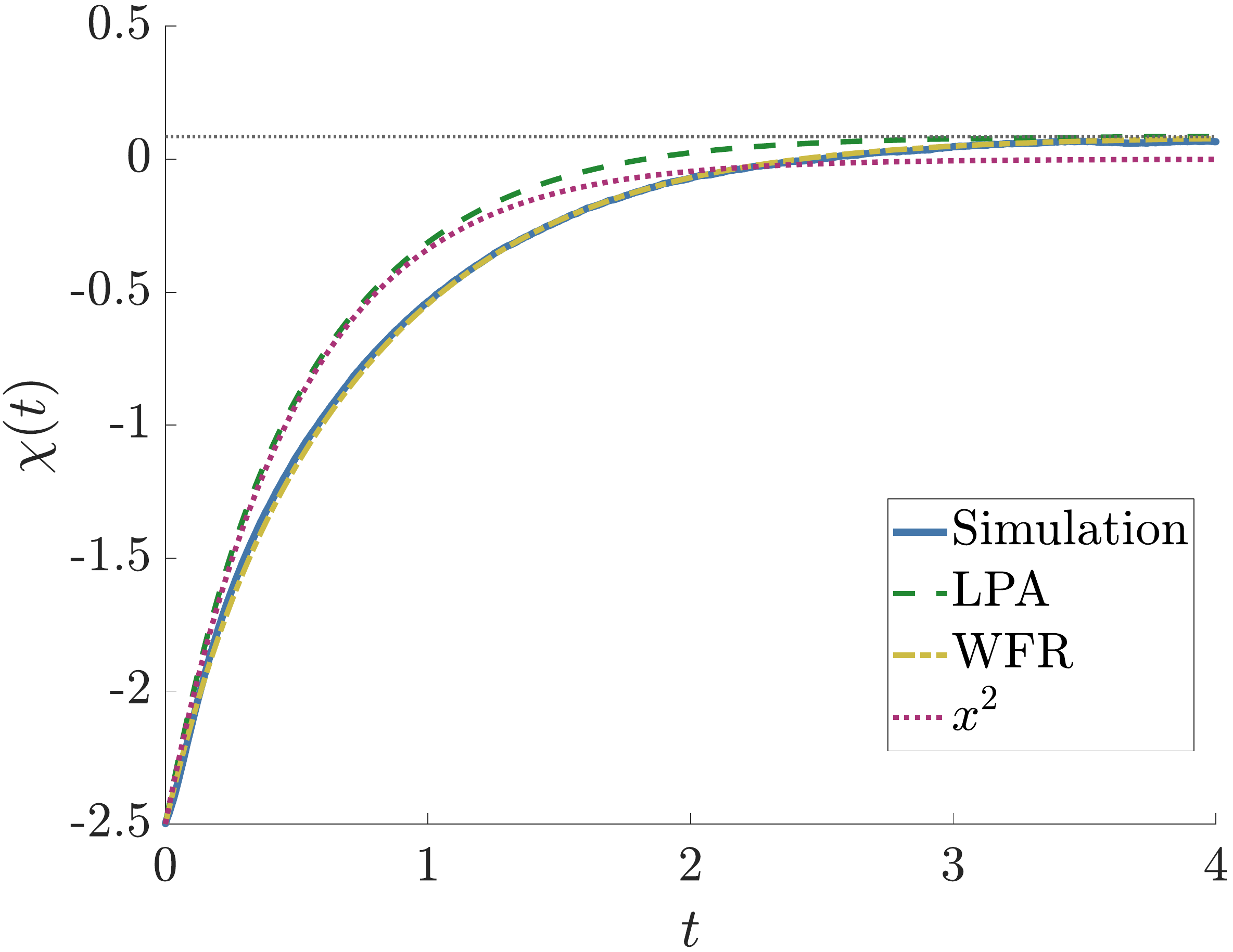}
\caption[Evolution of average position for a $x^2$ plus six bumps potential]{The trajectory of the average position $\chi$ for $x^2$ potential plus six Gaussian bumps/dips $\tilde{V}$ by direct simulation \& solving the \ac{EEOM} (\ref{eq:QEOM}) using \ac{LPA} and \ac{WFR} for $\Upsilon = 3$.}\label{fig: ChiT_X2GaussMany}
\end{figure}
We know from section \ref{sec:1-point function} that the \ac{EEOM} for average position is given by a simple first order differential equation (\ref{eq:QEOM}). Having solved the appropriate flow equations to obtain the dynamical effective potentials we can now perform dynamics in this effective potential. Given the dynamical effective potential $\tilde{V}$ it only takes a couple of seconds to obtain the full trajectory of $\chi$ from some initial position $x_i = \chi_i$ to the equilibrium position. For the polynomial potential we initialised the particle far away from the equilibrium position at $x =  1$. In Fig.~\ref{fig: Chi_Gies} we show how the average position of the particle changes with time using direct simulation of the Langevin equation (\ref{eq:langevindimless}) over 50000 runs, by numerically solving the \ac{F-P} equation (\ref{eq:FP1}) and as calculated by the evolution in the dynamical effective potentials $\tilde{V}$ given using the \ac{LPA} and \ac{WFR} methods at $\Upsilon = 10$. All four trajectories agree to a very high precision. This is perhaps not surprising as the polynomial potential we consider is rather simple. What is more surprising however is how well the \ac{FRG} works for the symmetric doublewell. In Fig.~\ref{fig: Chi_doublewell} we plot the four trajectories where the particle for each starts at the bottom of the right hand well ($x = \sqrt{2}$). Even the simple \ac{LPA} describes pretty well the evolution of $\chi (t)$ towards the equilibrium point at $\chi_{eq} = 0$. When \ac{WFR} is also included it matches the simulated trajectory very closely although not quite as closely as solving the \ac{F-P} equation (\ref{eq:FP1}). This is a non-trivial system and it is remarkable how well the \ac{FRG} does to capture the correct dynamics. \\

In Fig.~\ref{fig: Chi5T_ULJ} we plot the evolution of $\chi (t)$ for the unequal L-J potential where the particle begins in the smaller well at $x = - 1.878$ and moves towards its equilibrium position. We see as before that while the \ac{LPA} captures the behaviour relatively poorly, the \ac{WFR} curve closely matches the simulated trajectory. For this system we were unable to get convergent numerics for the evolution of the \ac{F-P} equation (\ref{eq:FP1}) showing that the \ac{FRG} can derive important quantities even in highly non-trivial systems where competing methods struggle. Similarly in Fig.~\ref{fig: Chi1T_X2Gauss} the particle is initialised to the left of one of the Gaussian bumps at $x = -1.5$. While the \ac{LPA} offers little/no improvement over the bare $x^2$ ``prediction" including \ac{WFR} offers excellent agreement with direct simulations and the \ac{F-P} solution. This ability of the \ac{FRG} to capture the non-trivial evolution of average position is also shown in Fig.~\ref{fig: ChiT_X2GaussMany} for the $x^2$ potential plus six bumps/dips which is a much more complex potential landscape. Here the \ac{LPA} trajectory offers improvement over the $x^2$ ``prediction" by converging to the correct equilibrium position and including \ac{WFR} more closely matches the true simulated trajectory. It is noteworthy that the \ac{FRG} is able to reasonably capture these difficult dynamics well in systems where the \ac{F-P} solution is difficult to obtain.

It is important to note the time advantage offered by the \ac{FRG} compared to direct numerical simulation or by solving the \ac{F-P} equation (\ref{eq:FP1}). Solving the \ac{FRG} flow equations is comparable in computation time to direct simulation while solving the \ac{F-P} equation (\ref{eq:FP1}) takes longer than both. However the latter two methods obtain solutions that are only valid for a single initial condition. A huge advantage of the \ac{FRG} is that once the dynamical effective potential $\tilde{V}$ is obtained it is trivial to solve the \ac{EEOM} (\ref{eq:QEOM}) in a couple of seconds for any initial position whereas for both direct numerical simulation of (\ref{eq:langevindimless}) and solving the \ac{F-P} equation (\ref{eq:FP1}) one has to start again from scratch. 

\subsection{Evolution of the variance}
\begin{figure}[t!]
\centering
\includegraphics[width=0.45\textwidth]{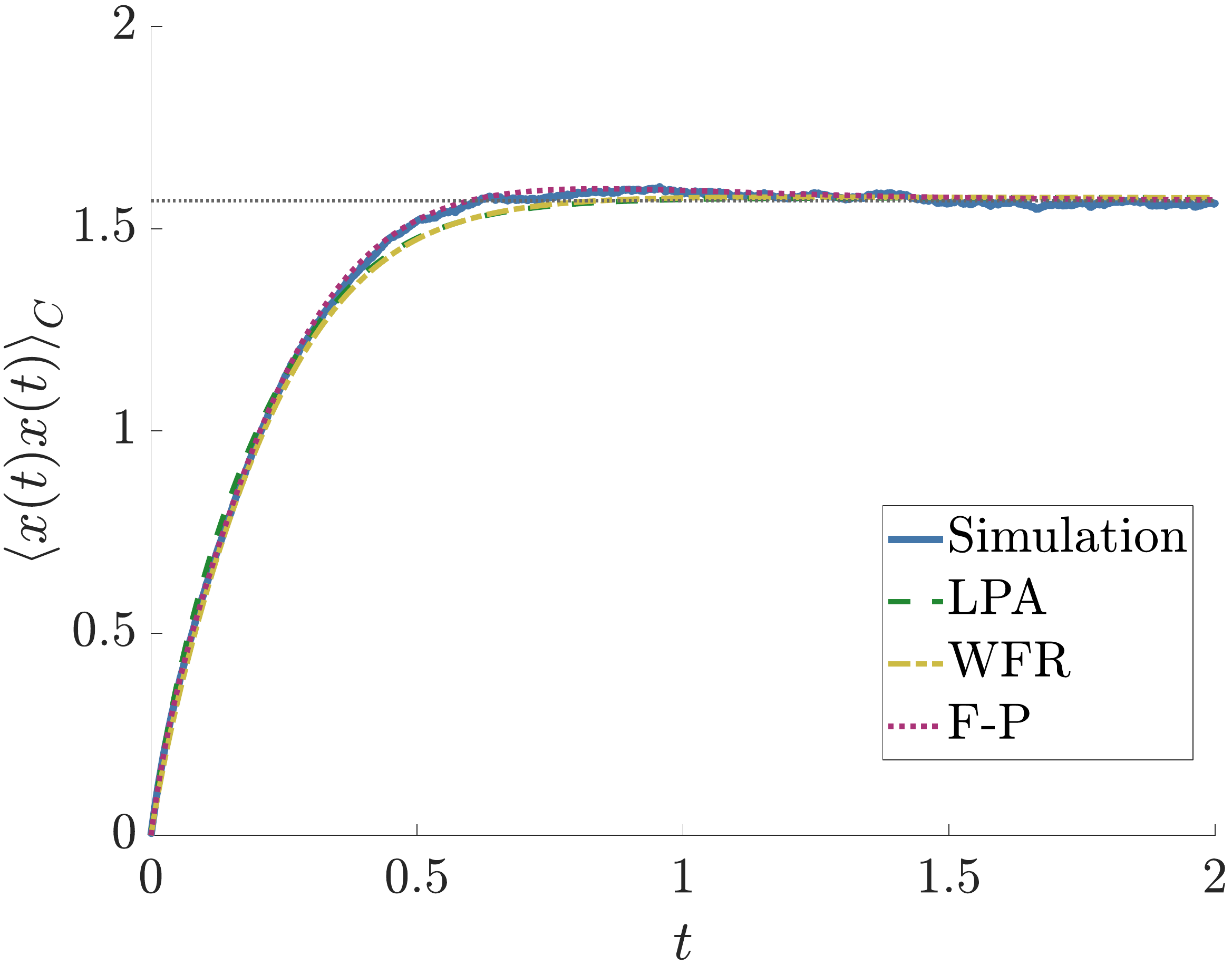}
\includegraphics[width=0.45\textwidth]{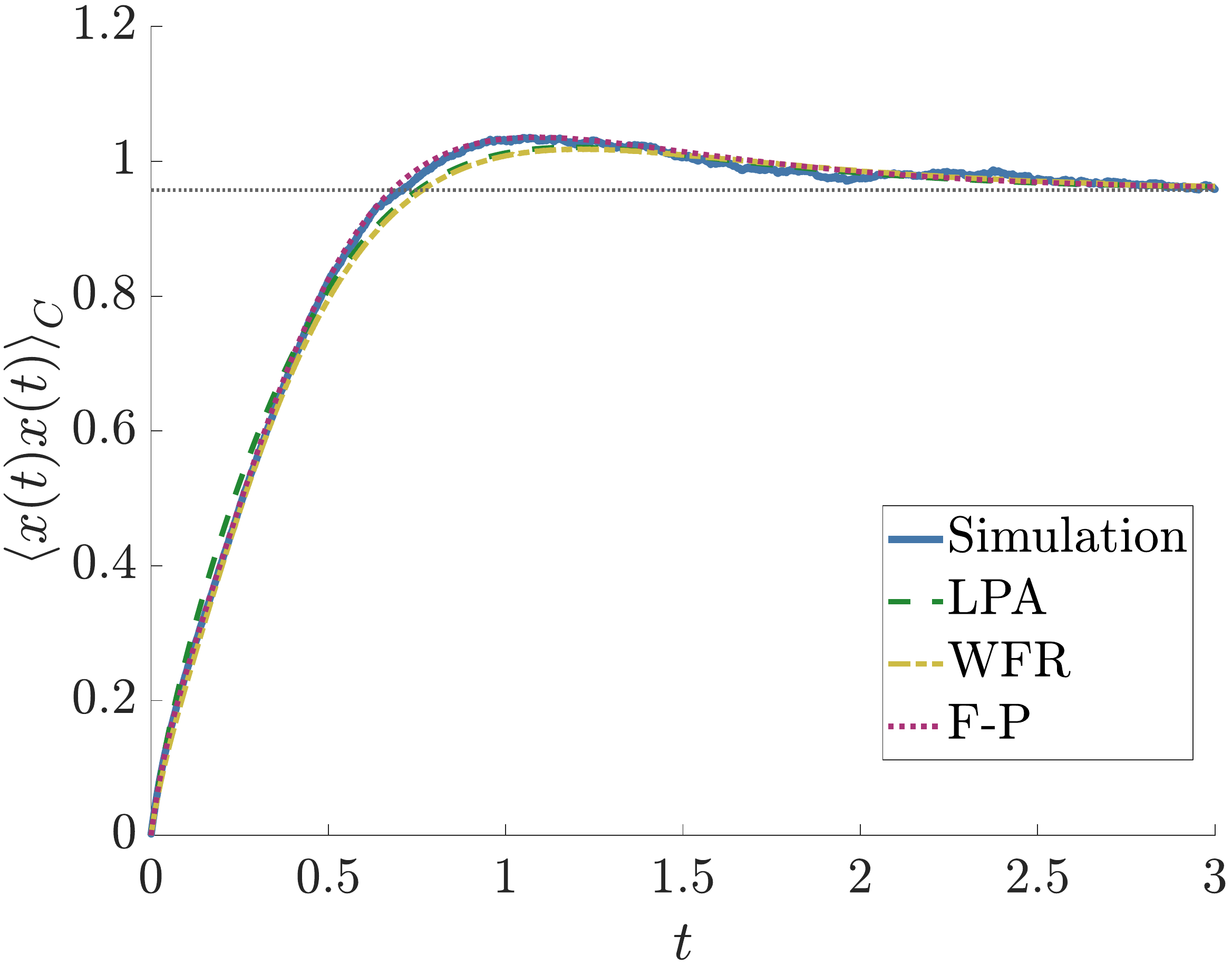}
\caption[Evolution of the variance in a polynomial potential]{The evolution of the variance \textbf{Var}(x) in a polynomial potential by direct simulation, solving the Fokker-Plank equation \& solving the \ac{EEOM} (\ref{eq:EEOM Variance}) for $\Upsilon = 10$ (left) and $\Upsilon = 4$ (right). The equilibrium variance as calculated from the Boltzmann distribution is also plotted with the horizontal dotted line.}\label{fig: VarT_Gies}
\end{figure}

\begin{figure}
    \centering
    \includegraphics[width=0.45\textwidth]{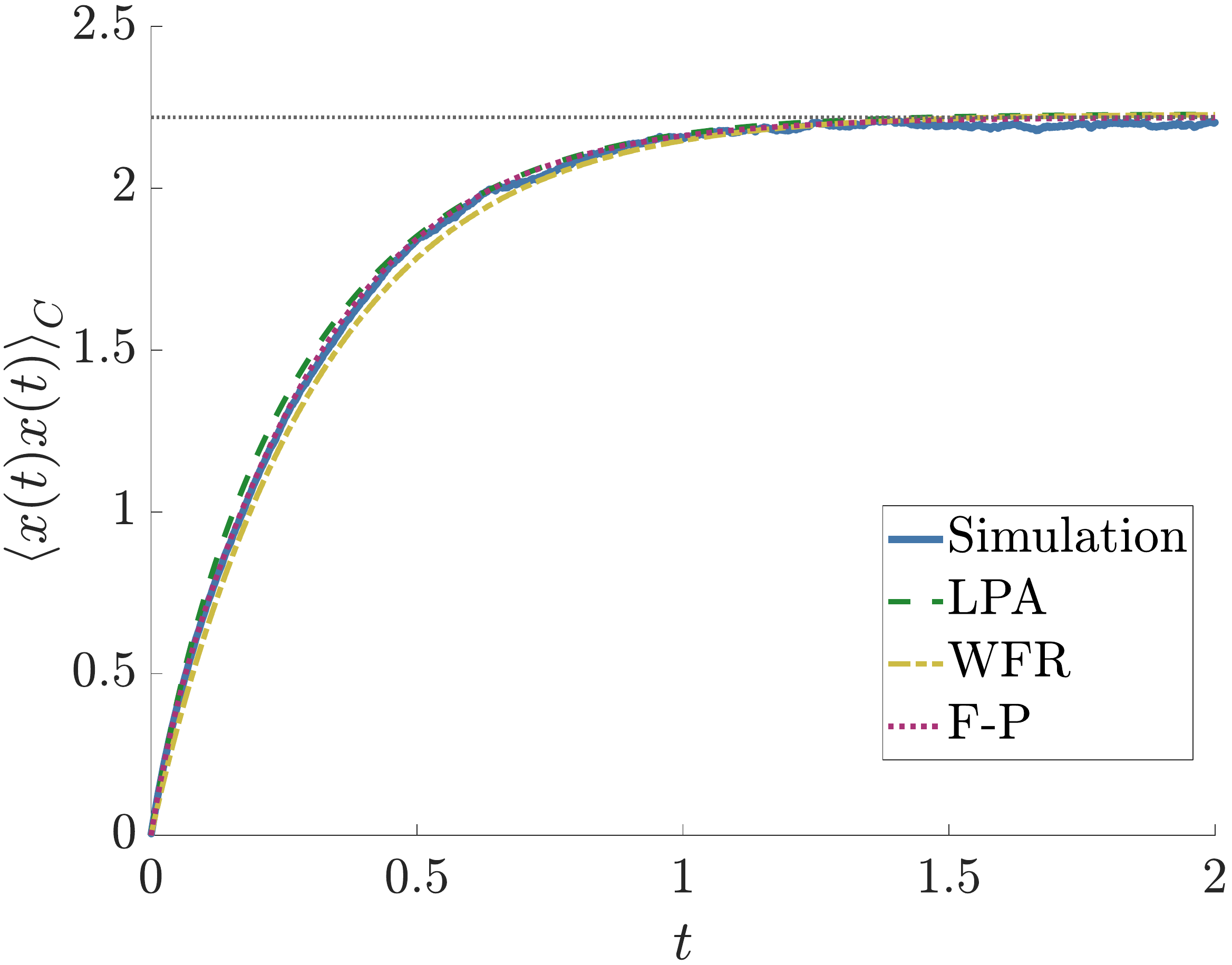}
    \caption[Evolution of the variance in a doublewell potential]{The evolution of the variance \textbf{Var}(x) in a doublewell potential by direct simulation \& solving the \ac{EEOM} (\ref{eq:EEOM Variance}) for $\Upsilon = 10$. The equilibrium variance as calculated from the Boltzmann distribution is also plotted with the horizontal dotted line.}
    \label{fig:VarT_DW}
\end{figure}

\begin{figure}[t!]
\centering
\includegraphics[width=0.45\textwidth]{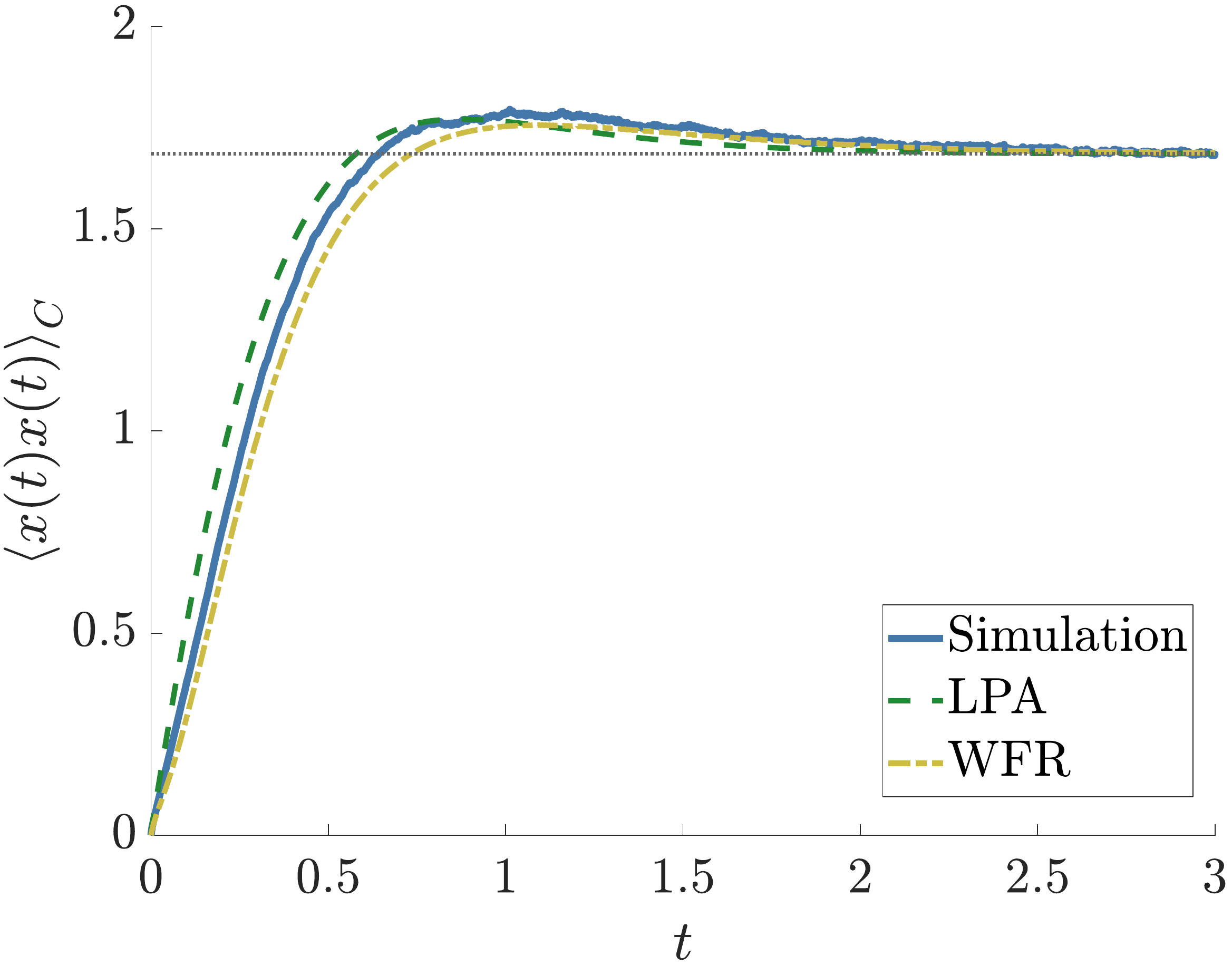}
\caption[Evolution of the variance in a Lennard-Jones type potential]{The evolution of the variance \textbf{Var}(x) in an unequal L-J potential by direct simulation \& solving the \ac{EEOM} (\ref{eq:EEOM Variance}) for $\Upsilon = 10$. The equilibrium variance as calculated from the Boltzmann distribution is also plotted with the horizontal dotted line.}\label{fig: Var5T_ULJ}
\end{figure}
\begin{figure}[t!]
	\centering
	\includegraphics[width=0.45\textwidth]{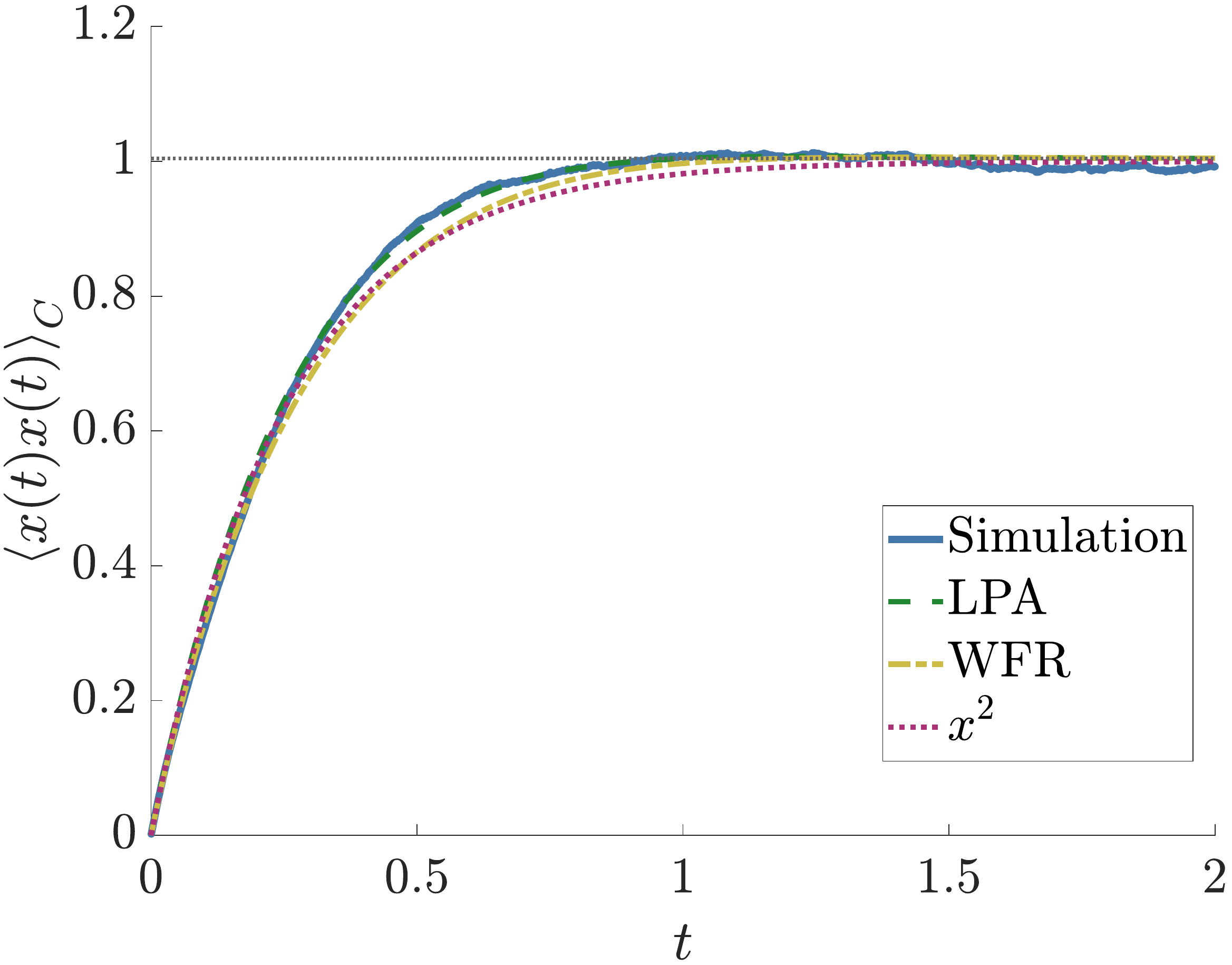}
	\includegraphics[width=0.45\textwidth]{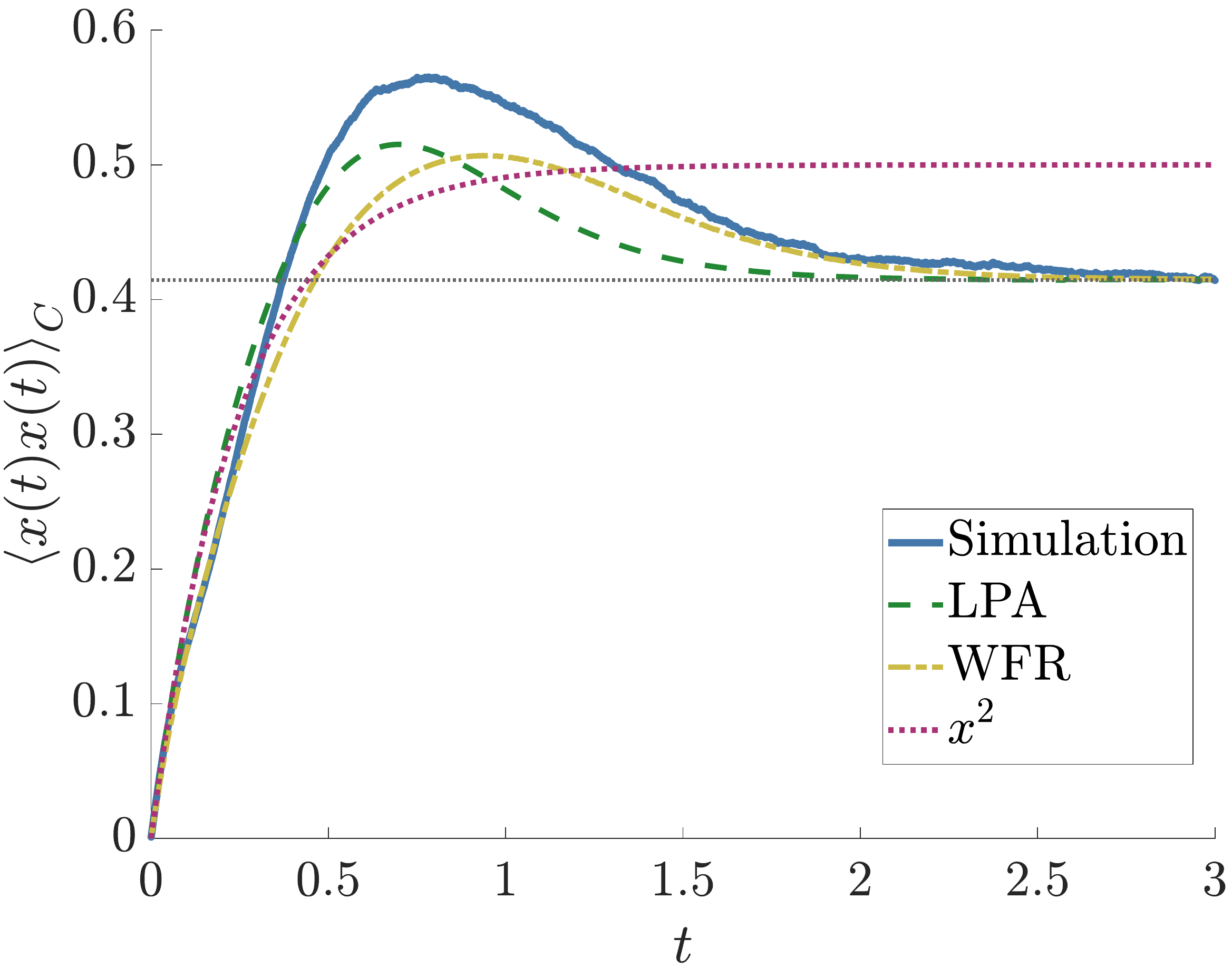}
	
	\caption[Evolution of the variance in a $x^2$ plus two Gaussian bumps potential]{The evolution of the (normalised) variance \textbf{Var}(x) by direct simulation \& solving the \ac{EEOM} (\ref{eq:EEOM Variance}) in a $x^2$ plus two Gaussian bumps potential for $\Upsilon = 4$ (left) and $\Upsilon = 2$ (right). The equilibrium variance as calculated from the Boltzmann distribution is also plotted with the horizontal dotted line.}
	\label{fig: VarT_X2Gauss.png}
\end{figure}
\begin{figure}[t!]
	\centering
	\includegraphics[width=0.45\textwidth]{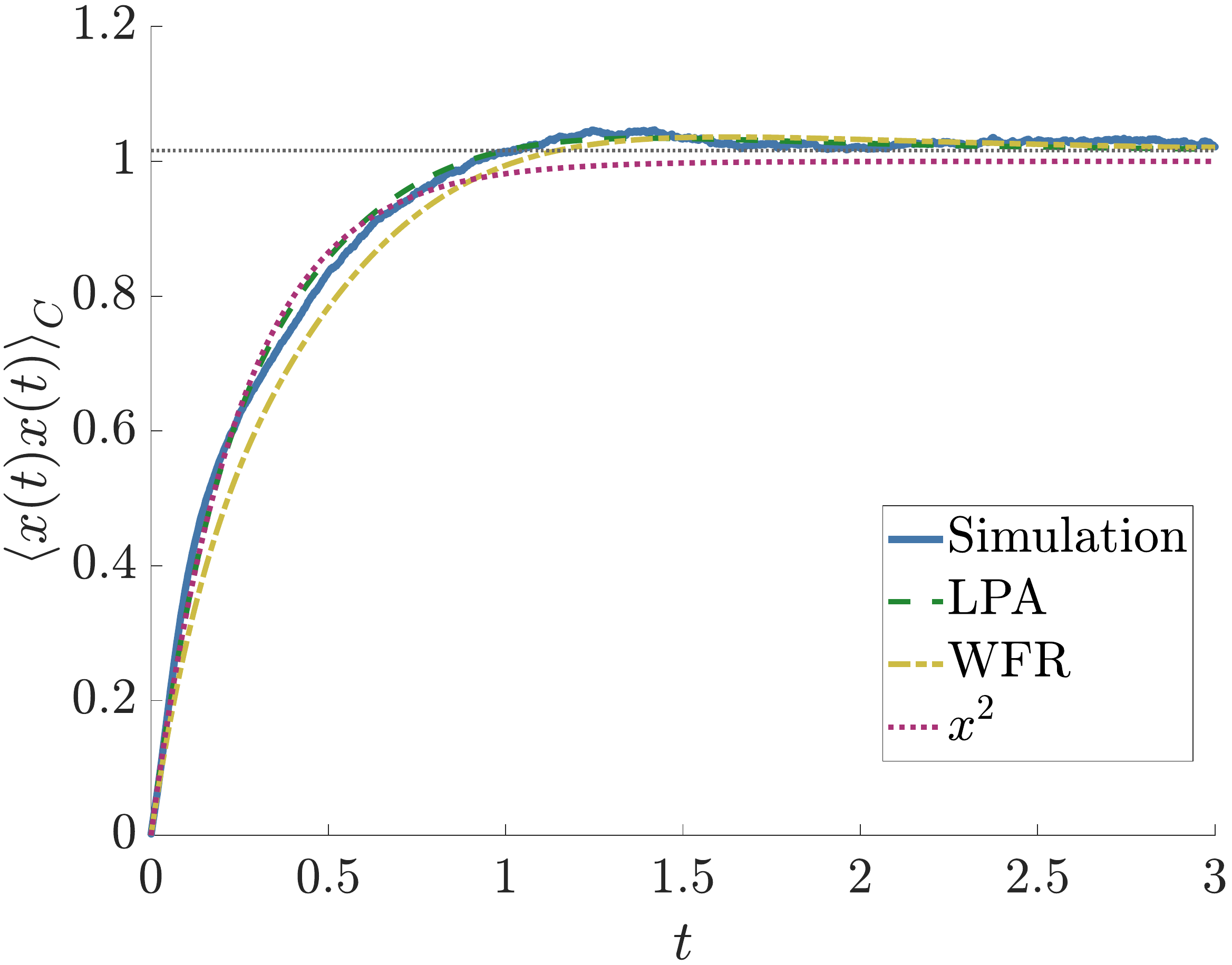}
	\includegraphics[width=0.45\textwidth]{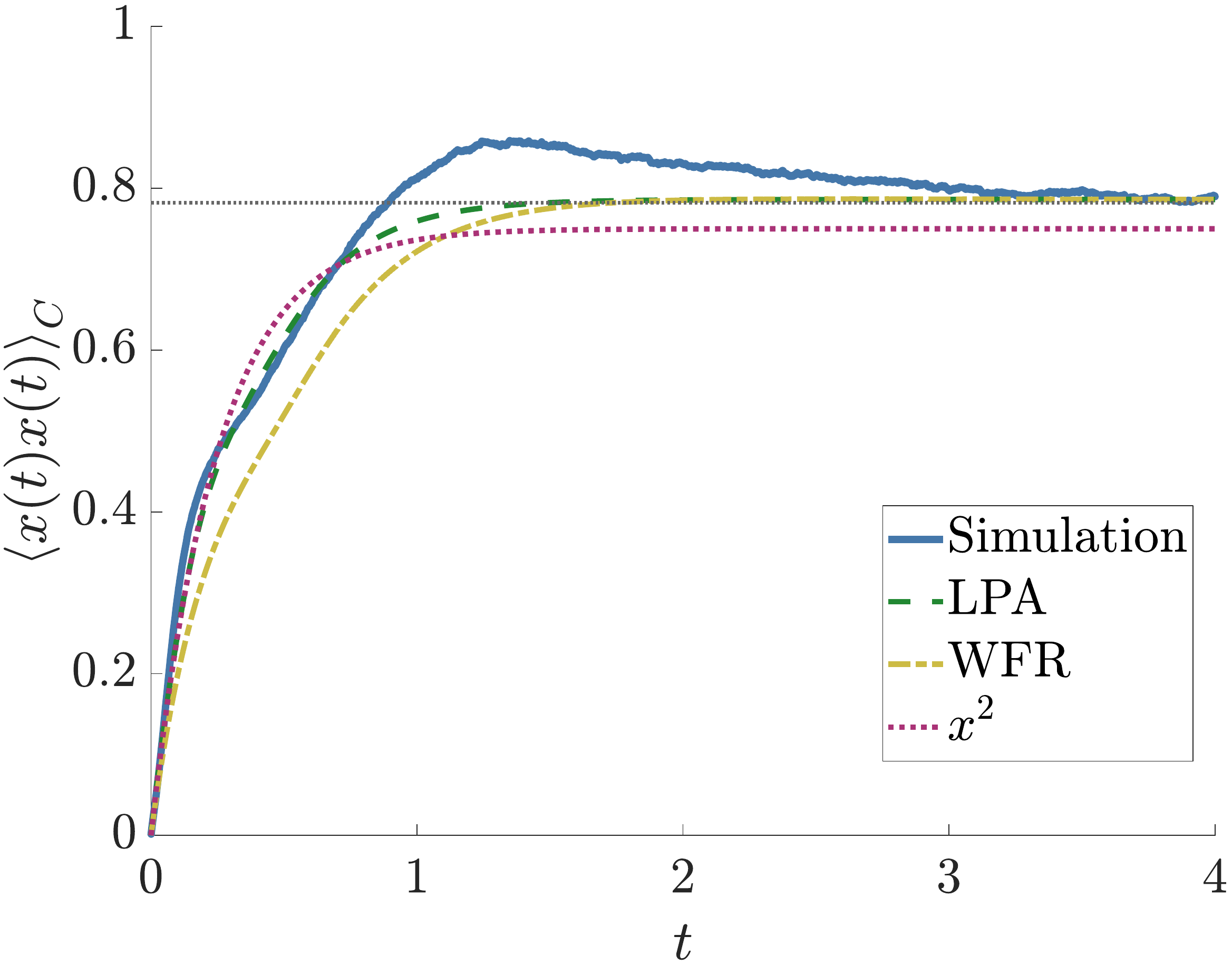}
	\includegraphics[width=0.45\textwidth]{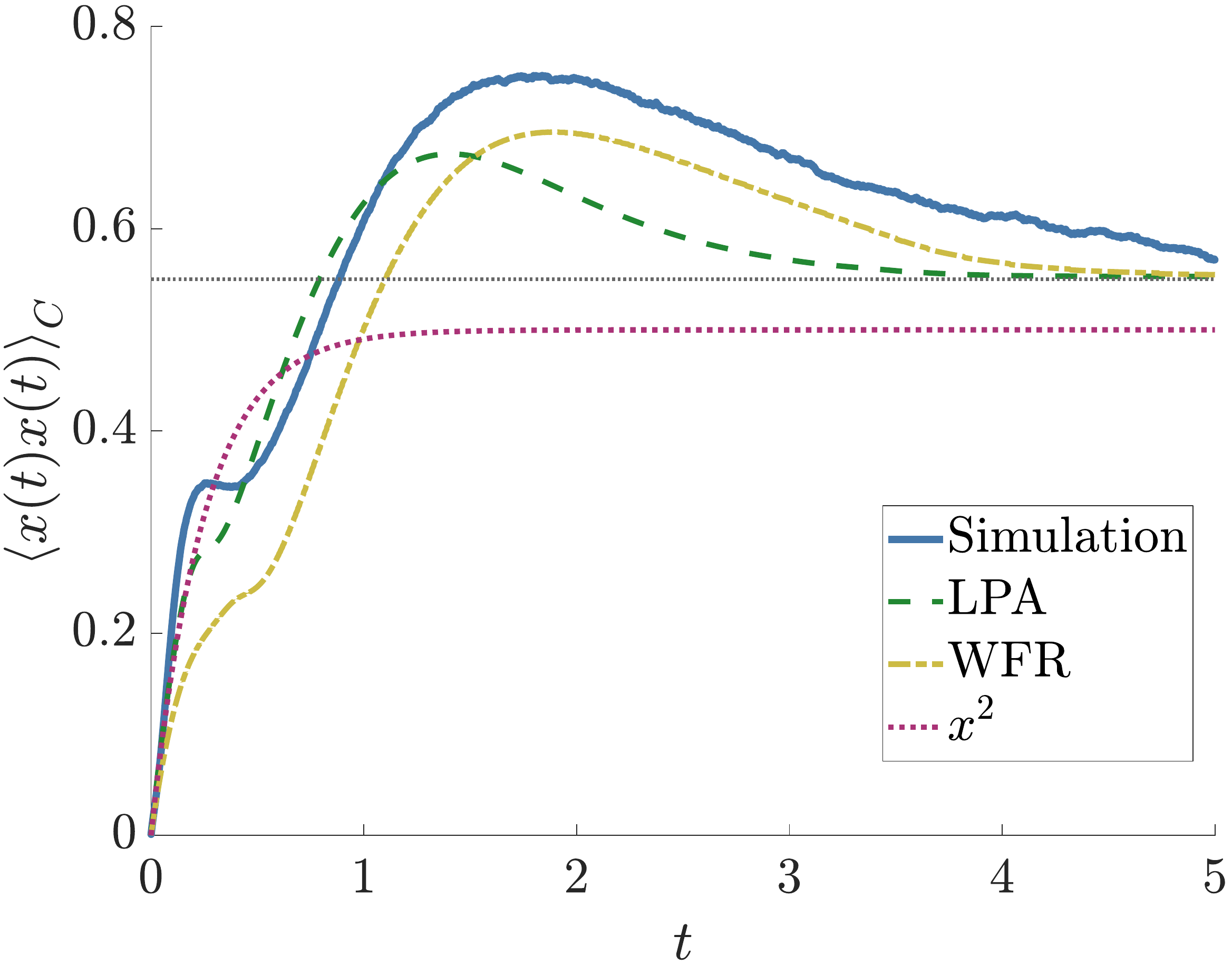}
	\caption[Evolution of the variance in a $x^2$ plus six Gaussian bumps/dips potential]{The evolution of the (normalised) variance \textbf{Var}(x) in a $x^2$ plus six Gaussian bumps/dips potential for $\Upsilon = 4$ (top left), $\Upsilon = 3$ (top right) and $\Upsilon = 2$ (bottom). The equilibrium variance as calculated from the Boltzmann distribution is also plotted with the horizontal dotted line.}
	\label{fig: VarT_X2Gaussmany.png}
\end{figure}

For our accelerated trajectories we initialised the particles at the exact same point every time. This means that at t = 0 the probability distribution of the particles had zero variance \textbf{Var}$(x)$ = 0. Using this as our initial condition we solved numerically the \ac{EEOM} for the variance (\ref{eq:EEOM Variance}), derived in section \ref{sec:EEOM_2point}.\\
In Fig.~\ref{fig: VarT_Gies} we show how the variance evolves with time for the polynomial potential for $\Upsilon = 10$ (left) and $\Upsilon = 4$ (right). For $\Upsilon = 10$ can see that the \ac{LPA} closely matches the numerical and \ac{F-P} evolution for the first 0.5 time units before departing slightly although it still tends towards the correct equilibrium distribution. In this case including \ac{WFR} offers no improvement. Solving the full \ac{F-P} equation (\ref{eq:FP1}) does however match very well with the simulated case. For $\Upsilon = 4$ we see some more interesting behaviour, the variance \textit{overshoots} the equilibrium value before asymptoting to it. In this case both \ac{LPA} and \ac{WFR} capture this non-trivial behaviour well and closely match the late time decay.

In Fig.~\ref{fig:VarT_DW} we show how the variance evolves with time for a symmetric doublewell potential for $\Upsilon = 10$. We can see  that the \ac{LPA} gives us incredible agreement with the simulated and \ac{F-P} evolution which is highly significant.

In Fig.~\ref{fig: Var5T_ULJ} we show how the variance evolves with time for an unequal Lennard-Jones potential at $\Upsilon = 10$. As with the one-point function the \ac{F-P} was unable to give sensible statistics however the \ac{LPA} is able to very well match the early simulated trajectory even capturing the overshooting of the variance. The \ac{WFR} on the other hand is better at capturing the late-time decay to equilibrium.

In Fig.~\ref{fig: VarT_X2Gauss.png} we show how the variance evolves for the $x^2$ plus two bumps potential at $\Upsilon = 4$ (left) and $\Upsilon = 2$ (right). Here we have also included the prediction for the variance evolution if the potential was simply $x^2$ (red dotted curve). In this way we can highlight the \ac{FRG}'s ability to capture non-trivial features. For $\Upsilon = 4$ it is clear that this $x^2$ prediction captures well the actual simulated evolution of the variance. It is clear therefore that the features do not significantly affect the dynamics at this high temperature. As temperature is lowered however they become more important. For $\Upsilon = 2$ we can see how the $x^2$ prediction is wildly incorrect and converges to the wrong value. Here -- as in Fig.~\ref{fig: Var5T_ULJ} -- we can see that both \ac{LPA} and \ac{WFR} capture the qualitative nature of the overshooting with the \ac{LPA} better position for early evolution and \ac{WFR} for the late time near-equilibrium evolution. 

Finally in Fig.~\ref{fig: VarT_X2Gaussmany.png} we show how the variance evolves for the $x^2$ plus six Gaussian bumps/dips potential at three different temperatures. As before, lowering the temperature decreases accuracy.  In the top right and bottom plots the \ac{FRG} once again clearly captures the overshooting which is a feature of the Gaussian bumps' existence; the bare $x^2$ evolution does not capture this behaviour and overall describes the evolution poorly, converging to the wrong equilibrium variance. Again as before the \ac{LPA} much better describes the early evolution while \ac{WFR} more accurately describes late time evolution, this is most notable in the top right plot. 

\subsection{\label{sec:comparspec}Comparison with the spectral expansion}
The above results for the change in the relative performance of \ac{LPA} + \ac{WFR} as temperature is lowered can be interpreted by resorting to the spectral expansion. In section \ref{sec:F-P} we recalled how all observables can be computed in a standard way from the Schr\"{o}dinger-like, \ac{F-P} equation (\ref{eq: rescaled F-P}) using an expansion in eigenfunctions and eigenenergies. \\
If we utilise the definition for $\lan f(x) g(x_0)\ran$ in equation (\ref{eq:spectral_correlation}) and given that we want to calculate the average position $\lan x(t)\ran$ we can identify $f(x) = x$ and $g(x_0) = 1$. We then obtain:
\begin{eqnarray}
\lan x(t)\ran = \sum_{n = 0}^{\infty} \int_{-\infty}^{\infty}\mathrm{d}x\lsb p_0(x)xp_n(x)e^{-E_n t}\rsb\int_{-\infty}^{\infty}\mathrm{d}\tilde{x}~\lsb p_n(\tilde{x})\tilde{P}(\tilde{x},0)\rsb
\end{eqnarray}
where $p_n$ are the normalised eigenfunctions, $E_n$ are the eigenvalues and $\tilde{P}(\tilde{x},0)$ is the rescaled probability distribution function evaluated at $t=0$. In all cases we consider our initial condition to be no initial variance, therefore the rescaled probability takes the form of a delta function: 
\begin{eqnarray}
\tilde{P}(\tilde{x},0) &=& \delta (\tilde{x} -x_0)e^{V(x_0)/\Upsilon}\\
\Rightarrow \lan x(t)\ran &=& \sum_{n = 0}^{\infty} p_n (x_0)e^{V(x_0)/\Upsilon}e^{-E_n t}\int_{-\infty}^{\infty}\mathrm{d}x\lsb p_0(x)xp_n(x)\rsb \\
\Rightarrow \lan x(t)\ran &\underbrace{=}_{(\ref{eq:spec_lowest_eigfunc})}& N^2 \int_{\infty}^{\infty}\mathrm{d}x\lsb e^{-2V(x)/\Upsilon}\rsb \nonumber \\
&&+ N\sum_{n = 1}^{\infty}p_n (x_0)e^{V(x_0)/\Upsilon}e^{-E_n t}\int_{-\infty}^{\infty}\mathrm{d}x\lsb e^{-V(x)/\Upsilon}xp_n(x)\rsb 
\end{eqnarray}
\begin{figure}[t!]
	\centering
	\includegraphics[width=0.55\textwidth]{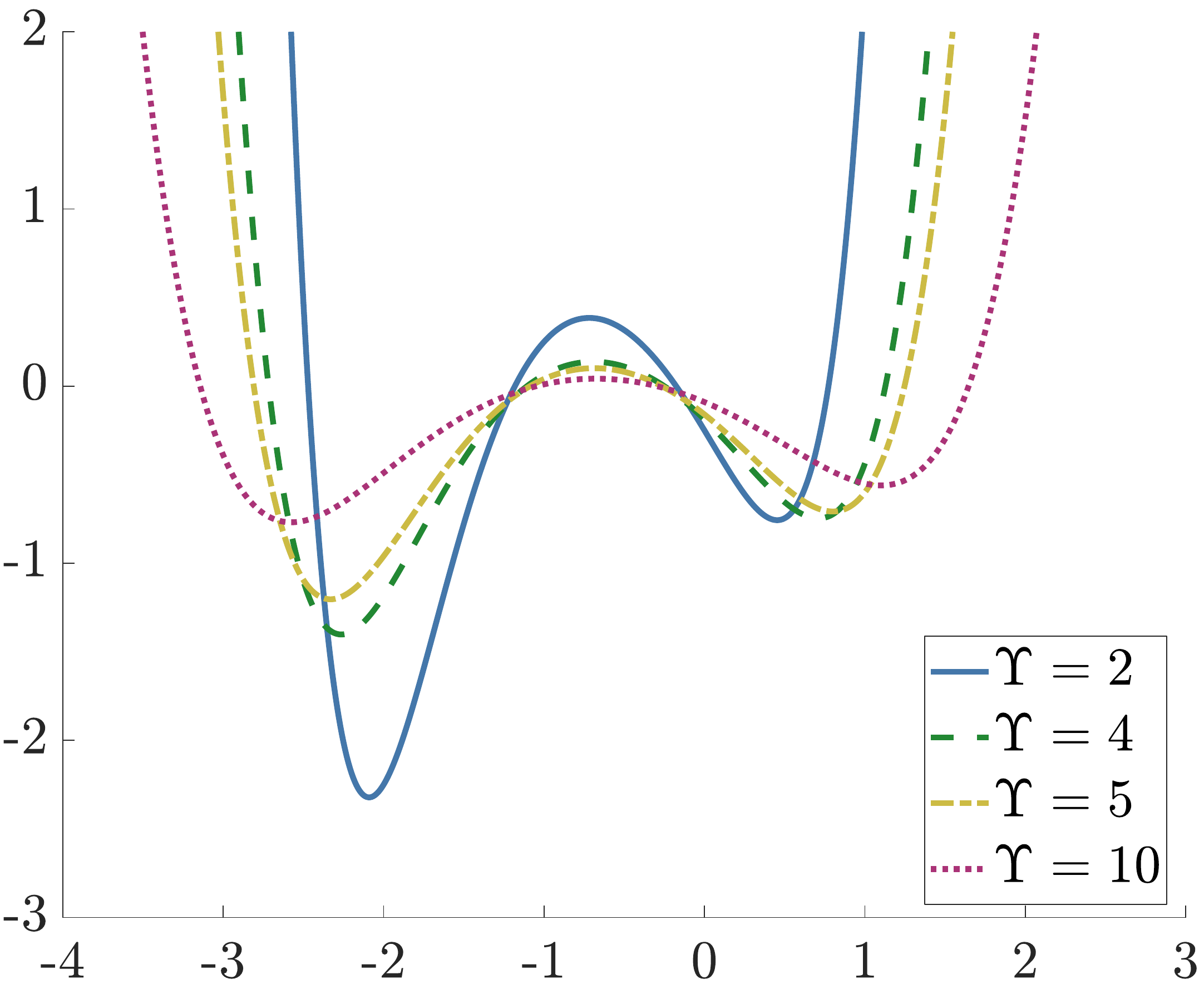}
	\caption[How the Schr\"{o}dinger potential depends on temperature $\Upsilon$]{The dependence of the Schr\"{o}dinger potential $-\bar{U}$ (\ref{eq: Ubar=}) on the temperature $\Upsilon $ for the polynomial potential.}
	\label{fig: multiT_Gies.png}
\end{figure}
If we identify the first term as the equilibrium position then we simply obtain:
\begin{eqnarray}
\left\langle x (t)\right\rangle = \chi_{eq} + \sum_{n=1}^{\infty} e^{-E_n t}e^{V(x_0)/\Upsilon}p_n(x_0)  \int_{-\infty}^{\infty}\mathrm{d}x~x~e^{-V(x)/\Upsilon}p_n(x)  \label{eq:spec-1-point}
\end{eqnarray} 
The story is very similar for computing the variance. We first compute the normal two point function $\left\langle x^2(t)\right\rangle$ by replacing $x$ with $x^2$ in the derivation above:
\begin{eqnarray}
\left\langle x^2(t)\right\rangle = \left\langle x^2\right\rangle_{eq} + \sum_{n=1}^{\infty} e^{-E_n t}e^{V(x_0)/\Upsilon}p_n(x_0) \int_{-\infty}^{\infty}\mathrm{d}x~x^2~e^{-V(x)/\Upsilon}p_n(x)  \label{eq:spec-2-point}
\end{eqnarray} 
and we can then straightforwardly obtain the variance by combining (\ref{eq:spec-1-point}) \& (\ref{eq:spec-2-point}) in the combination $\textbf{Var}(x) = \left\langle x^2(t)\right\rangle - \left\langle x(t)\right\rangle^2$. Obtaining the spectrum $E_n$ and $p_n(x)$ may be complicated  by the fact that the actual Schr\"{o}dinger potential $\bar{U}$ (\ref{eq: Ubar=}) can develop temperature dependent features as the temperature is decreased-- see Fig.~\ref{fig: multiT_Gies.png}. Even for the simple polynomial potential in the Langevin equation it is clear that at low temperatures $\bar{U}$ becomes non-trivial, developing highly asymmetrical trapping wells. The increasing energy gap between the two minima indicates that, for a fixed initial condition, higher order terms in the spectral expansion can become important as the temperature is lowered.

\begin{figure}[t!]
	\centering
	\includegraphics[width=0.45\textwidth]{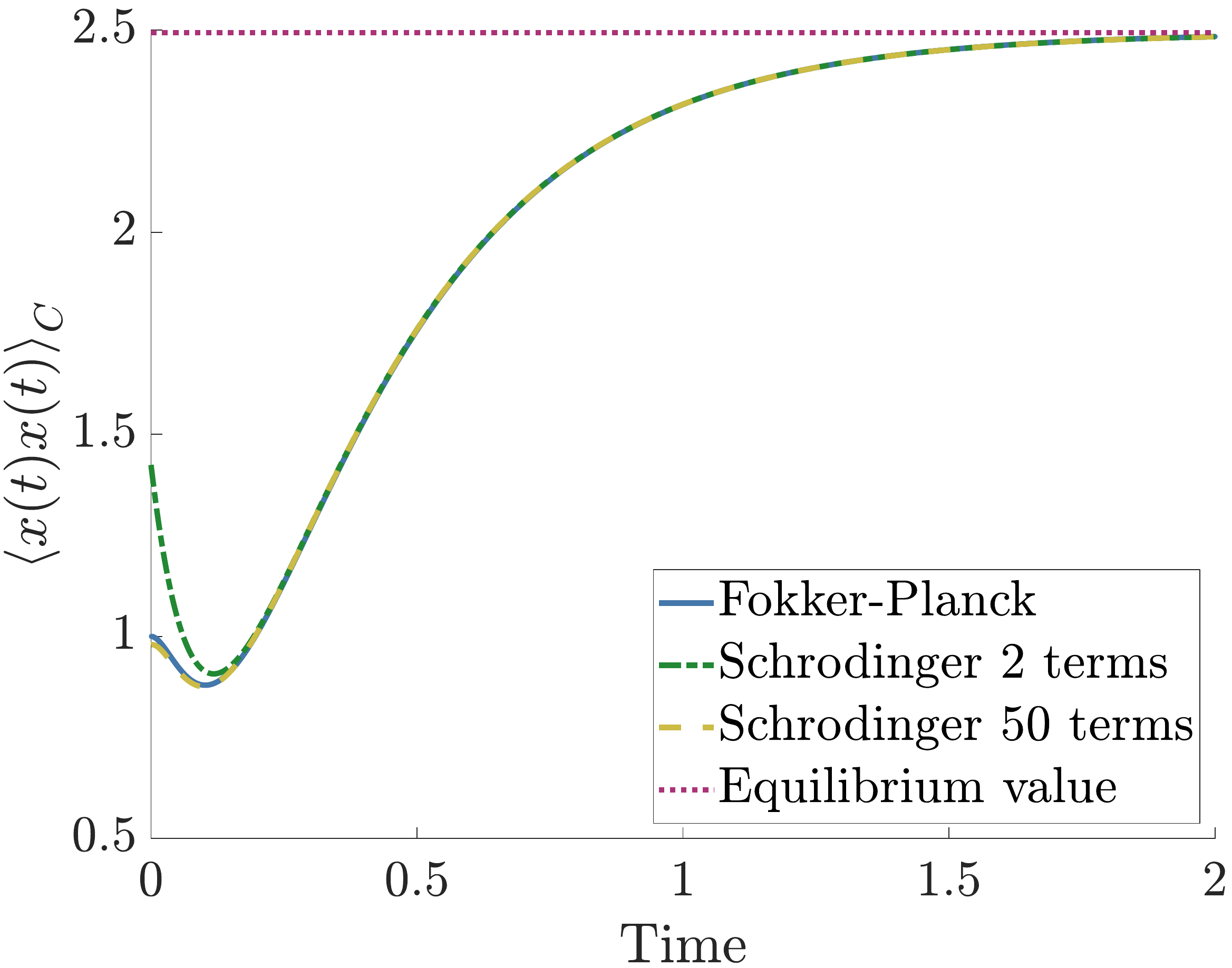}
	\includegraphics[width=0.45\textwidth]{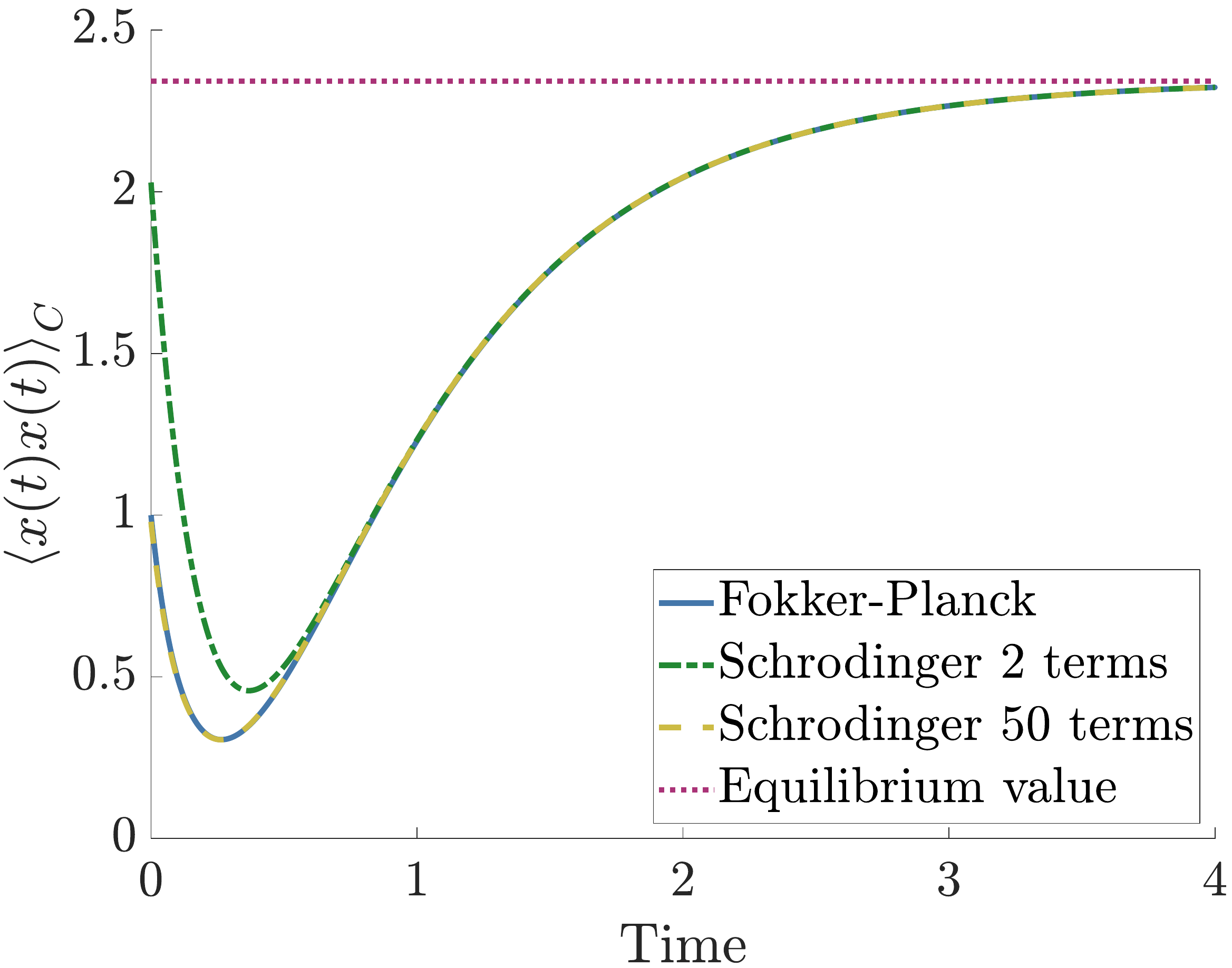}
	
	\caption[Spectral expansion method for the two point function in a polynomial potential]{The evolution of $\left\langle x^2(t)\right\rangle$ in the polynomial potential at $\Upsilon = 10$ (left) and $\Upsilon = 2$ (right). Included are the solutions from solving the whole \ac{F-P} equation numerically (\ref{eq: rescaled F-P}) as well as from a spectral expansion method (\ref{eq:spec-2-point}) keeping only the first 2 (red) and first 50 (blue) non-zero terms.}
	\label{fig: Gtt_Schro_5T_Gies.png}
\end{figure}

To illustrate the importance of these higher-order terms for the two-point function evolution in the polynomial potential we examine the accuracy of a finite truncation of the spectral expansion at two temperatures, $\Upsilon = 10$ and $\Upsilon = 2$, for the evolution of $\langle x^2(t)\rangle$, initialising trajectories at $x=1$: $P(x,t=0)=\delta(x-1)$. In Fig.~\ref{fig: Gtt_Schro_5T_Gies.png} we plot the evolution of $\langle x^2(t)\rangle$ by solving the \ac{F-P} equation (\ref{eq: rescaled F-P}) and by two different finite truncations of the spectral expansion method for $\Upsilon = 10$ (left) and $\Upsilon = 2$ (right). We can see at both temperatures that keeping only two terms in the spectral expansion method is still a very good description of the behaviour at all but early times. However in order to correctly describe the behaviour around the local minima more terms are required. As temperature is lowered we can see that the deviation at earlier times from the full \ac{F-P} trajectory is greater. \\
\begin{figure}[t!]
	\centering
	\includegraphics[width=0.55\textwidth]{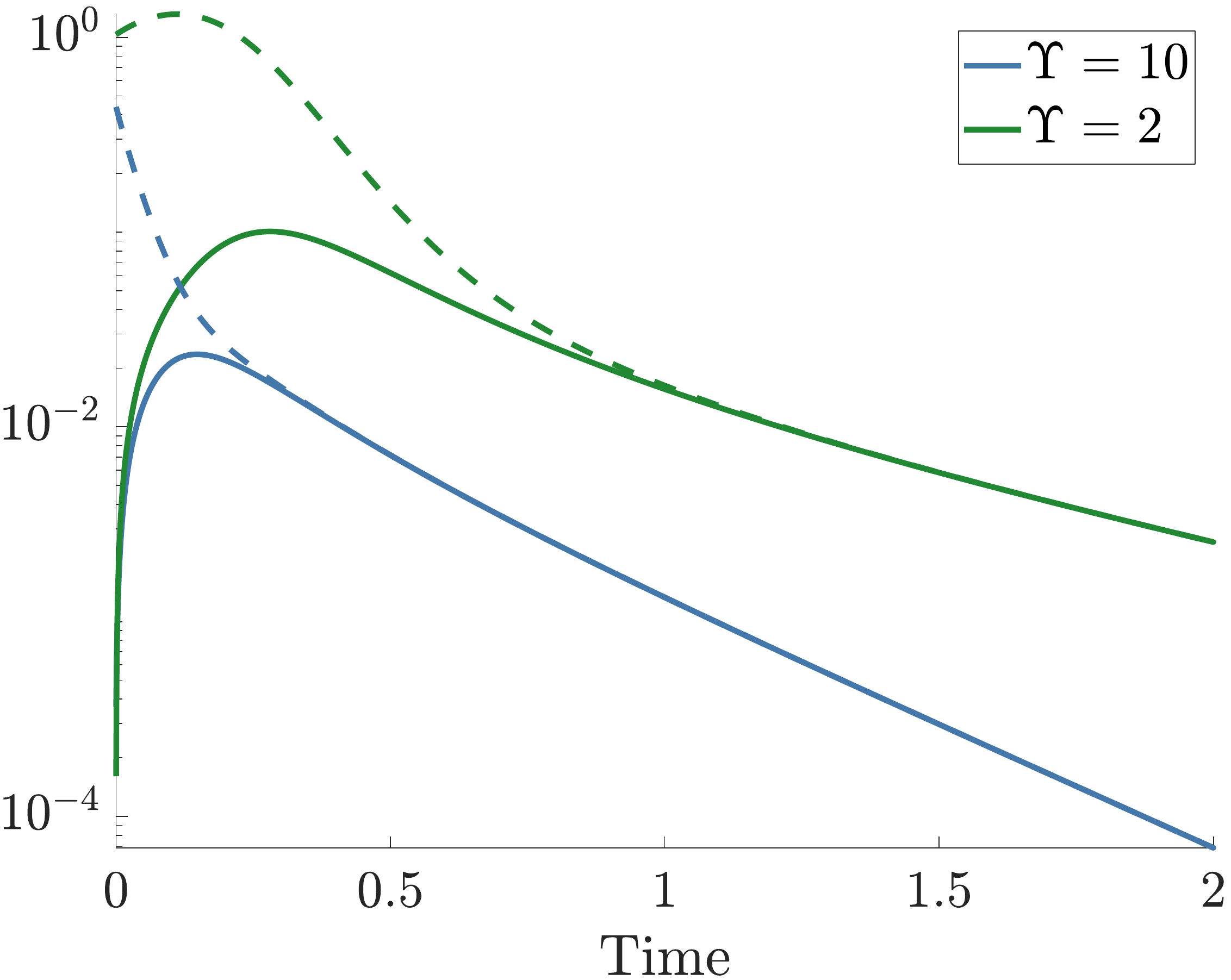}
	\caption[Error in the spectral expansion]{The error in $\left\langle x^2(t)\right\rangle$ for a finite truncation of the spectral expansion (\ref{eq:spec-2-point}) compared to the full \ac{F-P} equation (\ref{eq: rescaled F-P}) initialised at $\chi(0) = 1$ for the polynomial potential. The dashed (solid) lines correspond to keeping the first two (50) non-zero terms. }
	\label{fig: Gtt_Schro_AllT_Gies_error}
\end{figure}

In Fig.~\ref{fig: Gtt_Schro_AllT_Gies_error} we plot the error associated with a finite truncation of the spectral expansion, keeping only the first two (dashed line) or fifty (solid line) terms, at two different temperatures $\Upsilon=10$ (blue) or $\Upsilon=2$ (green). This error is computed by comparing the truncated expansion to the numerical solution of the \ac{F-P} equation (\ref{eq:FP1}). At early times, the error associated with keeping only two terms is larger than when 50 terms are kept, as one would expect, the discrepancy being more pronounced at lower temperatures. As the system relaxes, the contribution form the higher order terms decreases and the errors of the two truncations converge, until they are essentially indistinguishable at later times, as expected. This decay of the contribution from the higher eigenvalues occurs faster for the higher temperature, making the two-term truncation more accurate earlier. This observation reinforces our inference from the previous paragraph that as temperature is lowered, higher order terms in the spectral expansion become more important for accurately describing the evolution, at least for a fixed initial condition. Crucially, this offers an explanation for why the the \ac{LPA} + \ac{WFR} offers poorer agreement as temperature is lowered since it would be expected to most accurately describe circumstance where the lowest order terms in a spectral expansion dominate. The relation between the spectral expansion and the range of validity of the \ac{EA}'s derivative expansion is not entirely straightforward however, as it would also depend on the initial condition. 

\section{\label{sec:EEOM_conc}Conclusion}
\acresetall 
In this chapter we examined how the \ac{EA} $\Gamma$ allows one to derive \ac{EEOM} for the average position $\chi(t)\equiv \langle x(t) \rangle$ and variance $\langle x^2(t) \rangle_C$ in an analogous manner to the classical equations of motion, by taking variational derivatives. We used the \ac{LPA} and \ac{WFR} to compute the elements entering the \ac{EEOM}, for instance the \textit{dynamical effective potential}. We verified the accuracy of the equilibrium limit to these equations, further emphasising the physical significance of certain aspects of the effective potential $V_{{\kappa}=0}$: namely how the minimum of $V_{{\kappa}=0}$ corresponds to the equilibrium position and its second derivative evaluated at this point to the variance through equation (\ref{eq:equal2pt}). We noted here that while the \ac{LPA} reproduces these equilibrium quantities, the accuracy of the covariance's temporal evolution diminished as temperature was lowered.

Going beyond equilibrium, we examined how \ac{LPA}+\ac{WFR} handle relaxation towards it for the average position $\chi(t)$ in many complicated potentials including simple harmonic potentials with bumps on top. These potentials clearly demonstrates that the \ac{FRG} is capable of capturing the effect of the non-trivial local features. In fact, the \ac{FRG} still offers reasonable approximations even in those cases where the \ac{F-P} numerics failed to converge. We have also shown how the \ac{FRG} can closely match the relaxation of the variance $\langle x^2(t) \rangle_C$ to its equilibrium value: for asymmetric potentials, the \ac{LPA} variance has reasonable accuracy and still captures highly non-trivial behaviour such as the variance overshooting its equilibrium value before settling to it. This is in a system where numerically solving the \ac{F-P} equation failed to provide good results, at least using standard methods. Again, we find that accuracy decreases with decreasing temperature. 

A clear conclusion that can be drawn from the above investigations is that decreasing the temperature negatively impacts the accuracy of using the \ac{LPA} + \ac{WFR} derivative expansion for the \ac{FRG} to describe \ac{BM} in the potentials we examined. This appears to correlate with the increasing importance at lower temperatures of higher order terms in the spectral expansion; indeed, it is expected that the lowest order terms in the derivative expansion (\ac{LPA} + \ac{WFR}) are best placed to describe evolution dominated by the lowest non-zero eigenvalues of the \ac{F-P} spectral expansion. It would seem that the derivative expansion of the \ac{FRG} for studying thermal fluctuations has utility in the range from moderate temperatures (roughly when the classical force is comparable to the noise), up to the very high temperature regime where the small local features of the potential become less relevant.

\cleardoublepage 


\ctparttext{\color{black}Having examined stochastic processes and the Renormalisation Group in the context of mesoscale phenomena in part one we turn our attention to what might have happened billions of years ago. Cosmic Inflation couples the very small quantum phenomena with the very large cosmic expansion in a compelling way unique to \emph{The Early Universe}.} 

\part{\label{part:Cosmo}The Early Universe} 

\chapter{Inflationary Perturbations} 
\label{cha:Inflation} 
\acresetall 
\vspace{0.5cm}
\begin{flushright}{\slshape    
The truth is much too complicated to allow anything but approximations} \\ \medskip
--- John von Neumann \cite{JohnvonNeumann}
\end{flushright}
\vspace{0.5cm}

We now turn our attention away from the mesoscopic scale to the two extremes seen in the early universe; the very small quantum and the cosmically large. \\

We begin this chapter with a brief overview of what cosmic inflation is and why we need it in section \ref{sec:Inflation} where we also include a pedagogical discussion surrounding the horizon. We proceed in section \ref{sec:cosm_pert} to discuss the behaviour of perturbations away from homogeneous evolution. This is done both for linear perturbations as well as non-linear super-horizon perturbations. In particular we do a thorough treatise of the often neglected \ac{GR} momentum constraint. In section \ref{sec:stoch_infl} we consistently match these sub- and super-horizon perturbations and therefore successfully incorporate quantum backreaction on cosmic scales in a regime known as \textit{stochastic inflation}. We summarise our conclusions in section \ref{sec:infla_conc}. We defer many technical details of cosmological perturbations to Appendix \ref{app:Cosmo_perts}.\\

Much of this chapter will be familiar to those with a background in Cosmology. Section \ref{sec:Inflation} is very standard and has many excellent, more detailed, treatments in various textbooks and lecture notes, see e.g. \cite{Mukhanov2005,Baumann2009}. The cosmological perturbation theory covered in section \ref{sec:cosm_pert} is also rather standard with various treatments covered in e.g. \cite{Bardeen1980,Mukhanov1992, Arnowitt2008, Misner1964}. The details of stochastic inflation \cite{Starobinsky1988, Nambu1988,Nambu1989,Mollerach1991,Salopek1991,Habib1992, Linde1994,Starobinsky1994} covered in section \ref{sec:stoch_infl} however is not so standard and the non-expert is directed to the main results of this section listed below. The stochastic-$\delta \mathcal{N}$ formalism \cite{Enqvist2008, Fujita2013, Fujita2014, Vennin2015} covered in section \ref{sec:stochastic_deltaN} is even less well known and includes results from our paper \cite{Rigopoulos2021} which extends known results to be valid outside \ac{SR}. \\

Based on the expertise of the reader discussed above, they are directed to the main results of this chapter:
\begin{itemize}
    \item In section \ref{sec:Inflation} the main results are the first (\ref{eq:Friedmann_Inflation}) and second (\ref{eq: K-G eqn}) Friedmann equations for homogeneous inflation and the definition of the \ac{SR} parameters (\ref{eq:epsilon_multi_defn}). 
    \item In section \ref{sec:cosm_pert} the main results are the Mukhanov-Sasaki equation (\ref{eq:M-S equation}), the classical PDF (\ref{eq:PDF_varphiPi_squeezed}) that reproduces the inflationary correlators in the squeezed limit and the \ac{H-J} equation (\ref{eq: H-J equation}) with associated equation of motion (\ref{eq: dphidt HJ}).
    \item In section \ref{sec:stoch_infl} we refer the reader to Figs.~\ref{fig:Stochastic_fluctuations_kspace} \& \ref{fig:Stochastic_fluctuations_realspace} for a schematic overview of how stochastic inflation works. Equation (\ref{eq:deSitter_stochastic_equations}) is the main dynamical equation for stochastic inflation and from this one can obtain the coarse-grained curvature perturbation using equation (\ref{eq:Rcg_defn}).
\end{itemize}

\section{\label{sec:Inflation} What is Cosmic Inflation?}
In this section we will outline the basics needed to motivate an inflationary period as well as the background dynamics required to explain it mathematically.

\subsection{\label{sec:Homogeneous} Homogeneous equations of motion}
The evolution of the universe is determined by the Einstein equation:
\begin{eqnarray}
G_{\mu\nu} = \dfrac{1}{M_{\mathrm{p}}^2} T_{\mu\nu} \label{eq:Einstein_eqn}
\end{eqnarray}
which relates the curvature of spacetime through the Einstein tensor $G_{\mu\nu}$ to the matter content of the universe through the stress-energy tensor $T_{\mu\nu}$. N.B. we are writing quantities in terms of the \emph{reduced} Planck mass:
\begin{eqnarray}
M_{\mathrm{p}}^2 = \dfrac{8\pi G}{\hslash c} \label{eq:planck_mass_defn}
\end{eqnarray} 
but in practice from now on we set the speed of light $c = 1$. The Einstein Tensor is defined in terms of the Ricci tensor, $R_{\mu\nu}$, metric tensor, $g_{\mu\nu}$, and Ricci scalar $R$:
\begin{eqnarray}
G_{\mu\nu} &=& R_{\mu\nu}-\dfrac{1}{2}g_{\mu\nu}R \label{eq:Einstein_tensor}\\
R &=& R^{\lambda}_{~\lambda}\label{eq:Ricci scalar}\\
R_{\mu\nu} &=& R^{\lambda}_{~\mu \lambda \nu}\label{eq:Ricci tensory}\\
R^{\alpha}_{~\beta \mu\nu} &=& \partial_{\mu}\Gamma^{\alpha}_{~\nu\beta} - \partial_{\nu}\Gamma^{\alpha}_{~\mu\beta} +\Gamma^{\alpha}_{~\mu\lambda}\Gamma^{\lambda}_{~\beta\nu} - \Gamma^{\alpha}_{~\nu\lambda}\Gamma^{\lambda}_{~\mu\beta}  \label{eq:Riemann tensor}\\
\Gamma^{\alpha}_{~\mu\nu} &=& \dfrac{1}{2}g^{\alpha \lambda} \lb \partial_{\nu}g_{\lambda\mu} + \partial_{\mu}g_{\lambda\nu} - \partial_{\lambda}g_{\mu\nu} \rb\label{eq:Christoffel symbols}
\end{eqnarray}
The most general metric we can work with that respects homogeneity and isotropy is the FLRW metric:
\begin{eqnarray}
g_{\mu\nu}\mathrm{d}x^{\mu}\mathrm{d}x^{\nu} = -\mathrm{d}t^2 + a^2(t)\lb \dfrac{\mathrm{d}r^2}{1-Kr^2} + r^2 \lb \mathrm{d}\theta^2 + \sin^2 \theta \mathrm{d}\phi^2\rb\rb \label{eq:FLRW_full}
\end{eqnarray}
where $r$, $\theta$ and $\phi$ are the standard radial and spherical coordinates. $K$ describes the sign of the curvature on spatial hypersurfaces and is positive, zero or negative for positively curved, flat and negatively curved spatial hypersurfaces respectively. For reasons we will justify later we can assume that we are dealing with flat hypersurfaces to obtain the simpler, flat FLRW metric:
\begin{eqnarray}
g_{\mu\nu}\mathrm{d}x^{\mu}\mathrm{d}x^{\nu} = -\mathrm{d}t^2 + a^2(t)\delta_{ij}\mathrm{d}x^i\mathrm{d}x^j \label{eq:FLRW}
\end{eqnarray}
From this we can compute the non-zero components of the Einstein tensor $G^{\mu}_{\nu}$:
\begin{subequations}
\begin{align}
    G^{0}_{0} &= 3\lb\dfrac{\dot{a}}{a}\rb^2 \\
    G^{i}_{j} &= \lsb 2\dfrac{\ddot{a}}{a} + \lb\dfrac{\dot{a}}{a}\rb^2 \rsb \delta^{i}_{j}
\end{align} \label{eq:G for FLRW}
\end{subequations}
so we know what the left hand side of (\ref{eq:Einstein_eqn}) is. As for the right hand side we shall consider a perfect fluid:
\begin{eqnarray}
T^{\mu}_{\nu} = \lb \rho + P\rb U^{\mu}U_{\nu} - P\delta^{\mu}_{\nu} \label{eq:T for FLRW}
\end{eqnarray}
where $\rho$ and $P$ are the density and pressure of the fluid respectively and $U^{\mu}$ is the relative four-velocity between the fluid and the observer. While a perfect fluid can source an inhomogeneous spacetime in generality, for our current considerations spatial homogeneity implies that $\rho$, $P$ and $U^{\mu}$ can only depend on time\footnote{Isotropy imposes the weaker requirement that $U^{\mu}$ can only depend on time.}. For a comoving observer $U^{\mu} = (1,0,0,0)$. This means we can simply combine equations (\ref{eq:G for FLRW}) \& (\ref{eq:T for FLRW}) to get the well known \emph{Friedmann equations}:
\begin{subequations}
\begin{align}
    \lb\dfrac{\dot{a}}{a}\rb^2 &= \dfrac{1}{3M_{\mathrm{p}}^2}\rho \\
    \dfrac{\ddot{a}}{a} &= -\dfrac{1}{6M_{\mathrm{p}}^2}\lb \rho + 3P\rb 
\end{align} \label{eq:Friedmann equations}
\end{subequations}
The first Friedmann equation is usually written in terms of the Hubble rate, defined as:
\begin{eqnarray}
H \equiv \dfrac{\dot{a}}{a} \label{eq:H_homo-defn}
\end{eqnarray}
so that the first Friedmann equation (from the energy constraint) becomes:
\begin{eqnarray}
H^2 = \dfrac{1}{3M_{\mathrm{p}}^2}\rho \label{eq:Friedmann_theflat}
\end{eqnarray}
It is worth noting that there is one more equation that needs to be accounted for. Covariant conservation of the stress-energy tensor $\nabla_{\mu}T^{\mu}_{\nu} = 0$ gives us the \emph{continuity equation}:
\begin{eqnarray}
\dot{\rho} + 3H(\rho + P) \label{eq:continuity eqn} = 0
\end{eqnarray}
If we now parameterise the fluids independently in terms of the equation of state parameter $w = P/\rho$ then -- assuming $w$ is constant or slowly varying -- the solution to (\ref{eq:continuity eqn}) is:
\begin{eqnarray}
\rho &\propto &a^{-3\lb 1 + w\rb } \\
\Rightarrow \rho &\propto &
\begin{cases}
a^{-3}, \quad \text{for pure matter}\\
a^{-4}, \quad \text{for pure radiation}\\
a^{0}, \quad \text{for just a cosmological constant}\\
\end{cases}
\end{eqnarray}

\subsection{Horizons}
In an expanding spacetime the propagation of light is best examined using conformal time $\tau$ defined such that $a(\tau)\mathrm{d}\tau = \mathrm{d}t$. Then we can describe the evolution using the two-dimensional\footnote{As the spacetime is isotropic we can always define the coordinate system so that we can suppress the angular coordinates and the light travels purely in the radial direction.} line element:
\begin{eqnarray}
\mathrm{d}s^2 = a^2(\tau)\lsb \mathrm{d}\tau^2 -\mathrm{d}r^2\rsb
\end{eqnarray}
As light travels along null geodesics ($\mathrm{d}s^2 = 0$) their path is simply given by lines at $45^{\circ}$ in $r-\tau$ coordinates:
\begin{eqnarray}
\Delta r = \pm \Delta \tau
\end{eqnarray}
where the plus (minus) corresponds to outgoing (ingoing) photons. We can use this to describe different kinds of cosmological \emph{horizons}. \\

The \emph{particle horizon} is concerned by what events in the past \emph{could} have influenced an observer. Put another way, if the Big Bang `started' with the singularity at $t_i = 0$ then the greatest comoving distance from which an observer at time $t$ can receive signals is given by:
\begin{eqnarray}
r_{\mathrm{ph}}(\tau ) = \tau - \tau_i = \int_{t_i}^{t}\dfrac{\mathrm{d}t}{a(t)} \label{eq:particle_horizon}
\end{eqnarray}
This defines the comoving particle horizon. This means that for a comoving particle to have influenced an observer at $p$, the particle's worldline must have intersected the past lightcone of $p$. \\

The \emph{event horizon} in contrast focuses on the future and determines what future events will we be able to see eventually. To determine where this boundary is we consider the distance light can travel from the current time to the final time $t_f$:
\begin{eqnarray}
r_{\mathrm{eh}} (\tau) = \tau_f - \tau = \int_{t}^{t_f}\dfrac{\mathrm{d}t}{a(t)} \label{eq:event_horizon}
\end{eqnarray}
this defines the comoving event horizon. As the name suggests it is similar to the event horizon of black holes. \\

There is another quantity that is often (confusingly) called the horizon, this is the comoving \emph{Hubble radius} defined by:
\begin{eqnarray}
r_{\mathrm{H}} = \dfrac{1}{aH} \label{eq:Hubble_radius}
\end{eqnarray}
For a universe dominated by a fluid with constant equation of state $w\equiv P/\rho$ we find:
\begin{eqnarray}
(aH)^{-1} \propto a^{(1+3w)/2}
\end{eqnarray}
This means that for `normal' matter sources -- e.g. radiation, baryons -- which satisfy the strong energy condition, $1+3w > 0$, that the comoving Hubble radius increases as the universe expands and the particle horizon and Hubble radius are approximately equivalent i.e. $r_{\mathrm{ph}} \sim (aH)^{-1}$. This explains why the two are often used interchangeably and both unhelpfully referred to as ``the horizon". While both describe how different observers are causally connected there is a fundamental difference. If we consider two observers with a comoving separation $\lambda$ then the Hubble radius $(aH)^{-1}$ tells you whether they could communicate within the next \emph{Hubble time}\footnote{This is denoted by $H^{-1}$ and is the time in which the scale factor increases by a factor of $e$.} i.e. whether they are in causal contact \emph{right now}. On the other hand the particle radius tells us whether the two observers have ever been able to communicate. \\
\clearpage
Summarising all this we can say:
\begin{itemize}
    \item If $\lambda > r_{\mathrm{ph}}$, then the two observers could \emph{never have} communicated before now. 
    \item If $\lambda > r_{\mathrm{H}}$, then the two observers can't communicate \emph{within the next Hubble time}.
    \item If $\lambda > r_{\mathrm{eh}}$, then the two observers will \emph{never be able to} communicate in the future.
\end{itemize}

\subsection{\label{sec:why inflation}Why do we need inflation?}
The \ac{CMB} has demonstrated how incredibly homogeneous\footnote{Strictly speaking the \ac{CMB} only directly shows isotropy, To demonstrate homogeneity one has to also invoke the Copernican principle; namely that we are not in a ``privileged" position in the universe and our observations can be considered typical.} and isotropic the universe is on large scales. The deviations in temperature from the average value is of order $10^{-5}$ \cite{Akrami2020}. We will now explain why this poses a problem for the standard Big Bang evolution of the universe.

\subsubsection{The Horizon Problem}
In the standard evolution of Big Bang Cosmology the comoving Hubble radius decreases as you go back in time. This would suggest that if we pick two distant points we observe in the \ac{CMB} they would be causally disconnected. We demonstrate what we mean by this in the left plot of Fig.~\ref{fig:horizon_problem}. We can see that the past lightcones (given by the $45^{\circ}$ lines) of the distant points in the \ac{CMB} do not overlap. This means that $\lambda > r_{\mathrm{ph}}$ i.e. they lie outside each others particle horizon and could never have communicated. This would not be a problem if it weren't for the fact that these two regions are actually very similar! The whole \ac{CMB} is so uniform that the temperature deviation is given by $\delta T/T \sim 10^{-5}$ which means that the universe would have had to be incredibly uniform to start with. This introduces a fine-tuning problem to the start of the universe that we would like to avoid. 

\subsubsection{The Flatness Problem}
We have been explicitly working with a spatially flat universe but why are we justified in doing so? \ac{GR} famously curves space in response to matter in the universe so why should it be so well described by Euclidean space? To make this problem more concrete we consider the first Friedmann equation (\ref{eq:Friedmann_theflat}) but now with non-zero spatial curvature:
\begin{eqnarray}
M_{\mathrm{p}}^2H^2 = \dfrac{1}{3}\rho -\dfrac{K}{a^2}M_{\mathrm{p}}^2\label{eq:Friedmann_notflat}
\end{eqnarray}
which can be rewritten in terms of critical energy density $\rho_{crit} \equiv 3M_{\mathrm{p}}^2H^2$:
\begin{eqnarray}
1 - \Omega \equiv 1-\dfrac{\rho}{\rho_{crit}} = -\dfrac{K}{\lb aH\rb^2}
\end{eqnarray}
The critical energy density $\rho_{crit}$ corresponds to the energy density in a flat universe, therefore $1 - \Omega = 0$ corresponds to a flat universe. It is straightforward to show that:
\begin{eqnarray}
\dfrac{\mathrm{d} | \Omega - 1 |}{\mathrm{d}\ln a} = (1+3w)\Omega \lb \Omega -1 \rb \label{eq:dOmegadlna}
\end{eqnarray}
This tells us that $\Omega = 1$ is an unstable fixed point if the strong energy condition is satisfied:
\begin{eqnarray}
\dfrac{\mathrm{d} | \Omega - 1 |}{\mathrm{d}\ln a} > 0 \Leftrightarrow 1 + 3w > 0
\end{eqnarray}
In standard Big Bang Cosmology the strong energy condition is satisfied and extreme fine-tuning of the initial conditions are required for the observed value of $\Omega \approx 1$. 

\subsection{Inflation as the solution}
\begin{figure}[t!]
    \centering
    \includegraphics[width = 0.95\linewidth]{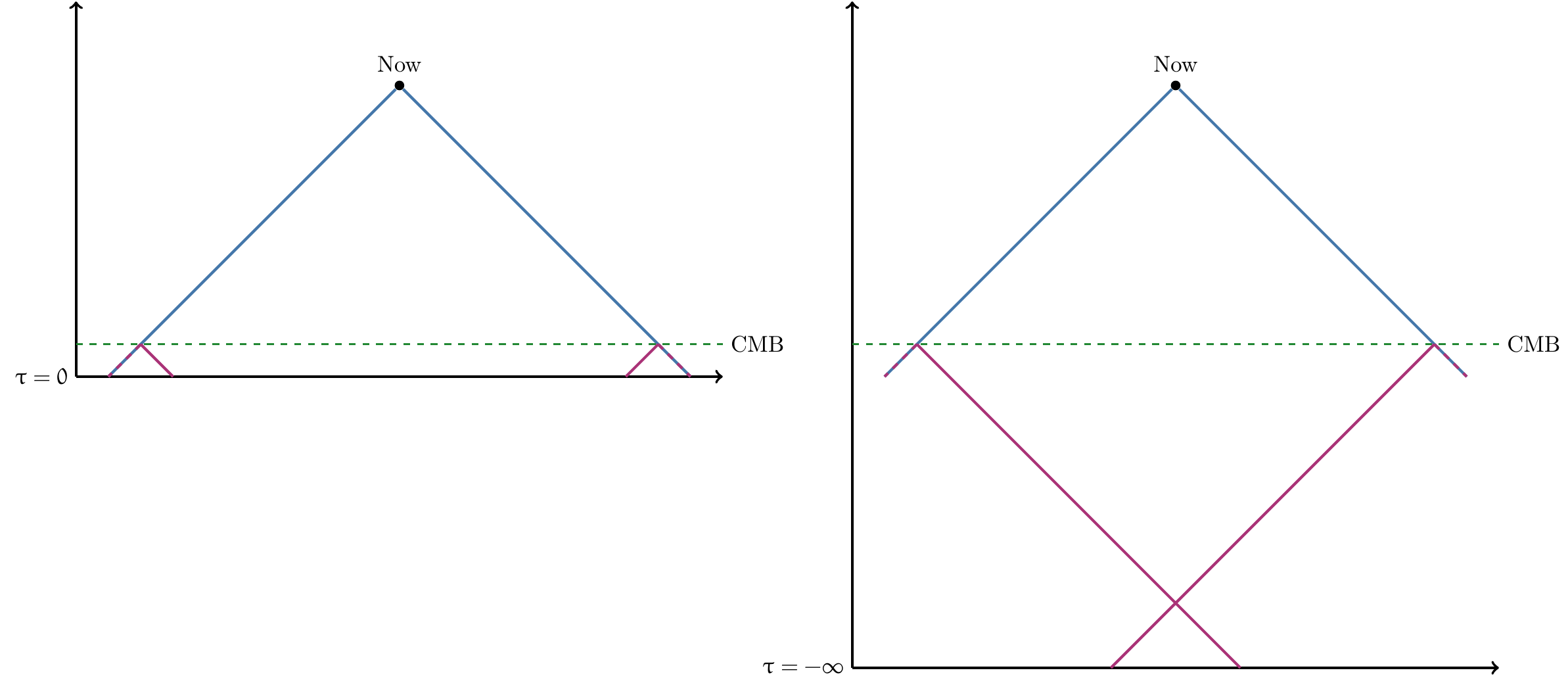}
    \caption[The Horizon Problem]{In the left plot we plot the past lightcone, in conformal time for an observer today, with blue lines at $45^{\circ}$ angles in the standard Big Bang Cosmology. We can see that distant points we observe in the \ac{CMB} have past lightcones, shown by the red lines, that do not intersect. These means that these distant regions we observe could not have communicated with each other. In the right plot we include an inflationary period prior to the standard Big Bang Cosmology which means we can extend the conformal time coordinate down to $\tau = -\infty$. We can see that this additional conformal time now means that the past lightcones of the distant regions we observe in the \ac{CMB} do intersect so that they were in causal contact with one another at some time in the past.}
    \label{fig:horizon_problem}
\end{figure}
We have seen the problems that need to be solved result from a conformal Hubble radius that is increasing. A natural way to try to solve the horizon problem is to consider a phase of a decreasing Hubble radius:
\begin{eqnarray}
\dfrac{\mathrm{d}}{\mathrm{d}t}(aH)^{-1} < 0
\end{eqnarray}
For this to be true we require a fluid that violates the strong energy condition i.e. $1+3w<0$. If this period of a decreasing Hubble radius occurs for long enough then the horizon problem is solved -- see right plot of Fig.~\ref{fig:horizon_problem}. Here we can see that adding this period of accelerated expansion essentially gives us more conformal time allowing the past lightcones of two seemingly disconnected regions in the \ac{CMB} to come into causal contact at finite time in the past. This means that our entire observable universe was, at some time, all causally connected -- i.e. it lay within the same Hubble sphere -- and so explains how the \ac{CMB} appears so uniform. Another way of looking at this is what although distant patches we observe in the \ac{CMB} were not in each others Hubble sphere at that time, they were in their particle horizon: $r_{\mathrm{H}} < \lambda < r_{\mathrm{ph}}$. In this way inflation is a mechanism to make the particle horizon larger than the Hubble radius $r_{\mathrm{ph}} > r_{\mathrm{H}}$. As we have had to violate the strong-energy condition we have also solved the flatness problem as $\Omega =1$ is now an attracter of (\ref{eq:dOmegadlna}).\\

We can define this inflationary period in several complementary ways:
\begin{itemize}
    \item \emph{Accelerated Expansion} \\
    \begin{eqnarray}
    \dfrac{\mathrm{d}}{\mathrm{d}t}(aH)^{-1} = -\dfrac{\ddot{a}}{\dot{a}^2}< 0 \Rightarrow \ddot{a} > 0
    \end{eqnarray}
    A shrinking comoving Hubble radius is equivalent to a period of accelerated expansion -- hence the name \emph{Inflation}.
    \item \emph{Slowly-varying $H$}\\
    If we define a quantity known as the first \ac{SR} parameter:
    \begin{eqnarray}
    \varepsilon_1 \equiv -\dfrac{\dot{H}}{H^2} = -\dfrac{\mathrm{d}\ln H}{\mathrm{d}\alpha}\label{eq:SR1_H}
    \end{eqnarray}
    where we have defined $\mathrm{d}\alpha = H\mathrm{d}t$ which measures the number of e-folds $\alpha$ of inflationary expansion. We then have the condition:
    \begin{eqnarray}
    \dfrac{\mathrm{d}}{\mathrm{d}t}(aH)^{-1} = -\dfrac{1}{a}\lb 1 - \varepsilon_1 \rb \Rightarrow \varepsilon_1 < 1
    \end{eqnarray}
    \item \emph{Negative pressure fluid}\\
    We can rewrite the condition on $\varepsilon_1$ in terms of the density, $\rho$, and pressure, $P$ of the cosmological fluid:
    \begin{eqnarray}
    \varepsilon_1 = \dfrac{3}{2}\lb 1 + \dfrac{P}{\rho}\rb < 1 \Rightarrow P < -\dfrac{\rho}{3}
    \end{eqnarray}
\end{itemize}
\subsection{Homogeneous Inflation}
While a cosmological constant might seem like the natural candidate to drive inflation it has some problems, mainly that inflation would last forever! Instead the natural thing to do is to consider a quantum scalar field $\varphi (t,\vec{x} )$ suggestively called the \emph{inflaton} evolving in a potential $V(\varphi )$. The stress-energy tensor for this field is:
\begin{eqnarray}
T_{\mu\nu} = \partial_{\mu}\partial_{\nu}\varphi - g_{\mu\nu}\lb \dfrac{1}{2}g^{\alpha\beta}\partial_{\alpha}\varphi \partial_{\beta}\varphi  -V(\varphi)\rb 
\end{eqnarray}
If we want consistency with the symmetries of FLRW spacetime -- i.e. homogeneity and isotropy -- then the inflaton can only depend on time, $\varphi = \varphi (t)$. This allows us to determine the density, $\rho_{\varphi}$, and pressure, $P_{\varphi}$, from the $T^{0}_{0}$ and $T^{i}_{j}$ components respectively:
\begin{eqnarray}
\rho_{\varphi} &=& \dfrac{1}{2}\dot{\varphi}^2 + V(\varphi)\\
P_{\varphi} &=& \dfrac{1}{2}\dot{\varphi}^2 - V(\varphi)
\end{eqnarray}
A field configuration that leads to inflation requires $P_{\varphi} < \rho_{\varphi}/3$:
\begin{eqnarray}
\dfrac{1}{2}\dot{\varphi}^2 < 2V(\varphi) \label{eq:inflation_condition_potential}
\end{eqnarray}
i.e. the potential energy dominates over the kinetic energy. Using all this we can rewrite the first Friedmann equation as:
\begin{empheq}[box = \fcolorbox{Maroon}{white}]{equation}
H^2 = \dfrac{1}{3M_{\mathrm{p}}^2}\lsb \dfrac{1}{2}\dot{\varphi}^2 + V(\varphi)\rsb \label{eq:Friedmann_Inflation}
\end{empheq}
and by taking a time derivative of this and substituting in the second Friedmann equation\footnote{N.B. that this can also be straightforwardly derived from the continuity equation (\ref{eq:continuity eqn})} we get the well known Klein-Gordon equation:
\begin{empheq}[box = \fcolorbox{Maroon}{white}]{equation}
\ddot{\varphi} + 3H\dot{\varphi} + \dfrac{\mathrm{d} V(\varphi)}{\mathrm{d}\varphi} = 0 \label{eq: K-G eqn}
\end{empheq}
Notice how the Klein-Gordon equation describes damped motion where the gradient of the potential acts like a force and the expansion of the universe, $H$, acts like friction. 
\subsubsection{Slow-Roll Inflation}
For inflation to take place we require that $\varepsilon_1 < 1$, which suggests -- see (\ref{eq:inflation_condition_potential}) -- that the kinetic energy is subdominant compared to the potential energy. If we take $\varepsilon_1 \ll 1$ then we can approximate the Friedmann equation (\ref{eq:Friedmann_Inflation}) as:
\begin{eqnarray}
H^2 \approx \dfrac{V}{3M_{\mathrm{p}}^2} \label{eq:Friedmann_SR}
\end{eqnarray}
Notice how this restricts the Hubble parameter to depend only on $\varphi$ i.e. $H = H(\varphi )$ without any explicit time dependence. To determine when we can simplify the Klein-Gordon equation (\ref{eq: K-G eqn}) we introduce the second \ac{SR} parameter:
\begin{eqnarray}
\varepsilon_2 \equiv -\dfrac{\ddot{H}}{H\dot{H}} + 2\dfrac{\dot{H}}{H^2} =-\dfrac{\ddot{H}}{H\dot{H}}  -2\varepsilon_1 \label{eq:SR2_H}
\end{eqnarray}
If this is also very small -- i.e. $\varepsilon_2 \ll 1$ -- then we can neglect the acceleration term in (\ref{eq: K-G eqn}) to obtain:
\begin{eqnarray}
3H\dot{\varphi} \approx -\dfrac{\mathrm{d} V}{\mathrm{d}\varphi} \label{eq:K-G_SR}
\end{eqnarray}
Equations (\ref{eq:Friedmann_SR}) \& (\ref{eq:K-G_SR}) are the \emph{slow-roll equations} which are valid for $\varepsilon_1, \varepsilon_2 \ll 1$. 

We can then define the first three Hubble \ac{SR} parameters, $\epsilon_1$, $\epsilon_2$ \& $\varepsilon_{3}$ in a pleasingly iterative way:
\begin{subequations}
\begin{empheq}[box = \fcolorbox{Maroon}{white}]{align}
\epsilon_1 &\equiv -\dfrac{\mathrm{d}~\text{ln} H}{\mathrm{d}\alpha} \label{eq: epsilon1 defn} \\
\epsilon_2 &\equiv -\dfrac{\mathrm{d}~\text{ln} \epsilon_1}{\mathrm{d}\alpha}  \label{eq: epsilon2 defn} \\
\epsilon_3 &\equiv -\dfrac{\mathrm{d}~\text{ln} \epsilon_2}{\mathrm{d}\alpha}  \label{eq: epsilon3 defn} 
\end{empheq} \label{eq:epsilon_multi_defn}
\end{subequations}
It is worth remembering that despite the name these \ac{SR} parameters are \textit{exact definitions} and make no a-priori assumption about the inflaton being in a \ac{SR} regime. \\
While this homogeneous picture is very nice and solves some of the problems with the standard Big Bang evolution its greatest strength is actually how it can seed the small deviations we observe from homogeneity. We discuss how to accomplish this in the next section. 

\section{\label{sec:cosm_pert} Cosmological Perturbations }
We have outlined in section \ref{sec:Inflation} what inflation is and how it solves problems with the Big Bang model such as the horizon and flatness problems. If this was all inflation was able to do however it wouldn't be very interesting. What is notable is that while inflation works as a method to make things homogeneous on large scales, it also offers a mechanism for creating the small deviations from homogeneity we observe in the \ac{CMB}. In this section we will outline how to do this using two different methods. In section \ref{sec:lin_pert} we will consider small, linear perturbations from homogeneity and isotropy and discuss issues surrounding choice of gauge. In section \ref{sec:nonlin_pert} we will outline a framework for computing large \emph{non-linear} perturbations, the sacrifice to be made here is that one can only consider the behaviour of perturbations on long-wavelengths. 
\subsection{\label{sec:lin_pert}Linear perturbations}
In this subsection we will outline how to compute small linear perturbations from homogeneity. While this section is fully self-contained many details are skipped, those wishing to see all the nitty gritty are referred to Appendix \ref{sec:App_lin_pert}. \\

We begin by perturbing the homogeneous metric and inflaton in the following way:
\begin{eqnarray}
g_{\mu\nu}(t,\vec{x}) = \bar{g}_{\mu\nu}(t) + \delta g_{\mu\nu}(t,\vec{x}), \quad \varphi(t,\vec{x}) = \bar{\varphi} (t) + \delta \varphi (t,\vec{x}) \label{eq:linear_split}
\end{eqnarray}
where an overbar will be used from here on in to refer to the homogeneous quantity which obeys the equations discussed in section \ref{sec:Inflation}. If we are only interested in scalar perturbations then we can consider the following perturbed flat, FLRW metric:
\begin{eqnarray}
\mathrm{d}s^2 =  -\lb 1+2A \rb \mathrm{d}t^2 +2a(t) \partial_i B\mathrm{d}x^i\mathrm{d}t + a^2(t)\lsb \lb 1-2C\rb \delta_{ij} +2\partial_i\partial_j E\rsb\mathrm{d}x^i\mathrm{d}x^j \label{eq:perturb line element_scalar}
\end{eqnarray}
written in terms of four different scalar perturbations $A$, $B$, $C$ and $E$. $A$ is often called the lapse, B the shift, $C$ the spatial curvature and E the shear. The naive thing to do would be to perturb the Einstein equation:
\begin{eqnarray}
\delta G_{\mu\nu} = \dfrac{1}{M_{\mathrm{p}}^2}\delta T_{\mu\nu} \label{eq:Einstein_eqn_perturbed}
\end{eqnarray}
and compute the left and right hand sides using (\ref{eq:linear_split}) \& (\ref{eq:perturb line element_scalar}) to determine how the scalars $A$, $B$, $C$ and $E$ evolve. It is at this moment we need to address the gauge problem. \ac{GR} allows coordinate changes so we are free to redefine the coordinates in (\ref{eq:perturb line element_scalar}) such that there is no spatial curvature $C = 0$ or there is no perturbation in time $A = 0$ and so on. Different choices of coordinates correspond to different gauges. While physics does not care about what gauge you are in, the equations for $A$, $B$, $C$ and $E$ do. We will therefore introduce \emph{gauge-invariant} quantities that are the same regardless of the gauge choice. Before we do that however we will cover a few different, popular gauges. \\
\begin{itemize}
    \item \emph{Synchronous Gauge}\\
    This is defined such that there are no perturbations in the time coordinate, $A = B = 0$. 
    \item \emph{Newtonian Gauge} \\
    This is defined so that the equations reduce to Newtonian gravity in the small-scale limit. For this $B = E = 0$.
    \item \emph{Uniform $\varphi $ Gauge}\\
    Also known more generally as the uniform density gauge, this corresponds to setting the fluctuations in the inflaton to zero $\delta \varphi = 0$. After this there is still one more gauge freedom so we will also take\footnote{\label{footnote:gauge_stuff}It is worth noting that another common choice for both uniform $\varphi$ and comoving gauges is to set $B = 0$ instead of $E$. } $E = 0$ as well. It is then standard practice to introduce the \emph{uniform-density curvature perturbation}, $\zeta = -A$. 
    \item \emph{Comoving Gauge}\\
    This gauge comoves with the inflaton such that it always takes the inflaton the same amount of time to travel between two points on the inflationary potential. As inflation ends at a fixed field value, $\varphi_e$, this is also called the uniform expansion gauge as inflation lasts the same amount of time everywhere. In many ways this is the `preferred' gauge as it is the one we live in, i.e. it is what we observe in the \ac{CMB}. This gauge is determined by requiring the scalar momentum density to vanish $q = 0$. We also set$^{\ref{footnote:gauge_stuff}}$ $E= 0$. It is common to then introduce the \emph{comoving curvature perturbation}, $\mathcal{R} = -C$.
    \item \emph{Spatially-flat Gauge}\\
    As the name suggests the perturbations in the spatial component of the metric are set to zero: $C = E = 0$. This gauge is useful as it enables you to focus directly on the fluctuations in the inflaton $\delta \varphi$.
\end{itemize}

\subsubsection{Gauge-invariant quantities}
\begin{figure}[t!]
    \centering
    \includegraphics[width = 0.45\linewidth]{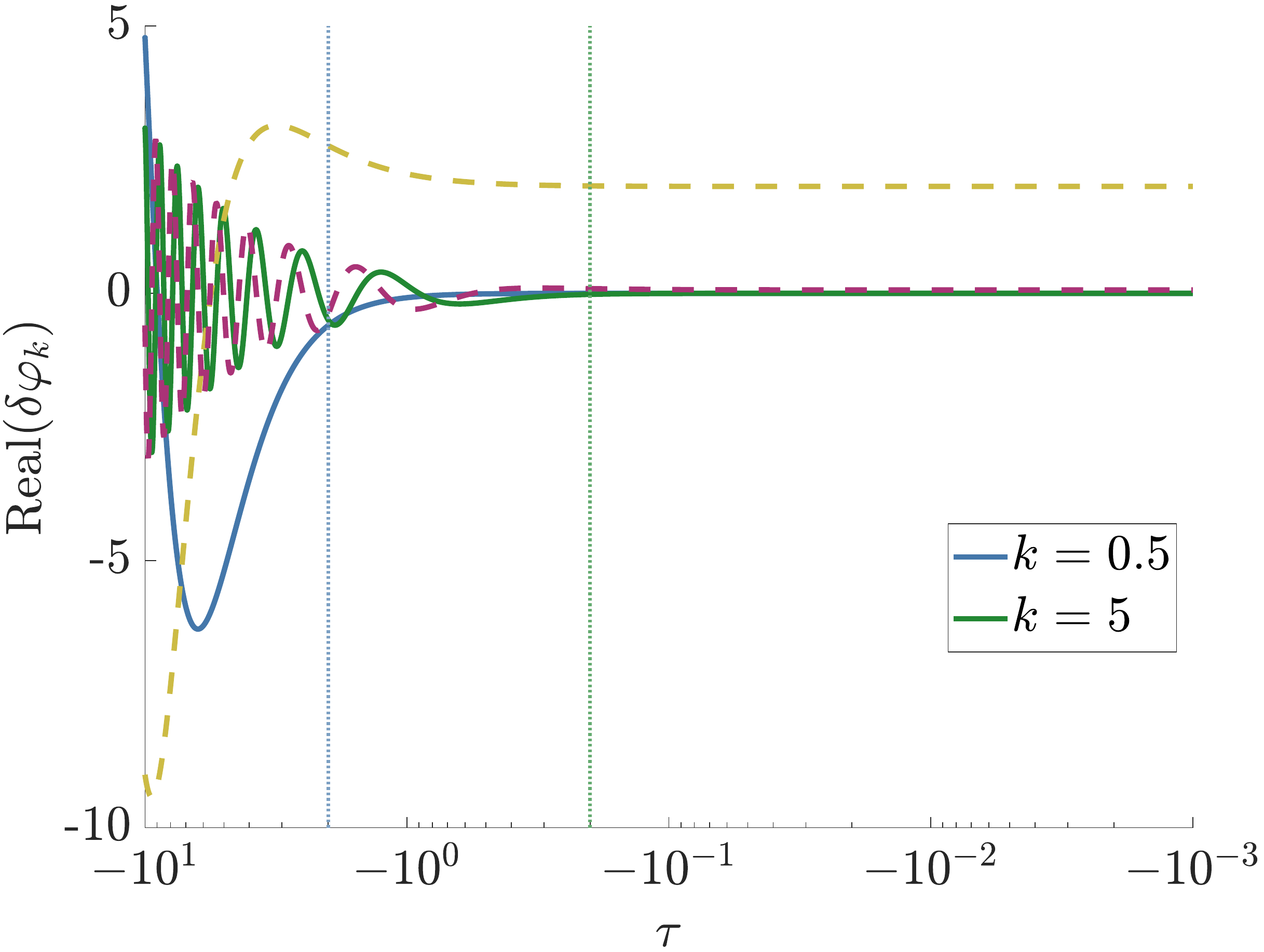}
    \includegraphics[width = 0.45\linewidth]{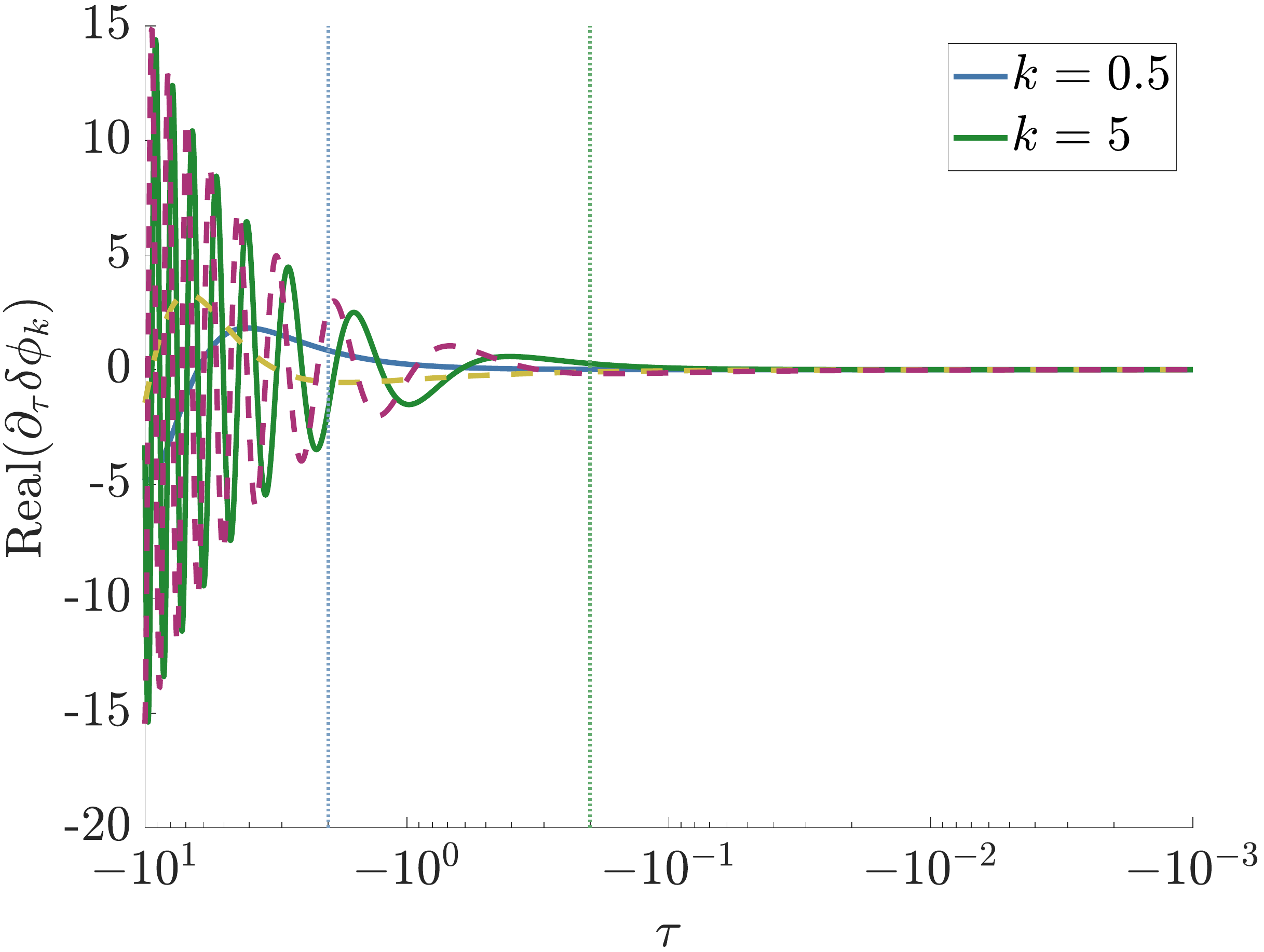}
    \includegraphics[width = 0.45\linewidth]{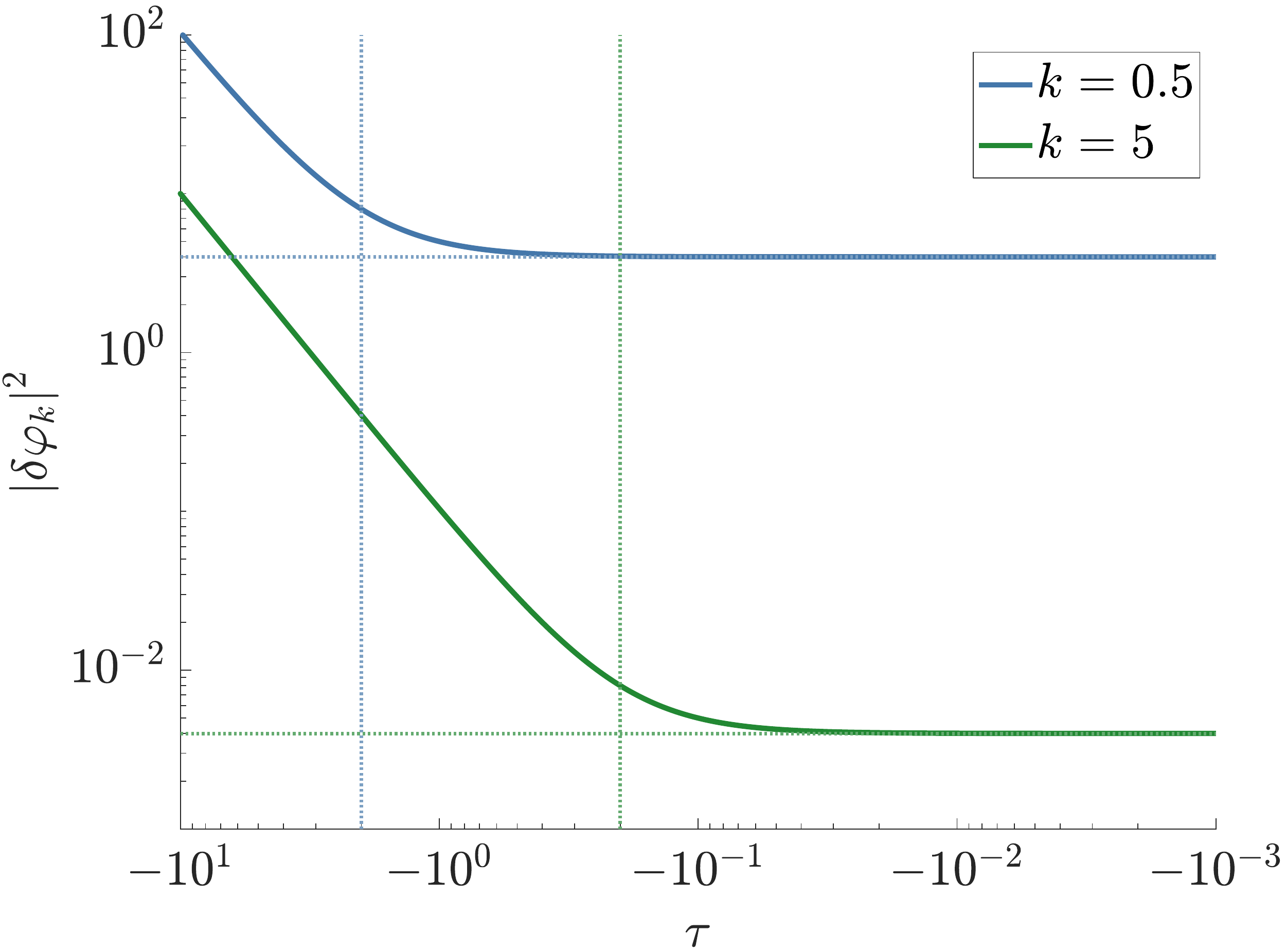}
    \includegraphics[width = 0.45\linewidth]{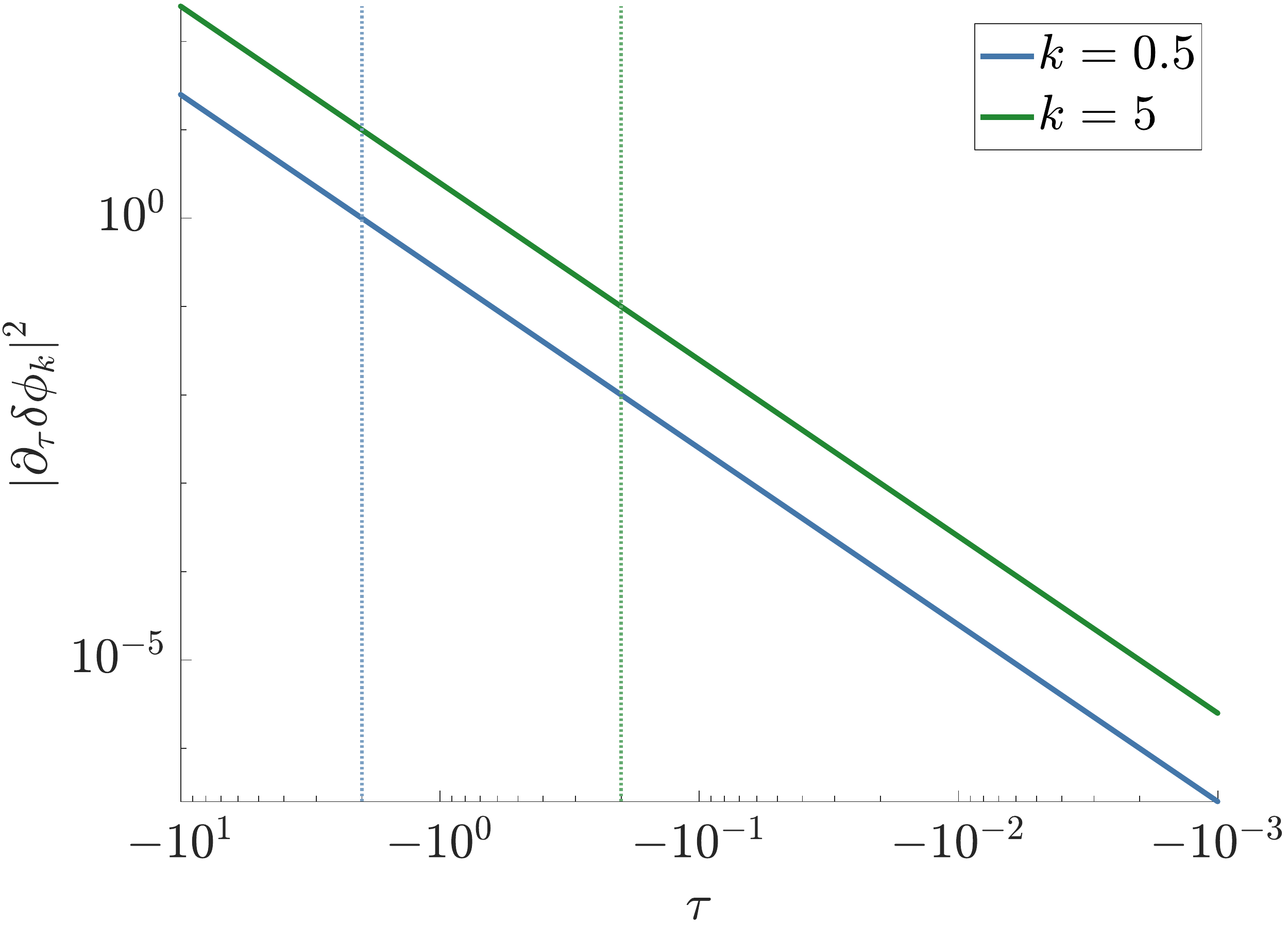}
    \caption[The solutions to the Mukhanov-Sasaki equation for de Sitter]{Evolution of the real components (top row) and absolute values (bottom row) of the the solutions to the Mukhanov-Sasaki equation (\ref{eq:M-S equation}) for $\delta\varphi_k$ (left) and $\partial_{\tau}\delta\varphi_k$ (right) for two different $k$ modes in units of $H = 1$. The vertical dotted lines correspond to the time when the mode ``exits" the horizon i.e. when $k = -1/\tau$.}
    \label{fig:M-S_deSitter}
\end{figure}

While we introduced two different curvature perturbations $\zeta$ and $\mathcal{R}$ for the uniform density and comoving gauge respectively, they can actually be computed in any gauge and are therefore gauge-invariant. 
For instance the comoving curvature perturbation can be computed in any gauge using:
\begin{eqnarray}
\mathcal{R} \equiv C + \dfrac{H}{\dot{\bar{\varphi}}}\delta \varphi
\end{eqnarray}
and the uniform-density curvature perturbation can similarly be computed in any gauge:
\begin{eqnarray}
\zeta \equiv C + \dfrac{H}{\dot{\bar{\varphi}}}\delta \rho_{\varphi}
\end{eqnarray}
Spatially-flat gauge by definition has no curvature perturbations $(C = 0)$ but there are density perturbations. We can also construct the gauge-invariant quantity known as the Mukhanov-Sasaki variable:
\begin{eqnarray}
Q \equiv \delta \varphi + \dfrac{\dot{\bar{\varphi}}}{H}C
\end{eqnarray}
which is valid in any gauge but corresponds to scalar fluctuations on spatially-flat hypersurfaces. 
Gauge-invariant variables are not unique and they aren't independent, e.g. :
\begin{eqnarray}
-\zeta = \mathcal{R} + \dfrac{k^2}{\lb aH\rb^2}\dfrac{2\bar{\rho}}{3\lb \bar{\rho} +\bar{P}\rb }\Psi_{B}
\end{eqnarray}
written in terms of the Bardeen variable $\Psi_B$ -- see \cite{Bardeen1980} and Appendix \ref{sec:App_lin_pert}. We can clearly see that on superhorizon scales, $k \ll aH$, so that $\zeta$ and $\mathcal{R}$ are equivalent on long wavelengths. During \ac{SR} $\zeta \approx \mathcal{R}$ so that the two are roughly equivalent on all scales. Because of this the two are often used interchangeably in the literature. \\
A nice thing about $\zeta$ is that it is naturally related to the perturbed expansion from the zero-curvature to uniform density time-slices:
\begin{eqnarray}
\delta N \equiv \delta (\ln a) = \zeta
\end{eqnarray}
At this stage it is worth remembering that fields in real space can be represented as an integral over Fourier modes. 
\begin{eqnarray}
\delta \varphi (t,\vec{x}) = \int \dfrac{\mathrm{d}^3k}{(2\pi)^3} ~\delta \varphi_{\vec{k}} (t) e^{i\vec{k}\cdot\vec{x}}
\end{eqnarray}
where the Fourier modes form a complete orthonormal basis:
\begin{eqnarray}
\int \mathrm{d}^3x ~e^{i\vec{k}_1\cdot\vec{x}}e^{i\vec{k}_2\cdot\vec{x}} = (2\pi)^3 \delta^{(3)} \lb \vec{k}_1 -\vec{k}_2\rb 
\end{eqnarray}
Expressing the Mukhanov-Sasaki variable $Q$ in terms of its fourier modes one can derive:
\begin{eqnarray}
\ddot{Q}_{\vec{k}} + 3H\dot{Q}_{\vec{k}} + \lsb \dfrac{k^2}{a^2} + V_{,\bar{\varphi}\bar{\varphi}} -\dfrac{1}{a^3M_{\mathrm{p}}^2} \dfrac{\mathrm{d}}{\mathrm{d}t}\lb \dfrac{a^3}{H}\dot{\bar{\varphi}}^2\rb\rsb Q_{\vec{k}}= 0 \label{eq:linear_pert_Q}
\end{eqnarray}
If we now introduce the variable $v_{\vec{k}} = a Q_{\vec{k}}$ we obtain the famous Mukhanov-Sasaki equation:
\begin{empheq}[box = \fcolorbox{Maroon}{white}]{equation}
v_{\vec{k}}'' + \lb k^2 - \dfrac{z''}{z}\rb v_{\vec{k}} = 0 \label{eq:M-S equation}
\end{empheq}
where, as before, a prime denotes a derivative with respect to conformal time $\tau$ and we have introduced $z = a\sqrt{2\varepsilon_1}M_{\mathrm{p}}$. This equation is particularly useful as $v_{\vec{k}} = z \mathcal{R}_{\vec{k}}$. \\
The Mukhanov-Sasaki equation (\ref{eq:M-S equation}) resembles that of a harmonic oscillator if we define the time dependent frequency:
\begin{eqnarray}
\omega^2 (\tau,k) = k^2 - \dfrac{z''}{z}
\end{eqnarray}
In full generality:
\begin{eqnarray}
\dfrac{z''}{z} = \mathcal{H}^2\lb 2-\varepsilon_1 - \dfrac{3\varepsilon_2}{2} + \dfrac{\varepsilon_1\varepsilon_2}{2} + \dfrac{\varepsilon_{2}^2}{4}+\dfrac{\varepsilon_2\varepsilon_3}{2}\rb 
\end{eqnarray}
where $\mathcal{H} = a'/a$ is the conformal Hubble parameter. To get a better sense of the behaviour of this equation we examine the case of exact de Sitter where $z''/z \rightarrow 2/\tau^2$ so that the solutions to the Mukhanov-Sasaki equation (\ref{eq:M-S equation}) are:
\begin{eqnarray}
v_{\vec{k}} = A\dfrac{e^{-ik\tau}}{\sqrt{2k}}\lb1-\dfrac{i}{k\tau}\rb + B\dfrac{e^{ik\tau}}{\sqrt{2k}}\lb 1+\dfrac{i}{k\tau} \rb \label{eq:M-S_soln_dS_gen}
\end{eqnarray}
where $A$ and $B$ are arbitrary constants. To determine them we need to impose the proper boundary conditions by \emph{quantising} the field. We do this in the standard way by promoting the Fourier components to operators:
\begin{eqnarray}
v_{\vec{k}} \rightarrow \hat{v}_{\vec{k}} = v_k \hat{a}_{\vec{k}} + v_{-k}^{*}\hat{a}^{\dagger}_{-\vec{k}}
\end{eqnarray}
written in terms of the creation and annihilation operators $\hat{a}^{\dagger}_{-\vec{k}}$ and $\hat{a}_{\vec{k}}$ which satisfy the canonical commutation relation:
\begin{eqnarray}
\lsb\hat{a}_{\vec{k}} , \hat{a}^{\dagger}_{\vec{k}'}\rsb = (2\pi )^3 \delta (\vec{k} -\vec{k}')
\end{eqnarray}
This condition suggests that the mode functions are normalised as follows:
\begin{eqnarray}
\left\langle v_k,v_k \right\rangle \equiv \dfrac{i}{\hslash} (v_{k}^{*}v_{k}' - v_{k}^{*}{}'v_{k}) = 1 \label{eq:mode_func_norm}
\end{eqnarray}
which provides one of the boundary conditions. To determine the other we must choose a vacuum state for the fluctuations so that:
\begin{eqnarray}
\hat{a}_{\vec{k}} \left| 0\right\rangle = 0
\end{eqnarray}
is satisfied. Specifying this together with (\ref{eq:mode_func_norm}) give us the two necessary boundary conditions to solve (\ref{eq:M-S equation}). The standard choice\footnote{Note that this is a choice and there are others with varying levels of motivation.} is to choose a vacuum state corresponding to the Minkowski vacuum of a comoving observer in the far past. This is well motivated as at these early times, $\tau \rightarrow -\infty$, the observer is well within the horizon and in effect does not ``see" the expansion of space. In this limit it transpires that the appropriate limit for the vacuum to the minimum energy state is given by the initial condition:
\begin{eqnarray}
\lim_{\tau \rightarrow-\infty}v_k = \dfrac{e^{-ik\tau}}{\sqrt{2k}} \label{eq:IC_for_BD}
\end{eqnarray}
This initial condition corresponds to choosing the \emph{Bunch-Davies} vacuum and is appropriate for all inflationary spacetimes, not just exact de Sitter. \\
Using (\ref{eq:IC_for_BD}) we can determine that $A =1$ and $B = 0$ in (\ref{eq:M-S_soln_dS_gen}) leading to the Bunch-Davies mode functions for a massless field in de Sitter:
\begin{eqnarray}
v_{k} = \dfrac{e^{-ik\tau}}{\sqrt{2k}}\lb1-\dfrac{i}{k\tau}\rb \label{eq:vk_dS_soln}
\end{eqnarray}
These can be straightforwardly rewritten to give the inflaton fluctuations in spatially flat gauge using $v_k = a\delta\varphi_k$:
\begin{eqnarray}
\delta\varphi_k = -H \dfrac{e^{-ik\tau}}{\sqrt{2k}}\lb \tau-\dfrac{i}{k}\rb \label{eq:delvarphi_dS_soln}
\end{eqnarray}
from which we straightforwardly obtain the derivative of the inflaton fluctuations with respect to conformal time:
\begin{eqnarray}
\partial_{\tau}\delta\varphi_k = H\tau\sqrt{\dfrac{k}{2}}ie^{-ik\tau} \label{eq:delvarphiprime_dS_soln}
\end{eqnarray}
In the top row of Fig.~\ref{fig:M-S_deSitter} we plot the real parts of (\ref{eq:delvarphi_dS_soln}) \& (\ref{eq:delvarphiprime_dS_soln}) in the left and right panels respectively. In both cases we can clearly see that on subhorizon scales there are damped oscillations, however after horizon exit the oscillations completely dissipate and asymptote to zero. What will be of greater relevance later is the modulus squared of these mode functions which for a massless field in de Sitter are:
\begin{eqnarray}
| \delta\varphi_k  |^2 &=& \dfrac{H^2}{2k^3}\lb k^2\tau^2 + 1\rb \underset{\tau \rightarrow 0^{-}}{\longrightarrow} \dfrac{H^2}{2k^3} \label{eq:mod_delvarphi_dS_soln}\\
| \partial_{\tau}\delta\varphi_k |^2 &=& \dfrac{k}{2}H^2\tau^2 \underset{\tau \rightarrow 0^{-}}{\longrightarrow} 0 \label{eq:mod_delvarphiprime_dS_soln}
\end{eqnarray}
Here we have also clearly indicated that while the derivative vanishes in the superhorizon limit, $| \delta\varphi_k  |^2$ tends to a constant. We have plotted this in the bottom row of Fig.~\ref{fig:M-S_deSitter}.  
\subsubsection{Power Spectrum in (quasi-)de Sitter}
We define the power spectrum $\mathcal{P}_{\mathcal{O}}(k)$ of an observable $\mathcal{O}$ in the standard way:
\begin{eqnarray}
\left\langle \mathcal{O}_{\vec{k}} \mathcal{O}_{\vec{k}'} \right\rangle = (2\pi)^3\delta (\vec{k} + \vec{k}')\mathcal{P}_{\mathcal{O}}(k)
\end{eqnarray}
The power spectrum has dimensions of [length$]^3$ so it is useful to introduce the \emph{dimensionless} power spectrum $\Delta_{\mathcal{O}}^2$ of an observable $\mathcal{O}$
\begin{eqnarray}
\Delta_{\mathcal{O}}^2 = \dfrac{k^3}{2\pi^2}\mathcal{P}_{\mathcal{O}}(k)
\end{eqnarray}
Using (\ref{eq:mod_delvarphi_dS_soln}) we can compute the power spectrum of the field:
\begin{eqnarray}
\left\langle\hat{\varphi}_{\vec{k}}(\tau)\hat{\varphi}_{\vec{k}'}(\tau)\right\rangle &=& (2\pi)^3\delta (\vec{k} + \vec{k}') | \delta\varphi_k  |^2 \\
\Rightarrow \left\langle\hat{\varphi}_{\vec{k}}(\tau)\hat{\varphi}_{\vec{k}'}(\tau)\right\rangle &=& (2\pi)^3\delta (\vec{k} + \vec{k}') \dfrac{H^2}{2k^3}\lb k^2\tau^2 + 1\rb \\
\Rightarrow \left\langle\hat{\varphi}_{\vec{k}}(\tau)\hat{\varphi}_{\vec{k}'}(\tau)\right\rangle & \underset{\tau \rightarrow 0^{-}}{\longrightarrow}& (2\pi)^3\delta (\vec{k} + \vec{k}') \dfrac{H^2}{2k^3} \\
\Rightarrow \Delta_{\varphi}^2 &=& \lb \dfrac{H}{2\pi}\rb^2 \label{eq:dS_temperature}
\end{eqnarray}
Where we find that the dimensionless power spectrum for the inflaton is given by the Hawking temperature of de Sitter space. This result can straightforwardly be extended outside of pure de Sitter. If we consider a period of \ac{SR} then we can compute the power spectrum of the comoving curvature perturbation $\mathcal{R} = H\delta\varphi/\dot{\bar{\varphi}}$ at horizon crossing $k = aH$:
\begin{eqnarray}
\Delta_{\mathcal{R}}^2 = \lb\dfrac{H_{\star}}{2\pi}\rb^2 \lb\dfrac{H_{\star}}{\dot{\bar{\varphi}}_{\star}}\rb^2 \label{eq:SR_powerspec}
\end{eqnarray}
where the $\star$ subscript indicates that a quantity has been evaluate at horizon crossing i.e. $k = aH$. We can evaluate $\mathcal{R}$ at horizon crossing because it approaches a constant on super-horizon scales. Different modes will exit the horizon at slightly different times when $a_{\star}H_{\star}$ has a slightly different value. The $H_{\star}/\dot{\bar{\varphi}}_{\star}$ factor compensates for this and allows us to extend the pure de Sitter result to be valid for \ac{SR}. For non \ac{SR} inflation the background needs to be tracked more carefully and the Mukhanov-Sasaki equation (\ref{eq:M-S equation}) will probably have to be integrated numerically. Wildly incorrect results can be obtained by naively applying \ac{SR} formula outside their regime of validity \cite{Kinney2005}.

\subsection{Squeezing, decoherence and classicalisation}
The decay of the perturbations in the momenta of the field $\Pi = \partial_t \varphi$ shown in Fig.~\ref{fig:M-S_deSitter} suggests a simplification of the system at hand. To make this more clear we quantise the field and its momenta by promoting them to operators in terms of creation and annihilation operators
\begin{eqnarray}
\hat{\varphi}_{\vec{k}} &\equiv & \varphi_{\vec{k}} \hat{a}_{\vec{k}}+ \varphi_{\vec{k}}^{*}\hat{a}_{-\vec{k}}^{\dagger} \\
\hat{\Pi}_{\vec{k}} &\equiv & \Pi_{\vec{k}} \hat{a}_{\vec{k}}+ \Pi_{\vec{k}}^{*}\hat{a}_{-\vec{k}}^{\dagger} 
\end{eqnarray}
where the creation, $\hat{a}_{\vec{k}}^{\dagger}$, and annihilation, $\hat{a}_{\vec{k}}$, operators obey the usual commutation relations:
\begin{eqnarray}
\lsb \hat{a}_{\vec{k}} , \hat{a}_{\vec{k}'}^{\dagger}\rsb = \delta (\vec{k}-\vec{k}'), \quad \lsb \hat{a}_{\vec{k}}, \hat{a}_{\vec{k}'} \rsb = \lsb \hat{a}_{\vec{k}}^{\dagger}, \hat{a}_{\vec{k}'}^{\dagger} \rsb = 0
\end{eqnarray}
and the vacuum state satisfies $\hat{a}_{\vec{k}}\left| 0\right\rangle$. We have also introduced the mode functions $\varphi_{\vec{k}}$ \& $\Pi_{\vec{k}}$ which we will discuss in more detail shortly. If we assume that we are dealing with linear perturbation theory then the mode functions $\varphi_{\vec{k}}$ \& $\Pi_{\vec{k}}$ can be solved based on the solutions to the Mukhanov-Sasaki equation (\ref{eq:M-S equation}) and because of the assumed background isotropy they again only depend on the norm of $\vec{k}$. A problem with the current operators is that they are non-Hermitian and mix the $\vec{k}$ and $-\vec{k}$ sectors. It is beneficial therefore to introduce e.g. the real components of the operators defined as:
\begin{subequations}
\begin{align}
\hat{\varphi}_{\vec{k}}^R &\equiv \dfrac{\hat{\varphi}_{\vec{k}} + \hat{\varphi}_{\vec{k}}^{\dagger}}{\sqrt{2}}    \\
    \hat{\Pi}_{\vec{k}}^R &\equiv \dfrac{\hat{\Pi}_{\vec{k}} + \hat{\Pi}_{\vec{k}}^{\dagger}}{\sqrt{2}} 
\end{align}
\end{subequations}
from which we can determine the following quantum correlators in terms of the mode functions:
\begin{subequations}
\begin{align}
    \gamma_{\scalebox{0.5}{$\varphi\varphi$}} &\equiv \left\langle 0 \right| \hat{\varphi}_{\vec{k}}^R\hat{\varphi}_{\vec{k}}^R\left| 0\right\rangle = 2|\varphi_k|^2\\
    \gamma_{\scalebox{0.5}{$\Pi\Pi$}} &\equiv \left\langle 0 \right| \hat{\Pi}_{\vec{k}}^R\hat{\Pi}_{\vec{k}}^R\left| 0\right\rangle = 2|\Pi_k|^2\\
    \gamma_{\scalebox{0.5}{$\varphi\Pi$}}^{+} &\equiv \dfrac{1}{2}\Big\langle 0 \Big| \lcb \hat{\varphi}_{\vec{k}}^R ,\hat{\Pi}_{\vec{k}}^R \rcb\Big| 0\Big\rangle = 2\text{Re}\lb \varphi_k \Pi_{k}^{*}\rb \\
    \gamma_{\scalebox{0.5}{$\varphi\varphi$}}^{-} &\equiv \dfrac{1}{2}\Big\langle 0 \Big| \lsb \hat{\varphi}_{\vec{k}}^R ,\hat{\Pi}_{\vec{k}}^R \rsb\Big| 0\Big\rangle = 2i~\text{Im}\lb \varphi_k \Pi_{k}^{*}\rb
\end{align}
\end{subequations}
These quantum correlators it transpires can be reproduced from the following classical probability distribution:
\begin{eqnarray}
P(\varphi,\Pi) = \dfrac{1}{2\pi \sqrt{\gamma_{\scalebox{0.5}{$\varphi\varphi$}}\gamma_{\scalebox{0.5}{$\Pi\Pi$}} - \gamma_{\scalebox{0.5}{$\varphi\Pi$}}^{+}\gamma_{\scalebox{0.5}{$\varphi\Pi$}}^{+}}}\exp \lb -\dfrac{\gamma_{\scalebox{0.5}{$\Pi\Pi$}}\varphi^2 + \gamma_{\scalebox{0.5}{$\varphi\varphi$}}\Pi^2 -2\gamma_{\scalebox{0.5}{$\varphi\Pi$}}^{+}\varphi \Pi}{2\lb\gamma_{\scalebox{0.5}{$\varphi\varphi$}}\gamma_{\scalebox{0.5}{$\Pi\Pi$}} - \gamma_{\scalebox{0.5}{$\varphi\Pi$}}^{+}\gamma_{\scalebox{0.5}{$\varphi\Pi$}}^{+}\rb}\rb \label{eq:PDF_varphiPi_full}
\end{eqnarray}
where we have dropped the $\vec{k}$ subscripts and $R$ superscripts to lighten the notation. To see how this distribution behaves as modes become superhorizon we consider the massless de Sitter solutions derived earlier -- see e.g. (\ref{eq:M-S_soln_dS_gen}). Then the correlators become:
\begin{subequations}
\begin{align}
     \gamma_{\scalebox{0.5}{$\varphi\varphi$}} &= \dfrac{H^2}{k^3}\lsb 1 + \lb \dfrac{k}{aH}\rb^2\rsb\\
    \gamma_{\scalebox{0.5}{$\Pi\Pi$}} &= \dfrac{H^4}{k^3}\lb \dfrac{k}{aH}\rb^4 \\
    \gamma_{\scalebox{0.5}{$\varphi\Pi$}}^{+} &= -\dfrac{H^3}{k^3}\lb \dfrac{k}{aH}\rb^2 \\
    \gamma_{\scalebox{0.5}{$\varphi\varphi$}}^{-} &= i\dfrac{H^3}{k^3}\lb \dfrac{k}{aH}\rb^3 
\end{align} \label{eq:deSitter_gamma_matrices}
\end{subequations}
Which suggests on long wavelengths ($k \ll aH $) that all correlators tend to zero apart from the inflaton correlator $\gamma_{\scalebox{0.5}{$\varphi\varphi$}}$. This means that the PDF (\ref{eq:PDF_varphiPi_full}) can be simplified to:
\begin{eqnarray}
P(\varphi,\Pi) \underset{k \ll aH}{\longrightarrow} \dfrac{1}{\sqrt{2\pi \gamma_{\scalebox{0.5}{$\varphi\varphi$}}}}\exp \lb -\dfrac{\varphi^2}{2\gamma_{\scalebox{0.5}{$\varphi\varphi$}}}\rb \delta (\Pi) \label{eq:PDF_varphiPi_dS}
\end{eqnarray}
which we can see essentially reduces the dimension of the problem as the momentum is forced to the line $\Pi = 0$ and all uncertainty is left in the inflaton itself which is described by a Gaussian distribution. In this way (\ref{eq:PDF_varphiPi_dS}) represents a highly squeezed state as the inescapable Heisenberg uncertainty which was initially spread between $\varphi$ and $\Pi$ has been squeezed into the inflaton. Phrased another way we have sacrificed knowing the inflaton's position with any certainty in order to know that the momentum is negligible. This means that the behaviour of the inflaton perturbations can be well described by a \emph{classical, stochastic process} which we will discuss in more detail in section \ref{sec:stoch_infl}. This ``classicalisation" has many other both heuristic and more formal derivations -- see e.g. \cite{Guth1985,Albrecht1994, Polarski1996, Martin2021} -- and has resulted in the notion of ``decoherence without decoherence". This is because although in our presentation there are no interactions to cause the perturbations to decohere, they appear to have effectively done so. However it is worth emphasising that in a closed environment there is \emph{no true decoherence} \cite{Hsiang2022, Martin2021}. A state being highly squeezed actually -- in some sense -- makes it \emph{more} quantum. If interactions are added to this picture it has been shown \cite{Martin2021} that there is a competition between the correlation build up induced by the squeezing of the perturbations and the interaction erasing quantum features. \\

While we have presented this squeezing in the case of massless de Sitter it is a generic feature of an inflationary spacetime that the perturbations will be squeezed on super-horizon scales and effectively classicalise, the caveat is that the line they are forced to might not correspond exactly with the $\Pi = 0$ but there will still only be one effective degree of freedom. It transpires -- see e.g. \cite{Figueroa2021a} -- that the large squeezing limit in general corresponds to $\gamma_{\scalebox{0.5}{$\varphi\Pi$}}^{+}\gamma_{\scalebox{0.5}{$\varphi\Pi$}}^{+} \rightarrow \gamma_{\scalebox{0.5}{$\varphi\varphi$}}\gamma_{\scalebox{0.5}{$\Pi\Pi$}}$ for which the PDF (\ref{eq:PDF_varphiPi_full}) can be simplified to:
\begin{empheq}[box = \fcolorbox{Maroon}{white}]{equation}
P(\varphi,\Pi) \underset{k \ll aH}{\longrightarrow} \dfrac{1}{\sqrt{2\pi \gamma_{\scalebox{0.5}{$\varphi\varphi$}}}}\exp \lb -\dfrac{\varphi^2}{2\gamma_{\scalebox{0.5}{$\varphi\varphi$}}}\rb \delta \lb\Pi - \text{sign}(\gamma_{\scalebox{0.5}{$\varphi\Pi$}}^{+})\sqrt{\dfrac{\gamma_{\scalebox{0.5}{$\Pi\Pi$}}}{\gamma_{\scalebox{0.5}{$\varphi\varphi$}}}}\varphi \rb \label{eq:PDF_varphiPi_squeezed}
\end{empheq}
In this way it is generically true that long-wavelength perturbations can be treated as \emph{effectively} classical in the sense they can be described by a stochastic process\footnote{There are some subtleties if the inflaton enters a period of \ac{USR} which we do not go into right now.}. A more thorough look into squeezing is done in Appendix \ref{app:squeezing}.

\subsection{\label{sec:nonlin_pert}Long-wavelength non-linear perturbations}
We have so far looked at the behaviour of perturbations from homogeneity at linear order. In principle this can be extended to higher orders in perturbation theory but the framework is still unable to handle large deviations from homogeneity. In this section we will instead examine a framework that can fully describe the non-linear perturbations, the payoff will be that it is only valid on superhorizon scales. \\

Let us consider the Arnowitt-Deser-Misner (ADM) parameterisation of the metric \cite{Arnowitt2008}. The idea is that spacetime is foliated by spacelike hypersurfaces with a normal vector $n^{\mu}$ which has components:
\begin{eqnarray}
n_0 = -N,\quad n_i = 0,\quad n^0 = N^{-1}, \quad n^i = N^{-1}N^i
\end{eqnarray}
where $N$ is the lapse function and $N^i$ is the shift vector. The four functions $N$ and $N^i$ are arbitrary and reflect the gauge freedom of \ac{GR}. The metric can then be defined component wise:
\begin{eqnarray}
g_{00} = -N^2 + \gamma^{ij}N_iN_j,\quad g_{0i} =-N_i,\quad g_{ij} = \gamma_{ij} \label{eq:ADM_metric}
\end{eqnarray}
with the inverse metric given by
\begin{eqnarray}
g^{00} = -N^{-2} ,\quad g^{0i} =-N^{-2}N^i,\quad g^{ij} = \gamma^{ij} - N^{-2}N^iN^j \label{eq:ADM_metric_inv}
\end{eqnarray}
where $\gamma_{ij}$ is the metric on the spatial hypersurfaces. The way these spatial hypersurfaces are embedded in the 4D geometry is parameterised by the extrinsic curvature tensor:
\begin{eqnarray}
K_{ij} = -\dfrac{1}{2N}\lb \tilde{\nabla}_j N_{i} + \tilde{\nabla}_iN_{j} + \dfrac{\partial}{\partial t}\gamma_{ij}\rb
\end{eqnarray}
where $\tilde{\nabla}_i$ corresponds to a covariant derivative with respect to the the spatial metric $\gamma_{ij}$.\\
To make progress we split the field into a smoothed long-wavelength, background or coarse-grained field $\varphi_{\scalebox{0.5}{$>$}}$ and a residual short wavelength field $\varphi_{\scalebox{0.5}{$<$}}$:
\begin{eqnarray}
\varphi (t,\vec{x} ) &=& \varphi_{\scalebox{0.5}{$>$}}(t,\vec{x} ) + \varphi_{\scalebox{0.5}{$<$}}(t,\vec{x} ) \\
\varphi_{\scalebox{0.5}{$>$}}(t,\vec{x} ) &\equiv &\int \mathrm{d}^3 x ~\mathcal{W}(t,\vec{x}-\vec{x}') \varphi (t,\vec{x}' )
\end{eqnarray}
where $\mathcal{W}$ is a window or smoothing function in the spatial coordinates whose Fourier transform falls off at high momentum. It is worth noting that this smoothing is gauge dependent and one must therefore be careful about relating quantities not computed in the same gauge this coarse-graining is performed in. For stochastic inflation the natural smoothing scale is (a multiple of) the comoving Hubble length $(aH)^{-1}$ and the natural hypersurfaces are those where $aH$ is constant. We will discuss the exact nature of this split in more detail when we cover stochastic inflation in \ref{sec:stoch_infl}, for now let us assume we can make this split. We will identify $\phi \equiv \varphi_{\scalebox{0.5}{$>$}}$ and therefore all quantities expressed in terms of $\phi$ will correspond to the coarse-grained, long-wavelength part of it -- e.g. $H(\phi)$ strictly speaking means $H_{\scalebox{0.5}{$>$}} (\varphi_{\scalebox{0.5}{$>$}})$.
It can be shown \cite{Salopek1990} -- see also Appendix \ref{sec:app_ADM_pert} -- that during inflation the long wavelength metric can be written as: 
\begin{eqnarray}
\mathrm{d}s^2 = -N^2(t,x^i)\mathrm{d}t^2 + e^{2\alpha(t,\textbf{x})}h_{ij}(\textbf{x})\mathrm{d}x^i\mathrm{d}x^j \label{eq:longwavemetric}
\end{eqnarray}
The shift vector, $N_i$, has been set to $0$ but coordinate freedom remains in the choice of the lapse function $N$. 
The local expansion rate is defined as:
\begin{eqnarray}
H(t,\textbf{x}) \equiv \dfrac{1}{N}\dfrac{\partial \alpha}{\partial t} \label{eq: inhom H expan rate}
\end{eqnarray}
while the dynamics of $h_{ij}(\textbf{x})$, describing volume-preserving deformations of the spatial geometry, can be ignored as a first approximation. By keeping the leading order in spatial gradients we can obtain a dynamical equation for the long wavelength modes of the inflaton field $\phi$:
\begin{eqnarray}
\Pi = \dfrac{1}{N}\dfrac{\partial \phi}{\partial t} \label{eq: mom defn} \\
\dfrac{1}{N}\dfrac{\partial \Pi}{\partial t} + 3H \Pi + \dfrac{\mathrm{d}V}{\mathrm{d}\phi} = 0 \label{eq: inhom K-G}
\end{eqnarray} 
and we also obtain the energy constraint equation:
\begin{eqnarray}
H^2 = \dfrac{1}{3 M_{\mathrm p}^2}\left( \dfrac{\Pi^2}{2} + V(\phi) \right) \label{eq: inhom energy constraint}
\end{eqnarray}
It important to realise that although equations (\ref{eq: inhom K-G}) and (\ref{eq: inhom energy constraint}) look identical to the homogeneous versions, (\ref{eq: K-G eqn}) and (\ref{eq:Friedmann_Inflation}), they are valid at each spatial point with \textit{a priori} different initial conditions. Equations (\ref{eq: inhom K-G}) and (\ref{eq: inhom energy constraint}) represent the \textit{separate universe evolution} as each spatial point independently follows its own homogeneous cosmology evolution. What has yet to be taken into account however is the \ac{GR} momentum constraint which must also be obeyed and we will see this restricts the separate universe picture.   
\subsubsection{The momentum constraint}
At leading order in spatial gradients, the \ac{GR} momentum constraint tells us that:
\begin{eqnarray}
\tilde{\nabla}_{i}H = -\dfrac{1}{2M_{\mathrm p}^2}\Pi \tilde{\nabla}_{i}\phi \label{eq: GR mom constraint}
\end{eqnarray}
Taking this additional constraint into account, one can show \cite{Salopek1990} -- see also Appendix \ref{sec:app_ADM_pert} -- that both $H$ and $\Pi$ are solely functions of $\phi$ with no explicit time dependence and are related through:
\begin{eqnarray}
\Pi(\phi) = -2 M_{\mathrm p}^2 \,\dfrac{\mathrm{d}H(\phi)}{\mathrm{d}\phi} \label{eq: Pi and H relation}
\end{eqnarray}
If (\ref{eq: Pi and H relation}) is inserted into the local energy constraint (\ref{eq: inhom energy constraint}) we obtain the \ac{H-J} equation for $H(\phi)$:
\begin{empheq}[box = \fcolorbox{Maroon}{white}]{equation}
\left( \dfrac{\mathrm{d}H}{\mathrm{d}\phi}\right)^2 = \dfrac{3}{2M_{\mathrm p}^2}H^2 - \dfrac{1}{2M_{\mathrm p}^4}V(\phi) \label{eq: H-J equation}
\end{empheq}
which can be solved for any given potential $V(\phi)$ to give a family of solutions $H(\phi, \mathcal{C})$. One of these solutions combined with:
\begin{eqnarray}
\dfrac{\mathrm{d}\phi}{\mathrm{d}t} = -2M_{\mathrm p}^2 \,N \, \dfrac{\mathrm{d}H}{\mathrm{d}\phi}
\end{eqnarray} 
and the evolution of the expansion rate (\ref{eq: inhom H expan rate}) offers the complete description of the long wavelength evolution of the inflaton field $\phi$ in the long wavelength metric (\ref{eq:longwavemetric}). \\
It is important at this stage to notice that the naive separate universe picture suggests that at each spatial point we can pick any initial value for the inflaton field and its momentum i.e. that $\mathcal{C} = \mathcal{C}(\textbf{x})$ has an explicit spatial dependence on the initial hypersurface. However this would violate the \ac{GR} momentum constraint (\ref{eq: GR mom constraint}) which restricts $\mathcal{C}$ to be a \textit{global constant} meaning that all spatial points must be placed along the same integral curve of (\ref{eq: H-J equation}). 
Therefore, once a particular solution of $H(\phi, \mathcal{C})$ of (\ref{eq: H-J equation}) has been obtained the field evolution is given by (for e-fold time $\alpha$):
\begin{empheq}[box = \fcolorbox{Maroon}{white}]{equation}
\dfrac{\mathrm{d}\phi}{\mathrm{d}\alpha} = -2M_{\mathrm p}^2 \, \dfrac{\partial \text{ln} H (\phi,\phi_{0})}{\partial \phi} \label{eq: dphidt HJ}
\end{empheq}
where $\phi_{0} = M_{\mathrm p}\sqrt{\frac{2}{3}}~\text{ln}~\mathcal{C}$ is a global constant whose physical significance will become clear later. We see that the inclusion of the \ac{GR} momentum constraint (\ref{eq: GR mom constraint}) has reduced the dynamics from second order (\ref{eq: inhom K-G}) to first order (\ref{eq: dphidt HJ}) massively reducing the difficulty of the problem. Of particular note this reduction to first order does not make any assumptions about whether the inflaton field is in the \ac{SR} (or any other) regime. Indeed for \ac{SR} the \ac{GR} momentum constraint (\ref{eq: GR mom constraint}) is trivially satisfied but the crucial detail is that the \ac{H-J} equation (\ref{eq: H-J equation}) is valid in any regime, including in the \ac{USR} regime which is significant for the formation of \ac{PBHs} as we will see in the next chapter. \\
Recalling the definitions of the first few \ac{SR} parameters we can express them in terms of $\phi$ derivatives: 
\begin{eqnarray}
\epsilon_1 &\equiv & -\dfrac{\mathrm{d}~\text{ln} H}{\mathrm{d}\alpha} = 2M_{\mathrm{p}}^2\dfrac{\tilde{H}_{,\phi}^2}{\tilde{H}^2} \label{eq: epsilon1_phi defn} \\
\epsilon_2 &\equiv & -\dfrac{\mathrm{d}~\text{ln} \epsilon_1}{\mathrm{d}\alpha} = 4\dfrac{\tilde{H}_{,\phi\phi}}{\tilde{H}} - \dfrac{2\epsilon_1}{M_{\mathrm{p}}^2} \label{eq: epsilon2_phi defn} \\
\epsilon_3 &\equiv & -\dfrac{\mathrm{d}~\text{ln} \epsilon_2}{\mathrm{d}\alpha} = 4M_{\mathrm{p}}^2\dfrac{\tilde{H}_{,\phi\phi\phi}\tilde{H}_{,\phi}}{\tilde{H}^2\varepsilon_2} - 4\dfrac{\tilde{H}_{,\phi\phi}\varepsilon_1}{\tilde{H}\varepsilon_2} -2\varepsilon_1 \label{eq: epsilon3_phi defn}
\end{eqnarray}
where $\tilde{H}^2 \equiv H_{\scalebox{0.5}{$>$}}^2/8\pi^2M_{\mathrm{p}}^{2}$. All of these equations describe the evolution of the long-wavelength modes, however modes which were initially subhorizon get stretched to superhorizon scales during inflation. So far we have assumed that when these initially short wavelength modes become superhorizon they do not affect the dynamics at all. This is a simplification too far and we will address how to adequately incorporate this backreaction when we discuss stochastic inflation in \ref{sec:stoch_infl}.
\clearpage
\section{\label{sec:stoch_infl} Stochastic Inflation}
Stochastic Inflation \cite{Starobinsky1988, Nambu1988,Nambu1989,Mollerach1991,Salopek1991,Habib1992, Linde1994,Starobinsky1994} has enjoyed much success as the leading framework to describe the evolution of non-linear perturbations and their backreaction on the dynamics of the inflaton. The basic idea is to split inflationary perturbations into short- and long-wavelength components. As discussed earlier the long-wavelength perturbations can be treated as effectively classical greatly simplifying the analysis. The initially short-wavelength quantum perturbations are stretched by the rapid inflationary expansion and can be consistently included as a classical random noise term on the dynamical equations which is a well established approximation for the the behaviour of IR quantum fields in inflationary spacetimes \cite{Tsamis2005,Finelli2009,Finelli2010,Garbrecht2014,Garbrecht2015,Moss2017} -- see \cite{Cable2021,Cable2022} however for how this picture breaks down for too massive test fields and see \cite{Cohen2021b} for \ac{NNLO} corrections to the standard stochastic framework. In this way it is clear that the stochastic framework can be imagined as an \ac{EFT} of the long-wavelength sector. We will make this idea more precise in the next section. 
\subsection{\label{sec: An EFT}An EFT of the long-wavelength sector}
\begin{sidewaysfigure}
    \includegraphics[width = 0.99\linewidth]{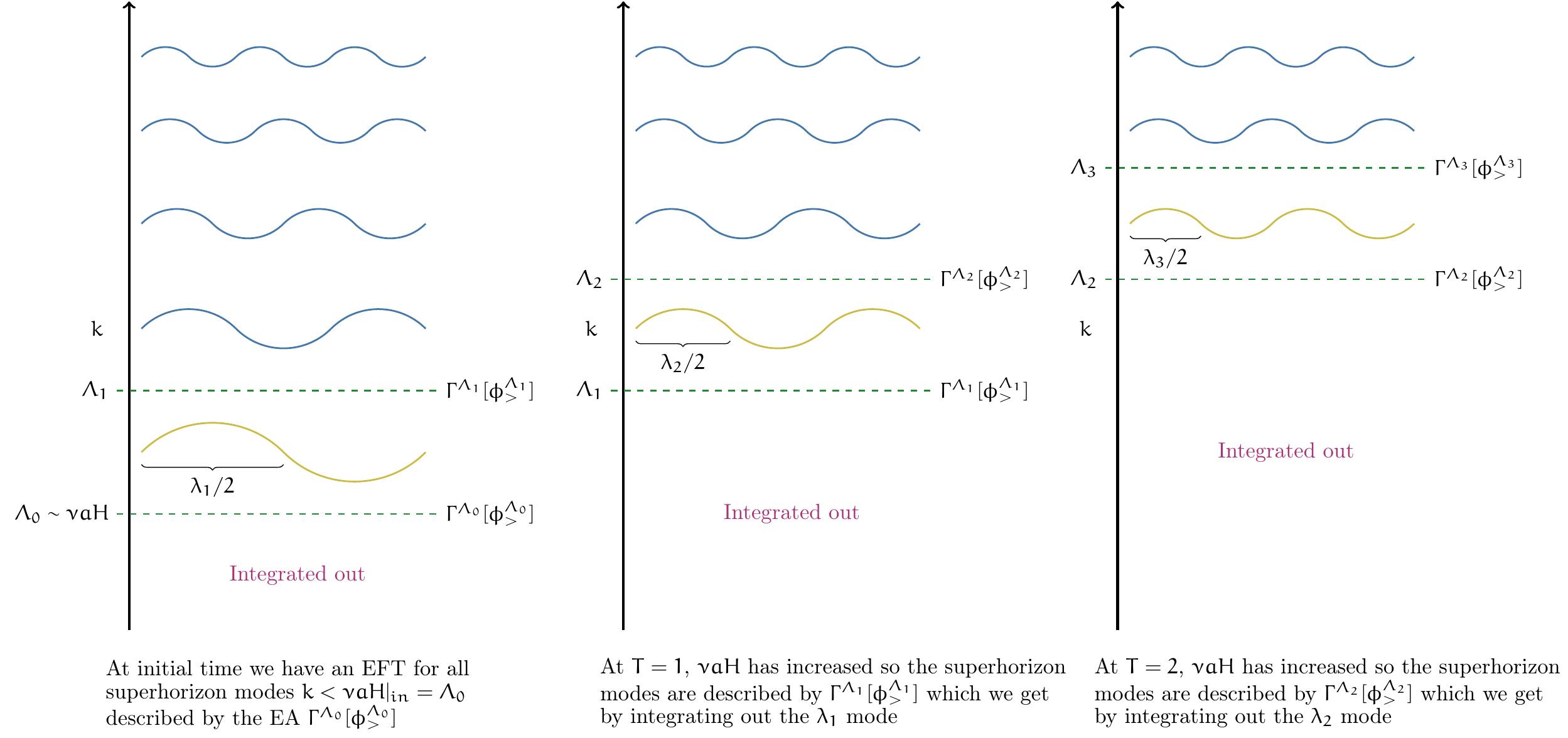}
    \caption[Stochastic Inflation as an \ac{EFT}]{Schematic for how stochastic inflation behaves like an \ac{EFT} for the long-wavelength sector. The long-wavelength modes are integrated out and contained within $\Gamma^{\Lambda}[\phi_{>}^{\Lambda}]$ defined at the coarse-graining scale $\Lambda \sim \nu aH$. However because this scale changes with time we find that the cutoff $\Lambda$ increases with time such that short wavelength modes need to be incorporated into the long-wavelength sector.}
    \label{fig:Stochastic_fluctuations_kspace}
\end{sidewaysfigure}
Let us be more precise about the splitting of long- and short-wavelength modes. We introduce a parameter $\nu < 1 $ which we identify as the coarse-graining parameter. We split the inflaton $\hat{\varphi}$ between long and short wavelengths at the coarse-graining scale $k_{\nu} = \nu aH$ so that:
\begin{eqnarray}
\hat{\varphi}(t,\vec{x}) &=& \hat{\phi}(t,\vec{x}) +  \delta\hat{\varphi} (t,\vec{x})\\
\hat{\phi}(t,\vec{x}) &\equiv &\int \dfrac{\mathrm{d}^3\vec{k}}{(2\pi)^{3/2}}\mathcal{W}\lb \dfrac{k_{\nu}}{k}\rb\lsb \hat{a}_{\vec{k}}\varphi_{\vec{k}}e^{i\vec{k}\cdot \vec{x}} + \hat{a}_{\vec{k}}^{\dagger}\varphi_{\vec{k}}^{*}e^{-i\vec{k}\cdot \vec{x}}\rsb  \\
\delta\hat{\varphi}(t,\vec{x}) &\equiv &\int \dfrac{\mathrm{d}^3\vec{k}}{(2\pi)^{3/2}}\mathcal{W}\lb \dfrac{k}{k_{\nu}}\rb\lsb \hat{a}_{\vec{k}}\varphi_{\vec{k}}e^{i\vec{k}\cdot \vec{x}} + \hat{a}_{\vec{k}}^{\dagger}\varphi_{\vec{k}}^{*}e^{-i\vec{k}\cdot \vec{x}}\rsb 
\end{eqnarray}
where $\mathcal{W}\lb k_{\nu}/k\rb$ is a window function in $k$-space that selects the the long wavelengths, $k < k_{\nu}$, for $\hat{\phi}$ and the short wavelengths, $k > k_{\nu}$, for $\delta\hat{\varphi}$. From now on we have also working in units where $M_{\mathrm{p}}^2 = 1$. Recall that $\varphi_{\vec{k}}$ corresponds to the Fourier modes of the full field $\hat{\varphi}$ and we have explicitly written out hats on all quantum operators -- at this stage no assumption has been made about any objects being \emph{effectively classical}. \\
If we were not in e.g. Minkowski space then this split would be valid at all times and we could treat the long-wavelength sector $\hat{\phi}$ independently of the short-wavelength sector. If we look at the left plot of Fig.~\ref{fig:Stochastic_fluctuations_kspace} you can imagine $\Lambda_0 = \nu aH$ acts as a cutoff between the long- and short-wavelength modes. Therefore, in complete analogy\footnote{The subtle difference is that here we are considering the Fourier modes of fluctuations in \emph{space} whereas in Part \ref{part:Meso} we were considering the Fourier modes of fluctuations in \emph{time}} with Fig.~\ref{fig:fRG_fluctations}, we can define an \ac{EFT} for the long-wavelength modes in terms of the \ac{EA}  $\Gamma^{\Lambda_0}[\hat{\phi}]$ for the cutoff $\Lambda_0$ which is computed by integrating out all modes with $k < \Lambda_0$. However as the comoving Hubble horizon, $(aH)^{-1}$, shrinks with time the coarse-graining scale, $k_{\nu}$, is implicitly a function of time and therefore modes that were originally in the short-wavelength regime enter the long-wavelength one. This means that $\Gamma^{\Lambda}[\phi]$ varies with time as $k_{\nu}$ increases with time. For instance looking back to Fig.~\ref{fig:Stochastic_fluctuations_kspace} we can see that the \ac{EFT} at $T = 0$ differs from the one at $T = 1$ by a single mode with wavelength $\lambda_1$. Therefore $\Gamma^{\Lambda_1}[\phi]$ is equivalent to $\Gamma^{\Lambda_0}[\phi]$ after integrating out the $\lambda_1$ mode -- see middle plot of Fig.~\ref{fig:Stochastic_fluctuations_kspace}. We can see in the right plot of Fig.~\ref{fig:Stochastic_fluctuations_kspace} how this procedure continues for $T=2$ and in principle carries on for the entirety of inflation. In this way stochastic inflation is an \ac{EFT} of the long-wavelength sector with a time dependent cutoff given by the coarse-graining scale $k_{\nu}$.
\subsection{Stochastic Equations of Motion}
We follow the work of \cite{Salopek1991} to derive the stochastic equations of motion in the ADM formalism. The equations of motion we derived in section \ref{sec:nonlin_pert} did not include the effect of short-wavelength modes entering the long-wavelength sector. We recall that the ADM formalism gives us the following equations:
\begin{subequations}
\begin{align}
    \dfrac{\mathrm{d}\Pi}{\mathrm{d}\alpha} &= -3\Pi -\dfrac{1}{H^2}\lb \tilde{\nabla}_i H\rb \lb \tilde{\nabla}^i \varphi \rb +\dfrac{1}{H} \tilde{\nabla}_i\tilde{\nabla}^i \varphi - \dfrac{1}{H}\dfrac{\mathrm{d}V(\varphi)}{\mathrm{d}\varphi} \\
    \dfrac{\mathrm{d}\varphi}{\mathrm{d}\alpha} &= \dfrac{\Pi}{H} \\ 
    3H^2 &= \dfrac{\Pi^2}{2}+ V(\varphi)  + \tilde{\nabla}_i \tilde{\nabla}^i \varphi - \dfrac{R_{(3)}}{2} \\
    \tilde{\nabla}_i H &= -\dfrac{1}{2}\Pi\tilde{\nabla}_i \varphi
\end{align} \label{eq:Full_ADM_beforesplit}
\end{subequations}
The next step is quantise the inflaton and its conjugate momentum:
\begin{eqnarray}
\hat{\varphi}_{\vec{k}} &\equiv & \varphi_{\vec{k}} \hat{a}_{\vec{k}}+ \varphi_{\vec{k}}^{*}\hat{a}_{-\vec{k}}^{\dagger} \\
\hat{\Pi}_{\vec{k}} &\equiv & \Pi_{\vec{k}} \hat{a}_{\vec{k}}+ \Pi_{\vec{k}}^{*}\hat{a}_{-\vec{k}}^{\dagger} 
\end{eqnarray}
where the creation, $\hat{a}_{\vec{k}}^{\dagger}$, and annihilation, $\hat{a}_{\vec{k}}$, operators obey the usual commutation relations:
\begin{eqnarray}
\lsb \hat{a}_{\vec{k}} , \hat{a}_{\vec{k}'}^{\dagger}\rsb = \delta (\vec{k}-\vec{k}'), \quad \lsb \hat{a}_{\vec{k}}, \hat{a}_{\vec{k}'} \rsb = \lsb \hat{a}_{\vec{k}}^{\dagger}, \hat{a}_{\vec{k}'}^{\dagger} \rsb = 0
\end{eqnarray}
and the vacuum state satisfies $\hat{a}_{\vec{k}}\left| 0\right\rangle$. We have also introduced the mode functions $\varphi_{\vec{k}}$ \& $\Pi_{\vec{k}}$ which we will discuss in more detail shortly. If we now split the field into long and short wavelengths, $\hat{\varphi} = \hat{\varphi}_{\scalebox{0.5}{$>$}} + \hat{\varphi}_{\scalebox{0.5}{$<$}}$:
\begin{subequations}
\begin{align}
\hat{\varphi}_{\scalebox{0.5}{$>$}}(t,\vec{x}) &\equiv \int \dfrac{\mathrm{d}^3\vec{k}}{(2\pi)^{3/2}}\mathcal{W}\lb \dfrac{k_{\nu}}{k}\rb  \lsb \hat{a}_{\vec{k}}\varphi_{\vec{k}}e^{-i\vec{k}\cdot \vec{x}} + \hat{a}_{\vec{k}}^{\dagger}\varphi_{\vec{k}}^{*}e^{+i\vec{k}\cdot \vec{x}}\rsb  \\
\hat{\varphi}_{\scalebox{0.5}{$<$}}(t,\vec{x}) &\equiv \int \dfrac{\mathrm{d}^3\vec{k}}{(2\pi)^{3/2}}\mathcal{W}\lb \dfrac{k}{k_{\nu}}\rb\lsb \hat{a}_{\vec{k}}\varphi_{\vec{k}}e^{-i\vec{k}\cdot \vec{x}} + \hat{a}_{\vec{k}}^{\dagger}\varphi_{\vec{k}}^{*}e^{+i\vec{k}\cdot \vec{x}}\rsb 
\end{align} \label{eq:varphi_split_defn}
\end{subequations}
and do the same for its momentum $\hat{\Pi} = \hat{\Pi}_{\scalebox{0.5}{$>$}} + \hat{\Pi}_{\scalebox{0.5}{$<$}}$: 
\begin{subequations}
\begin{align}
\hat{\Pi}_{\scalebox{0.5}{$>$}}(t,\vec{x}) &\equiv \int \dfrac{\mathrm{d}^3\vec{k}}{(2\pi)^{3/2}}\mathcal{W}\lb \dfrac{k_{\nu}}{k}\rb  \lsb \hat{a}_{\vec{k}}\Pi_{\vec{k}}e^{-i\vec{k}\cdot \vec{x}} + \hat{a}_{\vec{k}}^{\dagger}\Pi_{\vec{k}}^{*}e^{+i\vec{k}\cdot \vec{x}}\rsb  \\
\hat{\Pi}_{\scalebox{0.5}{$<$}}(t,\vec{x}) &\equiv \int \dfrac{\mathrm{d}^3\vec{k}}{(2\pi)^{3/2}}\mathcal{W}\lb \dfrac{k}{k_{\nu}}\rb\lsb \hat{a}_{\vec{k}}\Pi_{\vec{k}}e^{-i\vec{k}\cdot \vec{x}} + \hat{a}_{\vec{k}}^{\dagger}\Pi_{\vec{k}}^{*}e^{+i\vec{k}\cdot \vec{x}}\rsb 
\end{align} \label{eq:Pi_split_defn}
\end{subequations}
where the window function $\mathcal{W}$ suppresses all modes not in the relevant sector, i.e. it ensures no short wavelength modes are in $\hat{\varphi}_{\scalebox{0.5}{$>$}}$. The issue is that it is not known how to quantise the inflaton in an arbitrarily curved spacetime with backreaction so it is not clear what equations the mode functions $\varphi_{\vec{k}}$ \& $\Pi_{\vec{k}}$ should obey. Instead of worrying about the full mode functions we instead concentrate on the short-wavelength components. To make progress we will assume that the short-wavelength components can be described by linear perturbation theory. For this reason we will suggestively make the following identifications:
\begin{subequations}
\begin{align}
    \delta\hat{\varphi} &\equiv \hat{\varphi}_{\scalebox{0.5}{$<$}} \\
    \delta\hat{\Pi} &\equiv \hat{\Pi}_{\scalebox{0.5}{$<$}} 
\end{align}
\end{subequations}
so that they can be expressed in terms of the mode functions of linear perturbations $\delta\varphi_{\vec{k}}$ \& $\delta\Pi_{\vec{k}}$:
\begin{subequations}
\begin{align}
  \delta\hat{\varphi}  &= \int \dfrac{\mathrm{d}^3\vec{k}}{(2\pi)^{3/2}}\mathcal{W}\lb \dfrac{k}{k_{\nu}}\rb\lsb \hat{a}_{\vec{k}}\delta\varphi_{\vec{k}}e^{-i\vec{k}\cdot \vec{x}} + \hat{a}_{\vec{k}}^{\dagger}\delta\varphi_{\vec{k}}^{*}e^{+i\vec{k}\cdot \vec{x}}\rsb \\
   \delta\hat{\Pi} &= \int \dfrac{\mathrm{d}^3\vec{k}}{(2\pi)^{3/2}}\mathcal{W}\lb \dfrac{k}{k_{\nu}}\rb\lsb \hat{a}_{\vec{k}}\delta\Pi_{\vec{k}}e^{-i\vec{k}\cdot \vec{x}} + \hat{a}_{\vec{k}}^{\dagger}\delta\Pi_{\vec{k}}^{*}e^{+i\vec{k}\cdot \vec{x}}\rsb
\end{align}\label{eq:short-wav_window_defn}
\end{subequations}
We will assume that these mode functions obey the same linear perturbation equations as those derived earlier for a homogeneous background. The reasoning behind being able to make this simplification is that on the subhorizon scales for which $\delta\varphi_{\vec{k}}$ \& $\delta\Pi_{\vec{k}}$ exist the perturbations do not feel the large scale deviations from the flat FLRW metric but instead each Hubble sphere acts as its own little FLRW universe with background values given by $\varphi_{\scalebox{0.5}{$>$}}$, $\Pi_{\scalebox{0.5}{$>$}}$, $H_{\scalebox{0.5}{$>$}}$ smoothed over that patch. Using equation (\ref{eq:linear_pert_Q}) the mode functions of linear perturbations\footnote{Note that the mode functions now only depend on the norm $k$ rather than the vector $\vec{k}$. This is because isotropy is assumed to hold over the sub-Hubble scales the mode functions are defined over.} $\delta\varphi_{k}$ \& $\delta\Pi_{k}$ in spatially flat gauge obey:
\begin{subequations}
\begin{align}
\dfrac{\mathrm{d}\delta\varphi_{k}}{\mathrm{d}\alpha} &= \dfrac{\delta \Pi_k}{H_{\scalebox{0.5}{$>$}}}\\
\dfrac{\mathrm{d}\delta\Pi_{k}}{\mathrm{d}\alpha} &= -3 \delta \Pi_k -\omega_{k}^2\delta\varphi_k \\
\omega_{k}^{2} &\equiv \dfrac{1}{H_{\scalebox{0.5}{$>$}}}\lsb \dfrac{k^2}{a^2} + \dfrac{\mathrm{d}^2V(\phi)}{\mathrm{d}\phi^2} + 3\Pi_{\scalebox{0.5}{$>$}}^2 + 2\dfrac{\Pi_{\scalebox{0.5}{$>$}}}{H_{\scalebox{0.5}{$>$}}}\dfrac{\mathrm{d}V(\phi)}{\mathrm{d}\phi} - \dfrac{\Pi_{\scalebox{0.5}{$>$}}^4}{H_{\scalebox{0.5}{$>$}}^2}\rsb
\end{align} \label{eq:mode_eqns_short}
\end{subequations}
where we have identified $\phi \equiv \varphi_{\scalebox{0.5}{$>$}}$. The background quantities $\phi$, $\Pi_{\scalebox{0.5}{$>$}}$, $H_{\scalebox{0.5}{$>$}}$ are evaluated using the classical equations of motion for the long-wavelength sector derived in section \ref{sec:nonlin_pert}. We will further discuss this assumption shortly. \\
Having appropriately defined the mode functions we can then insert the splits (\ref{eq:varphi_split_defn}) \& (\ref{eq:Pi_split_defn}) into (\ref{eq:Full_ADM_beforesplit}) and -- after dropping second order gradient terms on long wavelengths since $\tilde{\nabla}_i\tilde{\nabla}^i \phi  \simeq 0$ -- we find:
\begin{subequations}
\begin{align}
    \dfrac{\mathrm{d}\Pi_{\scalebox{0.5}{$>$}}}{\mathrm{d}\alpha} &= -3\Pi_{\scalebox{0.5}{$>$}} -3\Pi_{\scalebox{0.5}{$<$}} - \dfrac{\mathrm{d}\delta \Pi}{\mathrm{d}\alpha} - \dfrac{1}{H_{\scalebox{0.5}{$>$}}}\dfrac{\mathrm{d}V(\phi)}{\mathrm{d}\phi} - \dfrac{\delta \varphi}{H_{\scalebox{0.5}{$>$}}}\dfrac{\mathrm{d}^2V(\phi)}{\mathrm{d}\phi^2} \nonumber \\
    &\quad + \dfrac{1}{H_{\scalebox{0.5}{$>$}}}\tilde{\nabla}_i\tilde{\nabla}^i \delta \varphi-\dfrac{1}{H_{\scalebox{0.5}{$>$}}^2}\lb \tilde{\nabla}_i H_{\scalebox{0.5}{$>$}}\rb \lsb\lb \tilde{\nabla}^i \phi \rb + \lb \tilde{\nabla}^i \delta \varphi \rb\rsb\\
    \dfrac{\mathrm{d}\phi}{\mathrm{d}\alpha} &= \dfrac{\Pi_{\scalebox{0.5}{$>$}}}{H_{\scalebox{0.5}{$>$}}} + \dfrac{\delta \Pi}{H_{\scalebox{0.5}{$>$}}} - \dfrac{\mathrm{d}\delta \varphi}{\mathrm{d}\alpha}
\end{align} \label{eq:Full_ADM_aftersplit}
\end{subequations}
We can compute the derivatives of the short-wavelength sector (e.g. $\partial_{\alpha}\delta \varphi$) using the definition (\ref{eq:short-wav_window_defn}) and the mode function equations (\ref{eq:mode_eqns_short}). The time derivative either hits the window function or it hits the mode functions themselves. If it hits the mode functions we find that it cancels exactly\footnote{This is not strictly true, they cancel exactly if one only includes the first two terms of $\omega_{k}^2$ in (\ref{eq:mode_eqns_short}). The remaining terms in $\omega_{k}^2$ arise from metric backreaction which cannot be included unless one appropriately perturbs the background quantities $H$ and $\alpha$ \cite{Rigopoulos2005}. One also has to set the $\tilde{\nabla}_i H_{\scalebox{0.5}{$>$}}$ term $= 0 $ by hand in (\ref{eq:Full_ADM_aftersplit}), the justification being that on short-wavelengths we treat $H_{\scalebox{0.5}{$>$}}$ as if it was constant over the whole Hubble sphere and on long-wavelengths the term is second order in spatial gradients so should also be dropped.} the other short-wavelength terms because of (\ref{eq:mode_eqns_short}) so the only terms that survive from the short-wavelength sector are those from the derivative acting on the window function. We can therefore rewrite (\ref{eq:Full_ADM_aftersplit}) like so:
\begin{subequations}
\begin{align}
    \dfrac{\mathrm{d}\Pi_{\scalebox{0.5}{$>$}}}{\mathrm{d}\alpha} &= -3\Pi_{\scalebox{0.5}{$>$}}  - \dfrac{1}{H}\dfrac{\mathrm{d}V(\phi)}{\mathrm{d}\phi} + \hat{\xi}_{\Pi}\\
    \dfrac{\mathrm{d}\phi}{\mathrm{d}\alpha} &= \dfrac{\Pi_{\scalebox{0.5}{$>$}}}{H} + \hat{\xi}_{\varphi}
\end{align} \label{eq:Split_ADM_withnoise}
\end{subequations}
where we have introduced the quantum noise terms $\hat{\xi}_{\varphi}$ \& $\hat{\xi}_{\Pi}$ defined as:
\begin{subequations}
\begin{align}
    \hat{\xi}_{\varphi} &= -  \int \dfrac{\mathrm{d}^3\vec{k}}{(2\pi)^{3/2}}\partial_{\alpha}\mathcal{W}\lb \dfrac{k}{k_{\nu}}\rb\lsb \hat{a}_{\vec{k}}\delta\varphi_{k}e^{-i\vec{k}\cdot \vec{x}} + \hat{a}_{\vec{k}}^{\dagger}\delta\varphi_{k}^{*}e^{+i\vec{k}\cdot \vec{x}}\rsb \\
    \hat{\xi}_{\Pi} & = -\int \dfrac{\mathrm{d}^3\vec{k}}{(2\pi)^{3/2}}\partial_{\alpha}\mathcal{W}\lb \dfrac{k}{k_{\nu}}\rb\lsb  \hat{a}_{\vec{k}}\delta\Pi_{k}e^{-i\vec{k}\cdot \vec{x}} + \hat{a}_{\vec{k}}^{\dagger}\delta\Pi_{k}^{*}e^{+i\vec{k}\cdot \vec{x}}\rsb
\end{align} \label{eq:quantum_noise_defn}
\end{subequations}
To determine the statistical properties of the quantum noises we consider their two-point correlation matrix:
\begin{eqnarray}
\Xi (\vec{x}_1,\alpha_1;\vec{x_2},\alpha_2) \equiv \begin{pmatrix}
\left\langle 0\right| \hat{\xi}_{\Pi}(\vec{x}_1,\alpha_1) \hat{\xi}_{\varphi}(\vec{x}_2,\alpha_2 )\left| 0\right\rangle  & \left\langle 0\right|\hat{\xi}_{\Pi}(\vec{x}_1,\alpha_1) \hat{\xi}_{\varphi}(\vec{x}_2,\alpha_2) \left| 0\right\rangle \\
\left\langle 0\right|\hat{\xi}_{\varphi}(\vec{x}_1,\alpha_1) \hat{\xi}_{\varphi}(\vec{x}_2,\alpha_2 )\left| 0\right\rangle & \left\langle 0\right|\hat{\xi}_{\Pi}(\vec{x}_1,\alpha_1) \hat{\xi}_{\Pi}(\vec{x}_2,\alpha_2 )\left| 0\right\rangle 
\end{pmatrix}
\end{eqnarray}
Introducing the notation $\Xi_{f_1,g_2} \equiv \left\langle 0\right|\hat{\xi}_{f}(\vec{x}_1,\alpha_1) \hat{\xi}_{g}(\vec{x}_2,\alpha_2 )\left| 0\right\rangle$ and allowing the annihilation and creation operators to act on the vacuum we find that:
\begin{eqnarray}
\Xi_{f_1,g_2} = \int \dfrac{\mathrm{d}^3k}{(2\pi)^3}\partial_{\alpha}\mathcal{W}\lb\dfrac{k}{k_{\nu}(\alpha_1)}\rb \partial_{\alpha}\mathcal{W}\lb\dfrac{k}{k_{\nu}(\alpha_2)}\rb f_{k}(\alpha_1)g_{k}^{*}(\alpha_2)e^{i\vec{k}\cdot(\vec{x}_2 - \vec{x}_1)}
\end{eqnarray}
As the mode functions only depend on the norm of $\vec{k}$ we can perform the angular integral such that:
\begin{eqnarray}
\Xi_{f_1,g_2} = \int \dfrac{k^2\mathrm{d}k}{2\pi^2}\partial_{\alpha}\mathcal{W}\lb\dfrac{k}{k_{\nu}(\alpha_1)}\rb \partial_{\alpha}\mathcal{W}\lb\dfrac{k}{k_{\nu}(\alpha_2)}\rb f_{k}(\alpha_1)g_{k}^{*}(\alpha_2)\dfrac{\sin (k|\vec{x_2} - \vec{x_1} |)}{k|\vec{x_2} - \vec{x_1} |}
\end{eqnarray}
To make anymore progress we now need to specify the window function. The easiest choice is a simple Heaviside function $\mathcal{W}(k/k_{\nu}) = \Theta (k/k_{\nu} -1)$ so that the time derivative gives a Dirac distribution. This leads to:
\begin{eqnarray}
\Xi_{f_1,g_2} = \dfrac{1}{6\pi^2}\dfrac{\mathrm{d}k^3_{\nu}(\alpha)}{\mathrm{d}\alpha}\Bigg\vert_{\alpha_1}f_{k_{\nu}}(\alpha_1)g_{k_{\nu}}^{*}(\alpha_1)\dfrac{\sin (k_{\nu}(\alpha_1)|\vec{x_2} - \vec{x_1} |)}{k_{\nu}(\alpha_1)|\vec{x_2} - \vec{x_1} |}\delta (\alpha_1 - \alpha_2)
\end{eqnarray}
We will be focusing on autocorrelation of the noises, i.e. $\vec{x}_1 = \vec{x}_2$, and as the noises are white -- due to the choice of window function -- the correlations are only non-zero at equal time. We therefore write $\Xi_{f_1,g_2} \equiv \Xi_{f,g}(\alpha_1)\delta (\alpha_1 - \alpha_2 )$ which we can write in terms of the power spectrum of quantum fluctuations:
\begin{eqnarray}
\Xi_{f,g}(\alpha) &=& \dfrac{\mathrm{d}\ln k_{\nu}(\alpha)}{\mathrm{d}\alpha} \Delta_{f,g} \\
\Delta_{f,g} &\equiv & \dfrac{k^3}{2\pi^2}f_k(\alpha)g_{k}^{*}(\alpha)
\end{eqnarray}
In general these solutions do not have a proper analytic form and must be obtained numerically, however to get a sense of what is going on we will consider the case of massless de Sitter. Then we can use the solutions obtained earlier -- see equations (\ref{eq:deSitter_gamma_matrices}) -- to compute the correlation matrix elements:
\begin{subequations}
\begin{align}
    \Xi_{\varphi,\varphi} &= \dfrac{H_{\scalebox{0.5}{$>$}}^2}{4\pi^2}\lsb 1 + \nu^2\rsb \\
    \Xi_{\Pi,\Pi} &= \dfrac{H_{\scalebox{0.5}{$>$}}^4}{4\pi^2} \nu^4\\
    \Xi_{\varphi,\Pi} + \Xi_{\Pi,\varphi} &= -\dfrac{H_{\scalebox{0.5}{$>$}}^3}{2\pi^2}\nu^2\\
    \Xi_{\varphi,\Pi} - \Xi_{\Pi,\varphi} &= \dfrac{iH_{\scalebox{0.5}{$>$}}^3}{2\pi^2}\nu^3
\end{align}\label{eq:noise_matrix_deSitter}
\end{subequations}
We can see here that nearly all terms are suppressed by powers of the coarse-graining scale $\nu \ll 1$, the notable exception being the noise in the inflaton $\Xi_{\varphi,\varphi}$ which approaches the de Sitter temperature (\ref{eq:dS_temperature}) computed earlier. It is worth noting that if the field is not exactly massless then the RHS of equations (\ref{eq:noise_matrix_deSitter}) will have an additional prefactor \cite{Nakao1988} of $\nu^{2M^2/3H^2}$ which suggests that $\nu \gg \exp (-3H^2/|M|^2)$. One in general should therefore be careful about choosing $\nu$ to be arbitrarily small. It is also worth pointing out that the whole point of this stochastic inflation business is to create a formalism that can handle non-linear perturbations. As the mode functions that source the stochastic noise terms are derived from linear perturbation theory, if we coarse-grain ``too late" we will miss the non-linear evolution we are trying to capture in the first place!\\
Equations (\ref{eq:noise_matrix_deSitter}) can also point to the ``classicalisation" of perturbations like so:
\begin{eqnarray}
\left\langle 0\right| \lsb \delta\hat{\varphi}, \delta\hat{\Pi}\rsb \left| 0\right\rangle &=& \dfrac{iH_{\scalebox{0.5}{$>$}}}{2\pi^2}\nu^3, \quad \left\langle 0\right|  \lcb \delta\hat{\varphi}, \delta\hat{\Pi}\rcb \left| 0\right\rangle = -\dfrac{H_{\scalebox{0.5}{$>$}}^3}{2\pi^2}\nu^2\\
\Rightarrow \dfrac{\left\langle 0\right|  \lsb \delta\hat{\varphi}, \delta\hat{\Pi}\rsb \left| 0\right\rangle}{\left\langle 0\right|  \lcb \delta\hat{\varphi}, \delta\hat{\Pi}\rcb \left| 0\right\rangle} &=& -i\nu \underset{\nu \rightarrow 0}{\longrightarrow} 0
\end{eqnarray}
i.e. the commutator of the operators can be neglected compared to the anti-commutator at the coarse-graining scale, provided it is chosen to be sufficiently small. While this result is for exact de Sitter, it is generically true that there will be a decaying mode that we can choose to neglect so that we essentially ignore the ``quantumness" of the fields on long wavelengths. We therefore identify the quantum noises with classical stochastic noises i.e. $\hat{\xi}_{f}\rightarrow \xi_{f}$ which also have vanishing mean and covariances given by $\Xi_{f,g}$. In this way we relate the vacuum expectation value with a stochastic average -- e.g. $\left\langle 0\right| \hat{\xi}_{f} \left| 0 \right\rangle \rightarrow \left\langle  \xi_{f} \right\rangle$. Restoring the Planck mass, $M_{\mathrm{p}}$, our complete stochastic equations therefore are:
\begin{subequations}
\begin{empheq}[box = \fcolorbox{Maroon}{white}]{align}
    \dfrac{\mathrm{d}\Pi_{\scalebox{0.5}{$>$}}}{\mathrm{d}\alpha} &= -3\Pi_{\scalebox{0.5}{$>$}}  - \dfrac{1}{H_{\scalebox{0.5}{$>$}}}\dfrac{\mathrm{d}V(\phi)}{\mathrm{d}\phi} + \xi_{\Pi}\\
    \dfrac{\mathrm{d}\phi}{\mathrm{d}\alpha} &= \dfrac{\Pi_{\scalebox{0.5}{$>$}}}{H_{\scalebox{0.5}{$>$}}} + \xi_{\varphi} \\
    \left\langle  \xi_{\varphi}(\alpha_1) \xi_{\varphi}(\alpha_2) \right\rangle &= \dfrac{1}{6\pi^2}\dfrac{\mathrm{d}(\nu aH_{\scalebox{0.5}{$>$}})^3}{\mathrm{d}\alpha}\Bigg\vert_{\alpha_1}\left| \delta \varphi_{k}\right|^2 \delta (\alpha_1 -\alpha_2)\\
    \left\langle  \xi_{\Pi} (\alpha_1) \xi_{\Pi} (\alpha_2) \right\rangle &= \dfrac{1}{6\pi^2}\dfrac{\mathrm{d}(\nu aH_{\scalebox{0.5}{$>$}})^3}{\mathrm{d}\alpha}\Bigg\vert_{\alpha_1}\left| \delta \Pi_{k}\right|^2\delta (\alpha_1 -\alpha_2)\\
    \left\langle  \xi_{\varphi}(\alpha_1)\xi_{\Pi} (\alpha_2)  \right\rangle &= \dfrac{1}{6\pi^2}\dfrac{\mathrm{d}(\nu aH_{\scalebox{0.5}{$>$}})^3}{\mathrm{d}\alpha}\Bigg\vert_{\alpha_1}\text{Re}(\delta\varphi_k \delta\Pi_k^{*}) \delta (\alpha_1 -\alpha_2)\\
    3H_{\scalebox{0.5}{$>$}}^2M_{\mathrm{p}}^2 &= \dfrac{\Pi_{\scalebox{0.5}{$>$}}^2}{2}+ V(\phi)  \\
    \tilde{\nabla}_i H_{\scalebox{0.5}{$>$}}M_{\mathrm{p}}^2 &= -\dfrac{1}{2}\Pi_{\scalebox{0.5}{$>$}}\tilde{\nabla}_i \phi
\end{empheq} \label{eq:Full_stochastic_equations}
\end{subequations}
which together with (\ref{eq:mode_eqns_short}) can -- in principle -- be straightforwardly solved numerically. However straightforward does not mean easy and it is common to neglect the noise in the momentum. This is easily justified for exact de Sitter -- see (\ref{eq:noise_matrix_deSitter}) -- but it is true in generality that $\left|\delta \Pi_{k}\right|^2 \ll \left|\delta \varphi_{k}\right|^2$ at the coarse-graining scale. We can therefore utilise the constraints to reduce the dynamics of the system and recover the equations of motion from section \ref{sec:nonlin_pert} with the addition of a stochastic noise term:
\begin{subequations}
\begin{empheq}[box = \fcolorbox{Maroon}{white}]{align}
    \dfrac{\mathrm{d}\phi}{\mathrm{d}\alpha} &= -2\dfrac{\mathrm{d}\ln H_{\scalebox{0.5}{$>$}}}{\mathrm{d}\phi} + \xi_{\varphi} \\
    \left\langle  \xi_{\varphi}(\alpha_1) \xi_{\varphi}(\alpha_2) \right\rangle &= \dfrac{1}{6\pi^2}\dfrac{\mathrm{d}(\nu aH_{\scalebox{0.5}{$>$}})^3}{\mathrm{d}\alpha}\Bigg\vert_{\alpha_1}\left| \delta \varphi_{k}\right|^2 \delta (\alpha_1 -\alpha_2)\\
    3H_{\scalebox{0.5}{$>$}}^2M_{\mathrm{p}}^2 &= 2\lb\dfrac{\mathrm{d}H_{\scalebox{0.5}{$>$}}}{\mathrm{d}\phi}\rb^2M_{\mathrm{p}}^4 + V(\phi) 
\end{empheq} \label{eq:first_order_stochastic_equations}
\end{subequations}
which simplifies even further if we assume the mode functions can be well approximated\footnote{See \cite{Figueroa2021} for potential problems with this.} by the de Sitter ones:
\begin{subequations}
\begin{empheq}[box = \fcolorbox{Maroon}{white}]{align}
    \dfrac{\mathrm{d}\phi}{\mathrm{d}\alpha} &= -2\dfrac{\mathrm{d}\ln H_{\scalebox{0.5}{$>$}}}{\mathrm{d}\phi} + \xi_{\varphi} \\
    \left\langle  \xi_{\varphi}(\alpha_1) \xi_{\varphi}(\alpha_2) \right\rangle &= \lb\dfrac{H}{2\pi}\rb^2\delta (\alpha_1 -\alpha_2)\\
    3H_{\scalebox{0.5}{$>$}}^2M_{\mathrm{p}}^2 &= 2\lb\dfrac{\mathrm{d}H_{\scalebox{0.5}{$>$}}}{\mathrm{d}\phi}\rb^2M_{\mathrm{p}}^4 + V(\phi) 
\end{empheq} \label{eq:deSitter_stochastic_equations}
\end{subequations}
Equations (\ref{eq:Full_stochastic_equations}), (\ref{eq:first_order_stochastic_equations}) \& (\ref{eq:deSitter_stochastic_equations}) are the main results of this subsection. 
\subsection{\label{sec:stochastic_deltaN}Stochastic-\texorpdfstring{$\delta \mathcal{N}$}{} formalism}
\begin{figure}
    \centering
    \includegraphics[width = 0.95\linewidth]{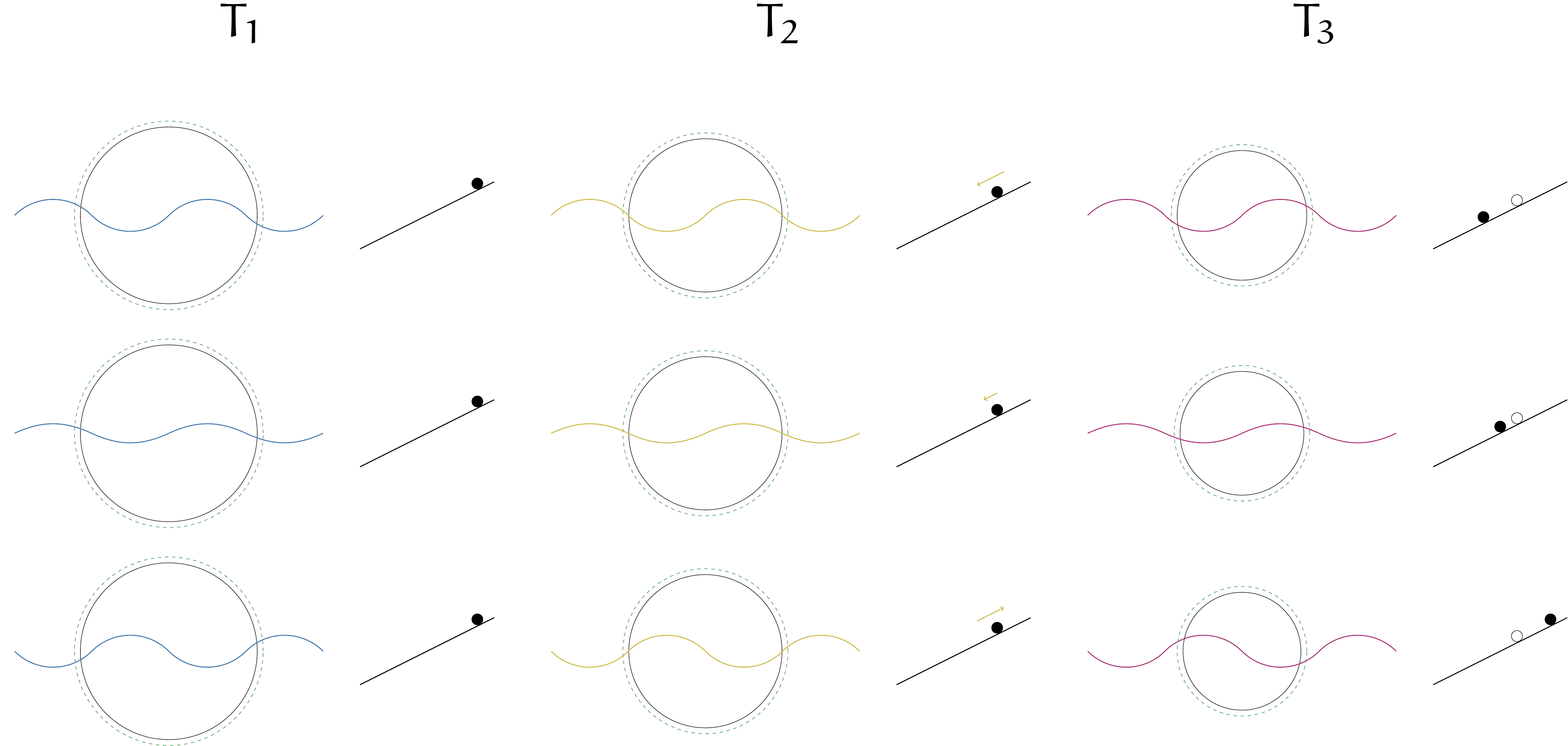}
    \caption[How fluctuations in real space affect the inflaton]{A schematic drawing for how perturbations of the same wavelength, but different amplitude, affect the inflaton trajectory. The the green dashed circles correspond to different coarse-grained patches which are slightly larger than the Hubble sphere (solid black circles). Next to each patch is a representation of how the inflaton slides down a linear potential in that patch. Each patch contains only a single perturbation with the same wavelength but different amplitude. At $T = 1$ the perturbation wavelength is shorter than the coarse-grained patch so is shown in blue. At $T = 2$ the perturbation wavelength is comparable and is shown in yellow. This is when it sources a stochastic kick on the inflaton (shown by a yellow arrow). At $T = 3$ the perturbation has entered the long-wavelength sector and is shown in red.}
    \label{fig:Stochastic_fluctuations_realspace}
\end{figure}
As we have described above, the stochastic formulation of inflation allows us to treat the long-wavelength modes of the inflaton, $\phi$, as a classical stochastic variable which obeys stochastic equations of motion -- in principle (\ref{eq:Full_stochastic_equations}) but in practice usually (\ref{eq:deSitter_stochastic_equations}). Because the equations of motion are now stochastic this means that the inflaton's evolution on the potential is more complicated. In Fig.~\ref{fig:Stochastic_fluctuations_realspace} we show a schematic for a toy model of this behaviour at three different moments in time, $T_1$, $T_2$, and $T_3$. At $T_1$ we have the setup of three different coarse-grained patches (shown by the green dashed-line) with a Hubble patch (solid line) also shown. Inside each of these patches is a perturbation of wavelength $\lambda < \nu aH$ but it has a different amplitude in each one. To the right is the inflaton slowly sliding down a slope. At this stage the perturbation has no effect on the long wavelength sector and is shown in blue. In the middle panel at $T_2$ the inflaton has slid a little bit down the slope and the coarse-graining scale has shrunk now so that $\lambda \sim \nu aH$ and is therefore show in yellow. The perturbation is now added to the long-wavelength sector as a stochastic kick and because the amplitude is different in each patch the size of the stochastic kick is different in each case. We can see the size and directions of these different stochastic kicks by the yellow arrows above the inflaton. At a later time $T_3$ we can see that these kicks have resulted in very different positions of the inflaton. The empty circle corresponds to where the inflaton would have been just due to classical drift. Because of these different positions the value of $H$ is slightly different and so the comoving coarse-graining scale -- and therefore Hubble sphere -- is also slightly different in each patch. Therefore each Hubble patch, corresponding to a different stochastic realisation, will evolve differently to other patches. As the perturbation is now in the long-wavelength sector it is plotted in red.\\

The stochastic nature of the dynamics means that the time taken (measured in e-folds) for the inflaton to reach the end of inflation, corresponding to $\phi_e$, is also a stochastic quantity, denoted by $\mathcal{N}$. We can imagine computing the average e-fold time taken, $\left\langle \mathcal{N}\right\rangle$, by averaging over many different realisations of (\ref{eq:deSitter_stochastic_equations}). This is useful because the stochastic $\delta \mathcal{N}$ formalism \cite{Enqvist2008, Fujita2013, Fujita2014, Vennin2015} allows one to compute the coarse-grained comoving curvature perturbation on uniform energy density time-slices\footnote{Phrased another way the formalism allows us to move between quantities computed in spatially flat gauge, the stochastic equations, to quantities in comoving or uniform expansion gauge.} $\mathcal{R}_{cg}$ through:
\begin{empheq}[box = \fcolorbox{Maroon}{white}]{equation}
\mathcal{N} - \left\langle \mathcal{N}\right\rangle = \mathcal{R}_{cg} = \dfrac{1}{(2\pi)^{3/2}}\int_{k_{in}}^{k_{end}}\mathrm{d}\vec{k}\mathcal{R}_{\vec{k}}e^{i\vec{k}\cdot\vec{x}} \label{eq:Rcg_defn}
\end{empheq}  
which -- as the name suggests -- is just the usual comoving curvature perturbation coarse-grained between scales $k_{in}$, the scale that crossed the Hubble radius at initial time, and $k_{end}$, the scale that crosses out the Hubble radius at final time. This reduces the problem of computing curvature perturbations to the one of performing first-passage time analysis on the stochastic equations of motion to obtain the PDF for exit time $\rho (\mathcal{N})$. This can be achieved for example by following the method outlined in \cite{Vennin2015} for \ac{SR} inflation which uses first passage time analysis on the stochastic differential equation:
\begin{subequations}
\begin{align}
\dfrac{\mathrm{d}\phi}{\mathrm{d}\alpha} &= -\dfrac{v_{,\phi}}{v} + \xi (\alpha) \label{eq:Venin  dphidalpha}   \\
\left\langle \xi (\alpha_1) \xi(\alpha_2) \right\rangle &= 2v \delta (\alpha_1 -\alpha_2)
\end{align} \label{eq:Vennin_both}
\end{subequations} 
where $v \equiv V/24\pi^2M_{\mathrm{p}}^{4}$ is the dimensionless potential. It is clear that equation (\ref{eq:Vennin_both}) is of the same form as (\ref{eq:deSitter_stochastic_equations})
\begin{subequations}
\begin{align}
\dfrac{\mathrm{d}\phi}{\mathrm{d}\alpha} &= - 2\dfrac{\partial_\phi( \tilde{H}^2)}{\tilde{H}^2} + \xi (\alpha) \\
\left\langle \xi (\alpha_1) \xi(\alpha_2) \right\rangle &= 2\tilde{H}^2 \delta (\alpha_1 -\alpha_2)
\end{align} \label{eq:Htilde  dphidalpha}
\end{subequations} 
if we make the identification $v \rightarrow \tilde{H}^2$ where, as before, $\tilde{H}^2 \equiv H_{\scalebox{0.5}{$>$}}^2/8\pi^2M_{\mathrm{p}}^{2}$ is the dimensionless Hubble expansion rate and where now $\phi$ is dimensionless . We can therefore utilise the \ac{SR} formulae given in \cite{Vennin2015} and rewrite them in terms of $\tilde{H}$ giving them full validity outside of the \ac{SR} regime. The average number of e-folds it takes to reach $\phi_e$ starting at $\phi_{*}$, $\left\langle\mathcal{N}\right\rangle$, is given by:
\begin{eqnarray}
\left\langle\mathcal{N}\right\rangle (\phi_{*}) = \int_{\phi_{e}}^{\phi_{*}}\dfrac{\mathrm{d}x}{M_{\mathrm{p}}}\int_{x}^{\bar{\phi}}\dfrac{\mathrm{d}y}{M_{\mathrm{p}}}\dfrac{1}{\tilde{H}^2(y)}\text{exp}\left[ \dfrac{1}{\tilde{H}^2(y)} - \dfrac{1}{\tilde{H}^2(x)} \right] \label{eq:mean N full}
\end{eqnarray}
where $\bar{\phi}$ is a reflective boundary set high up in the UV that can be necessary to regularise the integrals. So long as this is set large enough it typically does not change the results \cite{Vennin2015}. The variation in the number of e-folds, $\delta \mathcal{N}^2 = \left\langle\mathcal{N}^2\right\rangle - \left\langle\mathcal{N}\right\rangle^2$ is given by:
\begin{eqnarray}
\delta \mathcal{N}(\phi_{*})^2 = \int_{\phi_{e}}^{\phi_{*}}\mathrm{d}x\int_{x}^{\bar{\phi}}\mathrm{d}y \left[ \dfrac{\partial }{\partial y}\left\langle\mathcal{N}\right\rangle (y)\right]^2\text{exp}\left[ \dfrac{1}{\tilde{H}^2(y)} - \dfrac{1}{\tilde{H}^2(x)} \right] \label{eq:Var N full}
\end{eqnarray}
As $\left\langle\mathcal{N}\right\rangle$ and $\delta \mathcal{N}^2 $ are both functions of $\phi_{*}$ we can express them in terms of one another to obtain the power spectrum:
\begin{eqnarray}
\Delta_{\mathcal{R}}^2 = \dfrac{\mathrm{d}\delta\mathcal{N}^2}{\mathrm{d}\left\langle \mathcal{N}\right\rangle} \label{eq:Powerspecfull}
\end{eqnarray}
Similarly the local $f_{\scalebox{0.5}{$\mathrm{NL}$}}$ parameter is given by:
\begin{eqnarray}
f_{\scalebox{0.5}{$\mathrm{NL}$}} = \dfrac{5}{72}\lb\dfrac{\mathrm{d}\delta\mathcal{N}^3}{\mathrm{d}\left\langle \mathcal{N}\right\rangle}\rb^2 \left(\dfrac{\mathrm{d}\delta\mathcal{N}^2}{\mathrm{d}\left\langle \mathcal{N}\right\rangle}\right)^{-2} \label{eq:fNL full}
\end{eqnarray}
These results should reduce to the standard $\delta N$ formalism in the appropriate limit which we will call \emph{semi-classical}. By semi-classical we mean that the integrals above can be well approximated by the leading order contribution in the saddle-point approximation as in \cite{Vennin2015}. To ensure that this approximation is under control we introduce the classicality parameter, $\eta_{cl}$, as derived in \cite{Vennin2015} from the second order term in the expansion -- it being small ensures the validity of being in the semi-classical regime. In this sense it is a more sophisticated measure of classicality than simply the ratio of the quantum diffusion over classical drift $\delta \phi_{qu}/\delta \phi_{cl}$ as is often used. By performing this saddle point approximation we find that the \emph{semi-classical} formulae for the average e-fold time, $\left\langle \mathcal{N}\right\rangle$, the deviation from the average e-fold time (or variance), $\delta \mathcal{N}^2 = \left\langle \mathcal{N}^2\right\rangle - \left\langle \mathcal{N}\right\rangle^2$, the power spectrum of curvature perturbations, $\Delta_{\mathcal{R}}^2\vert_{cl}$, the local non-Gaussianity, $f_{\scalebox{0.5}{$\mathrm{NL}$}} $, the spectral tilt, $n_{\phi}$, and the classicality parameter, $\eta_{cl}$, are:
\begin{eqnarray}
\left\langle \mathcal{N}\right\rangle\vert_{cl} &=& \dfrac{1}{2}\int_{\phi_{end}}^{\phi}\dfrac{\mathrm{d}x}{M_{\mathrm{p}}^2}\dfrac{\tilde{H}(x)}{\tilde{H}_{,x}(x)} = \int_{\phi_{e}}^{\phi}\dfrac{\mathrm{d}x}{M_{\mathrm{p}}}\dfrac{1}{\sqrt{2\epsilon_1}(x)}  \label{eq: classical average e-fold} \\
\delta \mathcal{N}^2\vert_{cl} &=& \dfrac{1}{4}\int_{\phi_{end}}^{\phi}\dfrac{\mathrm{d}x}{M_{\mathrm{p}}^4}\dfrac{\tilde{H}^{5}(x)}{\tilde{H}_{,x}^{3}(x)} \label{eq: classical var e-fold} \\
\Delta_{\mathcal{R}}^2\vert_{cl} &=& \dfrac{1}{2}\dfrac{1}{M_{\mathrm{p}}^2}\dfrac{\tilde{H}^{4}(\phi)}{\tilde{H}_{,\phi}^{2}(\phi)} = \dfrac{\tilde{H}^{2}(\phi)}{\epsilon_{1}} \label{eq: classical power spectrum} \\
f_{\scalebox{0.5}{$\mathrm{NL}$}}\vert_{cl} &=& \dfrac{5}{24}\lsb 16\lb\dfrac{\tilde{H}_{,\phi}}{\tilde{H}}\rb^2 - 8\dfrac{\tilde{H}_{,\phi\phi}}{\tilde{H}}\rsb  = \dfrac{5}{24}\lb 4\varepsilon_1 -2\varepsilon_2\rb\label{eq:classical fNL}\\
n_{\phi} \vert_{cl} &=& 1-\lsb 8\lb\dfrac{\tilde{H}_{,\phi}}{\tilde{H}}\rb^2 - 4\dfrac{\tilde{H}_{,\phi\phi}}{\tilde{H}}\rsb = 1-2\varepsilon_1 +\varepsilon_2 \label{eq:classical ns} \\
\eta_{cl} &=& \left| \dfrac{3}{2}\tilde{H}^2 - \dfrac{\tilde{H}_{,\phi\phi}\tilde{H}^3}{2\tilde{H}_{,\phi}^{2}}\right| = \Delta_{\mathcal{R}}^2\vert_{cl} \left| \dfrac{7\epsilon_1}{2} - \dfrac{\epsilon_2}{4}\right| \label{eq: classicality criterion}
\end{eqnarray}
We can see that these all reduce to the standard formulae one obtains from the usual $\delta N$ formalism. For reasons we will explore in chapter \ref{cha:PBH} it is desirable to know the whole PDF of $\mathcal{R}_{cg}$. Pattison et.al \cite{Pattison2017} outline a program to compute the PDF of exit time, $\rho (\mathcal{N})$ using characterstic function techniques. As highlighted earlier, their formulae for \ac{SR} can be fully valid outside the \ac{SR} regime under the replacement $v \rightarrow \tilde{H}^2$. We will focus on the expansion of the characteristic function around the classical limit. At leading order every trajectory takes the same amount of time and there are no coarse-grained comoving curvature perturbations. One must go to the \ac{NLO} equation in \cite{Pattison2017} to obtain curvature perturbations with a Gaussian shape, which under the replacement $v \rightarrow \tilde{H}^2$ is simply:
\begin{eqnarray}
\rho_{\scalebox{0.5}{$\mathrm{NLO}$}}(\mathcal{R}_{cg}) &=& \dfrac{1}{\tilde{H}\sqrt{4\pi \gamma_{1}^{\scalebox{0.5}{$\mathrm{NLO}$}}}}\exp \lsb -\lb\dfrac{\mathcal{R}_{cg} }{{2\tilde{H}\sqrt{\gamma_{1}^{\scalebox{0.5}{$\mathrm{NLO}$}}}}} \rb^2\rsb \label{eq:rhoN_char_NLO}\\
\gamma_{1}^{\scalebox{0.5}{$\mathrm{NLO}$}} &\equiv & \dfrac{1}{8\tilde{H}^2M_{\mathrm{p}}^{4}}\int_{\phi_{end}}^{\phi_{in}}\mathrm{d}\phi~\dfrac{\tilde{H}^5}{\tilde{H}_{,\phi}^3} \label{eq:gamma1 NLO}
\end{eqnarray}
If we go to the \ac{NNLO} in \cite{Pattison2017} and again perform the replacement $v \rightarrow \tilde{H}^2$ we obtain non-Gaussianities:
\begin{eqnarray}
\rho_{\scalebox{0.5}{$\mathrm{NNLO}$}}(\mathcal{R}_{cg}) &=& \dfrac{1}{\tilde{H}\sqrt{4\pi \gamma_{1}^{\scalebox{0.5}{$\mathrm{NNLO}$}}}}\exp \lsb -\lb\dfrac{\mathcal{R}_{cg} }{{2\tilde{H}\sqrt{\gamma_{1}^{\scalebox{0.5}{$\mathrm{NNLO}$}}}}} \rb^2\rsb \nonumber \\
&&\times \lcb 1-\dfrac{\gamma_{2}^{\scalebox{0.5}{$\mathrm{NNLO}$}}  \mathcal{R}_{cg} }{8 \tilde{H}^2\lb \gamma_{1}^{\scalebox{0.5}{$\mathrm{NNLO}$}}\rb^3 } \lsb 6 \tilde{H}^2 \gamma_{1}^{\scalebox{0.5}{$\mathrm{NNLO}$}} -  \mathcal{R}_{cg}^2 \rsb \rcb \label{eq:rhoN_char_NNLO}\\
\gamma_{1}^{\scalebox{0.5}{$\mathrm{NNLO}$}} &\equiv & \dfrac{1}{16\tilde{H}^2M_{\mathrm{p}}^{4}}\int_{\phi_{end}}^{\phi_{in}}\mathrm{d}\phi \left[2\dfrac{\tilde{H}^5}{\tilde{H}_{,\phi}^3} + 7\dfrac{\tilde{H}^7}{\tilde{H}_{,\phi}^3} - 5\dfrac{\tilde{H}^8\tilde{H}_{,\phi\phi}}{\tilde{H}_{,\phi}^5}\right]\label{eq:gamma1 NNLO}\\
\gamma_{2}^{\scalebox{0.5}{$\mathrm{NNLO}$}} &\equiv & \dfrac{1}{16\tilde{H}^4M_{\mathrm{p}}^{6}}\int_{\phi_{end}}^{\phi_{in}}\mathrm{d}\phi~\dfrac{\tilde{H}^9}{\tilde{H}_{,\phi}^5} \label{eq:gamma2 NNNO}
\end{eqnarray}
In principle one could carry on continuing this expansion to arbitrary order. The problem with this approach is that this expansion is carried out around the \emph{peak} of the distribution i.e. around the average $\lan \mathcal{N}\ran$ which means it does a very poor job of capturing behaviour around the tail of the distribution.

\section{\label{sec:infla_conc} Conclusion}
\acresetall 
In this chapter we have reviewed the concept of an inflationary period and outlined its motivation, namely its ability to solve the horizon problem whilst simultaneously providing a mechanism for the perturbations observed in the \ac{CMB} that will go on to form all large-scale structure. We have reviewed how these inflationary perturbations can be described in both a linear and non-linear regime. We have discussed the notion of classicalisation of perturbations and emphasised the importance of incorporating \ac{GR} momentum constraint.\\

We have also included a pedagogical overview for the stochastic inflation formalism in phase space, paying particular care to emphasise the assumptions built into the standard framework. In particular the way the noise correlators are usually computed requires assuming that each Hubble patch can essentially be treated as its own (approximately) homogeneous and isotropic universe. We point out that if one wishes to include metric backreaction in the linear mode functions that one should also appropriately perturb the background quantities $H$ and $\alpha$ for the stochastic equations of motion to be consistent. We reviewed the stochastic $\delta \mathcal{N}$ formalism and note that \ac{SR} formulae previously computed in this framework can be made valid outside of the \ac{SR} regime by making the identification $v \rightarrow \tilde{H}^2$. We use this to modify the works of Pattison et.al \cite{Pattison2017} to outline how to compute the first-passage time problem at \ac{NLO} and \ac{NNLO} from the classical trajectory using characteristic function techniques. 
\chapter{Primordial Black Holes} 
\label{cha:PBH} 
\acresetall 
\vspace{0.5cm}
\begin{flushright}{\slshape    
If you hear a ``prominent" economist using the word `equilibrium,' or `normal distribution,' \\do not argue with him; just ignore him, or try to put a rat down his shirt.} \\ \medskip
--- Nassim Nicholas Taleb \cite{NassimNicholasTaleb}
\end{flushright}
\vspace{0.5cm}
What Nassim is alluding to is the phenomenon of a \emph{black swan}: an event that is deemed improbable and yet causes massive consequences. For over 1000 years\footnote{The earliest known use is from the 2nd-century Roman poet Juvenal's characterisation in his Satire VI of something being "rara avis in terris nigroque simillima cygno" -- "a rare bird in the lands and very much like a black swan" \cite{Puhvel1984}.} the existence of a black swan was considered so ludicrous it was a common expression as a statement of impossibility. However after Dutch explorers were the first Europeans to see black swans in Western Australia in 1697 a black swan came to refer to events so rare they were previously thought to be impossible\dots until they weren't. As the quote indicates therefore, one should be wary about soley focusing on the common events you can describe by a normal distribution, because a black swan may come and ruin the whole thing! In this chapter we will discuss the exceedingly rare events required to form \ac{PBHs}\footnote{Note that in spite of the perceived rarity of black swans they actually account for $\sim $ 1 in 10 of the global population whereas for \ac{PBHs} to make up all of dark matter their initial rarity was more like 1 in 10000. Perhaps the black swan phenomenon should be renamed to the Primordial Black Hole phenomenon!}.
\section{Introduction}
\label{sec:intro}
In the previous chapter we outlined how the stochastic formalism can be applied to the period of accelerated expansion in the early universe known as inflation. In this chapter we utilise this framework to describe the formation of the most extreme objects known to exist: black holes. Black holes created due to large inflationary fluctuations will have formed before all large-scale structure and are dubbed \textit{primordial} to distinguish them from those black holes created due to the collapse of stars -- \textit{stellar} black holes.   \\

This chapter is in part based on the publication \cite{Rigopoulos2021} and the outline is as follows. Section \ref{sec:PBH_overview} is background information and gives an overview of \ac{PBHs}, covering their status as a possible \ac{DM} candidate, how one can characterise their collapse and how their abundance can be computed from knowledge of the PDF for the comoving curvature perturbation. Our contribution begins in section \ref{sec: plateau} where we use the \ac{H-J} formalism to extend the results of \cite{Prokopec2021} to a plateau of finite width. We consider two main scenarios, Scenario A where the classical inflaton velocity is large enough to carry it through the plateau and Scenario B where the inflaton comes to a stop and is followed by a phase of free diffusion. In section \ref{sec: Inflection} we extend these results to a local inflection point. We summarise our results in section \ref{sec: PBHConclusion}. Some technical computations for the first-passage time are deferred to Appendix \ref{APP:FPT}.

The busy reader is directed to the main results of this chapter:
\begin{itemize}
    \item Fig.~\ref{fig:Inflationary_fluctuations_PBHs} shows how \ac{PBHs} can act as a probe of inflation below the scales observed in the \ac{CMB}.
    \item The mass fraction of \ac{PBHs} will be computed in this chapter predominantly from equation (\ref{eq:massfracdef_both}). 
    \item In Fig.~\ref{fig:Scenario A and B} we show the two possible scenarios for a plateau in the potential that we will examine.
    \item The mass fraction of \ac{PBHs} generated by a plateau in the potential can be computed exactly from equation (\ref{eq:massfrac_exactFULL}) with the result plotted in Fig.~\ref{fig:chi_sigma_forBeta_USRandSR}. This allows us to determine a pivot scale given by (\ref{eq:sigma pivot}) between under- and over-production of \ac{PBHs}.
    \item In Fig.~\ref{fig:Beta_Riotta_compare} we highlight how computing the mass fraction more accurately in terms of the density contrast would only serve to \emph{enhance} the abundance of \ac{PBHs}.
    \item In Fig.~\ref{fig:inflec_powerandbeta.pdf} we show the abundance of \ac{PBHs} generated from an inflection point using a semi-classical expansion. 
\end{itemize}

\section{\label{sec:PBH_overview} Primordial Black Holes -- an overview}
\begin{figure}[t!]
    \centering
    \includegraphics[width = 0.95\linewidth]{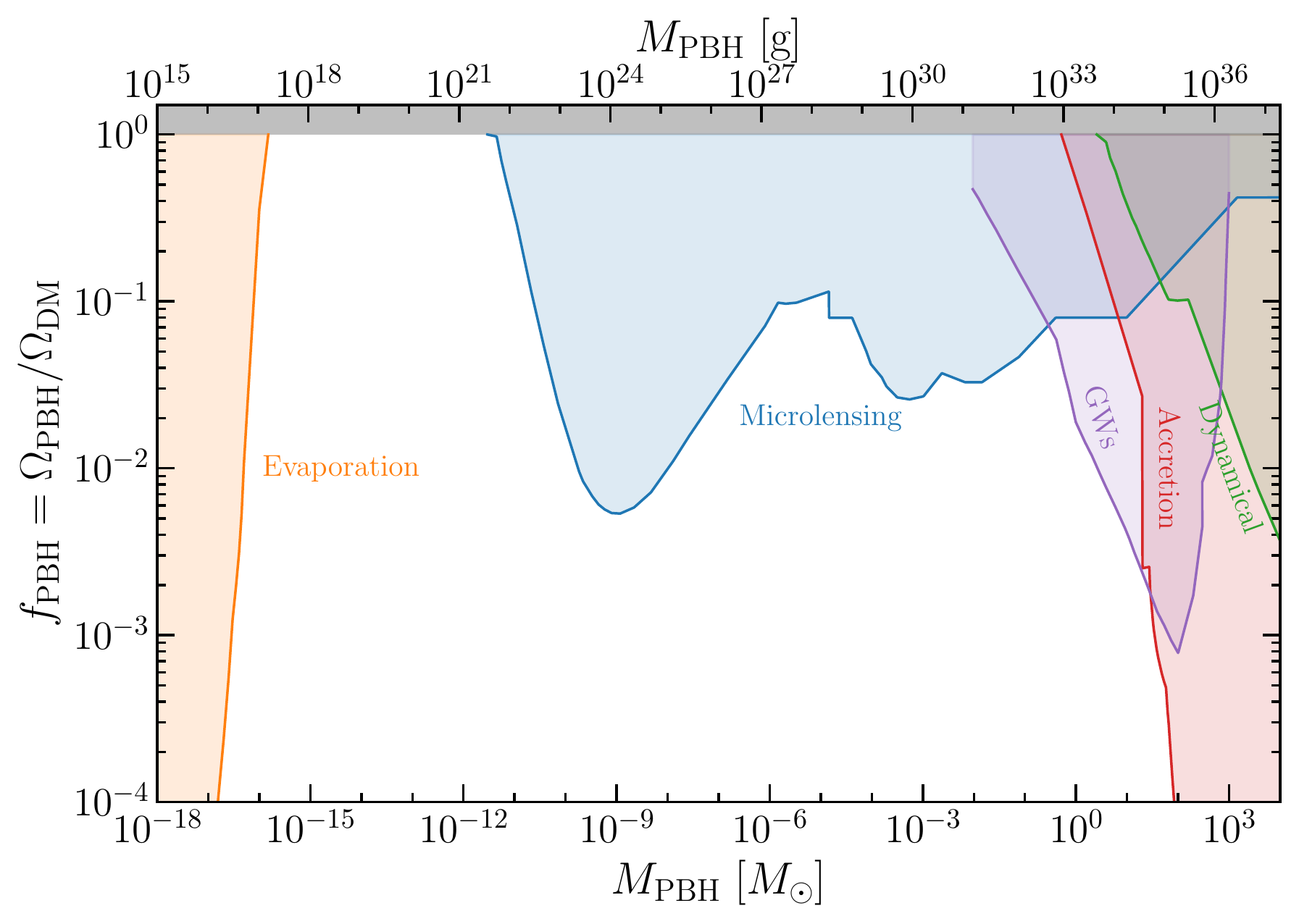}
    \caption[Review of various bounds on \ac{PBHs}]{Review of various bounds on \ac{PBHs} obtained using the open source code \href{https://github.com/bradkav/PBHbounds}{PBH Bounds} -- see \cite{Green2020}}
    \label{fig:PBHbounds_review}
\end{figure}

\begin{figure}[t!]
    \centering
    \includegraphics[width = 0.95\linewidth]{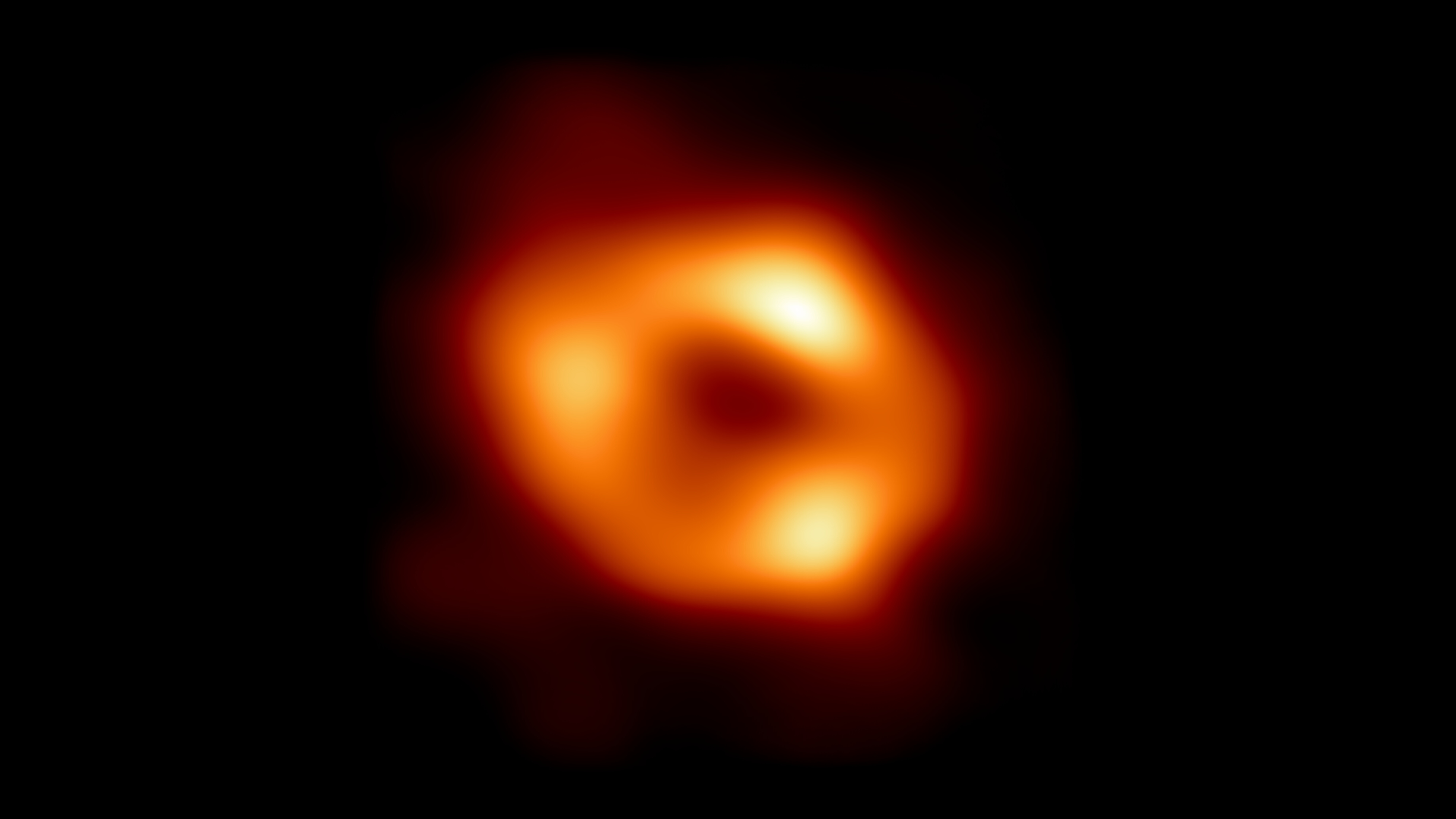}
    \caption[First image of Sagittarius A*]{First image of Sagittarius A*, the supermassive black hole at the centre of our galaxy -- credit to the Event Horizon Telescope collaboration \cite{EventHorizonTelescopeCollaboration2022}.}
    \label{fig:Sagittarius A*}
\end{figure}

\ac{PBHs} were first theorised in the $60$s and $70$s \cite{1967SvA....10..602Z, Hawking1971} and it was soon realised that they could be a \ac{DM} candidate \cite{Hawking1971, CHAPLINE1975}. Interest in \ac{PBHs} has been renewed in the wake of the LIGO-VIRGO detection of the merger of intermediate mass black holes \cite{Abbott2016} which could be primordial rather than astrophysical in origin \cite{Bird2016, Sasaki2016, Clesse2017}. There are numerous constraints on the abundance of \ac{PBHs} from lots of different effects, many of which are shown in Fig.~\ref{fig:PBHbounds_review} -- for a comprehensive review of these see e.g. \cite{Carr2020} or for a shorter pedagogical overview see e.g. \cite{Green2020}. These constraints are expressed in terms the \ac{DM} fraction of \ac{PBHs} $f_{PBH}$ defined intuitively:
\begin{eqnarray}
f_{PBH} \equiv \dfrac{\Omega_{PBH}}{\Omega_{DM}}
\end{eqnarray}
so that if all of \ac{DM} is compromised of \ac{PBHs} then $f_{PBH} = 1$. We can see from Fig.~\ref{fig:PBHbounds_review} that $f_{PBH} = 1$ is ruled out at most masses, but there is a window for all of \ac{DM} to be \ac{PBHs} of around asteroid mass. It is worth pointing out that the constraints in Fig.~\ref{fig:PBHbounds_review} assume a monochromatic mass function, i.e. that all \ac{PBHs} form at the same mass. As we will later see it is unphysical that all \ac{PBHs} would form at exactly the same mass although it is a reasonable approximation so long as the power spectrum of curvature perturbations is not too ``wide". There are other assumptions built into all of the other constraints shown in Fig.~\ref{fig:PBHbounds_review} which might be partially evaded -- even for monochromatic mass functions -- if e.g. \ac{PBHs} are clustered in a particular way when they are formed. Regardless, even if \ac{PBHs} are not all of \ac{DM}, their abundance (be it small or large) can act as an invaluable probe of the inflationary potential outside of the narrow \ac{CMB} window. While very large \ac{PBHs} are ruled out as a \ac{DM} candidate it is possible that they could act as the seed for the supermassive black holes at the centre of galaxies -- see Fig.~\ref{fig:Sagittarius A*} for an image of the black hole at the centre of our galaxy, Sagittarius A*.\\
The current density of \ac{PBHs}, $\Omega_{PBH}$ is related to the mass fraction of \ac{PBHs} of mass $M$, at the time they are formed, $\beta (M) = \rho_{PBH}(M)/\bar{\rho}$, through \cite{Carr1975}:
\begin{eqnarray}
\Omega_{PBH} \sim \dfrac{10^{18}}{\sqrt{\beta \lb\dfrac{M}{10^{15}g}\rb }} \label{eq:Omega_Beta relation}
\end{eqnarray}
Note that this equation is only valid for $M > 10^{15}g$ as \ac{PBHs} lighter than this would have evaporated by the present time. Equation (\ref{eq:Omega_Beta relation}) is very rough as the current density of \ac{PBHs} depends on the entire cosmic evolution since the time when the \ac{PBHs} were formed and any modifications to the standard $\Lambda$CDM evolution would alter the relationship. The constraints on the abundance of \ac{PBHs} of mass, $M_{PBH}$, in the range $(10^{9}-10^{50})$g limit the upper bound of $\beta$, to $10^{-24}$ – $10^{-17}$ for $10^9$g $ \ll M_{PBH} \ll 10^{16}$g and $10^{-11}$ – $10^{-5}$ for $10^{16}$g $ \ll M_{PBH} \ll 10^{50}$g with a transition in constraints for $M_{PBH} \sim 10^{16}$g. While lighter \ac{PBHs} evaporate so quickly they don't even survive to Big Bang Nucleosynthesis and therefore can't play any part in structure formation, constraints can still be imposed on their mass fraction. It has been shown \cite{Papanikolaou2020} that there is an upper bound on $\beta$, in the range of $10^{-4}$ – $10^{-2}$ for $10$g $ < M_{PBH} < 10^9$g. 

\subsection{How to form a Primordial Black Hole}
\begin{figure}[t!]
    \centering
    \includegraphics[width = 0.95\linewidth]{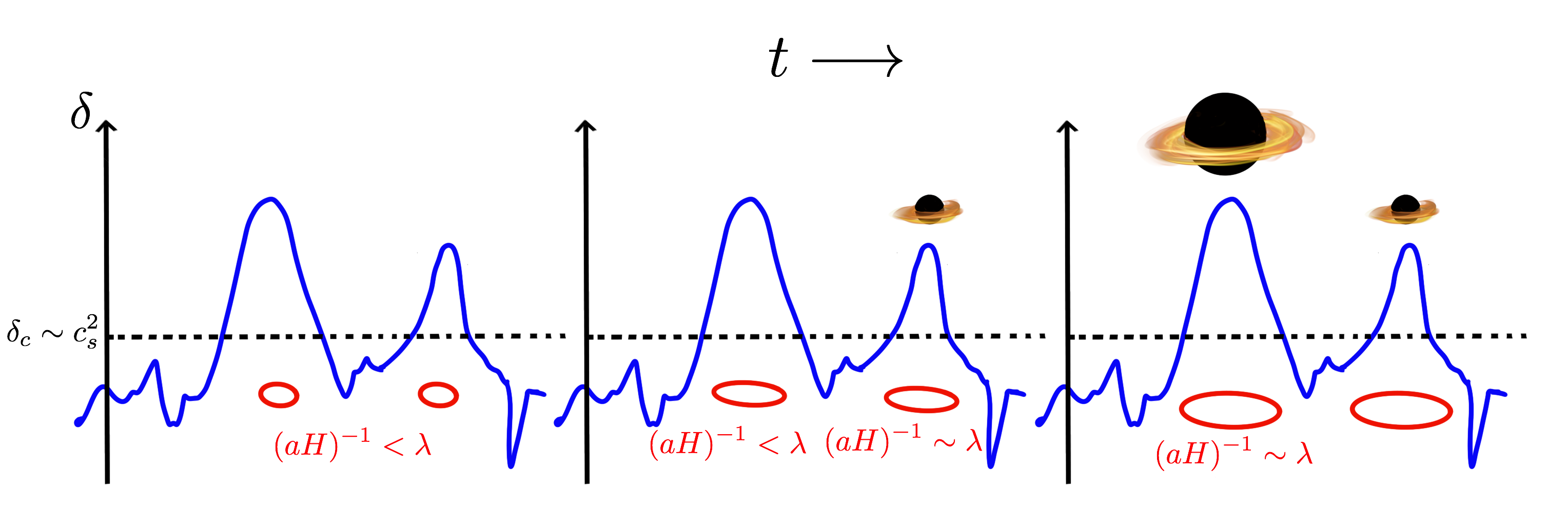}
    \caption[Schematic of the formation of \ac{PBHs}]{Schematic of the formation of \ac{PBHs} from overdensities at three successive moments in time. Taken from \cite{Villanueva-Domingo2021}.}
    \label{fig:PBH_formation}
\end{figure}
Forming a black hole is not an easy feat, requiring an \emph{overdense} region of space such that gravity is stronger than the pressure of the matter involved. Notice that it is not simply enough for matter to be dense, for if it is very dense \emph{everywhere} there is no large gravity gradient in any one direction. Therefore the criterion for the collapse of a black hole should be in terms of \emph{deviations} from background values. An intuitive parameter to consider is the density contrast $\delta$ defined as the density deviation from the background value $\bar{\rho}$:
\begin{eqnarray}
\delta \equiv \dfrac{\delta\rho}{\bar{\rho}} = \dfrac{\rho - \bar{\rho}}{\bar{\rho}} \label{eq:density_contrast_defn}
\end{eqnarray}
On the \ac{CMB} scales $\delta \sim \mathcal{R} \sim 5\times 10^{-5}$ which is way too small to lead to production of \ac{PBHs}. Bernard Carr \cite{Carr1975} in the '70s estimated, using Newtonian gravity, that an overdensity would collapse if the density contrast -- evaluated at horizon crossing -- exceeded the sound speed of perturbations\footnote{One should caveat this by noting that if the universe is in a phase dominated by pressureless matter (such as the Cold \ac{DM} scenario) where $c_{s}^2 = 0$ it is not true that any perturbation will collapse to form a black hole. Instead the main barrier to collapse is deviations from spherical symmetry. In matter domination the mass fraction of \ac{PBHs} can be shown to be \cite{Khlopov1980}:
\begin{eqnarray}
\beta (M) = 0.02\delta_H (M)^5
\end{eqnarray}
which is much larger than the exponentially suppressed fraction we will see for the radiation dominated case.}:
\begin{eqnarray}
\delta \vert_{k=aH} > \delta_c = c_{s}^2
\end{eqnarray}
However \ac{PBHs} can only form when the scale of the overdensity is comparable to the Hubble horizon\footnote{As discussed in the previous chapter, this is because the Hubble horizon indicates regions that are in causal contact within the next Hubble time.}. In Fig.~\ref{fig:PBH_formation} we show a schematic for how this might happen. At the initial time (leftmost panel) the two perturbations large enough to form \ac{PBHs}, $\delta > \delta_c$, have wavelength larger than the Hubble horizon and so cannot collapse yet to form a black hole. At the next time (middle panel) the horizon has grown sufficiently such that the right perturbation has wavelength comparable to the horizon. This region therefore collapses to form a black hole with mass approximately equal to the horizon mass -- more on this later. At the next instant (rightmost panel) the horizon has grown enough for the wavelength of the left perturbation to be comparable in size and so this region will also collapse to form a black hole. Notice that because this collapse happened later, when the horizon was larger and therefore the horizon mass was bigger, the black hole formed will be more massive. 

During radiation domination the sound speed is $c_s = 1/\sqrt{3}$ which suggests that $\delta_c = 1/3$. More accurate \ac{GR} simulations \cite{Yoo2020, Escriva2022} have revised this to be $\delta_c \simeq 0.45$ which is remarkably close to Carr's original estimate. In reality the precise value of $\delta_c$ depends not only on the equation of state at horizon re-entry but also the shape of the perturbation itself \cite{Musco2020}. The density contrast $\delta$ can be written for spherically symmetric peaks -- see e.g. \cite{Musco2019,Young2022} -- in terms of the curvature perturbation $\mathcal{R}$ like so:
\begin{eqnarray}
\delta (r,t) = -\dfrac{4}{9}\lb \dfrac{1}{aH}\rb^2 e^{-2\mathcal{R}(r)}\lb \mathcal{R}''(r) + \dfrac{2}{r}\mathcal{R}'(r) + \dfrac{1}{2}\mathcal{R}'(r)^2\rb \label{eq:delta_R_nonlinear}
\end{eqnarray}
the linear component of which can be simply written in Fourier space as:
\begin{eqnarray}
\delta_l (k) = \dfrac{4}{9}\lb \dfrac{k}{aH}\rb^2 \mathcal{R}(k) \label{eq:delta_R_linear}
\end{eqnarray}
so we can see that at horizon crossing the two are equivalent at linear order up to an order unity factor. 
\subsubsection{Smoothing and window functions}
\begin{figure}[t!]
    \centering
    \includegraphics[width = 0.95\linewidth]{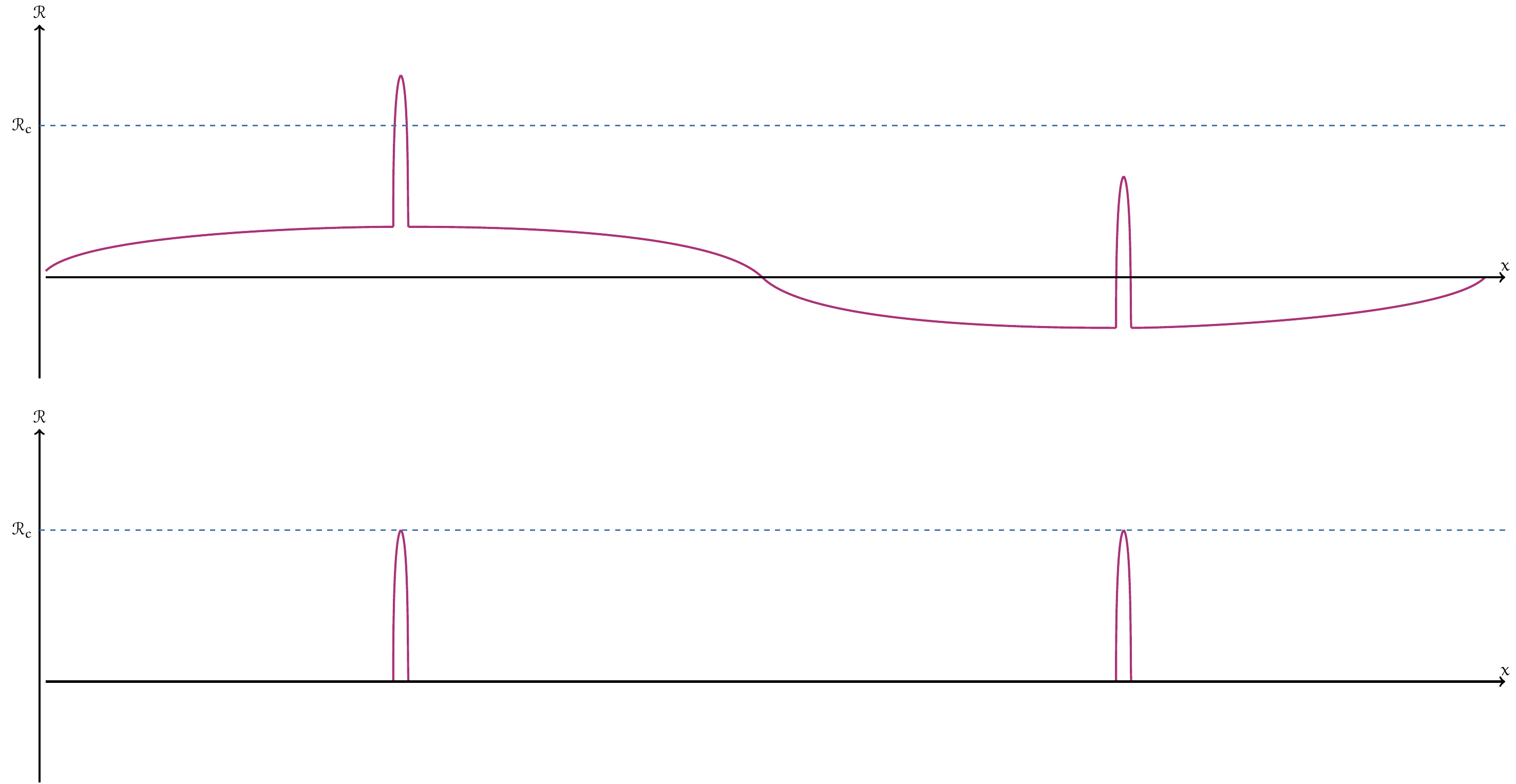}
    \caption[How perturbations across many scales can confuse things]{How large perturbations across large scales can confuse things. Adapted from \cite{Young2020}.}
    \label{fig:zeta_issue}
\end{figure}
In reality things are more complicated than the simple picture we have outlined. Consider for instance that we use the linear order relation (\ref{eq:delta_R_linear}) to determine the abundance of \ac{PBHs} in terms of the curvature perturbation $\mathcal{R}$. In Fig.~\ref{fig:zeta_issue} we show in the top panel a large scale small perturbation with two smaller scale perturbations on top. We can see because of the large scale perturbation that only one of the small scale perturbations would seem to cross the threshold and thus form a black hole. The problem with this is that black holes form due to \emph{local} overdensities and so these small scale overdensities should be compared to the \emph{local} value of $\mathcal{R}$ rather than the global average. In the bottom panel we have subtracted off the large scale perturbation to see that both regions do in fact reach the criteria so both regions should collapse to form \ac{PBHs}. This means that even small, large scale deviations can give incorrect predictions for whether a black hole forms. Therefore the use of the curvature perturbation, $\mathcal{R}$, to compute the mass fraction of \ac{PBHs} is heavily criticised in the literature \cite{Musco2019, Young2019, Germani2020, Young2020, Biagetti2021, DeLuca2022}. It is customary then to \emph{smooth} quantities like the density contrast using a smoothing function $W$ like so:
\begin{eqnarray}
\delta_{sm} = \int_{0}^{\infty}\mathrm{d}r ~ 4\pi r^2 W(r,R)\delta(r)
\end{eqnarray}
where the smoothing function is typically either a real-space top hat, Fourier-space top hat or Gaussian window function. The choice of window function is non-trivial as the abundance of \ac{PBHs} is sensitive to it \cite{Young2020} and care should be taken to not compare quantities computed using one window function with those computed with another. The window function will smooth things on sub-horizon scales, and therefore prevent contamination in the analysis of modes that had previously re-entered the horizon. It is clear that to get the most accurate result one should instead work with e.g. the smoothed density contrast $\delta_{sm}$ which is related to $\mathcal{R}$ in a highly non-linear way -- see (\ref{eq:delta_R_nonlinear}). However, such non-linear effects are expected to only reduce $\beta$ at most by a factor of a few, eg $\sim 2$ according to \cite{Young2019}. This is only true if the power spectrum is very peaked, as it will be for us; more generic shapes of the power spectrum are analysed in \cite{Germani2020}. 
\clearpage
\subsubsection{The compaction function}
With all these issues in mind it transpires that the most appropriate parameter to determine whether a perturbation will collapse to form a black hole is the compaction function $\mathcal{C}$ \cite{Musco2019, Young2014, Young2019} defined as the excess of the Misner-Sharp mass $\delta M$ in the spherical region with radius r:
\begin{subequations}
\begin{align}
\mathcal{C}(\vec{x},r) &\equiv \dfrac{\delta M}{4\pi R} = \dfrac{M_{MS} - M_b}{4\pi R}    \label{eq:compaction_function_defn}\\
M_{MS} &\equiv 4\pi \int_{0}^{R} \rho \tilde{R}^2\mathrm{d}\tilde{R}\label{eq:MS_mass_defn}\\
M_b &\equiv \dfrac{4\pi}{3}\bar{\rho}R^3\label{eq:Mb_defn}\\
R &\equiv a(t)re^{\mathcal{R}(t,\vec{x} + r)}\label{eq:areal_radius_defn}
\end{align}
\end{subequations}
where $M_{MS}$ is the Misner-Sharp mass, $M_b$ is the background mass defined in terms of the background energy density $\bar{\rho}$, $R$ is the \emph{areal} radius defined in terms of the curvature perturbation $\mathcal{R}$ which in turn is defined in the usual way on spatial hypersurfaces:
\begin{eqnarray}
\mathrm{d}s_{(3)}^2 = a^2(t)e^{2\mathcal{R}(\vec{x})}\gamma_{ij}\mathrm{d}x^i\mathrm{d}x^j
\end{eqnarray}
The parameter $\vec{x}$ is essentially an index for different Hubble patches (i.e. varying $\vec{x}$ means moving between the origin of different Hubble patches) and $r$ is the radial distance from the centre of each Hubble patch. The areal radius $R$ therefore measures the radial distance from the centre of each Hubble patch while incorporating the effect of local curvature. It is worth emphasising that while individual components of the compaction function are time dependent, the compaction function itself is not. The formation of \ac{PBHs} corresponds to rare peaks in the curvature perturbation $\mathcal{R}$ which can be assumed to be spherically symmetric \cite{Bardeen1986}. Under this assumption it is possible to express the compaction function in terms of its linear component $\mathcal{C}_l$:
\begin{eqnarray}
\mathcal{C}(\vec{x},r) &=& -\dfrac{4}{3}r\mathcal{R}'(r)\lb 1 + \dfrac{1}{2}r\mathcal{R}'(r)\rb \\
&=& \mathcal{C}_l (\vec{x},r) -\dfrac{3}{8}\mathcal{C}_l (\vec{x},r)^2
\end{eqnarray}
It transpires from this that there is a maximum value\footnote{Note that this is the maximum value $\mathcal{C}$ can take for \emph{any} perturbation, for most perturbations the compaction function will be smaller.} for the compaction $\mathcal{C}_{max} = 2/3$ which corresponds to $\mathcal{C}_l =4/3$. Perturbations are therefore split into Type I and Type II corresponding to perturbations with $\mathcal{C}_l < 4/3$ and $\mathcal{C}_l > 4/3$ respectively. Type I perturbations are the best understood and will form a black hole if above some threshold value $\mathcal{C}_{c}$ with a mass spectrum that observes the scaling law of critical collapse \cite{Musco2019}:
\begin{eqnarray}
M_{PBH} = KM_{H}(\mathcal{C}-\mathcal{C}_{c})^{\gamma}
\end{eqnarray}
where $M_{H}$ is the horizon mass at the time when the horizon scale is equal to the smoothing scale used and $\gamma $ depends on the nature of the background fluid. In general for a radiation dominated fluid $\gamma \simeq 0.36$. The parameters $K$ and $C_{c}$ depend on the exact shape of the profile and the window function used\footnote{While there might not appear to be a window function used in the definition of the compaction function (\ref{eq:compaction_function_defn}), $\mathcal{C}$, it can be understood as the average of the comoving density contrast.}. However in general for a real space top-hat smoothing function $K \simeq 4$, $\mathcal{C}_{c} \simeq 0.55$ and for a Gaussian window function $K \simeq 10$, $C_{c} \simeq 0.25$. Type I perturbations are relatively well understood and are the ones we will focus on in this work. 

Type II perturbations on the other hand are a lot more strange as the areal radius $R$ does not increase monotonically with $r$. It was thought that such perturbations did not actually form \ac{PBHs} -- instead forming separate universes -- however it has been shown \cite{Kopp2011} that such perturbations \emph{always} form \ac{PBHs} but it is not clear at what mass. In any case Type II perturbations are probabilistically suppressed compared to Type I perturbations so while interesting academically are typically ignored when computing the abundance of \ac{PBHs} and we will do so here. In this way focusing only on Type I perturbations will generically \emph{underestimate} the total abundance of \ac{PBHs}. 

An important point is $r_m$ which corresponds to the innermost maximum of the compaction function $\mathcal{C}$. $R(r_m)$ is then understood as the proper ``size" of the overdensity and the compaction function should be evaluated at this point in order to determine whether a black hole will form after its horizon re-entry when $R(r_m)H =1$.

\subsubsection{\label{sec:P-S vs Peaks}Press-Schecter versus Peaks theory}
The concept of the Press-Schecter approach \cite{Press1974} is to say that \ac{PBHs} form in regions where the compaction function at a given point, $r_m$, is above the threshold value. In this way the Press-Schecter approach \emph{computes the volume of the universe above a certain threshold.} The mass fraction, $\beta$ of \ac{PBHs} is then given by integrating the PDF for the compaction function from the threshold value up to $2/3$ as we are only considering Type I perturbations:
\begin{eqnarray}
\beta = \int_{\mathcal{C}_{c}}^{2/3}\mathrm{d}\mathcal{C}~ \dfrac{M(r_m)}{M_{H}}P(\mathcal{C}(r_m)) \label{eq:type1_pert}
\end{eqnarray}
where we note that all relevant quantities should be evaluated at the innermost maximum $r_m$. Here $M_{H}$ is the mass enclosed within a Hubble sphere or simply the Hubble mass. We can write (\ref{eq:type1_pert}) more neatly in terms of the linear part of the compaction function $\mathcal{C}_l$:
\begin{eqnarray}
\beta = \int_{\mathcal{C}_{l,c}}^{4/3}\mathrm{d}\mathcal{C}_1~ K\lsb \mathcal{C}_l(r_m) - \dfrac{3}{8}\mathcal{C}_l(r_m)^2 - \mathcal{C}_{l,c} \rsb^{\gamma} P(\mathcal{C}_l(r_m))
\end{eqnarray}
From our perspective however it is more desirable to work directly with the curvature perturbation $\mathcal{R}$ as this is the object inflation naturally computes for us. If we consider Gaussian fluctuations so that the power spectrum $\Delta_{\mathcal{R}}^2$ fully describes the fluctuations we can consider a standard parameterisation of the curvature profile:
\begin{eqnarray}
\mathcal{R}(r) = \mathcal{R}_0\exp \lsb -\lb \dfrac{r}{r_m}\rb^{2\gamma}\rsb
\end{eqnarray}
where $\mathcal{R}_0$ is the value at the centre of the perturbation. This enables to write the linear compaction function and thus the mass fraction like so:
\begin{eqnarray}
\mathcal{C}_1 (r_m) &=& \dfrac{8}{3}\dfrac{\gamma}{e}\mathcal{R}_0\\
M &=& KM_H \lsb \dfrac{8\gamma}{3e}\lb \mathcal{R}_0 -\dfrac{\gamma}{e}\mathcal{R}_{0}^2 -\mathcal{R}_{c} + \dfrac{\gamma}{e}\mathcal{R}_{c}^2\rb  \rsb^\gamma
\end{eqnarray}
For simplicity we will assume a delta function power spectrum $\Delta_{\mathcal{R}}^2 = \delta (k-k_{*})\sigma_{\mathcal{R}}^2$. Then -- assuming a monochromatic mass function -- the mass fraction $\beta$ simply becomes\footnote{The eagle eyed viewer will note that the upper limit of integration seems to include Type II perturbations as well. Under the monochromatic mass function assumption it is customary in Press-Schecter to then integrate over all perturbations above the threshold as it is known that Type II perturbations will always form \ac{PBHs}.}:
\begin{eqnarray}
\beta = \dfrac{1}{\sqrt{2\pi}\sigma_{\mathcal{R}}}\int_{\mathcal{R}_c}^{\infty}\mathrm{d}\mathcal{R}~\exp (-\dfrac{1}{2}\lb \dfrac{\mathcal{R}}{\sigma_{\mathcal{R}}}\rb^2) = \dfrac{1}{2}\text{erfc}\lb \dfrac{\mathcal{R}_c}{\sqrt{2}\sigma_{\mathcal{R}}}\rb 
\end{eqnarray}
suggesting that the mass fraction is in general exponentially sensitive to the threshold value $\mathcal{R}_c$. \\

Peaks theory is more accurate than Press-Schecter as it adds the condition that \ac{PBHs} will form at locations where the compaction function is at a maximum. In this way the central object of peaks theory is the number density of peaks $n(C)$ which for a delta function power spectrum and assuming Gaussian perturbations is \cite{Bardeen1983}:
\begin{eqnarray}
n(\mathcal{C}) = \dfrac{1}{3^{3/2}(2\pi)^2}k_{p}^6\lb\dfrac{\mathcal{C}}{\sigma_\mathcal{C}}\rb^3\exp \lb -\dfrac{\mathcal{C}^2}{2\sigma_{\mathcal{C}}^2}\rb 
\end{eqnarray}
where $k_p$ is the pivot scale of interest. The mass fraction is then given by:
\begin{eqnarray}
\beta = (2\pi)^{3/2}r^3 \int \mathrm{d}\mathcal{C} \dfrac{M_{PBH}(\mathcal{C})}{M_{H}}n(\mathcal{C})
\end{eqnarray}
where the integration is performed over the Type I perturbations above the appropriate threshold as discussed above. In this way it can be said that peaks theory calculates the number of peaks above the critical value. Perhaps counter-intuitively peaks theory will generically predict the formation of a greater number of \ac{PBHs} than the Press-Schecter approach -- for a demonstration of this see e.g. Appendix A of \cite{Kitajima2021}. For the rest of this thesis however we will neglect many of these complications and in the next subsection we will outline a more straightforward method for obtaining the mass fraction directly from quantities computed during inflation. 

\subsection{\label{sec: PBHs from USR}Seeding Primordial Black Holes from a period of Ultra Slow-Roll inflation}
\begin{sidewaysfigure}
    \includegraphics[width=0.99\linewidth]{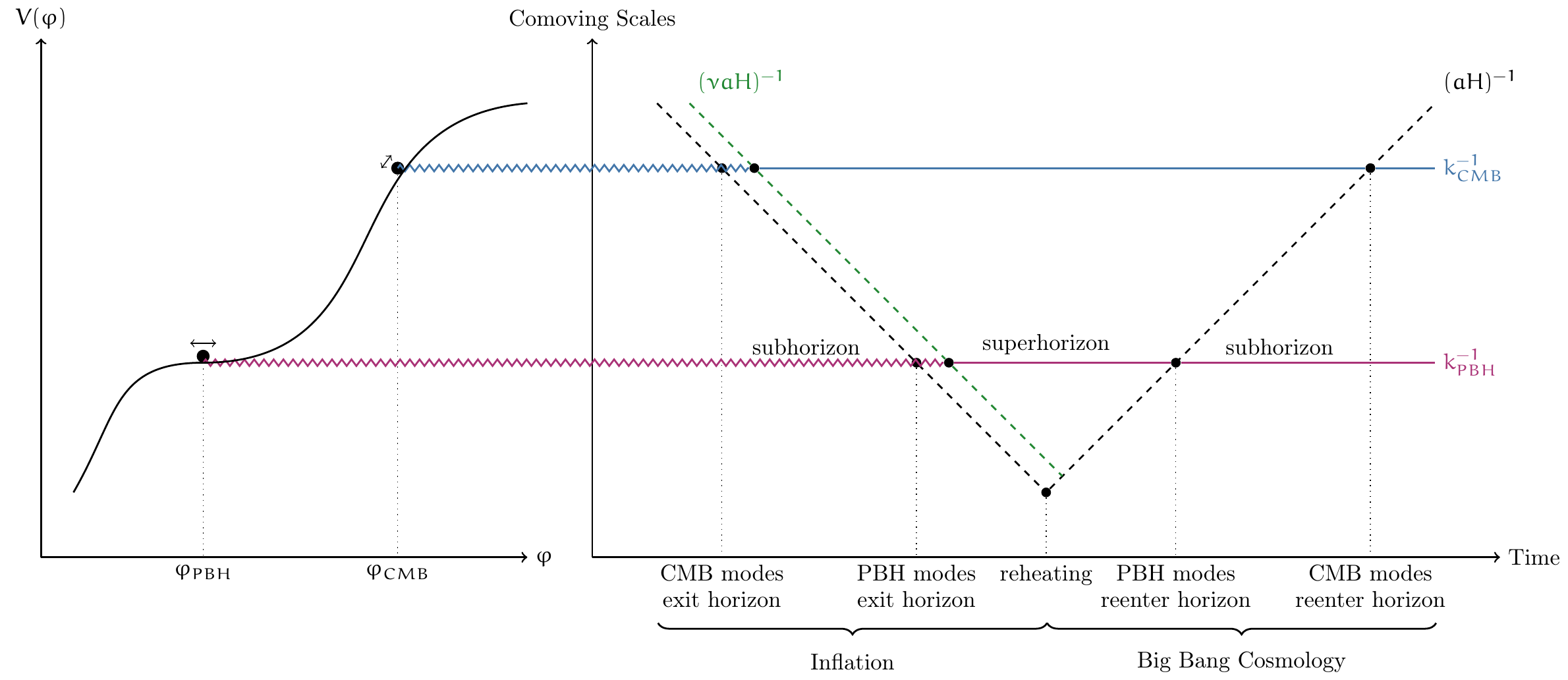}
    \caption[Schematic for inflationary perturbations]{How inflaton perturbations can be related to modes we observe in the \ac{CMB} and the formation of \ac{PBHs}. The initially quantum fluctuations (blue and red lines) exit the Hubble horizon (black dashed line) during inflation. A short time later they exit the coarse-graining scale (green dashed line) at which point they can be treated as \emph{effectively classical}. This is why the the line goes from a wavy to a straight line. After inflation ends the Hubble horizon grows again and the fluctuations will \emph{re-enter} the Hubble horizon. We can see on the left hand side how fluctuations sourced earlier during inflation (higher up the potential) correspond to modes later observed in the \ac{CMB} while modes that exit later during inflation (lower down the potential) correspond to modes that might collapse to form \ac{PBHs}.}
    \label{fig:Inflationary_fluctuations_PBHs}
\end{sidewaysfigure}
While \ac{PBHs} can be formed from bubble collisions \cite{Crawford1982, Hawking1982, Kodama1982}, cosmic strings \cite{Hawking1989, Polnarev1991} or the collapse of domain walls \cite{Rubin2000, Rubin2001} to name a few, we will focus here on curvature perturbations generated from a period of inflation.

We draw the reader's attention to Fig.~\ref{fig:Inflationary_fluctuations_PBHs} where we relate inflaton perturbations to observables. On the left we show the inflaton evolving in a potential and on the right we show the evolution of curvature perturbations throughout cosmic history. The comoving Hubble horizon $(aH)^{-1}$ is depicted by a black dashed line and the coarse-graining scale $(\nu aH)^{-1}$ is shown by a dashed green line. We can see that modes we later observe in the \ac{CMB} correspond to fluctuations generated when the inflaton is relatively high up the potential and exit the horizon approximately 60 e-folds before the end of inflation. These quantum fluctuations are initially depicted with a wavy blue line until they reach the coarse-graining scale $(\nu aH)^{-1}$ outside the horizon. At this stage -- as discussed in section \ref{sec:stoch_infl} -- we can neglect the quantum aspect of the fluctuations and treat their long-wavelength evolution in a classical, stochastic manner. When they eventually re-enter the horizon they will seed the perturbations observed in the \ac{CMB}. The story is very similar for those perturbations that will go on to form \ac{PBHs}. We have shown these in red and because \ac{PBHs} must form well before the \ac{CMB} their perturbations must have exited the horizon during inflation much later than the \ac{CMB} perturbations did. In this way \ac{PBHs} are a natural probe of what happened during inflation much later than $\approx 60$ e-folds before its end that the \ac{CMB} probes. In section 2.2. of \cite{Ozsoy:2023ryl} they outline how to relate the number of e-folds, $\Delta N$, between the \ac{CMB} mode exiting the horizon and the mode that will source the \ac{PBHs} exiting the horizon. We will briefly summarise this computation now to give a sense of when these modes should exit the horizon. The mass of the black hole at the time it is formed can be approximately\footnote{In order to obtain this relation one must assume the horizon mass at matter-radiation equality is given by $M_{H} \sim 2.8 \times 10^{17} M_{\odot}$ in line with \cite{Nakama2017} and that the relativistic degrees of freedom in energy density and entropy are equivalent.} given by:
\begin{eqnarray}
    M_{PBH}(k_{PBH}) \simeq \lb\dfrac{k_{PBH}}{3.2\times 10^5 ~Mpc^{-1}}\rb^{-2}30M_{\odot} \label{eq: PBHmasswith solar}
\end{eqnarray}
where $k_{PBH}$ is the wavenumber of the mode that forms a black hole. In order to express (\ref{eq: PBHmasswith solar}) in terms of e-folds we note that:
\begin{eqnarray}
    \dfrac{k_{PBH}}{k_{CMB}} = \dfrac{\lb aH\rb_{PBH}}{\lb aH\rb_{CMB}} \label{eq:kPBHandCMB}
\end{eqnarray}
and if we assume that the first slow roll parameter $\varepsilon_1$ is slowly varying we can express the scale factor and Hubble expansion rate like so:
\begin{eqnarray}
 \lb aH\rb_{PBH} = \lb aH\rb_{CMB}e^{\Delta N\lb 1-\varepsilon_1\rb}   \label{eq:aH for PBH}
\end{eqnarray}
We can then combine equations (\ref{eq: PBHmasswith solar}), (\ref{eq:kPBHandCMB}) \& (\ref{eq:aH for PBH}) to obtain the mass of \ac{PBHs} in terms of the number of e-folds its mode exited the horizon after the \ac{CMB} modes did\footnote{Assuming $k_{CMB} = 0.002 ~Mpc^{-1}$.}:
\begin{empheq}[box = \fcolorbox{Maroon}{white}]{equation}
M_{PBH}(\Delta N) \sim  7.7\times 10^{17}M_{\odot} e^{-2\Delta N\lb 1- \varepsilon_1 \rb } \label{eq:PBH_efolds}
\end{empheq}
Inverting equation (\ref{eq:PBH_efolds}) yields:
\begin{eqnarray}
\Delta N \sim 20 + \log_{10} \lb \dfrac{M_{PBH}}{M_{\odot}}\rb
\end{eqnarray}
which suggests that the modes sourcing \ac{PBHs} of asteroid mass would have exited the horizon $\approx 35$ e-folds after the \ac{CMB} modes did. \\

If one wants to generate an appreciable number of \ac{PBHs} from single-field inflation then one generally needs to go beyond the \ac{SR} regime into a so-called period of \ac{USR} inflation \cite{Tsamis2004, Kinney2005, Namjoo2013, Martin2013, Dimopoulos2017, Salvio2018, Pattison2018}. A period of \ac{USR} is characterised by a negligible gradient in the potential $V_{,\phi} \sim 0$ or, equivalently, the second \ac{SR} parameter $\varepsilon_2 \sim 6$\footnote{Different conventions will define this slightly differently, for instance $\varepsilon_2 \sim \pm 6$ or $\eta \sim \pm 3$}, this is why in Fig.~\ref{fig:Inflationary_fluctuations_PBHs} we have shown these modes being sourced from a local inflection point in the potential. Going beyond \ac{SR} is in general necessary to get significant non-Gaussianity which will significantly enhance the abundance of \ac{PBHs} as compared to the Gaussian case with the same power spectrum. There have been many works examining the effects of primordial non-Gaussianity on the abundance of \ac{PBHs} -- see e.g \cite{Byrnes2012a} where this is done for Press-Schecter and more recently for peaks theory \cite{Yoo2019, Young2022a}. However, the local non-Gaussianity contained in objects such as $f_{NL}$ is not the whole story. In fact it is now known to be a generic feature of the quantum backreaction in stochastic inflation that there will always be exponential tails in the distribution \cite{Ezquiaga2020}, even for those which appear very Gaussian at their peak. There has been recent effort to incorporate these exponential tails in a rigorous manner within the Press-Schecter formalism \cite{Biagetti2021,DeLuca2022} and peaks theory \cite{Kitajima2021}.

While there has been a lot of work done on generating \ac{PBHs} from a period of \ac{USR} inflation \cite{Germani2017, Pattison2017, Biagetti2018, Ezquiaga2018, Firouzjahi2019, Passaglia2019, Ezquiaga2020, Figueroa2020, Biagetti2021, Figueroa2021a, DeLuca2022}, until very recently most work focused on the large velocity -- e.g. \cite{Firouzjahi2019} -- or negligible velocity/diffusion dominated regime -- e.g \cite{Pattison2017, Ezquiaga2020}. There have been strong efforts to describe both limits in the same framework \cite{Pattison2021}, but there isn't good control over the transition period between the two regimes. Direct numerical simulation of the stochastic equations of motion is usually prohibitively expensive to get accurate values for the mass fraction of \ac{PBHs} -- see \cite{Figueroa2020,Figueroa2021} for a treatment of this problem and \cite{Jackson2022} for a possible workaround using importance sampling.  

Based on our discussions in section \ref{sec:P-S vs Peaks} we will use the Press-Schecter formalism for computing the mass fraction assuming a near monochromatic peak in the power spectrum. This will provide an \emph{underestimate} of the true abundance of \ac{PBHs}. The mass fraction of \ac{PBHs}, $\beta$, can then be computed from the probability distribution function (PDF) of the coarse-grained\footnote{There are some subtleties involved with using the coarse-grained curvature perturbation rather than the standard comoving curvature perturbation. The main issue is that $\mathcal{R}_{cg}$ is actually typically coarse-grained at the scale of the end of inflation hypersurface according to (\ref{eq:Rcg_defn}) whereas the smoothing required for to accurately compute the abundance of \ac{PBHs} is on the scale of the perturbation itself given by (\ref{eq:Rcg defn_PBH}) and the two are not necessarily equivalent. There have been some attempts to relate the coarse-grained comoving curvature perturbation directly to quantities like the density contrast \cite{Tada2022} however these have relied on equating quantities computed using different window functions and as discussed before \cite{Young2020} this eliminates any gain in accuracy this procedure would hope to achieve.} scalar curvature perturbation $\mathcal{R}_{cg}$:
\begin{subequations}
\begin{empheq}[box = \fcolorbox{Maroon}{white}]{align}
\beta (M) &= 2 \int_{\mathcal{R}_{c}}^{\infty} P(\mathcal{R}_{cg})~\mathrm{d}\mathcal{R}_{cg} \label{eq:massfracdef} \\
\mathcal{R}_{cg}(\textbf{x}) &\equiv  (2\pi)^{-3/2} \int_{k > aH_{form}}\mathrm{d}\textbf{k}\mathcal{R}_{\textbf{k}}e^{i\textbf{k}\cdot\textbf{x}} \label{eq:Rcg defn_PBH}
\end{empheq} \label{eq:massfracdef_both}
\end{subequations}
so that the mass fraction $\beta$ represents the area under the curve\footnote{Multiplied by a factor of 2 to account for the under-counting in Press-Schecter theory \cite{Press1974}.} of the PDF above some critical value, $\mathcal{R}_c$. Recall from (\ref{eq:Rcg_defn}) that the coarse-grained curvature perturbation is given by $\mathcal{R}_{cg}(\textbf{x}) = \mathcal{N} - \lan \mathcal{N}\ran$ where each ``point $x$" is a Hubble-sized\footnote{Strictly speaking a coarse-grained sized patch will be larger than a Hubble-sized patch by a factor of $1/\nu^3$.} patch whose field value is represented by one of the trajectories in the random walk of the stochastic inflation equation (\ref{eq:stochastic H-J}). 

\section{\label{sec: plateau}Primordial Black Holes from a plateau in the potential}
\begin{figure}[t!]
\centering 
\includegraphics[width=.95\textwidth]{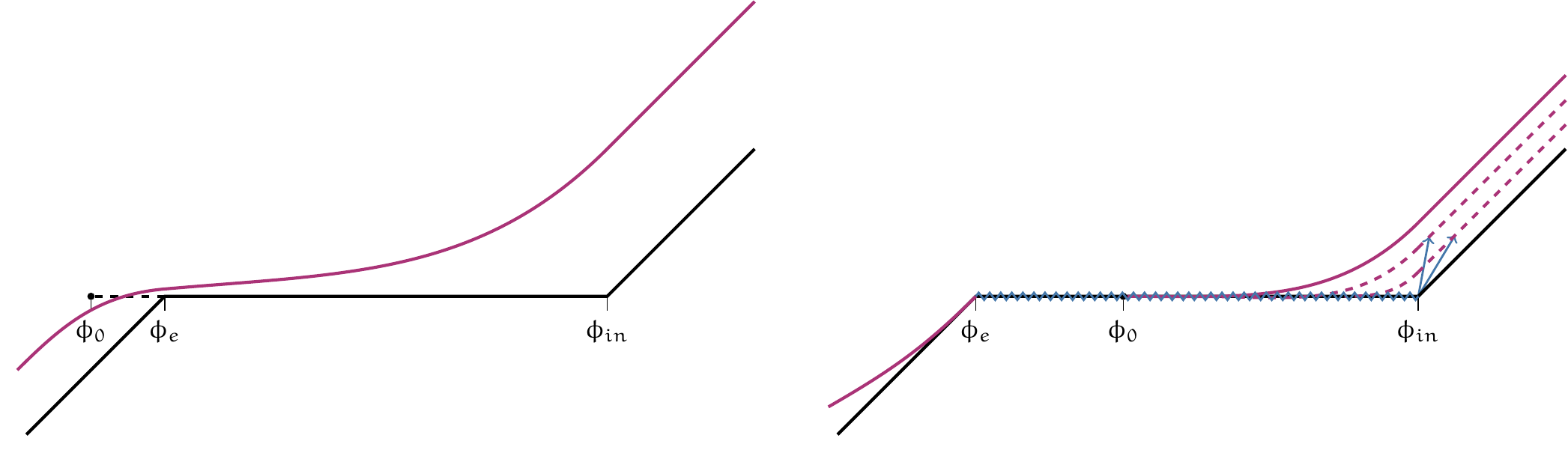}
\caption[Schematic of scenarios A and B on a plateau]{\label{fig:Scenario A and B} Scenario A (left panel) corresponds to a plateau shorter than $\Delta \phi_{cl}$, the \ac{H-J} trajectory is plotted in red. Scenario B (right panel) corresponds to a plateau longer than $\Delta \phi_{cl}$, the free diffusion is plotted in blue and the \ac{H-J} trajectory in red as before. When the diffusion reaches $\phi_{in}$ the field jumps onto a new \ac{H-J} trajectory and will follow it to a new $\phi_0$ before freely diffusing again.}
\end{figure}

If we imagine that the inflaton enters a plateau region (i.e. $V_{,\phi} = 0$) of the potential of width $\Delta \phi_{pl} \equiv \phi_{in} - \phi_{e}$ from the right hand side with some initial (negative) velocity $\Pi_{in}$\footnote{Entering from the left hand side is equivalent up to a few irrelevant sign changes.}, then the \ac{H-J} equation (\ref{eq: H-J equation}) can be solved exactly \cite{Prokopec2021}:
\begin{eqnarray}
H(\phi) &= &
\begin{cases}
H_0~\cosh\left( \sqrt{\dfrac{3}{2}} \dfrac{\phi - \phi_0}{M_{\mathrm p}}\right), & \text{for } \Pi_{in} \neq 0\\[10pt]
H_0 = M_{\mathrm p}^{-1}\sqrt{\dfrac{V_0}{3}}, & \text{for } \Pi_{in} = 0
\end{cases} \label{eq: H exact flat}
\end{eqnarray}
where $\phi_0$ represents the field value the inflaton asymptotes to. In this sense $\Delta \phi_{cl} \equiv \phi_{in} - \phi_{0}$ represents the distance the classical drift will carry the inflaton as it enters a plateau with finite initial velocity. As we are imagining that the field will enter the plateau from some non-negligible gradient in the potential it will enter with some initial velocity i.e. $\Pi_{in} \neq 0$. The total distance that can be travelled due to the classical velocity is then:
\begin{eqnarray}
\Delta \phi_{cl} \equiv \phi_{in} - \phi_{0} = M_{\mathrm p} \, {\mathrm arc}{\sinh} \left(-\dfrac{\Pi_{in}}{\sqrt{2V_0}}\right) = M_{\mathrm p}\,{\mathrm arc}{\sinh} \left(  \sqrt{\dfrac{\varepsilon_{in}}{3-\varepsilon_{in}}} \right) \label{eq: Delta phicl definition}
\end{eqnarray}
where $\varepsilon_{in}$ the first Hubble \ac{SR} parameter -- defined in equation (\ref{eq: epsilon1 defn}) -- as the field enters the plateau. As shown explicitly later in (\ref{eq:class e-folds}) it actually takes an infinite amount of time to reach $\phi_0$ in the absence of stochastic noise. We see that the range over which the field can slide on the plateau is solely determined by the \ac{SR} parameter associated with the injection velocity. Further assuming \ac{SR} to hold prior to entering the plateau, $\varepsilon_{in} \ll 1$, we have
\begin{eqnarray}
\Delta \phi_{cl} \simeq  M_{\mathrm p} \sqrt{\dfrac{\varepsilon_{in}}{3}}
\end{eqnarray} 
This naturally gives rise to two scenarios represented in Fig. \ref{fig:Scenario A and B}. Scenario A (left panel) corresponds to a plateau width $\Delta \phi_{pl} \leq \Delta \phi_{cl}$ i.e. the field enters the region with sufficient velocity to carry it all the way through. The inflaton therefore stays on the \ac{H-J} trajectory for all times. Scenario B (right panel) corresponds to $\Delta \phi_{pl} > \Delta \phi_{cl}$, meaning the inflaton cannot be carried all the way through by classical drift. Therefore, once the inflaton has crossed $\phi_0$ by a stochastic kick it undergoes free diffusion. If the field arrives at the exit point to the plateau, $\phi_{e}$, then the gradient of the potential will start to dominate the evolution and it will have joined a new \ac{H-J} trajectory. If however the field reaches $\phi_{in}$, i.e. the edge of the plateau where it originally entered from, then its evolution is more complicated. The field will jump onto a new \ac{H-J} curve; the momentum constraint will not be violated when neighbouring spatial points also lie on this new \ac{H-J} curve and the whole region is surrounded by a zero deterministic velocity boundary. The field will then re-enter the plateau with a different initial velocity, arriving at a new $\phi_0$ before freely diffusing. Scenario B is therefore a highly complicated system to describe. Fortunately -- as we will demonstrate -- realising scenario B is in general forbidden as it leads to an overproduction of \ac{PBHs}.
\clearpage
If the classical velocity of the field $\Pi \neq 0$ then it follows the \ac{H-J} evolution described by (\ref{eq: H-J equation}). Incorporating the short-wavelength quantum fluctuations results in the addition of a stochastic noise term to (\ref{eq: dphidt HJ}):
\begin{eqnarray}
\dfrac{\mathrm{d}\phi}{\mathrm{d}\alpha} &=& -2 M^2_{\mathrm p}\,\dfrac{\partial \text{ln} H (\phi,\phi_{0})}{\partial \phi} + \dfrac{H(\phi,\phi_{0})}{2\pi}\xi (\alpha) \label{eq:stochastic H-J} \\
\left\langle \xi (\alpha)\xi (\alpha ')\right\rangle &=& \delta (\alpha - \alpha ') \label{eq: xi defn HJ}
\end{eqnarray}
Where $H (\phi,\phi_{0})$ represents a particular solution to the \ac{H-J} equation (\ref{eq: H-J equation}). If we are in a region of the potential where $V_{,\phi} \neq 0$ or $V_{,\phi} = 0$ but $\phi$ has not yet reached $\phi_{0}$ then $\phi$ still lies on the \ac{H-J} trajectory and equations (\ref{eq:stochastic H-J}) and (\ref{eq: xi defn HJ}) are the appropriate dynamical equations to use. \\

If the classical velocity $\Pi = 0$ then the field must be on a plateau portion of the potential and have either reached $\phi_{0}$ from a previous \ac{H-J} trajectory or have started in the region with $\Pi = 0$. The problem is therefore equivalent to pure de Sitter with $H = H_{0} = \sqrt{V_{0}/3}$. The field evolution is then simply given by:
\begin{eqnarray}
\dfrac{\mathrm{d}\phi}{\mathrm{d}\alpha} &=& \dfrac{H_{0}}{2\pi}\xi (\alpha) \label{eq: stochastic deSitter}\\
\left\langle \xi (\alpha)\xi (\alpha ')\right\rangle &=& \delta (\alpha - \alpha ') \label{eq: xi defn deSitter}
\end{eqnarray}
This means that for a plateau of width $\Delta \phi_{pl} > \Delta \phi_{cl}$ that once $\phi = \phi_{0}$ is reached, the inflaton is injected into the $\Pi= 0$ de Sitter trajectory and freely diffuses along the plateau. What happens at the boundaries is what we cover next.

We can insert the $\Pi_{in} \neq 0$ solution of (\ref{eq: H exact flat}) into the classical equation of motion (\ref{eq: dphidt HJ}) to find the classical number of e-folds, $\Delta \alpha_{cl}$, it takes to reach $\phi$ having started at $\phi_{in}$ for Scenario A:
\begin{eqnarray}
\Delta \alpha_{cl} = -\dfrac{1}{3}\text{ln}\left\lbrace \dfrac{\sinh\left[\sqrt{\frac{3}{2}}\dfrac{\phi-\phi_0}{M_{\mathrm p}} \right]}{\sinh\left[\sqrt{\frac{3}{2}}\dfrac{\phi_{in}-\phi_0}{M_{\mathrm p}} \right]}\right\rbrace \label{eq:class e-folds}
\end{eqnarray}
Note however that classically it takes an infinite number of e-folds to reach $\phi_0$ and thus for the classical field velocity $\Pi$ to reach zero. Therefore the $\Pi_{in} = 0$ and  $\Pi_{in} \neq 0$ solutions are completely distinct and there is no way to go between them classically.\\
We can describe this problem in terms of two dimensionless parameters, $\Omega$ and $\mu$. $\Omega$ is defined in terms of the classical drift distance, $\Delta \phi_{cl}$ - equivalently the \ac{SR} parameter $\varepsilon_{1}$ of the prior \ac{SR} region - and the dimensionless plateau height $v_0 = V_0/24\pi^2M_{\mathrm{p}}^{4}$:
\begin{eqnarray}
\Omega \equiv  \sqrt{\dfrac{3}{2v_0}}\dfrac{\Delta \phi_{cl}}{M_{\mathrm{p}}} 
\simeq\sqrt{\dfrac{\varepsilon_{in}}{2v_0}} \label{eq: chi_in defn}
\end{eqnarray}
$\mu$ parameterises how wide the plateau is relative to the classical drift distance $\Delta \phi_{cl}$:
\begin{eqnarray}
\mu \equiv \dfrac{\Delta \phi_{cl} - \Delta \phi_{pl} }{\Delta \phi_{cl}} \label{eq:sigma defn}
\end{eqnarray}
Notice that $\mu = 0$ corresponds to $\Delta \phi_{cl} = \Delta \phi_{pl}$ and that the limit $\mu \rightarrow 1$ corresponds to $\Delta \phi_{pl} \rightarrow 0$. Scenario A corresponds to $0 \leq \mu < 1$ and Scenario B to $ \mu <0$. Using these parameters we can straightforwardly compute the semi-classical observables on the plateau\footnote{These parameters are evaluated at the exit point $\phi_e$. It has been argued \cite{Kinney2005} that parameters like the power spectrum during \ac{USR} should be computed at the end of the \ac{USR} phase as opposed to horizon crossing.} as given by equations (\ref{eq: classical average e-fold}) - (\ref{eq: classicality criterion}):
\begin{eqnarray}
\varepsilon_1 &\approx & 3\tilde{H}_{0}^2\Omega^2\mu^2 \approx \dfrac{3}{2}\varepsilon_{in}^2 \mu^2\label{eq:SR1_flat} \\
\varepsilon_2 &\approx & 6\lb 1 - \tilde{H}_{0}^2\Omega^2\mu^2\rb \approx 3 \lb 2 - \varepsilon_{in}^2 \mu^2\rb \label{eq:SR2_flat} \\
\left\langle \mathcal{N}\right\rangle\vert_{cl} &\approx & -\dfrac{1}{3}\ln (\mu) \label{eq: classical average e-fold_flat} \\
\delta \mathcal{N}^2\vert_{cl} &\approx & \dfrac{1}{18}\dfrac{1-\mu^2}{\Omega^2\mu^2} \label{eq: classical var e-fold_flat} \\
\mathcal{P}_{\mathcal{R}}\vert_{cl} &\approx & \dfrac{1}{3\Omega^2\mu^2}\label{eq: classical power spectrum_flat} \\
f_{\scalebox{0.5}{$\mathrm{NL}$}}\vert_{cl} &\approx & -\dfrac{5}{2}\lb 1 - 2\tilde{H}_{0}^2\Omega^2\mu^2\rb \approx -\dfrac{5}{2}\lb 1 - \varepsilon_{in}^2 \mu^2 \rb \label{eq:classical fNL_flat}\\
n_{\phi} \vert_{cl} - 1&\approx & 6\lb 1 - 2\tilde{H}_{0}^2\Omega^2\mu^2 \rb \approx 6\lb 1 - \varepsilon_{in}^2 \mu^2 \rb\label{eq:classical ns_flat} \\
\eta_{cl} &\approx & \dfrac{\tilde{H}_{0}^2}{2}\left\vert 8 - \dfrac{1}{\tilde{H}_{0}^2\Omega^2\mu^2} \right\vert \approx \dfrac{1}{2\Omega^2\mu^2}\label{eq: classicality criterion_flat}
\end{eqnarray}

\subsection{The case \texorpdfstring{$\Delta \phi_{pl} \leq \Delta \phi_{cl}$ ($0 \leq \mu <1$)}{}}
We start by examining how many \ac{PBHs} are produced in Scenario A where the inflaton's classical velocity when entering the plateau is enough to carry it all the way through, i.e. $\Delta \phi_{pl} \leq \Delta \phi_{cl}$. In \cite{Prokopec2021} it was shown how (\ref{eq:stochastic H-J}) can be reformulated in terms of a \ac{F-P} equation and thus using heat kernel techniques the PDF of e-fold time spent in the plateau $\rho(\mathcal{N})$ can be obtained. Assuming that the inflaton enters the plateau from a previous \ac{SR} phase at the same e-fold number $N_{in}$ in every stochastic realisation of its trajectory -- see Appendix \ref{sec:SR_HJ_deriv} where we drop this assumption -- then the PDF $\rho (\mathcal{N})$ for time taken to reach $\phi_{e}$ can be expressed in terms of the difference $\Delta \mathcal{N} \equiv \mathcal{N} - N_{in}$ like so \cite{Prokopec2021}\footnote{See also Appendix \ref{sec:H-J_phase deriv}.}:  
\begin{eqnarray}
\rho (\mathcal{N}) &=& -\dfrac{3}{\sqrt{\pi}}\exp \lsb -(n+1)\bar{U}^2\rsb \lsb n\sqrt{n+1}\Omega\mu -\sqrt{n}(n+1)\Omega \rsb \nonumber \\
&& +\dfrac{3}{\sqrt{\pi}}\exp \lsb -n\bar{V}^2\rsb \lsb \sqrt{n}(n+1)\Omega-n\sqrt{n+1}\lb 2-e^{-6\Delta \mathcal{N}}\rb\Omega\mu \rsb e^Y \nonumber \\
&& -3\Omega\mu\lsb \Omega\mu e^{-6\Delta\mathcal{N}}-n(2n+3)\Omega e^{-3\Delta\mathcal{N}}\rsb e^Y \text{erfc}\lsb \sqrt{n}\bar{V}\rsb \label{eq:rhoN HJ} \\
n &\equiv &  \dfrac{1}{e^{6\Delta \mathcal{N}} - 1} \\
Y &\equiv & -\dfrac{2n+1}{n+1}\Omega^2\mu^{2} -2n\Omega\mu \Omega e^{-3\Delta \mathcal{N}}  \label{eq: Y defn} \\
\bar{U} &\equiv & \Omega\mu - \Omega e^{-3\Delta \mathcal{N}} \label{eq: U defn}\\
\bar{V} &\equiv & \Omega - \Omega\mu e^{-3\Delta \mathcal{N}} \label{eq: V defn}
\end{eqnarray}
where erfc$(x)$ is the standard complementary error function. As discussed in Appendix \ref{sec:H-J_phase deriv}, the PDF (\ref{eq:rhoN HJ}) is not normalised to 1 for values of $ 0<\Omega\mu < 3$. However even in this interval the deviation is small -- see equation (\ref{eq:chisigma_norm}) and Fig.~\ref{fig:chisig_norm} -- so we do not write out this small correction explicitly in this section but it is accounted for in all graphs. \\

\begin{figure}[t!]
    \centering
    \includegraphics[width = 0.45\linewidth]{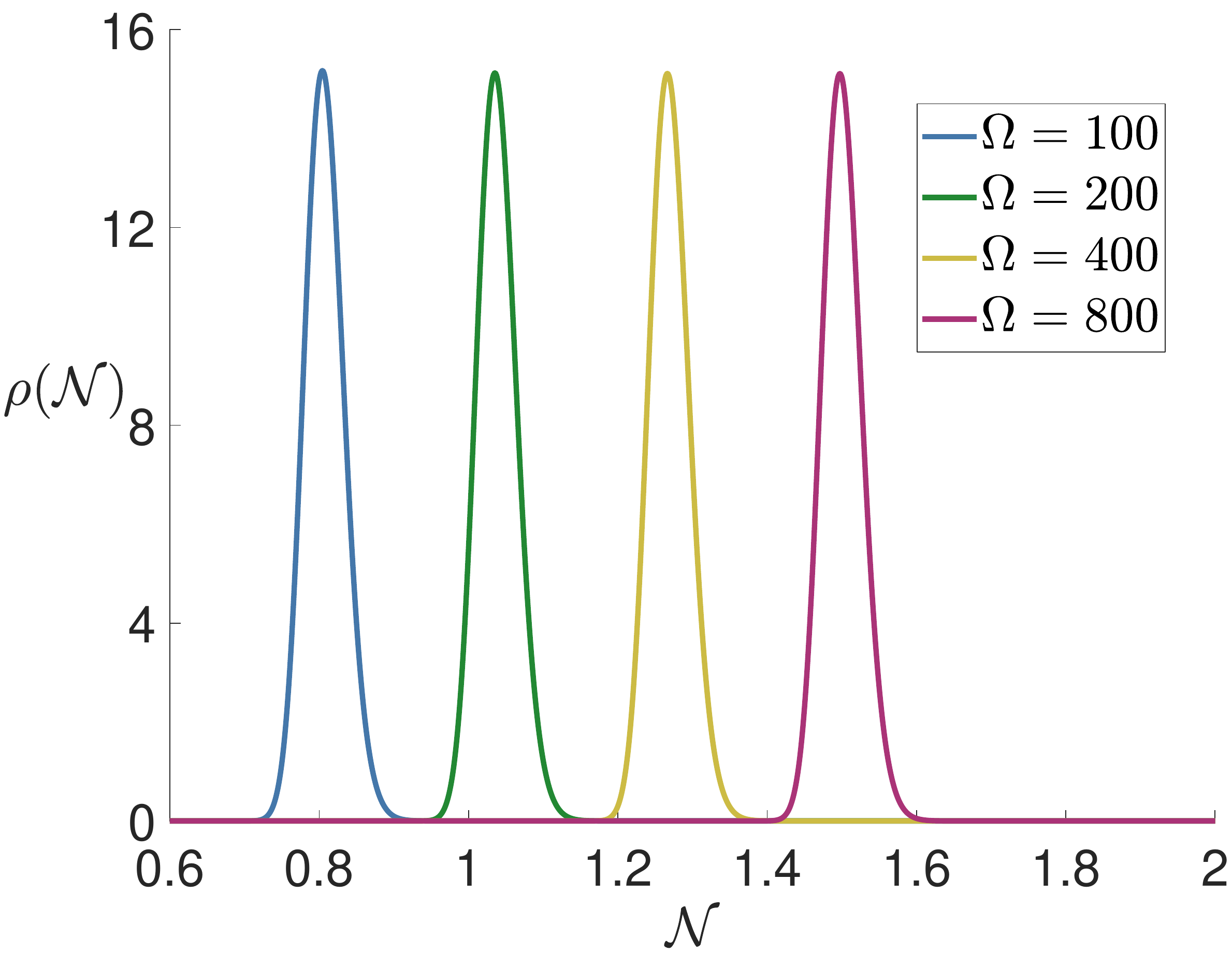}
    \includegraphics[width = 0.45\linewidth]{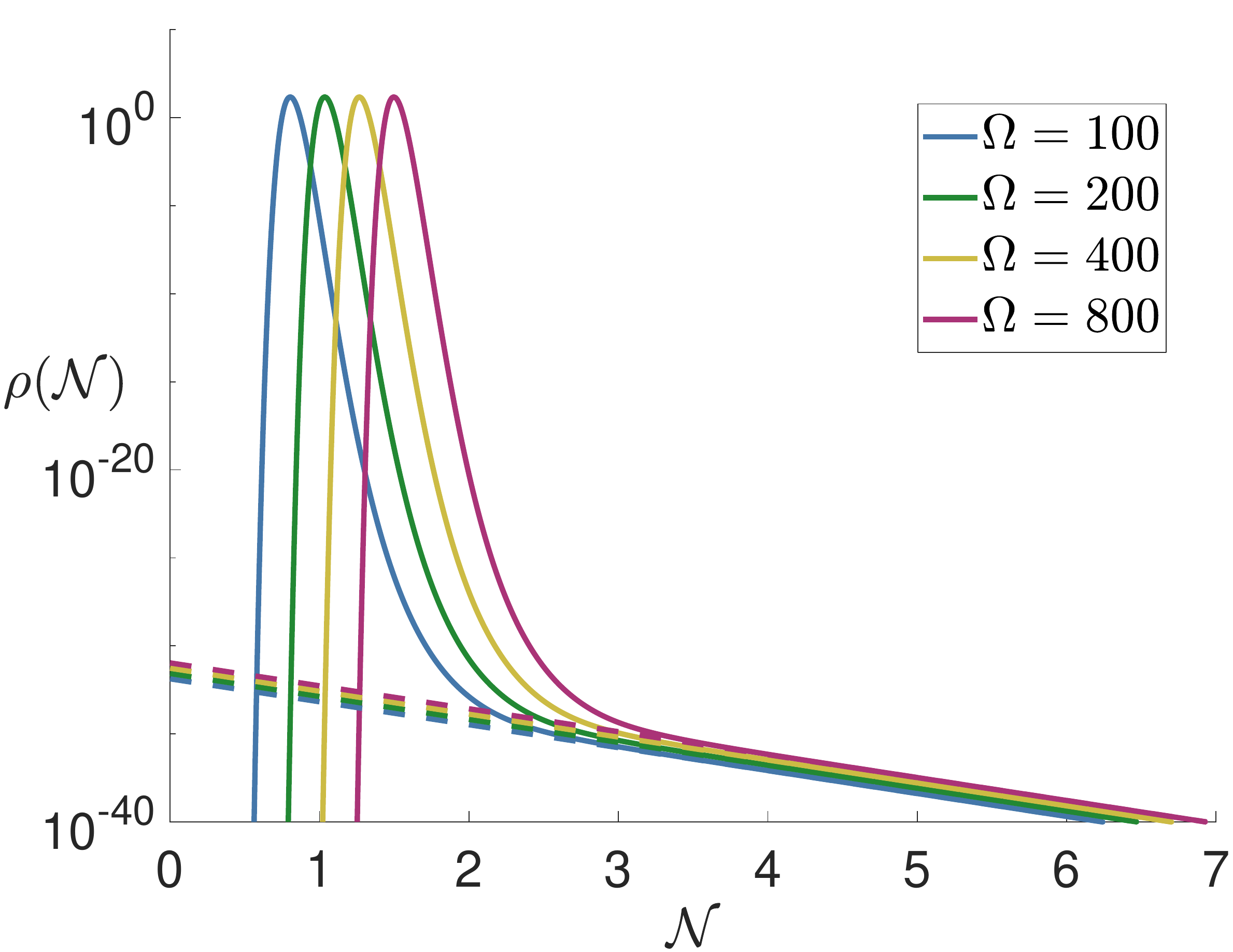}
    \includegraphics[width = 0.45\linewidth]{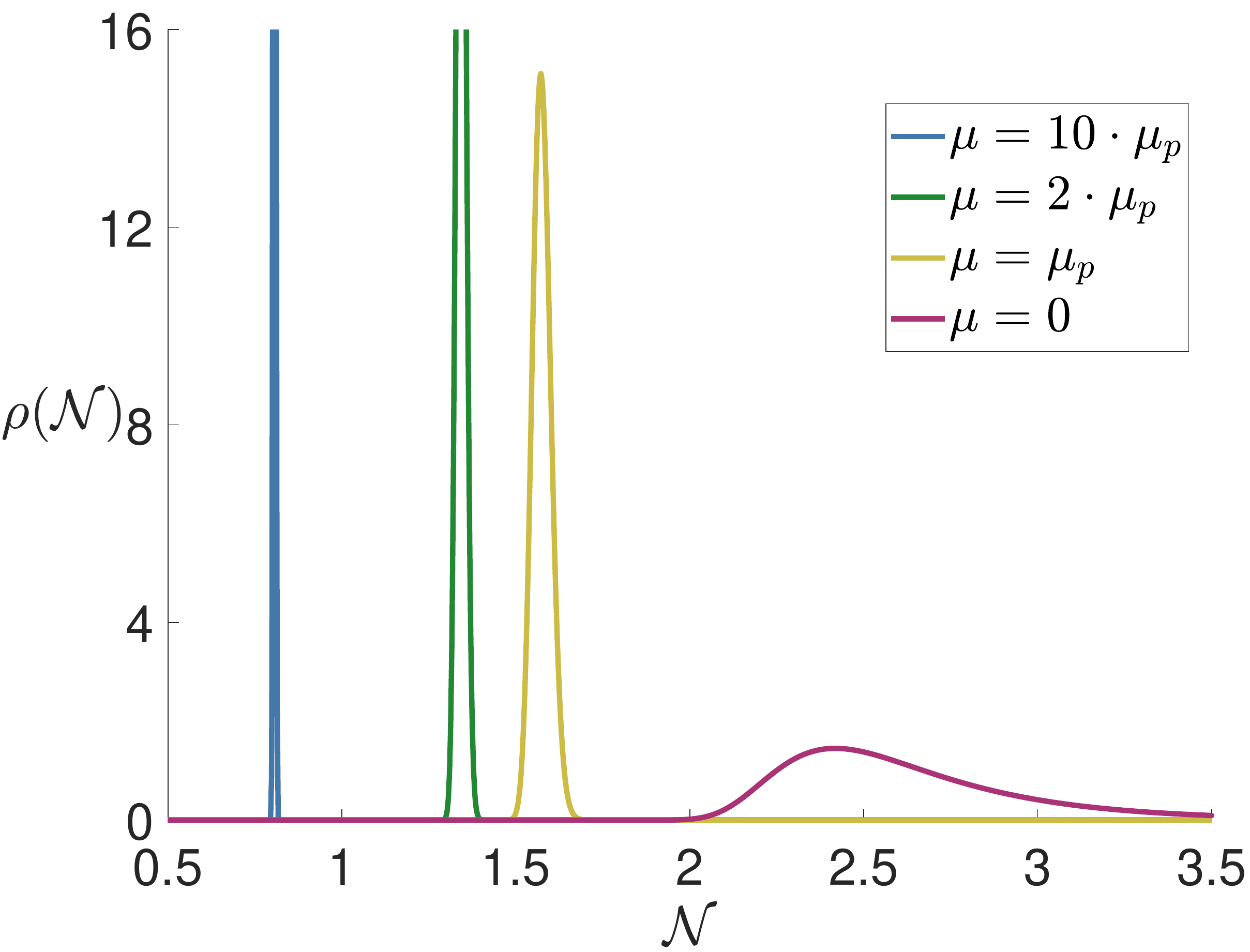}
    \includegraphics[width = 0.45\linewidth]{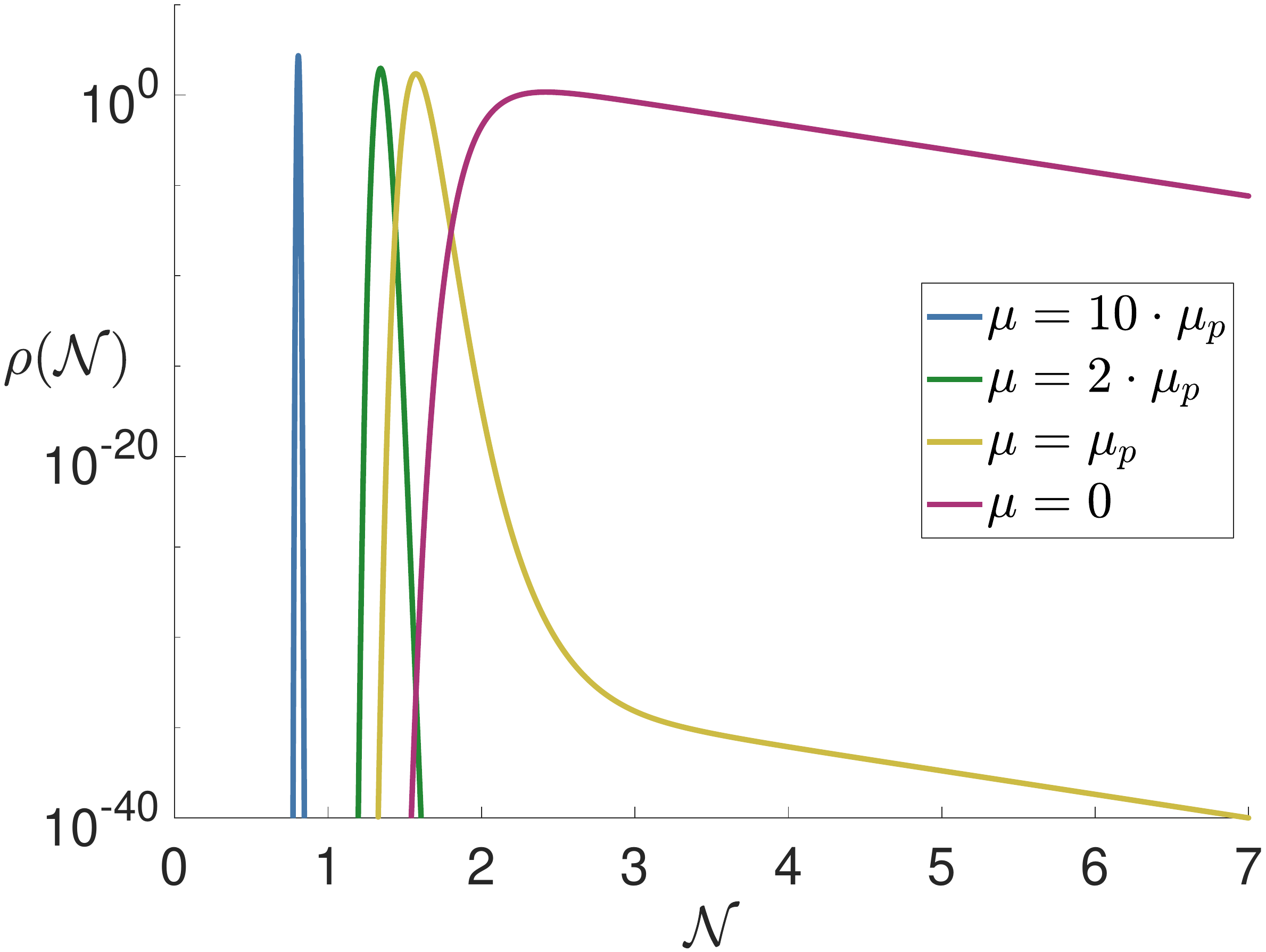}
    \caption[PDF for exit time on a plateau while varying $\Omega$ and $\mu$.]{The top row shows $\rho (\mathcal{N})$ with linear (left) and log (right) scale on a plateau for $\mu = \mu_p$ (\ref{eq:sigma pivot}). This means that the quantity $\Omega\mu$ is constant for each curve. The dashed lines correspond to the asymptotic tail given by (\ref{eq: rhoN HJ LARGE N}). The bottom row is the same but for $\Omega = 1000$.}
    \label{fig:rhoN_USR_chi_and_sigma}
\end{figure}

To aid understanding of how (\ref{eq:rhoN HJ}) behaves as we vary $\Omega$ and $\mu$ we first plot $\rho (\mathcal{N})$ in the top row of Fig.~\ref{fig:rhoN_USR_chi_and_sigma} for four different values of $\Omega$, choosing $\mu$ in each case such that $\Omega\mu$ is constant. Recall that increasing $\Omega$ can be viewed as decreasing the inflationary scale -- and hence both the noise and the friction -- or increasing the velocity of the inflaton as it enters the plateau\footnote{Up to a limit imposed by $\varepsilon_{in} < 1$ required for inflation to be taking place.}. Looking at the PDF in linear scale (left top plot) we can clearly see how varying $\Omega$ does not affect the shape of the PDF but merely translates it such that the average e-fold time $\lan \mathcal{N}\ran $ varies. While in linear scale these PDFs look very Gaussian we can see by examining the right top plot that in fact they have a highly non-Gaussian exponential tail -- shown by dashed lines. These exponential tails are a known feature for stochastic dynamics on plateaus \cite{Pattison2019, Pattison2021} and it is the presence of this tail that will ultimately greatly enhance the abundance of \ac{PBHs} beyond a naive estimate. In fact the value of $\Omega\mu$ chosen in the top row of Fig.~\ref{fig:rhoN_USR_chi_and_sigma} is such that the abundance of \ac{PBHs} starts to violate constraints -- see discussion around (\ref{eq:sigma pivot}). \\

To examine more explicitly the dependence on the value of $\mu$ we plot $\rho (\mathcal{N})$ for four different values of $\mu$ for a value of $\Omega = 1000$ in the bottom row of Fig.~\ref{fig:rhoN_USR_chi_and_sigma}. We have plotted $\mu$ in terms of the pivot scale $\mu_p$ -- given by (\ref{eq:sigma pivot}) -- which corresponds to the transition between under- and over-production of \ac{PBHs}. We find that the shape of the PDF is highly sensitive to this value -- even for a value of $\mu$ that is only twice as large we see how the tail disappears\footnote{It is worth emphasising that for these stochastic processes there will always be an exponential tail in the distribution. However if the parameters aren't chosen carefully the tail will be so small the distribution is essentially indistinguishable from a Gaussian.} and the distribution is highly Gaussian. On the other hand as $\mu \rightarrow 0$, corresponding to $\Delta \phi_{pl} = \Delta \phi_{cl}$, we can see the tail becomes so large that even without going to log scale it is visible. We will deal with this case fully in section \ref{sec:sigma_equal_zero}.\\

In spite of its rather inelegant form it transpires that the PDF (\ref{eq:rhoN HJ}) can be integrated exactly and so using the mass fraction definition (\ref{eq:massfracdef}) and the coarse-grained curvature perturbation relation (\ref{eq:Rcg_defn}), $\mathcal{R}_{cg} = \mathcal{N} - \lan \mathcal{N}\ran$, we obtain the mass fraction as:
\begin{empheq}[box = \fcolorbox{Maroon}{white}]{equation}
\beta = \lsb \text{erfc}(\Omega\mu)-e^{-\Omega^2\mu^2}-\text{erfc}(\bar{U}_c) + e^{Y_{c}}\text{erfc}(\bar{V}_c)\rsb\underbrace{e^{\frac{9}{4}\Omega^{-2}}}_{SR} \label{eq:massfrac_exactFULL}
\end{empheq}
where the quantities $Y_c$, $\bar{U}_c$ \& $\bar{V}_c$ correspond to the parameters defined in equations (\ref{eq: Y defn}), (\ref{eq: U defn}) \& (\ref{eq: V defn}) evaluated at $n_c$ which in turn is given by:
\begin{eqnarray}
n_{c} \equiv \dfrac{1}{e^{6(\mathcal{R}_c + \left\langle \mathcal{N}\right\rangle - N_{in})} -1} \label{eq:nc defn} 
\end{eqnarray}
We have also included the contribution from a previous \ac{SR} phase -- see Appendix \ref{sec:SR_HJ_deriv} -- underlined as SR in the above formula. This \ac{SR} contribution is only important for $\Omega \ll 1$ where it very quickly forces $\beta$ to unrealistically large values. It is worth noting that $\Omega \ll 1$ would correspond to either super-Planckian inflationary scales or (using the \ac{CMB} bounds) $\varepsilon_{in} \ll v_0 \leq 10^{-10}$.\\

To better analyse the behaviour of the mass fraction let us first consider the large $\Delta \mathcal{N}$ limit i.e. deep in the tail that we observe in Fig.~\ref{fig:rhoN_USR_chi_and_sigma}. Then the PDF (\ref{eq:rhoN HJ}) simplifies to:
\begin{eqnarray}
\rho (\mathcal{N}) \simeq \dfrac{6}{\sqrt{\pi}}\Omega ~e^{-\Omega^{2}\mu^2}e^{-3\Delta \mathcal{N}} \label{eq: rhoN HJ LARGE N}
\end{eqnarray}
Using the mass fraction definition (\ref{eq:massfracdef}) under the large $\Delta \mathcal{N}$ limit (\ref{eq: rhoN HJ LARGE N}) we obtain:
\begin{eqnarray}
\beta (M) \simeq \dfrac{4}{\sqrt{\pi}}\Omega ~e^{-\Omega^{2}\mu^2}~e^{-3(\mathcal{R}_c + \left\langle \mathcal{N}\right\rangle - N_{in})}\underbrace{e^{\frac{9}{4}\Omega^{-2}}}_{SR} \label{eq: Beta chie > 0}
\end{eqnarray} 
where again we have included the contribution from a previous \ac{SR} phase. For a Gaussian PDF the mass fraction depends on on the combination $\mathcal{R}_c /\delta \mathcal{R}$ where $\delta \mathcal{R}$ is the variance of the perturbations. However the exponential tail of (\ref{eq: rhoN HJ LARGE N}) means that the mass fraction does not depend on this simple combination. To see how the mass fraction does depend on the variance we note that the classical power spectrum -- given in (\ref{eq: classical power spectrum}) -- in this case is $\mathcal{P}_{\mathcal{R}}\vert_{cl} \sim 1/ 3\Omega^2\mu^2$ suggesting that the variance of perturbations is given by $\delta \mathcal{R} \sim 1/\Omega^2\mu^2$ while the evolution is classically dominated. 
\begin{figure}[t!]
\centering 
\includegraphics[width=.75\textwidth]{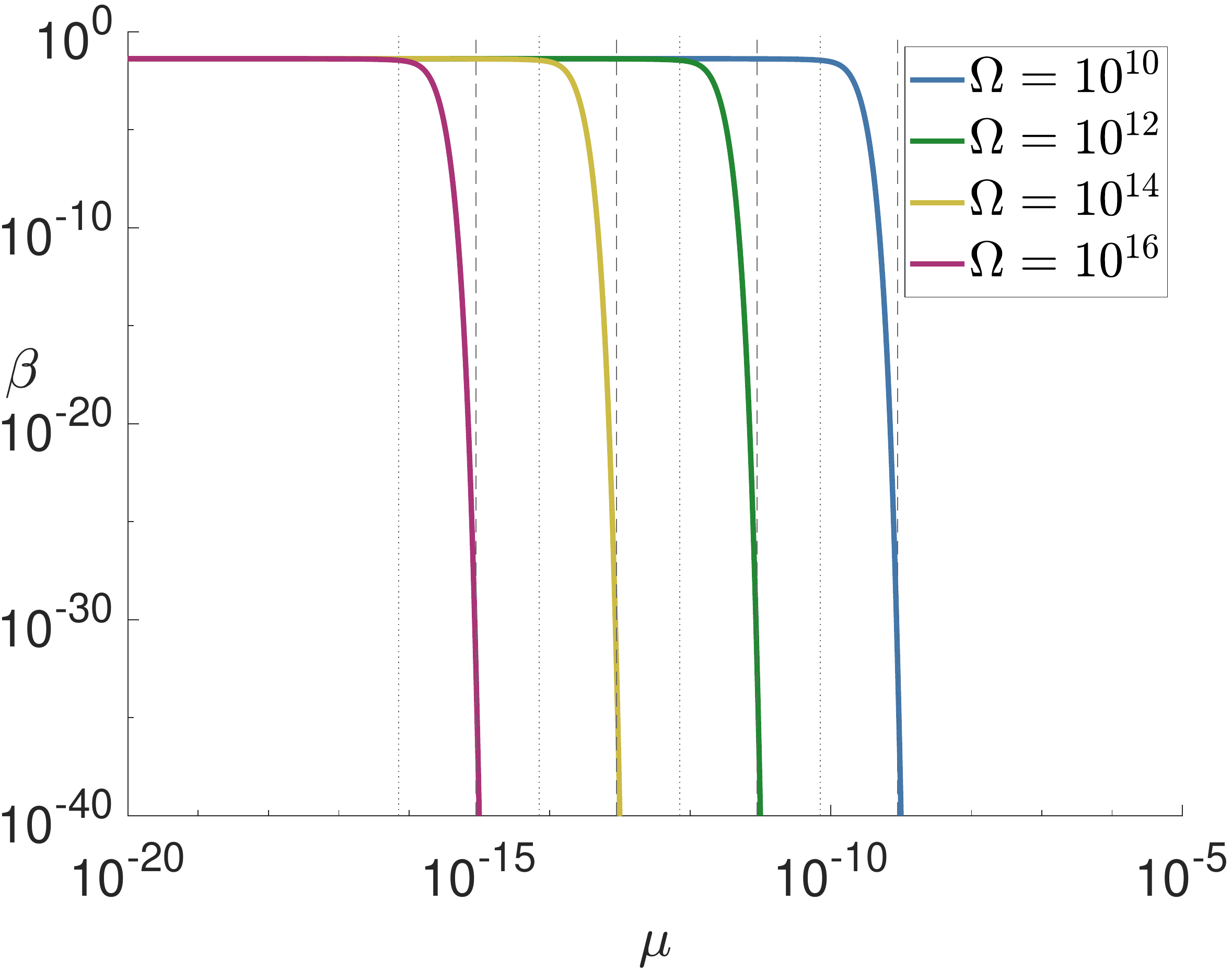}
\caption[$\beta $ for $\Delta \phi_{pl} < \Delta \phi_{cl}$]{\label{fig:chi_sigma_forBeta_USRandSR} The mass fraction of \ac{PBHs}, $\beta$, as a function of $\mu$ for four values of $\Omega$ computed using (\ref{eq:massfrac_exactFULL}) for $\mathcal{R}_c = 1$. The dotted lines represent the scale when the classical approximation fails, $\mu_{cl}$, given by (\ref{eq:sigma classical}). The dashed lines represent the pivot scale to overproduce \ac{PBHs}, $\mu_p$, predicted by (\ref{eq:sigma pivot}). It is clear that values of $\mu$ significantly smaller than $\mu_p$ overproduced \ac{PBHs} whereas values of $\mu$ larger than this value correspond to negligible production.}
\end{figure}
Looking at equation (\ref{eq: Beta chie > 0}) it is clear for large values of $\Omega$ that the $e^{-\Omega^{2}\mu^2}$ factor forces $\beta$ to be incredibly small unless $\mu$ is very small. Assuming we can approximate the average e-fold time by its classical value $\left\langle \mathcal{N}\right\rangle \simeq \left\langle \mathcal{N}\right\rangle\vert_{cl}$ which can be computed from (\ref{eq: classical average e-fold}) as:
\begin{eqnarray}
\left\langle \mathcal{N}\right\rangle\vert_{cl} \simeq -\dfrac{1}{3}\text{ln}(\mu)
\end{eqnarray}
This allows us to rewrite (\ref{eq: Beta chie > 0}) as:
\begin{eqnarray}
\beta (M) &\simeq &\dfrac{4}{\sqrt{\pi}}\Omega \mu ~e^{-\Omega^{2} \mu^{2}}~e^{-3\mathcal{R}_c} \label{eq: Beta chie > 0 chisigma}\\
&\simeq &\dfrac{4}{\sqrt{\pi}}\Omega\mu ~e^{-3\mathcal{R}_c} \label{eq: Beta chie > 0 small sigma}
\end{eqnarray}
where the second approximation uses the fact that $x~e^{-x^2} \sim x$ for small $x$.
We can use (\ref{eq: Beta chie > 0 small sigma}) to define a scale for $\mu$ where the mass fraction is important. We therefore find that:
\begin{empheq}[box = \fcolorbox{Maroon}{white}]{equation}
\begin{split}
\mu &\ll \dfrac{\sqrt{\pi}}{4\Omega}~e^{3\mathcal{R}_c} \equiv \mu_{p}\Leftrightarrow \beta \text{ Violates constraints} \\
\mu &\gg \dfrac{\sqrt{\pi}}{4\Omega}~e^{3\mathcal{R}_c} \equiv \mu_{p} \Leftrightarrow \beta \text{ Negligible} 
\end{split} \label{eq:sigma pivot}
\end{empheq}
where we have identified the pivot scale $\mu_{p}$. This scale is verified in Fig.~\ref{fig:chi_sigma_forBeta_USRandSR} where we plot, using the full expression (\ref{eq:massfrac_exactFULL}),  the dependence of the mass fraction, $\beta$ on $\mu$ for a few values of $\Omega$. We can see that the behaviour of $\beta$ almost looks like a step function with a sharp drop off as $\mu$ is increased.  The dashed lines -- corresponding to the scale predicted by (\ref{eq:sigma pivot}) -- accurately describes where this sharp dropoff takes place and marks the separation between overproduction of \ac{PBHs} and negligible production. 

We can verify our use of $\left\langle \mathcal{N}\right\rangle \simeq \left\langle \mathcal{N}\right\rangle\vert_{cl}$ by computing the value of $\mu$ for which the classicality criterion is violated, $\eta_{cl}=1$. Using equation (\ref{eq: classicality criterion_flat}), we find that the classicality parameter evaluated at $\phi_e$ is given by:
\begin{eqnarray}
\eta_{cl}(\phi_{e}) \simeq \left| \dfrac{3}{2}\tilde{H}_{0}^{2} - \dfrac{1}{2\Omega^2\mu^2}\right| \simeq \dfrac{1}{2\Omega^2\mu^2} \label{eq:etacl for sig > 0}
\end{eqnarray}
which suggests that smaller (bigger) values of the combination $\Omega^2\mu^2$ correspond to being in the quantum (classical) regime. This also justifies the use of the second approximation in (\ref{eq: Beta chie > 0 small sigma}) as the exponential dependence on $\Omega^2\mu^2$ which massively suppresses the formation of \ac{PBHs} also corresponds to being deep in the classical regime. This transition from classically dominated to quantum diffusion dominated dynamics takes place when $\eta_{cl}=1$ which we substitute into (\ref{eq:etacl for sig > 0}):
\begin{empheq}[box = \fcolorbox{Maroon}{white}]{equation}
\mu_{cl} = \dfrac{1}{\sqrt{2}\Omega} \label{eq:sigma classical_thisone} 
\end{empheq} 
identifying the transition:
\begin{empheq}[box = \fcolorbox{Maroon}{white}]{equation}
\begin{split}
\mu & < \mu_{cl} \Leftrightarrow \text{ The inflaton dynamics has a diffusion dominated regime}\\
\mu & > \mu_{cl} \Leftrightarrow \text{ Inflaton evolution is always classically dominated}  
\end{split}
\label{eq:sigma classical}
\end{empheq}
As $\mu_{cl} < \mu_{p}$, we are consistent in using $\left\langle \mathcal{N}\right\rangle\vert_{cl}$ to evaluate $\mu_p$. However,  this result has a more significant consequence. As can be clearly shown by plotting $\mu_{cl}$ with the dotted lines in Fig.~\ref{fig:chi_sigma_forBeta_USRandSR} one only enters the diffusion dominated regime once the mass fraction of \ac{PBHs}, $\beta$, is prohibitively high. This value is given by substituting $\mu_{cl}$ into $\beta$:
\begin{eqnarray}
\beta_{\mu_{cl}} \sim \dfrac{4}{\sqrt{2\pi e}}e^{-3\mathcal{R}_c} \sim 0.4289 \cdot  \dfrac{4}{\sqrt{\pi}}e^{-3\mathcal{R}_c} \label{eq:beta_plat_classical}
\end{eqnarray}
In other words the classically dominated evolution will already overproduce \ac{PBHs} before the inflaton even enters the diffusion dominated regime -- we will expand on this point in section \ref{sec:sigma_equal_zero}. 

\subsubsection{Perturbative expansion around classical solution}
\begin{figure}[t!]
    \centering
    \includegraphics[width = 0.45\linewidth]{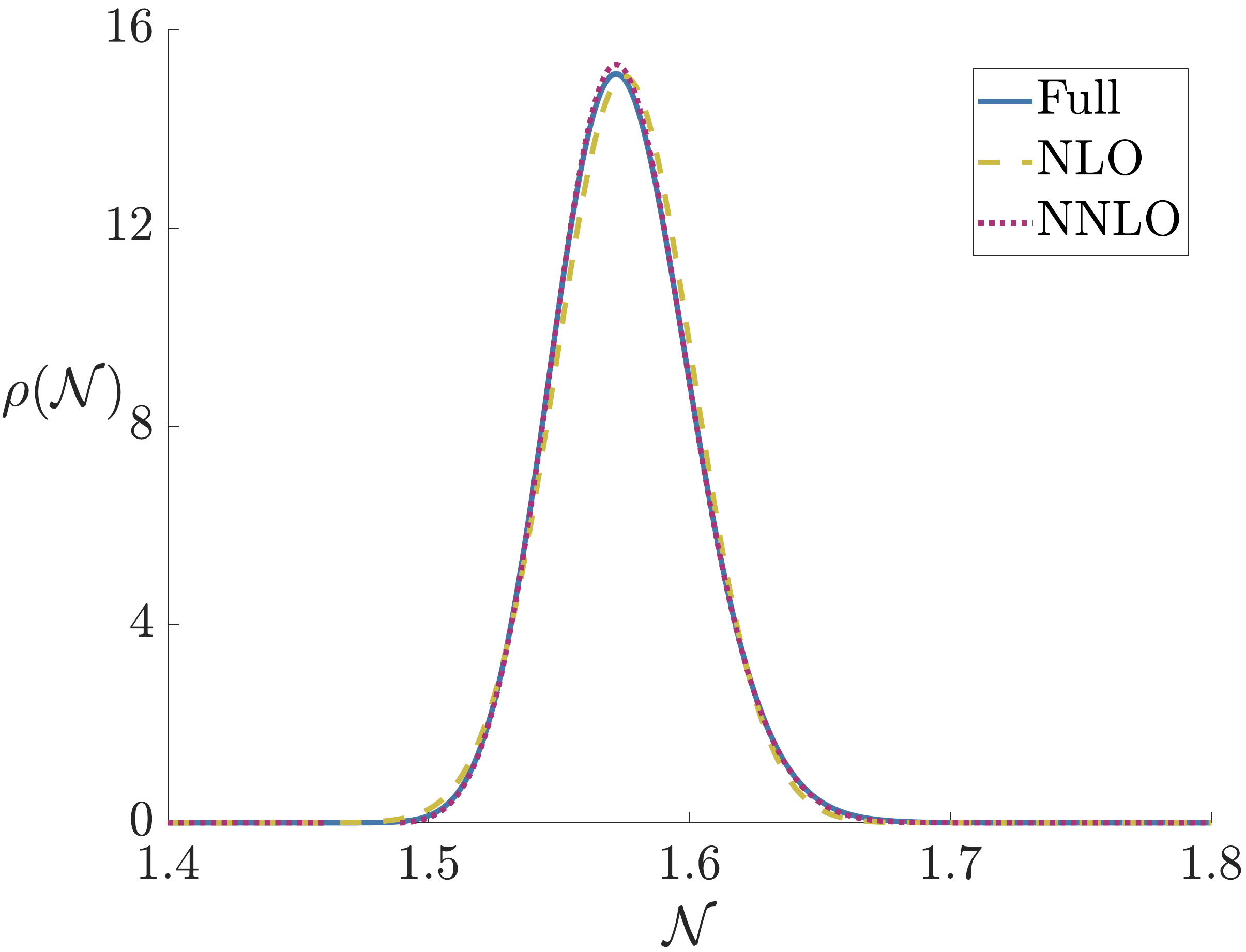}
    \includegraphics[width = 0.45\linewidth]{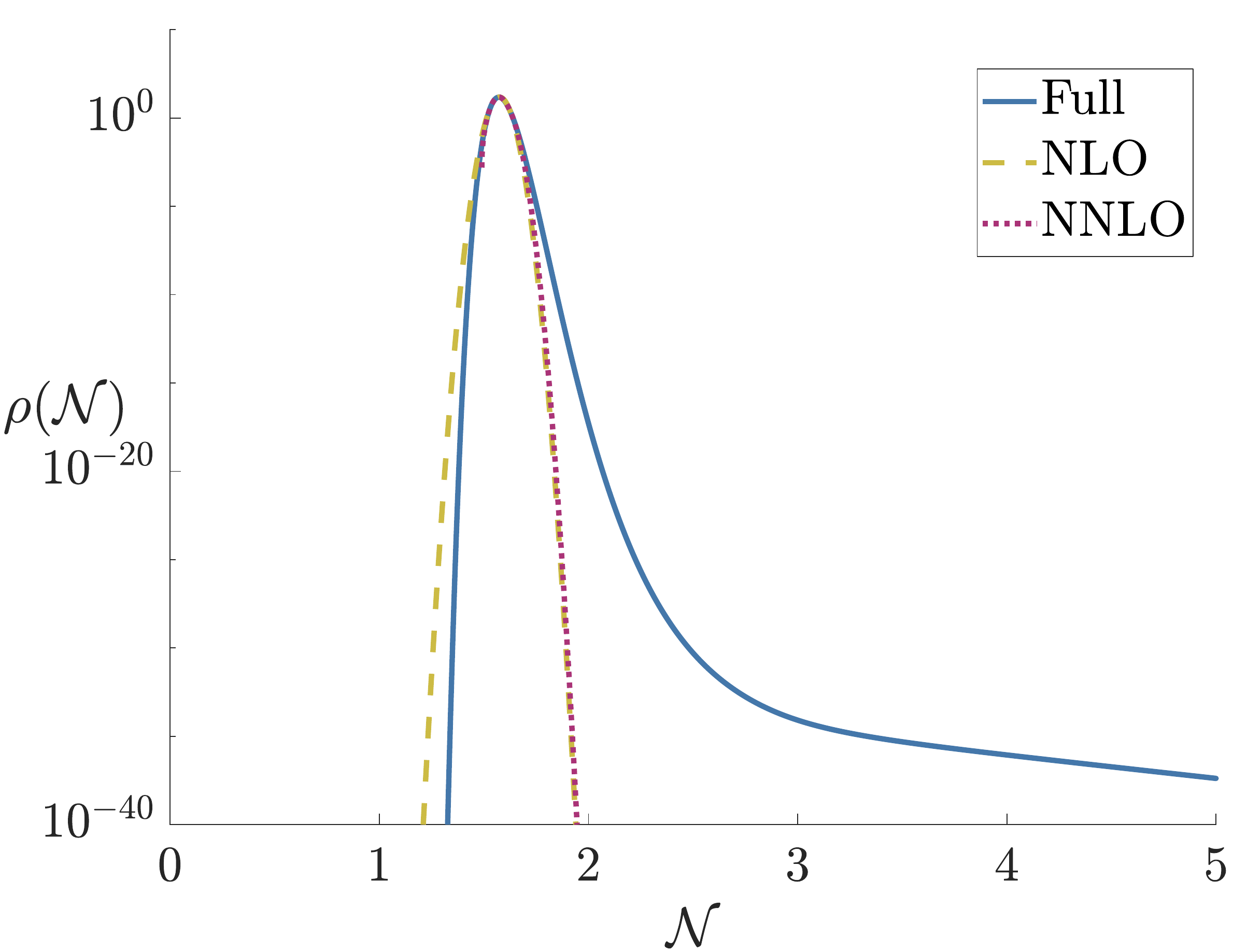}
    \caption[Full $\rho (\mathcal{N})$ as compared to semi-classical expansion on a plateau]{$\rho (\mathcal{N})$ with normal (left) and log (right) scale on a plateau for $\Omega = 1000$ and $\mu = \mu_p$ as computed by \ac{NLO}, \ac{NNLO} methods as well as the full PDF (\ref{eq:rhoN HJ}). }
    \label{fig:PDF_NNLO_flat}
\end{figure}
\begin{figure}[t!]
    \centering
    \includegraphics[width = 0.45\linewidth]{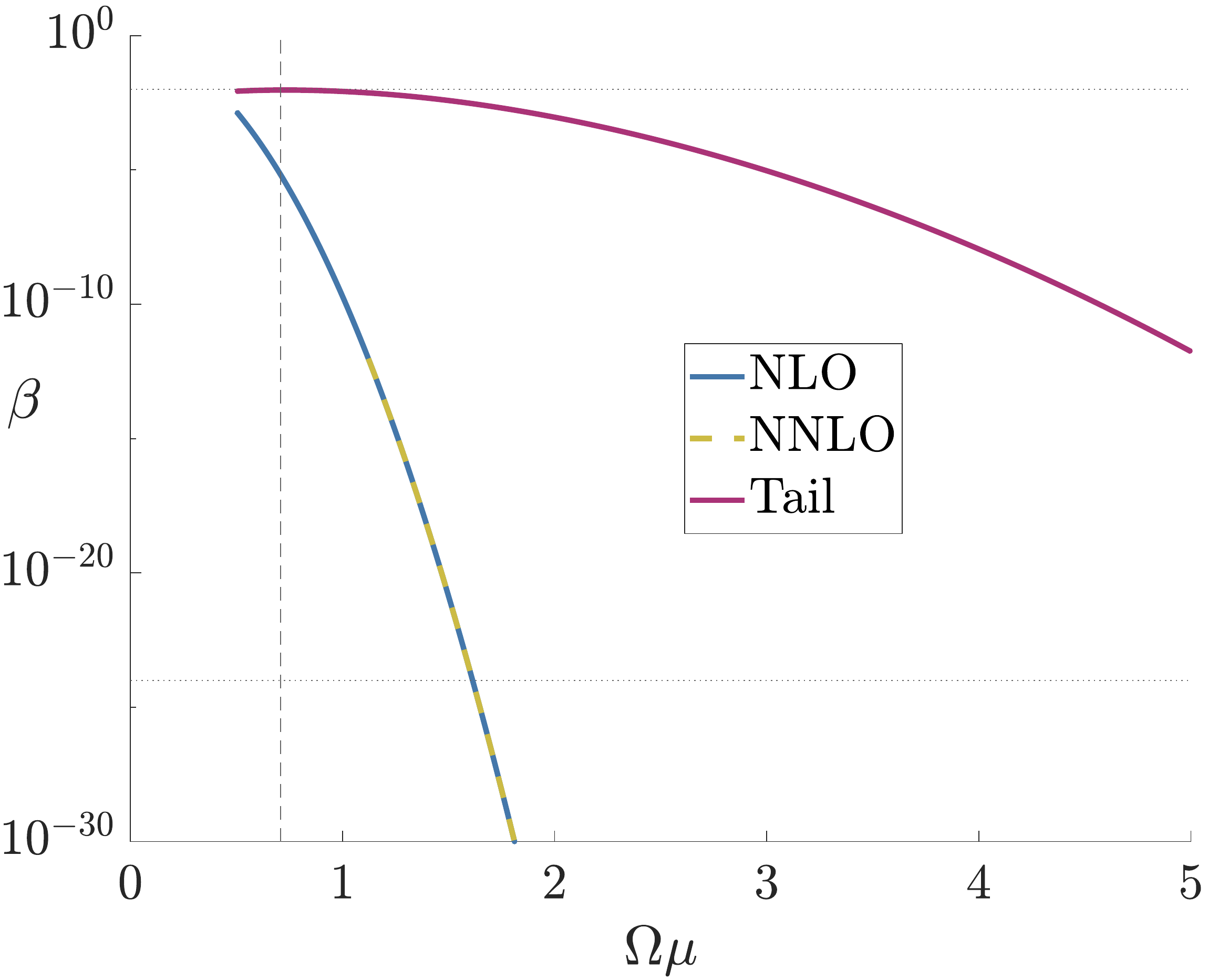}
    \includegraphics[width = 0.45\linewidth]{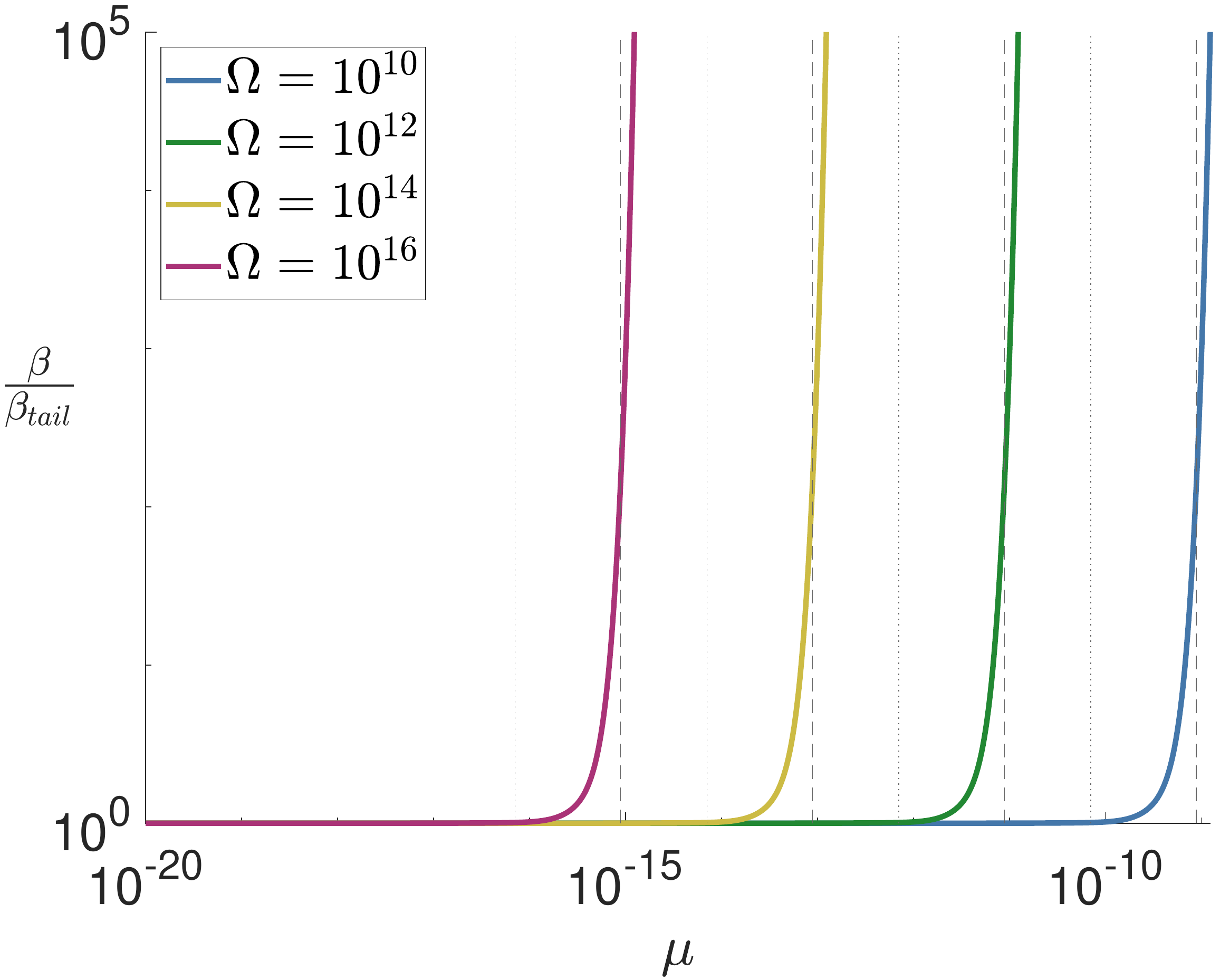}
    \caption[Full $\beta $ as compared to expansion around classical limit and expansion in the tail on a plateau]{Mass fraction $\beta $ on a plateau as computed by different methods for $\mathcal{R}_c = 1$. In the left panel we compare the \ac{NLO}, \ac{NNLO} expansion around the classical limit as well as the simple tail expression (\ref{eq: Beta chie > 0 chisigma}). The horizontal dotted lines correspond to the weakest \cite{Papanikolaou2020} and strongest \cite{Carr2020} bounds on the abundance of \ac{PBHs} and the vertical dashed line is when the classicality criterion is violated. In the right panel we compare the enhancement that the full expression (\ref{eq:massfrac_exactFULL}) has over the expansion in the tail (\ref{eq: Beta chie > 0}). The vertical dotted line corresponds to $\mu = \mu_{cl}$ (\ref{eq:sigma classical}) and the vertical dashed line to $\mu = \mu_{p}$ (\ref{eq:sigma pivot}).}
    \label{fig:Beta_NNLO_flat}
\end{figure}

As discussed in section \ref{sec:stochastic_deltaN} Pattison et.al \cite{Pattison2017} outlined a procedure to obtain the PDF using characterstic function techniques. We demonstrated how these formulae can be extended to be valid outside of \ac{SR} resulting in a Gaussian PDF at \ac{NLO} (\ref{eq:rhoN_char_NLO}) and a slightly non-Gaussian PDF at \ac{NNLO} (\ref{eq:rhoN_char_NNLO}). We can therefore straightforwardly obtain the following predictions for the mass fraction at \ac{NLO}:
\begin{eqnarray}
\beta_{\scalebox{0.5}{$\mathrm{NLO}$}} &=& \text{erfc}\left( \dfrac{\mathcal{R}_c}{2\tilde{H}\sqrt{\gamma_{1}^{\scalebox{0.5}{$\mathrm{NLO}$}}}} \right)\label{eq:beta NLO}
\end{eqnarray}
and at \ac{NNLO}:
\begin{eqnarray}
\beta_{\scalebox{0.5}{$\mathrm{NNLO}$}} &=& \text{erfc}\left( \dfrac{\mathcal{R}_c}{2\tilde{H}\sqrt{\gamma_{1}^{\scalebox{0.5}{$\mathrm{NNLO}$}}}} \right) + \dfrac{\gamma_{2}^{\scalebox{0.5}{$\mathrm{NNLO}$}}\left[\mathcal{R}_{c}^{2}-2\gamma_{1}^{\scalebox{0.5}{$\mathrm{NNLO}$}}\tilde{H}^2\right]}{4\tilde{H}\sqrt{\pi (\gamma_{1}^{\scalebox{0.5}{$\mathrm{NNLO}$}})^5}}\text{exp}\left(-\dfrac{\mathcal{R}_{c}^{2}}{4\tilde{H}^2\gamma_{1}^{\scalebox{0.5}{$\mathrm{NNLO}$}}}\right) \nonumber \\ \label{eq:beta NNLO} 
\end{eqnarray}
If we substitute our solution for $\tilde{H}$ on a plateau (\ref{eq: H exact flat}) we can determine the parameters $\gamma_{1}^{\scalebox{0.5}{$\mathrm{NLO}$}}$, $\gamma_{1}^{\scalebox{0.5}{$\mathrm{NNLO}$}}$ \& $\gamma_{2}^{\scalebox{0.5}{$\mathrm{NNLO}$}}$:
\begin{eqnarray}
\gamma_{1}^{\scalebox{0.5}{$\mathrm{NLO}$}} & \approx & \dfrac{1}{36\Omega^2\mu^2\tilde{H}_{0}^{2}}\\
\gamma_{1}^{\scalebox{0.5}{$\mathrm{NNLO}$}} &\approx & \dfrac{4\Omega^2\mu^2 -5}{144\Omega^4\mu^4\tilde{H}_{0}^2}\\
\gamma_{2}^{\scalebox{0.5}{$\mathrm{NNLO}$}} & \approx & \dfrac{1}{216\Omega^4\mu^4\tilde{H}_{0}^{4}}
\end{eqnarray}
where we have neglected higher order terms in powers of $\tilde{H}_0$. We have plotted the corresponding PDFs for $\Omega = 1000$, $\mu = \mu_p$ in Fig.~\ref{fig:PDF_NNLO_flat} as well as the full result. Here we can see that on a linear scale (left plot) the \ac{NLO} and \ac{NNLO} PDFs closely match the full result capturing well quantities like $\lan \mathcal{N}\ran$ and the variance. However if we look at a log scale (right plot) we can see that the \ac{NLO} and \ac{NNLO} fail spectacularly at capturing the tail behaviour. This reinforces the well known result that perturbative expansions around the mean fail to accurately capture the tail of the distribution \cite{Pattison2017}. Unsurprisingly this has drastic consequences for the abundance of \ac{PBHs} predicted by these methods. In the left plot of Fig.~\ref{fig:Beta_NNLO_flat} we plot the mass fraction as computed using the \ac{NLO} and \ac{NNLO} method as well as the simplified tail expression (\ref{eq: Beta chie > 0 chisigma}) for $\mathcal{R}_c = 1$. It is obvious that the perturbative methods drastically underestimate the abundance by many orders of magnitude. This emphasises that although the distribution is never dominated by quantum diffusion effects, as $\eta_{cl} \ll 1$, these diffusion effects are still crucially important for correctly resolving the tail of the distribution and therefore getting the correct abundance of \ac{PBHs}. On the other hand an expansion in the tail is much more accurate. In the right plot of Fig.~\ref{fig:Beta_NNLO_flat} we plot the enhancement the full mass fraction (\ref{eq:massfrac_exactFULL}) has over just the tail expression (\ref{eq: Beta chie > 0}). We can see that for $\mu \ll \mu_p$ -- i.e. when the abundance becomes significant -- that the tail expression is a very good approximation of the full result. 
\subsection{\label{sec:sigma_equal_zero}The case \texorpdfstring{$\Delta \phi_{pl} = \Delta \phi_{cl}$ ($\mu=0$)}{} }

\begin{figure}[t!]
\centering 
\includegraphics[width=.48\textwidth]{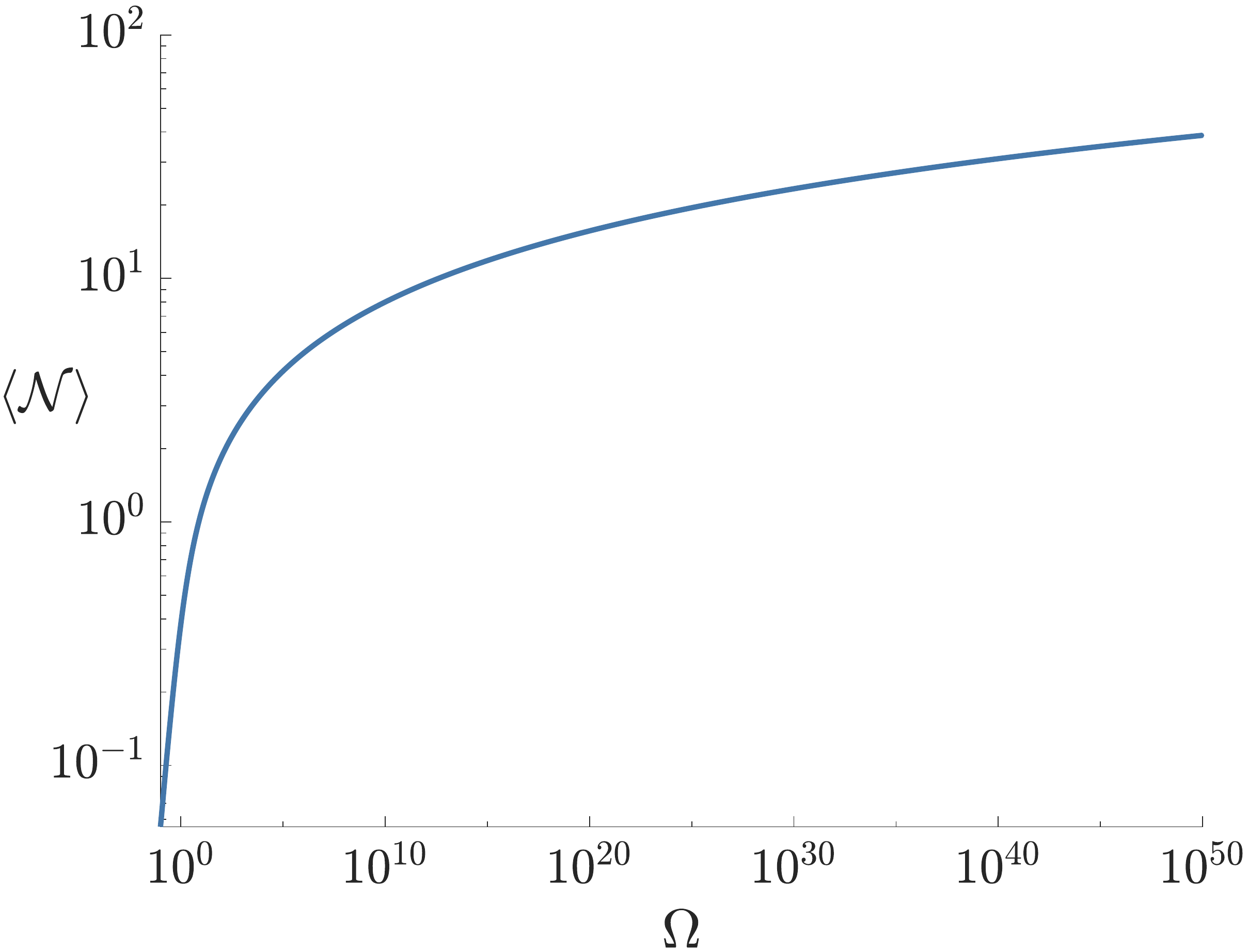}
\includegraphics[width=.48\textwidth]{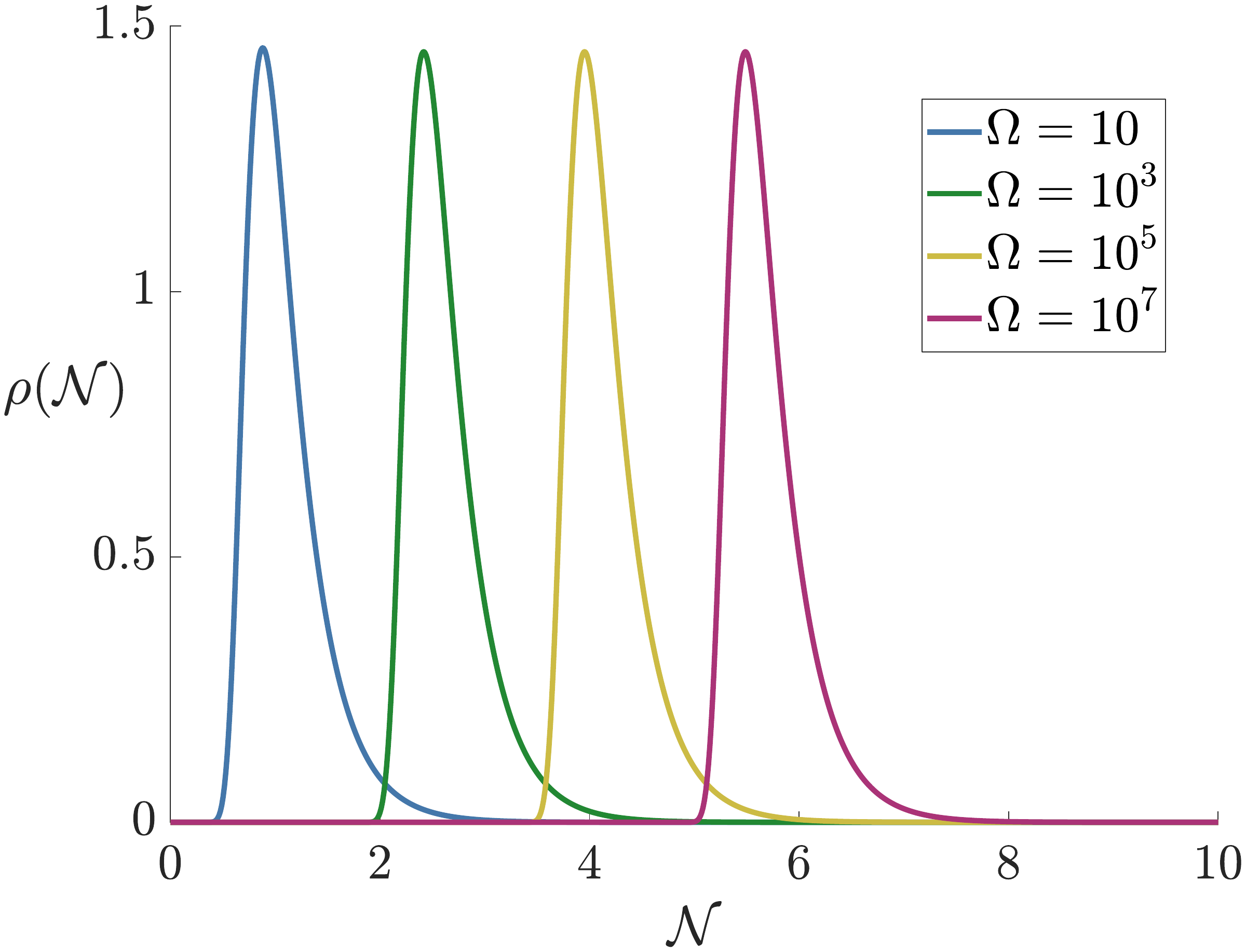}
\caption[$\left\langle \mathcal{N}\right\rangle$ and $\rho (\mathcal{N})$ for $\Delta \phi_{cl} = \Delta \phi_{pl}$]{\label{fig:rhoN_chi_e_zero.pdf} The dependence of average number of e-folds $\left\langle \mathcal{N}\right\rangle$ realised on the plateau on $\Omega$ (left) and the PDF $\rho (\mathcal{N})$ of e-fold time spent in the plateau for four values of $\Omega$ (right). Both for $\Delta \phi_{cl} = \Delta \phi_{pl}$.}
\end{figure}

In the limit $\Delta \phi_{pl} = \Delta \phi_{cl}$ (equivalently $\mu = 0$), (\ref{eq:rhoN HJ}) simplifies to:
\begin{eqnarray}
\rho(\mathcal{N}) = \dfrac{6}{\sqrt{\pi}} \sqrt{n}(n+1)\Omega ~e^{-n\Omega^{2}}  \label{eq:rho chie = 0}
\end{eqnarray}
If we consider the large $\Delta \mathcal{N}$ limit of (\ref{eq:rho chie = 0}) to examine the behaviour in the tail we obtain:
\begin{eqnarray}
\rho(\mathcal{N})\sim \dfrac{6}{\sqrt{\pi}}\Omega ~e^{-3\Delta\mathcal{N}} \label{eq:rho chie = 0 approx}
\end{eqnarray}
which corresponds to (\ref{eq: rhoN HJ LARGE N}) in the $\mu \rightarrow 0$ limit as it should. Importantly therefore it also exhibits the same non-Gaussian exponential tail $e^{-3\Delta\mathcal{N}}$. The average number of e-folds $\left\langle \mathcal{N}\right\rangle$ realised can be computed exactly from (\ref{eq:rho chie = 0}) and is given by:
\begin{eqnarray}
\left\langle \mathcal{N}\right\rangle = \dfrac{\pi}{6}\text{erfi}(\Omega)-\dfrac{\Omega^{2}}{3}\,\, {}_2F_2 \left( \lbrace 1,1\rbrace, \left\lbrace \dfrac{3}{2},2\right\rbrace ,\Omega^{2}\right) 
\end{eqnarray}
where erfi is the imaginary error function and $_2F_2$ is a generalised hypergeometric function. Note in practice that for very large values of $\Omega$ it is usually more practical numerically to compute $\left\langle \mathcal{N}\right\rangle$ directly from the PDF.

In Fig.~\ref{fig:rhoN_chi_e_zero.pdf} we plot both the dependence of average e-fold time spent in the plateau, $\left\langle \mathcal{N}\right\rangle$, on $\Omega$ (left panel) and the PDF $\rho(\mathcal{N})$ for four values of $\Omega$ (right panel). In the left panel we see how the average number of e-folds $\left\langle \mathcal{N}\right\rangle $ grows with $\Omega$ very quickly initially before growing logarithmically at very large values of $\Omega$. At $\Omega \sim 10^{50}$ the average time spent in the plateau, $\left\langle \mathcal{N}\right\rangle $, is comparable to the total duration of inflation.\footnote{Strictly speaking we mean the average time spent in the plateau, $\left\langle \mathcal{N}\right\rangle $, is longer than the allowed number of e-folds between \ac{CMB} modes exiting the horizon and inflation ending.} 
Looking at the right panel we can see how $\rho(\mathcal{N})$ is non-Gaussian with a deep tail and that this shape does not change noticeably as $\Omega$ is increased by several orders of magnitude. Indeed, the only noticeable impact of  increasing $\Omega$ is to translate the whole PDF to the right. This suggests that even if the inflaton spends a large amount of time on the plateau it is still reasonably localised in time around its average value and we can reasonably assign a time for these modes to exit the horizon. 

The mass fraction of \ac{PBHs}  for the PDF (\ref{eq:rho chie = 0}) can be calculated exactly as:
\begin{empheq}[box = \fcolorbox{Maroon}{white}]{equation}
\beta (M) = 2~\text{erf}(\sqrt{n_{c}}\Omega)\times \underbrace{e^{\frac{9}{4}\Omega^{-2}}}_{\text{SR}}  \label{eq: Massfrac chie = 0}
\end{empheq}
where $n_c$ is given by equation (\ref{eq:nc defn}) and again we have included the contribution from the previous \ac{SR} phase.
Note that if we consider the large $\Delta \mathcal{N}$ limit of (\ref{eq: Massfrac chie = 0}) then it reduces to:
\begin{eqnarray}
\beta (M) \sim \dfrac{4}{\sqrt{\pi}}\Omega~e^{-3(\mathcal{R}_c + \left\langle \mathcal{N}\right\rangle - N_{in})}\times e^{\frac{9}{4}\Omega^{-2}}  \label{eq: massfrac chie = 0 approx}
\end{eqnarray}
We see that this is the $\mu \rightarrow 0$ limit of equation (\ref{eq: Beta chie > 0}) confirming the two results are consistent with each other. It is worth appreciating that, like in the analysis of \cite{Pattison2017}, the PDF $\rho(\mathcal{N})$, average number of e-folds $\left\langle \mathcal{N}\right\rangle$ and mass fraction of \ac{PBHs} $\beta$ only depends on a single parameter $\Omega$\footnote{In \cite{Pattison2017} their parameter is $\mu \equiv \Delta \phi_{pl}/\sqrt{v_0}$ which for $\mu = 0$ is related to $\Omega$ through $\Omega = \mu\sqrt{3/2}$.}. However unlike in \cite{Pattison2017} this computation fully takes into account the velocity of the inflaton as it enters the plateau albeit in the restricted case where $\Delta \phi_{pl} = \Delta \phi_{cl}$. \\

\begin{figure}[t!]
\centering 
\includegraphics[width=.75\textwidth]{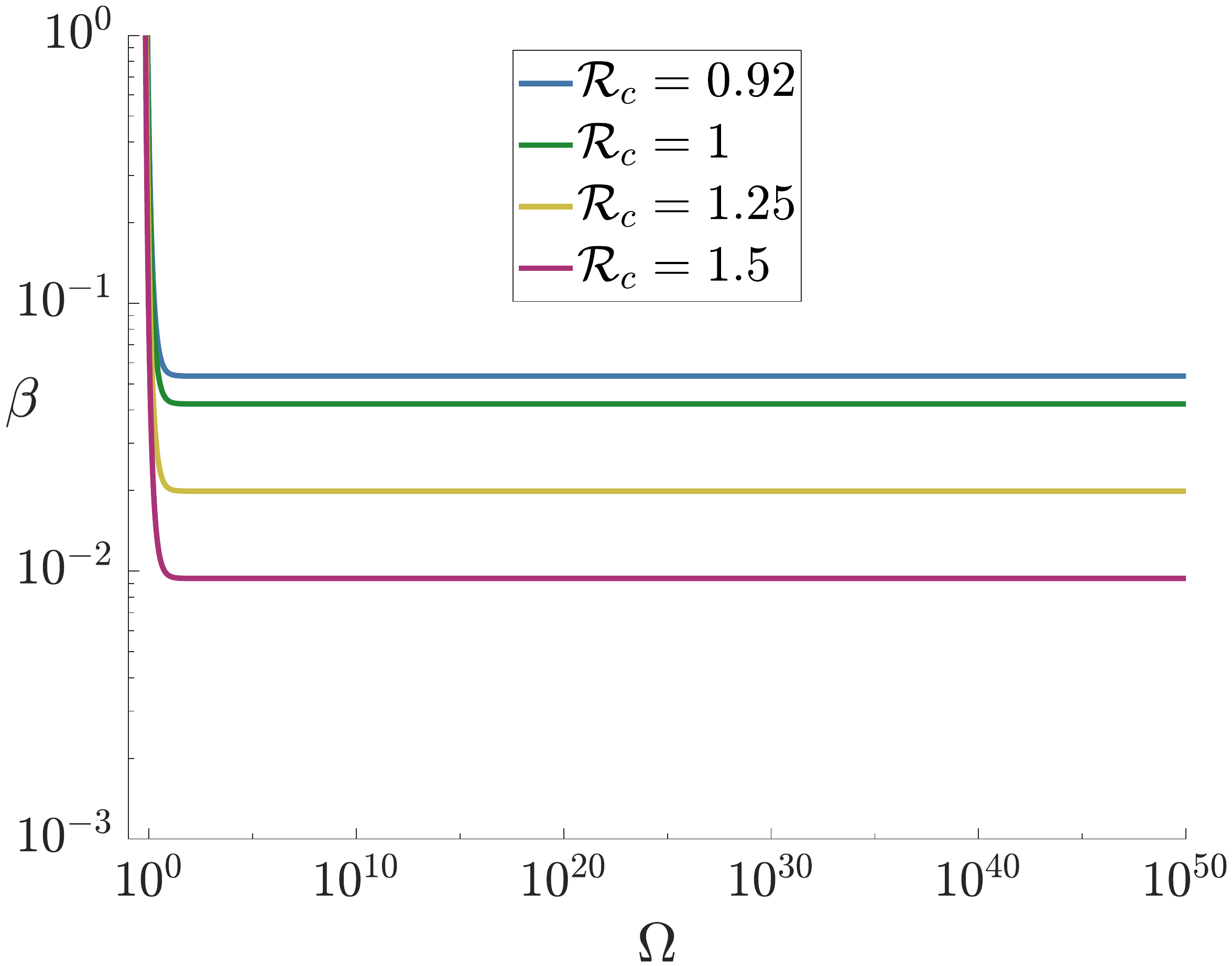}
\caption[$\beta$ for $\Delta \phi_{cl} = \Delta \phi_{pl}$]{\label{fig:Beta_USR_chi_e_zero.pdf} Dependence of mass fraction $\beta$ on $\Omega$ using (\ref{eq: Massfrac chie = 0}) for four different values of the cutoff $\mathcal{R}_c$ for  $\Delta \phi_{cl} = \Delta \phi_{pl}$. The sudden spike in $\beta$ for small values of $\Omega$ is the contribution from the previous \ac{SR} phase}
\end{figure}

In Fig.~\ref{fig:Beta_USR_chi_e_zero.pdf} we plot the dependence of the mass fraction, $\beta$, on $\Omega$ for four different values of the cutoff, $\mathcal{R}_c$, between the lower and upper limits of $\sim$ 0.92 and 1.5 permitted \cite{Musco2019}. We can see that -- apart from the sharp spike at small $\Omega$ due to the previous \ac{SR} phase -- the mass fraction is constant for all values of $\Omega$ and approximately lies in the range $10^{-1}$ - $10^{-2}$. The mass fraction, $\beta$, is therefore in excess of all the upper limits imposed in the possible mass ranges for \ac{PBHs}.\\

This constant value occurs because for large $\Omega$ the quantity $\Omega ~e^{-3(\left\langle \mathcal{N}\right\rangle -N_{in})} \approx 0.3747$. We can therefore say that the mass fraction converges very quickly for $\Omega \gg 1$ to:
\begin{eqnarray}
\beta (M) \sim 0.3747\times\dfrac{4}{\sqrt{\pi}}~e^{-3\mathcal{R}_c} \label{eq: massfrac chie = 0 const approx}
\end{eqnarray}
This equation very accurately describes the horizontal lines displayed in Fig.~\ref{fig:Beta_USR_chi_e_zero.pdf}.\\

It is therefore clear from the analysis of this section that for any plateau of equal width to the classical drift distance that one will generically overproduce \ac{PBHs} for any remotely realistic inflationary potential. This means that Scenario B in Fig.~\ref{fig:Scenario A and B} is completely ruled out as a subsequent phase of free diffusion would only \textit{enhance} the curvature perturbation, producing even more \ac{PBHs} -- we verify this in section \ref{sec:ScenB}. Not only that but as shown in Fig.~\ref{fig:chi_sigma_forBeta_USRandSR} even a diffusion dominated regime with non-zero classical drift is forbidden. We therefore arrive at the main result of this section:  \\
\textit{Any period of quantum diffusion dominated dynamics on a plateau will overproduce \ac{PBHs}}\\

\subsection{\label{sec:ScenB}The case \texorpdfstring{$\Delta \phi_{pl} > \Delta \phi_{cl}$ ($ \mu < 0$)}{}}
\begin{figure}[t!]
\centering 
\includegraphics[width=.48\textwidth]{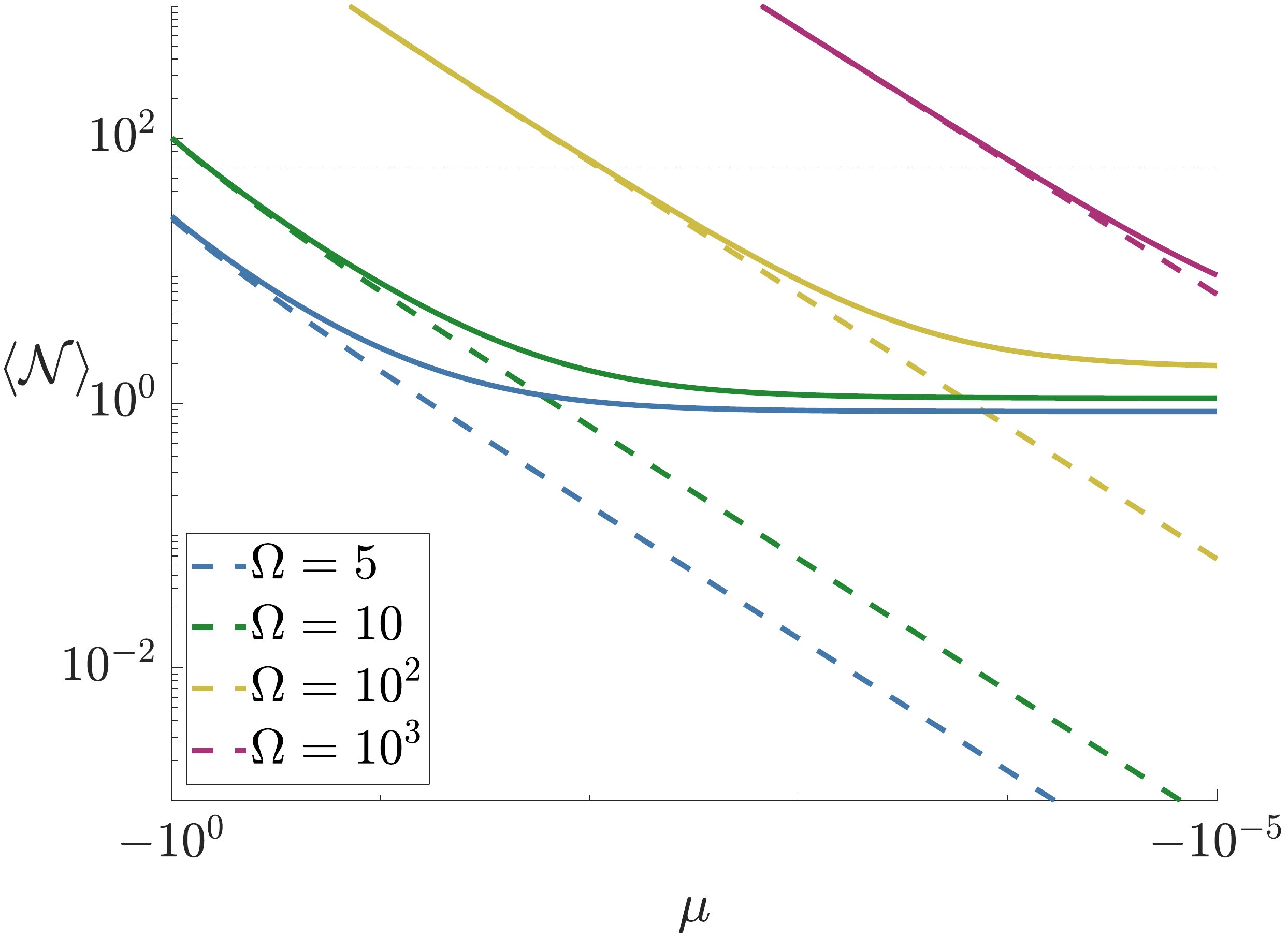}
\includegraphics[width=.48\textwidth]{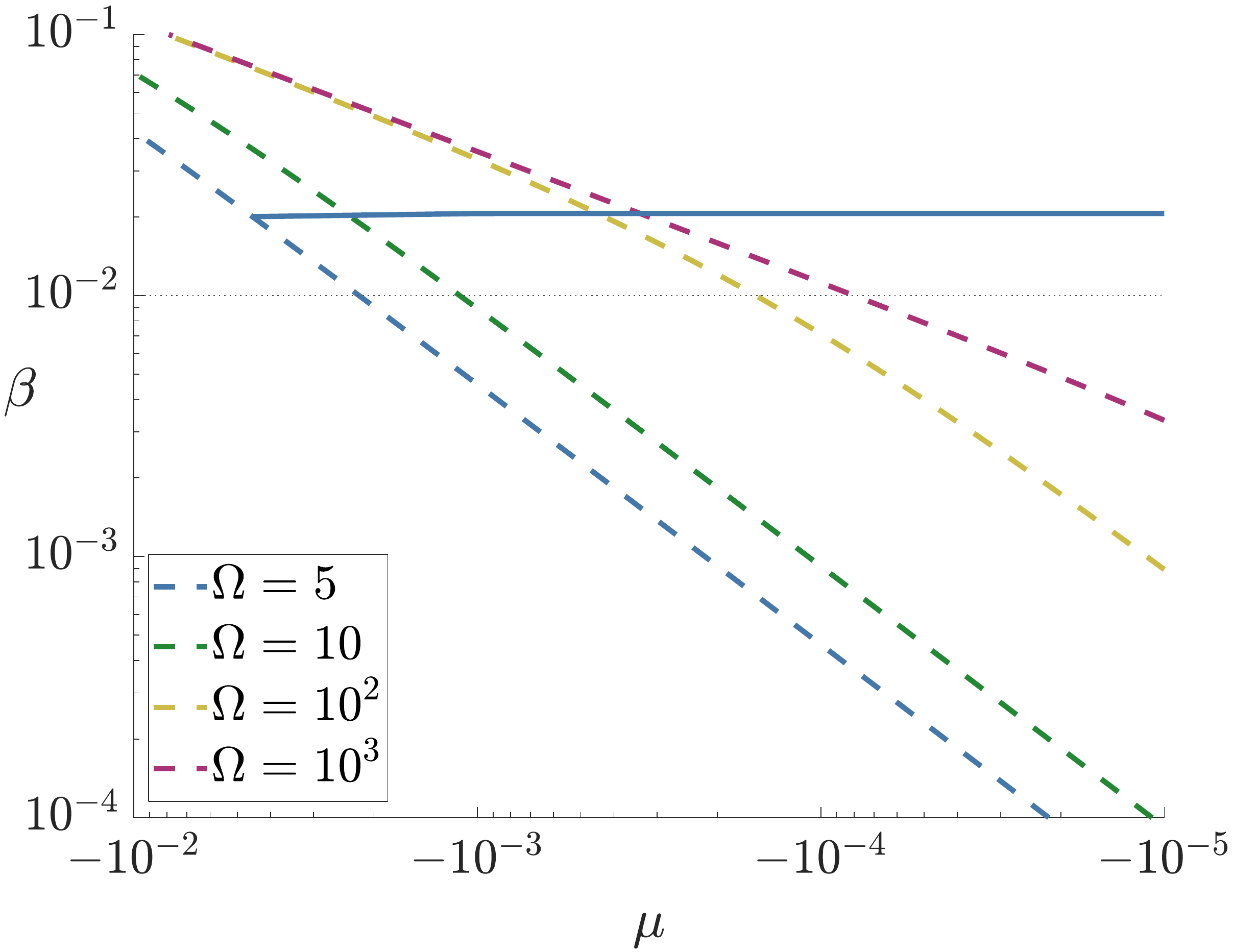}
\caption[$\left\langle \mathcal{N}\right\rangle$ and $\beta$ for $\Delta \phi_{pl} > \Delta \phi_{cl}$]{\label{fig:Pattison muN and beta} The dependence on $\mu$ of the average number of e-folds $\left\langle \mathcal{N}\right\rangle$ (left) and the mass fraction $\beta$ (right) given as a function of $\mu<0$ ($\Delta \phi_{pl} > \Delta \phi_{cl}$). The dashed lines correspond to the free diffusion value only and the solid lines to the total of the \ac{H-J} plus free diffusion phase. For small $\mu$, $\left\langle \mathcal{N}\right\rangle$ in the \ac{H-J} computation levels off to the number of e-folds required for the field to slide on the plateau via its initial velocity with free diffusion making a very small contribution. The horizontal dotted line on the left plot corresponds to 60 e-folds. The mass fraction accounting for both phases given by (\ref{eq:rhoHJ+dS}) is shown for $\Omega = 5$ by the solid blue line in the right plot where the horizontal dotted line corresponds to the weakest ($10^{-2}$) bound on the abundance of \ac{PBHs} \cite{Papanikolaou2020}.}
\end{figure}

If the plateau is wider than the classical drift distance, the field enters a period of free diffusion as described by scenario B in Fig.~\ref{fig:Scenario A and B}. When the field reaches $\phi_0$ it exits the \ac{H-J} trajectory and enters a period of free diffusion with zero drift velocity. If we assume -- for now -- that the distribution enters as a delta function then we can use the results of Pattison \textit{et.~al} \cite{Pattison2017} to describe this second phase. The PDF for exit time, $\rho_{dS}$, average time spent during free diffusion, $\left\langle  \mathcal{N}\right\rangle_{dS}$, and the mass fraction, $\beta_{dS}$, for this pure de Sitter phase are given by \cite{Pattison2017}:
\begin{eqnarray}
\rho_{dS}(\mathcal{N}) &=& \dfrac{3\pi}{\Omega^2 (1-\mu)^2}\sum_{n=0}^{\infty}\lb n+\scalebox{0.5}{$\dfrac{1}{2}$}\rb\text{sin}\left[ \dfrac{\mu \lb n+\scalebox{0.5}{$\dfrac{1}{2}$}\rb}{\mu-1}\pi\right]~\text{exp}\left[ -\dfrac{3\pi^2\lb n+\scalebox{0.5}{$\dfrac{1}{2}$}\rb^2}{2\Omega^2 (1-\mu)^2} \mathcal{N}\right]\label{eq:PDF de Sitter}\\
\left\langle \mathcal{N} \right\rangle_{dS} &=& \dfrac{2\Omega^2}{3}\mu \left( \dfrac{\mu}{2} -1\right)\label{eq:muN dS}\\
\beta_{dS} &=& \dfrac{4}{\pi}\sum_{n=0}^{\infty}\dfrac{\text{sin}\left[ \dfrac{\mu\lb n+\scalebox{0.5}{$\dfrac{1}{2}$}\rb}{\mu-1}\pi\right]}{  n+\scalebox{0.5}{$\dfrac{1}{2}$}} ~\text{exp}\left[ -\dfrac{3\pi^2\lb n+\scalebox{0.5}{$\dfrac{1}{2}$}\rb^2}{2\Omega^2 (1-\mu)^2} (\mathcal{R}_c + \left\langle \mathcal{N}_3 \right\rangle )\right]\label{eq:massfrac de Sitter}
\end{eqnarray}
expressed in terms of our parameters $\Omega$ and $\mu$. $\left\langle \mathcal{N} \right\rangle_{dS}$ and $\beta_{dS}$ are plotted as dashed lines on the left and right plots of Fig.~\ref{fig:Pattison muN and beta} respectively as functions of $\mu$ for different values of $\Omega$. We see that unless $\mu$ is very small in absolute value, the average number of e-folds realised throughout the plateau can easily exceed the number of e-folds needed between the \ac{CMB} and the end of inflation, at least for inflationary scales not too close to $M_{\mathrm{p}}$ (i.e. for large $\Omega$). The mass fraction given by (\ref{eq:massfrac de Sitter}) places tighter bounds on $\mu$, forcing it to be small in absolute value to not violate constraints. All these conclusions are drawn from the pure diffusion computation.
\begin{figure}[t!]
\centering 
\includegraphics[width=.48\textwidth]{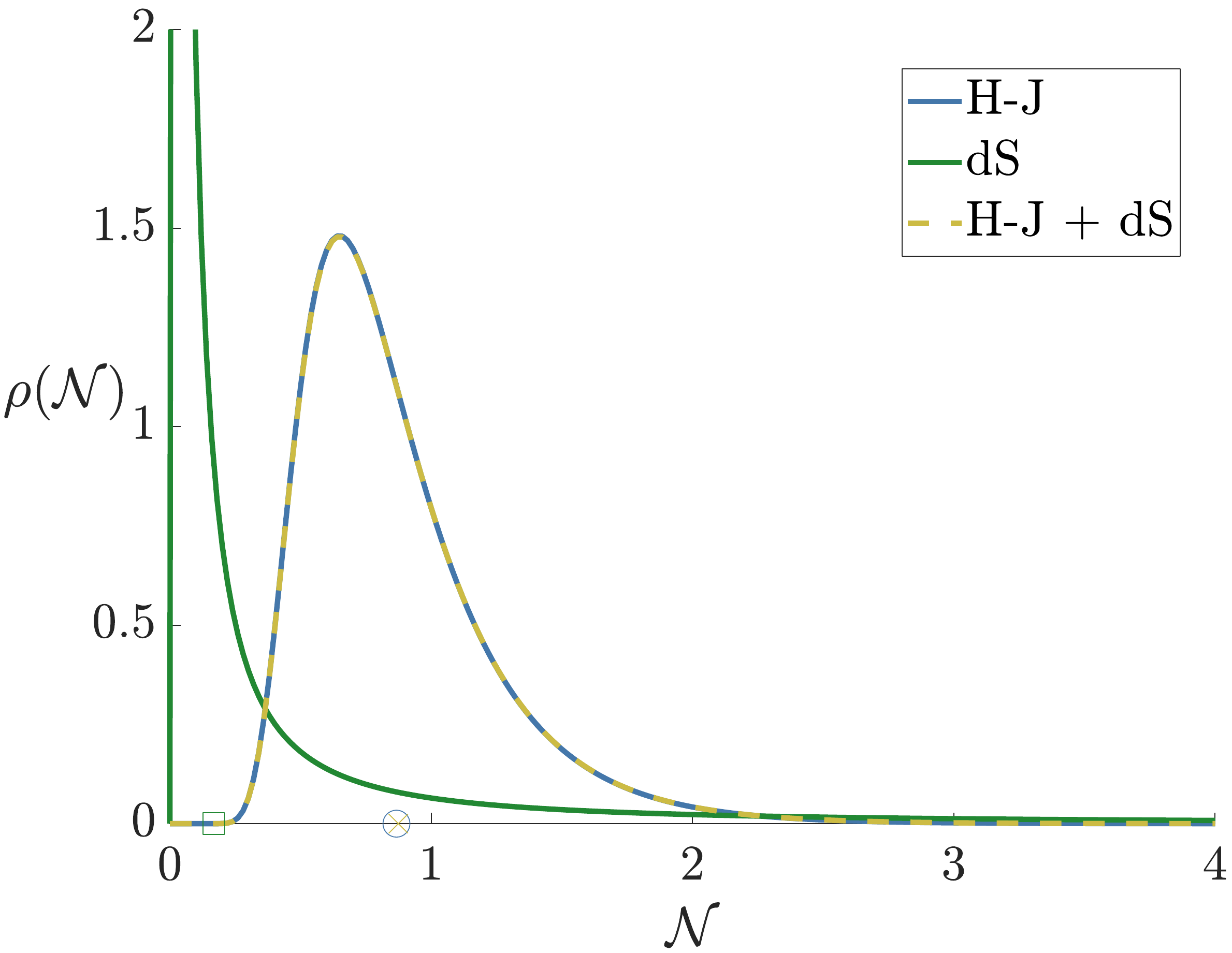}
\includegraphics[width=.48\textwidth]{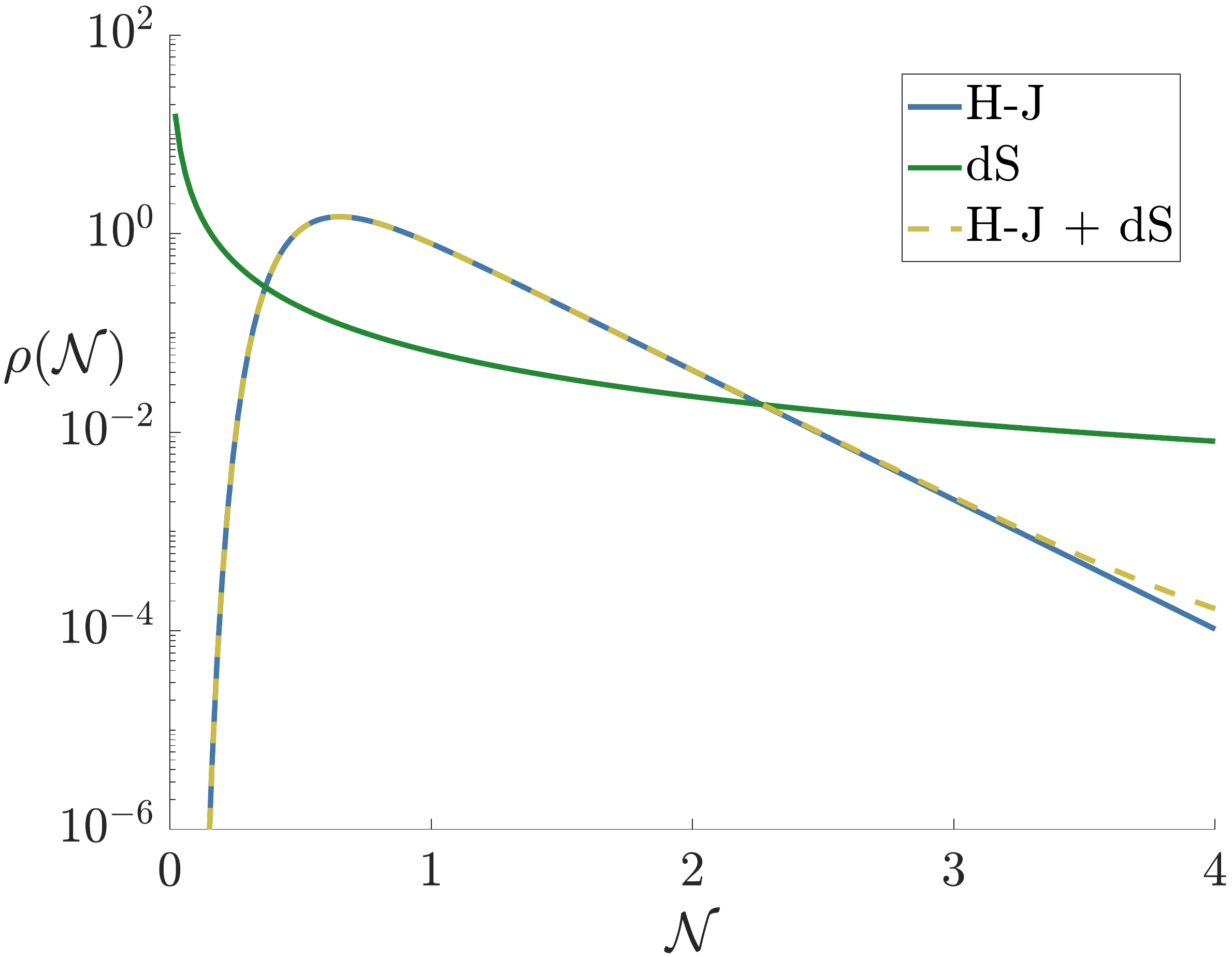}
\caption[$\rho (\mathcal{N})$ for $\Delta \phi_{pl} > \Delta \phi_{cl}$]{\label{fig:rhoN scenarioB} The PDF for exit times, $\rho (\mathcal{N})$, in scenario B where $\Omega = 5$, $\mu = -10^{-2}$ plotted on a linear (left) and logarithmic scale (right). The green solid line corresponds to the pure free diffusion result (\ref{eq:PDF de Sitter}) and the solid blue to the \ac{H-J} solution (\ref{eq:rho chie = 0}). The dashed line is the full \ac{H-J} plus free diffusion phase (\ref{eq:rhoHJ+dS}). The circle, square and cross on the left plot correspond to the average number of e-folds, $\left\langle \mathcal{N}\right\rangle$, realised during the \ac{H-J} phase, free diffusion only and \ac{H-J} + free diffusion phase respectively - the circle and cross are essentially identical. Adding free diffusion to the \ac{H-J} result only enhances the tail slightly as seen in the right plot. Note that both the average number of e-folds, the variance and the general shape of the PDF in the \ac{H-J} computation differ significantly from those in the pure diffusion computation.}
\end{figure}
However, these computations have assumed that the field starts the free diffusion phase at a fixed time and localised on the plateau. This is clearly not true as is evident e.g. from the right plot of Fig.~\ref{fig:rhoN_chi_e_zero.pdf} where a similar looking PDF would determine the starting time of the post-\ac{H-J} free diffusion phase. To include the effect on the mass fraction of the prior slide of the field on the plateau, one should instead do the convolution of the \ac{H-J} phase (\ref{eq:rhoN HJ}) and the free diffusion phase (\ref{eq:PDF de Sitter}), and then compute the mass fraction from this convoluted PDF. While in principle this can be done numerically in a similar way as the procedure outlined in Appendix \ref{sec:SR_HJ_deriv} we find that the more illuminating method is to modify the procedure presented in \cite{Prokopec2021} to account for a finite width plateau. In \cite{Prokopec2021} the PDF was computed for a free diffusion phase on an infinitely wide plateau with one absorbing boundary. In Appendix \ref{sec:dS+HJ_deriv} we adapt this computation to account for a finite plateau by including a reflecting boundary\footnote{As discussed earlier the right boundary is a lot more complicated than a simple reflection but this will suffice for our purposes.} at one side in spirit with the computation in \cite{Pattison2017}. The final result can be written as:
\begin{empheq}[box = \fcolorbox{Maroon}{white}]{equation}
\begin{split}
\rho_{\mu < 0}(\mathcal{N}) &= \dfrac{2\sqrt{\pi}}{\Omega(1-\mu)^2}\int_{0}^{\mathcal{N}}\mathrm{d}u~\text{sinh}(3u)^{\scalebox{0.5}{$-\dfrac{3}{2}$}} ~\text{exp}\left[\dfrac{3}{2}u -\dfrac{1}{2}\left( \text{coth}(3u ) -1\right)\Omega^2\right]  \\
&\times \sum_{n=0}^{\infty}\left(n+\scalebox{0.5}{$\dfrac{1}{2}$} \right)\Bigg\lbrace \text{sin}\left[\lb n+\scalebox{0.5}{$\dfrac{1}{2}$}\rb \dfrac{\pi\mu}{\mu-1}\right] ~\text{exp}\left[-\lb n+\scalebox{0.5}{$\dfrac{1}{2}$}\rb^2\dfrac{2\pi^2 (\mathcal{N} -u)}{3\Omega^2 (1-\mu)^2}  \right] \Bigg\rbrace
\end{split}\label{eq:rhoHJ+dS}
\end{empheq}

The analytical evaluation of the integral in (\ref{eq:rhoHJ+dS}) and the summation of the series is challenging, but a numerical evaluation is feasible. We show what the PDF looks like in Fig.~\ref{fig:rhoN scenarioB} for $\mu=-10^{-2}$, where the corresponding free diffusion PDF (\ref{eq:PDF de Sitter}) is also shown for comparison. As expected, for the very small values of $\mu$ allowed, the total PDF resembles very closely the \ac{H-J} one and increasing $\mu$ only serves to slightly enhance the tail. From this PDF we can compute the total mass fraction of \ac{PBHs} accounting for both the \ac{H-J} and free diffusion phase. This is shown by the solid line for $\Omega = 5$ on the right plot of Fig.~\ref{fig:Pattison muN and beta}. The other values of $\Omega$ were not shown due to being indistinguishable graphically. The mass fraction is largely unchanged from its $\mu = 0$ value for the plotted $\mu$ range. Evaluating it for $|\mu|>5 \times 10^{-2}$ is time-consuming and unnecessary  and the result would simply follow the dashed free diffusion line. The conclusion to be drawn however is clear: From the right plot of Fig.~\ref{fig:Pattison muN and beta} we see that $\beta$ in the $\mu <0$ case is always above the allowed value - allowing for any period of free diffusion always overproduces black holes according to the \ac{H-J} computation.   

\subsection{Density contrast versus curvature perturbation}
\begin{figure}[t!]
    \centering
    \includegraphics[width = .45\linewidth]{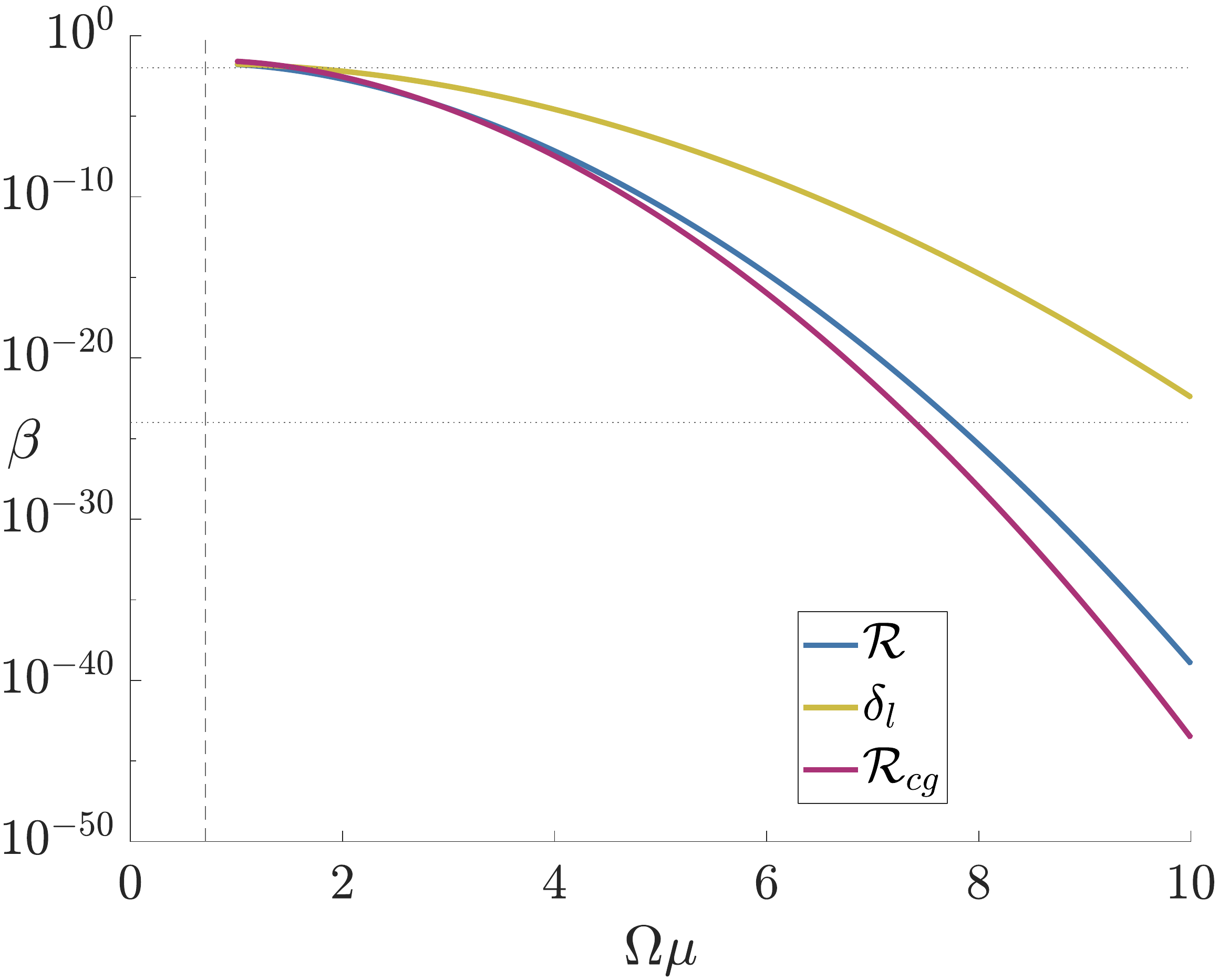}
    \includegraphics[width = .45\linewidth]{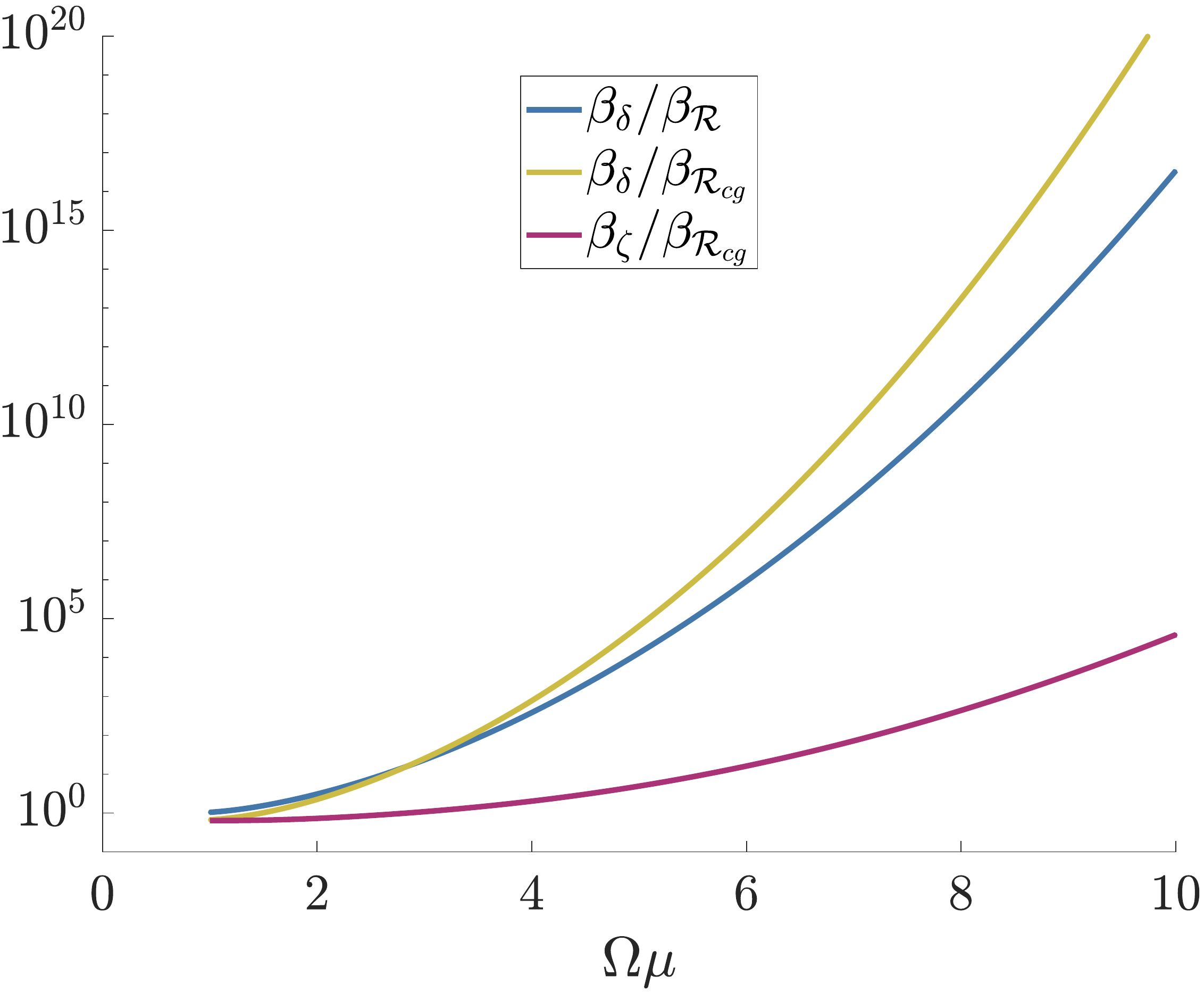}
    \caption[Density contrast formation comparison]{Comparison of the mass fraction (left) from different theoretical methods and the error between them (right). The horizontal dashed lines correspond to the weakest \cite{Papanikolaou2020} and strongest \cite{Carr2020} constraints on \ac{PBHs} and the vertical dashed line is when the classicality criterion is violated. }
    \label{fig:Beta_Riotta_compare}
\end{figure}
We now compare our results to those obtained using the $\delta N$ formalism for a period of \ac{USR} by Biagettia et.al \cite{Biagetti2021} and Luca \& Riotto \cite{DeLuca2022} as they have claimed to unambiguously compute the correct mass fraction for this period. As discussed earlier, it is clear that there are issues with using a universal threshold for the curvature perturbation $\mathcal{R}$ and it should instead be defined in terms of e.g. the (linear) density contrast $\delta_l$. In these works \cite{Biagetti2021, DeLuca2022} they compute $\mathcal{R}$ using the $\delta N$ formalism for a period of \ac{USR}. As they do not account for the momentum constraint we would disagree with their expression for $\rho (\mathcal{R})$:
\begin{eqnarray}
\rho (\mathcal{R}) = \dfrac{1}{\sqrt{2\pi}\sigma_{\mathcal{R}_g}}\exp \lsb -\dfrac{1}{18\sigma_{\mathcal{R}_g}^2}\lb e^{-3\mathcal{R}} -1\rb^2 -3\mathcal{R} \rsb \label{eq:Riotto_pdf}
\end{eqnarray}
defined in terms of the variance of the Gaussian curvature perturbation component $\sigma_{\mathcal{R}_g}^2$. In \cite{Biagetti2021, DeLuca2022} this is defined as the variance of the Gaussian inflaton fluctuations divided by the classical velocity at exit squared. If we look to modify (\ref{eq:Riotto_pdf}) however we find that this ratio does not correspond to the Gaussian curvature perturbation component which is instead given by (\ref{eq: classical power spectrum_flat}), $\sigma_{\mathcal{R}_g}^2 \approx 1/3\Omega^2\mu^2$. Using this identification we plot the mass fraction in the left plot of Fig.~\ref{fig:Beta_Riotta_compare} for (\ref{eq:Riotto_pdf}) in blue and (\ref{eq: Beta chie > 0 chisigma}) in red. We can see that the two curves are similar but there is some disagreement, especially as $\Omega\mu$ increases -- this enhancement is shown by the red curve on the right plot. More importantly however -- modifying the equations in \cite{Biagetti2021, DeLuca2022} to account for $\sigma_{\mathcal{R}_g}^2 \approx 1/3\Omega^2\mu^2$ -- we have also plotted the mass fraction as computed from the density contrast by the yellow line in the left plot of Fig.~\ref{fig:Beta_Riotta_compare}. It is clear that accounting for the density contrast generically \textit{enhances} the abundance of \ac{PBHs} formed as compared to using the curvature perturbation, sometimes by many orders of magnitude at larger $\Omega\mu$ -- see yellow and blue curves in the right plot. \\
It is therefore safe to say that while correctly accounting for the non-linear relationship between the density contrast and the curvature perturbation is critical for getting the \textit{correct} abundance of \ac{PBHs}, neglecting it only strengthens our conclusions about there being no period of diffusion dominated dynamics. This is because including the non-linear effects of the density contrast would only seek to \textit{enhance} the formation of \ac{PBHs}, not suppress it. 
\section{\label{sec: Inflection}Primordial Black Holes from a local inflection point}
\begin{figure}[t!]
    \centering
    \includegraphics[width=.75\textwidth]{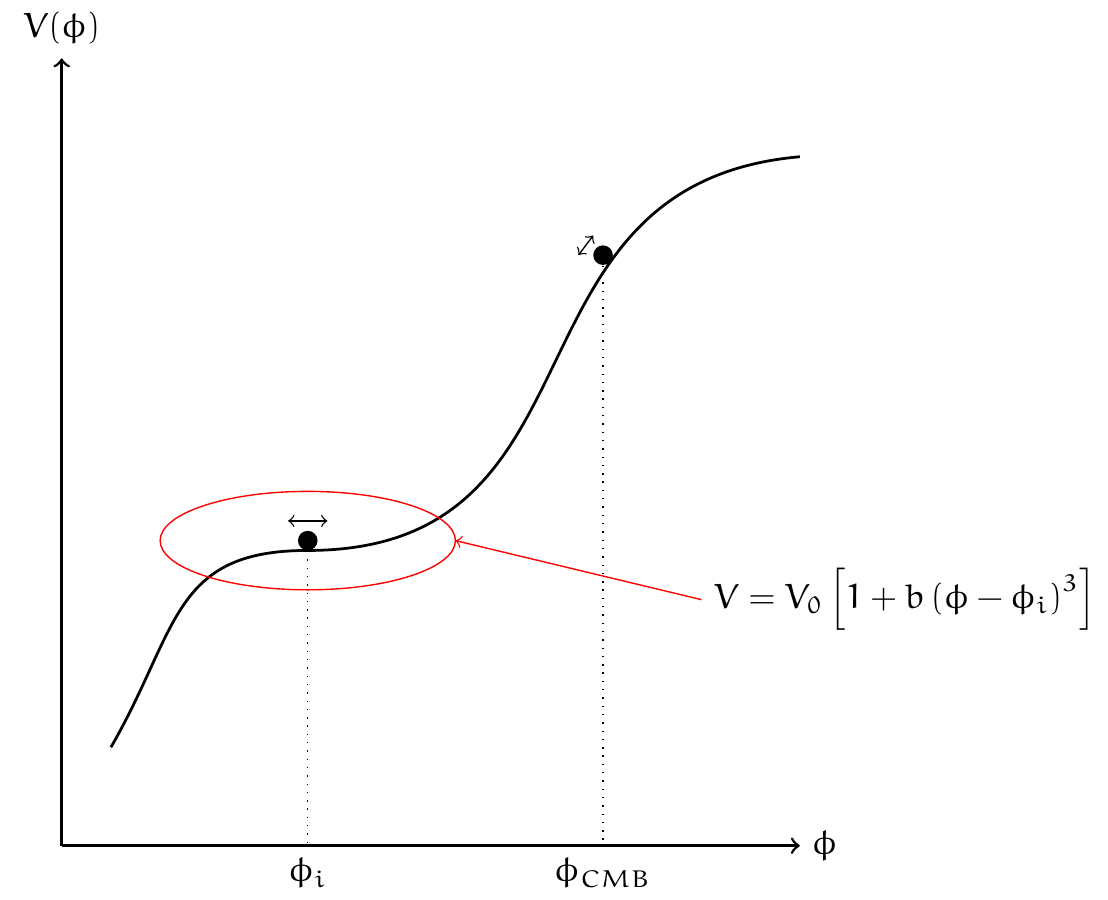}
    \caption{Expanding around an inflection point in the potential}
    \label{fig:Inflationary_Potential_PBHs_inflection}
\end{figure}

We now consider a smoother entry into a \ac{USR} regime which we will approximate locally as an inflection point. In contrast to other work on stochastic inflation and inflection points \cite{Ezquiaga2020, Pattison2021}, we will not define an ``effectively flat'' region around the inflection point and use our plateau results. Instead we will solve the \ac{H-J} equation exactly and see if this gives us qualitatively different results than our conclusions for a plateau.  Concretely, we can imagine Taylor expanding around the inflection point, $\phi_{i}$, -- see Fig.~\ref{fig:Inflationary_Potential_PBHs_inflection} -- to obtain:
\begin{eqnarray}
V = V_0\left[ 1 + b\left( \phi-\phi_{i}\right)^3\right] \label{eq: inflection potential}
\end{eqnarray}
with $V_0$ corresponding to the height of the inflection point. While we cannot obtain an analytic solution for $H$ using this potential, it is straightforward enough to obtain numerically from (\ref{eq: H-J equation}) subject to an initial condition. We will parametrise a family of initial conditions in terms of the value of the first slow parameter, $\varepsilon_1$, evaluated as the field enters our inflection point potential approximation (\ref{eq: inflection potential}). As we are looking to maximise stochastic effects we will choose the largest possible value of $V_0$ as allowed by the \ac{CMB} -- see e.g. \cite{Akrami2020} -- which is determined by $v_0 = 10^{-10}$.\\

\begin{figure}[t!]
\centering 
\includegraphics[width=.48\textwidth]{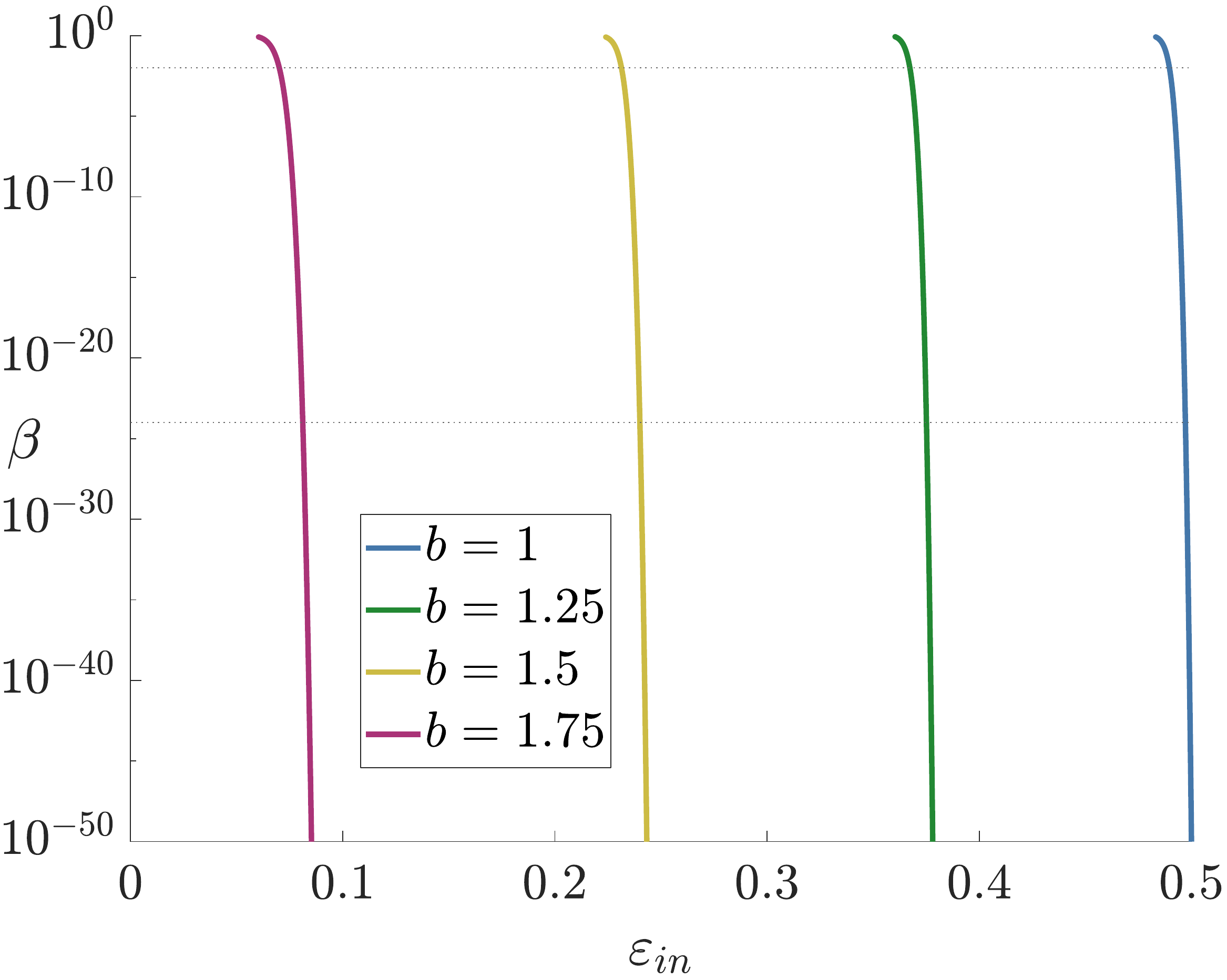}
\includegraphics[width=.48\textwidth]{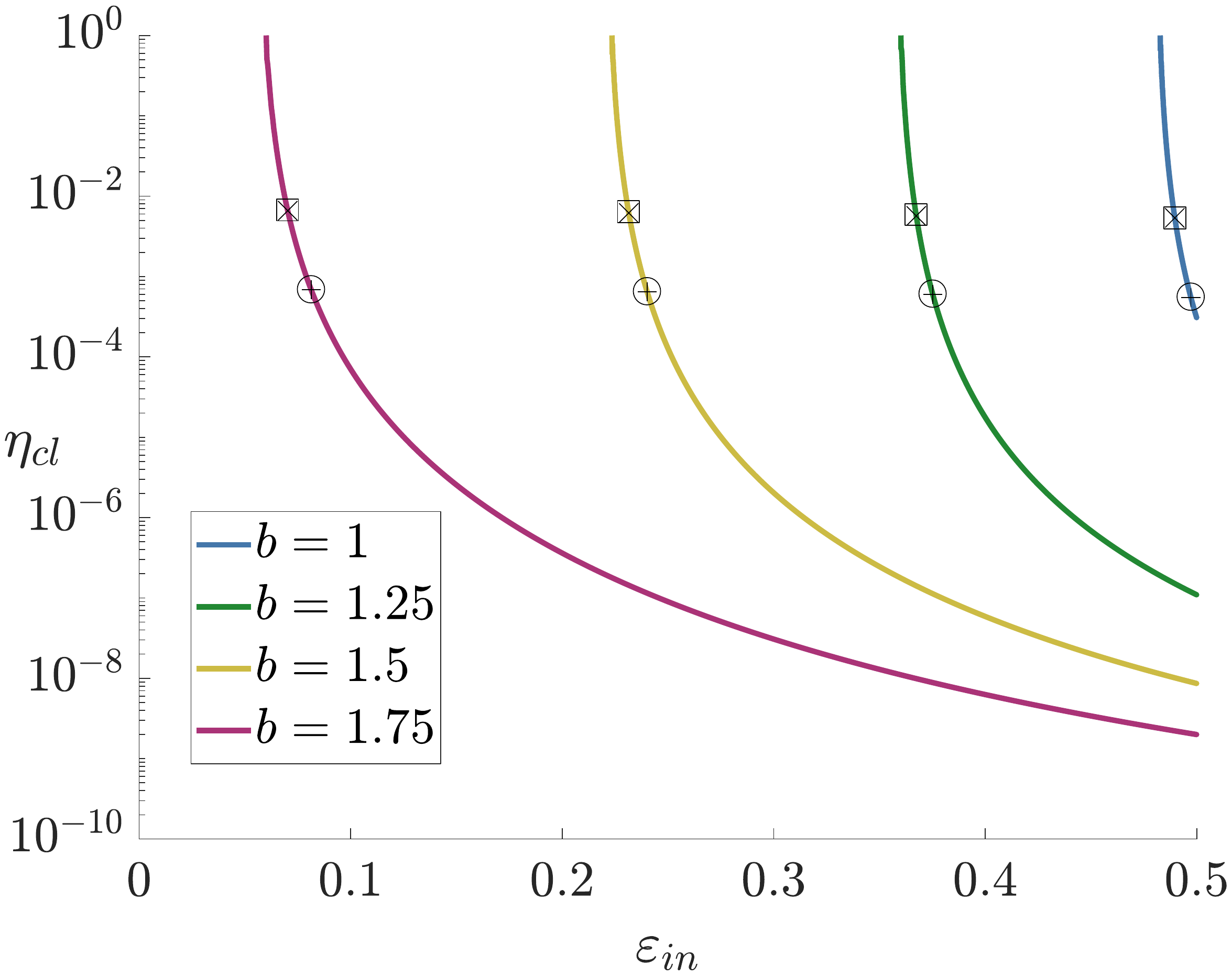}
\caption[$\beta $ and $\eta_{cl}$ for an inflection point]{\label{fig:inflec_powerandbeta.pdf} The left panel shows the mass fraction, $\beta$, as a function of the first Hubble \ac{SR} parameter as the inflaton enters the inflection point region for four different values of $b$. The solid and dashed lines represent the \ac{NLO} (\ref{eq:beta NLO}) and \ac{NNLO} (\ref{eq:beta NNLO}) approximations respectively, the dashed lines are indistinguishable from the solid lines as the \ac{NNLO} does not substantially differ from the \ac{NLO} in this case. The dotted horizontal lines show the weakest, $10^{-2}$, and strongest, $10^{-24}$ , constraints. The right panel shows the value of the classicality parameter $\eta_{cl}$ in the same parameter space. The boxes (circles) and crosses (pluses) correspond to the \ac{NLO} and \ac{NNLO} approximations respectively for $\beta$ being equal to the weakest (strongest) constraint $10^{-2}$ ($10^{-24}$). }
\end{figure}

The mass fraction, $\beta$, in both the \ac{NLO} and \ac{NNLO} approximations is plotted on the left panel of Fig.~\ref{fig:inflec_powerandbeta.pdf}. We can see that $\beta_{\scalebox{0.5}{$\mathrm{NLO}$}}$ and $\beta_{\scalebox{0.5}{$\mathrm{NNLO}$}}$ are essentially indistinguishable in this regime.\\

As we are expanding around the classical limit we require that the classicality parameter (\ref{eq: classicality criterion}) $\eta_{cl}\ll 1$ for these formulae to be valid. In the right panel of Fig.~\ref{fig:inflec_powerandbeta.pdf} we plot the classicality parameter and show that for both the \ac{NLO} and \ac{NNLO} approximation the weakest bounds on $\beta$ are violated at around $\eta_{cl} \sim 10^{-2}$ and the strongest bounds are violated at around $\eta_{cl} \sim 10^{-3}$. In both cases $\eta_{cl} \ll 1$ which corroborates the conclusions of the previous section's plateau analysis. This demonstrates that \emph{\ac{PBHs} will be generically overproduced before the inflaton can enter a quantum diffusion dominated regime} which corresponds to $\eta_{cl} \geq 1$. This is not to say that quantum diffusion effects aren't of significance and we expect them to play an important role in enhancing the tail of the distribution \cite{Ezquiaga2020}. While accounting for these effects is undoubtedly crucial to get a precise value of the mass fraction, we would only expect these effects to \textit{enhance} the abundance of \ac{PBHs} from the \ac{NNLO} computation -- see e.g. Fig.~\ref{fig:Beta_NNLO_flat} -- and therefore our statement that the inflaton never enters a quantum diffusion dominated regime is still valid.
\section{\label{sec: PBHConclusion}Conclusion}
\acresetall 
We begun this chapter by reviewing \ac{PBHs} concluding that they are still a viable \ac{DM} candidate in the asteroid mass range and can act as a valuable probe of inflation below \ac{CMB} scales. We discussed the formation of \ac{PBHs} and highlighted issues with the use of a universal threshold for the curvature perturbation in generality. However for our purposes in this chapter we demonstrated that the effects we were neglecting would only serve to enhance the number of \ac{PBHs} formed which further supports the main result of this chapter: \\

\noindent\textit{The inflaton cannot enter a period of quantum diffusion dominated dynamics without first overproducing Primordial Black Holes}. \\

\noindent A semi-classical approximation seems to always be adequate for observationally allowed inflationary dynamics. We demonstrated this by considering the evolution of the inflaton entering both a finite width plateau as well as a local inflection point from a previous \ac{SR} phase. We have updated the results of \cite{Prokopec2021} to obtain the probability density function of e-fold exit times, $\mathcal{N}$, in the stochastic inflation formalism (which is valid beyond \ac{SR}) for evolution on a finite width plateau, or more generally an \ac{USR} phase, taking into account the velocity of the field as it slides in the \ac{USR} region. \\

We showed that for a classical drift distance, $\Delta \phi_{cl}$, larger than the plateau width, $\Delta \phi_{pl}$, corresponding to the scenario where the classical inflaton momentum carries the field all the way through the plateau, the mass fraction of \ac{PBHs}, $\beta$, closely resembles a step function. Unless $\Delta \phi_{cl} \approx  \Delta \phi_{pl}$, corresponding to extremely small values of $\mu < 10^{-9}$, the mass fraction of \ac{PBHs} produced is negligible. On the contrary if $\mu$ is too small then \ac{PBHs} will be overproduced violating observational and theoretical constraints. This very sharp transition between negligible production and over-production of \ac{PBHs} occurs around $\mu \sim \dfrac{\sqrt{\pi}}{4\Omega}~e^{3\mathcal{R}_c} \simeq e^{3\mathcal{R}_c}\sqrt{\dfrac{\pi v_0}{8\varepsilon_{in}}}$. This would indicate that \ac{PBHs} will always be overproduced before $\mu$ can reach $0$, meaning that the inflaton is observationally forbidden from getting stranded on the plateau and exploring it via pure quantum diffusion. Furthermore, \ac{PBHs} are overproduced even before $\mu = \mu_{cl}$, corresponding to when quantum diffusion effects would be dominant even for an inflaton that is still classically drifting. \\

We examined the robustness of these constraints by taking the explicit case of $\Delta \phi_{cl} = \Delta \phi_{pl}$, or $\mu = 0$, and varying the cutoff $\mathcal{R}_c$. In this case, the mass fraction $\beta$ generically lies in the range $10^{-1}$ - $10^{-2}$ for all realistic values of the cutoff $\mathcal{R}_c$ the mass fraction remained constant as $\Omega$ was varied-- the exception to this is when $\Omega$ is very small, corresponding to super-Planckian inflationary energies, where $\beta \rightarrow 1$. We can therefore say that the case where the classical field momentum carries the field right to the edge of the plateau, $\Delta \phi_{cl} = \Delta \phi_{pl}$, will always overproduce \ac{PBHs}. Consequently, the scenario where $\Delta \phi_{cl} < \Delta \phi_{pl}$, corresponding to a period of free diffusion, is also forbidden as this subsequent phase would only serve to \textit{enhance} the curvature perturbations. We verified this assumption by extending the free diffusion results of \cite{Prokopec2021} to a finite width plateau which confirmed that the $\Delta \phi_{cl} < \Delta \phi_{pl}$ case always overproduces \ac{PBHs}. We also verified that accounting for the non-linear relationship between the curvature perturbation and the density contrast would only serve to \emph{enhance} the abundance of \ac{PBHs}. We therefore arrive at the conclusion stated above, namely that there can be no period of free diffusion during inflation without overproducing \ac{PBHs}.\\

When examining the more general setup of an inflection point we found that even in the Gaussian case, \ac{PBHs} are overproduced before the classicality criterion $\eta_{cl} < 1$ is violated. This agrees with the plateau result and further suggests that the distortion of the classical relationship between field values and wavenumbers explored in \cite{Ando2020} is never realised and that a late period of quantum diffusion which spoils the \ac{CMB} power spectrum is already ruled out by \ac{PBHs}. 

\chapter{Stochastic Spectator Fields and the Functional Renormalisation Group} 
\label{cha:RG_spect} 
\acresetall 
\vspace{0.5cm}
\begin{flushright}{\slshape    
Do you know in 900 years of time and space,\\ I've never met anyone who wasn't important before} \\ \medskip
--- The Doctor
\end{flushright}
\vspace{0.5cm}
\section{\label{sec:RGSpec_intro} Introduction}

In this chapter we pull together the \ac{FRG} techniques developed in chapters \ref{cha:fRG} \& \ref{cha:fRG EEOM} with the behaviour of scalar fields in the early universe using stochastic techniques as chapters \ref{cha:Inflation} \& \ref{cha:PBH} did for the inflaton field. In this chapter however we will not be considering the inflaton directly but instead the behaviour of another light scalar field in a de Sitter background as the \ac{FRG} techniques were developed for systems with a constant noise amplitude. \\

This chapter is all new research and since the original submission of this thesis has been published in JCAP \cite{Rigopoulos:2022gso}. We begin in section \ref{sec:Stoch_Spec} by reviewing the concept of a stochastic spectator in the early universe and outline how this behaviour can be related to the path integral formulation for stochastic behaviour described in chapter \ref{cha:stochastic_processes}. In section \ref{sec:RG_Stoch_Spec} we adapt the \ac{EEOM} developed in chapter \ref{cha:fRG EEOM} and introduce the \ac{EEOM} for the third central moment. In section \ref{sec:Cosmo_Spec} we discuss how we can obtain cosmological observables like the power spectrum and spectral tilt from \ac{FRG} computed quantities. In section \ref{sec:FPT_spec} we show how the \ac{FRG} can solve the \ac{FPT} problem for a spectator field and can predict quantities such as $\lan \mathcal{N} \ran $ and $\delta \mathcal{N}^2$ for several potentials. \\

The busy reader is directed to the main results of this chapter:
\begin{itemize}
    \item Equation (\ref{eq:FRG_threepoint}) and Fig.~\ref{fig:Three_point} for the \ac{FRG} prediction for the third central moment $\lan \sigma(\alpha)^3\ran_{C}$.
    \item Equation (\ref{eq:Power_spec_sig}) and Fig.~\ref{fig:spectral_tilt_compare} for how the \ac{FRG} can predict the spectral tilt and how features on the spectator's potential complicate inferences from observations.
    \item Equation (\ref{eq:rhoN_spect}) and Figs.~\ref{fig:FPT5H2_DW}, \ref{fig:FPTALL_mean} \& \ref{fig:FPTALL_Var} for how the \ac{FRG} can solve the \ac{FPT} for a spectator field.
\end{itemize}

\section{\label{sec:Stoch_Spec} The Stochastic Spectator}
In chapters \ref{cha:Inflation} \& \ref{cha:PBH} we focused on the behaviour of a light scalar field called the inflaton that could drive a period of accelerated expansion known as inflation. We assumed that the inflaton comprised the total energy budget of the universe. However there is reason to suspect that other scalar fields would be present during the inflationary period. For instance string theory predicts the presence of many extra light moduli fields \cite{Turok1988, Damour1996,Kachru2003} and unless we are dealing with Higgs inflation -- see e.g. \cite{Martin2014} -- we would expect the Higgs field to be present also. \\

To be concrete we introduce another scalar field $\sigma$ evolving in a potential $U(\sigma)$. Then we find that the first Friedmann equation (\ref{eq:Friedmann_Inflation}) must be appropriately modified:
\begin{eqnarray}
3M_{p}^2 H^2 = \dfrac{1}{2}\dot{\varphi}^2 + V(\varphi) + \dfrac{1}{2}\dot{\sigma}^2 + U(\sigma) \label{eq:Friedmann_sigma}
\end{eqnarray}
so that there are also contributions from the kinetic and potential energy of $\sigma$. We also obtain a Klein-Gordon equation for $\sigma$:
\begin{eqnarray}
\ddot{\sigma} +3H\dot{\sigma} + \dfrac{\mathrm{d}U(\sigma)}{\mathrm{d}\sigma} = 0 \label{eq:K-G_sigma}
\end{eqnarray}
If the energy scale of $\sigma$ is comparable to $\varphi$ -- i.e. $V(\varphi) \sim U(\sigma)$ -- then both fields are relevant for the dynamics of inflation and we are in a multi-field inflation scenario. Inflation will then proceed along a direction in the $(\varphi, \sigma)$ field space -- see \cite{Pinol2021} for an example of how to deal with this in the stochastic inflation approach. In this chapter we will consider the much more straightforward scenario where $U(\sigma) \ll V(\varphi)$ such that the full Friedmann equation (\ref{eq:Friedmann_sigma}) is well described by (\ref{eq:Friedmann_Inflation}) and the field $\sigma$ does not affect the inflationary dynamics. It is therefore referred to as a \emph{spectator} field. \\

At this stage the reader might be inclined to ask what the point of investigating the behaviour of a spectator field is, if (by definition) it cannot affect the behaviour of the inflaton. In the \emph{curvaton} scenario \cite{Linde1997,Moroi2001,Lyth2002,Moroi2002,Lyth2003,Lyth2005,Vennin2015a} the inflaton produces a subdominant contribution to the primordial density perturbation and the spectator field is the main contribution to the curvature perturbation hence the name. This is typically achieved by having the inflaton decay into radiation before the curvaton decays so that there is a period where the curvaton is the dominant contribution to the energy budget. In some cases this can even drive a short second period of inflation -- see \cite{Vennin2015a} for a full breakdown of all the possible configurations. A curvaton field could also be used as a means of measuring the duration of inflation \cite{Torrado2018}. Even if the spectator field is not the dominant contribution to the curvature perturbation observed in the \ac{CMB} a spectator field could still form \ac{PBHs} from field bubbles \cite{Maeso2022}. An inflationary period also affects the dynamics of any spectator field and if this field becomes important later on (e.g. the Higgs) it is useful to know how inflation sets the initial conditions for these spectator fields after inflation is over. \\

Having hopefully motivated that spectator fields are of interest we will proceed to outline how to deal with them in the stochastic formalism.

\subsection{Coarse-graining a spectator field}
The arguments outlined in section \ref{sec:stoch_infl} for the inflaton still (largely) hold for a spectator field. We can still split the spectator into long, $\sigma_{\scalebox{0.5}{$>$}}$, and short, $\sigma_{\scalebox{0.5}{$<$}}$, wavelength modes and inflation will still force the short wavelength modes to backreact on the long wavelength modes. The key difference is that as this field is a pure spectator this backreaction does not modify the \emph{geometry} of the background spacetime. Assuming we are dealing with overdamped motion (i.e. the \ac{SR} limit) it is straightforwardly shown from our previous arguments -- see also the original treatments \cite{Starobinsky1988,Starobinsky1994} -- that the equation of motion for $\sigma_{\scalebox{0.5}{$>$}}$ is:
\begin{eqnarray}
\dfrac{\mathrm{d}\sigma_{\scalebox{0.5}{$>$}}}{\mathrm{d}\alpha} &=& -\dfrac{1}{3H^2}\dfrac{\partial U(\sigma_{\scalebox{0.5}{$>$}})}{\partial \sigma_{\scalebox{0.5}{$>$}}} + \eta (\alpha) \label{eq:Langevin_sigma}\\
\lan \eta (\alpha) \eta (\alpha ')\ran &=& \dfrac{H^2}{4\pi^2} \delta (\alpha - \alpha ')\label{eq:sigma_noise}
\end{eqnarray}
where we have set $M_{\mathrm{p}}^2 =1$. In principle the value of the Hubble parameter will vary with time depending on the inflationary potential. We will choose the background inflationary potential to be of the plateau type so that $H$ is roughly constant and we can therefore assume that the spectator field exists in an exact de Sitter background. It has also been shown recently \cite{Cable2021,Cable2022} that the noise term for a scalar field in a de Sitter background is not given by (\ref{eq:sigma_noise}) unless the field is exactly massless. We will assume here that the field is sufficiently light such that (\ref{eq:sigma_noise}) is a good approximation, in any case the procedure we outline in this chapter is easily adapted to incorporate different values of the noise. To lighten the notation we will drop the subscript on $\sigma_{\scalebox{0.5}{$>$}}$ going forward and $\sigma$ can be assumed to refer to the coarse-grained long-wavelength field.\\

Ideally we would like to link (\ref{eq:Langevin_sigma}) with the dimensionless Langevin equation (\ref{eq:langevindimless}) from section \ref{sec:langevin_eqn} so we can utilise the \ac{FRG} formulae derived in Part \ref{part:Meso}. We could introduce the dimensionless parameters $v$ and $\tilde{H}$ from chapters \ref{cha:Inflation} \& \ref{cha:PBH} however this uses the Planck mass $M_{pl}$ as a reference scale which is much larger than the scales we are interested in. Instead, as in section \ref{sec:langevin_eqn}, we will introduce a reference Hubble scale $H_0$\footnote{Not to be confused with the value of the Hubble parameter today which is also often called $H_0$.} to define the dimensionless Hubble parameter $\hat{H}$ and in turn the other terms in equation (\ref{eq:Langevin_sigma}):
\begin{subequations}
\begin{align}
H &=  H_0 \hat{H} \label{eq:H_dimless}\\
	\sigma &=  \dfrac{H_0}{2\pi}\hat{\sigma} \label{eq:sigma_dimless} \\
	U(\sigma) &=  \dfrac{3\hat{H}^2H_0^4}{4\pi^2}\hat{U}(\hat{\sigma}) \label{eq:V_dimless}\\
	\eta (\alpha) &=  \dfrac{H_0}{2\pi}\hat{\eta} (\alpha )  \label{eq:eta_dimless}
\end{align}\label{eq:all_sigma_dimensionless}   
\end{subequations}
Notice that as the number of e-folds $\alpha$ is already dimensionless we do not need to rescale it. Also worth commenting on is that the dimensionless potential $\hat{U}$ depends on the dimensionless Hubble $\hat{H}$. This is because, unlike in the more simple case dealt with in Part \ref{part:Meso}, the friction coefficient and amplitude of the noise are both determined by the same parameter $H$ hence why our dimensionless potential effectively depends on the \textit{temperature} of the system. We will deal with this more when we come to section \ref{sec:RG_Stoch_Spec}. All this will give us the following dimensionless Langevin equation:
\begin{eqnarray}
\dfrac{\mathrm{d}\hat{\sigma}}{\mathrm{d}\alpha} &=& -\dfrac{\partial \hat{U}(\hat{\sigma})}{\partial \hat{\sigma}} + \hat{\eta} (\alpha) \label{eq:Langevin_sigma_dimless}\\
\lan \hat{\eta} (\alpha) \hat{\eta} (\alpha ')\ran &=&\hat{H}^2 \delta (\alpha - \alpha ')\label{eq:sigma_noise_dimless}
\end{eqnarray}
which we can readily identify with the thermal dimensionless Langevin equation (\ref{eq:langevindimless}) from part \ref{part:Meso} by making the transformation $\hat{H}^2 \rightarrow \Upsilon$. We will now drop the hat on $\sigma $ for notational simplicity. 
\subsection{The Spectator Path Integral}
As discussed in section \ref{sec:BM-PI} we can turn this stochastic problem into a path integral. Modifying the \ac{BPI}, (\ref{eq:BM_path_int}), to be in terms of our new dimensionless parameters yields:
\begin{subequations}
\begin{align}
\mathcal{P}(\sigma_{f}\vert \sigma_{i}) &= \int\mathcal{D}\sigma\mathcal{D}\tilde{\sigma} \mathcal{D}c\mathcal{D}\bar{c} \text{ exp}\left[-\mathcal{S}_{Spect}(\sigma,\tilde{\sigma},\bar{c},c)\right] \label{eq:TransProb_Sigma} \\
\mathcal{S}_{Spect}(\sigma,\tilde{\sigma},\bar{c},c) &= \int \text{d}\alpha\bigg[ \frac{\hat{H}^2}{2}\tilde{\sigma}^2 - i\tilde{\sigma}(\dot{\sigma}+ \hat{U}_{,\sigma}) - \bar{c}\left( \partial_{\alpha} +  \hat{U}_{,\sigma\sigma} \right)c \bigg] \label{eq: PDF action_Sigma}    
\end{align}
\end{subequations} 
where we have again introduced the response field $\tilde{\sigma}$ to $\sigma$ and the anticommuting variable $c$ and $\bar{c}$. We can then in analogy with (\ref{eq:LPAidentifications}) identify this with the \ac{SUSY} action (\ref{eq:SUSYClass})\footnote{N.B. that here $\varphi$ does not refer to the inflaton.}:
\begin{subequations}
\begin{align}
\sigma(\alpha) &\equiv  \hat{H}\,\varphi(\alpha) \\
	\hat{U}(\sigma) &\equiv  {\hat{H}^2} \, W(\varphi)  \\
	\tilde{\sigma} &\equiv  \dfrac{1}{\hat{H}}\,(i\dot{\varphi} - \tilde{F}) \\
	\bar{c}c &\equiv  i\bar{\rho}\rho   
\end{align}\label{eq:sigma_identifications}   
\end{subequations}
and as before construct the \ac{REA} at the cutoff ${\kappa} = \Lambda$ for \ac{LPA} like so:
\begin{eqnarray}
\Gamma_{\kappa}[\Sigma,\tilde{\Sigma},C,\bar{C}] = \int\mathrm{d}\alpha~\dfrac{\hat{H}^2}{2}\tilde{\Sigma}^2 -i\tilde{\Sigma}\lb \dot{\Sigma} + \partial_{\Sigma}U_{{\kappa}}\rb  -\bar{C}\dot{C} -\bar{C}C\partial_{\Sigma\Sigma}U_{\kappa} \label{eq:Gamma_k_LPA_Spect}
\end{eqnarray}
Which is written in terms of the mean fields 
\begin{subequations}
\begin{align}
\Sigma &= \dfrac{\delta \mathcal{W}[\mathcal{J}]}{\delta J_{\sigma}} = \lan \sigma\ran_{J_{\sigma}} \\
\tilde{\Sigma} &= \dfrac{\delta \mathcal{W}[\mathcal{J}]}{\delta J_{\tilde{\sigma}}} = \lan \tilde{\sigma}\ran_{J_{\tilde{\sigma}}} \\
C &= \dfrac{\delta \mathcal{W}[\mathcal{J}]}{\delta \vartheta} = \lan c \ran_{\vartheta} \\
\bar{C} &= \dfrac{\delta \mathcal{W}[\mathcal{J}]}{\delta \bar{\vartheta}} = \lan \bar{c}\ran_{\bar{\vartheta}}    
\end{align}   \label{eq:Spect_mean_fields} 
\end{subequations}
which in turn depend on the currents:
\begin{equation}
 \int \mathrm{d}\alpha \,{\mathcal{J}} \Vec{\Sigma} \equiv \int \mathrm{d}\alpha \left(  J_\sigma \sigma + J_{\tilde{\sigma}} \tilde{\sigma} + \bar{c}\vartheta + \bar{\vartheta}c    \right) \label{eq:currents_spect}
\end{equation}
We now have the appropriate ingredients to apply the \ac{FRG} technology to the problem at hand. 

\section{\label{sec:RG_Stoch_Spec} The Effective Equations of Motion for a Spectator Field}
Having successfully linked the dynamics of the spectator field with \ac{SUSY} QM we can again utilise the appropriate flow equations as derived in sections \ref{sec:LPA_BM_FLOW} \& \ref{sec:WFR_BM_FLOW} for the \ac{LPA} and \ac{WFR} approximations respectively:
\begin{eqnarray}\label{eq:dVsigma/dktilde2}
\partial_{{{\kappa}}}\bar{U}_{{{\kappa}}}(\Sigma) &=& \dfrac{3\hat{H}^4}{4}\cdot\dfrac{1}{{{\kappa}}+ \partial_{\Sigma \Sigma}\bar{U}_{{{\kappa}}}(\Sigma)} \\
\partial_{{{\kappa}}}\zeta_{,\Sigma} &=& \dfrac{3\hat{H}^6}{2}\cdot\dfrac{\mathcal{P}}{\zeta_{,\Sigma}\cdot\mathcal{D}^2 } \label{eq:dzetax_sigma/dktilde}\\
\mathcal{D} &\equiv & \bar{U}_{,\Sigma\Sigma} + {\kappa}\,\zeta_{,\Sigma}^{2}, \quad \mathcal{P} \equiv \dfrac{4\zeta_{,\Sigma\Sigma} \bar{U}_{,\Sigma\Sigma\Sigma}}{\mathcal{D}} - \left( \zeta_{,\Sigma\Sigma}\zeta_{\Sigma}\right)_{,\Sigma} - \dfrac{3\zeta_{,\Sigma}^{2}\bar{U}_{,\Sigma\Sigma\Sigma}^{2}}{4\mathcal{D}^2} 
\end{eqnarray}
where we have rescaled the potential like so:
\begin{eqnarray}
\bar{U}(\Sigma ) = 3\hat{H}^2\hat{U} (\Sigma) \label{eq:Ubar_relation}
\end{eqnarray}
and $\kappa \in [0, ~ 3\hat{H}^2\Lambda]$. Equations (\ref{eq:dVsigma/dktilde2}) and (\ref{eq:dzetax_sigma/dktilde}) can be solved as outlined in section \ref{sec:FRG SOL} and we will do so for a few different potentials. Namely the $\sigma^2$ plus two bumps (\ref{eq:x^2_2bumpsdefn}), the doublewell (\ref{eq:Vdoublewell}) and the polynomial (\ref{eq:Vpoly}) potentials:
\begin{eqnarray}
&& \sigma^2 \text{ plus two bumps: }  \bar{U}(\sigma) = \sigma^2 + \dfrac{3}{2}\lcb \exp \lsb-\dfrac{\lb \sigma - 1\rb^2}{0.06} \rsb + \exp \lsb -\dfrac{\lb \sigma + 1\rb^2}{0.06} \rsb \rcb \label{eq:varsigma^2_2bumpsdefn}\\
&& \text{ Doublewell: }   \bar{U}(\sigma) = -\sigma^2 + \dfrac{\sigma^4}{4}   \label{eq:sigma_DW_defn} \\
    && \text{ Polynomial: }   \bar{U}(\sigma) = \sigma + \dfrac{\sigma^2}{2} + \dfrac{2\sigma^3}{3} + \dfrac{\sigma^4}{4}   \label{eq:sigma_poly_defn}
\end{eqnarray}
\subsection{Equilibrium}
As before the equilibrium position $\Sigma_{eq}$ is given by the minimum of the effective potential $\bar{U}_{{\kappa}=0}$. The connected two point function is straightforwardly modified from section \ref{sec:equi_2point} so that the appropriate solution to (\ref{eq:2pointfuncequi}) providing the connected correlation function at equilibrium is  
\begin{eqnarray}
\textbf{Cov}_{eq}(\sigma(\alpha_1)\sigma(\alpha_2)) = G_{eq}(\alpha_1, \alpha_2)  &=& \dfrac{3\hat{H}^4}{2\bar{U}_{,\Sigma\Sigma}\vert} e^{-\lambda|\alpha_1-\alpha_2|} \label{eq:2-pointfunc_sigma} \\
\textbf{Var}_{eq}(\sigma (\alpha)) = G_{eq}(\alpha, \alpha) &=& \dfrac{3\hat{H}^4}{2\bar{U}_{,\Sigma\Sigma}\vert}
\label{eq:equal2pt_sigma}
\end{eqnarray}
where $\lambda$ is given by 
\begin{eqnarray}\label{eq:lambdadef_sigma}
\lambda^2 \equiv  
\begin{cases}
\dfrac{\bar{U}_{,\Sigma\Sigma}^{2}\vert}{9\hat{H}^4}, & \text{for \ac{LPA}}\\[10pt]
\dfrac{\bar{U}_{,\Sigma\Sigma}^{2}\vert}{9\hat{H}^4\zeta_{,\Sigma}^{4}\vert}, & \text{for \ac{WFR}}
\end{cases} 
\end{eqnarray}
and $\vert$ means the quantity has been evaluated at $\kappa = 0$ and the equilibrium value of $\Sigma$. Note how the non-connected correlators follow the same behaviour:
\begin{eqnarray}
\lan \sigma (\alpha_1) \sigma ( \alpha _2) \ran_{eq} &=& G_{eq}(\alpha_1 , \alpha_2) + \lan \sigma (\alpha_1)\ran_{eq} \lan \sigma (\alpha_2)\ran_{eq} \\
\Rightarrow \lan \sigma (\alpha_1) \sigma ( \alpha _2) \ran _{eq} &=& \dfrac{3\hat{H}^4}{2\bar{U}_{,\Sigma\Sigma}\vert}  e^{-\lambda|\alpha_1-\alpha_2|} + \Sigma_{eq}^2 \label{eq:2pointfunc_equi_spect_dimless}
\end{eqnarray}
If we rearrange (\ref{eq:lambdadef_sigma}) so that\footnote{N.B. for \ac{LPA} $\zeta_{,\Sigma}^{2}$ is unity.} $\bar{U}_{,\Sigma\Sigma} = 3\hat{H}^2\lambda \zeta_{,\Sigma}^{2}\vert $ and then restore (\ref{eq:2pointfunc_equi_spect_dimless}) to the true physical parameters we obtain:
\begin{eqnarray}
\lan \sigma (\alpha_1) \sigma ( \alpha _2) \ran _{eq} &=&  \dfrac{H^2}{8\pi^2\lambda\zeta_{,\hat{\Sigma}}\vert}  e^{-\lambda|\alpha_1-\alpha_2|}+ \Sigma_{eq}^2 \label{eq:2pointfunc_equi_spect_dimfull}
\end{eqnarray}

\subsection{Non-Equilibrium}
Given that it can take many e-folds for the system to relax to the de Sitter equilibrium it is worth examining the non-equilibrium behaviour of the system at hand. Returning to dimensionless parameters, as before we can determine the evolution of the average field value $\Sigma$ by a simple first order differential equation:
\begin{eqnarray}
\dot{\Sigma} = -\tilde{U}_{,\Sigma}(\Sigma) \label{eq:EEOM_sigma}
\end{eqnarray} 
where again we have introduced the \textit{effective dynamical potential} $\tilde{U}$ defined by 
\begin{eqnarray}
\tilde{U}_{,\Sigma}(\Sigma) \equiv 
\begin{cases}
\dfrac{\bar{U}_{,\Sigma}({\kappa} = 0, \Sigma)}{3\hat{H}^2}, & \text{ for \ac{LPA}}  \\[10pt]
\dfrac{\bar{U}_{,\Sigma}({\kappa} = 0, \Sigma)}{3\hat{H}^2\zeta_{,\Sigma}^{2}({\kappa} = 0, \Sigma)}, & \text{ for \ac{WFR}} 
\end{cases}\label{eq: Vtilde_sigma}
\end{eqnarray}
We know from equation (\ref{eq:EEOM Covariance}) that the covariance is written in terms of a normalised function $\tilde{Y}_2$ which we identify with $f(\alpha)$. We can therefore write the non-connected two point function like so:
\begin{eqnarray}
\lan \sigma (\alpha_1) \sigma ( \alpha _2) \ran = G(\alpha_1,\alpha_1)f(\alpha_2 - \alpha_1 ) + \Sigma (\alpha_1) \Sigma(\alpha_2)
\end{eqnarray}

\subsection{Three-point function}
In part \ref{part:Meso} we only derived the \ac{FRG} predictions for the one-point and connected two-point function. Here we will extend our results to the three point function by use of the formula -- see e.g. pages 381-382 \cite{Peskin:1995ev}:
\begin{eqnarray}
\lan \sigma_a \sigma_b \sigma_c \ran_{C} = \int \mathrm{d}u\mathrm{d}v\mathrm{d}w  ~\dfrac{\delta^3\Gamma [\Sigma]}{\delta \Sigma_u \delta \Sigma_v\delta \Sigma_w}~G_{au}G_{bv}G_{cw}
\end{eqnarray}
where subscripts indicate the argument and $G_{ab} \equiv \lan \sigma(a) \sigma(b) \ran_{C}$. The third functional derivative of the \ac{EA} can be computed from (\ref{eq:Gamma_k_LPA_Spect}) as:
\begin{eqnarray}
\dfrac{\delta^3\Gamma [\Sigma]}{\delta \Sigma_u \delta \Sigma_v\delta \Sigma_w} &=& \dfrac{1}{\hat{H}^2}\lsb 2\zeta_{,\Sigma}\zeta_{,\Sigma\Sigma} \partial_{ww} - \mathcal{W}(\Sigma)\rsb \delta (w-u) \delta (w-v) \\
\mathcal{W}(\Sigma) &\equiv & 3\dfrac{\hat{U}_{,\Sigma\Sigma}\hat{U}_{,\Sigma\Sigma\Sigma}}{\zeta_{,\Sigma}^2} + \dfrac{\hat{U}_{,\Sigma}\hat{U}_{,\Sigma\Sigma\Sigma\Sigma}}{\zeta_{,\Sigma}^2} - 6\dfrac{\hat{U}_{,\Sigma\Sigma}^2 \zeta_{,\Sigma\Sigma}}{\zeta_{,\Sigma}^3} \nonumber \\
&& -6\dfrac{\hat{U}_{,\Sigma}\hat{U}_{,\Sigma\Sigma\Sigma}\zeta_{,\Sigma\Sigma}}{\zeta_{,\Sigma}^3} -9\dfrac{\hat{U}_{,\Sigma}\hat{U}_{,\Sigma\Sigma}\zeta_{,\Sigma\Sigma\Sigma}}{\zeta_{,\Sigma}^3} -2\dfrac{\hat{U}_{,\Sigma}^2\zeta_{,\Sigma\Sigma\Sigma\Sigma}}{\zeta_{,\Sigma}^3} \nonumber \\
&&+18\dfrac{\hat{U}_{,\Sigma}\hat{U}_{,\Sigma\Sigma}\zeta_{,\Sigma\Sigma}^2}{\zeta_{,\Sigma}^4} + 9 \dfrac{\hat{U}_{,\Sigma}^2 \zeta_{,\Sigma\Sigma}\zeta_{,\Sigma\Sigma\Sigma}}{\zeta_{,\Sigma}^4} -12\dfrac{\hat{U}_{,\Sigma}^2\zeta_{,\Sigma\Sigma}^3}{\zeta_{,\Sigma}^5}
\end{eqnarray}
If we modify appropriately the initial conditions outlined in section \ref{sec:equi_2point} we can write the connected two point function like so:
\begin{eqnarray}
G_{ab} = \Theta (b-a) G_{aa}\dfrac{G_{0b}}{G_{0a}}
\end{eqnarray}
which we can combine to give the following \ac{EEOM} for the third central moment:
\begin{empheq}[box = \fcolorbox{Maroon}{white}]{equation}
\begin{split}
\lan \sigma (\alpha)^3\ran_{C} = \lb\dfrac{G_{\alpha\alpha}}{G_{0\alpha}}\rb^3 \int_{\alpha}^{\infty}\mathrm{d}\tilde{\alpha}~ \Bigg\{ &\dfrac{6\zeta_{\Sigma}\zeta_{\Sigma\Sigma}}{\hat{H}^2}\lsb \lb G_{0\tilde{\alpha}}\rb^2 \partial_{\tilde{\alpha}\tilde{\alpha}}G_{0\tilde{\alpha}} + 2G_{0\tilde{\alpha}}\lb \partial_{\tilde{\alpha}}G_{0\tilde{\alpha}}\rb^2 \rsb  \\
&  - \dfrac{\lb G_{0\tilde{\alpha}}\rb^3}{\hat{H}^2}\mathcal{W}(\Sigma (\tilde{\alpha}))\Bigg\}  \label{eq:FRG_threepoint}
\end{split}
\end{empheq}

\begin{figure}[t!]
    \centering
    \includegraphics[width = 0.45\linewidth]{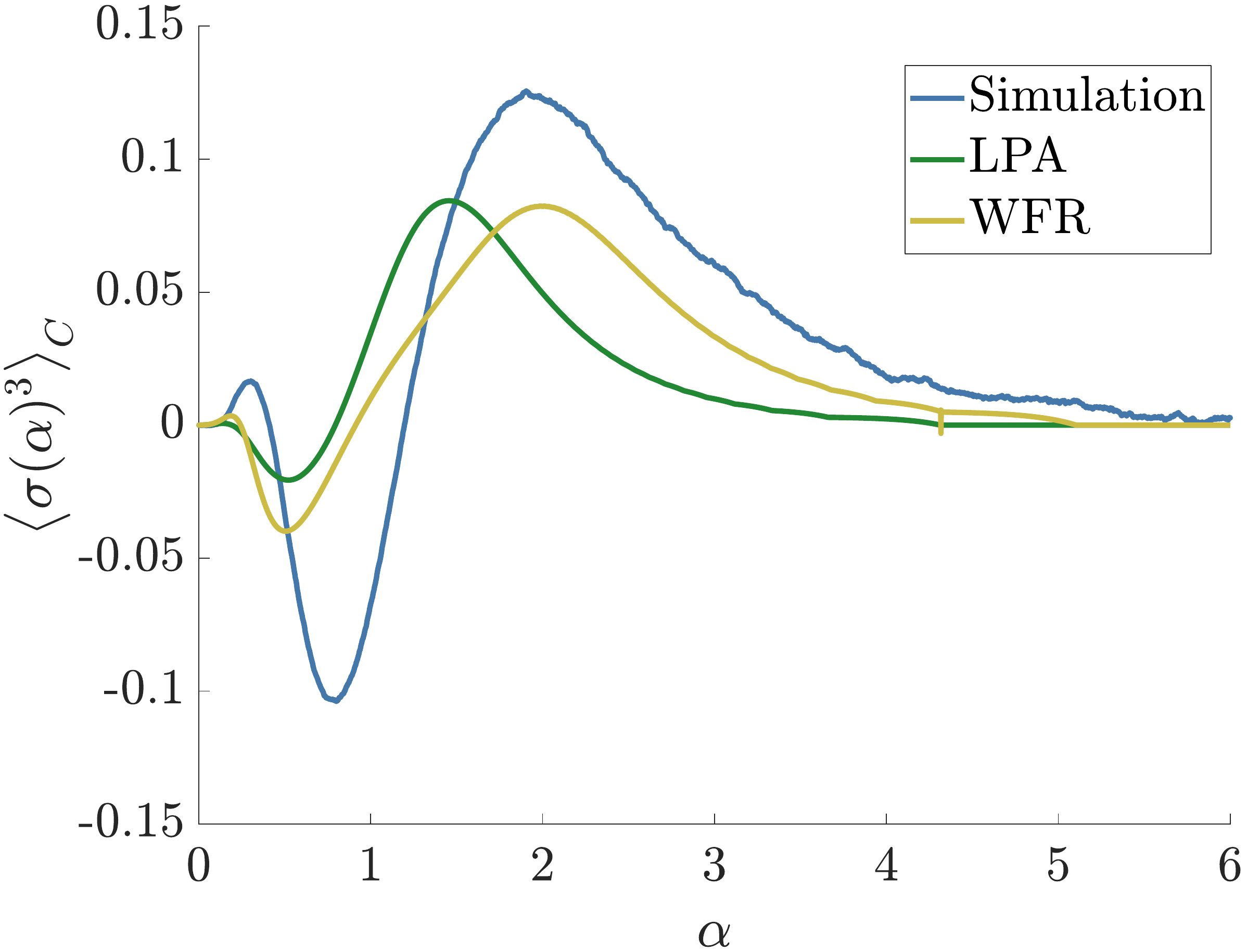}
    \includegraphics[width = 0.45\linewidth]{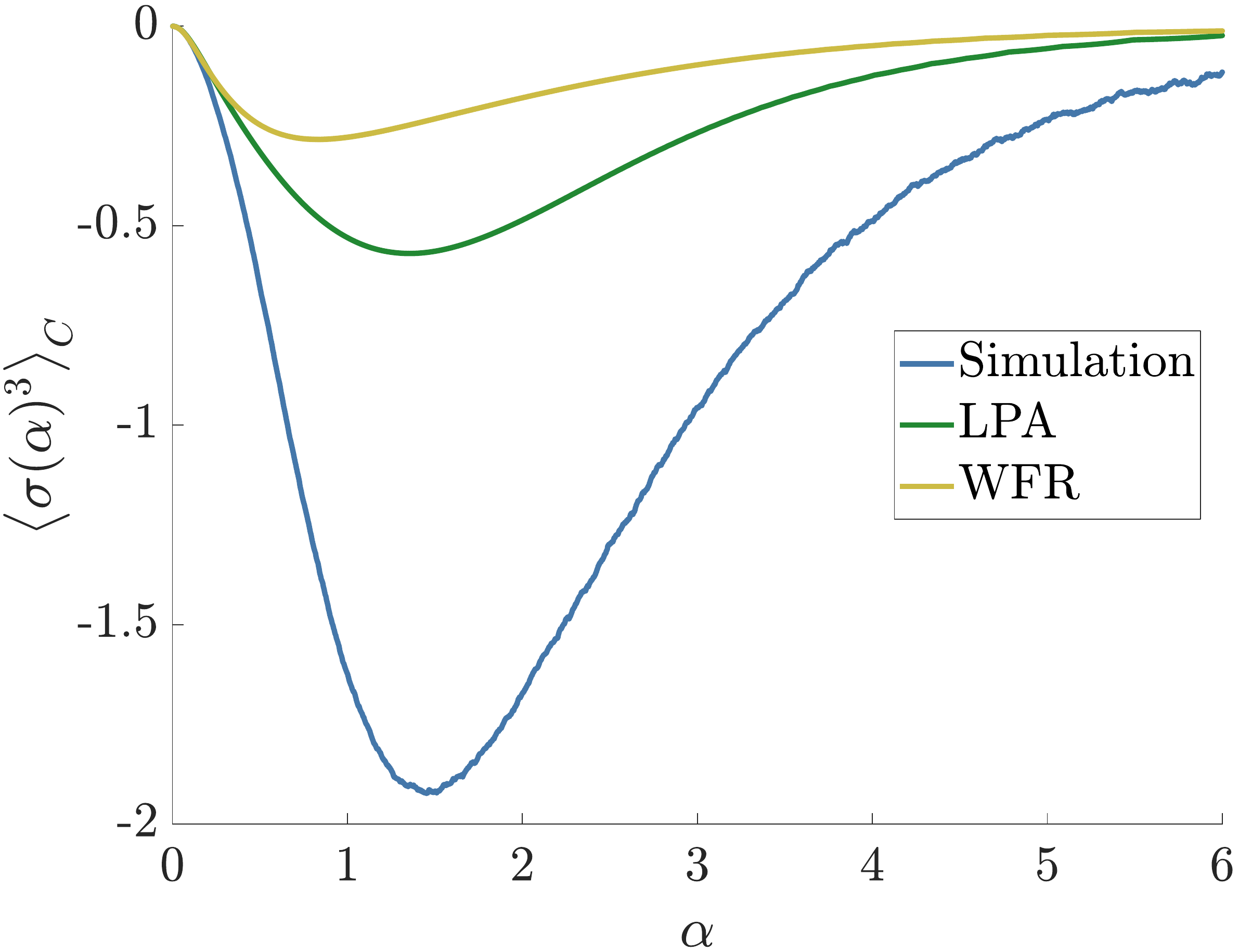}
    \caption[Evolution of the third central moment for the doublewell and quadratic potential with bumps]{Evolution of the third central moment $\lan \sigma (\alpha)^3\ran$ in a $\sigma^2$ plus two bumps potential for $\hat{H}^2 = 1.5$ (left) and the doublewell potential at $\hat{H}^2 = 5$ (right) as computed by direct numerical simulation and from the \ac{FRG} \ac{EEOM} (\ref{eq:FRG_threepoint}).}
    \label{fig:Three_point}
\end{figure}
In Fig.~\ref{fig:Three_point} we plot the solution to (\ref{eq:FRG_threepoint}) for the $\sigma^2$ plus two bumps potential (left plot) and doublewell (right plot) for favourable choices of $\hat{H}^2$ and compare to direct numerical simulations. It is clear that the \ac{FRG} can capture the qualitative evolution of the third central moment reasonably well, however it is not very precise and we were unable to improve the accuracy for different choices of initial conditions or $\hat{H}^2$. It would therefore seem that we have reached the limit of reasonable accuracy that the current \ac{FRG} procedure can provide. It is possible that one needs to go to higher order in the derivative expansion to get accurate results for the third central moment, or perhaps one should instead look at a vertex expansion \cite{Morris1994} of the \ac{FRG} approach instead.  \\
The equilibrium  limit is much simpler and assuming that $t_4 \geq t_3 \geq t_2 \geq t_1$ can be written as:
\begin{eqnarray}
\left\langle \sigma(\alpha_1)\sigma(\alpha_2)\sigma(\alpha_3) \right\rangle_{C}  &=& \left\langle \sigma(\alpha_1)^3 \right\rangle_{C} e^{-\lambda (2\alpha_3 -\alpha_2-\alpha_1)} \label{3-pointfunc}\\
\left\langle \sigma(\alpha_1)^3 \right\rangle_{C} & = & -\left\langle \sigma(\alpha_1)^2 \right\rangle_{C}^{3}\dfrac{\mathcal{W}(\Sigma_{eq})}{3\lambda}
\end{eqnarray}
with the potential $\mathcal{W}$ evaluated at the equilibrium point $\chi_{eq}$.


\section{\label{sec:Cosmo_Spec} Cosmological Observables}

Because of de Sitter invariance \cite{Starobinsky1994} any correlator of a scalar observable $\mathcal{O}(\sigma)$ can only depend on the de Sitter invariant quantity:
\begin{eqnarray}
y = \text{cosh} \lb \alpha_1 - \alpha_2  \rb -\dfrac{H^2}{2}\exp \lb \alpha_1 + \alpha_2  \rb \vert \vec{r}_1 - \vec{r}_2\vert^2
\end{eqnarray}
where $\vec{r}_1$ and $\vec{r}_2$ are comoving position vectors. Provided $\vert y\vert \gg 1$ then both time-like and space-like separations can be expressed like so:
\begin{eqnarray}
\lan \mathcal{O}(\sigma (\vec{r}_1, \alpha_1)) \mathcal{O}(\sigma (\vec{r}_2, \alpha_2))\ran = \lan \mathcal{O}(\sigma(0)) \mathcal{O}(\sigma (H^{-1}\ln \vert 2y -1\vert  )) \ran
\end{eqnarray}
where we can see that the right hand side is the spatial coincidence $(\vec{r}_1 = \vec{r}_2)$, temporal correlation function that can in principle be obtained through the stochastic approach i.e. equation (\ref{eq:Langevin_sigma}). To simplify things we consider correlators at equal time for the observable $\mathcal{O} = \sigma$:
\begin{eqnarray}
\lan \sigma (\vec{x}_1,\alpha) \sigma (\vec{x}_2,\alpha)\ran = \lan \sigma(0) \sigma  (2\ln \lb \left| \vec{x}_1 - \vec{x}_2\right| H\rb /H) \ran \label{eq:equal_time=equal_space}
\end{eqnarray}
which is valid at distances $\vert \vec{x}_1 -\vec{x}_2\vert \gg 1/H$ and $\vec{x} = a\vec{r}$ is the physical, non-comoving coordinate and $sigma$ is now \emph{dimensionful}. As discussed in chapter \ref{cha:Inflation} in cosmology equal-time correlation functions are often described by their power spectrum:
\begin{eqnarray}
\mathcal{P}_{\sigma}(k) = \dfrac{k^3}{2\pi^2} \int \mathrm{d}^3x~e^{-i\Vec{k}\cdot \Vec{x} }\lan \sigma (\Vec{x}_1,\alpha) \sigma (\Vec{x}_2,\alpha)\ran \label{eq:power_spec_sigma_defn}
\end{eqnarray}
where here $k$ is the Fourier transform of position. The question now is how the \ac{FRG} can compute the RHS of (\ref{eq:equal_time=equal_space}) and therefore the power spectrum. 

\subsection{Power Spectrum from an equilibrium distribution}
\begin{table}[t!]
	\centering
		\begin{tabular}{l l l l l}
			\toprule
			$\hat{H}^2$ & \ac{LPA} & \ac{WFR} & Sim & Bare\\
			\midrule
			5 & 0.7902  & 0.7874 & 0.7841 & 0.8\\
			2 & 1.9916 & 1.8427 &  1.8703 & 2\\
			1 & 6.4109 & 5.6368 & 5.3684 & 4\\
			0.5 & 9.3241 & 9.2108 & 9.2849 & 8\\
			\bottomrule 
		\end{tabular}
	 \caption[Value of spectral tilt as computed by \ac{FRG} and by direct simulation for a harmonic potential with Gaussian bumps.]{Value of the (shifted) spectral tilt $n_{\sigma} - 1$ as computed by the \ac{LPA}, \ac{WFR} and by direct numerical simulation for the $\sigma^2$ plus two bumps potential. The simulation values were generated by averaging over 50,000 runs. In the final column we have included, for comparison, what the prediction from the underlying harmonic potential would be without the Gaussian bumps.}
	  	 \label{tabel:spectral_tilt}
\end{table}
In equilibrium the \ac{FRG} predicts that the two point function follows a simple exponential decay (\ref{eq:2pointfunc_equi_spect_dimfull}) -- for simplicity we assume a symmetric potential such that $\Sigma_{eq} = 0$. We can substitute this into the RHS of (\ref{eq:equal_time=equal_space}) to obtain:
\begin{eqnarray}
\lan \sigma (x_1,\alpha) \sigma (x_2,\alpha)\ran = \dfrac{H^2}{8\pi^2\lambda\zeta_{,\hat{\Sigma}}\vert}  \dfrac{1}{\lb \left| \Vec{x}_1 - \Vec{x}_2\right| H\rb^{2\lambda}}
\end{eqnarray}
Which suggests a power law form:
\begin{eqnarray}
\lan \sigma (\Vec{x}_1,\alpha) \sigma (\Vec{x}_2,\alpha)\ran &=& \dfrac{A_{\sigma}}{\lb \left| \Vec{x}_1 - \Vec{x}_2\right| H\rb^{n_{\sigma} -1}} \\
A_{\sigma} &=& \dfrac{H^2}{4\pi^2}\dfrac{1}{\zeta_{,\hat{\Sigma}}\vert \lb n_{\sigma} -1\rb}  \label{eq:powerspec_ampli_sigma}\\
n_{\sigma} &=& 1 + 2\lambda \label{eq:spectral_tilt_sigma}
\end{eqnarray}
Using the definition of the power spectrum (\ref{eq:power_spec_sigma_defn}) we obtain\footnote{It is worth noting that this simple form assumes that $\vert n_{\sigma} -1 \vert \ll 1$ otherwise the power spectrum is more generally given by:
\begin{eqnarray}
\mathcal{P}_{\sigma}(k) &=& \dfrac{2}{\pi}A_{\sigma}\Gamma \lsb 2 - 2\lambda \rsb \sin \lb \pi\lambda\rb \lb\dfrac{k}{H}\rb^{n_{\sigma} -1} 
\end{eqnarray}}
\begin{eqnarray}
\mathcal{P}_{\sigma}(k) = A_{\sigma}(n_{\sigma} -1) \lb\dfrac{k}{H}\rb^{n_{\sigma} -1} 
\end{eqnarray}
which suggests that $A_{\sigma}$ and $n_{\sigma}$ are the amplitude of the power spectrum and the spectral tilt respectively for $\sigma$. In terms of \ac{FRG} quantities the power spectrum is given by:
\begin{empheq}[box = \fcolorbox{Maroon}{white}]{equation}
\mathcal{P}_{\sigma}(k) =   \dfrac{H^2}{4\pi^2\zeta_{,\hat{\Sigma}}\vert}  \lb\dfrac{k}{H}\rb^{2\lambda}\label{eq:Power_spec_sig}
\end{empheq}
\begin{figure}[t!]
    \centering
    \includegraphics[width = 0.45\linewidth]{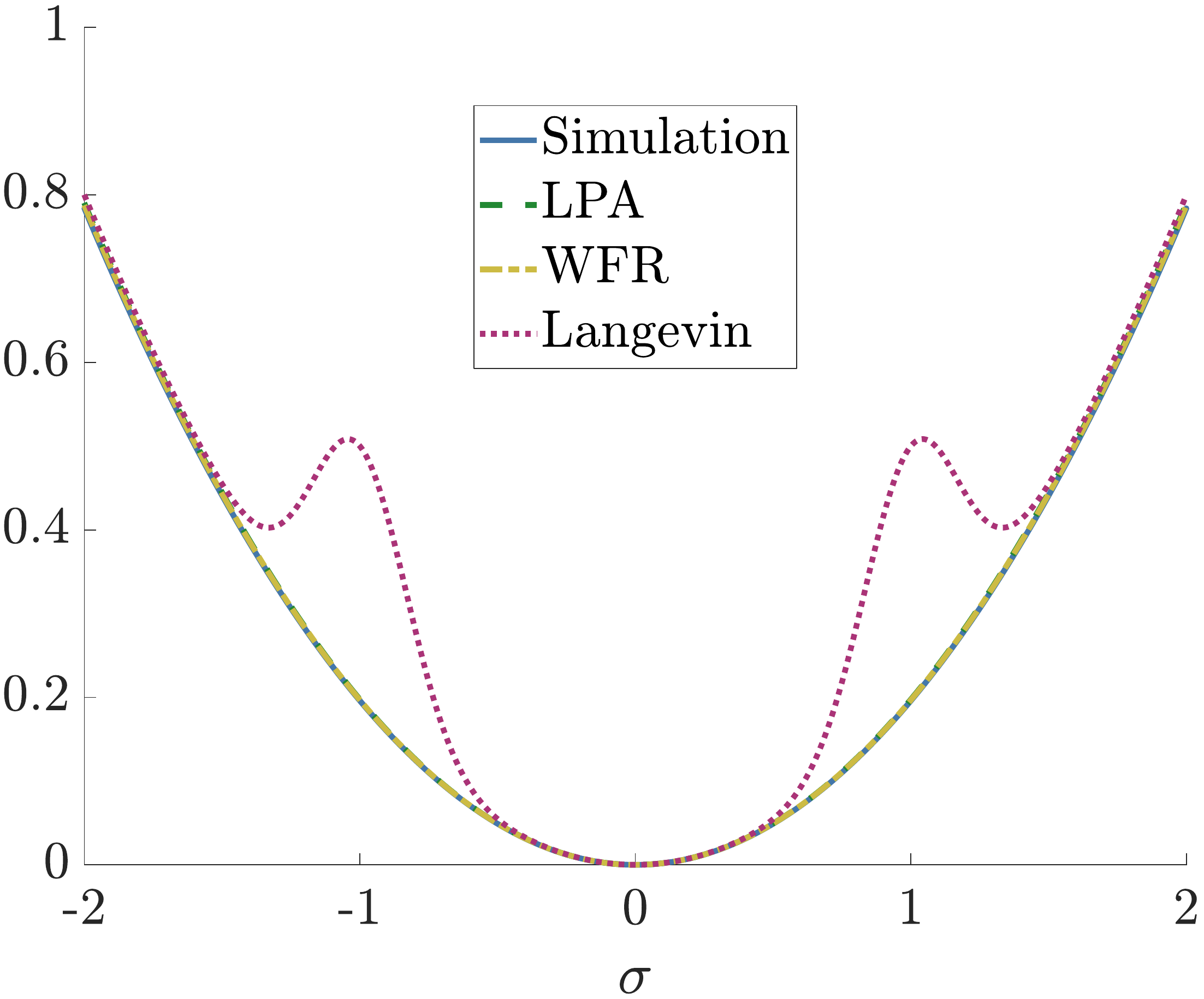}
    \includegraphics[width = 0.45\linewidth]{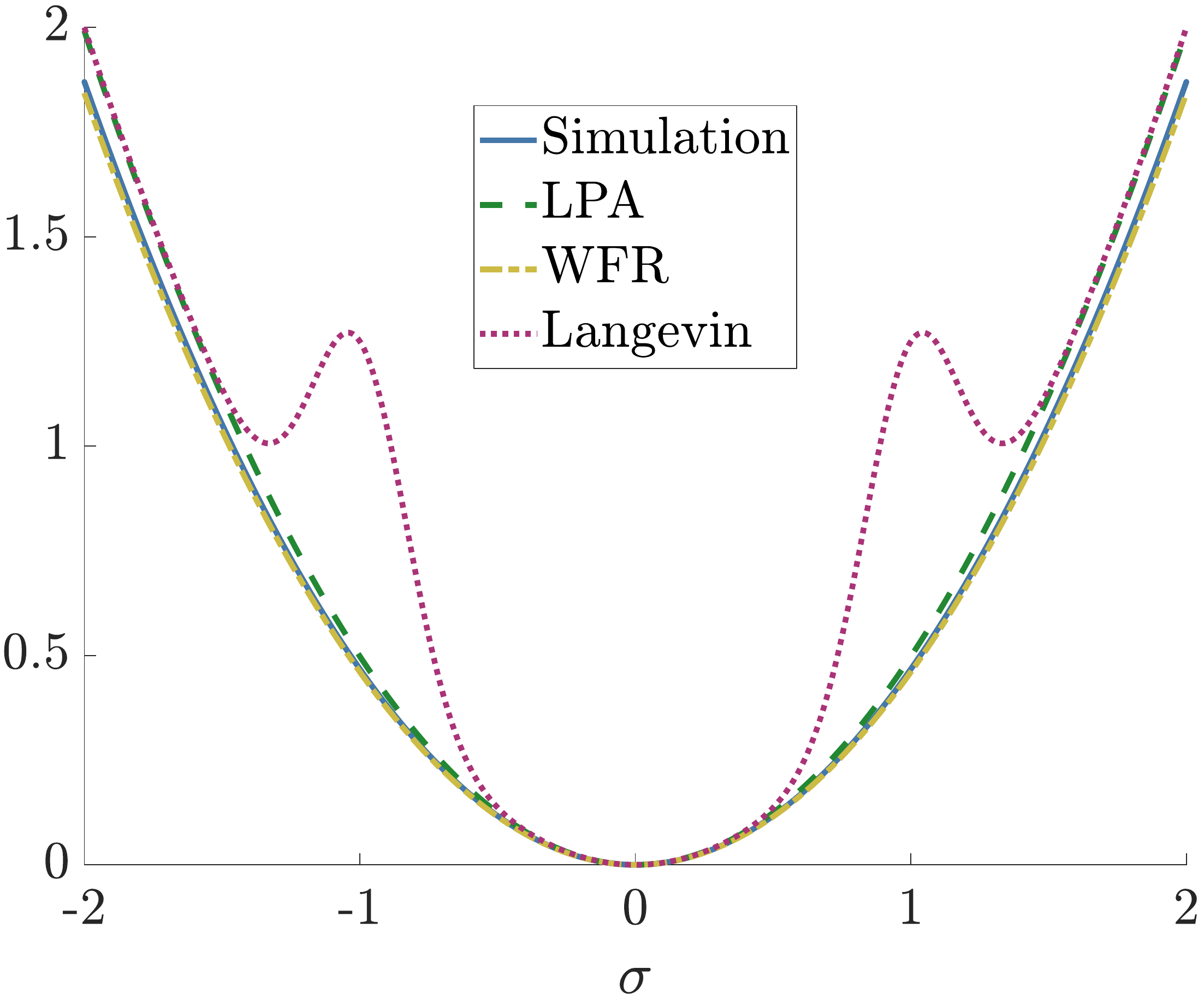}
    \includegraphics[width = 0.45\linewidth]{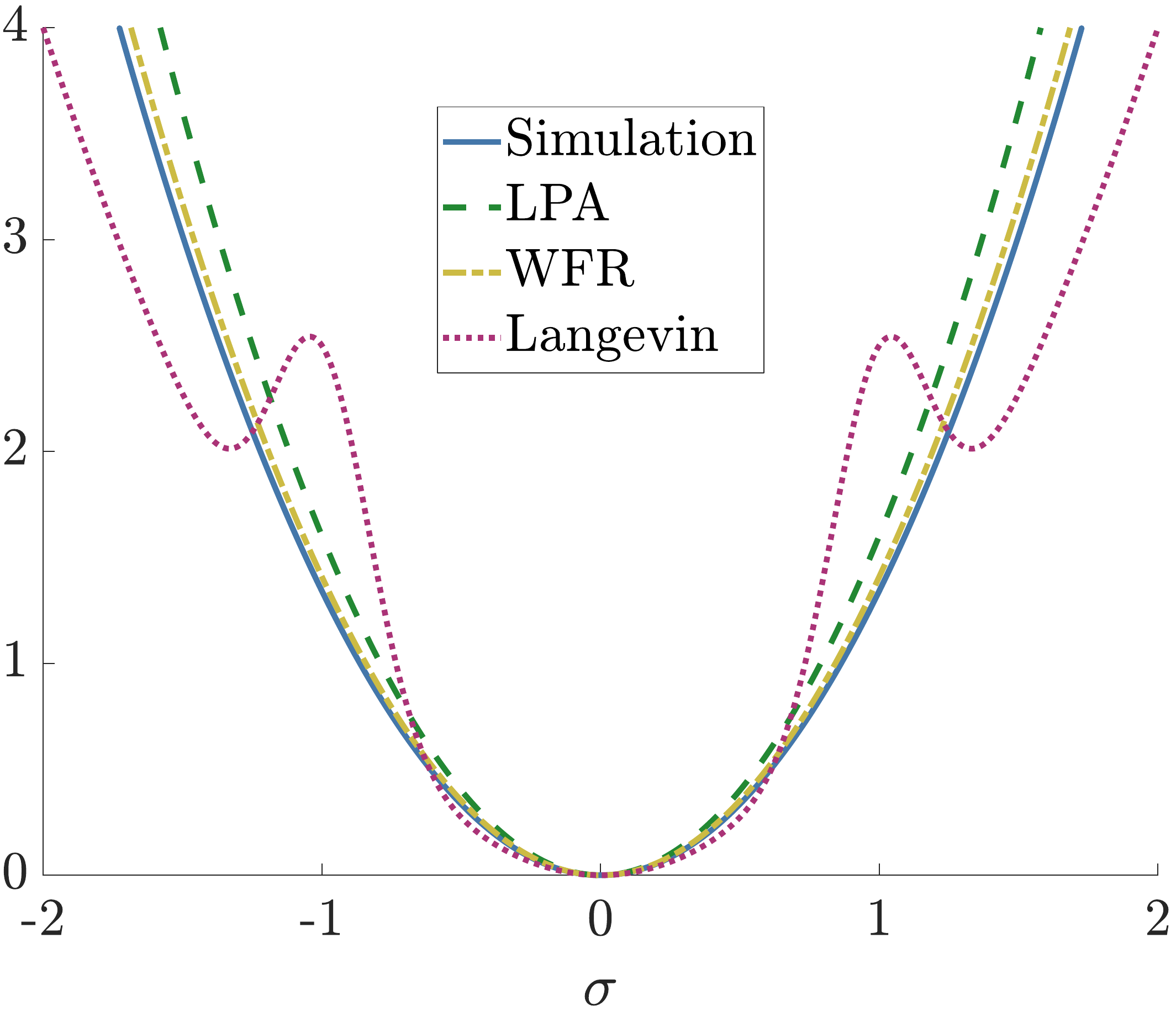}
    \includegraphics[width = 0.45\linewidth]{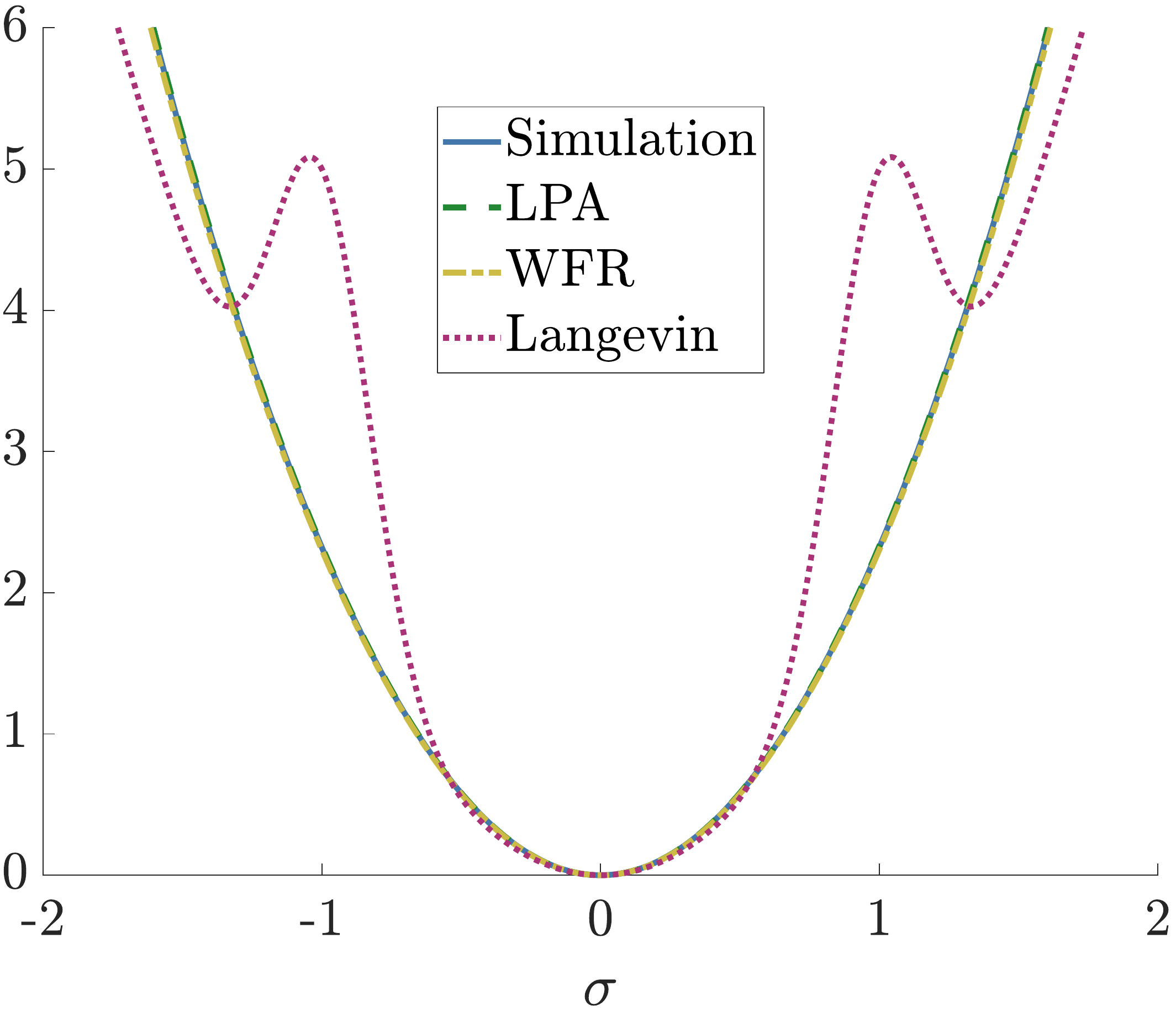}
    \caption[Different potentials predicting the same spectral tilt]{Various harmonic potentials $U(\sigma) \propto \sigma^2$ that give the same prediction for $\lambda$ and therefore the spectral tilt $n_{\sigma}$ as computed by direct simulation, the \ac{LPA} and \ac{WFR} for the $\sigma^2$ plus bumps potential -- shown here by the dotted red line. The parameters for each subplot are: $\hat{H}^2 = 5$ (top left), $\hat{H}^2 = 2$ (top right), $\hat{H}^2 = 1$ (bottom left)  and $\hat{H}^2 = 0.5$ (bottom right). }
    \label{fig:spectral_tilt_compare}
\end{figure}
 The values for $\lambda$ computed in Table \ref{tabel:effective mass for different potentials and methods} therefore take on a new interpretation, they tell us how accurately the \ac{FRG} can predict the power spectrum of a spectator field $\sigma$. In \cite{Markkanen2019,Markkanen2020} the stochastic spectral expansion -- see section \ref{sec:comparspec} -- is used to obtain the amplitude of the power spectrum and the spectral tilt for a standard fourth order polynomial and doublewell potential. Here we will use \ac{FRG} techniques to compute the spectral tilt for the $\sigma^2$ plus two bumps potential (\ref{eq:varsigma^2_2bumpsdefn}) while varying $\hat{H^2}$. \\
 
It is straightforward to solve the appropriate flow equations to obtain $\lambda$ from the \ac{FRG} and then using (\ref{eq:spectral_tilt_sigma}) obtain the (shifted) spectral tilt $n_{\sigma} -1$. In Table~\ref{tabel:spectral_tilt} we compare these computations to the result from direct numerical simulation as well as what the spectral tilt would be for the simple underlying harmonic potential in the absence of Gaussian bumps. In line with our results from part \ref{part:Meso} we can see good agreement using \ac{FRG} techniques with \ac{WFR} offering improvement over the \ac{LPA} result. It is also clear that we are capturing non-trivial effects as the deviation from the bare potential prediction is significant. However this does point to a degeneracy in our results, and theoretical predictions for observations in general. \\
The values we obtained in Table~\ref{tabel:spectral_tilt} could have just as easily been obtained from a harmonic potential with a suitably modified coefficient. In this way many \emph{different} potentials can give \emph{identical} predictions for the spectral tilt. To make this point more transparent we plot in Fig.~\ref{fig:spectral_tilt_compare} the harmonic potentials $U(\sigma) \propto \sigma^2$ that would reproduce the spectral tilt predictions in Table~\ref{tabel:spectral_tilt} for the \ac{FRG} methods and direct numerical simulation. We can see that at $\hat{H}^2 = 5$ that these harmonic potentials closely match the original Langevin potential which makes sense from the results in Table~\ref{tabel:spectral_tilt}. However as $\hat{H}^2$ decreases we can see that the deviations becomes more significant so that it does not resemble the original Langevin potential at all. In this way it is clear that one should be careful about making inferences about the potential from values of the spectral tilt. At high $\hat{H}^2$ one could very easily add features like bumps that would negligibly change the spectral tilt but make the potential look very different. At lower $\hat{H}^2$, added features will modify the spectral tilt significantly, but would still match the prediction from a suitably modified harmonic potential. At very low $\hat{H}^2$ however the equilibrium distribution will be heavily contained near the equilibrium point and features further away will again have little impact on the spectral tilt. \\
The values for the spectral tilt we show in Table~\ref{tabel:spectral_tilt} are too high to correspond to the curvature perturbation so the potentials we consider here could not correspond to the \emph{curvaton} scenario.


\subsection{Information retention}
Another interesting thing to capture is the information retention from the initial conditions. These computations  rely heavily on the work of Hardwick et.al \cite{Hardwick2017}. We can measure the relative information between two different distributions $P_1$ and $P_2$ using the Kullback-Leibler divergence, $D_{KL}$, \cite{Kullback1951}:
\begin{eqnarray}
D_{KL}(P_1 ||P_2) \equiv \int_{-\infty}^{\infty}\mathrm{d}\sigma P_1 (\sigma) \log_2 \lsb \dfrac{P_1 (\sigma)}{P_2 (\sigma)}\rsb \label{eq:DKL_defn} 
\end{eqnarray}
If we now consider two initial distributions separated by an amount of information $\delta D_{KL}^{in}$ which then leads to two distributions at a later time separated by $\delta D_{KL}^{f}$ we can define the information retention criterion by:
\begin{eqnarray}
\mathcal{I} \equiv \dfrac{\delta D_{KL}^{f}}{\delta D_{KL}^{in}} \label{eq:inform_reten_defn}
\end{eqnarray}
The value of $\mathcal{I}$ can tell us how dependent on the initial conditions the later distributions are. For instance $\mathcal{I} > 1$ indicates that the initial conditions are amplified and any initially small deviation becomes larger at later times. On the other hand $\mathcal{I} < 1$ indicates that any initial information is \textit{smoothed} out and the later state does not depend strongly on it. For instance if our final distribution corresponds to the equilibrium distribution $P_{eq}(\sigma)$ we would expect $\mathcal{I} = 0$ as any initial condition should reach this eventually for our system. \\

If we assume Gaussian distributions -- as this is what the \ac{FRG} can tell us -- then equation (\ref{eq:inform_reten_defn}) becomes:
\begin{eqnarray}
\mathcal{I} = \lb \dfrac{\partial G(\mathcal{N}_{f},\mathcal{N}_{f})}{\partial G(\mathcal{N}_{in},\mathcal{N}_{in})} \rb^2 \lb \dfrac{G (\mathcal{N}_{in},\mathcal{N}_{in})}{G (\mathcal{N}_{f},\mathcal{N}_{f})}\rb^2
\end{eqnarray}
where the variance $G(\mathcal{N},\mathcal{N})$ has been evaluated at the initial (in) and final (f) distribution appropriately. If we adapt the \ac{EEOM} for variance (\ref{eq:EEOM Variance}) for our spectator parameters our information retention becomes:
\begin{eqnarray}
\mathcal{I} &=& \lb \dfrac{\zeta_{,\Sigma}^2(\mathcal{N}_{in})}{\zeta_{,\Sigma}^2(\mathcal{N}_f)}f^2(\mathcal{N}_f - \mathcal{N}_{in})\rb^2 \lb \dfrac{G (\mathcal{N}_{in})}{G (\mathcal{N}_{f})}\rb^2 \\
&=& \lb \dfrac{\zeta_{,\Sigma}^2(\mathcal{N}_{in})}{\zeta_{,\Sigma}^2(\mathcal{N}_f)} \dfrac{G^{2} (\mathcal{N}_f, \mathcal{N}_{in} )}{G (\mathcal{N}_{in},\mathcal{N}_{in})G (\mathcal{N}_{f},\mathcal{N}_{f})}\rb^2 
\end{eqnarray}
where in the second line we have been able to rewrite it in terms of the ratio between the covariance squared $G^{2} (\mathcal{N}_f, \mathcal{N}_{in} )$ and the variance at the final and initial times. \\
Unsurprisingly we find the information retention rapidly drops to zero within half an e-fold for reasonable initial conditions. Due to the presence of a slow-roll attracter in the equations of motion it makes sense that there would be a rapid \emph{erasure} of initial conditions but it is reassuring that the \ac{FRG} can robustly predict this. 

\section{\label{sec:FPT_spec} First-Passage Time prediction}
\begin{figure}[t!]
    \centering
    \includegraphics[width = 0.45\linewidth]{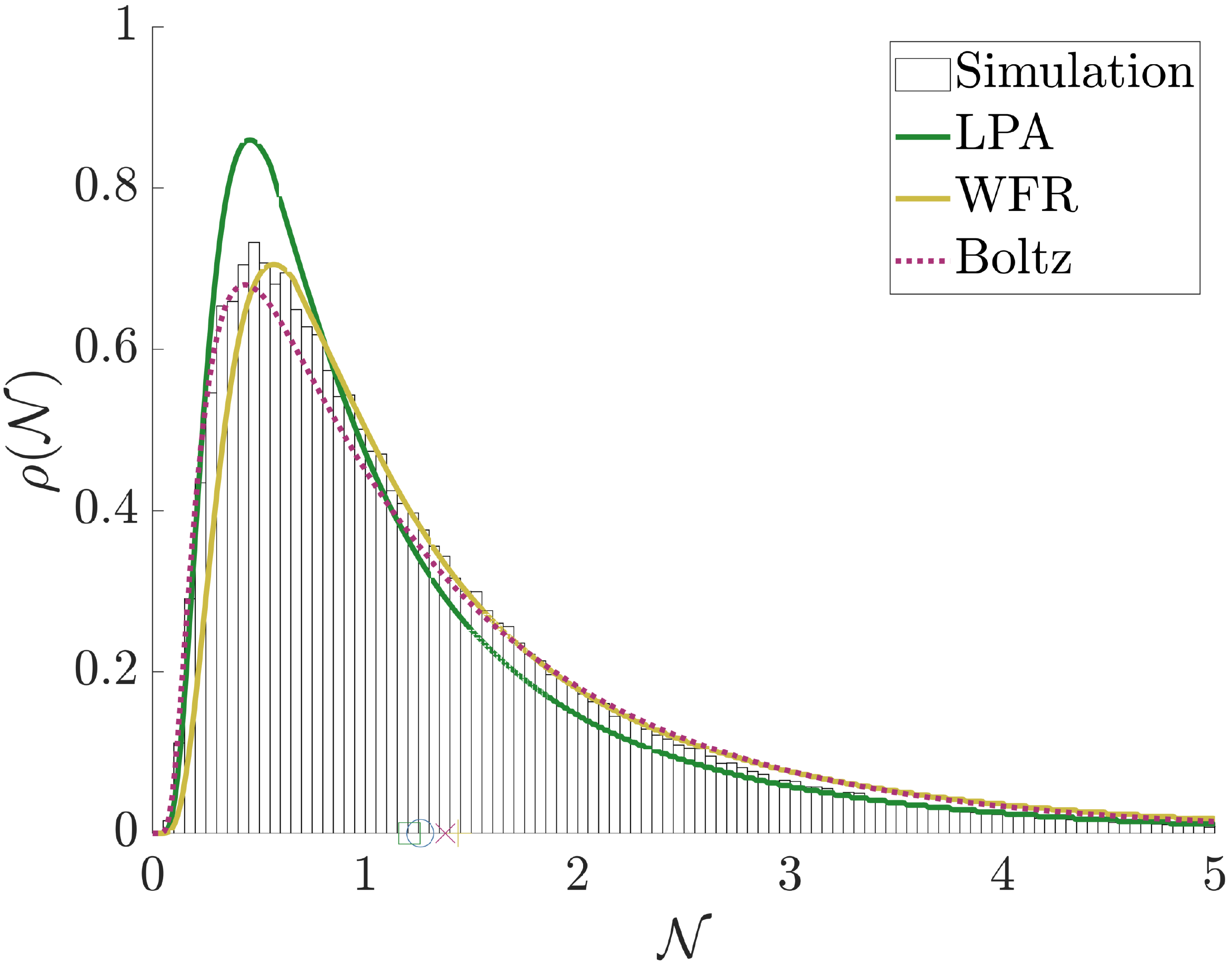}
    \includegraphics[width = 0.45\linewidth]{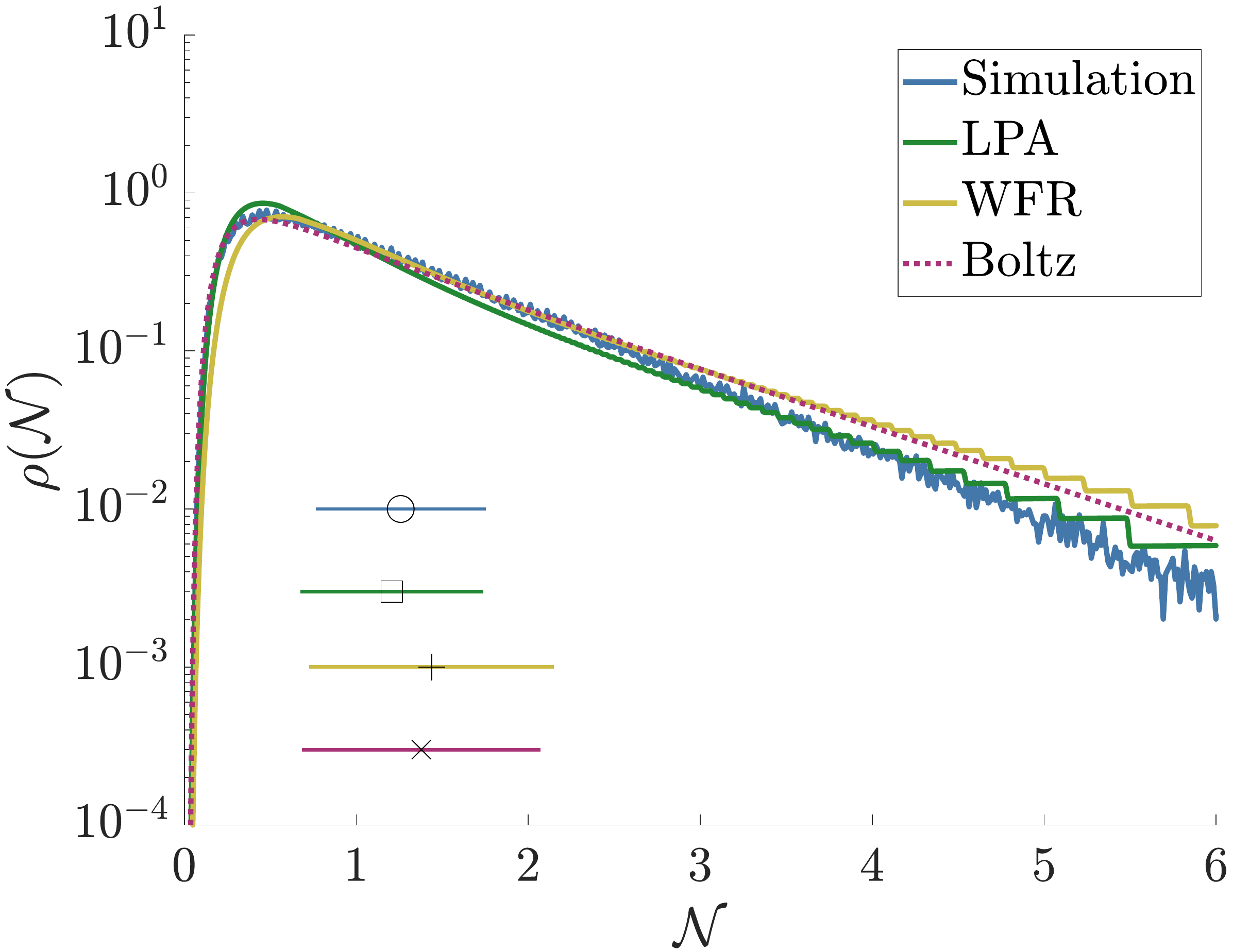}
    \includegraphics[width = 0.45\linewidth]{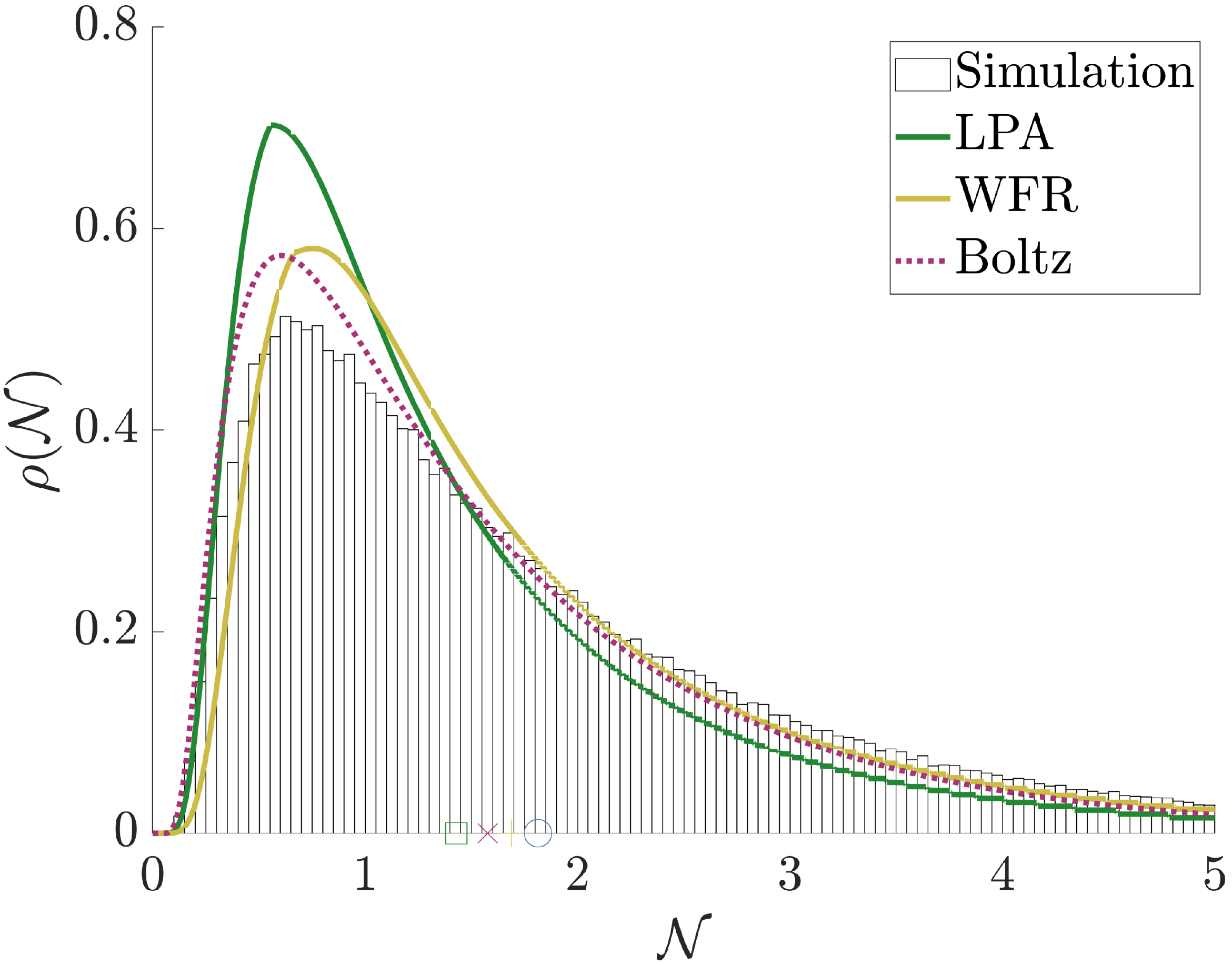}
    \includegraphics[width = 0.45\linewidth]{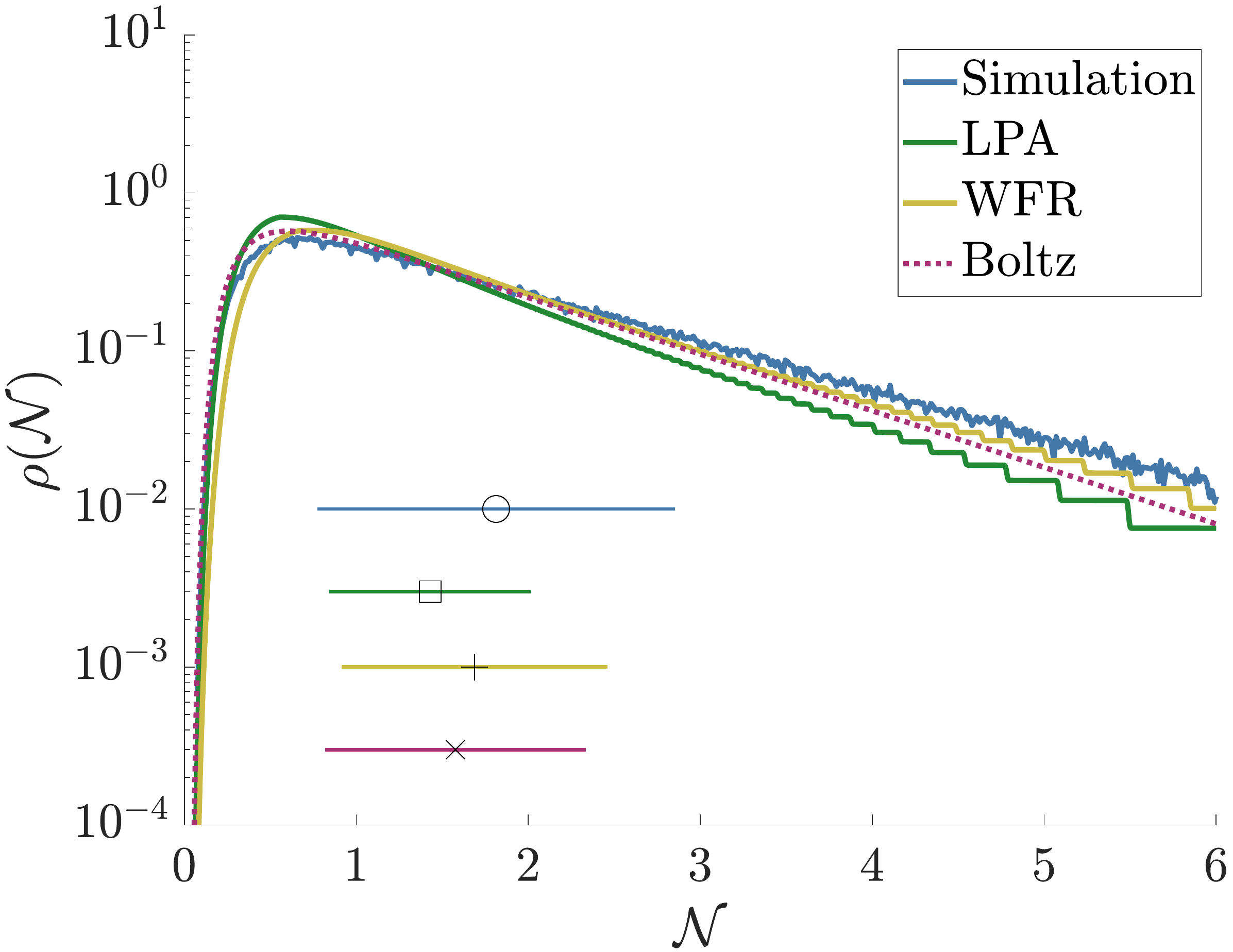}
    \caption[First-Passage Time PDF for the doublewell]{The PDF for time taken for the spectator to reach $\sigma = 0.5$ (top row) and the equilibrium $\sigma = 0$ (bottom row) in the doublewell potential for $\hat{H}^2 = 5$ using linear (left) and log (right) scales. The circle, box, plus and cross symbols in the plots represent the mean time taken $\lan \mathcal{N} \ran$ as computed by direct numerical simulation, by the \ac{LPA}, by \ac{WFR} and assuming a Boltzmann type potential respectively. The horizontal lines in the log plots correspond to the respective variances $\delta \mathcal{N}^2 = \lan \mathcal{N}^2\ran - \lan \mathcal{N}\ran^2$. The initial conditions were a Gaussian distribution centred at $\sigma = 3$ with variance $ = 0.05$.}
    \label{fig:FPT5H2_DW}
\end{figure}
The \ac{FRG} naturally gives us a prediction for the evolution of $\Sigma$ and the connected two point function of the field $G$ and we showed in section \ref{sec:Cosmo_Spec} how this can be turned into predictions for cosmological observables such as the power spectrum. In this section we will instead examine the \ac{FPT} problem in line with our computations for the inflaton in chapters \ref{cha:Inflation} \& \ref{cha:PBH}. In particular what we wish to know is the probability distribution, $\rho (\mathcal{N})$ for number of e-folds it takes to reach a field value $\sigma_2$ given it was initially at $\sigma_1$ at some initial time $\mathcal{N}_i$ which for simplicity we identify with $0$. For a spectator field $\sigma$ that obeys the Langevin equation (\ref{eq:Langevin_sigma_dimless}) we can straightforwardly write down the \ac{F-P} equation that the PDF $P(\sigma, \alpha)$ obeys:
\begin{eqnarray}
	\dfrac{\partial P(\sigma,\alpha)}{\partial \alpha} = \partial_{\sigma}(P(\sigma,t)\partial_{\sigma} U) + \dfrac{\hat{H}^2}{2}\partial_{\sigma\sigma} P(\sigma,\alpha) \label{eq: F-P_spect}
\end{eqnarray}
The question now is how one can compute the solution to (\ref{eq: F-P_spect}) using \ac{FRG} techniques.
\subsection{Normal Distribution}
It is true that in general the solution to (\ref{eq: F-P_spect}) is not a normal distribution, however we will assume it is as the \ac{FRG} is able to accurately predict the evolution of the average position $\Sigma (\alpha)$ and the variance $G(\alpha )$ -- now denoted with a single argument for notational brevity. This gives us the following ansatz:
\begin{eqnarray}
P(\sigma, \alpha) = \dfrac{1}{\sqrt{2\pi G (\alpha)}} \exp \lsb -\dfrac{1}{2}\dfrac{\lb \sigma -\Sigma (\alpha)\rb^2}{G(\alpha)}\rsb \label{eq:pdf_gaus_spect}
\end{eqnarray}
We now wish to compute the probability $\rho (\mathcal{N})$ that $\sigma_2$ is reached between $\mathcal{N}$ and $\mathrm{d}\mathcal{N}$ e-folds. This can be related to (\ref{eq:pdf_gaus_spect}) using equation (\ref{eq:FPT_FP_pdf_equivalence}) reproduced here for clarity:
\begin{eqnarray}
\int_{\mathcal{N}}^{\infty} \rho (\alpha) \mathrm{d}\alpha &=& \int_{\sigma_2}^{\infty}P(\sigma,\mathcal{N})\mathrm{d}\sigma \label{eq:FPT_FP_pdf_equivalence_spect_int}\\
\Rightarrow \rho (\mathcal{N}) &=& -\dfrac{\partial}{\partial \mathcal{N}} \int_{\sigma_2}^{\infty}P(\sigma,\mathcal{N})\mathrm{d}\sigma \label{eq:FPT_FP_pdf_equivalence_spect}
\end{eqnarray}
As discussed in Appendix \ref{APP:FPT} for (\ref{eq:FPT_FP_pdf_equivalence_spect}) to hold exactly one must impose an absorbing boundary condition at $\sigma_2$ otherwise you overestimate the number of runs that have not yet reached $\sigma_2$. An absorbing boundary condition is naturally imposed for the \ac{F-P} equation for the inflaton -- as it corresponds to inflation ending -- but this is not so for a spectator field and the PDF (\ref{eq:pdf_gaus_spect}) is not endowed with such boundary conditions. Instead the PDF that enters (\ref{eq:FPT_FP_pdf_equivalence_spect}) will be different to the one that solves (\ref{eq: F-P_spect}). We can easily modify the ansatz (\ref{eq:pdf_gaus_spect}) to include an absorbing boundary condition by adding another Gaussian solution that cancels at $\sigma_2$:
\begin{eqnarray}
    P(\sigma, \alpha) = \dfrac{A}{\sqrt{2\pi G (\alpha)}} \lcb \exp \lsb -\dfrac{1}{2}\dfrac{\lb \sigma -\Sigma (\alpha)\rb^2}{G(\alpha)}\rsb - \exp \lsb -\dfrac{1}{2}\dfrac{\lb 2\sigma_2 -\sigma -\Sigma (\alpha)\rb^2}{G(\alpha)}\rsb\rcb \label{eq:pdf_gaus_spect_absorb}
\end{eqnarray}
Where $A$ is a normalisation factor to be determined. Substituting (\ref{eq:pdf_gaus_spect_absorb}) into (\ref{eq:FPT_FP_pdf_equivalence_spect}) yields:
\begin{eqnarray}
\rho (\mathcal{N}) = -A\dfrac{\partial}{\partial \mathcal{N}}\lsb  \text{erfc}\lb \dfrac{\sigma_2 -\Sigma (\mathcal{N})}{\sqrt{2G(\mathcal{N})}}\rb \rsb \end{eqnarray}
Which can be straightforwardly evaluated to obtain one of the main results of this chapter:
\begin{empheq}[box = \fcolorbox{Maroon}{white}]{equation}
\rho (\mathcal{N}) = \dfrac{A}{\sqrt{2\pi G (\mathcal{N})}}\lsb \dfrac{\lb \Sigma (\mathcal{N}) - \sigma_2\rb\partial_{\mathcal{N}}G(\mathcal{N})}{2G(\mathcal{N})} - \partial_{\mathcal{N}}\Sigma (\mathcal{N})\rsb \exp \lsb -\dfrac{1}{2} \dfrac{\lb\sigma_2 -\Sigma (\mathcal{N})\rb^2}{G(\mathcal{N})}\rsb \label{eq:rhoN_spect}
\end{empheq}
The normalisation condition, integrating between the initial e-fold time $\mathcal{N}_{in}$ and $\mathcal{N} \rightarrow \infty$ yields:
\begin{eqnarray}
\dfrac{1}{A} =   \text{erf}\lb \dfrac{\sigma_2 -\Sigma_{in}}{\sqrt{2G_{in}}} \rb - \text{erf}\lb \dfrac{\sigma_2 -\Sigma_{eq}}{\sqrt{2G_{eq}} }\rb  \label{eq:norm_spect}
\end{eqnarray}
where subscripts $in$ and $eq$ indicate quantities evaluated at the initial e-fold time $\mathcal{N}_{in}$ and at equilibrium respectively. It is worth noting that if our initial condition corresponds to a delta function then (\ref{eq:norm_spect}) simplifies to:
\begin{eqnarray}
\dfrac{1}{A} = 1 -  \text{erf}\lb \dfrac{\sigma_2 -\Sigma_{eq}}{\sqrt{2G_{eq}} }\rb    \label{eq:norm_spect_delta}
\end{eqnarray}
and if $\sigma_2$ is the equilibrium point the norm can be further simplified to $A = 1$. Equations (\ref{eq:rhoN_spect}) \& (\ref{eq:norm_spect}) are the main results of this section.\\

We also recall the Boltzmann potential (\ref{eq: BoltztildeV}):
\begin{eqnarray}
\tilde{U}_{Boltz}(\Sigma) = \dfrac{\Upsilon}{4G_{eq}}\left( \Sigma - \Sigma_{eq}\right)^2 \label{eq: BoltztildeU}
\end{eqnarray}
which suggests the effective dynamical potential is simply a harmonic potential centred at the equilibrium point with mass determined by the equilibrium variance. A potential of this form gives the following simple predictions for the average position and variance:
\begin{eqnarray}
\Sigma_{Boltz} (\alpha) &=& \Sigma_{in}\exp \lb -\dfrac{\Upsilon}{2G_{eq}} \alpha\rb + \Sigma_{eq}\\
G_{Boltz} (\alpha ) &=& \lsb G_{in} - G_{eq} \rsb \exp \lb -\dfrac{\Upsilon}{G_{eq}} \alpha\rb + G_{eq}
\end{eqnarray}
these equations can be substituted into (\ref{eq:rhoN_spect}) to give a prediction for \ac{FPT} quantities and will act as a benchmark for the \ac{FRG}.\\

In Fig.~\ref{fig:FPT5H2_DW} we plot the PDF for the \ac{FPT} to reach $\sigma = 0.5$ (top row) and $\sigma = 0$ (bottom row) for the doublewell potential at $\hat{H}^2 =5$. We have compared the results from simulations with the \ac{FRG} from (\ref{eq:rhoN_spect}) as well as the Boltzmann potential prediction. We can readily see -- as expected -- that \ac{WFR} offers an improvement over \ac{LPA} with matching the \ac{FRG}. What is more surprising is how well the Boltzmann potential prediction also does even when the final position is not the equilibrium point. \\
\begin{figure}[t!]
    \centering
    \includegraphics[width = 0.45\linewidth]{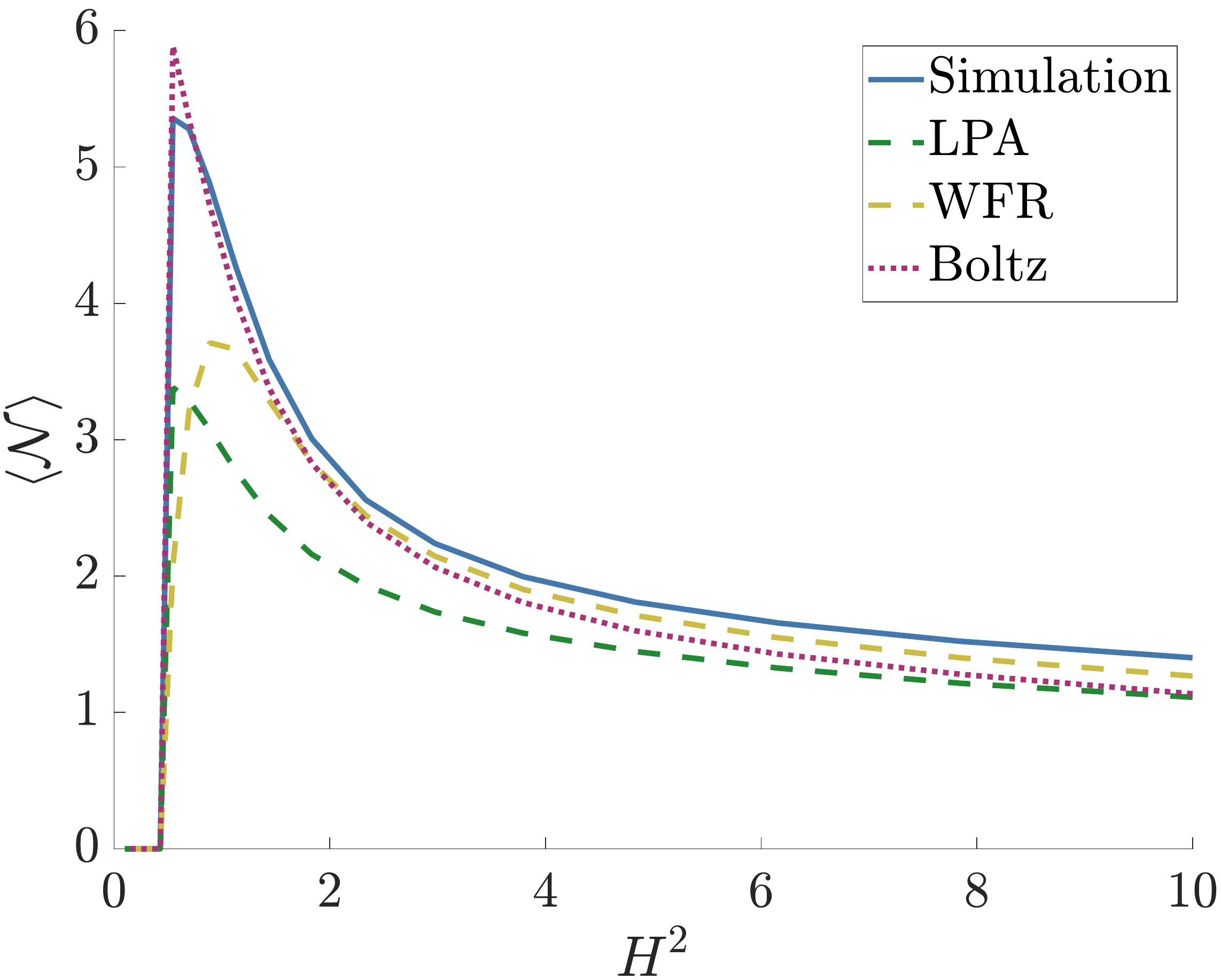}
    \includegraphics[width = 0.45\linewidth]{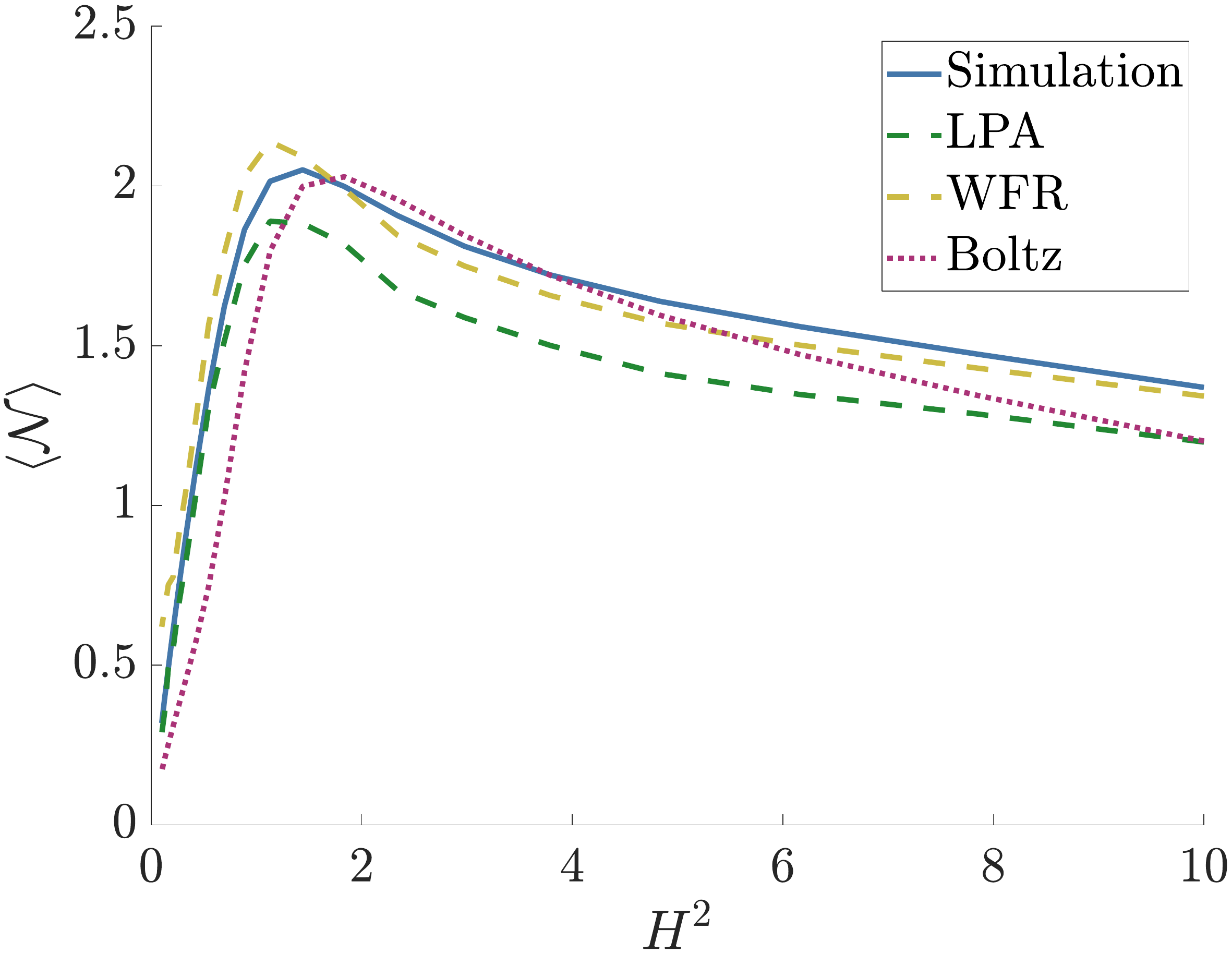}
    \caption[Average time taken to reach equilibrium for a spectator field]{Dependence of the average time taken to reach the equilibrium point, $\lan \mathcal{N} \ran$, on $\hat{H}^2$ for the doublewell potential (left) and polynomial (right) as computed by different approaches.}
    \label{fig:FPTALL_mean}
\end{figure}

To get a more general sense of how well the \ac{FRG} does at predicting \ac{FPT} quantities we plot the predictions for the average time taken, $\lan \mathcal{N} \ran $ to reach the equilibrium point for the doublewell and polynomial potentials in Fig.~\ref{fig:FPTALL_mean} over a range of $\hat{H}^2$. We can see that the \ac{FRG} does a good job at correctly predicting how $\lan \mathcal{N} \ran $ changes as the value of $\hat{H}^2$ is varied with \ac{WFR} in particular offers good agreement with the result from direct numerical simulation. \\

\begin{figure}[t!]
    \centering
    \includegraphics[width = 0.45\linewidth]{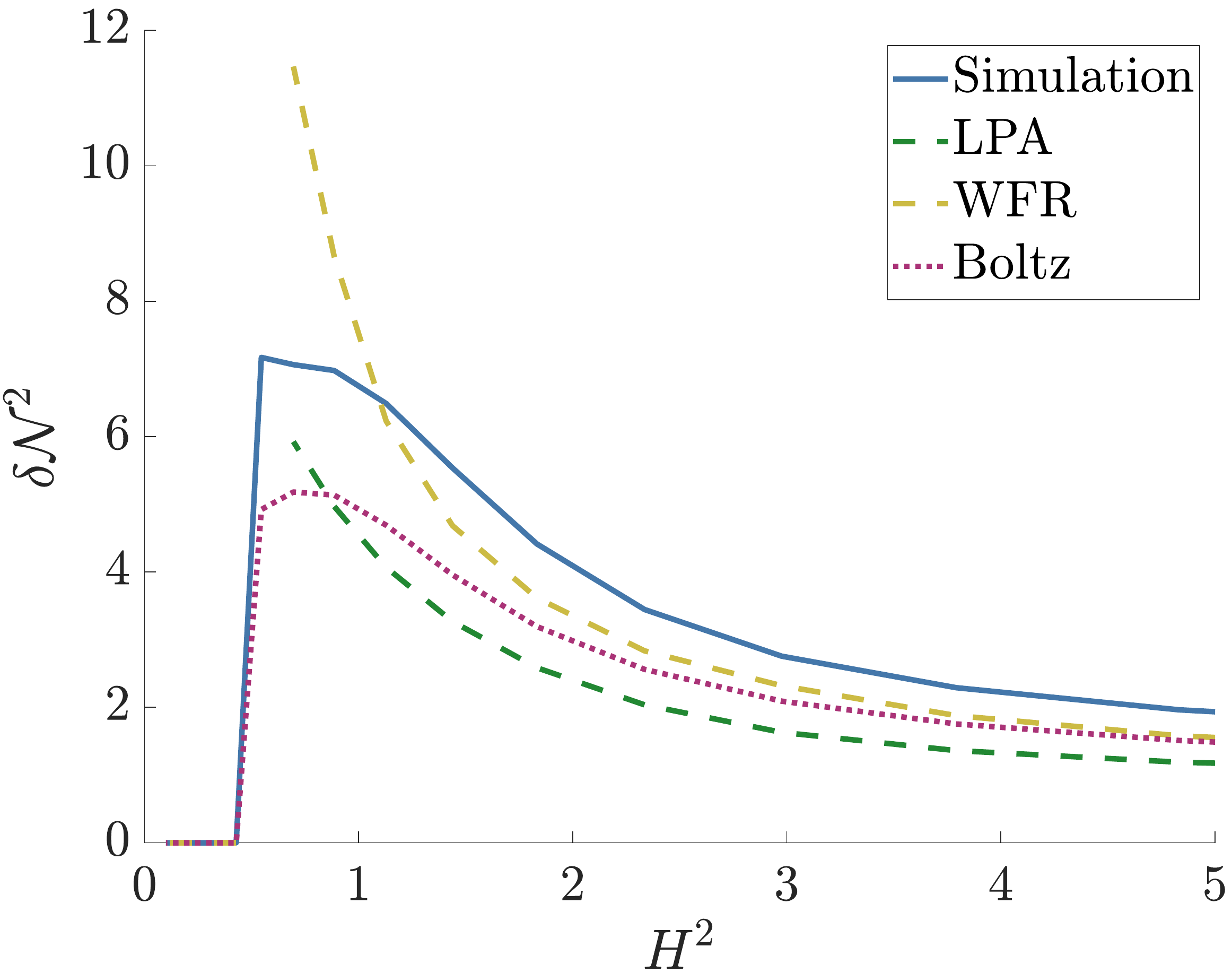}
    \includegraphics[width = 0.45\linewidth]{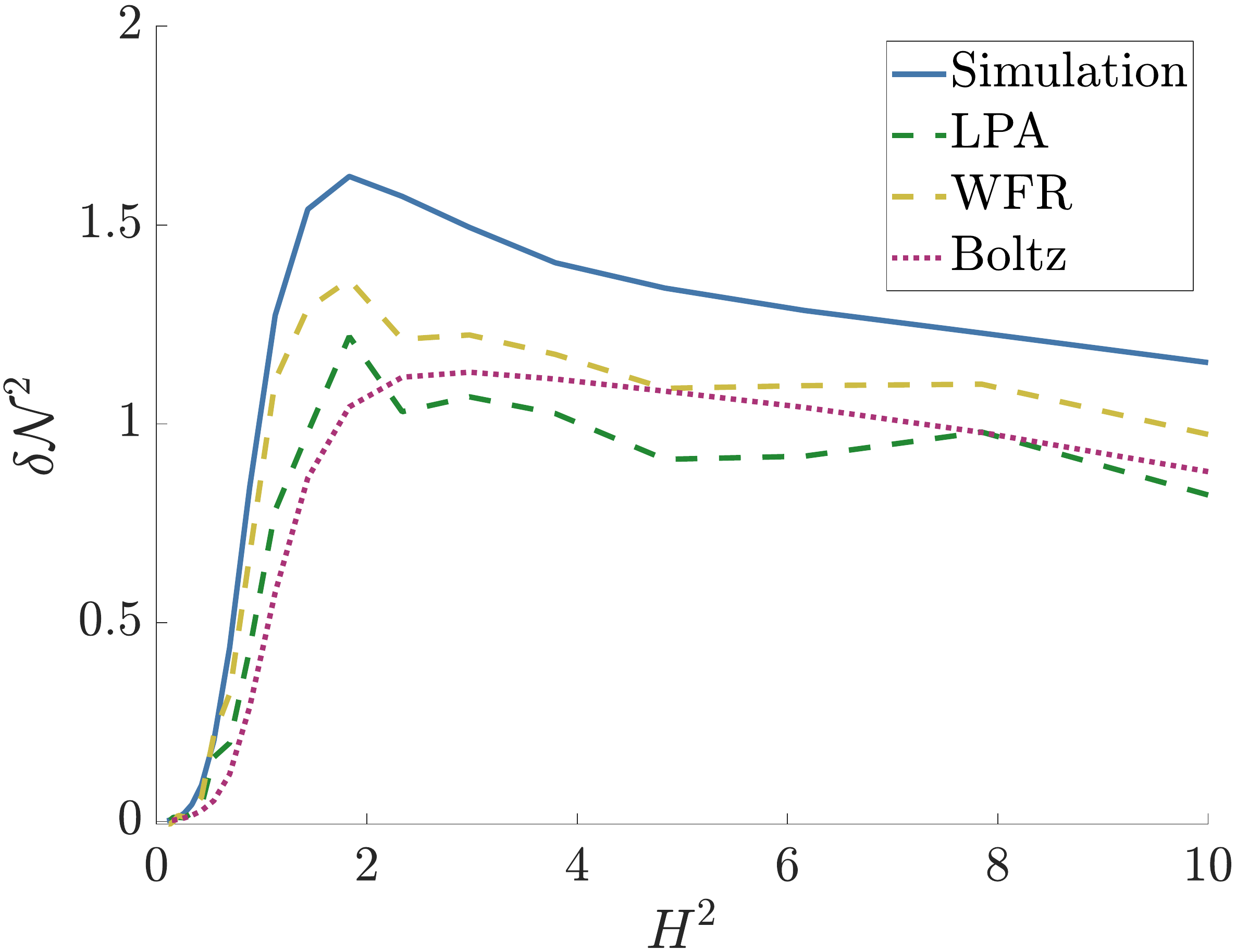}
    \caption[Variance in time taken to reach equilibrium for a spectator field]{Dependence of the variance in time taken to reach the equilibrium point, $\delta \mathcal{N}^2 = \lan \mathcal{N}^2\ran - \lan \mathcal{N} \ran^2 $, on $\hat{H}^2$ for the doublewell potential (left) and polynomial (right) as computed by different approaches.}
    \label{fig:FPTALL_Var}
\end{figure}
We have also plotted in Fig.~\ref{fig:FPTALL_Var} how the variance in time taken to reach equilibrium $\delta \mathcal{N}^2 = \lan \mathcal{N}^2\ran - \lan \mathcal{N} \ran^2 $ changes as the value of $\hat{H}^2$ is varied. We can see that while the \ac{FRG} does not match as well as it does for $\lan \mathcal{N} \ran $  it still offers good agreement and improvement over the Boltzmann prediction. It is remarkable given the number of assumptions that had to be taken to achieve this result -- derivative expansion of the \ac{REA}, simple regulator, normal solution to the \ac{F-P} equation -- that the \ac{FRG} agrees as well as it does. \\

\subsection{Skew-Normal Distribution}
In principle one can go beyond an initially normal distribution and introduce skewness through the third central moment $\lan \sigma(\alpha)^3\ran_{C}$. There are many different distributions with skew but here we will assume a skew-normal distribution given by:
\begin{eqnarray}
P(\sigma, \alpha) = \dfrac{1}{\sqrt{2\pi B (\alpha)}} \lsb 1 + \text{erf}\lb \dfrac{C(\alpha)\lb \sigma - A(\alpha) \rb}{\sqrt{2B(\alpha )}}\rb \rsb \exp \lsb -\dfrac{1}{2}\dfrac{\lb \sigma -A (\alpha)\rb^2}{B(\alpha)}\rsb 
\end{eqnarray}
where the time dependent parameters $A (\alpha)$, $B(\alpha)$ and $C(\alpha)$ are related to the mean, $\Sigma (\alpha)$, variance, $G(\alpha)$, and the third central moment, $\lan \sigma (\alpha)^3\ran_{C}$, in the following way:
\begin{subequations}
\begin{align}
   D(\alpha) &= \sqrt{\dfrac{\pi}{2}}\dfrac{u^{1/3}}{\sqrt{1 + u^{2/3}}}, \quad u \equiv \dfrac{2}{4-\pi}\dfrac{\lan \sigma (\alpha)^3\ran_{C}}{G(\alpha)^{3/2}}\\
   C(\alpha) &= \dfrac{D}{\sqrt{1-D^2}}\\
   B(\alpha) &= \dfrac{G(\alpha)}{1-2D^2/\pi}\\
   A(\alpha) &=  \Sigma (\alpha) - D\sqrt{\dfrac{2G(\alpha)}{\pi}}
\end{align}\label{eq:ABCD_relations_append}
\end{subequations}
Then we can proceed as in the normal case:
\begin{eqnarray}
\rho (\mathcal{N}) &=& -\dfrac{\partial}{\partial \mathcal{N}}\int_{\sigma_2}^{\infty}\dfrac{\mathrm{d}\sigma}{\sqrt{2\pi B (\mathcal{N})}} \lsb 1 + \text{erf}\lb \dfrac{C(\mathcal{N})\lb \sigma - A(\mathcal{N}) \rb}{\sqrt{2B(\mathcal{N} )}}\rb \rsb \exp \lsb -\dfrac{\lb \sigma -A (\mathcal{N})\rb^2}{2B(\mathcal{N})}\rsb  \nonumber\\
&& \\
&=& -\dfrac{\partial}{\partial \mathcal{N}}\lsb  \dfrac{1}{2} \text{erfc}\lb \dfrac{\sigma_2 - A(\mathcal{N})}{\sqrt{2B(\mathcal{N})}} \rb  + 2T \lb  \dfrac{\sigma_2 - A(\mathcal{N})}{B(\mathcal{N})}, C(\mathcal{N}) \rb \rsb \\
&=& \exp \lsb - \dfrac{\lb\sigma_2 -A (\mathcal{N})\rb^2}{2B(\mathcal{N})}\rsb \Bigg\lbrace  - \dfrac{\partial_{\mathcal{N}}C(\mathcal{N})}{2\pi\lb 1 + C(\mathcal{N})^2\rb}\exp \lsb - \dfrac{C(\mathcal{N})^2\lb \sigma_2 - A(\mathcal{N})\rb^2}{2B(\mathcal{N})}\rsb \nonumber \\
&& +\dfrac{1}{\sqrt{2\pi B (\mathcal{N})}}\lsb \dfrac{\lb A (\mathcal{N}) - \sigma_2\rb\partial_{\mathcal{N}}B(\mathcal{N})}{B(\mathcal{N})} - \partial_{\mathcal{N}}A (\mathcal{N})\rsb  \text{erfc} \lb \dfrac{\sigma_2 - A(\mathcal{N})}{\sqrt{2B(\mathcal{N})}}\rb \Bigg\rbrace \label{eq:rhoN_spect_skew}
\end{eqnarray}
where we have used Owen's T function \cite{Owen1956} defined as:
\begin{eqnarray}
T(x,a) \equiv \dfrac{1}{2\pi}\int_{0}^{a}\mathrm{d}y~\dfrac{\exp \lsb -x^2 \lb 1 + y^2\rb /2\rsb}{1+y^2}
\end{eqnarray}
so that the norm is simply given as:
\begin{eqnarray}
\lan 1 \ran &=&  \dfrac{1}{2} \text{erf}\lb \dfrac{\sigma_2 - A_{eq}}{\sqrt{2B_{eq}}} \rb - \dfrac{1}{2} \text{erf}\lb \dfrac{\sigma_2 - A_{in}}{\sqrt{2B_{in}}} \rb \nonumber \\
&&+ 2T \lb  \dfrac{\sigma_2 - A_{in}}{B_{in}}, C_{in} \rb   - 2T \lb  \dfrac{\sigma_2 - A_{eq}}{B_{eq}}, C_{eq} \rb \label{eq:norm_spect_skeq}
\end{eqnarray}
However as we previously indicated the \ac{FRG} poorly predicts the third central moment -- and thus the skewness -- so we find that going beyond a Gaussian distribution actually \emph{worsens} our \ac{FRG} predictions for the PDF. 

\section{\label{sec:RGSpec_conc}Conclusions}
\acresetall 
In this chapter we have successfully applied the \ac{FRG} techniques developed in the earlier parts of the thesis for thermally driven stochastic dynamics to stochastic scalar fields during an inflationary period in the early universe. Having outlined how our \ac{FRG} equations derived in part \ref{part:Meso} can be successfully applied to a spectator scalar field during inflation we derived \ac{EEOM} for the third central moment in an attempt to go beyond Gaussian statistics. Unfortunately while the \ac{FRG} is capable of correctly describing the qualitative nature of the third central moment it does not offer sufficient quantitative accuracy. We surmised that this is probably the limit of the accuracy of the derivative expansion of the \ac{REA} and that going to higher orders or focusing on a vertex expansion might offer better results. \\

We went on to discuss what cosmological observables could be predicted from an \ac{FRG} approach. In the curvaton scenario the spectator field could provide the dominant contribution to the primordial curvature perturbation and therefore one may wish to compute the power spectrum and spectral tilt of $\sigma$. We reviewed how de Sitter invariance allows us to relate correlations in space -- what we observe in the \ac{CMB} -- to correlations in time -- what can be computed in a stochastic approach. As the \ac{FRG} predicts that (in equilibrium) the covariance follows a simple exponential in time, i.e. $\lan \sigma_0 \sigma_{\alpha}\ran \propto e^{-\lambda \alpha}$, the real space correlator follows a simple power law form. This means that the spectral tilt is simply given by $n_{\sigma} - 1 = 2\lambda$, and we showed for a $\sigma^2$ plus bumps potential that the \ac{FRG} (and in particular \ac{WFR}) can accurately compute the spectral tilt. We showed in Fig.~\ref{fig:spectral_tilt_compare} how this creates a degeneracy in predictions such that potentials with features, like Gaussian bumps, give the same predictions for the spectral tilt as an appropriately scaled harmonic potential. One should therefore be wary about making inferences about the potential from observational measurements like the spectral tilt. We also used the \ac{FRG} to confirm the erasure of initial condition dependence of the spectator field during inflation as to be expected by the presence of the \ac{SR} attracter. \\

We finished this chapter by an examination of the \ac{FPT} problem for a spectator field. In particular we derived an analytic formula for the PDF for time taken to traverse between two points assuming a normal distribution (\ref{eq:rhoN_spect}) and skew-normal distribution (\ref{eq:rhoN_spect_skew}). As the \ac{FRG} is able to predict the evolution of average position $\Sigma$ and its variance with time this meant the \ac{FRG} could make predictions for \ac{FPT} quantities. We showed that the \ac{FRG} captured the shapes of the PDFs well and commented on the surprising robustness of the simple Boltzmann equilibrium prediction even far from equilibrium. We showed that even assuming a normal distribution that the \ac{FRG} is able to accurately predict the average time taken to traverse between two points $\lan \mathcal{N} \ran $ and the variance in the time taken $\delta \mathcal{N}^2 = \lan \mathcal{N}^2\ran - \lan \mathcal{N} \ran^2 $. This represents a first, crucial step towards using \ac{FRG} techniques to compute \ac{FPT} quantities for the inflaton. 
\chapter{Summary} 
\label{cha:Summary} 
\acresetall 
\vspace{0.5cm}
\begin{flushright}{\slshape    
In literature and in life we ultimately pursue, not conclusions, but beginnings.} \\ \medskip
--- Sam Tanenhaus \cite{Tanenhaus1986}
\end{flushright}
\vspace{0.5cm}
In this work we have covered a wide range of topics and it would therefore be easy to ``miss the wood for the trees". To this end we will briefly summarise the key results of this thesis. \\

\noindent Part \ref{part:Meso} was concerned with examining the behaviour of the one-dimensional overdamped Langevin equation (\ref{eq:langevindimless}) which describes the \ac{BM} of a particle moving in a thermal bath. \\

\noindent In chapter \ref{cha:stochastic_processes} we introduced the notion of a path integral and demonstrated how \ac{BM} could be expressed in terms of the \ac{BPI}, (\ref{eq:BM_path_int}), which is equivalent to the path integral for Euclidean SuperSymmetric Quantum Mechanics. We also reviewed the concepts of generating functionals from \ac{QFT} and discussed how the \ac{EA}, (\ref{eq:Gamma_S_relation}), resembles the classical action $\mathcal{S}$ but with all the thermal fluctuations integrated out. We also included a derivation of the \ac{F-P} equation which is most illuminating in the form that resembles the Schr\"{o}dinger equation (\ref{eq:FP1}). \\

\noindent In chapter \ref{cha:fRG} we introduced the concept of an \ac{EFT} and how the \ac{FRG} can be used to obtain an \ac{EFT} for the coarse-grained \emph{in time} theory for \ac{BM}. This is achieved by the introduction of a regulator in frequency space that interpolates from the classical action $\mathcal{S}$ down to the \ac{EA} $\Gamma$. This interpolation is controlled by the renormalisation scale $\kappa$, where $\kappa = 0$ corresponds to the \ac{EA} with all fluctuations integrated out. We reviewed the derivative expansion approach and introduced the \ac{LO} and \ac{NLO} approximations known more commonly as the \ac{LPA} and \ac{WFR}. In terms of these approximations we derived the flow equations (\ref{eq:dV/dk}) and (\ref{eq:WFR_flow_eqn}) which are PDEs describing how the \emph{effective potential} and \ac{WFR} parameter $\zeta_x$ vary with $\kappa$. We solved these flow equations numerically for several non-trivial potentials such as the doublewell and harmonic potential with multiple Gaussian bumps added. This reinforces the \ac{FRG}'s ability to obtain non-perturbative results. \\

\noindent In chapter \ref{cha:fRG EEOM} we derived the \ac{EEOM} for the \ac{BM} problem for the first time. These \ac{EEOM} allow one to relate the static $\kappa = 0$ quantities obtained by the \ac{FRG} in chapter \ref{cha:fRG} to dynamical objects of interest. We focused on the \ac{EEOM} for the average position of the particle $\lan x(t)\ran$ (\ref{eq:QEOM}) as well as the \ac{EEOM} for the variance $\lan x(t)x(t)\ran_{C}$ (\ref{eq:EEOM Variance}) and the covariance $\lan x(t_1)x(t_2)\ran_{C}$ (\ref{eq:EEOM Covariance}). We verified that these \ac{EEOM} reduce to the appropriate static quantities in equilibrium, resulting in the \ac{FRG} passing a significant consistency check. We then computed these \ac{EEOM} as the system relaxes towards equilibrium in several non-trivial potentials and compared the results to direct numerical simulations of the Langevin equation (\ref{eq:Langevin_sigma_dimless}). We found that the \ac{FRG} is capable of accurately describing this relaxation and can capture non-trivial dynamics such as the variance overshooting its equilibrium value. We found the accuracy of the \ac{FRG} generically decreases as the temperature of the thermal bath is decreased. \\

\noindent Part \ref{part:Cosmo} was concerned with examining the behaviour of stochastic processes in the early universe. In contrast to part \ref{part:Meso} the appropriate Langevin equations, (\ref{eq:deSitter_stochastic_equations}) \& (\ref{eq:Langevin_sigma_dimless}), described the evolution of scalar fields rather than particles and the noise wasn't due to a thermal bath but instead inherently quantum fluctuations stretched to large scales that could be treated as \emph{effectively} thermal. \\

\noindent In chapter \ref{cha:Inflation} we reviewed a period of cosmic inflation in the early universe driven by a scalar field called the inflaton. Paying particular care to the ADM formalism for non-linear, superhorizon perturbations we obtained the \ac{H-J} equation (\ref{eq: H-J equation}) for the Hubble expansion rate $H(\phi)$ which together with the inflaton evolution equation (\ref{eq: dphidt HJ}) fully describes the dynamics in the absence of quantum backreaction. To incorporate the effect of initially short wavelength quantum fluctuations being stretched to super-horizon scales and backreacting on the inflaton dynamics we reviewed the stochastic inflation formalism. Assuming the short-wavelength mode functions can be well approximated by those of de Sitter we arrived at the Langevin equation for the inflaton (\ref{eq:deSitter_stochastic_equations}). We also reviewed the stochastic-$\delta \mathcal{N}$ formalism and obtained the semi-classical expression for the coarse-grained curvature perturbation. \\

\noindent In chapter \ref{cha:PBH} we examined the prospect of forming \ac{PBHs} from these inflationary perturbations. After reviewing how large perturbations can form \ac{PBHs} due to gravitational collapse when they re-enter the horizon in the post-inflationary era, we expressed the mass fraction of \ac{PBHs} at the time they are formed in terms of the coarse-grained curvature perturbation in equation (\ref{eq:massfracdef_both}). We then applied the \ac{H-J} formalism to a plateau region in the potential corresponding to a period of \ac{USR}. We found an exact expression for the abundance of \ac{PBHs} formed due to a plateau region, (\ref{eq:massfrac_exactFULL}), and surprisingly found that before quantum backreaction is the dominant effect on the inflaton that \ac{PBHs} will be overproduced. Phrased another way, one can produce enough \ac{PBHs} during inflation to satiate observational and theoretical constraints while the classical drift is still the dominant effect on the inflaton. This is in contrast with the assumptions of previous works on the subject. We demonstrated that the approximations we made served to generically \emph{underestimate} the abundance of \ac{PBHs} therefore reinforcing our conclusions. We also examined an inflationary period characterised by an inflection point which also found overproduction of \ac{PBHs} while dynamics were still semi-classical. \\

\noindent In chapter \ref{cha:RG_spect} we turn our attention to a spectator field in an inflationary background. As the spectator field does not directly influence the inflationary dynamics its Langevin equation (\ref{eq:Langevin_sigma_dimless}) resembles the \ac{BM} overdamped Langevin equation (\ref{eq:langevindimless}) we examined in part \ref{part:Meso}. We were therefore able to modify the \ac{FRG} machinery developed for \ac{BM} and applied it to the spectator field. In addition to the straightforward modifications to the \ac{EEOM} for the one- and two-point functions from the \ac{BM} versions we derived the \ac{EEOM} for the third central moment $\lan \sigma(\alpha)^3\ran_{C}$, (\ref{eq:FRG_threepoint}). We solved this for a couple of different potentials and while we were able to qualitatively capture the correct behaviour we were unable to found good quantitative agreement. We therefore surmised that we were at the limits of the \ac{NLO} approximation of the \ac{FRG} derivative expansion. We discussed how cosmological observables such as the power spectrum and spectral tilt could be obtained from quantities computed by the \ac{FRG} resulting in equation (\ref{eq:Power_spec_sig}). We highlighted how this means that wildly different potentials, e.g. a simple harmonic compared to one with loads of bumpy features, will give the same prediction for the spectral tilt. This means that one should be careful when making inferences about the exact form of the spectator potential from cosmological observables. We went on to compute \ac{FPT} quantities like the average time, and variance in time taken to reach equilibrium for a spectator field from the \ac{FRG} \ac{EEOM} and demonstrated remarkable agreement for non-trivial potentials with barriers. This represents an important first step towards using \ac{FRG} techniques to compute \ac{FPT} quantities for the inflaton which, as explained before, yields the coarse-grained curvature perturbation. \\

\noindent To conclude we have examined stochastic processes across a wide range of scales focusing on the \ac{BM} of both particles and scalar fields in the early universe. We have been able to utilise non-perturbative techniques from \ac{QFT} to derive \ac{EEOM} for the one-, two- and three-point function of a particle in a thermal bath and a spectator field during inflation. These \ac{EEOM} are a direct alternative to solving directly either the Langevin equations, (\ref{eq:langevindimless}) \& (\ref{eq:Langevin_sigma}), or the \ac{F-P} equations (\ref{eq:FP1}) \& (\ref{eq: F-P_spect}). As well as having obvious value in and of themselves, these \ac{EEOM} can be used to describe \ac{FPT} quantities with clear applications to e.g. barrier escape and relaxation time. We also worked directly with the inflaton and, using the \ac{H-J} formulation of stochastic inflation, were able to compute the abundance of \ac{PBHs} from a period of \ac{USR} inflation. We arrived at the surprising conclusion that inflation must always be semi-classical to not overproduce \ac{PBHs}.\\

\noindent In spirit with the quote at the start of this chapter we will finish, not with conclusions but with speculations of new beginnings. An obvious future direction would be to examine the behaviour of two- and three-dimensional Langevin equations and generalise the \ac{FRG} results accordingly. Perhaps more pertinent would be to applied the vertex expansion of the \ac{FRG} to \ac{BM} and see if this yields more accurate \ac{EEOM} at lower temperatures. A different but equally pertinent extension would be to consider a Langevin equation with \emph{multiplicative} noise, i.e. noise that depends on the position of the particle/field as well as time. Successfully applying the \ac{FRG} to this scenario would enable one to derive \ac{EEOM} for the inflaton and so derive the primordial curvature perturbation through the stochastic-$\delta N$ formalism. There are also important, unanswered questions surrounding the formation of \ac{PBHs}. An important next step would be to relate the mass fraction as computed in this work using the coarse-grained curvature perturbation to the mass fraction using the density contrast to get a more precise value. This extension could also potentially incorporate going beyond Press-Schecter to peaks theory. It would also be desirable to resolve accurately the full PDF for the the coarse-grained curvature perturbation outside for more generic periods of \ac{USR}. There are also unanswered questions about the validity of the use of the de Sitter mode functions in the noise term of stochastic inflation. Going beyond this approximation in the context of \ac{USR} while utilising the \ac{H-J} is another exciting research direction. 

\vspace{0.5cm}
\begin{center}{\slshape    
No, this is not the beginning of a new chapter in my life; this is the beginning of a new book!\\ That first book is already closed, ended, and tossed into the seas;\\ this new book is newly opened, has just begun!\\ Look, it is the first page! \\And it is a beautiful one!} \\ \medskip
--- C. JoyBell C 
\end{center}
\vspace{0.5cm}

\cleardoublepage 

\ctparttext{\color{black}If brevity is the soul of wit then prepare to read the least intelligent part of this thesis as we show in much greater detail computations and results omitted from the main chapters.} 
\appendix

\part{Appendices} 
\chapter{\label{app:Cosmo_perts}Cosmological Perturbation Theory}
\acresetall 
In this appendix we will outline in more detail how to treat perturbations from homogeneity and isotropy in both the linear and non-linear regimes. We will also offer a more technical and detailed look at squeezing than presented in the main body of the text.
\section{\label{sec:App_lin_pert}Linear perturbations around FLRW}
In standard cosmological perturbation theory we define perturbations around the homogeneous background solutions for the metric $\bar{g}_{\mu\nu}(t)$ -- given by the FLRW metric (\ref{eq:FLRW}) -- and the inflaton $\bar{\varphi}(t)$ perturbations on top:
\begin{eqnarray}
g_{\mu\nu}(t,\vec{x}) = \bar{g}_{\mu\nu} + \delta g_{\mu\nu}(t,\vec{x}), \quad \varphi(t,\vec{x}) = \bar{\varphi} (t) + \delta \varphi (t,\vec{x})
\end{eqnarray}
i.e. we can split the perturbations into those in the gravity sector, metric perturbations, and those in the stress-energy tensor, inflaton perturbations. 
\subsection{Metric Perturbations} 
We can write the perturbed FLRW metric like so:
\begin{eqnarray}
\mathrm{d}s^2 =  -\lb 1+2\Phi \rb \mathrm{d}t^2 +2a(t) B_i\mathrm{d}x^i\mathrm{d}t + a^2(t)\lsb \lb 1-2\Psi\rb \delta_{ij} +2E_{ij}\rsb\mathrm{d}x^i\mathrm{d}x^j \label{eq:perturb line element}
\end{eqnarray}
where $\Phi$ is often called the lapse function and $B_i$ the shift vector. \\
In real space, the Scalar-Vector-Tensor decomposition of the metric allows us to split these perturbations -- $\Phi$, $B_i$, $\Psi$ \& $E_{ij}$ -- into linear combinations of scalar, vector and tensor components. As we will only be interested in scalar perturbations in this work we can define those perturbations not already manifestly as scalars like so:
\begin{eqnarray}
B_{i} \equiv \partial_i B, \quad E_{ij} \equiv \partial_i\partial_j E
\end{eqnarray}
This means that the intrinsic Ricci scalar curvature of constant time hypersurfaces is:
\begin{eqnarray}
R_{(3)} = \dfrac{4}{a^2}\nabla^2\Psi
\end{eqnarray}
which is why $\Psi$ is often called the curvature perturbation. \\
Scalar fluctuations considered here are not gauge invariant. Consider the gauge transformation:
\begin{subequations}
\begin{align}
t &\rightarrow t + \alpha \\
x^i &\rightarrow x^i + \delta^{ij} \beta_j 
\end{align}   \label{eq:gauge_transform} 
\end{subequations}
which tells us that the scalar metric perturbations transform as:
\begin{subequations}
\begin{align}
\Phi &\rightarrow \Phi - \dot{\alpha}\\
B &\rightarrow B + \alpha /a -a\dot{\beta} \\
E &\rightarrow E - \beta \\
\Psi &\rightarrow \Psi + H\alpha 
\end{align}   \label{eq:scalar_gauge_transform} 
\end{subequations}
This means that depending on the gauge one would compute different values of e.g. $B$ which isn't very helpful. We therefore note two important gauge-invariant quantities known as the Bardeen variables \cite{Bardeen1980}:
\begin{subequations}
\begin{align}
\Phi_B &\equiv \Phi - \dfrac{\mathrm{d}}{\mathrm{d}t}\lsb a^2\lb \dot{E}-B/a\rb\rsb\\
\Psi_B &\equiv \Psi + a^2 H\lb \dot{E} - B/a\rb
\end{align}   \label{eq:Bardeen_vars} 
\end{subequations}
\subsection{Stress-Energy Perturbations}
It is not just metric perturbations we must consider but perturbations in the fluid itself. Density and pressure perturbations transform under temporal gauge transformations as :
\begin{subequations}
\begin{align}
\delta \rho &\rightarrow \delta \rho \dot{\bar{\rho}}\alpha\\
\delta p &\rightarrow \delta p - \dot{\bar{p}}\alpha
\end{align}   \label{eq:densityandpressure_pert} 
\end{subequations}
Adiabatic pressure perturbations are defined as:
\begin{eqnarray}
\delta p_{\mathrm{ad}} \equiv \dfrac{\dot{\bar{p}}}{\dot{\bar{\rho}}}\delta\rho
\end{eqnarray}
which means the entropic -- more commonly called the non-adiabatic -- part of the pressure perturbations is gauge-invariant
\begin{eqnarray}
\delta p_{en}\equiv \delta p - \dfrac{\dot{\bar{p}}}{\dot{\bar{\rho}}}\delta\rho
\end{eqnarray}
If we also consider the scalar part of the 3-momentum density $(\delta q)_{,i}$ this transforms as:
\begin{eqnarray}
\delta q \rightarrow \delta q + (\bar{\rho} + \bar{p})\alpha
\end{eqnarray}
which allows us to define the gauge-invariant comoving density perturbation:
\begin{eqnarray}
\delta \rho_m \equiv \delta\rho -3H\delta q
\end{eqnarray}
We can then define two important gauge-invariant quantities which are formed from combinations of these fluid and metric perturbations. The first is the \textit{curvature perturbation on uniform density hypersurfaces}:
\begin{eqnarray}
-\zeta \equiv \Psi + \dfrac{H}{\dot{\bar{\rho}}}\delta\rho
\end{eqnarray}
and the \textit{comoving curvature perturbation}:
\begin{eqnarray}
\mathcal{R} \equiv \Psi - \dfrac{H}{\bar{\rho} + \bar{p}}\delta q
\end{eqnarray}
\subsection{Perturbed Einstein Equation}
To relate the metric and stress-energy perturbations we consider the perturbed Einstein equations\footnote{In units where $\hslash = c = 1$.}:
\begin{eqnarray}
\delta G_{\mu\nu} = 8\pi G \delta T_{\mu\nu}
\end{eqnarray}
At linear order this leads to the energy and momentum constraints respectively:
\begin{eqnarray}
3H\lb \dot{\Psi} + H\Phi\rb + \dfrac{k^2}{a^2}\lsb \Psi + H\lb a^2 \dot{E} -aB\rb\rsb &=& -4\pi G \delta \rho \\
\dot{\Psi} + H\Phi &=& -4\pi G \delta q
\end{eqnarray}
which can be combined to yield the gauge invariant Poisson equation:
\begin{eqnarray}
\dfrac{k^2}{a^2}\Psi_{B} = -4\pi G \delta \rho_m
\end{eqnarray}
The Einstein equations also yield two dynamical evolution equations:
\begin{eqnarray}
\ddot{\Psi} + 3H\dot{\Psi} + H\dot{\Phi} + (3H^2 + 2\dot{H})\Phi &=& 4\pi G \lb \delta p - \dfrac{2}{3}k^2 \delta\Sigma\rb \\
\lb\partial_t + 3H \rb \lb \dot{E} - B/a\rb &=& 8\pi G \delta \Sigma
\end{eqnarray}
The last of which can be rewritten in terms of the Bardeen variables:
\begin{eqnarray}
\Psi_{B} - \Phi_{B} = 8\pi G a^2\delta \Sigma
\end{eqnarray}
which shows that in the absence of anisotropic stress ($\delta \Sigma = 0$) that $\Psi_{B} = \Phi_{B}$.\\
Conservation of energy-momentum , $\nabla_{\mu} T_{\mu\nu} = 0$, gives the continuity equation and the Euler equation:
\begin{eqnarray}
\dot{\delta \rho} + 3H\lb \delta \rho + \delta p\rb &=& \dfrac{k^2}{a^2}\delta q + \lb \bar{\rho} + \bar{p}\rb \lsb 3\dot{\Psi} + k^2 \lb \dot{E} + B/a\rb\rsb \label{eq:perturb_continuity}\\
\dot{\delta q} + 3H\delta q &=& -\delta p + \dfrac{2}{3}k^2 \delta \Sigma - (\bar{\rho} + \bar{p})\Phi \label{eq:perturb_Euler}
\end{eqnarray}
The continuity equation (\ref{eq:perturb_continuity}) can be expressed in terms of the curvature perturbation on uniform-density hypersurfaces $\zeta$:
\begin{eqnarray}
\dot{\zeta} &=& -H \lcb \dfrac{\delta p_{en}}{\bar{\rho} + \bar{p}} + \dfrac{k^2}{3a^2H^2}\lsb \zeta - \Psi_{B}\lb 1 - \dfrac{2\bar{\rho}}{9\lb \bar{\rho} + \bar{p}\rb}\dfrac{k^2}{a^2H^2} \rb  \rsb\rcb \\
&\rightarrow &-H\dfrac{\delta p_{en}}{\bar{\rho} + \bar{p}}
\end{eqnarray}
where in the second line we have considered the superhorizon limit $k/(aH) \ll 1$. It is therefore clear that for adiabatic perturbations ($\delta p_{en}$) that the curvature perturbation $\zeta$ is constant on super-horizon scales.
\subsection{Different Gauges}
There are many widely used gauges each with its own advantages. Here we will briefly define some and write out their appropriate equations.
\subsubsection{Synchronous Gauge}
This is defined so that their is no perturbation in the time coordinate:
\begin{eqnarray}
\Phi = B = 0 \label{eq:Synchro_defn}
\end{eqnarray}
so that
\begin{eqnarray}
\mathrm{d}s^2 =  - \mathrm{d}t^2 + a^2(t)\lsb \lb 1-2\Psi\rb \delta_{ij} +2E_{ij}\rsb\mathrm{d}x^i\mathrm{d}x^j \label{eq:Synchro_line element}
\end{eqnarray}
With Einstein Equations:
\begin{subequations}
\begin{align}
3H\lb \dot{\Psi} + H\Phi\rb + \dfrac{k^2}{a^2}\lsb \Psi + Ha^2 \dot{E}\rsb &= -4\pi G \delta \rho \\
\dot{\Psi}  &= -4\pi G \delta q \\
\ddot{\Psi} + 3H\dot{\Psi} &= 4\pi G \lb \delta p - \dfrac{2}{3}k^2 \delta\Sigma\rb \\
\lb\partial_t + 3H \rb \dot{E}  &= 8\pi G \delta \Sigma
\end{align}   \label{eq:Sychro_Einstein} 
\end{subequations}
Also with continuity equations:
\begin{subequations}
\begin{align}
\dot{\delta \rho} + 3H\lb \delta \rho + \delta p\rb &= \dfrac{k^2}{a^2}\delta q + \lb \bar{\rho} + \bar{p}\rb \lsb 3\dot{\Psi} + k^2 \dot{E} \rsb \\
\dot{\delta q} + 3H\delta q &= -\delta p + \dfrac{2}{3}k^2 \delta \Sigma 
\end{align}   \label{eq:Sychro_conserv} 
\end{subequations}

\subsubsection{Newtonian Gauge}
As the name suggests this gauge reduces to Newtonian gravity in the small-scale limit. It is defined by 
\begin{eqnarray}
B = E = 0 \label{eq:Newt_defn}
\end{eqnarray}
so that
\begin{eqnarray}
\mathrm{d}s^2 =  -\lb 1+2\Phi \rb \mathrm{d}t^2  + a^2(t)\lsb \lb 1-2\Psi\rb \delta_{ij}\rsb\mathrm{d}x^i\mathrm{d}x^j \label{eq:Newt_line element}
\end{eqnarray}
With Einstein Equations:
\begin{subequations}
\begin{align}
3H\lb \dot{\Psi} + H\Phi\rb + \dfrac{k^2}{a^2} \Psi  &= -4\pi G \delta \rho \\
\dot{\Psi} + H\Phi &= -4\pi G \delta q \\
\ddot{\Psi} + 3H\dot{\Psi} + H\dot{\Phi} + (3H^2 + 2\dot{H})\Phi &= 4\pi G \lb \delta p - \dfrac{2}{3}k^2 \delta\Sigma\rb \\
\dfrac{\Psi - \Phi}{a^2} &= 8\pi G \delta \Sigma
\end{align}   \label{eq:Newt_Einstein} 
\end{subequations}
Also with continuity equations:
\begin{subequations}
\begin{align}
\dot{\delta \rho} + 3H\lb \delta \rho + \delta p\rb &= \dfrac{k^2}{a^2}\delta q + 3\lb \bar{\rho} + \bar{p}\rb \dot{\Psi} \\
\dot{\delta q} + 3H\delta q &= -\delta p + \dfrac{2}{3}k^2 \delta \Sigma - (\bar{\rho} + \bar{p})\Phi
\end{align}   \label{eq:Newt_conserv} 
\end{subequations}

\subsubsection{Uniform density gauge}
As the name suggests it is defined by:
\begin{eqnarray}
\delta \rho = 0 
\end{eqnarray}
This is actually not enough so we also take $E = 0$. Noticing that $\zeta = -\Psi$:
\begin{eqnarray}
\mathrm{d}s^2 =  -\lb 1+2\Phi \rb \mathrm{d}t^2 +2a(t) B_i\mathrm{d}x^i\mathrm{d}t + a^2(t)\lsb \lb 1+2\zeta\rb \delta_{ij}\rsb\mathrm{d}x^i\mathrm{d}x^j \label{eq:Uniform_line element}
\end{eqnarray}
With Einstein Equations:
\begin{subequations}
\begin{align}
3H\lb -\dot{\zeta} + H\Phi\rb - \dfrac{k^2}{a^2}\lsb \zeta + aHB\rsb &= 0 \\
-\dot{\zeta} + H\Phi &= -4\pi G \delta q \\
-\ddot{\zeta} - 3H\dot{\zeta} + H\dot{\Phi} + (3H^2 + 2\dot{H})\Phi &= 4\pi G \lb \delta p - \dfrac{2}{3}k^2 \delta\Sigma\rb \\
\lb\partial_t + 3H \rb B/a  + \dfrac{\zeta + \Phi}{a^2}&= -8\pi G \delta \Sigma
\end{align}   \label{eq:Uniform_Einstein} 
\end{subequations}
Also with continuity equations:
\begin{subequations}
\begin{align}
3H \delta p &= \dfrac{k^2}{a^2}\delta q + \lb \bar{\rho} + \bar{p}\rb \lsb -3\dot{\zeta} + k^2 B/a \rsb \\
\dot{\delta q} + 3H\delta q &= -\delta p + \dfrac{2}{3}k^2 \delta \Sigma - (\bar{\rho} + \bar{p})\Phi
\end{align}   \label{eq:Uniform_conserv} 
\end{subequations}

\subsubsection{Comoving gauge}
This is defined by the vanishing of the scalar momentum density:
\begin{eqnarray}
\delta q = 0, \quad E = 0
\end{eqnarray}
and we notice we can set $-\Psi = \mathcal{R}$. Then:
\begin{eqnarray}
\mathrm{d}s^2 =  -\lb 1+2\Phi \rb \mathrm{d}t^2 +2a(t) B_i\mathrm{d}x^i\mathrm{d}t + a^2(t)\lsb \lb 1+2\mathcal{R}\rb \delta_{ij}\rsb\mathrm{d}x^i\mathrm{d}x^j \label{eq:comov_line element}
\end{eqnarray}
With Einstein Equations:
\begin{subequations}
\begin{align}
3H\lb -\dot{\mathcal{R}} + H\Phi\rb - \dfrac{k^2}{a^2}\lsb \mathcal{R}  + aHB\rsb &= -4\pi G \delta \rho \\
-\dot{\mathcal{R}} + H\Phi &= 0 \\
-\ddot{\mathcal{R}} - 3H\dot{\mathcal{R}} + H\dot{\Phi} + (3H^2 + 2\dot{H})\Phi &= 4\pi G \lb \delta p - \dfrac{2}{3}k^2 \delta\Sigma\rb \\
\lb\partial_t + 3H \rb B/a + \dfrac{\mathcal{R} + \Phi}{a^2} &= -8\pi G \delta \Sigma
\end{align}   \label{eq:comov_Einstein} 
\end{subequations}
Also with continuity equations:
\begin{subequations}
\begin{align}
\dot{\delta \rho} + 3H\lb \delta \rho + \delta p\rb &= \dfrac{k^2}{a^2}\delta q + \lb \bar{\rho} + \bar{p}\rb \lsb -3\dot{\mathcal{R}} + k^2 B/a \rsb \\
0 &= -\delta p + \dfrac{2}{3}k^2 \delta \Sigma - (\bar{\rho} + \bar{p})\Phi
\end{align}   \label{eq:comov_conserv} 
\end{subequations}
The two equations = 0 can be combined into:
\begin{eqnarray}
\Phi = \dfrac{-\delta p + 2k^2\delta \Sigma /3}{\bar{\rho} + \bar{p}}, \quad kB = \dfrac{4\pi Ga^2\delta \rho -k^2 \mathcal{R}}{aH}
\end{eqnarray}

\subsubsection{Spatially-flat gauge}
A convenient gauge for inflationary perturbation is the spatially-flat gauge defined as:
\begin{eqnarray}
\Psi = E = 0 \label{eq:spat-flat_defn}
\end{eqnarray}
so that
\begin{eqnarray}
\mathrm{d}s^2 =  -\lb 1+2\Phi \rb \mathrm{d}t^2 +2a(t) B_i\mathrm{d}x^i\mathrm{d}t + a^2(t) \delta_{ij}\mathrm{d}x^i\mathrm{d}x^j \label{eq:spat-flat_line element}
\end{eqnarray}
During inflation therefore all scalar perturbations are described by $\delta \varphi$.\\
With Einstein Equations:
\begin{subequations}
\begin{align}
3H^2\Phi - \dfrac{k^2}{a^2}\lb aHB\rb &= -4\pi G \delta \rho \\
 H\Phi &= -4\pi G \delta q \\
H\dot{\Phi} + (3H^2 + 2\dot{H})\Phi &= 4\pi G \lb \delta p - \dfrac{2}{3}k^2 \delta\Sigma\rb \\
\lb\partial_t + 3H \rb B/a  + \dfrac{\Phi}{a^2}&= -8\pi G \delta \Sigma
\end{align}   \label{eq:spat-flat_Einstein} 
\end{subequations}
Also with continuity equations:
\begin{subequations}
\begin{align}
\dot{\delta \rho} + 3H\lb \delta \rho + \delta p\rb &= \dfrac{k^2}{a^2}\delta q + \lb \bar{\rho} + \bar{p}\rb \lsb 3\dot{\Psi} + k^2 B/a\rsb \\
\dot{\delta q} + 3H\delta q &= -\delta p + \dfrac{2}{3}k^2 \delta \Sigma - (\bar{\rho} + \bar{p})\Phi
\end{align}   \label{eq:spat-flat_conserv} 
\end{subequations}
\section{\label{app:squeezing}Squeezing and classicalisation of perturbations}
\begin{figure}[t!]
    \centering
    \includegraphics[width = 0.45\linewidth]{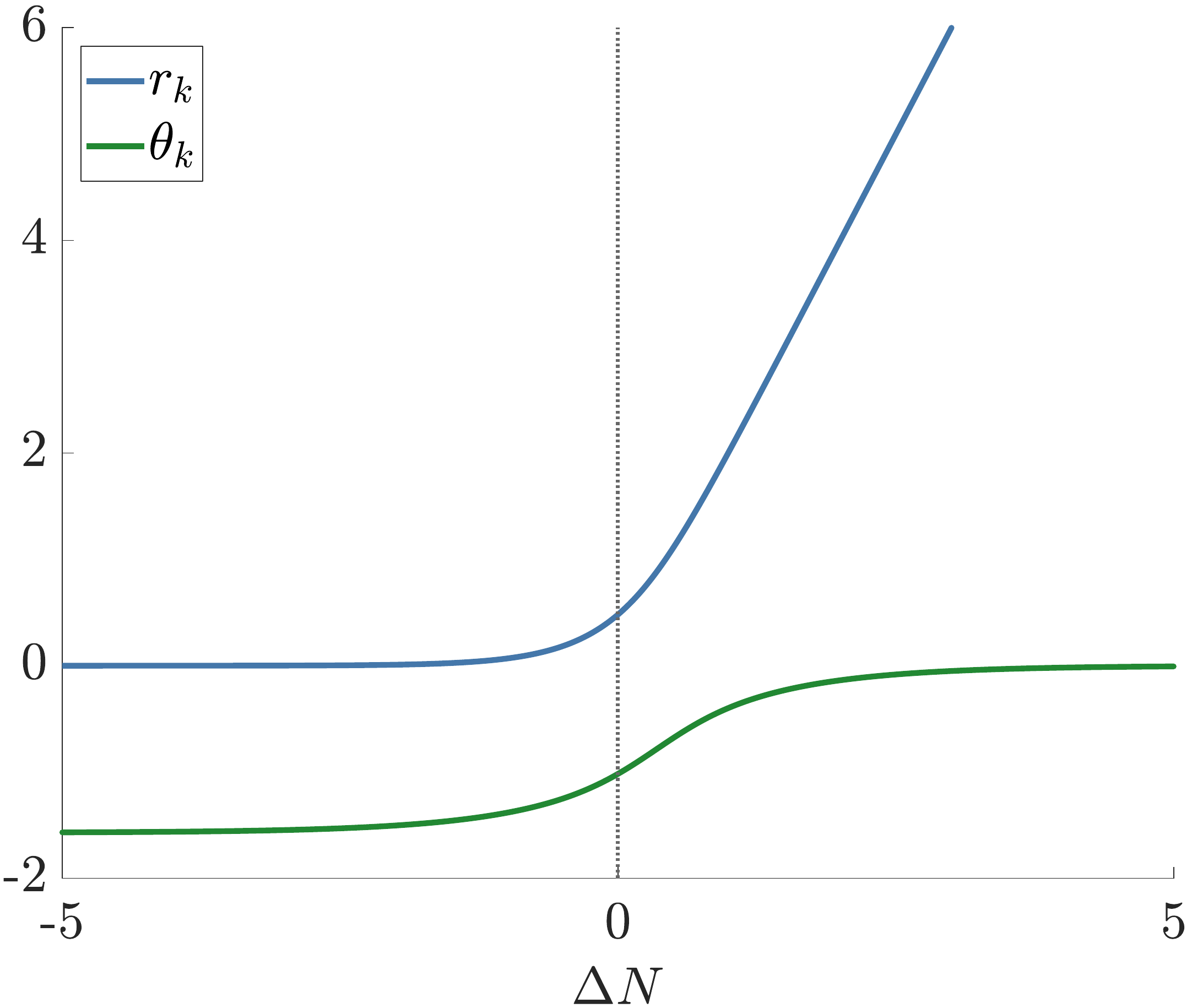}
    \includegraphics[width = 0.45\linewidth]{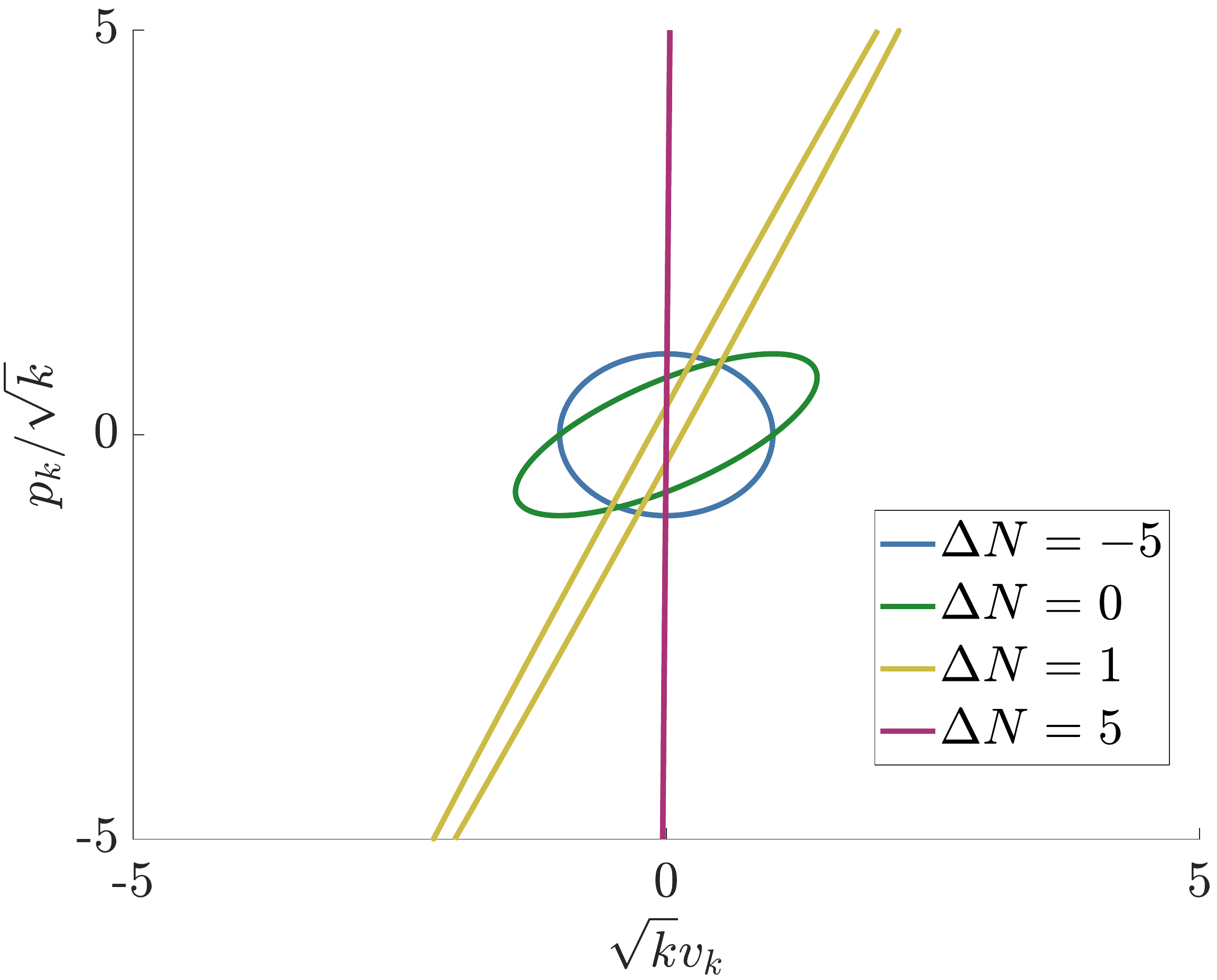}
    \caption[Evolution of squeezing parameters and phase space contours for de Sitter]{How de Sitter space squeezes perturbations. In the left plot we show how $r_k$ and $\theta_k$ vary as the perturbation leaves the horizon. In the right plot we show the $\sqrt{2}\sigma$ contour plot of the Wigner function for various values $\Delta N$. }
    \label{fig:squeezing_ellipse_dS}
\end{figure}

So far we have discussed how to compute linear perturbations from a homogeneous inflationary background and shown how for de Sitter space that while the perturbations in the inflaton asymptote to a constant on super-horizon scales, the perturbations in its momenta rapidly decay -- see right column of Fig.~\ref{fig:M-S_deSitter}. In this section we will more formally discuss how these perturbations are placed into a two-mode squeezed state and how this means these perturbations can be treated as effectively classical. This section is adapted from \cite{Martin2021}.\\
More concretely we consider the following quantum Hamiltonian in terms of a field $\hat{v}_{\vec{k}}$ and its conjugate momentum $\hat{p}_{\vec{k}}$:
\begin{eqnarray}
\hat{H} = \int_{\mathbb{R}^{3+}}\mathrm{d}\vec{k}\lsb \hat{p}_{\vec{k}}\hat{p}_{\vec{k}}^{\dagger} + \omega^2 \hat{v}_{\vec{k}}\hat{v}_{\vec{k}}^{\dagger}\rsb \label{eq:Hamiltonian_v_p}
\end{eqnarray}
where $\hat{v}_{\vec{k}}$ \& $\hat{p}_{\vec{k}}$ obey the following commutation relations:
\begin{eqnarray}
\lsb \hat{v}_{\vec{k}}, \hat{p}_{\vec{k}}^{\dagger}\rsb = i\delta (\vec{k}-\vec{k}'), \quad \lsb\hat{v}_{\vec{k}}, \hat{v}_{\vec{k}}^{\dagger} \rsb = \lsb\hat{p}_{\vec{k}}, \hat{p}_{\vec{k}}^{\dagger} \rsb = 0
\end{eqnarray}
One can therefore use Hamilton's equations to derive the Mukhanov-Sasaki equation for perturbations (\ref{eq:M-S equation}) from (\ref{eq:Hamiltonian_v_p}). In the helicity basis one can decompose the fields $\hat{v}_{\vec{k}}$ \& $\hat{p}_{\vec{k}}$ onto creation and annihilation operators as:
\begin{eqnarray}
\hat{v}_{\vec{k}} = \dfrac{1}{\sqrt{2k}}\lb \hat{c}_{\vec{k}} + \hat{c}_{-\vec{k}}^{\dagger}\rb, \quad  \hat{p}_{\vec{k}} = -i \sqrt{\dfrac{k}{2}}\lb \hat{c}_{\vec{k}} - \hat{c}_{-\vec{k}}^{\dagger}\rb \label{eq:v,p_ladderopers}
\end{eqnarray}
which obey the commutation relation $[\hat{c}_{\vec{k}} ,\hat{c}_{\vec{k}}^{\dagger}] = \delta (\vec{k}-\vec{k}') $. It is worth noting at this stage that the fields $\hat{v}_{\vec{k}}$ \& $\hat{p}_{\vec{k}}$ are not Hermitian and it is therefore worthwhile to split them into their real and imaginary parts:
\begin{eqnarray}
\hat{v}_{\vec{k}}^{R} = \dfrac{\hat{v}_{\vec{k}} + \hat{v}_{\vec{k}}^{\dagger}}{\sqrt{2}}, \quad \hat{v}_{\vec{k}}^{I} = \dfrac{\hat{v}_{\vec{k}} - \hat{v}_{\vec{k}}^{\dagger}}{i\sqrt{2}},\quad \hat{p}_{\vec{k}}^{R} = \dfrac{\hat{p}_{\vec{k}} + \hat{p}_{\vec{k}}^{\dagger}}{\sqrt{2}},\quad \hat{p}_{\vec{k}}^{I} = \dfrac{\hat{p}_{\vec{k}} - \hat{p}_{\vec{k}}^{\dagger}}{i\sqrt{2}}
\end{eqnarray}
which are Hermitian. This split to obtain Hermitian operators is not unique but what is nice is it splits the system into two independent subspaces, i.e. the Hamiltonian is sum separable:
\begin{eqnarray}
\hat{H} = \dfrac{1}{2}\int_{\mathbb{R}^{3+}} \mathrm{d}^3\vec{k}\sum_{s=R,I}\lsb (\hat{p}_{\vec{k}}^s)^2 + \omega^2 (\hat{v}_{\vec{k}}^s)^2\rsb 
\end{eqnarray}
Going forward we will focus on the behaviour of one of these partitions e.g. the real parts of the field $\hat{v}_{\vec{k}}^{R}$ \& $\hat{p}_{\vec{k}}^{R}$ but drop the explicit R. At this stage we note that because the Hamiltonian is quadratic, the dynamics it generates are linear and admits Gaussian solutions. Such states are therefore completely defined by their two-point function. With this in mind the introduce the covariance matrix elements:
\begin{subequations}
\begin{align}
   \gamma_{11} &= 2k\lan (\hat{v}_{\vec{k}}^{R})^2\ran = k\lan \lcb \hat{v}_{\vec{k}},\hat{v}_{\vec{k}}^{\dagger}\rcb\ran\\
  \gamma_{12} = \gamma_{21}  &= \lan \hat{v}_{\vec{k}}^{R}\hat{p}_{\vec{k}}^{R} + \hat{p}_{\vec{k}}^{R}\hat{v}_{\vec{k}}^{R}\ran = \lan  \hat{v}_{\vec{k}}\hat{p}_{\vec{k}}^{\dagger} + \hat{p}_{\vec{k}}\hat{v}_{\vec{k}}^{\dagger} \ran\\
  \gamma_{22}  &= \dfrac{2}{k} \lan (\hat{p}_{\vec{k}}^{R})^2\ran = \dfrac{1}{k} \lan \lcb \hat{p}_{\vec{k}},\hat{p}_{\vec{k}}^{\dagger}\rcb\ran
\end{align}
\end{subequations}
The creation and annihilation operators can also be rewritten in terms of these matrix elements:
\begin{subequations}
\begin{align}
    \lan \lcb \hat{c}_{\vec{k}}, \hat{c}_{\vec{k}}^{\dagger}\rcb\ran &= \lan \lcb \hat{c}_{-\vec{k}}, \hat{c}_{-\vec{k}}^{\dagger}\rcb\ran = \dfrac{\gamma_{11}+ \gamma_{22}}{2} \\
    \lan \lcb \hat{c}_{\vec{k}}, \hat{c}_{-\vec{k}}\rcb\ran &= \dfrac{\gamma_{11}-\gamma_{22}}{2} + i\gamma_{12}\\
    \lan \lcb \hat{c}_{\vec{k}}^{\dagger}, \hat{c}_{-\vec{k}}\rcb\ran &= \dfrac{\gamma_{11}-\gamma_{22}}{2} -i\gamma_{12}
\end{align}
\end{subequations}
and all others vanish. Note that these covariance matrix elements contain all information about the state. For instance the purity is given by:
\begin{eqnarray}
\hat{\rho} = \dfrac{1}{\gamma_{11}\gamma_{22}-\gamma_{12}^2}
\end{eqnarray}
In the Heisenberg picture the equation of motion for the ladder operators can be obtained by substituting (\ref{eq:v,p_ladderopers}) into (\ref{eq:Hamiltonian_v_p}) and in matricial form they are:
\begin{eqnarray}
\dfrac{\mathrm{d}}{\mathrm{d}\tau} \begin{pmatrix}
\hat{c}_{\vec{k}} \\
\hat{c}_{-\vec{k}}^{\dagger}  
\end{pmatrix} =  \begin{pmatrix}
-i\dfrac{k}{2} \lsb \dfrac{\omega^2}{k^2} +1\rsb & -i\dfrac{k}{2} \lsb \dfrac{\omega^2}{k^2} -1\rsb\\
i\dfrac{k}{2} \lsb \dfrac{\omega^2}{k^2} -1\rsb  & i\dfrac{k}{2} \lsb \dfrac{\omega^2}{k^2} +1\rsb
\end{pmatrix} \begin{pmatrix}
\hat{c}_{\vec{k}} \\
\hat{c}_{-\vec{k}}^{\dagger}  
\end{pmatrix} \label{eq:diff_eqn_ladders}
\end{eqnarray}
As this system is linear it can be solved with a linear transformation known as a Bogoliubov transformation:
\begin{eqnarray}
\begin{pmatrix}
\hat{c}_{\vec{k}} (\tau) \\
\hat{c}_{-\vec{k}}^{\dagger}  (\tau)
\end{pmatrix} =  \begin{pmatrix}
u_{\vec{k}}(\tau ) & w_{\vec{k}}(\tau )\\
w_{-\vec{k}}^{*}(\tau )  & u_{-\vec{k}}^{*}(\tau )
\end{pmatrix} \begin{pmatrix}
\hat{c}_{\vec{k}} (\tau_{in})\\
\hat{c}_{-\vec{k}}^{\dagger}  (\tau_{in})
\end{pmatrix}
\end{eqnarray}
where $u_{\vec{k}}$ \& $w_{\vec{k}}$ are the two complex Bogoliubov coefficients. In order to ensure that $[\hat{c}_{\vec{k}} ,\hat{c}_{\vec{k}}^{\dagger}] = \delta (\vec{k}-\vec{k}') $ is satisfied at all times, the Bogoliubov coefficients satisfy:
\begin{eqnarray}
|u_{\vec{k}} |^2 -|w_{-\vec{k}} |^2 = 1
\end{eqnarray}
Solving the problem at hand therefore simply reduces to computing the Bogoliubov coefficients which also satisfy the differential equation (\ref{eq:diff_eqn_ladders}) and have initial conditions $u_{\pm\vec{k}}(\tau_{in}) = 1$ and $w_{\pm\vec{k}}(\tau_{in}) = 0$. Note that because of statistical isotropy the Bogoliubov coefficients depend only on the norm of $\vec{k}$ so from now on they will be expressed in terms of the norm only. The two first order differential equations can be reformulated in terms of a single second order differential equation:
\begin{eqnarray}
\dfrac{\mathrm{d}^2}{\mathrm{d}\tau^2}\lb u_k + w_{k}^{*}\rb + \omega^2 (u_k + w_{k}^{*}) = 0 
\end{eqnarray}
with initial conditions $\lb u_k + w_{k}^{*}\rb (\tau_{in}) = 1$ and $\lb u_k + w_{k}^{*}\rb ' (\tau_{in}) = -ik$. This means that the combination $\lb u_k + w_{k}^{*}\rb$ essentially follows the Mukhanov-Sasaki equation and allows us to identify it with the Mukhanov-Sasaki mode function like so $\lb u_k + w_{k}^{*}\rb = \sqrt{2k} v_k$\footnote{The $\sqrt{2k}$ factor comes from the difference in initial conditions}. The covariance matrix elements can then be written in terms of the Bogoliubov coefficients and mode functions like so:
\begin{subequations}
\begin{align}
    \gamma_{11} &= |u_k + w_{k}^{*} |^2 = 2k|v_k|^2\\
    \gamma_{12} &= 2\text{Im}\lsb u_kw_k\rsb = 2\text{Re}[v_kp_{k}^{*}]\\
    \gamma_{22} &= |u_k - w_{k}^{*} |^2 = \dfrac{2}{k}|p_k|^2
\end{align}
\end{subequations}
which suggests the initial conditions $\gamma_{11}(\tau_{in}) = \gamma_{22}(\tau_{in}) = 1$ and $\gamma_{12}(\tau_{in}) = 0$. We can then rewrite all this in terms of squeezing parameters\footnote{Note that we have denoted the squeezing angle here as $\theta_k$ which is sometimes called $\varphi_k$. This is distinct from the rotation angle of the vacuum which we have neglected here as it is not of dynamical interest due to not appearing in the matrix elements $\gamma$.} $(r_k,\theta_k)$:
\begin{subequations}
\begin{align}
    u_k &= \cosh r_k \\
    w_k &= -e^{2i\theta_k}\sinh r_k
\end{align}
\end{subequations}
where $r_k$ measures the \emph{amount} of squeezing and $\theta_k$ measures the direction of it in phase space. We can therefore express the covariance matrix elements in terms of these squeezing parameters:
\begin{subequations}
\begin{align}
  \gamma_{11}  &= \cosh (2r_k) - \cos (2\theta_k)\sinh(2r_k)\\
  \gamma_{12}  &= - \sin (2\theta_k)\sinh(2r_k)\\
  \gamma_{22}  &= \cosh (2r_k) + \cos (2\theta_k)\sinh(2r_k)
\end{align}
\end{subequations}
\begin{figure}[t!]
    \centering
    \includegraphics[width = 0.75\linewidth]{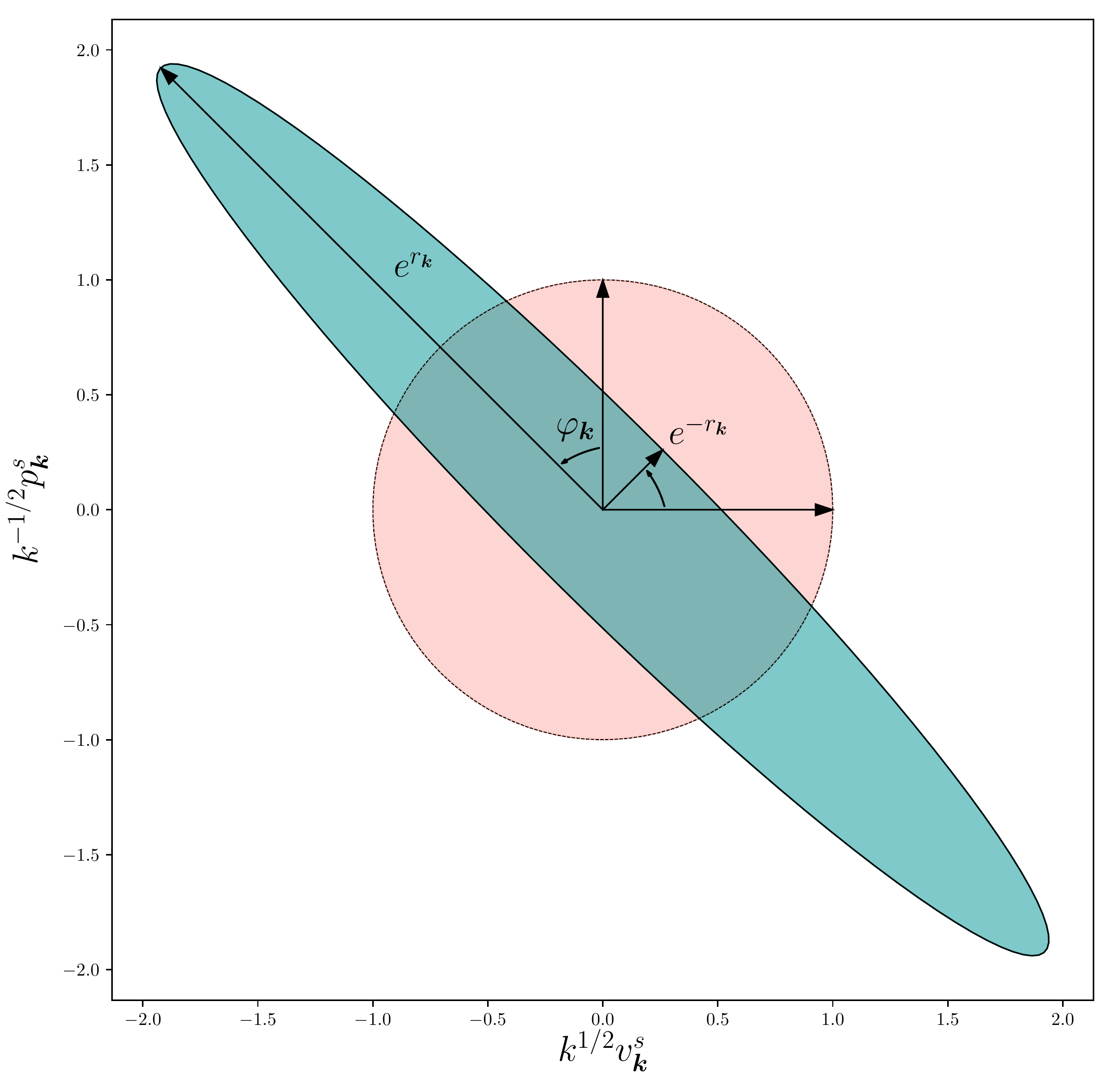}
    \caption[How squeezing affects the shape of the Wigner function]{Phase space representing the $\sqrt{2}\sigma$ contour level of the Wigner function for $\theta_k = \pi/4$, $r_k =1$ (green ellipse) compared to a pink circle corresponding to a vacuum state with no squeezing. Taken from \cite{Martin2021}.}
    \label{fig:Wigner_ellipse_martin}
\end{figure}
To make the geometrical nature of this squeezing more transparent we compute an object known as the Wigner function \cite{Simon1987, Simon1987a, Case2008} which for our Gaussian state simply reads:
\begin{eqnarray}
W = \dfrac{1}{\pi^2\lb\gamma_{11}\gamma_{22}-\gamma_{12}^2\rb}\exp \lsb -\dfrac{1}{\gamma_{11}\gamma_{22}-\gamma_{12}^2}\lb \gamma_{22}k v_{k}^2 + \dfrac{\gamma_{11}}{k}p_{k}^2 -2\gamma_{12}v_k p_k\rb \rsb 
\end{eqnarray}
For the initial, vacuum, state with no squeezing the Wigner function has no favoured direction and the $\sqrt{2}\sigma$ contour is simply given by a circle in phase space -- see Fig.~\ref{fig:Wigner_ellipse_martin}. However as the state is squeezed with time we can see that the circle is forced into an ellipse in a direction dictated by the angle $\theta_k$ and with length on the semi-major and semi-minor axes given by $e^{r_k}$ and $e^{-r_k}$ respectively. In this way we can see what squeezing does for us, it forces the uncertainty into one direction in phase space. To make this clearer we consider the case of perturbations in exact de Sitter. Then the covariance matrix elements are given by:
\begin{subequations}
\begin{align}
  \gamma_{11}  &= 1 + \dfrac{1}{k^2\tau^2}\\
  \gamma_{12}  &= - \dfrac{1}{k^3\tau^3}\\
  \gamma_{22}  &= 1 - \dfrac{1}{k^2\tau^2} + \dfrac{1}{k^4\tau^4}
\end{align}
\end{subequations}
From which it is straightforward to determine the dependence of the squeezing parameters in terms of conformal time $\tau$. To make the behaviour more transparent we describe the evolution in terms of the e-fold difference $\Delta N$ from when a mode exits the horizon at $\tau = -1/k$. In the left plot of Fig.~\ref{fig:squeezing_ellipse_dS} we plot the evolution of $r_k$ and $\theta_k$ as a function of $\Delta N$ in de Sitter. While $\theta_k$ asymptotes to 0 (aligning the Wigner function purely in the $p_k$ direction) the squeezing parameter $r_k$ grows without bound. In the right plot of Fig.~\ref{fig:squeezing_ellipse_dS} we show the effect this has on the $\sqrt{2}\sigma$ contours of the Wigner function. We see that even at horizon crossing $\Delta N = 0$ the perturbations have started to be squeezed and this squeezing only becomes more pronounced as $\Delta N$ is increased. We can see that at 5 e-folds over horizon exit the ellipse resembles a line in the $p_k$ direction. 

As discussed in the main body of the text we can see how the uncertainty in the perturbations are forced into one direction essentially reducing the dimensionality of the problem and allowing it to be described by a classical stochastic process. Notice however that the purity of the state is always equal to 1 regardless of the amount of squeezing meaning that there is \emph{no decoherence} taking place. The subdominant modes we neglect in our classical stochastic description are precisely what is needed to maintain that purity is one. In the squeezed limit $\gamma_{12}^2 \rightarrow \gamma_{11}\gamma_{22}$ we artificially take the purity = 0. This is because we have lost the quantum information in the system when we make this simplification. 
\section{\label{sec:app_ADM_pert}Perturbations in the ADM formalism}
Having described linear perturbations we move onto the ADM formalism which is better suited to accommodate nonlinear perturbations. Much of this is based on the seminal work of Salopek and Bond \cite{Salopek1990} although in a (hopefully) more pedagogical presentation.
\subsection{The ADM equations}
If we now look at fluctuations in the ADM formalism \cite{Arnowitt2008} where spacetime is sliced into three-dimensional hypersurfaces:
\begin{eqnarray}
\mathrm{d}s^2 = -N^2\mathrm{d}t^2 + \gamma_{ij}(\mathrm{d}x^i + N^i\mathrm{d}t)(\mathrm{d}x^j + N^j\mathrm{d}t) \label{eq:gen_ADM_metric}
\end{eqnarray}
where $\gamma_{ij}$ is the three-dimensional metric on slices of constant $t$. The lapse $N$ and the shift vector $N_i$ contain the same information\footnote{Unfortunately for students everywhere the lapse function and shift vector in the ADM formalism do not correspond one-to-one with $\Phi$ and $B_i$ in linear perturbation theory even though they share a name.} as the metric perturbations $\Phi$ and $B_i$ in (\ref{eq:perturb line element}). However they were chosen in such a way so as to be non-dynamical Lagrange multipliers in the action which becomes:
\begin{eqnarray}
\mathcal{S} &=& \int \mathrm{d}^4x~N\sqrt{\gamma} \Bigg\{ \dfrac{M_{p}^2}{2}\lb R_{(3)} + K_{ij}K^{ij} -K^2 \rb  + \dfrac{1}{2}\dfrac{1}{N^2}\lb \dot{\varphi}^2 -N^i \tilde{\nabla}_i \varphi\rb -\dfrac{1}{2} \tilde{\nabla}_i \varphi \tilde{\nabla}^i \varphi \Bigg\}\nonumber \\ \label{eq:ADM_action}
\end{eqnarray}
We have introduced a few new objects here. First is the intrinsic Ricci scalar curvature of constant time hypersurfaces $R_{(3)}$ associated with the 3-metric $\gamma_{ij}$. We have also denoted the metric of $\gamma_{ij}$ with $\gamma $ and defined three space covariant derivatives $\tilde{\nabla}_i$ with connection coefficients determined by $\gamma_{ij}$. The other object of interest is the extrinsic curvature three-tensor:
\begin{eqnarray}
K_{ij} = \dfrac{1}{2N}\lb \tilde{\nabla}_j N_i + \tilde{\nabla}_i N_j -\dfrac{\partial \gamma_{ij}}{\partial t}\rb \label{eq:ADM K defn}
\end{eqnarray}
The traceless part of a tensor will be denoted with an overbar\footnote{Not to be confused with the overbar denoting the homogenous solution in the previous section.} which for the extrinsic curvature looks like:
\begin{eqnarray}
\bar{K}_{ij} = K_{ij} -\dfrac{1}{3}K\gamma_{ij}, \quad K = K_{i}^{i}
\end{eqnarray}
where the trace $K$ is a generalisation of the Hubble parameter that appears in isotropic cosmologies. We will return to this in a moment. If we now vary the ADM action (\ref{eq:ADM_action}) with respect to the lapse $N$ and shift $N_i$ we obtain the energy and momentum constraint equations respectively:
\begin{eqnarray}
\bar{K}_{ij}\bar{K}^{ij} - \dfrac{2}{3}K^2 - R_{(3)} + \dfrac{2}{M_{p}^2}\mathcal{E} &=& 0  \label{eq:ADM_energy_constraint}\\
\tilde{\nabla}_j \bar{K}_{i}^{j} - \dfrac{2}{3} \tilde{\nabla}_i K + \dfrac{1}{M_p^2}\Pi \tilde{\nabla}_i \varphi &=& 0 \label{eq:ADM_momentum_constraint}
\end{eqnarray}
Where we have introduced the scalar field momentum:
\begin{eqnarray}
\Pi = \dfrac{1}{N}\lb \dot{\varphi} - N^i \tilde{\nabla}_i \varphi\rb^2 \label{eq:ADM_scalar_mom}
\end{eqnarray}
and the energy density on a constant time slice:
\begin{eqnarray}
\mathcal{E} = \dfrac{1}{2}\Pi^2 + \dfrac{1}{2} \tilde{\nabla}_i \varphi\tilde{\nabla}^i \varphi + V(\varphi) \label{eq:ADM_energydensity}
\end{eqnarray}
If we first note that the stress three-tensor is:
\begin{eqnarray}
S_{ij} = T_{ij} = \tilde{\nabla}_i \varphi\tilde{\nabla}^i \varphi + \gamma_{ij}\lsb\dfrac{1}{2}\Pi^2 - \dfrac{1}{2}\tilde{\nabla}_i \varphi\tilde{\nabla}^i \varphi - V(\varphi) \rsb 
\end{eqnarray}
Then variation with respect to $\gamma_{ij}$ yields the dynamical gravitational-field equations:
\begin{eqnarray}
\dfrac{\partial K}{\partial t} - N^i \tilde{\nabla}_i K &=& -\tilde{\nabla}_i\tilde{\nabla}^i N + \dfrac{3}{4}N\bar{K}_{ij}\bar{K}^{ij} + \dfrac{1}{2}NK^2 + \dfrac{1}{4}N R_{(3)} + \dfrac{1}{2}\dfrac{N}{M_{p}^2}S \nonumber \\\label{eq:ADM_K evoln}\\
\dfrac{\partial \bar{K}_{j}^{i}}{\partial t} + \bar{K}_{j}^{l}\tilde{\nabla}_l N^i -  N^l \tilde{\nabla}_l \bar{K}_{j}^{i} &=& -\tilde{\nabla}_i\tilde{\nabla}_j N + \dfrac{1}{3} \delta^{i}_{j}\tilde{\nabla}^l\tilde{\nabla}_l N + NK \bar{K}^{i}_{j} + N\bar{R}_{(3)j}^{i} -\dfrac{N}{M_p^2}\bar{S}^{i}_{j} \nonumber \\ \label{eq:ADM_Kbar evoln}
\end{eqnarray}
and finally variation with respect to $\varphi$ gives the scalar-field equations of motion:
\begin{eqnarray}
\dfrac{1}{N}\lb \dfrac{\partial \Pi}{\partial t} - N^i \tilde{\nabla}_i \Pi\rb - K\Pi - \dfrac{1}{N}\tilde{\nabla}_i N \tilde{\nabla}^i \varphi  - \tilde{\nabla}_i\tilde{\nabla}^i \varphi + \dfrac{\partial V(\varphi)}{\partial \varphi} \label{eq:ADM_phi evoln}
\end{eqnarray}
\subsection{Spatial gradient expansion of the ADM formalism}
What is nice about the terms in the previous subsection is that they are fully non-linear, the bad news is that they are basically impossible to solve. In order to have non-linear perturbations we can actually compute we must sacrifice something else, in particular we resolve to only examine long-wavelengths such that spatial gradients that are second order or higher can be neglected. Concretely we split the field into a smoothed long-wavelength, background or coarse-grained field $\varphi_{\scalebox{0.5}{$>$}}$ and a residual short wavelength field $\varphi_{\scalebox{0.5}{$<$}}$:
\begin{eqnarray}
\varphi (t,\vec{x} ) &=& \varphi_{\scalebox{0.5}{$>$}}(t,\vec{x} ) + \varphi_{\scalebox{0.5}{$<$}}(t,\vec{x} ) \\
\varphi_{\scalebox{0.5}{$>$}}(t,\vec{x} ) &\equiv &\int \mathrm{d}^3 x ~\mathcal{W}(t,\vec{x}-\vec{x}') \varphi (t,\vec{x}' )
\end{eqnarray}
where $\mathcal{W}$ is a window or smoothing function in the spatial coordinates whose Fourier transform falls off at high momentum. It is worth noting that this smoothing is gauge dependent and one must therefore be careful about relating quantities not computed in the same gauge this coarse-graining is performed in. For stochastic inflation the natural smoothing scale is (a multiple of) the comoving Hubble length $(aH)^{-1}$ and the natural hypersurfaces are those where $aH$ is constant. \\
If we now convolve the ADM equations with the smoothing function we can obtain equations in terms of the coarse-grained field $\varphi_{\scalebox{0.5}{$>$}}$. As we are working with quantities on long-wavelengths we shall only keep terms which are at most first order in spatial gradients. Finally we choose a gauge where the shift $N_i = 0$. This simplifies the equations massively and the evolution of $\bar{K}^{i}_{j}$ (\ref{eq:ADM_Kbar evoln}) is massively simplified:
\begin{eqnarray}
\dfrac{\partial\bar{K}^{i}_{j}}{\partial t} = N K \bar{K}_{j}^{i} 
\end{eqnarray}
Noting from the definition (\ref{eq:ADM K defn}) that $K$ can be expressed as:
\begin{eqnarray}
K = -\dfrac{\partial \ln \lb \sqrt{\gamma }\rb}{\partial t}
\end{eqnarray}
then we have a solution for $\bar{K}^{i}_{j}$:
\begin{eqnarray}
\bar{K}^{i}_{j} \propto \dfrac{1}{\sqrt{\gamma} } \equiv \dfrac{1}{a^3}\label{eq:ADM Kbar decay}
\end{eqnarray}
where we have noted how the scale factor is related to the determinant of the metric. Equation (\ref{eq:ADM Kbar decay}) tells us therefore that any non-zero anisotropic expansion rate, $\bar{K}^{i}_{j}$, will decay to zero extremely quickly during a period of accelerated expansion. The most general form of the metric is therefore:
\begin{eqnarray}
\mathrm{d}s^2 =  -N^2 \mathrm{d}t^2 + \exp \lsb 2\alpha (t,\vec{x}) \rsb\delta_{ij}\mathrm{d}x^i\mathrm{d}x^j \label{eq:gen_ADM_metric_simpler}
\end{eqnarray}
where we have introduced the spatially dependent e-fold $\alpha$ in terms of the spatially dependent expansion parameter $\alpha (t,\vec{x})\equiv \ln a (t,\vec{x}) $. This tells us that the trace of the extrinsic curvature can be simply expressed in terms of $\alpha$:
\begin{eqnarray}
K =  -3\dfrac{1}{N}\dfrac{\partial \alpha}{\partial t}
\end{eqnarray}
Which naturally allows us to define the spatially dependent Hubble parameter:
\begin{eqnarray}
H_{\scalebox{0.5}{$>$}} (t,\vec{x} ) \equiv \dfrac{1}{N}\dfrac{\partial \alpha}{\partial t} = -\dfrac{K}{3} \label{eq:ADM H defn}
\end{eqnarray}
With this in mind the energy constraint (\ref{eq:ADM_energy_constraint}) becomes:
\begin{eqnarray}
H_{\scalebox{0.5}{$>$}}^2 (t,\vec{x} ) = \dfrac{1}{3M_{p}^2}\lb \dfrac{1}{2} \Pi_{>} ^2 (t,\vec{x} ) + V(\varphi_{\scalebox{0.5}{$>$}} (t,\vec{x})) \rb \label{eq:ADM_energy constraint_simpler}
\end{eqnarray}
and the momentum constraint (\ref{eq:ADM_momentum_constraint}) becomes:
\begin{eqnarray}
2\tilde{\nabla}_i H_{\scalebox{0.5}{$>$}}(t,\vec{x} ) = -\dfrac{1}{M_{p}^2}\Pi_{>} (t,\vec{x} ) \tilde{\nabla}_i\varphi_{\scalebox{0.5}{$>$}} (t,\vec{x} )  \label{eq:ADM_momentum constraint_simpler}
\end{eqnarray}
In general $H_{\scalebox{0.5}{$>$}}$ is a function of the scalar field values and of time:
\begin{eqnarray}
H_{\scalebox{0.5}{$>$}}(t,\vec{x} ) \equiv H_{\scalebox{0.5}{$>$}} (\varphi_{\scalebox{0.5}{$>$}}(t,\vec{x} ), t ) \label{eq:ADM H as func phi}
\end{eqnarray}
which if we combine with (\ref{eq:ADM_momentum constraint_simpler}) allows us to relate the field momentum with the Hubble expansion rate:
\begin{eqnarray}
\Pi_{>} = -2M_{p}^2 \lb \dfrac{\partial H_{\scalebox{0.5}{$>$}}}{\partial \varphi_{\scalebox{0.5}{$>$}}}\rb_{t} \label{eq:ADM Pi = partial H}
\end{eqnarray}
The subscript on the bracket of the partial derivative indicates what is being held constant -- in this case time. We can now verify that there is no explicit dependence on time for $H_{\scalebox{0.5}{$>$}}$. We see from the original ADM equation (\ref{eq:ADM_K evoln}):
\begin{eqnarray}
\lb \dfrac{\partial H_{\scalebox{0.5}{$>$}}}{\partial x}\rb_{\vec{x}} = -\dfrac{1}{2}\Pi_{>}^2
\end{eqnarray}
and if we compare this with taking a time derivative of (\ref{eq:ADM H as func phi}):
\begin{eqnarray}
\lb \dfrac{\partial H_{\scalebox{0.5}{$>$}}}{\partial x}\rb_{\vec{x}} &=&  \dfrac{1}{N}\lb \dfrac{\partial \varphi_{\scalebox{0.5}{$>$}}}{\partial t}\rb_{\vec{x}} \lb \dfrac{\partial H_{\scalebox{0.5}{$>$}}}{\partial \varphi_{\scalebox{0.5}{$>$}}}\rb_{t}   + \dfrac{1}{N} \lb \dfrac{\partial t}{\partial t}\rb_{\vec{x}}\lb \dfrac{\partial H_{\scalebox{0.5}{$>$}}}{\partial t}\rb_{\varphi_{\scalebox{0.5}{$>$}}}\\
&=& -\underbrace{2M_{p}^2 \lb \dfrac{\partial H_{\scalebox{0.5}{$>$}}}{\partial \varphi_{\scalebox{0.5}{$>$}}}\rb_{t}^2}_{\Pi_{>}^2 /2} + \dfrac{1}{N} \lb \dfrac{\partial H_{\scalebox{0.5}{$>$}}}{\partial t}\rb_{\varphi_{\scalebox{0.5}{$>$}}}
\end{eqnarray}
then clearly
\begin{eqnarray}
\lb \dfrac{\partial H_{\scalebox{0.5}{$>$}}}{\partial t}\rb_{\varphi_{\scalebox{0.5}{$>$}}} = 0 \Rightarrow H_{\scalebox{0.5}{$>$}}(t,\vec{x}) = H_{\scalebox{0.5}{$>$}}(\varphi_{\scalebox{0.5}{$>$}} (t,\vec{x}) )
\end{eqnarray}
i.e. $H_{\scalebox{0.5}{$>$}}$ only depends on time and space through its dependence on the field $\varphi_{\scalebox{0.5}{$>$}}$. A similar argument can be made for the field momentum $\Pi_{>}$. The equation of motion for the scalar (\ref{eq:ADM_phi evoln}) becomes:
\begin{eqnarray}
\dfrac{1}{N}\dfrac{\partial\Pi_{\scalebox{0.5}{$>$}}}{\partial t} + 3H_{\scalebox{0.5}{$>$}}\Pi_{>} + \dfrac{\partial V}{\partial \varphi_{\scalebox{0.5}{$>$}}} = 0 \label{eq:ADM_phi evoln_simpler}
\end{eqnarray}
If we combine the energy (\ref{eq:ADM_energy constraint_simpler}) and momentum (\ref{eq:ADM Pi = partial H}) constraints we obtain the \ac{H-J} equation for $H_{\scalebox{0.5}{$>$}}(\varphi_{\scalebox{0.5}{$>$}} )$:
\begin{eqnarray}
\lb \dfrac{\partial H_{\scalebox{0.5}{$>$}}}{\partial \varphi_{\scalebox{0.5}{$>$}}}\rb^2 = \dfrac{3}{2}H_{\scalebox{0.5}{$>$}}^2 (\varphi_{\scalebox{0.5}{$>$}}) -\dfrac{1}{2}V(\varphi_{\scalebox{0.5}{$>$}}) \label{eq:ADM_H-J equation}
\end{eqnarray}
which as a first order differential equation admits for any given potential $V(\varphi_{\scalebox{0.5}{$>$}})$ a family of solutions $H_{\scalebox{0.5}{$>$}} = H_{\scalebox{0.5}{$>$}} (\varphi_{\scalebox{0.5}{$>$}} , \mathcal{C})$ where different solutions are parameterised by an arbitrary constant $\mathcal{C}$. This is equivalent to specifying the momentum of the field $\Pi_{\scalebox{0.5}{$>$}} $ at some value of $\varphi_{\scalebox{0.5}{$>$}}$ i.e. fixing $H_{\scalebox{0.5}{$>$}}$ to a point ($\varphi_{\scalebox{0.5}{$>$}}^{\mathcal{C}}, \Pi_{\scalebox{0.5}{$>$}}^{\mathcal{C}} )$ in the $\varphi_{\scalebox{0.5}{$>$}} ~$--$ ~\Pi_{\scalebox{0.5}{$>$}}$ phase space. Just looking at (\ref{eq:ADM_H-J equation}) one could conclude that $\mathcal{C} = \mathcal{C}(\vec{x})$ i.e. for every point $\vec{x}$ on the specified hypersurface ($\varphi_{\scalebox{0.5}{$>$}}^{\mathcal{C}}, \Pi_{\scalebox{0.5}{$>$}}^{\mathcal{C}} )$ there would be a separate integration constant $\mathcal{C}(\vec{x})$. This would mean that every point in space would therefore encode a different solution, $H_{\scalebox{0.5}{$>$}}(\varphi_{\scalebox{0.5}{$>$}}, \mathcal{C})$, for the H-J (\ref{eq:ADM_H-J equation}). If $\mathcal{C} = \mathcal{C}(\vec{x})$ then we can obtain:
\begin{eqnarray}
\tilde{\nabla} H_{\scalebox{0.5}{$>$}} (\varphi_{\scalebox{0.5}{$>$}}, \mathcal{C}) = (\partial_{\mathcal{C}}H_{\scalebox{0.5}{$>$}})\tilde{\nabla} \mathcal{C} + (\partial_{\varphi}H)\tilde{\nabla} \varphi_{\scalebox{0.5}{$>$}} \neq -\dfrac{1}{2}\Pi_{\scalebox{0.5}{$>$}} \tilde{\nabla} \varphi_{\scalebox{0.5}{$>$}}
\end{eqnarray}
which as indicated by the inequality does not match the required momentum constraint (\ref{eq:ADM Pi = partial H}). For this to be satisfied either $(\partial_{\mathcal{C}}H_{\scalebox{0.5}{$>$}}) = 0$ or $\tilde{\nabla} \mathcal{C} = 0$. If $(\partial_{\mathcal{C}}H_{\scalebox{0.5}{$>$}}) = 0$ then we must be in exact de Sitter with no field momentum i.e. $V = V_0$, $\Pi_{\scalebox{0.5}{$>$}} = 0 \Rightarrow H_{\scalebox{0.5}{$>$}}^2 = V_0/3 M_{p}^2$. If however this is not the case then the value of $H_{\scalebox{0.5}{$>$}}$ depends on $\mathcal{C}$ and we are restricted to $\tilde{\nabla} \mathcal{C} = 0$ i.e. $\mathcal{C} \neq \mathcal{C}(\vec{x})$. Phrased another way, $\mathcal{C}$ must be a \textit{global} constant on the initial hypersurface ($\varphi_{\scalebox{0.5}{$>$}}^{\mathcal{C}}, \Pi_{\scalebox{0.5}{$>$}}^{\mathcal{C}} )$ with no spatial dependence. Altogether this means that $H_{\scalebox{0.5}{$>$}}$ and $\Pi_{\scalebox{0.5}{$>$}}$ can only get their inhomogeneities through $\varphi$ rather than any explicit dependence on $\vec{x}$. \\
One of these solutions to (\ref{eq:ADM_H-J equation}) together with:
\begin{eqnarray}
\dfrac{1}{N}\dfrac{\mathrm{d}\varphi_{\scalebox{0.5}{$>$}}}{\mathrm{d}t} = -2\dfrac{\mathrm{d}H_{\scalebox{0.5}{$>$}}}{\mathrm{d}\varphi_{\scalebox{0.5}{$>$}}}
\end{eqnarray}
provides a complete description of the inhomogeneous, long-wavelength scalar field configuration $\varphi_{\scalebox{0.5}{$>$}}$. The metric is then recovered from:
\begin{eqnarray}
\dfrac{1}{N}\dfrac{\partial \alpha}{\partial t} = H_{\scalebox{0.5}{$>$}}
\end{eqnarray}

\subsection{The lack of a spatially varying constant of integration}
Suppose we induced\footnote{We ignore whether or not it is actually possible to do this by quantum backreaction in this section.} some perturbation away from the \ac{H-J} trajectory such that at a particular value of $\varphi_{\scalebox{0.5}{$>$}}$, $H_{\scalebox{0.5}{$>$}}$ is different i.e. $\mathcal{C}$ is different.\\
We start by taking a derivative of the \ac{H-J} equation (\ref{eq: H-J equation}) with respect to $\mathcal{C}$:
\begin{eqnarray}
2\dfrac{\partial H_{\scalebox{0.5}{$>$}}}{\partial \varphi_{\scalebox{0.5}{$>$}}}\dfrac{\partial}{\partial \varphi_{\scalebox{0.5}{$>$}}}\lb \dfrac{\partial H_{\scalebox{0.5}{$>$}}}{\partial \mathcal{C}}\rb &=& 3H \dfrac{\partial H}{\partial \mathcal{C}}\\
\Rightarrow 2\dfrac{\partial H_{\scalebox{0.5}{$>$}}}{\partial \varphi_{\scalebox{0.5}{$>$}}}\dfrac{1}{\partial_{\mathcal{C}}H_{\scalebox{0.5}{$>$}}}\dfrac{\partial}{\partial \varphi_{\scalebox{0.5}{$>$}}}\lb \dfrac{\partial H_{\scalebox{0.5}{$>$}}}{\partial \mathcal{C}}\rb &=& 3H_{\scalebox{0.5}{$>$}} \\
\Rightarrow 2\dfrac{\partial H_{\scalebox{0.5}{$>$}}}{\partial \varphi_{\scalebox{0.5}{$>$}}} \partial_{\varphi_{\scalebox{0.5}{$>$}}}\ln \left| \dfrac{\partial H_{\scalebox{0.5}{$>$}}}{\partial \mathcal{C}}\right| &=& 3H_{\scalebox{0.5}{$>$}} \\
\underbrace{\Rightarrow}_{(\ref{eq:ADM Pi = partial H})} - \Pi_{\scalebox{0.5}{$>$}} \partial_{\varphi}\ln \left| \dfrac{\partial H_{\scalebox{0.5}{$>$}}}{\partial \mathcal{C}}\right| &=& 3H_{\scalebox{0.5}{$>$}} \\
\underbrace{\Rightarrow}_{(\ref{eq:ADM_scalar_mom})} -\dfrac{1}{N}\dfrac{\partial \varphi_{\scalebox{0.5}{$>$}}}{\partial t} \partial_{\varphi_{\scalebox{0.5}{$>$}}}\ln \left| \dfrac{\partial H_{\scalebox{0.5}{$>$}}}{\partial \mathcal{C}}\right| &=& 3H_{\scalebox{0.5}{$>$}} \\
\underbrace{\Rightarrow}_{(\ref{eq:ADM H defn})} \dfrac{1}{N} \partial_{t}\ln \left| \dfrac{\partial H_{\scalebox{0.5}{$>$}}}{\partial \mathcal{C}}\right| &=& -\dfrac{3}{N}\dfrac{\partial\alpha}{\partial t} \\
\Rightarrow \ln \left| \dfrac{\partial H_{\scalebox{0.5}{$>$}}}{\partial \mathcal{C}}\right| &\propto &-3\alpha \\
\Rightarrow \dfrac{\partial H_{\scalebox{0.5}{$>$}}}{\partial \mathcal{C}} &\propto & e^{-3\alpha} = a^{-3}
\end{eqnarray}
Indicating that $\partial_{\mathcal{C}}H_{\scalebox{0.5}{$>$}}$ decays rapidly during inflation:
\begin{eqnarray}
\Delta H_{\scalebox{0.5}{$>$}} \approx \partial_{\mathcal{C}}H_{\scalebox{0.5}{$>$}} \Delta \mathcal{C} \propto e^{-3\alpha} = a^{-3}
\end{eqnarray}
This means that any variation away from the \ac{H-J} trajectory corresponds to the transient, decaying mode that is ever-present in inflationary perturbation theory. It is precisely the decaying of this mode that allows us to treat the truly quantum nature of the backreaction on long-wavelengths as an effective stochastic theory. If one wishes to explore a regime where this mode should be retained then there is no obvious way how it could be included in a stochastic description of inflation. 
\chapter{\label{APP:FPT}Computation of First-Passage Time Formulae}
\acresetall 
In this appendix we will explicitly show some computations related to the \ac{FPT} problems considered in part \ref{part:Cosmo}. In particular we will be exploiting the fact that the probability distribution $\rho (\mathcal{N})$ for number of e-folds to reach a field value $\phi_e$ can be obtained from the knowledge of the probability $P(\phi,\alpha)$ that enters the corresponding \ac{F-P} equation. These are related through \cite{VanKampen2007}:
\begin{eqnarray}
\int_{\mathcal{N}}^{\infty} \rho (\alpha) \mathrm{d}\alpha &=& \int_{\phi_e}^{\infty}P(\phi,\mathcal{N})\mathrm{d}\phi  \label{eq:FPT_FP_int_equivalence} \\
\Rightarrow \rho (\mathcal{N}) &=& -\dfrac{\partial}{\partial \mathcal{N}} \int_{\phi_e}^{\infty}P(\Phi,\mathcal{N})\mathrm{d}\phi \label{eq:FPT_FP_pdf_equivalence}
\end{eqnarray}
To see why this is the case consider that the LHS of (\ref{eq:FPT_FP_int_equivalence}) is simply the probability that it takes longer than $\mathcal{N}$ e-folds for the field to reach $\phi_e$. If this field is the inflaton then this is simply the probability that inflation last longer than $\mathcal{N}$, $P($ inflation $ > \mathcal{N})$. The RHS is the area under the \ac{F-P} PDF above the exit point $\phi_e$ at the time $\mathcal{N}$. Provided there is an absorbing boundary condition at $\phi_e$ then this area is the fraction of trajectories that have not yet reached $\phi_e$ at time $\mathcal{N}$. This means that all of these trajectories will take longer than $\mathcal{N}$ to reach $\phi_e$ so this area does indeed equal $P($ inflation $ > \mathcal{N})$ so the LHS = RHS. If there is not an absorbing boundary condition then the RHS will be larger than the LHS. This is because the RHS will now include contributions from trajectories that have reached $\phi_e$ previously but are now at $\phi > \phi_e$ meaning we are no longer computing a true \emph{first}-passage time quantity. Therefore if one does not include an absorbing boundary condition, a computation of the RHS would \emph{overestimate} the number of trajectories that have yet to reach $\phi_e$ and the prediction for $\rho (\mathcal{N})$ would have a fatter tail than the true value.

\section{\label{sec:H-J_phase deriv}Hamilton-Jacobi phase}
\begin{figure}[t!]
    \centering
    \includegraphics[width = 0.75\linewidth]{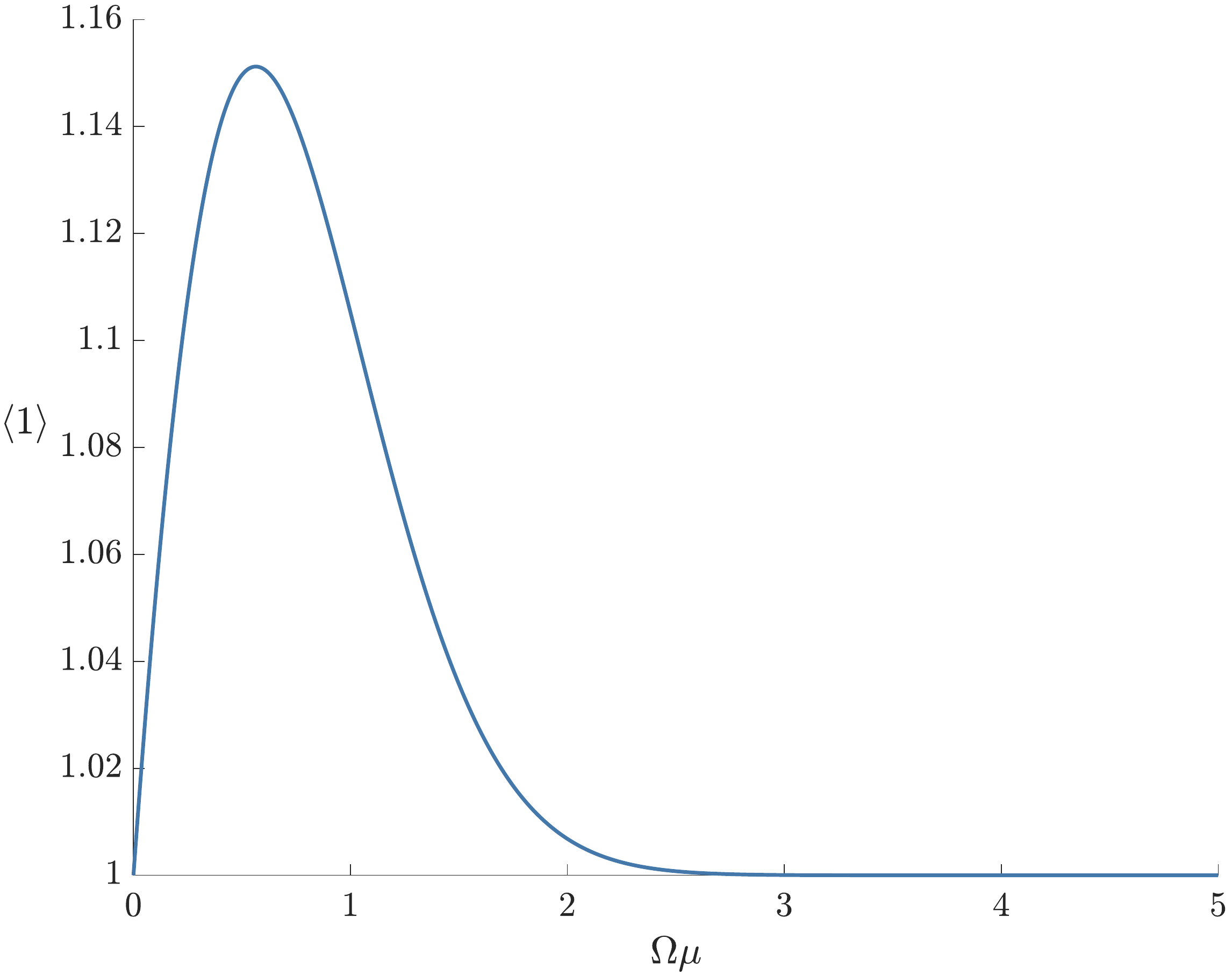}
    \caption[Normalisation condition for Hamilton-Jacobi trajectory]{The dependence of $\lan 1\ran$ on $\Omega \mu$}
    \label{fig:chisig_norm}
\end{figure}

This section is largely based on the computations in \cite{Prokopec2021} although with some small mistakes corrected.\\
The \ac{H-J} trajectory obeys the following PDF:
\begin{eqnarray}
\dfrac{\partial P_{HJ}}{\partial \alpha} \simeq 3 \dfrac{\partial}{\partial \chi} (\chi P_{HJ})+ \dfrac{3}{2}\dfrac{\partial^2 P_{HJ}}{\partial \chi^2}
\end{eqnarray}
where 
\begin{eqnarray}
\chi \equiv \sqrt{\dfrac{3}{2}}\dfrac{\phi - \phi_0}{\tilde{H}_0}
\end{eqnarray}
If we perform the transformation 
\begin{eqnarray}
P_{HJ} (\chi,\alpha) = C\exp \lb \dfrac{3}{2}\alpha - \dfrac{1}{2}\chi^2 \rb \Psi (\chi,\alpha) \label{eq:PHJ_transform}
\end{eqnarray}
where $C$ is some constant and $\Psi (\chi,\alpha)$ obeys:
\begin{eqnarray}
-\dfrac{1}{3}\dfrac{\partial\Psi}{\partial\alpha} = \dfrac{1}{2}\lb -\dfrac{\partial^2}{\partial \chi^2} + \chi^2\rb \Psi
\end{eqnarray}
which reduces the problem to the quantum mechanical kernel for the simple harmonic oscillator. The free propagator, which for stochastic processes is known as the \textit{Mehler heat kernel} \cite{Pauli2000}, is given by:
\begin{eqnarray}
K(\chi,\alpha;\chi_{in},\alpha_{in}) &=& \dfrac{1}{\sqrt{2\pi A(\Delta \alpha)}}\exp \lb -\dfrac{1}{2}B(\Delta \alpha)\lb\chi^2 + \chi_{in}^2\rb + \dfrac{\chi\chi_{in}}{A(\Delta \alpha)}\rb \\
A(\Delta \alpha) &\equiv &\sinh \lb 3\Delta \alpha\rb \\
B(\Delta \alpha) &\equiv &\coth \lb 3\Delta \alpha\rb \\
\Delta \alpha &\equiv & \alpha - \alpha_{in}
\end{eqnarray}
However because we have an absorbing boundary at $\phi_e$, $\Psi$ is not given by the free kernel. To accommodate the absorbing boundary we use the method of images to add another free kernel mirrored so as to cancel at the boundary $\chi_e$:
\begin{eqnarray}
\Psi (\chi, \alpha) &=& K(\chi,\alpha;\chi_{in},\alpha_{in}) - K(2\chi_e -\chi,\alpha;\chi_{in},\alpha_{in}) \\
&=& \dfrac{1}{\sqrt{2\pi A(\Delta \alpha)}} \Bigg\lbrace \exp \lb -\dfrac{1}{2}B(\Delta \alpha)\lb\chi^2 + \chi_{in}^2\rb + \dfrac{\chi\chi_{in}}{A(\Delta \alpha)}\rb \nonumber \\
&&- \exp \lb -\dfrac{1}{2}B(\Delta \alpha)\lsb (2\chi_e -\chi)^2 + \chi_{in}^2\rsb + \dfrac{(2\chi_e -\chi)\chi_{in}}{A(\Delta \alpha)}\rb \Bigg\rbrace
\end{eqnarray}
We can then use (\ref{eq:PHJ_transform}) to obtain $P_{HJ}$:
\begin{eqnarray}
P_{HJ} (\chi,\alpha) &=& \sqrt{\dfrac{3}{2v_0}}\exp \lb \dfrac{3}{2}\Delta \alpha - \dfrac{1}{2} (\chi^2 - \chi_{in}^2) \rb \Psi (\chi, \alpha )\\
&=&\dfrac{\sqrt{3(n+1)}}{\sqrt{2\pi v_0}}e^{-n\chi_{in}^{2}}\Bigg\lbrace \exp \lsb -(n+1)\chi^2 + 2(n+1)\chi \chi_{in}e^{-3\Delta \alpha}\rsb \nonumber \\
&&-\exp \lsb -2(2n+1)\chi_{e}^{2} +4(n+1)\chi_e\chi_{in}e^{-3\Delta \alpha}\rsb \nonumber \\
&& \times \exp \lsb  -(n+1)\chi^2 + \chi\lb 2(2n+1)\chi_e -2(n+1)\chi_{in}e^{-3\Delta \alpha}\rb\rsb \Bigg\rbrace \label{eq:P_HJ}
\end{eqnarray}
where we were able to determine the constant $C = \sqrt{3/2v_0}$ by use of the initial condition:
\begin{eqnarray}
P_{HJ}(\chi, \alpha \rightarrow \alpha_{in}) = C\delta (\chi - \chi_{in}) = C\sqrt{\dfrac{2v_0}{3}}\delta (\phi - \phi_{in})
\end{eqnarray}
We will look to generalise the initial condition to a normal distribution in section \ref{sec:SR_HJ_deriv}. We have also used the definitions of the hyperbolic trig functions so that we can rewrite $A$ and $B$ in terms of a new parameter $n$:
\begin{eqnarray}
n(\Delta \alpha) &\equiv &\dfrac{1}{e^{6\Delta {\alpha}}-1} \\
\dfrac{1}{A(\Delta \alpha)} &=& 2\lsb n(\Delta \alpha) + 1\rsb e^{-3\Delta {\alpha}}\\
B(\Delta \alpha) &=& 2n(\Delta \alpha) + 1 
\end{eqnarray}
Equation (\ref{eq:P_HJ}) looks horribly complex and too difficult to integrate according to (\ref{eq:FPT_FP_pdf_equivalence}), fortunately however if you peer at it long enough you realise that it is actually just the sum of Gaussian integrals so that:
\begin{eqnarray}
\rho_{HJ} (\mathcal{N}) &=& -\dfrac{\partial}{\partial \mathcal{N} }\lcb \dfrac{1}{2}\text{erfc} \lsb\sqrt{n + 1}\bar{U} \rsb  - e^Y\dfrac{1}{2}\text{erfc} \lsb \sqrt{n} \bar{V}\rsb \rcb \\
Y &\equiv & -\dfrac{2n+1}{n+1}\chi_{e}^{2} -2n\chi_e \chi_{in}e^{-3\Delta \mathcal{N}} \\
\bar{U} &\equiv & \chi_e - \chi_{in}e^{-3\Delta \mathcal{N}}\\
\bar{V} &\equiv & \chi_{in} - \chi_e e^{-3\Delta \mathcal{N}}
\end{eqnarray}
where $n = n(\Delta \mathcal{N})$. Evaluating the derivative we obtain:
\begin{eqnarray}
\rho_{HJ} (\mathcal{N}) &=& \dfrac{3}{\sqrt{\pi}}\exp \lsb -(n+1)\bar{U}^2\rsb \lsb n\sqrt{n+1}\chi_e -\sqrt{n}(n+1)\chi_{in} \rsb \nonumber \\
&& -\dfrac{3}{\sqrt{\pi}}\exp \lsb -n\bar{V}^2\rsb \lsb \sqrt{n}(n+1)\chi_{in}-n\sqrt{n+1}\lb 2-e^{-6\Delta \mathcal{N}}\rb\chi_e \rsb e^Y \nonumber \\
&& +3\chi_e\lsb \chi_e e^{-6\Delta\mathcal{N}}-n(2n+3)\chi_{in}e^{-3\Delta\mathcal{N}}\rsb e^Y \text{erfc}\lsb \sqrt{n}\bar{V}\rsb \label{eq:rhoN HJ_append}
\end{eqnarray}
If we now identify $\chi_{in} \rightarrow \Omega$ \& $\chi_e \rightarrow \Omega\mu$ according to the definitions (\ref{eq: chi_in defn}) \& (\ref{eq:sigma defn}) respectively we recover the PDF quoted in the main body of the text (\ref{eq:rhoN HJ}). \\
It is worth commenting at this stage that (\ref{eq:rhoN HJ_append}) is not actually normalised:
\begin{eqnarray}
\lan 1\ran = \dfrac{1}{2}\lsb 1 + \text{erf} (\Omega \mu) + e^{-\Omega^2\mu^2}\rsb \label{eq:chisigma_norm}
\end{eqnarray}
We have plotted this in Fig.~\ref{fig:chisig_norm} where we can see for $\Omega\mu > 3$ and $\Omega\mu = 0$ the deviation from 1 is negligible. Even around the peak, located at $\Omega\mu = 1/\sqrt{\pi}$, the deviation is very small. For this reason we will neglect (\ref{eq:chisigma_norm}) from equations in the main body of the text, however the graphs in chapter \ref{cha:PBH} do include the small contribution from this factor. 

\section{\label{sec:SR_HJ_deriv}Slow-Roll prior to Hamilton-Jacobi}
If we wish to add more general initial condition to the derivation in Appendix \ref{sec:H-J_phase deriv} we could modify the constant $C$ to account for this. However for our purposes, where we are focusing on the tail of the distribution, we can more simply compute the probability density for exit time given a \ac{SR} phase followed by an \ac{USR} phase by the convolution of the two PDFs:
\begin{eqnarray}
\rho_{SR+USR}(\mathcal{N}_2) = \int_{-\infty}^{\infty} d \mathcal{N}_1~\rho_{SR}(\mathcal{N}_1)\rho_{USR}(\mathcal{N}_2 -\mathcal{N}_1)
\end{eqnarray}
where the individual PDFs are given by:
\begin{eqnarray}
\rho_{SR}(\mathcal{N}_1) &=&  \dfrac{1}{\sigma_{SR} \sqrt{2\pi}}\text{exp}\left[ -\dfrac{1}{2}\dfrac{(\mathcal{N}_1 - \left\langle  \mathcal{N}_1\right\rangle)^2}{\sigma_{SR}^{2}}\right] \\
\rho_{USR}(\mathcal{N}_2 -\mathcal{N}_1) &=& \dfrac{6}{\sqrt{\pi}}\Omega ~e^{-\Omega^{2}\mu^2}e^{-3\Delta \mathcal{N}_{12}}
\end{eqnarray}
where $\Delta \mathcal{N}_{12} \equiv \mathcal{N}_2 - \mathcal{N}_1$ and $\sigma_{SR} = \sqrt{v_0/\epsilon_{in}} \approx 1/\sqrt{2}\Omega$. Performing the integration over $\mathcal{N}_1$ and realising that the lower bound of integration is restricted to $\left\langle \mathcal{N}_1 \right\rangle$ as $\rho_{USR}$ has zero weight below this:
\begin{eqnarray}
\rho_{tot}(\mathcal{N}_2) &=& \dfrac{3}{\sqrt{\pi}}\Omega ~e^{-\Omega^{2}\mu^2} e^{-3(\mathcal{N}_2 - \left\langle \mathcal{N}_2\right\rangle )}e^{-3(\left\langle \mathcal{N}_2\right\rangle - \left\langle \mathcal{N}_1\right\rangle )} e^{\frac{9}{4}\Omega^{-2}} \nonumber \\
&& \times ~\text{erfc}\left[ -\dfrac{3}{\sqrt{2}\Omega} -\Omega(\sqrt{2} -1)\left\langle \mathcal{N}_1\right\rangle \right] \\
&\approx &  \dfrac{6}{\sqrt{\pi}}\Omega ~e^{-\Omega^{2}\mu^2} e^{-3(\zeta_c + \left\langle \mathcal{N}_2\right\rangle - \left\langle \mathcal{N}_1\right\rangle )} e^{\frac{9}{4}\Omega^{-2}}
\end{eqnarray}
where in the second line we have approximated the complementary error function as 2 which is valid as the argument is generically large and negative. We can then use (\ref{eq:massfracdef}) to obtain the convolved mass fraction of \ac{PBHs}:
\begin{eqnarray}
\beta_{SR+USR}(M) &=& \dfrac{4}{\sqrt{\pi}}\Omega ~e^{-\Omega^{2}\mu^2}~e^{-3(\zeta_c +  \left\langle  N_2\right\rangle -\left\langle  N_1\right\rangle)} \times e^{\frac{9}{4}\Omega^{-2}} \\
&=& \beta_{USR} \times e^{\frac{9}{4}\Omega^{-2}} \label{eq:betaUSR+SR}
\end{eqnarray}
So we can see that a previous \ac{SR} phase enhances the \ac{USR} calculation by a factor of $e^{\frac{9}{4}\Omega^{-2}}$. This factor is clearly negligible for $\Omega \gg 1$ and is only relevant for $\Omega < 1$ where it significantly enhances the mass fraction $\beta$.

\section{\label{sec:dS+HJ_deriv}de Sitter phase preceded by Hamilton-Jacobi}
To deal with a period of free diffusion phase we recall that the injected current $J(\alpha)$ into the free diffusion branch from the \ac{H-J} phase is given by \cite{Prokopec2021}:
\begin{eqnarray}
J(\alpha) = \dfrac{6\pi\chi}{\left[ 2\pi ~\text{sinh}(3\Delta \alpha)\right]^{3/2}}~\text{exp}\left[\dfrac{3}{2}\Delta\alpha -\dfrac{1}{2}\left( \text{coth}(3\Delta \alpha ) -1\right)\chi^2\right] \label{eq:J current}
\end{eqnarray}
This will allow us to compute the probability distribution for $\phi$ on the free diffusion branch, $P_{v_0}$, through:
\begin{eqnarray}
P_{v_0}(\phi,\alpha) = \int_{\alpha_{in}}^{\alpha}\mathrm{d}u ~G(\phi-\phi_0, \alpha - u )J(u) \label{eq:Pv0}
\end{eqnarray}
where $G$ is the solution to the diffusive Green's function with exit boundary condition at $\phi_{e}$ and reflecting boundary condition at $\phi_{in}$:
\begin{subequations}
\begin{align}
\partial_{\alpha}G &= \tilde{H}_{0}^{2}\partial_{\phi\phi}G \\
G(\phi = \phi_e , \alpha) &= \partial_{\phi}G(\phi = \phi_{in} , \alpha) = 0 \\
G(\phi, \Delta\alpha \rightarrow 0) &= \delta (\phi-\phi_0)
\end{align}\label{eq:greenseqn} 
\end{subequations}
which has the solution:
\begin{eqnarray}
G(\phi,\Delta \alpha) &=& \dfrac{1}{\sqrt{2}\Delta\phi_{pl}}\sum_{n=0}^{\infty}\text{exp}\left[-\left(n + \dfrac{1}{2}\right)^2\dfrac{\tilde{H}_{0}^{2}}{\Delta\phi_{pl}^{2}}\pi^2\Delta\alpha \right]\nonumber \\
&\times &  \Bigg\lbrace \text{cos}\left[ \left(n + \dfrac{1}{2}\right)\dfrac{\pi (\phi-\phi_0)}{\Delta\phi_{pl}}\right] - \text{cos}\left[ \left(n + \dfrac{1}{2}\right)\dfrac{\pi (\phi+\phi_0-2\phi_e)}{\Delta\phi_{pl}}\right]\Bigg\rbrace \nonumber \\
\label{eq:G soln}
\end{eqnarray}
We can use this solution and equation (\ref{eq:Pv0}) to obtain the PDF for exit times, $\rho_{HJ+dS}(\mathcal{N})$, for a \ac{H-J} phase followed by a free diffusion phase using the relation:
\begin{eqnarray}
\rho_{HJ+dS}(\mathcal{N}) = \tilde{H}_{0}^{2}\dfrac{\partial P_{v_0}(\phi,\mathcal{N})}{\partial \phi}\Bigg\vert_{\phi_{e}} \label{eq:rhoHJ+dS defn}
\end{eqnarray}
yielding
\begin{equation}
\begin{split}
\rho_{\mu < 0}(\mathcal{N}) &= \dfrac{2\sqrt{\pi}}{\Omega(1-\mu)^2}\int_{0}^{\mathcal{N}}\mathrm{d}u~\text{sinh}(3u)^{\scalebox{0.5}{$-\dfrac{3}{2}$}} ~\text{exp}\left[\dfrac{3}{2}u -\dfrac{1}{2}\left( \text{coth}(3u ) -1\right)\Omega^2\right]  \\
&\times \sum_{n=0}^{\infty}\left(n+\scalebox{0.5}{$\dfrac{1}{2}$} \right)\Bigg\lbrace \text{sin}\left[\lb n+\scalebox{0.5}{$\dfrac{1}{2}$}\rb \dfrac{\pi\mu}{\mu-1}\right] ~\text{exp}\left[-\lb n+\scalebox{0.5}{$\dfrac{1}{2}$}\rb^2\dfrac{2\pi^2 (\mathcal{N} -u)}{3\Omega^2 (1-\mu)^2}  \right] \Bigg\rbrace
\end{split}\label{eq:rhoHJ+dS_append}
\end{equation}
which is the main result of this subsection.



\cleardoublepage

\label{app:bibliography} 

\manualmark 
\markboth{\spacedlowsmallcaps{\bibname}}{\spacedlowsmallcaps{\bibname}} 
\refstepcounter{dummy}

\addtocontents{toc}{\protect\vspace{\beforebibskip}} 
\addcontentsline{toc}{chapter}{\tocEntry{\bibname}}

\printbibliography 



\end{document}